%% file: main.tex
\input{preamble}
\begin{document}

\newcommand{\google}{\affiliation{Google Quantum AI, Venice, CA 90291, United States}
}

\newpage

\title{A Denser Planar Surface Code}

\author{Guang Hao Low}
\email{guanghaolow@google.com}\google
\author{William J. Huggins}\google
\author{Dominic W. Berry}
\affiliation{School of Mathematical and Physical Sciences,
Macquarie University, Sydney, NSW 2109, Australia}
\author{Tanuj Khattar}\google
\author{Alec F. White}\google
\author{Nicholas C. Rubin}\google
\author{Ryan Babbush}\google

\begin{abstract}
We present a quantum code implementable on a regular $2$D hex grid
with an estimated encoding rate up to $4.5\times$ of that of a rotated surface code patch using circuit-level noise in a one- and two-qubit $10^{-3}$ error uniform depolarizing model.
Our approach is based on yoking a dense packing of surface code twist defects, enabled by new stabilizer measurement cycles with an optimal four layers of nearest-neighbor two-qubit gates, almost no distance-reducing hook errors, and efficient decoding.
We demonstrate a space-efficient architecture for computing on densely packed logical qubits, including new padding-free lattice surgery protocols in an optimal bounding box of $2d^2$ data and measurement qubits per patch.
Assuming a $1\mu$s surface code cycle time and a $10\mu$s reaction time, these developments enable chemically accurate ground state phase estimation of a broad class of `utility-scale' electronic structure simulation problems such as the $108$ spin-orbital FeMoco-based nitrogen fixation catalyst in under a month with $89$k noisy superconducting qubits. We elucidate a Pareto frontier of space-time trade-offs and find a minimum physical quantum volume of $1.3$ mega-qubit-hours.
These correspond to a $36\times$ space and $6.6\times$ spacetime improvement, respectively, over our previous state-of-the-art minimum-Toffoli resource estimates (Phys. Rev. X 15, 041016).
\end{abstract}
\maketitle

\tableofcontents

\section{Introduction}
Scalable quantum computation requires fault-tolerant quantum error correction as elementary quantum gates are noisy.
The surface code~\cite{Kitaev2002Topological} is a topological quantum error correcting code that has been extensively studied over the past three decades, and is notable for a high threshold, polynomial-time decoding with good performance, and being experimentally feasible under a minimal set of hardware assumptions.
A modern dynamical implementation~\cite{Hastings2021Floquet} only requires nearest-neighbor connectivity on a degree-$3$ regular hex grid~\cite{McEwen2023RelaxingHardware}.

Unfortunately, the surface code has an extremely low encoding rate of logical qubits into physical qubits: The standard one-logical-qubit rotated patch~\cite{Bombin2007RotatedSC} achieves code distance $d$ using $2d^2-1$ physical data and measurement qubits~\cite{Fowler2018LowOverhead}, which is further padded to $2(d+1)^2$ for computation based on lattice surgery~\cite{Fowler2018LowOverhead}.
This becomes a severe bottleneck when applied to the broad class of promising `utility-scale'~\cite{Castaldo2026Utilityscale} electronic structure simulation problems~\cite{goings2022reliably,PhysRevResearch.3.033055} with more than a hundred spin-orbitals.
For the common benchmark of chemically accurate ground state energy estimation of the nitrogen fixation FeMo-cofactor~\cite{Reiher2017Elucidating}, realistic future superconducting qubit parameters suggest that  distances up to $27$ are required~\cite{PRXQuantum.2.030305}, which corresponds to at least $169$k physical qubits simply to store the $108$-qubit fermionic wavefunction, plus another $3$ million to complete the computation with a minimum number of Toffoli gates~\cite{low2025fast}.

\begin{figure}
    \centering
    
    \includegraphics[scale=0.8,valign = m]{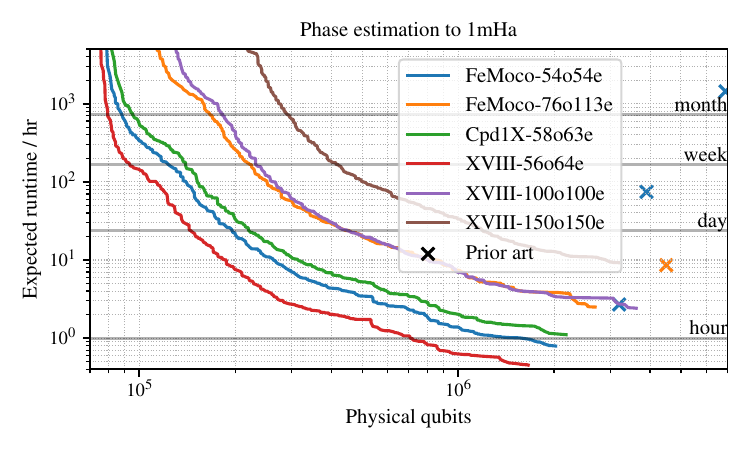}
    \begin{tabular}{c|c|ccc}
    \hline\hline
    \multicolumn{2}{c|}{FeMoco-54}&Physical & Expected & Volume/$10^6$ 
    \\ 
    \multicolumn{2}{c|}{resources}& qubits & runtime & qubit-hrs
         \\
         \hline
    \multicolumn{2}{c|}{DF~\cite{PhysRevResearch.3.033055,Beverland2020NonClifford}}& $5.9\times 10^6$ & 2 months &8500
         \\
          \multicolumn{2}{c|}{THC~\cite{PRXQuantum.2.030305}}& $3.9\times 10^6$ & 3 days & 280 
         \\
         \multicolumn{2}{c|}{DFTHC~\cite{low2025fast}} & $3.2\times {10^6}^\text{a}$& 3 hours &8.6
         \\
         \hline
          \multirow{4}{*}{\rotatebox{90}{This work}}& 
         \multirow{1}{*}{Compact} & $8.9\times 10^4$ & $<1$ month  &$60$
         \\
         & \multirow{1}{*}{Balanced} & $2.0\times 10^5$ & $<1$ day&$4.5$
         \\
         & \multirow{1}{*}{Efficient} &  $1.1\times 10^6$ & $1.1$ hours &$1.3$
         \\
         & \multirow{1}{*}{Fast} &  $2.0\times 10^6$ & $48$ minutes &$1.6$
         \\
         \hline\hline
         \multicolumn{5}{l}{\tiny $^\text{a}$Using the explicit layout of~\cite{PRXQuantum.2.030305}.}
    \end{tabular}
    \caption{Pareto frontier of physical resources for chemically-accurate ground state phase estimation of `utility-scale' benchmark molecules with $54$ to $150$ orbitals, assuming quantum hardware with nearest-neighbor connectivity, a $10^{-3}$ physical error rate, a surface code cycle time of $1\mu$s, and a reaction time of $10\mu$s.
    Shown are estimates for the FeMo-cofactor for nitrogen fixation, the XVIII configuration of a Ruthenium-based catalyst for carbon fixation, and the Cpd1X configuration of a metabolic cofactor p450.
    We present new space-time tradeoffs in quantum circuits and quantum error correction for compiling the second-quantized Double-Factorized-Tensor-HyperContracted (DFTHC) electronic structure representation~\cite{low2025fast} with spectrum amplification~\cite{King2025SOSSA}, and obtain order-of-magnitude improvements over prior art.
    Selected point estimates for FeMoco-54 include: `Compact', using $36\times$ fewer physical qubits than prior art; `Balanced', using $16\times$ less space and $1.9\times$ less volume; `Efficient', using $6.6\times$ less volume, and `Fast', using $3.4\times$ less time.
    }
    \label{fig:intro_pareto}
\end{figure}

In this work, we demonstrate that utility-scale quantum simulation can be achieved in a fraction of the previously assumed footprint, without additional hardware assumptions.
Assuming a $10^{-3}$ physical error rate in a one- and two-qubit uniform depolarizing noise model, a $1\mu$s surface code cycle time, and a $10\mu$s reaction time, we show that chemically accurate ground state energy estimation of the $108$ spin-orbital FeMoco system can be realized with as few as $89$k noisy physical qubits in under a month of runtime.
We also showcase the path towards utility-scale efficiency in~\cref{fig:intro_pareto} with a Pareto frontier of spacetime tradeoffs, and find a minimum spacetime volume of $1.3$ mega-qubit-hours.
These represent a $36\times$ space and $6.6\times$ spacetime improvement respectively over our previous resource estimates focused on minimizing Toffoli gate count.
We also obtain similar results for other benchmarks, including a Ruthenium-based catalyst~\cite{PhysRevResearch.3.033055} for carbon fixation, and a metabolic cofactor p450~\cite{goings2022reliably}.


\subsection{Summary of contributions}
\begin{figure}
    \centering
\fbox{
\begin{tabular}{cccc}
    \includegraphics[width=0.24\linewidth,valign=m]{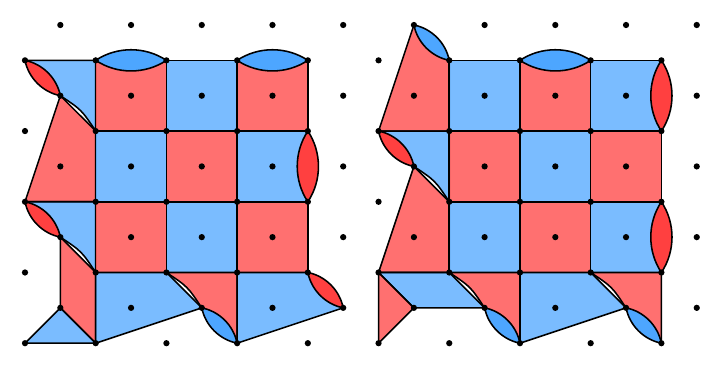}
    &
    \includegraphics[width=0.24\linewidth,valign=m]{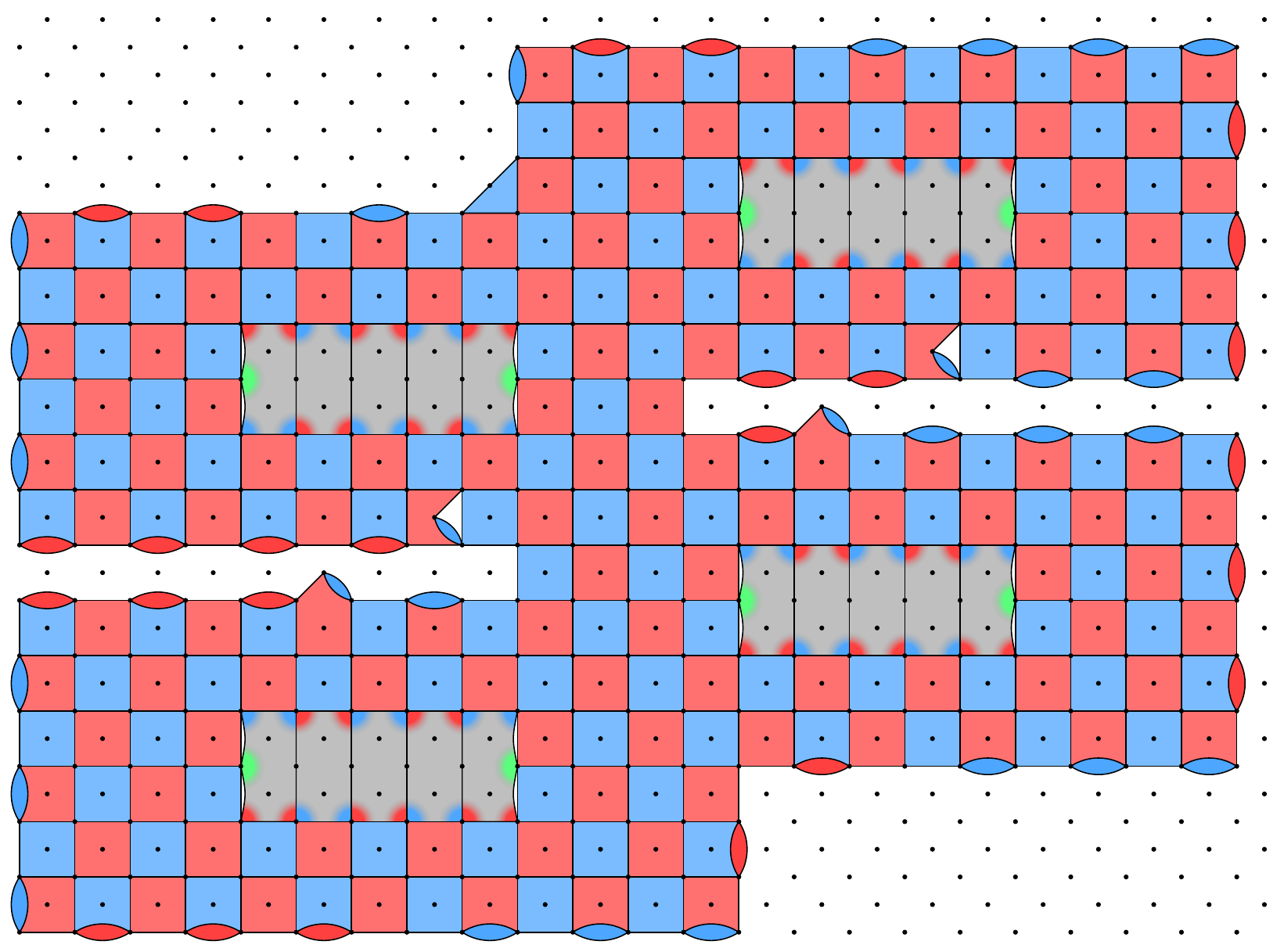}
    &
    \includegraphics[width=0.24\linewidth,valign=m]{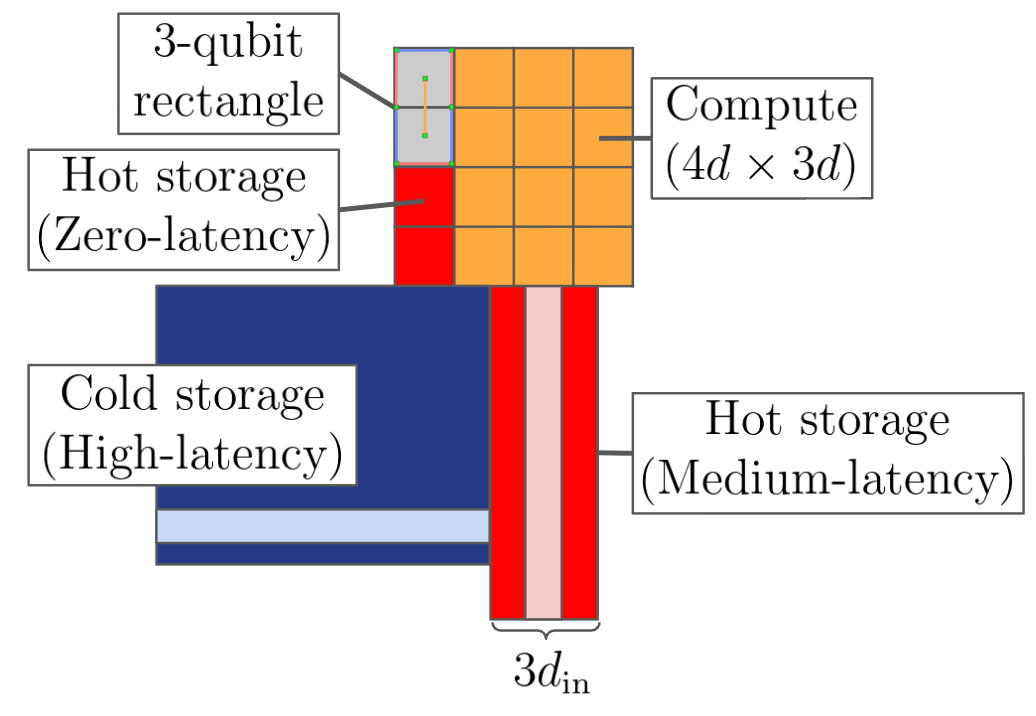}
    &
    \includegraphics[width=0.24\linewidth,valign=m]{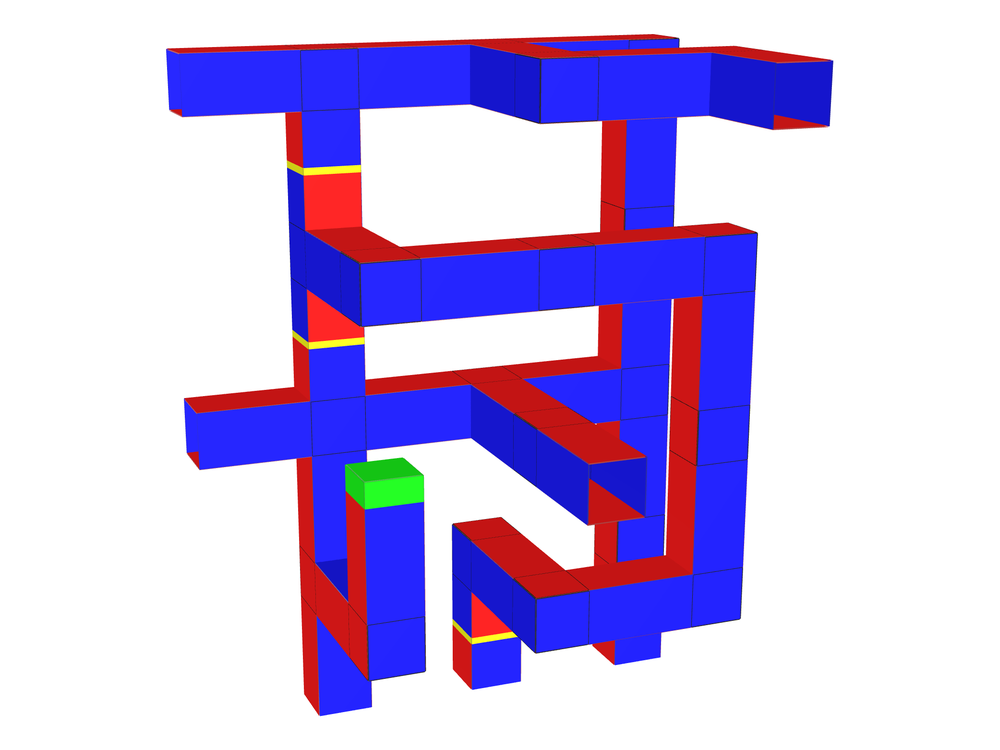}
    \\\parbox{0.24\linewidth}{\cref{sec:compact_patch}: Compact rotated surface code and padding-free lattice surgery}
&\parbox{0.24\linewidth}{\cref{sec:dense_memory}: Twist defect dense packing and yoking}
&\parbox{0.24\linewidth}{\cref{sec:architectures}: Mixed-latency architecture and layout}
&\parbox{0.24\linewidth}{\cref{sec:lattice_surgery_compilation}:  Optimized lattice surgery compilation of lookup tables and rotations}
    \\
    \multicolumn{2}{c}{\includegraphics[width=0.46\linewidth,valign=m]{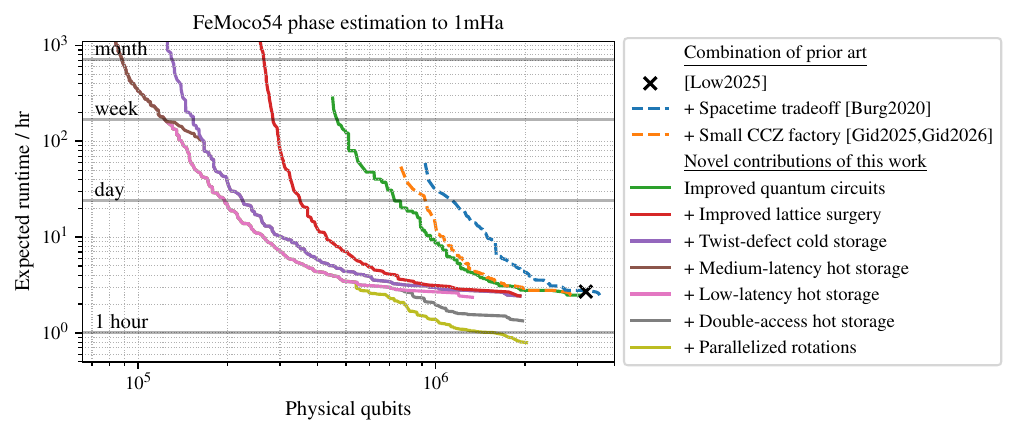}} &
    \multicolumn{2}{c}{\includegraphics[width=0.46\linewidth,valign=m]{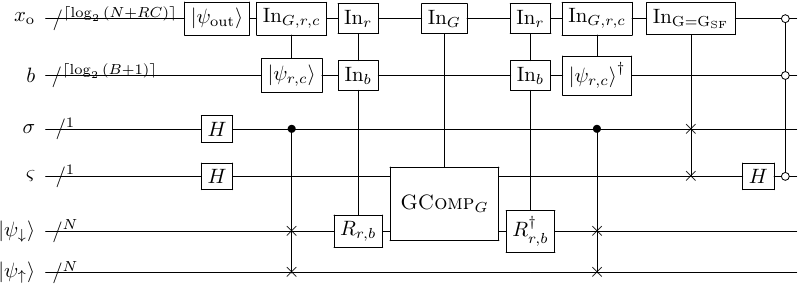}}
    \\
    \multicolumn{2}{c}{\cref{sec:quantum_simulation}: Application to electronic structure} &
    \multicolumn{2}{c}{\cref{sec:sec:appendix_tradeoffs}: Electronic structure quantum circuit optimization
}
    \\
\end{tabular}
}
    \caption{Summary of our key contributions.
    \cref{sec:compact_patch}: We reduce the lattice surgery bounding box of a distance $d$ rotated surface code patch from $2(d+1)^2$ to $2d^2$ physical qubits.
    \cref{sec:dense_memory}: We introduce an optimal-depth, hook-error-free, and efficiently decodable implementation of surface code twist defects using nearest-neighbor gates on a degree-3 hex grid. 
    This enables a packing of three logical qubits into a rectangular patch with a $1.5\times$ encoding rate. 
    By merging boundaries of three-qubit rectangles, we obtain a dense packing of logical qubits in twist defects with asymptotically a $2\times$ encoding rate.
    By yoking this dense packing of twist defects, we obtain logical qubit ``cold storage'' with an encoding rate to $4.5\times$ of a rotated surface code patch.
    \cref{sec:architectures}: We describe the architecture used to organize our quantum computation, including new variations of ``hot storage'' that are denser than surface code patches and support lattice surgery operations with almost the same throughput, and prove quantum communication complexity results showing that highly nonlocal Hamiltonians can be block-encoded with extremely few queries to high-density but slow high-latency cold storage.
    \cref{sec:lattice_surgery_compilation}: We compile common quantum circuit routines to highly optimized lattice surgery operations, such as the control logic for quantum lookup tables to $\approx 10\times$ less space, which dominates cost in the minimum-space regime, and parallelized Givens rotations, which dominates cost in the minimum-time regime.
    \cref{sec:quantum_simulation}: We apply our results to second quantized ground state phase estimation. Together with new spacetime-efficient quantum circuits in~\cref{sec:sec:appendix_tradeoffs} for block-encoding Double-Factorized-Tensor-Hyper-Contracted (DFTHC) chemistry with spectrum amplification, we elucidate a pareto space-time tradeoff spanning more than three orders of magnitude in time. 
    }
    \label{fig:improvements}
\end{figure}

We achieve these results by novel contributions across the entire quantum stack shown in~\cref{fig:improvements}. 
These are broadly split into two categories that are of interest to different audiences: Improvements in quantum error correction primitives, summarized in~\cref{sec:intro_denser}, and improvements in quantum circuits and compilation summarized in~\ref{sec:intro_electronic}. 

\subsubsection{Denser planar surface code}\label{sec:intro_denser}


The massive footprint of the surface code has motivated many alternatives.
Color codes~\cite{Bombin2006ColorCode}, tile codes~\cite{Steffan2025TileCodes}, codes on hyperbolic geometries~\cite{Breuckmann2017HyperbolicSC,Higgott2024Hyperbolic}, and LDPC codes~\cite{Bravyi2024BicycleB,cain2026shorsalgorithmpossible10000,tripier2026faulttolerantquantumcomputingtrapped} all feature higher encoding rates.
However, these invariably trade off some favorable practical aspect of the surface code, such as connectivity, decodability, speed, or threshold.
Even within the class of planar $2$D codes with nearest-neighbor connectivity, there are options that, in theory, almost quadruple~\cite{Kesselring2018BoundariesTwistDefects} the surface code encoding rate, but have geometries that are experimentally infeasible or inconducive to a scalable architecture. 
They generally also require higher weight stabilizers, whose measurement requires some combination of additional ancilla qubits, more layers, and increased qubit connectivity: factors that significantly degrade logical error rate in practice, but are not always studied in detail.
Actual code performance is most accurately estimated from noisy circuit-level simulations, which can deviate considerably from formal code parameters  $[[n,k,d]]$. A prominent example is surface code hook error in the $N$-\reflectbox{Z} gate sequence, which effectively halves distance if not carefully managed.

Due to its simplicity, the surface code is foundational in quantum error correction, and in the future, may still remain relevant for its speed, and especially in light of recent magic-state cultivation protocols~\cite{Gidney2025Cultivation}.
To maximize the viability of the surface code without relying on more complex alternative geometries, in~\cref{sec:compact_patch}, we review the surface code and lattice surgery, and as a warm-up, demonstrate full-distance surface code lattice surgery on a hex grid, and walking operations on a square grid, all within a bounding box of $2d^2$ physical qubits per patch, in an optimal depth of $6$ (reset layer; 4 $\textsc{CX}$ layers; measurement layer), and with comparable error scaling with prior approaches in~\cref{tab:surface_code_results}.
\begin{table}
\begin{tabular}{c|c|c|c|c|c|c|c}
\hline\hline
\multirow{2}{*}{Surface code gate sequence} & \multirow{2}{*}{Distance} &Bounding & \multirow{2}{*}{Grid} & \multirow{2}{*}{Layers} & \multicolumn{3}{c}{Logical error rate (empirical fit)} \\
\cline{6-8}
& & Box& & & Memory & Surgery & Walking \\
\hline
$N$-\reflectbox{Z}~\cite{Horsman2012LatticeSurgery,Fowler2018LowOverhead}&\multirow{4}{*}{$d$}&\multirow{4}{*}{$2(d+1)^2$}&Square& $6$\footnote[1]{$7$ to control hook errors in lattice surgery.}& $4.0^{-d}/10$~\cite{Gidney2025Yoked} &-& $3.5^{-d}/40$~\cite{McEwen2023RelaxingHardware} \\
Alternating ($X$-top only)~\cite{McEwen2023RelaxingHardware}&&&Hex& $6$& $4.0^{-d}/10$~\cite{Gidney2025Yoked}& $3.5^{-d}/40$~\cite{Gidney2025Factoring} & No\\
Diagonal~\cite{Kishony2026OffTheHook} &&&Square& $6$\footnote[2]{$8$ if measurement and reset are not interleaved with gates.} & - &-& No\\
Alternating ($X$- and $Z$-top)~\cite{Yuga2026NoHook} &&&Hex& $6$ & $4.0^{-d}/10$ & $3.5^{-d}/15$& No\\
\hline
\multirow{2}{*}{This work ($X$- and $Z$-top)}&\multirow{2}{*}{$d$}& \multirow{2}{*}{$2d^2$} &\multirow{2}{*}{Hex}&\multirow{2}{*}{$6$}& $4.0^{-d}/10$ & $3.5^{-d}/30$& $3.5^{-d}/20$\footnote[3]{Requires square grid and $\textsc{CXSwap}$ gates on the boundary.}\\
&&&&&\cref{fig:square_patch_alt}&\cref{fig:h_bridge}&\cref{fig:square_patch_alt}\\
\hline\hline
\end{tabular}
    \caption{Our work demonstrates surface code memory, lattice surgery, and walking all within a $2d^2$ bounding box and with the minimum number of gate layers ($4$ for $\textsc{CX}$ and counting measurement and reset each as one layer). For the purposes of comparison with prior art, we provide benchmarks of the empirical logical error rate per round ($6$ layers) per logical qubit in a standard $10^{-3}$ gate-level one- and two-qubit uniform depolarizing noise model~\cite{McEwen2023RelaxingHardware}, decoded by the sparse-blossom implementation of minimum weight perfect matching~\cite{Higgott2025SparseBlossom}.}
    \label{tab:surface_code_results}
\end{table}

Subsequently in~\cref{sec:dense_memory}, we demonstrate that the surface code on a planar regular hex grid can support an encoding rate up to $4.5\times$ that of an idling single-qubit rotated surface code patch.
Our higher rate codes are enabled by a new, practical implementation of surface code twist defects~\cite{Bombin2010Twists} under the same minimal hardware assumption of degree-$3$ nearest-neighbor-connectivity on a regular 2D planar lattice of qubits.
Standard surface codes are CSS codes comprised solely of $X$ and $Z$ stabilizers that are weight $4$ in the bulk.
In contrast, twist defects utilize non-CSS (odd) weight-$5$ stabilizers with a $Y$ Pauli operator, such as $XXYZZ$.
In condensed matter physics, they are extrinsic topological defects found at the end points of a domain wall that breaks parity conservation of Ising anyons~\cite{Brown2017PokingHoles}.

The utility of twist defects has long been known, such as enabling the Clifford $S$ gate~\cite{Brown2017PokingHoles} through non-Abelian braiding~\cite{Litinski2018LatticeTwist} and $YY$ measurements~\cite{Geher2025Tangling} between surface code patches.
However, these benefits have been offset by the experimental challenge of measuring their higher weight stabilizers.
Typically, more ancillae and gate layers are required, which lowers code threshold and significantly deteriorates real-world performance.
Although this can be overcome in principle by long-ranged connections and higher qubit degree in non-uniform lattices, engineering such devices is considerably more challenging than the simple nearest neighbor hex-grid required by the standard surface code.
When restricted to a regular square lattice, entangling stabilizer measurement gate schedules~\cite{Geher2025Tangling} provide another lower-depth option but halves code distance along the domain wall and lose polynomial-time decoding.
Our proposed implementation simultaneously achieves all desirable properties: Optimal depth in $6$ layers enabling a logical error rate scaling similarly to an idling surface code patch, the minimum qubit degree of $3$, no significant distance loss due to circuit-level hook errors, and efficient decoding by minimum weight perfect matching. 

\begin{figure}
        \includegraphics[height=2.5cm]{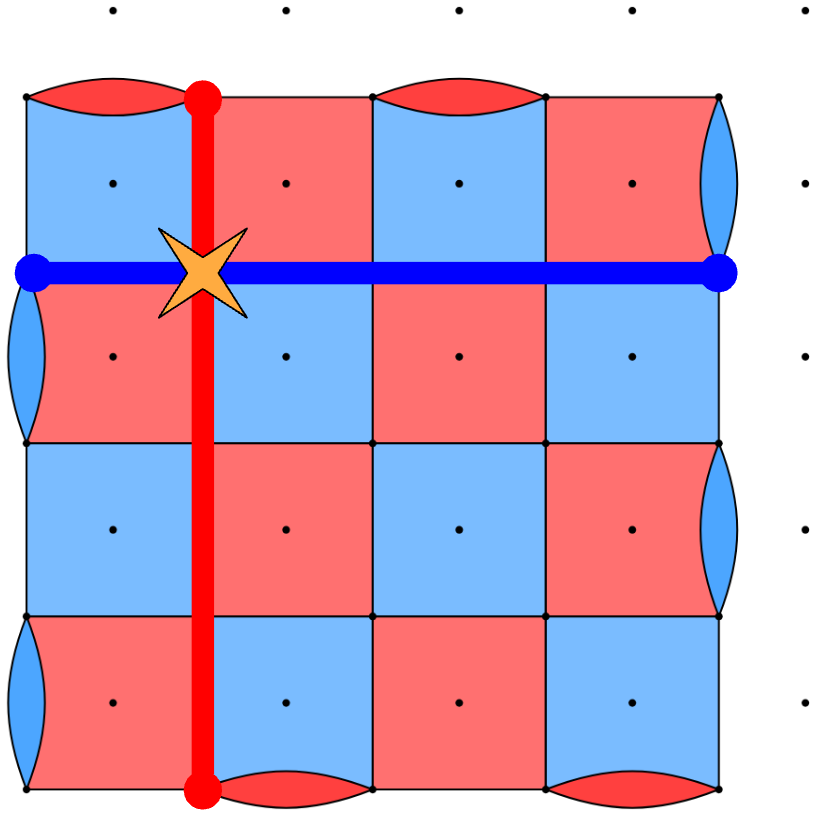}
        \hspace{0.5cm}
        \includegraphics[height=2.5cm]{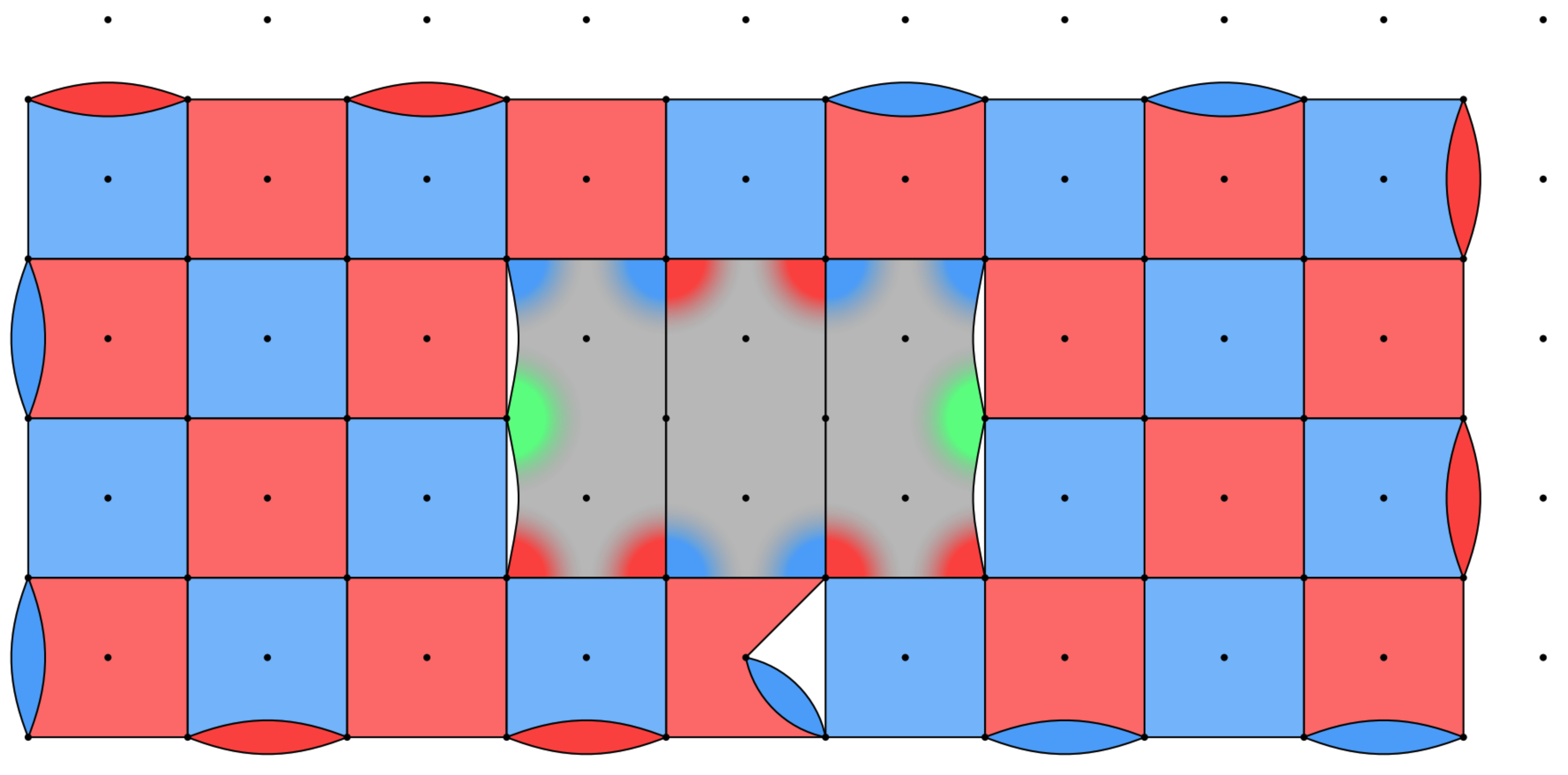}
        \hspace{0.5cm}
        \includegraphics[height=2.5cm]{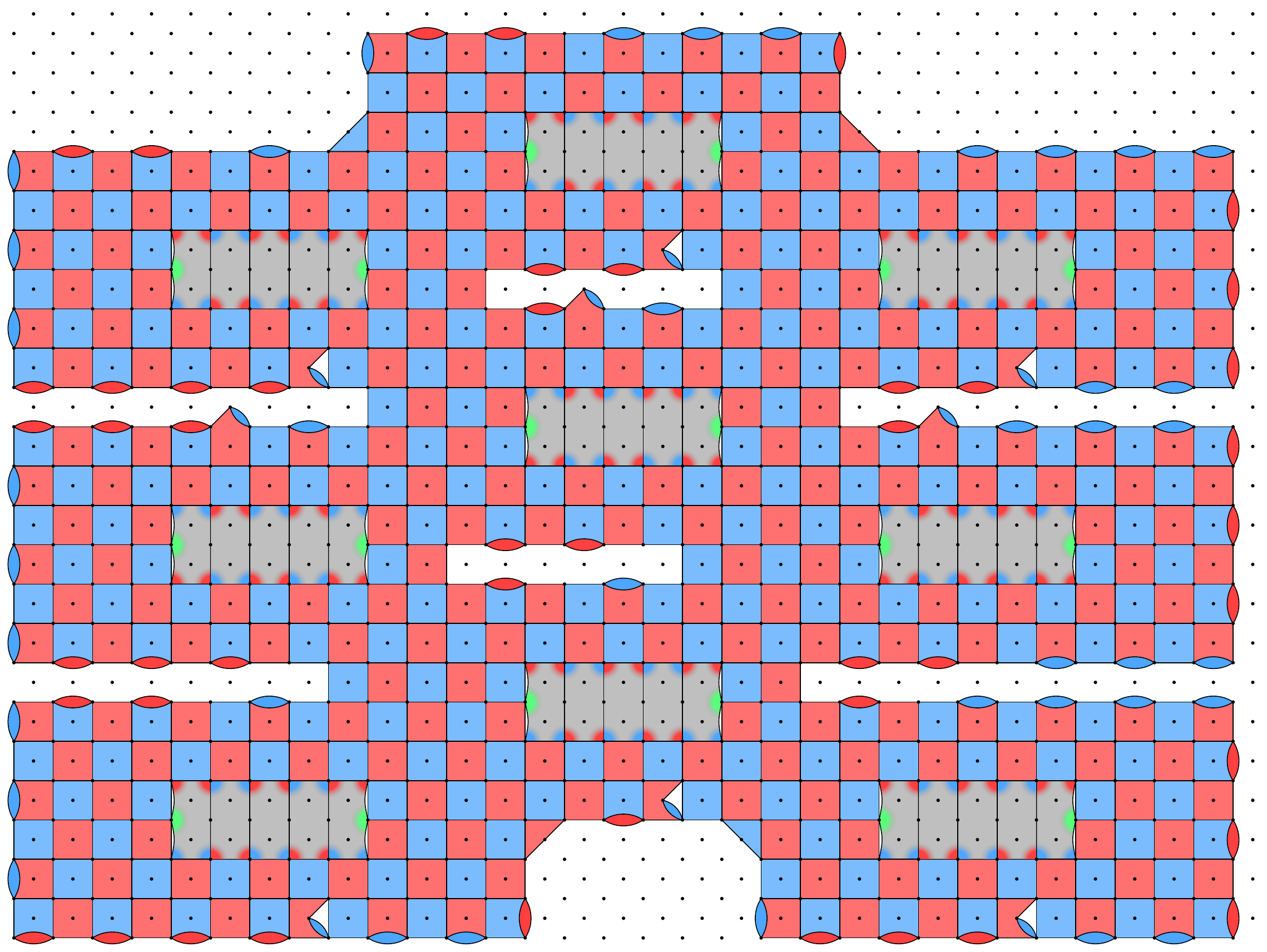}
    \caption{(Left) Rotated surface code patch with weight-4 Pauli-$Z$ (blue) and $X$ (red) stabilizers occupying a footprint of $d^2$ data qubits and $d^2-1$ measurement qubits. A logical qubit is defined by a pair of anti-commuting logical operators (bold lines) that commute with all stabilizers. The star marks where the logical operators anticommute. 
    (Middle) Rectangular surface code patch encoding three logical qubits with two weight-5 twist defects featuring a Pauli-$Y$ (green) at the ends of a domain wall. (Right) Boundaries of the $3$-qubit rectangle can be merged to realize a denser packing, such as this $n_\text{row}\times n_\text{col}=27$ logical qubit example with $n_\text{rows}=3$. The precise positions of twists and boundaries are carefully optimized for logical error rate.
    }
    \label{fig:surface_code_patch_intro}
\end{figure}

Our claimed encoding rates are obtained by a careful placement of twist defects.
Our dense quantum memories inherit all the favorable properties of our twist defect implementation, leading to significantly better performance than alternate twist-defect packing schemes and other 2D planar codes summarized in~\cref{tab:twist_defect_results}.
The three distinct dense-packing configurations we consider are:
\begin{enumerate}
    \item A rectangular patch shown in~\crefpos{fig:surface_code_patch_intro}{middle} encoding three logical qubits in a footprint of two $2d^2$-qubit rotated surface code patches with $1.5\times$ the encoding rate of a rotated surface code patch.
\item A dense-packing of twist defects shown in~\crefpos{fig:surface_code_patch_intro}{right} encoding a grid of $n_\text{rows}$ rows of $3m_\text{col}$ columns of logical qubits with an asymptotic density of one logical qubit per $d(d-1)$ physical qubits, or $2\times$ the encoding rate of a rotated surface code patch.
\item A concatenation of each column of $n_\text{row}$ logical qubits in the dense-packing with a $[[n_\text{row},n_\text{row}-2,2]]$ quantum parity check (iceberg~\cite{Self2024IcebergCode}) code, which doubles the code distance, or in practical situations, up to $4.5\times$ the encoding rate of a rotated surface code patch.
\end{enumerate}
Universal quantum computation on high-rate logical qubits is then achieved by unloading into rotated surface code patches to apply standard methods of Clifford gates by lattice surgery and non-Clifford gates by magic $\ket{T}$ and $\ket{CCZ}$ state cultivation and distillation~\cite{Gidney2019AutoCCZ,gidney2018factoringn2cleanqubits}.
Moreover, our twist defect implementation enables new lattice surgery primitives, such as an almost $S$ gate. 

In the following, our results for the compact rotated surface code, the rectangular patch, and dense-packing are validated by estimating the empirical logical error rate per round ($6$ layers) per logical qubit in a standard gate-level one- and two-qubit uniform depolarizing noise model~\cite{McEwen2023RelaxingHardware} defined in~\cref{sec:noise_model}, using $\approx0.9$ million core-hours of Monte-Carlo sampling implemented in Stim~\cite{Gidney2021Stim} and decoding by the sparse-blossom implementation of two-pass correlated minimum weight perfect matching~\cite{Higgott2025SparseBlossom}.
We choose the uniform depolarizing noise model, at a standard $10^{-3}$ error rate representing middle-to-far-term quantum hardware with an error suppression factor of $\Lambda\approx12$~\cite{Google2025SurfaceCodeThreshold}, in order to apply prior benchmarks on magic state cultivation~\cite{gidney2024magic} and to make our results directly comparable other resource estimates such as for factoring~\cite{Gidney2025Factoring}.
However, we do not perform an explicit simulation of yoking.
Unlike yoking of rotated surface code patches in prior art~\cite{Gidney2025Yoked}, logical operators of the dense packing lie in the interior in general, and may only be moved to the boundary by a complicated sequence of twist-defect lattice surgery operations that we describe, but is beyond the scope of this work to simulate.
Instead, we assume an ideal outer decoder with a logical error rate fitted to the rate of distance $\approx 2d$ errors in our explicit simulations of the dense packing, and show that that all correlated logical errors of distance $\lesssim 2d$ can be uniquely resolved.
\begin{table}
\begin{tabular}{lc|c|c|c|c|c|c}
\hline\hline
&\multirow{2}{*}{Error correcting code}&\multirow{2}{*}{Lattice}& Stabilizer&\multirow{2}{*}{Layers} &Encoding rate & Circuit-level distance & Logical error rate \\
&&& weight&&$2kd^2/n$  & evaluated & ($10^{-3}$ error fit)
\\
\hline 
&Rotated surface code~\cite{McEwen2023RelaxingHardware}  & Hex & 4 & 6 &1 & Yes & $4^{-d}/10$\footnote{$3.5^{-d}/30$ for lattice surgery.}
\\
&Surface code with a twist~\cite{Yoder2017SurfaceCodeTwist}& Irregular & $4$ & $8$ & $\frac{4}{3}$ & Yes & -
\\
&Qubit hotels~\cite{Gidney2025Factoring,Fujiu2025DensePacking}& Square & $4$  & $7$ & $<\frac{4}{3}$ & Yes & -\\
&Circle-packing codes~\cite{Sarkar2025Twists} & Irregular&$5$&$8$&$<2$& No & -
\\
&1D yoked surface code~\cite{Gidney2025Yoked,Gidney2025Factoring}& Square & $4$ & 6 & $\lesssim 2.5\footnotemark[2]$ & Yes & $n_{\text{row}}T_{\text{yoke}}\frac{15^{-d}}{100}$
\\
\hline
&\multirow{2}{*}{Stellated color codes~\cite{Kesselring2018BoundariesTwistDefects}}& Singularity & $6$ & $-$ & $<\frac{8}{3}$ & No & -\\
&& at origin & $8$ & $-$ & $<4$ & No & -\\
\hline
\multirow{4}{*}{\rotatebox{90}{\parbox{1.3cm}{This\\ work}}}&3-qubit rectangular patch& Hex&$5$ & $6$ &$1.5$ & Yes & $3.5^{-d}/10$\\
&Twist-defect dense-packing & Hex&$5$  & $6$ & $<2$& Yes & $3.5^{-d}/10$ \\
&1D yoked dense-packing & Hex & $5$ & $6$ & $\lesssim 4.5$\footnote{In the limit of large $k$ and $d$}& Yes & $n_{\text{row}}T_{\text{yoke}}\frac{3.5^{-2d}}{100}$ \\
\hline\hline
\end{tabular}
    \caption{Comparison of error correcting codes embeddable in $2$D planar surface. 
    Codes are specified by the number of physical data and measurement qubits $n$, the number of encoded logical qubits $k$, and the code distance $d$.
    Encoding rates with $<$ are achieved in the limit of large $k$.
    Our dense packing of logical qubits matches the highest known encoding rate and exhibits all properties favorable for experimental implementation, including embedding in a regular lattice with degree $3$ connectivity and an optimal number of gate layers, counting measurement and reset each as one layer. 
    We provide benchmarks of the empirical idling logical error rate per round ($6$ layers) per logical qubit in a standard $10^{-3}$ gate-level one- and two-qubit uniform depolarizing noise model~\cite{McEwen2023RelaxingHardware}, decoded by correlated minimum weight perfect matching~\cite{Higgott2025SparseBlossom}.}
    \label{tab:twist_defect_results}
\end{table}

\subsubsection{Application to electronic structure}\label{sec:intro_electronic}

We evaluate the practical impact of our results by describing a hybrid architecture \cref{sec:architectures} for performing computations on our high-latency-high-rate codes.
As such logical qubits in ``cold storage'' may only be accessed sequentially and loaded/unloaded slowly, a potential concern is that its high latency could dominate the overall runtime of an algorithm if access is too frequent.
We show that this is not an issue for quantum algorithms based on block-encoding highly nonlocal Hamiltonian~\cite{Low2016HamSim,low2019hamiltonian,GSLW2019QSVT}.
We prove that any $l$-local $n$-qubit or fermion Hamiltonian can be block-encoded with $\mathcal{O}(ln)$ quantum communication complexity between a large but slow $n$-qubit memory and a small but fast $\mathcal{O}(l\log(n))$-qubit workspace, independent of the number of terms that can be as large as $\mathcal{O}(3^l\binom{n}{l})$.
Towards maximizing the efficiency of the surface code, we also introduce variations of ``hot storage'' yoked with a single parity check instead of two parity checks~\cite{Gidney2025Yoked} that support high-rate lattice surgery operations at almost the same speed as a regular rotated surface code patch, with up to $1.4\times$ the density, or up to $1.9\times$ less spacetime volume.
The detailed layout of hot storage turns out to be essential to our organization of highly-optimized lattice surgery compilation of key quantum circuit bottnecks in~\cref{sec:lattice_surgery_compilation}, which focuses on quantum lookup tables using $10\times$ less space for the control logic than prior art~\cite{Gidney2019AutoCCZ}, a classical absorption technique that reduces the reaction depth of lookup tables from $\Theta(X)$ to $\mathcal{O}(1)$ by deferring all $\ket{CCZ}$ injection corrections to a streaming classical update of the lookup data, and parallelizing multiplexed rotations that are later used to implement single-particle basis rotations.

The primary application we consider in~\cref{sec:quantum_simulation} is chemically accurate ground state energy estimation~\cite{King2025SOSSA} of second quantized electronic structure. In particular, we use as benchmarks the FeMo-cofactor~\cite{Reiher2017Elucidating} for nitrogen fixation, a Ruthenium-based catalyst~\cite{PhysRevResearch.3.033055} for carbon fixation, and a metabolic cofactor p450~\cite{goings2022reliably}.
These examples have $2N> 100$ active space spin-orbitals, are far out of reach of exact exponential-time-scaling classical solvers, and they have the qualitative features of problems where approximate classical methods are expected to break down.
Although chemical accuracy of $1\text{kcal/mol}\approx 1.6$mHa by carefully targeted heuristic classical methods was recently demonstrated~\cite{Zhai2026Femoco}, it remains an open question whether such classical methods can remain generic and practical as $N$ increases. Unlike the quantum phase estimation method described here, the heuristics in Ref~\cite{Zhai2026Femoco} come with no performance guarantees and the best variational DMRG calculations still differ significantly from the extrapolated energy estimate.
Following a long series of work~\cite{berry2019qubitization,PhysRevResearch.3.033055,PRXQuantum.2.030305,Rocca2024SymmetryCompressedDF,Oumarou2024acceleratingquantum,caesura2025faster,low2025fast}, quantum algorithms can now simulate these in general with highly efficient block-encoding-based methods using $\mathcal{O}(N)$ logical qubits and $\mathcal{O}(N^2)$ Toffoli gates.

We optimize these block-encoding quantum circuits further in~\cref{sec:sec:appendix_tradeoffs} to realize generic space improvements, as well as an extremely an broad logical-qubit-Toffoli-count trade-off, down to as few as $2N+\mathcal{O}(\log N)$ logical qubits.
We then describe how the overall phase estimation and block-encoding algorithm is organized across distinct hybrid architecture layouts of ``minimum-space'', ``minimum-spacetime'', and ``minimum-time''.
We describe our procedure for enumerating over a wide range of quantum circuits to obtain logical-level resource estimates, and then how these are converted into physical-level resource estimates, and end with a review of key assumptions made across this work to obtain~\cref{fig:intro_pareto}.
\subsubsection{Remaining paper organization}
%
%
Finally, we discuss the impact of different architectural assumptions and conclude in~\cref{sec:discussion}.
\cref{sec:noise_model} defines the uniform depolarizing noise model used in this work.
\cref{sec:compacy_patch_appendix} provides additional surface code benchmark and lattice surgery gate sequences.
\cref{sec:lookup_appendix} and \cref{sec:multiplex_rotations_appendix} present additional lattice surgery pipe diagrams for the lookup table and multiplexed rotations.

\section{Compact rotated surface code}\label{sec:compact_patch}
\input{1_compact_patch/compact_patch}

\subsection{Lattice surgery without padding}\label{sec:dense_lattice_surgery}
\input{1_compact_patch/lattice_surgery}

\section{Dense surface code memory}\label{sec:dense_memory}
\input{2_denser_surface_code/denser_surface_code}

\section{Architecture}\label{sec:architectures}
\input{3_architecture/architecture}

\section{Lattice surgery compilation}\label{sec:lattice_surgery_compilation}
\input{4_compilation/quantum_circuits}

\subsection{Space-efficient multiplexed rotations}\label{sec:multiplex_rotations}
\input{4_compilation/multiplexed_rotations}

\subsection{Space-time-efficient multiplexed rotations}\label{sec:adder_multiplex_rotations}
\input{4_compilation/adder_multiplexed_rotations}

\section{Quantum simulation of electronic structure}\label{sec:quantum_simulation}
\input{5_simulation/quantum_simulation}

\section{Discussion and Conclusion}\label{sec:discussion}
In this work, we obtain a surface code encoding rate up to $4.5\times$ that of a rotated surface code patch under the most restrictive hardware assumptions of nearest-neighbor connectivity on a degree $3$ hex grid, and validate our results with circuit-level noisy simulations.
We describe a general hybrid architecture combining high-latency-high-rate codes with fast-low-rate codes to realize utility-scale electronic structure simulation in either as few as $89$k physical qubits in under a month, or with at least a $6.6\times$ improvement in spacetime volume over prior art.
Our chosen benchmark of phase estimation on block-encoded second quantized electronic structure is a representative test of general-purpose quantum computing capabilities as the optimal spacetime volume compilation has roughly equal contributions from magic state production and distillation, serial operations from quantum lookup tables, parallel operations from multiplexed Givens rotations, and routing overheads.
Our work greatly advances state-of-the-art and forms a robust baseline that allows the utility of future improvements and other quantum simulation applications to be accurately evaluated.

We may also substitute different quantum codes into the hybrid architecture.
For instance, the minimum-space compilation at $89$k physical qubits is dominated by $46$k for cold storage of $112$ logical qubits.
Cold storage based on a conjectured comparable-distance $[[432,12,24]]$ three-gross LDPC code~\cite{Bravyi2024BicycleB} would require only $8.6$k physical qubits, not including ancilla systems for lattice surgery.
However, we note that substituting high-rate codes into the compute region or even hot storage is still an active area of research.
The minimum spacetime volume can still be comparable to a surface code architecture due to different contributions of roughly equal importance: 
1) Clifford volume, specifically of arbitrary multi-target logical \textsc{CX} gates in large quantum lookup tables, that, on average, are applied to half of all logical qubits selected at random, almost all the time, and will need new high-rate lattice surgery~\cite{Zheng2025HighRate} protocols.
2) High-rate magic state cultivation and distillation directly into high-rate codes for parallelizing a large number of controlled-angle rotations, instead of the paradigm of first doing so in the surface code anyway.

We expect that our planar code results can still be advanced in various ways.
For instance, we perform a majority of layout and lattice surgery compilation by hand and used SAT solvers~\cite{Tan2024Scalpel} only for small segments due to a lack of automated tools~\cite{Tan2024Scalpel} suitable for large-scale compilation.
More advanced compilation techniques could be used, such as halving the lattice surgery logical timestep using complementary gap fallback~\cite{Akahoshi2025SoftLatticeSurgery} assuming a shorter reaction time,
or larger block distillation protocols in contrast to our focus on the $8\ket{T}\rightarrow\ket{CCZ}$ or catalyzed $\ket{CCZ}\rightarrow 2\ket{T}$ factory.
Although we now better understand how twist defects benefit quantum memory, we have not yet explored whether non-abelian braiding may improve the volume of lattice surgery compilations in practice.
Our reported error suppression factors $\Lambda$ may also be improved using advanced decoders~\cite{Beni2025Tesseract}, and we conjecture that 2D yoking of twist defects would yield an encoding rate $\lesssim6$ if the many more mechanisms for correlated logical errors can be decoded.
Finally, there remains an open problem on the maximum encoding rate of $2$D planar codes with higher-weight stabilizers, which could be significantly larger than $4$, and whether that could be realized under circuit-level noise.

A very-far-future superconducting qubit platform could also have an improved $10^{-4}$ error rate.
In this regime, the logical error rate per round of our compact surface code ($Z$-top orientation) is $\approx\frac{1}{20}3.5^{-2d}$, or $\Lambda\approx150$.
To a first approximation, a halving of patch distance like $24\rightarrow12$ reduces the total number of physical qubits by a factor of four.
So long as the cycle time $d\times 1\mu\text{s}$ remains larger than the $10\mu$s reaction time, the execution time for lookup tables will also be halved using our low-reaction depth modification. 
However, reaction time can remain a limiting factor in the $\ket{CCZ}$ factory and serial adders.
To a second approximation, the lower error rate also allows us to produce logical $\ket{T}$ states at the requisite error for all the electronic structure instances we consider using solely magic state cultivation.
This significantly reduces the spacetime overhead of parallelization in time-optimal compilation, and might generally favor multiplexed rotation synthesis based on the Matsumoto-Amano form rather than with phase gradient adders.
One could also consider implementing Toffoli gates directly with $4\ket{T}$ states as the $\textsc{CCX}$ pipe diagram~\cite{Huggins2025Flasq} is smaller than the $\ket{CCZ}$ factory used in this work, which has topological volume $4d\times3d\times 5d$.
In any case, color codes, which have a lower threshold but high encoding rates, will likely outperform surface codes in this regime.

\subsection{Code Availability}
Code to reproduce the results of this work will be made public in a Zenodo repository at a later date.

\begin{acknowledgments}
We thank Craig Gidney for developing the software tools of Stim~\cite{Gidney2021Stim} and Crumble that made this project possible. We thank Noah Shutty and Oscar Higgott for developing the pymatching library~\cite{Higgott2025SparseBlossom}.
Dominic W. Berry worked on this project under a sponsored research agreement with Google Quantum AI.
This project is supported by Australian Research Council Discovery Projects DP220101602 and DP260102543.
\end{acknowledgments}

\bibliography{main}
\clearpage
\appendix

\makeatletter
\addtocontents{toc}{\protect\renewcommand{\protect\l@subsection}[2]{}}
\addtocontents{toc}{\protect\renewcommand{\protect\l@subsubsection}[2]{}}
\makeatother

\section{Uniform depolarizing noise model}\label{sec:noise_model}
\input{1_compact_patch/noise_model}

\clearpage
\section{Additional surface code benchmarks and gate sequences}\label{sec:compacy_patch_appendix}
\input{1_compact_patch/lattice_surgery_appendix}

\clearpage
\section{Additional lookup table lattice surgery compilation details}
\label{sec:lookup_appendix}
\input{4_compilation/lookup_table_appendix}

\clearpage
\subsection{Constant reaction depth compilation of lookup tables}\label{sec:deferred_corrections_appendix}
\input{4_compilation/low_reaction_depth_lookup_appendix}

\clearpage
\section{Additional multiplexed rotations compilation details}\label{sec:multiplex_rotations_appendix}
\input{4_compilation/multiplexed_rotations_appendix}

\section{Additional electronic structure block-encoding details}\label{sec:sec:appendix_tradeoffs}
\input{5_simulation/circuit_tradeoffs}

\end{document}

%% file: preamble.tex
\documentclass[superscriptaddress,aps,prx,eqsecnum,nofootinbib,notitlepage,10pt,longbibliography]{revtex4-2}
\pdfoutput=1

\usepackage[caption=false]{subfig}
\usepackage{graphicx}
\usepackage{amsmath}
\usepackage{amsfonts}
\usepackage{amssymb}
\usepackage{amsthm}
\usepackage{mathtools}
\usepackage{booktabs}
\usepackage{array}
\usepackage[export]{adjustbox}
\usepackage{placeins}
\usepackage{multirow}
\usepackage{bbm}
\usepackage[colorlinks]{hyperref}
\hypersetup{
	pdfstartview={FitH},
	pdfnewwindow=true,
	colorlinks=true,
	linkcolor=blue,
	citecolor=blue,
	filecolor=blue,
	urlcolor=blue}
\usepackage{tikz}
\usepackage{verbatim}
\usetikzlibrary{arrows}
\usepackage{qcircuit}
\usepackage{enumerate}

\usepackage{listings}
\usepackage{color}
\usepackage[utf8]{inputenc}
\usepackage{nicematrix} 
\usepackage[table]{xcolor}
\usetikzlibrary{calc,positioning}

\usepackage[capitalise]{cleveref}
\newcommand{\crefpos}[2]{\hyperref[#1]{\cref*{#1} (#2)}}
\definecolor{codegreen}{rgb}{0,0.6,0}
\definecolor{codegray}{rgb}{0.5,0.5,0.5}
\definecolor{codepurple}{rgb}{0.58,0,0.82}
\definecolor{backcolour}{rgb}{0.95,0.95,0.92}

\usepackage{xcolor,colortbl}
\newtheorem{theorem}{Theorem}
\newtheorem{lemma}{Lemma}
\newtheorem{remark}{Remark}
\newtheorem{proposition}[theorem]{Proposition}

\crefname{definition}{Definition}{Definitions}
\newtheorem{definition}[theorem]{Definition}

\lstdefinestyle{mystyle}{
  backgroundcolor=\color{backcolour},   commentstyle=\color{codegreen},
  keywordstyle=\color{magenta},
  numberstyle=\tiny\color{codegray},
  stringstyle=\color{codepurple},
  basicstyle=\footnotesize,
  breakatwhitespace=false,
  breaklines=true,
  captionpos=b,
  keepspaces=true,
  numbers=left,
  numbersep=5pt,
  showspaces=false,
  showstringspaces=false,
  showtabs=false,
  tabsize=2
}

\def\bra#1{\mathinner{\langle{#1}|}}
\def\ket#1{\mathinner{|{#1}\rangle}}

\newcommand{\proj}[1]{\ket{#1}\!\bra{#1}}

\newcommand{\bits}[1]{\left\lceil\log_2 #1\right\rceil}

\newcommand{\identity}{\mathbb{I}}

\def\Be{\textsc{{Be}}}
\def\one{{\mathchoice {\rm 1\mskip-4mu l} {\rm 1\mskip-4mu l} {\rm
1\mskip-4.5mu l} {\rm 1\mskip-5mu l}}}

\makeatletter
\renewcommand{\tableofcontents}{%
    \vspace{-1em}

    \twocolumngrid 
    
    \@starttoc{toc} 
    
    \onecolumngrid 
}
\makeatother

\makeatletter
\def\l@subsubsection#1#2{}
\def\l@f@section{}%
\def\toc@@font{\footnotesize \sffamily}%
\makeatother

\usepackage{subfiles}

%% file: 1_compact_patch/compact_patch.tex
\begin{figure}
    \centering
    \includegraphics[width=\linewidth]{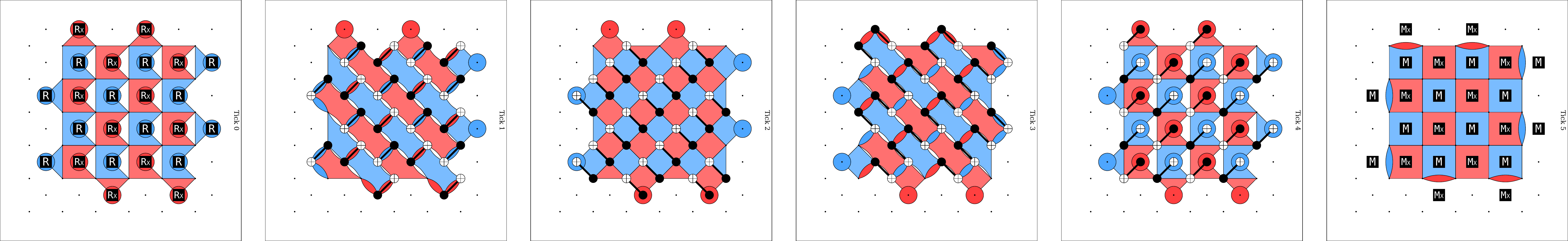}
    \caption{The $N$-\reflectbox{Z} gate sequence implementing stabilizer measurements of a distance $d=5$ rotated surface code patch encoding one logical qubit. Observe the `\reflectbox{Z}' pattern of \textsc{CX} gates controlled by the measurement qubits reset by $R_X$ to $\ket{+}$. Similarly, an `$N$' pattern is seen for \textsc{CX} gates targeted by measurement qubits reset by $R$ to $\ket{0}$.}
    \label{fig:basic_surface_code_patch_gate_sequence}
\end{figure}

The surface code~\cite{Kitaev2002Topological} is a topological quantum error correction code that has been extensively studied over the past three decades, and is notable for a high threshold and polynomial-time decoding by minimum weight perfect matching.
The stabilizers of the surface code are shown in~\cref{fig:surface_code_patch_intro} (left) where the blue and red squares represent weight-$4$ $X$ and $Z$ stabilizers acting on four data qubits respectively.
Also shown is a logical qubit defined by a pair of anti-commuting logical operators that commute with all stabilizers.
These stabilizers are the end-cycles of a detecting region, or detector~\cite{McEwen2023RelaxingHardware}, that makes explicit the sequence of quantum gates and additional qubits needed for their measurement.
The detecting region of the traditional $N$-\reflectbox{Z} gate sequence~\cite{fowler12} for a rotated surface code patch, illustrated in~\cref{fig:basic_surface_code_patch_gate_sequence}, is seen to be a periodic sequence of $6$ layers: 1 each for reset and measurement, and $4$ layers of \textsc{CX} gates.
Although a patch with side-length and distance $d$ is implemented using $d^2$ data qubits and $d^2-1$ measurement qubits, these qubits are not densely packed at the boundaries, with the empty gaps representing an inefficient use of physical qubits.
In a large-scale architecture, lattice surgery computation requires each patch to occupy a larger bounding box enclosing $2(d+1)^2$ physical qubits to either avoid overlapping measurements or breaking the checkerboard pattern of alternating $X$ and $Z$ squares.

\begin{figure}
    \centering
    \begin{minipage}[b]{0.48\linewidth}
        \centering
        \raisebox{-0.5\height}{\includegraphics[width=0.35\linewidth]{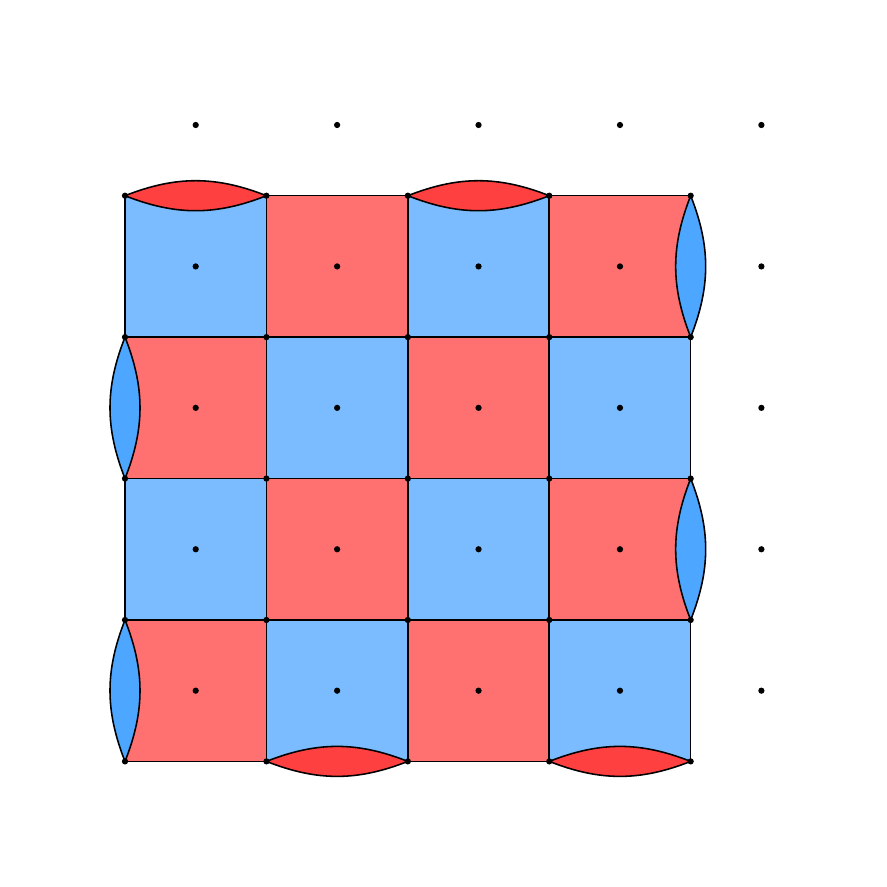}}
        \hfill
        $\begin{aligned}
            &\xrightarrow{G_X} \\
            &\xleftarrow{{G_X^{-1}}}
        \end{aligned}$
        \hfill
        \raisebox{-0.5\height}{\includegraphics[width=0.35\linewidth]{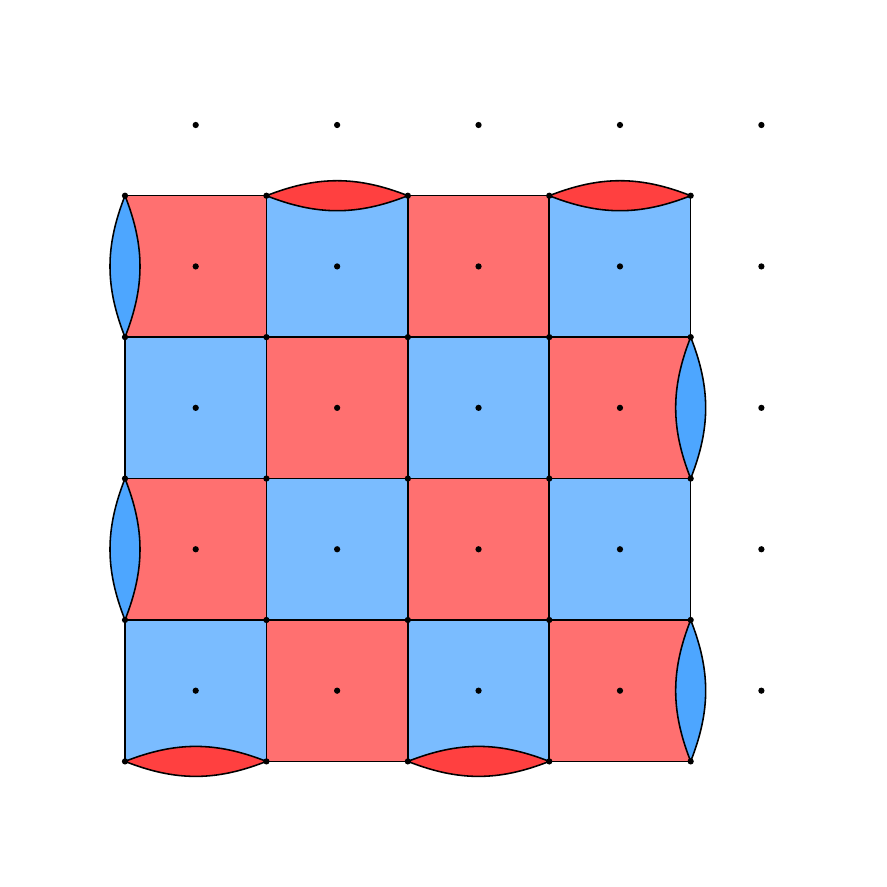}}
        \label{fig:compact_patch_X}
    \end{minipage}
    \hfill 
    \begin{minipage}[b]{0.48\linewidth}
        \centering
        \raisebox{-0.5\height}{\includegraphics[width=0.35\linewidth]{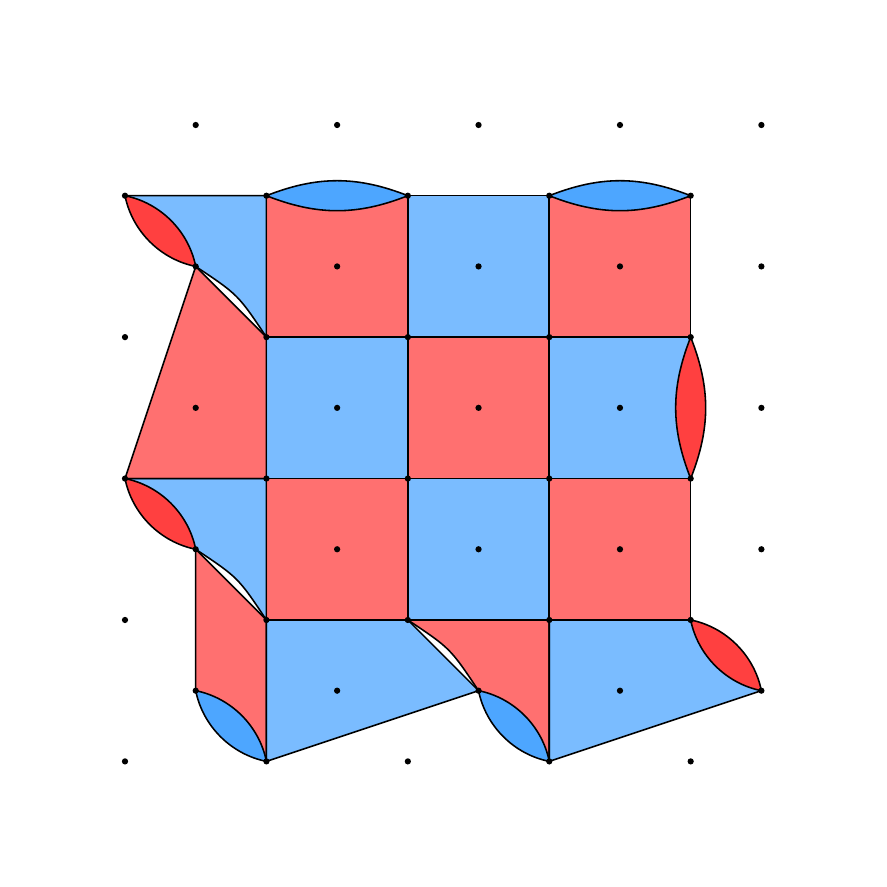}}
        \hfill
        $\begin{aligned}
            &\xrightarrow{G_Z} \\
            &\xleftarrow{{G_Z^{-1}}}
        \end{aligned}$
        \hfill
        \raisebox{-0.5\height}{\includegraphics[width=0.35\linewidth]{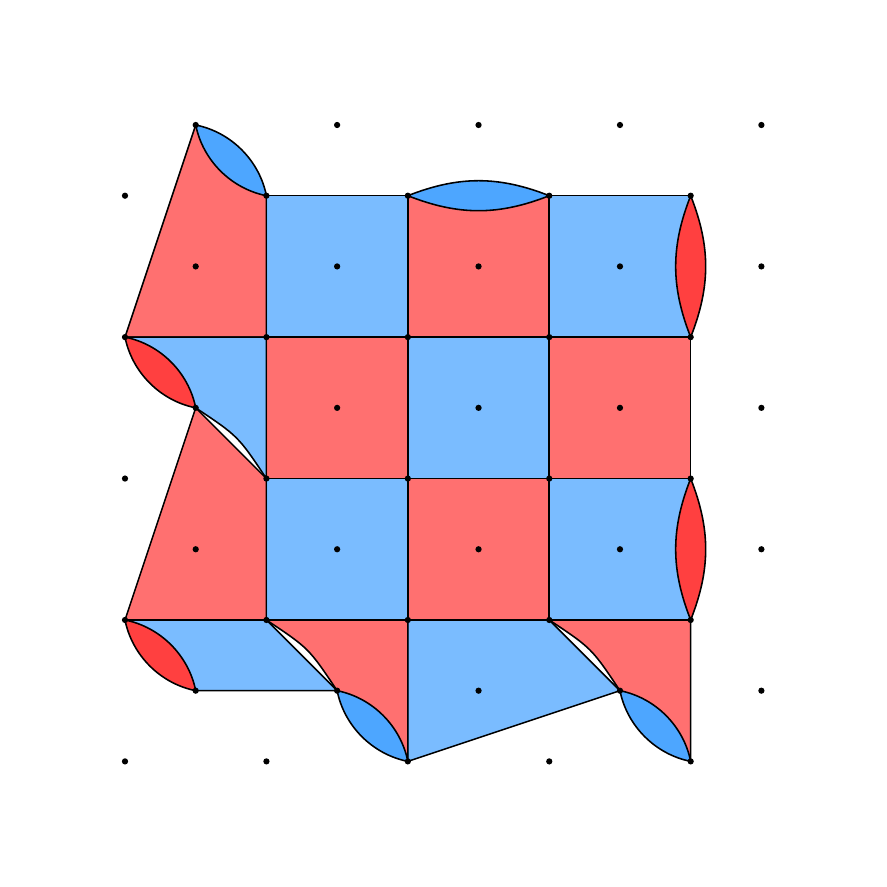}}
        \label{fig:compact_patch_Z} 
    \end{minipage}

    \vspace{0.5em} 
    
    \begin{minipage}[b]{0.48\linewidth}
        \centering
        \includegraphics[width=\linewidth]{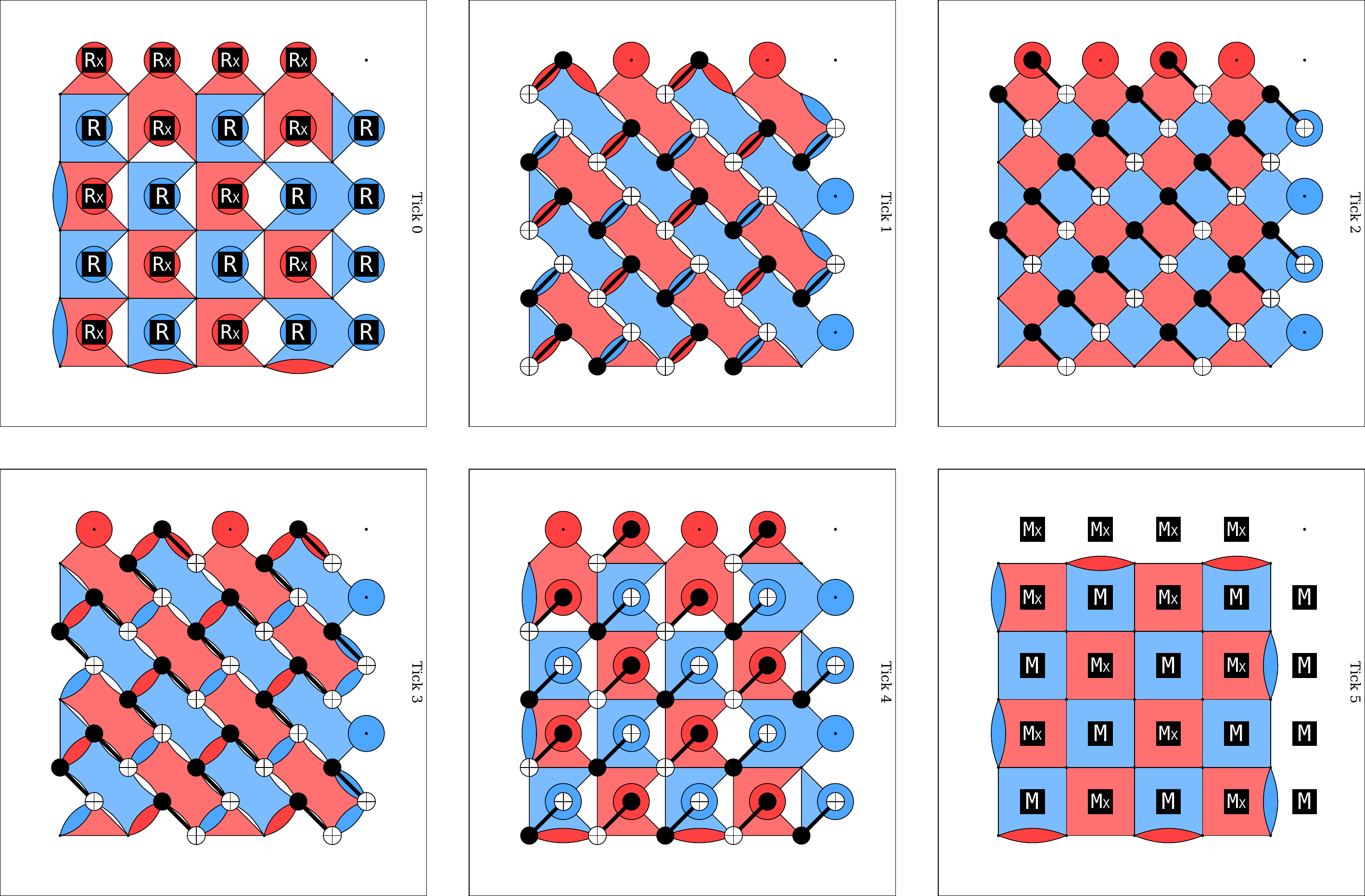}
        (c) Detecting regions of gate sequence $G_X$.
        \label{fig:square_patch_GX}
    \end{minipage}
    \hfill 
    \begin{minipage}[b]{0.48\linewidth}
        \centering
        \includegraphics[width=\linewidth]{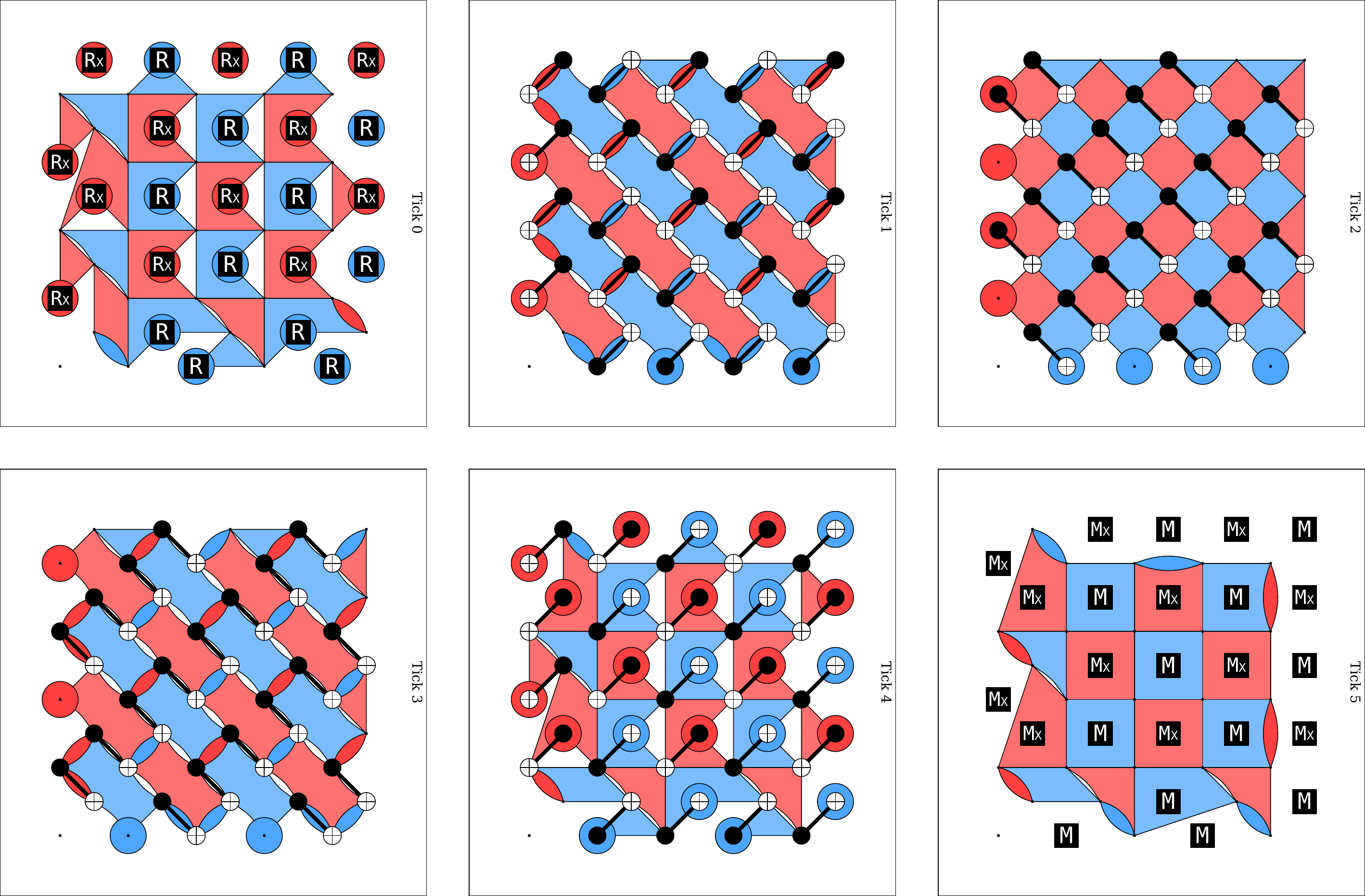}
        
        \vspace{0.2cm}
        (d) Detecting regions of gate sequence $G_Z$.
        \label{fig:square_patch_GZ}
    \end{minipage}
    \caption{Surface code patch encoding one logical qubit using $Z$ (blue) and $X$ (red) stabilizers with a footprint of $d^2$ data qubits and $d^2-1$ measurement qubits for (left) the case previously known $X$-top configuration. All stabilizers are measured by preparing either a $\ket{+}$ or $\ket{0}$ state followed by a sequence $G_X$ of four layers of $\textsc{CX}$ gates. The alternating sequence $G_X,G_X^{-1}$ of~\cite{McEwen2023RelaxingHardware} is a Floquet code that alternates between two end-cycles and achieves circuit-level distance $d$~\cite{Yuga2026NoHook}. In this work, we demonstrate (right) a new $Z$-top configuration that permits lattice surgery operations within a bounding box of $2d^2$ qubits per patch. The gate sequence (bottom right) $G_Z$ is identical to $G_X$ in the bulk but differs on the boundary.}
    \label{fig:compact_patch_main}
\end{figure}

In this work, we modify the surface code boundaries to eliminate unnecessary space.
We build upon the `alternating' gate sequence of prior art~\cite{McEwen2023RelaxingHardware}, which studied the case shown in~\cref{fig:compact_patch_main} (left) with vertically-oriented logical-$X$ and horizontally-oriented logical-$Z$ operators. 
We call this the $X$-top configuration due to the $X$-type boundary stabilizers above the patch. 
As seen in~\cref{fig:compact_patch_main} (left) the alternating sequence has a larger periodicity, leading to stabilizer end-cycles that alternate between two possible states, akin to a Floquet code~\cite{Hastings2021Floquet}.
Moreover, all data and measurement qubits neatly pack within the optimal $2d^2$-sized bounding box, and all two-qubit gates map to a hex, rather than square grid.
However, generic lattice surgery operations such as in patch rotations and our later dense packing of twist defect, also require patches with a $Z$-top boundary configuration with the same gate sequence in the bulk, but with an exchanged orientation of logical operators.
In~\cref{fig:compact_patch_main} (right), we provide the gate sequence for the $Z$-top configuration, also within a $2d^2$-sized bounding box, and with the same qubit connectivity, in contrast to prior art that always required the larger $2(d+1)^2$ bounding box~\cite{Yuga2026NoHook}, which we illustrate in~\cref{fig:square_patch_big} for completeness.
Due to using fewer qubits and quantum gates, our smaller $Z$-top configuration also has roughly half the logical error rate.
This space reduction is most significant for early fault-tolerant implementations, or at very low physical error rates -- for instance, this enables a $d=5$ patch with $30\%$ less space.

The non-square shape of our $Z$-top end-cycle also poses no impediment to lattice surgery.
For instance, a square-shaped boundary can still be recovered by conjugating the offset reset and measurement gates on the boundary with a \textsc{Swap} gate.
This is shown later in~\cref{fig:hadamard}, where some boundary \textsc{CX} gates are converted to $\textsc{CXSwap}$ gates.
We may therefore design lattice surgery protocols with the usual straight boundaries of a square-shaped end-cycle, where all reset and measurement qubits lie on the same horizontal and vertical lines, then obtain a pure \textsc{CX} construction by replacing all $\textsc{CXSwap}$ gates and shifting the resets and measurements that they conjugate.
The alternating gate sequence exhibits advantages over the traditional $N$-\reflectbox{Z} schedule.
Specifically, it was recently demonstrated that the alternating schedule is insensitive to hook errors~\cite{Yuga2026NoHook} - a distance $d$ qubit patch achieves a circuit-level distance $d$, regardless of the orientation of logical operators.
This allows the same $6$ layer gate sequence to be used uniformly throughout, in contrast to the $N$-\reflectbox{Z} schedule which requires numerous special-case gate sequences to manage hook errors, also executing in more layers.

Hook errors may be understood by the concept of detecting regions, or detectors~\cite{McEwen2023RelaxingHardware}.
In the absence of errors, the parity of detector measurement, will always be known in advance.
Single-qubit Pauli errors that anti-commute with the detecting region, which includes state-preparation and measurement errors cause a flip in detector parity and are therefore detectable. 
The code distance is then the minimum weight of any logical operator, which represents the number of single-qubit faults required to enact an erroneous logical operation.
However, as the stabilizers of each detecting region slice are generated by a quantum gate sequence, a single-qubit fault is transformed by the same sequence and can easily propagate into a multi-qubit error.
These multi-qubit hook errors could, in the worst-case, halve the distance of the quantum code.
For the $N$-\reflectbox{Z} gate sequence, two-qubit hook errors occur in horizontal or vertical lines, and are carefully managed by always aligning logical operators in a perpendicular direction, which leads to a $7$-layer implementation in general.

The set of all possible errors and their probabilities, known in advance through the noise model, generates a Detector Error Model (DEM) decoding graph~\cite{McEwen2023RelaxingHardware}.
Vertices of the DEM are detectors (and logical observables are sets of vertices), and individual error channels generate edges and hyperedges, weighted by the probability of their occurrence, between the detectors whose measurement outcome they flip. 
A key feature of the surface code is \textit{matchability}: Any string of $X$ ($Z$) errors is guaranteed to flip only two detectors of the opposite type, except at the boundaries where only one detector is flipped.
Hyperedges can also be generated by $Y$ or $YY$ errors, but these are in general decomposed into additional edges that flip $X$ and $Z$ detectors with the same probability.
This guarantees that the location of errors in the surface code can be decoded on independent $X$-only and $Z$-only DEMs in polynomial time by Minimum Weight Perfect Matching (MWPM) on the DEM.
Decoding by decomposing $Y$ errors leads to worst-case performance as it over-counts the number of actual errors.
Correlated MWPM can achieve better performance in multiple passes of adding edges between detectors of the joint DEM where $X$- and $Z$-matchgraph predict errors in the same locations.
Circuit-level distance is the minimum number of errors needed to cause an undetectable logical error.
On the decoding graph, it corresponds to the shortest non-trivial path of edges and hyperedges that anti-commutes with any logical operator and connects boundary vertices.

However, even circuit-level distance is only a heuristic for actual code performance as hyperedges are weighted, and the shortest path could rely on a particularly rare combination of errors.
Hence, the most reliable estimate of logical error rates is obtained by Monte Carlo sampling of DEM errors and a decoding with a specific choice of decoder.
In the interests of generality, we assume a standard one- and two- qubit uniform depolarizing noise model~\cite{McEwen2023RelaxingHardware}, with a physical gate set comprised of single-qubit Clifford operations, measurement and reset in the $X$ or $Z$ basis, and two-qubit $\textsc{CX}$ or $\textsc{CZ}$ gates. 
This noise model favors gate sequences with minimum depth, unlike other noise models such as for superconducting qubits that permit different optimizations, such as allowing for low-error idling, and also native $\textsc{CXSwap}$ gates in a single layer.
We validate the logical error rate of our approach with circuit-noise-level benchmarks implemented in \textsc{Stim}~\cite{Gidney2021Stim}. 
The results are summarized in~\cref{tab:surface_code_results}, where we also demonstrate walking~\cite{McEwen2023RelaxingHardware} in a $2d^2$ bounding box, assuming $\textsc{CXSwap}$ gates, which is of independent interest and not used in our later resource estimates.
Except for walking, all $\textsc{CXSwap}$ gates in this work can be replaced by \textsc{CX} gates without penalty.

We define a round, or surface code cycle, as the time taken to implement the $6$ layer gate sequence.
A logical timestep, is the time taken to implement $d$ rounds.
A distance-$d$ topological unit of volume is depicted as a $d\times d\times d$ box that represents a $2d^2$ physical qubit footprint executing a surface code gate sequence for one logical time step.
Our later resource estimates rely on the following fits of the logical error rate per round per logical qubit as a function of distance $d$:
\begin{align}\label{eq:compact_patch_error_formula}
\text{Memory}:\; p_\text{static}(d)\doteq\frac{1}{10}4.0^{-d},
\quad\text{Lattice surgery}:\; p_L(d)\doteq\frac{1}{3}p_\text{dynamic}(d),
\quad\text{Walking}:\; p_W(d)\doteq\frac{1}{2}p_\text{dynamic}(d),
\end{align}
where $p_\text{dynamic}(d)\doteq 3.5^{-d}/10$.
For example, we assume that a quantum algorithm occupying $n$ distance $d$ patches and executing $t$ logical timestep will have an overall failure probability upper bounded by
\begin{align}
p\le1-(1-p_L(d))^{dnt}.
\end{align}
For consistency with prior resource estimates on superconducting architectures we assume that each surface code cycle takes $1\mu$s~\cite{PRXQuantum.2.030305}.

\subsection{Benchmarks}
\subsubsection{Memory}\label{sec:memory}
The baseline operation in quantum algorithms is idling, where no logical operations are performed on a logical qubit, but its quantum state needs to be preserved regardless.
A key feature of the surface code is that, through Pauli frame tracking, single-qubit errors do not need to be explicitly corrected.
Whenever the decoder predicts that a single-qubit $X$ or $Z$ error has flipped the predicted eigenvalue of a stabilizer, it suffices to simply record this single-bit flip.
When the logical qubit is measured in a Pauli basis at the end, the parity of all flips that affect the logical Pauli string is merged into the result.
This introduces the concept of reaction time, which is the latency of decoding and resolving these parities.
Reaction time is important when lattice surgery operations are classically controlled by logical qubit measurement outcomes.

We benchmark the logical error rate per round per logical qubit for idling a distance $d$ patch, in memory experiments for $2d$ noisy rounds.
In order to simultaneously benchmark all logical observables $X,Y,Z$, we use the trick equivalent to forming a noiseless bell pair between the logical qubit with a single physical qubit.
This requires the assumption of noiseless initial state preparation and final state measurement.
This leads to a mild underestimate compared to including noisy state preparation and measurement or running the memory experiment run longer for, say, $20d$ rounds, but should still be an upper bound compared to using more sophisticated decoders~\cite{Beni2025Tesseract}, and will be essential later in studying correlated logical errors in patches encoding multiple logical qubits.

\begin{figure}
    \centering
        \includegraphics[scale=0.8]{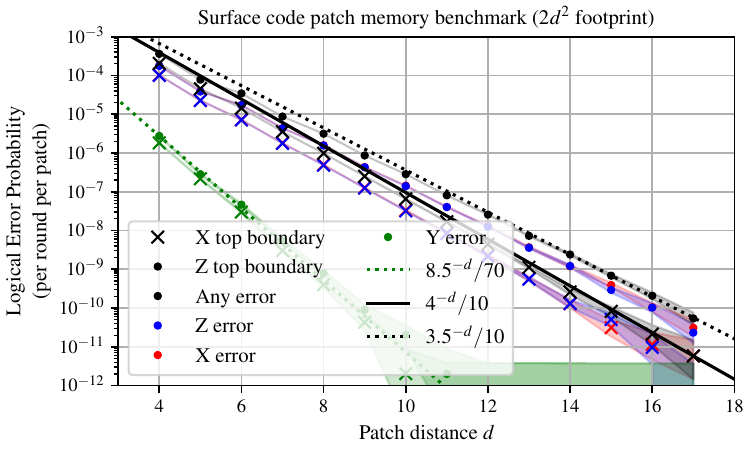}
        \includegraphics[scale=0.8]{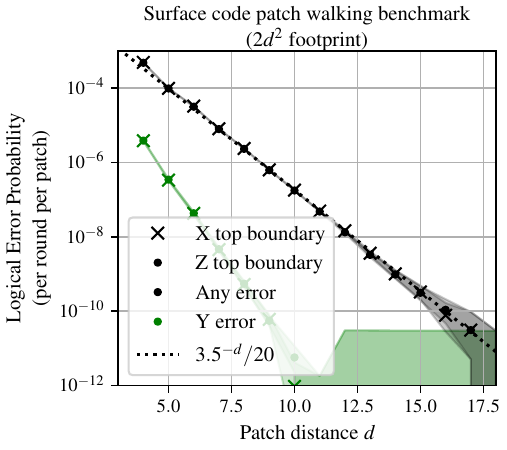}
    \caption{(Left) Memory benchmark of alternating gate sequence~\cref{fig:compact_patch_main}. The $X$-top configuration has better error scaling than the $Z$-top configuration. With each configuration, the probability of an erroneous logical-$X$ and $Z$ is identical, whereas the probability of both occurring in a logical $Y$ is suppressed. (Right) Benchmark of the walking gate sequence by~\cref{fig:walking}, which uses $2$ cycles ($12$ gate layers) to translate a patch horizontally or vertically by one lattice spacing on a rotated square grid, shows no bias between $X$-top and $Z$-top configuration and is fit by the average of the memory benchmark.
    }
    \label{fig:square_patch_alt}
\end{figure}

The results shown in~\cref{fig:square_patch_alt} demonstrate a clear bias in error rates.
The $X$-top configuration has an error fit of 
$p_\text{static}(d)=4.0^{-d}/10$, whereas the $Z$-top configuration is fit by $p_\text{dynamic}(d)=3.5^{-d}/10$.
This difference is accounted for by studying the detecting regions of the logical operators.
In the former case, the logical $X$ ($Z$) Pauli string is oriented vertically (horizontally), and their detecting regions do not include any measurement or reset gates.
In contrast, the latter $Z$-top case exchanges the orientation of the logical $X$ ($Z$) Pauli strings and their detecting regions include adjacent measurements and resets to their top-right.
The larger detecting region indicates greater sensitivity to errors, and hence a different scaling of logical error rate with $d$.
We note that the higher error rates of dynamically updated logical operators appear in previous memory benchmarks such as Ref.~\cite{Gidney2025Yoked} reporting a $4^{-d}/10$ logical error rate, and Ref.~\cite{Gidney2025Cultivation} reporting $3.5^{-d}/40$. These have exponents consistent with our fits for $X$-top and $Z$-top respectively.
For comparison, we also benchmark in~\cref{fig:square_patch_big} the $Z$-top error rate of prior art~\cite{Yuga2026NoHook} which requires a larger $2(d+1)^2$ bounding box and has roughly twice the error.

The error rate of memory can be significantly improved by a variety of methods.
In a typical quantum algorithm compiled to lattice surgery, the orientation of all patches will be known in advance.
Hence, we may always ensure that the $X$-top configuration dominates by reflecting the gate sequences of every single patch about the `$\backslash$' diagonal.
This leads to a worst-case error rate that is the average of both orientations and well-approximated by $\frac{1}{2}p_\text{dynamic}(d)$.
We instead choose to make this reflection local: a ($Z$-top) orientation is transformed into an ($X$-top) orientation whenever a surface code patch is to be placed in an idling state.
Observe in~\cref{fig:compact_patch_main} that the mid-cycle (tick $2$) of the detecting regions are mirror images of each other along the `$\backslash$' diagonal.
When the ($Z$-top) orientation reaches this layer, we continue with the $X$-top gate sequences reflected along the `$\backslash$' diagonal.
This transforms the $Z$-top to a reflected $X$-top orientation with the better error rate of $p_\text{static}(d)$, which we report in~\cref{tab:surface_code_results}, where we also assume that a similar technique can be applied to the larger $Z$-top orientation of~\cite{Yuga2026NoHook}.
When leaving the idling state, we restore the original $Z$-top orientation by merging the reflected $X$-top and the original $Z$-top gate sequence at the same mid-cycle.
Note that this local reflection is incompatible with lattice surgery, which requires a consistent bulk gate sequence across all involved surface code patches, and will require a different method of error reduction.

\subsubsection{Transversal Hadamard}
\begin{figure}
    \centering
    \includegraphics[width=0.48\linewidth,valign = m]{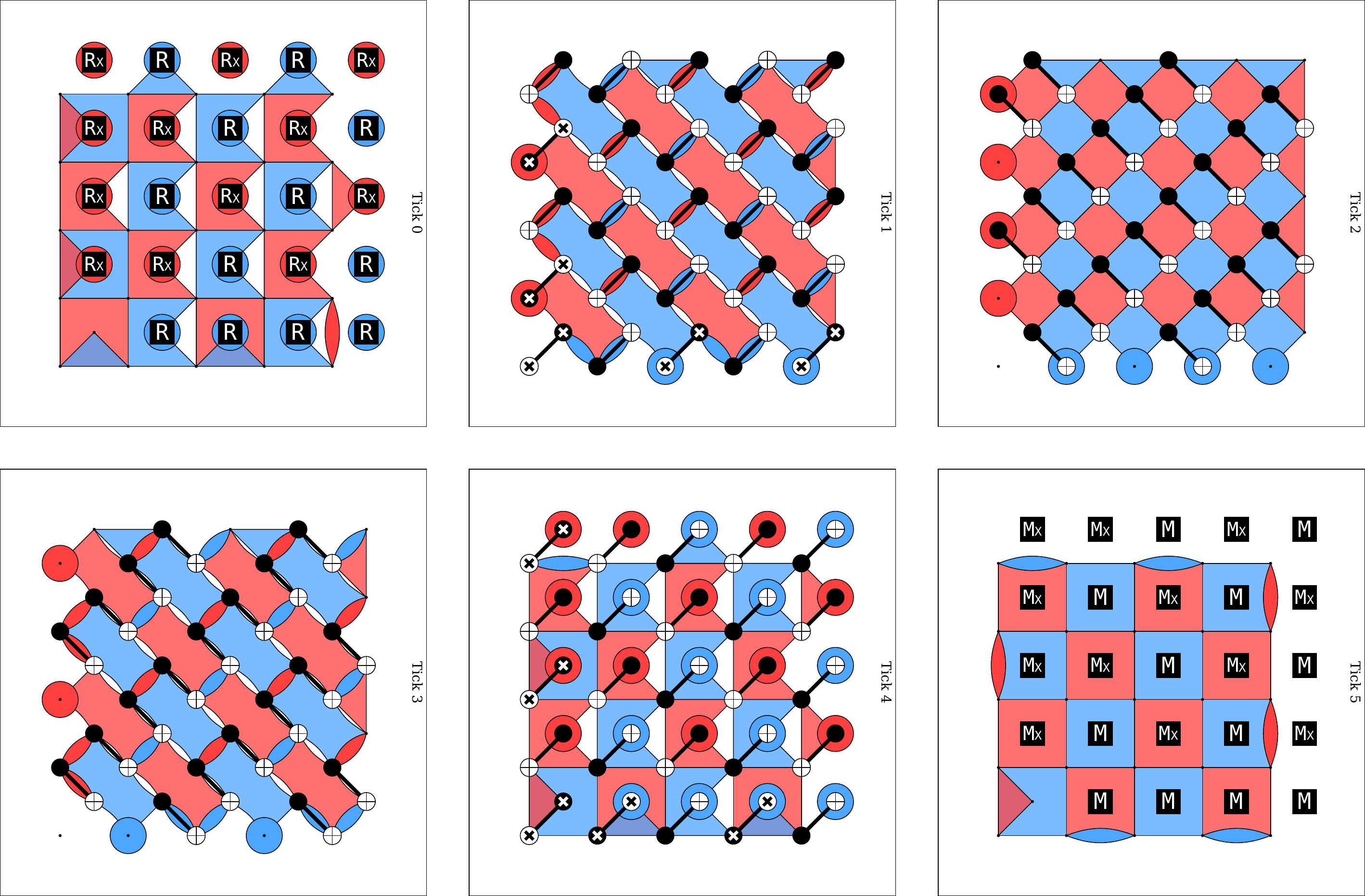}
    \includegraphics[scale=0.8,valign = m]{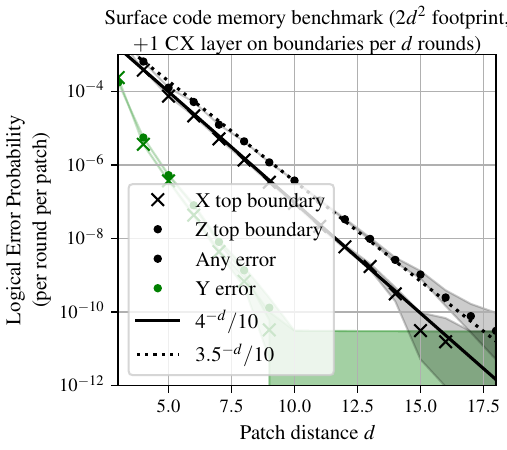}
    \caption{(Left) The $Z$-top gate sequence is isomorphic to the square-shaped end cycle using $\textsc{CXSwap}$ gates on two boundaries. Note that the bottom-left boundary stabilizer is weight three solely to produce a square shape. It can be made weight-two by removing the optional \textsc{CXSwap}. (Right) Logical error rate of Transversal Hadamard by synchronizing detecting regions by implementing $\textsc{CXSwap}$ in one round with one additional layer of boundary $\textsc{CX}$ gates.}
    \label{fig:hadamard}
\end{figure}

Another key operation is the logical Hadamard gate. 
In prior art, this is particularly simple as logical Hadamard is transversal and the end-cycle states for both $X$-top and $Z$-top orientations are identical.
In our case, logical Hadamard is still transversal, but inverting the Paulis of the stabilizers does not lead to detecting region slices that align.
This misalignment is easy to resolve with a larger bounding box, but in this work, we propose two options.
\begin{itemize}
    \item Add one additional layer of $\text{CX}$ gates on the boundary without additional space.
The key observation is that due to the alternating gate sequence, all measurements and resets are conjugated by a $\text{CX}$ gate. 
If in the $Z$-top orientation, we additionally conjugate the lower and left measurements and resets with a $\textsc{Swap}$ gate, we obtain $\textsc{CXSwap}$ gates that, as seen in~\crefpos{fig:hadamard}{left} produces a square-shaped end-cycle that can now be aligned with the $X$-top gate sequence after a transversal Hadamard.
Although applying an additional layer of gates would be extremely detrimental to the logical error rate if applied every round, we only need this layer once every $6d$ layers, and only if a Hadamard gate is required.
In~\cref{fig:hadamard}, we benchmark the impact of this additional layer, and find negligible impact to the overall logical error rate.
We note that in a superconducting architecture, this layer can be omitted by applying $\textsc{CXSwap}$ gates directly.
\item Alternatively, we first apply the error-reducing reflection trick if necessary so that the patch, regardless of whether it is $X$-top or $Z$-top, has a square-shaped end-cycle.
At any one of these end-cycles, say tick $5$, we apply a transversal Hadamard, and continue with the gate sequence reflected again about the `$\backslash$' diagonal.
If we originally started in the $Z$-top orientation, this completes the timelike logical Hadamard.
Otherwise, we wait until the next midcycle to continue with gate sequence reflected yet again about the `$\backslash$' diagonal. This completes the timelike logical Hadamard.
This approach should be preferred as it always executes in $6$ cyles without any reliance on \textsc{CXSwap} gates.
\end{itemize}

\subsubsection{Walking}\label{sec:walking}
We also demonstrate that a distance $d$ surface code patch can be translated one step to the right every $2d$ rounds within a bounding box of $2d^2$ qubits. However, our construction shown in~\cref{fig:walking} requires square grid connectivity and the use of $\textsc{CXSwap}$ gates on the boundary.
In contrast, prior art achieves this using only $\textsc{CX}$ gates, but within a bounding box of $2(d+1)^2$ qubits. 
We note that the resulting error rate achieved is essentially the average of the $X$-top and $Z$-top configurations, which equates to roughly half of the non-walking $Z$-top configuration.


%% file: 1_compact_patch/lattice_surgery.tex
Measurement-based computation~\cite{Briegal2009MBQC} between logical qubits in surface code patches is implemented by lattice surgery~\cite{Horsman2012LatticeSurgery}. 
Lattice surgery measures logical operators oriented along surface code boundaries.
For instance, the logical \textsc{CX} gate can be implemented by logical $XX$ or $ZZ$ measurements with a logical ancilla qubit.
Universal multi-qubit Clifford gates can be realized by such measurements, together with the single-qubit logical Clifford Hadamard and $S$ gates.
Universal quantum computation is then achieved by assuming magic resource states, such as a noisy logical $\ket{T}$, prepared by magic state cultivation~\cite{Gidney2025Cultivation}. 
Multiple copies of noisy magic states distributed across multiple surface code patches can then be distilled, also using lattice surgery, to obtain a lower-error $\ket{T}$ state or other useful magic states like the $\ket{\textsc{CCZ}}$ state~\cite{Gidney2025Factoring}.
 In quantum algorithms dominated by classical data input, such as those using  quantum lookup tables~\cite{Low2024Trading}, Clifford operations dominate spacetime volume. 
Hence, any improvements to lattice surgery lead to a proportional reduction in the overall spacetime volume of a quantum algorithm.

In this section, we demonstrate lattice surgery using the tightest possible bounding box of $2d^2$ qubits in an optimal-depth 6-layer implementation that maintains full code distance. 
We focus on logical $ZZ$ and $XX$ measurements --- more sophisticated forms of lattice surgery involving twist defects~\cite{Brown2017PokingHoles,Geher2025Tangling,Litinski2018LatticeTwist} enable direct measurement of other logical operators such as $YY$.
Later in~\cref{sec:dense_memory}, we also describe an improved implementation of twist defects that should enable these operations in the same bounding box.
Besides the $S$ gate and in-place single-qubit logical $Y$ measurements~\cite{Gidney2024Inplace}, which we assume can be performed in the same bounding box with similar error scaling, we do not use lattice surgery based on twist defect braiding and defer a detailed benchmark of twist-based lattice surgery to future studies.

\begin{figure}
    \centering
    \raisebox{-0.5\height}{\includegraphics[width=0.25\linewidth]{1_compact_patch/figs/merge/h_bridge_d5_Z_h_merge_end_cycles_A.pdf}}
        \hfill
        $\begin{aligned}
            \xrightarrow{{A}}
        \end{aligned}$
        \hfill
        \raisebox{-0.5\height}{\includegraphics[width=0.25\linewidth]{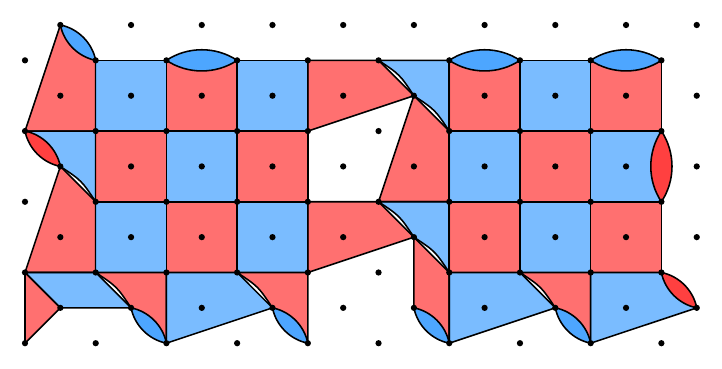}}
        \hfill
        $\begin{aligned}
            \xrightarrow{B}
        \end{aligned}$
        \hfill
        \raisebox{-0.5\height}{\includegraphics[width=0.25\linewidth]{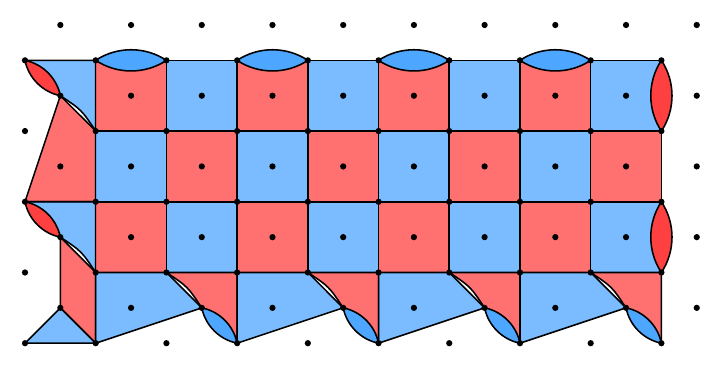}}
    \raisebox{-0.5\height}{\includegraphics[width=0.25\linewidth]{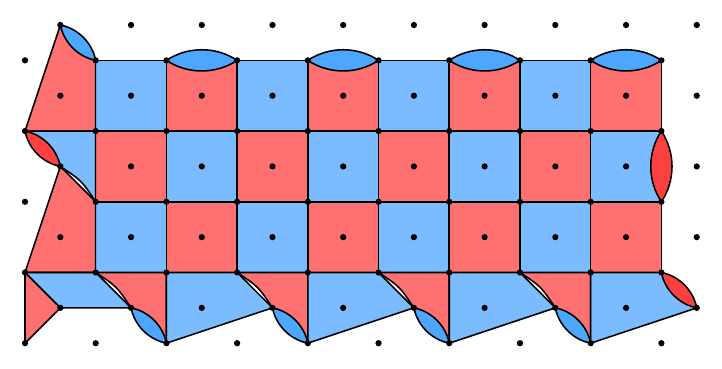}}
        \hfill
        $\begin{aligned}
            \xrightarrow{{A}}
        \end{aligned}$
        \hfill
        \raisebox{-0.5\height}{\includegraphics[width=0.25\linewidth]{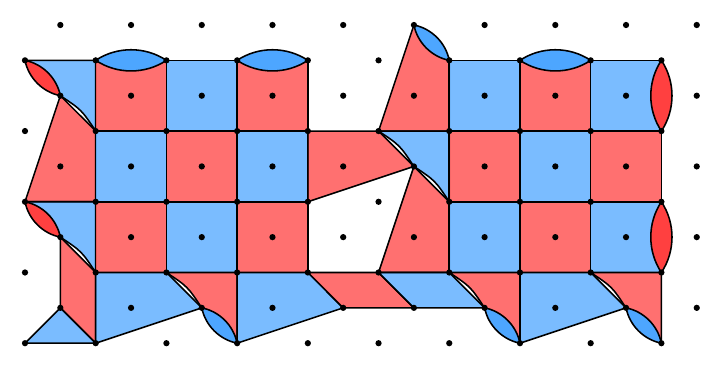}}
        \hfill
        $\begin{aligned}
            \xrightarrow{B}
        \end{aligned}$
        \hfill
        \raisebox{-0.5\height}{\includegraphics[width=0.25\linewidth]{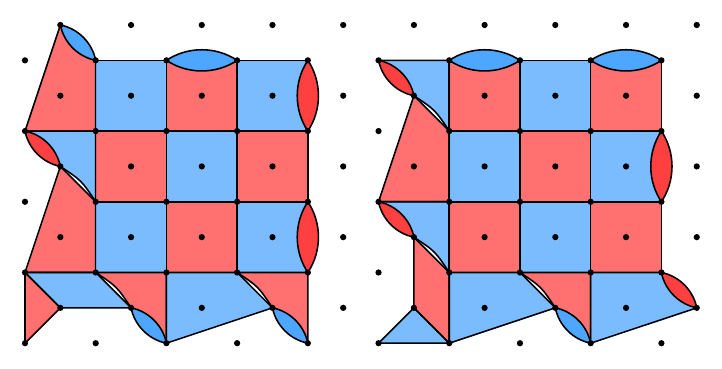}}
    \caption{(Top) Lattice surgery merge operation. Step $A$ measures all stabilizers that are the same color as the boundary. Step $B$ measures all remaining stabilizers to form a combined patch. 
    (Bottom) Lattice surgery split operation. Step $A$ omits stabilizers of the opposite color to the boundary. Step $B$ omits all remaining stabilizers to form two patches. See~\cref{fig:lattice_surgery_merge} and~\cref{fig:lattice_surgery_split} for the full gate sequence of all possible patch orientations, which also shows that all possible combinations of boundaries with different shapes and be merged and split in the optimum number of $6$ gate layers. 
    Note that boundaries are changing shape at each end-cycle as we implement a Floquet-like code.
    Moreover, the bottom-left weight-three boundary stabilizers can be made weight-two following the discussion in~\cref{fig:hadamard}.
}
    \label{fig:lattice_surgery_merge_split}
\end{figure}

Logical $XX$ and $ZZ$ measurements are realized through the merging and splitting of surface code patches along the same boundary type, which we illustrate in~\cref{fig:lattice_surgery_merge_split}.
We now outline the general steps~\cite{Fowler2018LowOverhead} of how a combined merge and split operation that measures logical $ZZ$ is completed in $d$ rounds --- the case of $XX$ is analogous.
\begin{enumerate}
    \item Select two patch boundaries of the same type, in this case one $Z$ boundary from each patch.
    \item Reset all inter-boundary uninitialized qubits with the $R_X$ gate to $\ket{+}$.
    These resets are concurrent with the other resets of the patch gate sequence and so have an amortized time cost of $0$ layers.
    \item In the first round, measure all inter-boundary $X$ stabilizers only, and in the second round, measure all inter-boundary stabilizers. This completes the merge.
    \item Repeat for a total of $d$ rounds to fault-tolerantly detect reset or measurement errors --- this could be less than $d$ by decoding with soft information in systems with a small reaction time~\cite{Akahoshi2025SoftLatticeSurgery}, or with temporally encoded lattice surgery~\cite{Chamberland2021Temporally}, but $d$ is sufficient in the standard uniform depolarizing model.
    \item Construct a detecting region in the form of a loop by choosing an appropriate product of $Z$ measurement gates.
    This loop should be enclosed on all sides by either the logical $Z$ operators of the patches or $Z$ boundary stabilizers.
    \item The parity of these measurement results from the selected gates, after decoding for errors on the DEM, reports the outcome of the logical $ZZ$ measurement.
    \item In the next-to-last round, only measure inter-boundary $X$ stabilizers, and in the last round, measure no inter-boundary stabilizers, and perform a $M_X$ measurement on all inter-boundary data qubits.
    Split the logical $X$ operator into two by updating either one of them with decoded inter-boundary $M_X$ measurement results.
    This completes the split.
\end{enumerate}

\begin{figure}
    \centering
    \includegraphics[scale = 0.8]{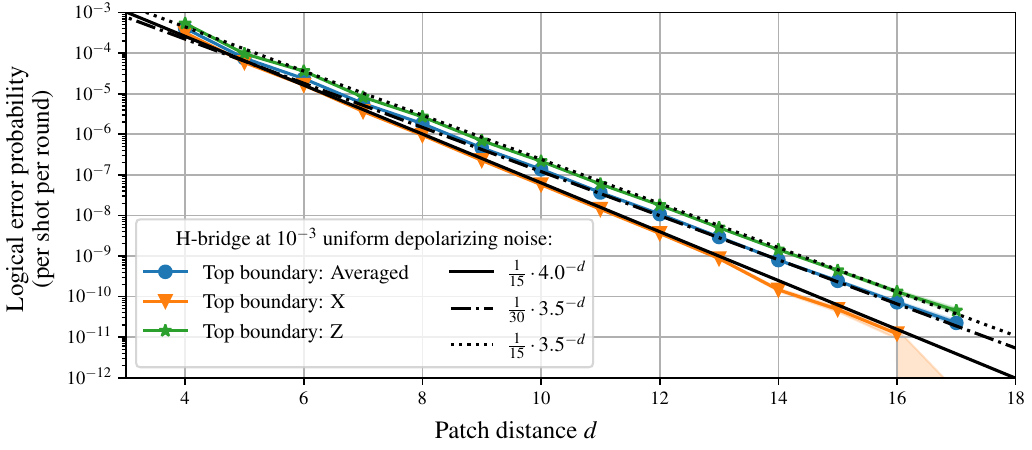}
    \caption{The logical error rate per round per patch of H-bridge lattice surgery between distance $d$ patches in a $10^{-3}$ one- and two-qubit uniform depolarizing noise model in the $\{Z,X\}$-top orientation, averaged over all combinations in~\cref{fig:h_bridge_breakdown} of a horizontal or vertical bridge orientation, and a bridge distance of $\{0,d,2d,3d\}$.
    The average of all configurations is well-approximated by $\frac{1}{30}3.5^{-d}$.}
    \label{fig:h_bridge}
\end{figure}

A standard benchmark for lattice surgery combines merge and split operations~\cref{fig:lattice_surgery_merge_split} to perform either a $ZZ$ or $XX$ measurement between two patches, and resembles an H-shaped bridge when visualized in $3$D spacetime.
We benchmark H-shaped bridge logical error rates over all possible patch orientations ($\{X,Z\}$-top), bridge directions $\{\text{horizontal},\text{vertical}\}$, and inter-patch bridge distances $\{0,d,2d,3d\}$ corresponding to zero, one, two, or three logical ancilla qubits.
We normalize the logical error rate by the topological volume of the H-bridge. 
To isolate the error contributions of the split and merge operations, we pad the start and end of the H-bridge with a minimum number of idling cycles.
Specifically, perfect state preparation is assumed for the two input patches, which allows us to simultaneously benchmark all logical operators, followed by $2$ rounds of noisy syndrome extraction.
We then perform a merge operation with $d$ rounds of syndrome extraction, followed by a split, $2$ rounds of noisy syndrome extraction, and finally perfect measurement of the two output patches.
As logical operators parallel to the short side of the bridge have at least double the distance, they effectively contribute no error. 
Hence, we find that the logical error rate fits in~\cref{fig:h_bridge} are very well approximated by half the memory logical error rate, times a combinatorial factor of $\frac{3}{2}$ for the additional error paths from one boundary to the other; additionally, it appears that noisy state preparation and measurement have little effect compared to a memory experiment in this noise model.

Similar to the memory experiments, the $Z$-top orientations have a worse logical error scaling than $X$-top orientations, owing to the static vs. dynamically updated properties of their logical operators.
On average, we expect all H-bridge orientations to be equally likely. 
Hence, we report the average of all logical error rates as $3.5^{-d}/30$ per round per logical qubit.
In practice, we can strongly bias lattice surgery compilation towards the lower-error $X$-top configuration, and we do so in~\cref{sec:hot_storage} to maximize the efficiency of quantum storage.
For instance, the same trick used to bias memory to the lower-error configuration (\cref{sec:memory}) may not be applied globally to large lattice surgery operations as those might involve an equal mixture of orientations, but can be applied locally to small lattice surgery segments, which are more likely to be predominantly one orientation that can be flipped.
Furthermore, one may apply timelike or spacelike Hadamard gates liberally to convert segments of $Z$-top configurations to $X$-top configurations, with the caveat that our spacelike Hadamards are later shown to require at least one unit of distance between patches. 
Otherwise, this trick could be applied to essentially all lattice surgery operations to obtain the lower error rate of $4.0^{-d}/15$.
One option would be to work with patches of dimension $d\times(d+1)$, and assume the better $X$-top error rates throughout, but we leave a careful study of this optimization to future work.

Combining the lattice surgery primitives of merge and split with ancilla patches realizes more complicated Clifford gates, such as logical $\textsc{CX}$, with an example gate sequence in the optimal bounding box provided in~\cref{fig:CNOT_X_Z}.
More general configurations of lattice surgery operations will involve the movement of boundaries across various interior and exterior corners.
We have been able to find distance-preserving $2d^2$ bounding box implementations of the most challenging configurations, such as examples with simultaneous horizontal and vertical zero-padding interior boundaries in~\cref{fig:straight_y_defects}, $Y$ defects oriented along straight boundaries, and other interior and exterior corners used in the next section for dense packing of defects. 
However, a future study should enumerate and benchmark all such possibilities.

%% file: 2_denser_surface_code/denser_surface_code.tex

In this section, we demonstrate that the surface code can support an encoding rate up to $4.5\times$ that of a one-qubit rotated surface code patch.
Notably, we achieve this without changing the minimal hardware assumption of degree $3$ qubit connectivity that is already needed to realize the surface code.
Moreover, we attain logical error scaling similar to the one-qubit rotated surface code patch using a quantum circuit with an optimal number of layers -- $4$ layers of controlled-NOT ($\textsc{CX}$) or controlled-phase ($\textsc{CZ}$) gates and $1$ layer each of measurement and reset.
We account for the effects of potential distance-reducing hook errors with extensive circuit-level noise benchmarks in a standard one- and two-qubit uniform depolarizing noise model~\cite{McEwen2023RelaxingHardware}.

\begin{figure}
        \centering
        \includegraphics[height=2.5cm]{2_denser_surface_code/figs/end_cycles/3qubit_d4_endcycle.png}
        \hspace{1cm}
        \includegraphics[height=2.5cm]{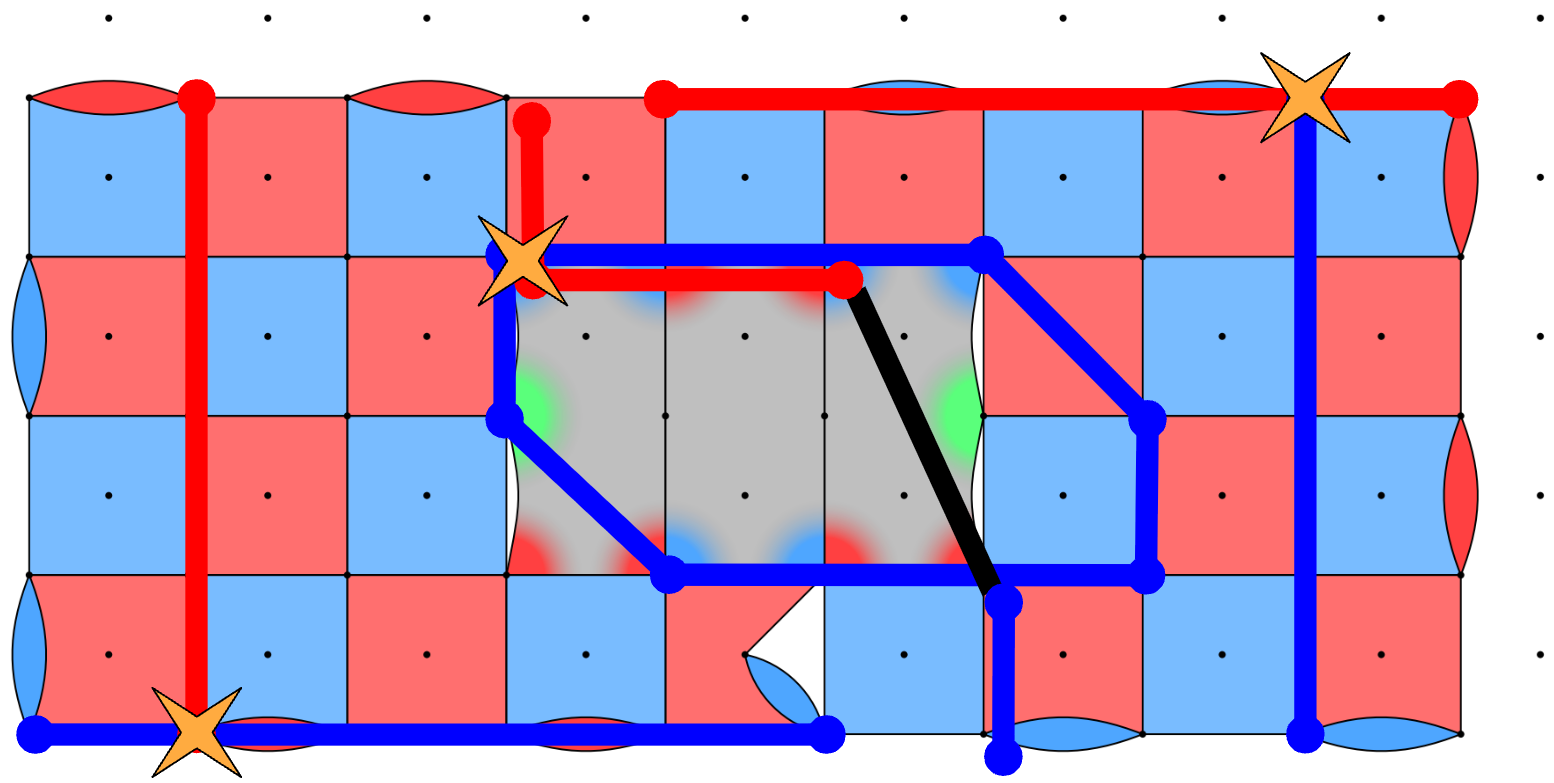}
        \hspace{1cm}
        \includegraphics[height=2.5cm]{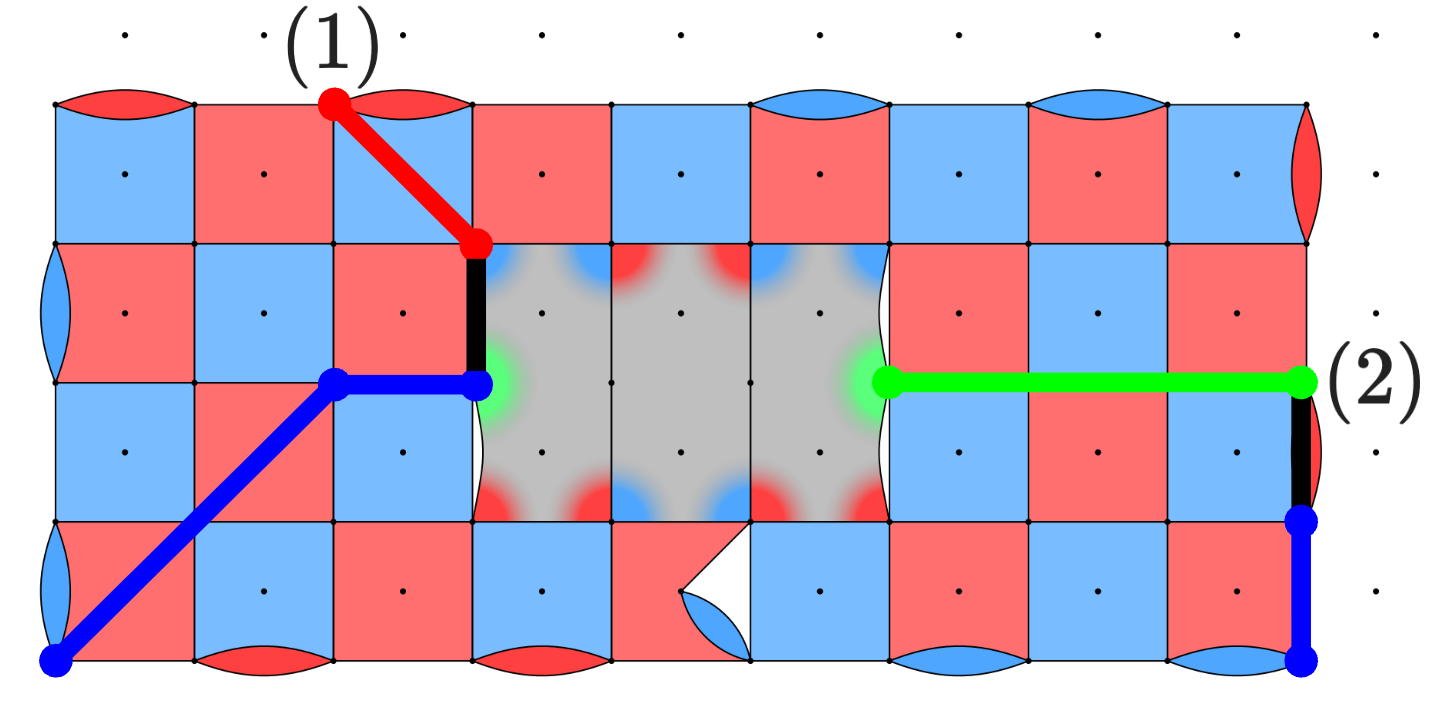}
    \caption{(Left) Rectangular surface code patch with two weight-5 twist defects featuring a Pauli-$Y$ (green) at the ends of a domain wall. 
    This end-cycle state has code parameters $[[49,3,5]]$. However, as discussed in~\cref{sec:3_qubit_patch}, both code-level and circuit-level distance are neither upper nor lower bounds on actual code performance.
    (Middle) One possible definition of logical operators that encode three logical qubits. Note the Pauli flip $X\leftrightarrow Z$ of the logical operator across the domain wall. That the $X$ and $Z$ chains form a single logical Pauli is indicated by a bold black line.
    (Right) Examples of other logical operators, or error mechanisms, that lead to undetectable logical errors as they anti-commute with logical qubit operators. 1) A mixed $X$-$Z$ error string that anti-commutes the loop logical $Z$ operator of the right figure. 2) A $Y$-$Z$ error string that anti-commutes with the loop and right logical $Z$ operators.
    We optimize the logical error rate by a precise positioning of twists and boundaries.
    }
    \label{fig:twist_patch}
\end{figure}

We accomplish this by a new quantum gate sequence for surface code twist defects.
Whereas the standard surface code~\cref{fig:compact_patch_main} is a CSS code with only weight $4$ $X$- and $Z$- stabilizers, codes with twist defects are not CSS.
Twist defect stabilizers have odd weight and contain an explicit $Y$ Pauli operator (see~\crefpos{fig:twist_patch}{left}) at the end-points of domain wall stabilizers in the bulk.
Similar to how point $Y$ defects also arise where $Z$ and $X$ boundaries intersect, these introduce new endpoints for various logical operators shown in~\crefpos{fig:twist_patch}{middle}.
In principle, this enables a greater density of logical operators in the form of loops and other Pauli strings mixing $\{X,Z,Y\}$ operators.
However, they also introduce new mechanisms for undetectable logical errors such as those shown in~\crefpos{fig:twist_patch}{right} that potentially reduce code distance and require larger patch dimensions to mitigate.
As logical operators flip $X\leftrightarrow Z$ across a domain wall, braiding and fusing twists enable a richer variety of logical operations such as the $S$ gate~\cite{Brown2017PokingHoles} or direct $YY$ measurements~\cite{Geher2025Tangling} between surface code patches.

However, known quantum gate sequences implementing twist defects typically suffer from some combination of distance-reducing hook errors, require higher-degree connectivity between physical qubits, require additional gate layers, or lose the polynomial-time decoding guarantee.
Although the utility of twist defects has long been known in principle, these practical limitations have resulted in their limited adoption in resource estimation architectures for compiling large-scale quantum algorithms to surface code operations, or in experimental demonstrations. 
Compared to prior art~\cref{tab:twist_defect_gate_sequence}, our proposed implementation simultaneously attains all desirable properties.
To avoid direct high-weight parity checks, we implement the twist defect stabilizer measurement via a subsystem code approach. 
We measure lower-weight gauge operators that commute with the stabilizer group but mutually anti-commute with one another. 
Although individual gauge measurement outcomes are random, their product yields the target high-weight stabilizer.
We take the combined parity of two gauge measurement outcomes to reliably reconstruct each twist defect and rectangle-shaped domain wall stabilizer.
We then add a smaller weight-$2$ domain wall to remove matchable interior boundaries.
This allows us to complete all stabilizer measurements within $6$ layers with the minimum hex-grid connectivity.
Moreover, we use an alternating cycle with nearly no distance loss from circuit-level hook errors and obtain good decoding performance in polynomial time by correlated MWPM.
\begin{table}
\begin{tabular}{c|c|c|c|c|c|c|c}
\hline\hline
\multirow{2}{*}{Twist defect implementation} & \multirow{2}{*}{Lattice}&{Surface code} &\multirow{2}{*}{Layers} &Qubit&\multicolumn{2}{c|}{Code distance loss at domain wall} & \multirow{2}{*}{Matchable}\\
&&bulk schedule&&Degree &Parallel&Perpendicular&\\
\hline
Hardware-unspecific~\cite{Litinski2018LatticeTwist} &Square& $N$-\reflectbox{Z}& $7$ & $\ge6$ & $-1$ &$0$& Yes
\\
Next-nearest-neighbor interactions~\cite{Chamberland2022Twist}  &Square&$N$-\reflectbox{Z}& $7$ & $\ge 6$ &$0$&$-1$ & Yes
\\
Nearest-neighbor interactions~\cite{Geher2025Tangling}  &Square&$N$-\reflectbox{Z}& $7$ & $4$ & $\times \frac{1}{2}$ & $0$ & No
\\
\hline
This work  &Hex&Alternating & $6$ & $3$ & $0$ &$-1$ (0 with $8$ layers) & Yes \\
\hline\hline
\end{tabular}
    \caption{Comparison of gate sequences that implement twist defects on a square lattice of qubits. Prior art uses standard `$N$-\reflectbox{Z}' gate sequence~\cite{Fowler2018LowOverhead} to measure surface code bulk stabilizers, leading to a gate depth at least $7$ layers (counting measurement and reset each as one layer) to control hook error propagation. Our construction is based on the alternating pattern~\cite{McEwen2023RelaxingHardware}, which loses at most one unit of distance in a $6$-layer gate sequence. 
    We note concurrent work that implements twist defects on a square grid with nearest neighbor interactions but with $>1$ distance loss: 
    In~\cite{Lensky2026Majorana}, the theory of dense memory with a design conceptually similar to~\cref{fig:twist_patch} and braiding gates based on the concept of dynamic Majorana codes~\cite{Lensky2023Majorana} is introduced and implemented using $>6$ layers to explicitly remove distance-halving $YY$ error chains discussed in~\cref{sec:dense_packing} and account for constraints of some superconducting architectures~\cite{Google2025SurfaceCodeThreshold}, including a requirement of measurement on all alternating columns; In~\cite{Hiari2026Sgate}, an $S$ gate occupying two square patches in a larger $4(d+2)^2$ bounding box is implemented using a mixture of \textsc{CXSwap} and \textsc{CX} gates.
    }
    \label{tab:twist_defect_gate_sequence}
\end{table}
\begin{figure}
\includegraphics[scale=0.8]{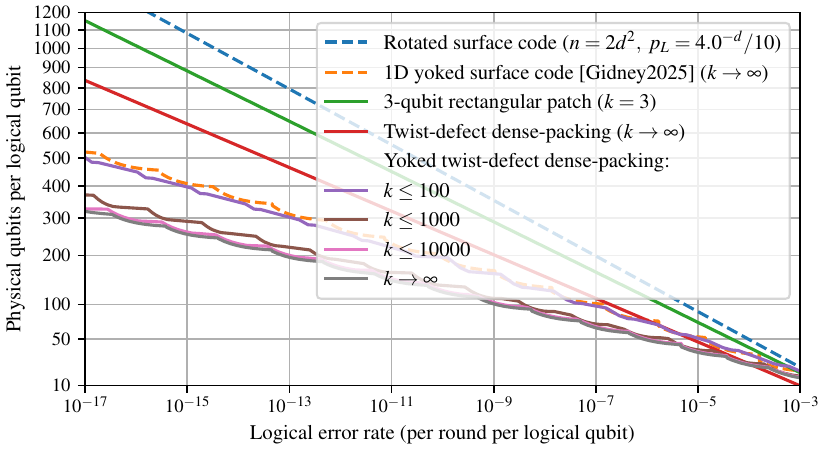}
\includegraphics[scale=0.8]{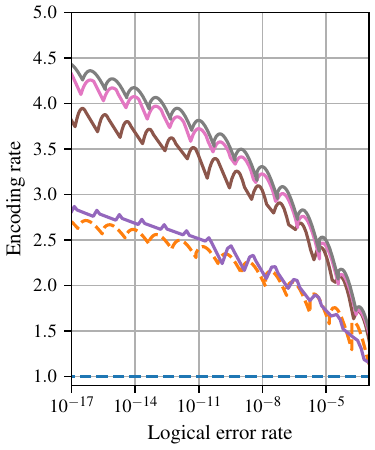}
    \caption{We present various dense packings of twist defects that enable up to a $4.5\times$ reduction in physical qubits needed to encode each logical qubit compared to the one-qubit rotated surface code patch. We use the empirical logical error rate $n_{\text{row}}T_{\text{yoke}}p^2_\text{dynamic}(d)$ to determine the minimum density of encoded logical qubits for finite-sized instances of up to $k\in\{100,1000,10000\}$ logical qubits. The density of 1D yoked surface codes is computed assuming the $2d^2$ footprint of this work instead of $2(d+1)^2$ in the original work.}
    \label{fig:dense_packing_density}
\end{figure}

The $6$-layer detectors that may be pieced together to form a twist defect at the endpoints of the domain wall are presented in~\cref{fig:twist_alternating_gate_sequence}.
Note that there are many equivalent realizations of these detecting regions, such as by conjugating various parts with single-qubit Clifford gates, which could substitute, for instance, the $Y$ measurement gates or replace all \textsc{CZ} gates with \textsc{CX} gates if desired.
We computed the graphlike distance of the twist defect including $YY$-error hyperedges and determined that two twist-defects $d$-rectangles apart (the example ~\cref{fig:twist_alternating_gate_sequence} is $5$-apart) has a distance $d$ $Y$-logical operator between the endpoints, and logical operators crossing the domain wall lose one unit of distance. 
In~\cref{fig:twist_8_cycles} we present an alternate $8$-layer construction with more detecting regions, leading to no distance reduction across the domain wall, that may be suitable for hardware architectures with low idling error rates, such as superconducting qubits.
Due to the symmetry of the `brick-work' pattern of the mid-cycle state at ticks $1$ and $4$, we may reverse the control and targets of the \textsc{CX} gates to translate the end-cycle of all stabilizers diagonally by one lattice spacing.
This trick was previously known for walking surface code patches~\cite{McEwen2023RelaxingHardware}, and we find that by alternating this flipping with a reflection about either the horizontal or vertical axis, we may also walk a twist defect right, left, up, or down by one lattice spacing every $2$ rounds, as shown in~\cref{fig:twist_walk_cycles} for the example of rightwards walking.
\begin{table}
\centering
\begin{tabular}{r|l|}
\hline\hline
&Detecting region and gate sequence\\
\hline
     \multirow{2}{*}{\rotatebox{90}{Twist defect (left)}}& \adjustbox{valign=t}{\includegraphics[width=0.9\linewidth]{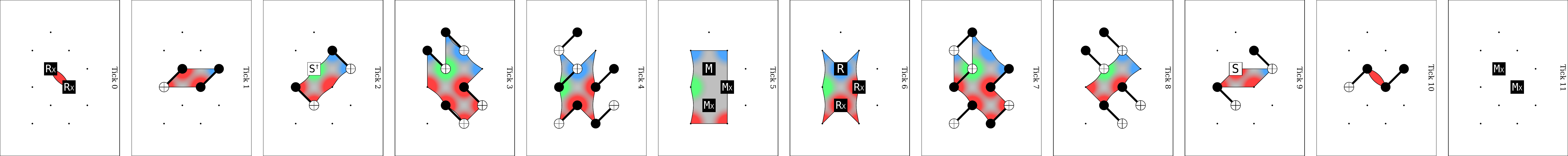}} \\
     & \adjustbox{valign=m}{\includegraphics[width=0.9\linewidth]{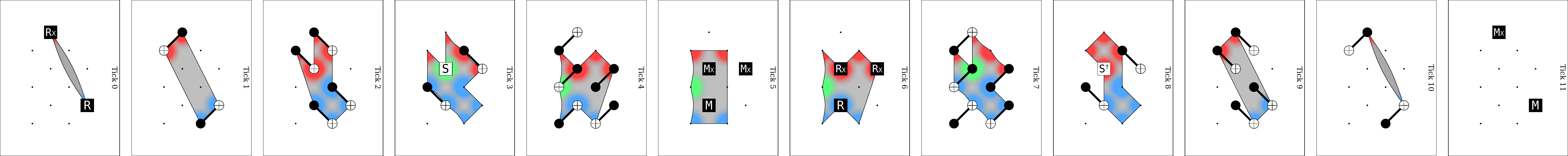}} \\
     \hline
     \multirow{2}{*}{\rotatebox{90}{Twist defect (right)}}& \adjustbox{valign=t}{\includegraphics[width=0.9\linewidth]{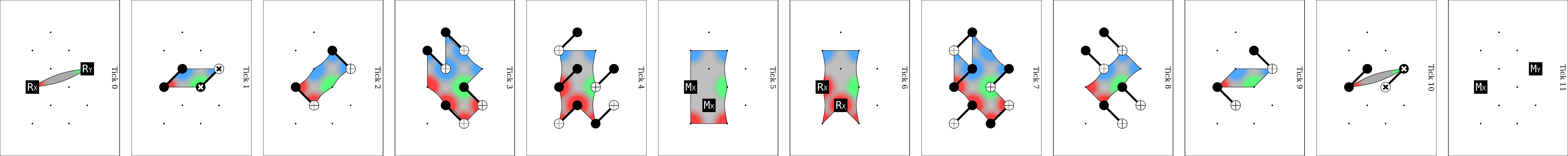}} \\
     & \adjustbox{valign=m}{\includegraphics[width=0.9\linewidth]{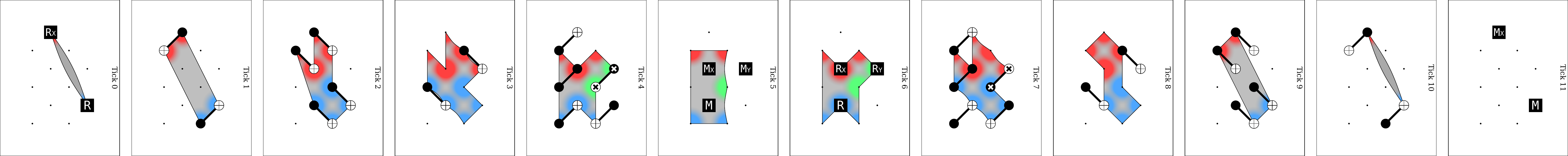}} \\
     \hline
     \multirow{3}{*}{\rotatebox{90}{Domain wall}}& \adjustbox{valign=m}{\includegraphics[width=0.9\linewidth]{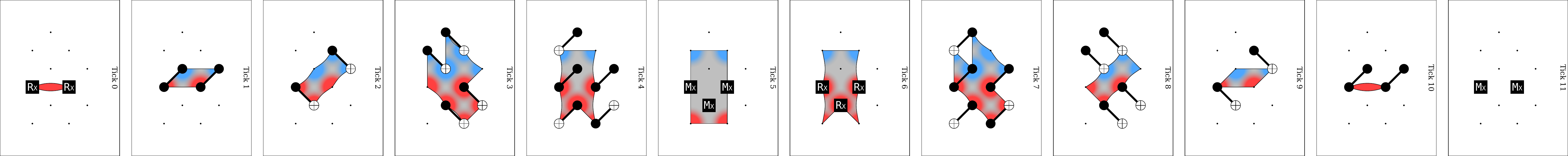}} \\
     & \adjustbox{valign=m}{\includegraphics[width=0.9\linewidth]{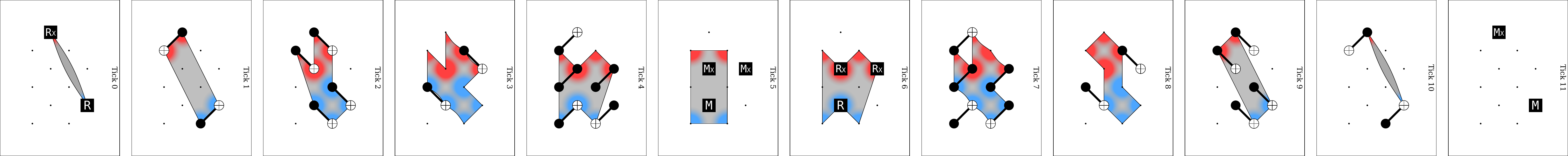}} \\
     & \adjustbox{valign=m}{\includegraphics[width=0.45\linewidth]{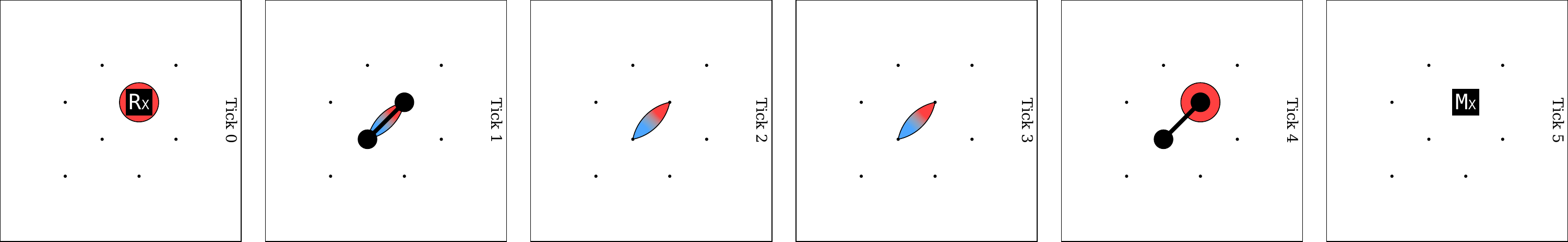}} \\
     \hline
     \rotatebox{90}{Full example}& \adjustbox{valign=m}{\includegraphics[width=0.9\linewidth]{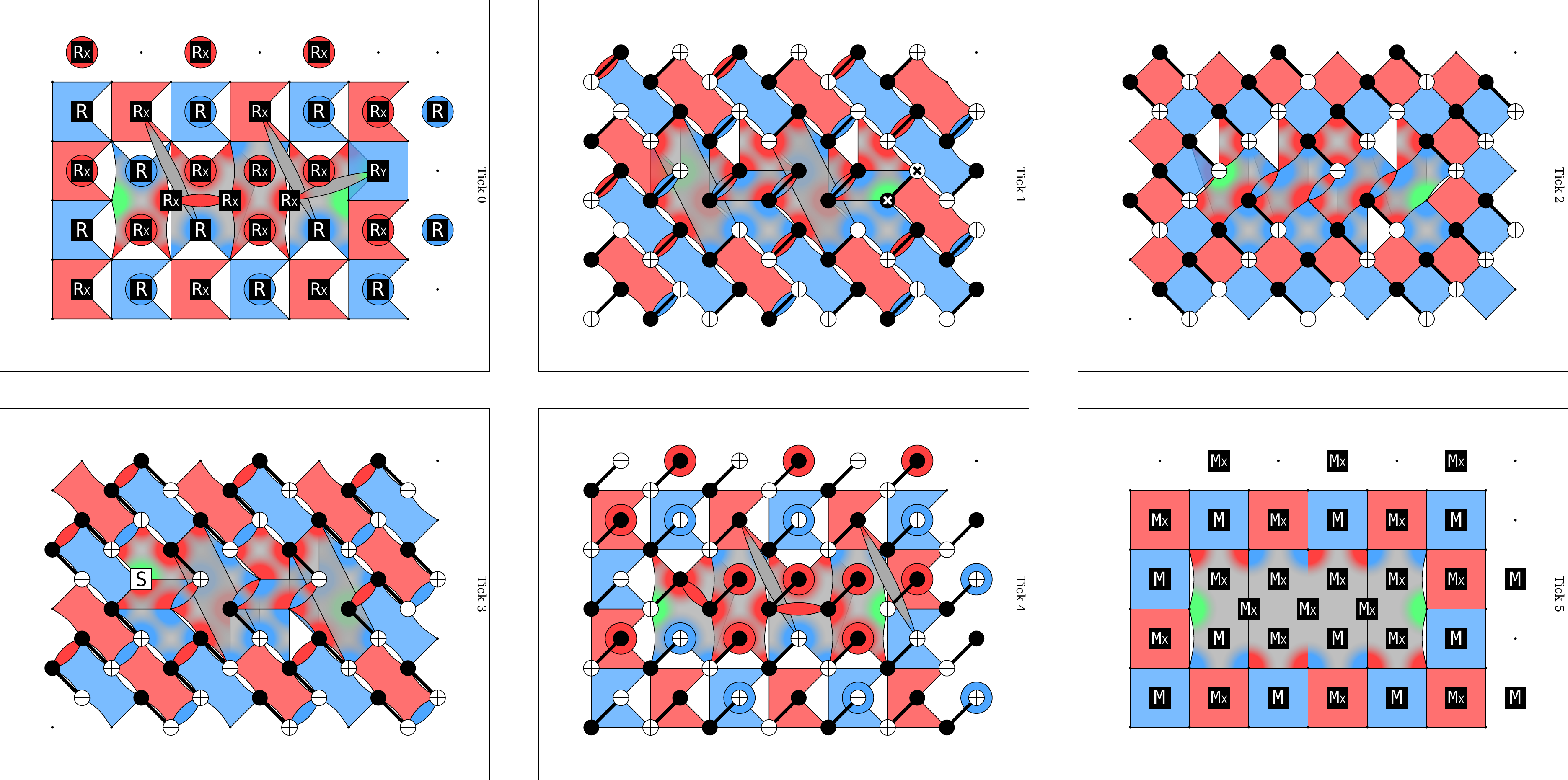}} \\
     \hline\hline
\end{tabular}
    \caption{Gate sequences implementing key detecting regions of the twist defect, where the first half is a depth-6 sequence $G$, and the second half is $G^{-1}$, where we define reset and measurement gates as inverses of each other. Note that \textsc{CXSwap} gates are used only to create a square-shaped end-cycle. 
    As the \textsc{CXSwap} gates conjugate a measurement and reset, and $\textsc{CXSwap}=\textsc{CX}\cdot\textsc{Swap}$, they can always be replaced by a single \textsc{CX}, and by shifting the location of the measurement and reset diagonally by one step.
    We also provide in~\cref{fig:domain_wall_bend_gate_sequence} gate sequences that implement vertically-oriented domain walls with all possible right-angle bends.}
    \label{fig:twist_alternating_gate_sequence}
\end{table}

By careful placement of twist defects, we greatly reduce the number of physical qubits encoding each logical qubit in the surface code as shown in~\cref{fig:dense_packing_density}.
We validate the logical error rate of our approach with a variety of circuit-noise-level benchmarks implemented in Stim~\cite{Gidney2021Stim}. 
We propose three different dense-memory configurations that we compared with other $2$D planar high-rate  alternatives in~\cref{tab:twist_defect_results}: 
\begin{enumerate}
    \item A rectangular patch encoding three logical qubits in a footprint of two $2d^2$-qubit rotated surface code patches with $\frac{3}{2}\times$ the encoding rate of a rotated surface code patch.
\item A dense-packing of twist defects with an asymptotic density of one logical qubit per $d(d-1)$ physical qubits, or $2\times$ the encoding rate of a rotated surface code patch.
\item A concatenation of each column of the dense-packing with a $[[n_\text{row},n_\text{row}-2,2]]$ quantum parity check code, which roughly doubles the code distance, leading up to $4.5\times$ the encoding rate of a rotated surface code patch.
\end{enumerate}
Our twist defect and domain wall gate sequences also enable space-efficient versions of other common logical operations, including spatial Hadamard, and an almost in-place $S$ gate with a footprint of $\approx 1.25$ patches and spacetime volume of $\approx 1.25 d^3$.

In~\cref{sec:dense_packing}, we describe our dense arrangement of twist defects, and how they enable new space-efficient patch operations.
In~\cref{sec:compute_dense_packing}, we describe how quantum computation may be executed on densely packed logical qubits through a sequence of steps that load and unload memory qubits into regular surface code patches.
In~\cref{sec:dense_packing_yoking}, we describe how we may concatenate our dense memory with a quantum parity check code to roughly double the density of logical qubits.

\subsection{Twist-defect dense packing}\label{sec:dense_packing}
Twist defects provide end points for new patterns of logical errors.
For instance~\crefpos{fig:twist_patch}{right} highlights how in the end-cycle state, $X$ and $Z$ errors propagate diagonally, with distance scaling like the $L_\infty$ norm, whereas $Y$ errors propagate horizontally or vertically, with distance scaling like the $L_1$ norm of a vector in~\cref{fig:twist_patch}, where lattice points on a horizontal or vertical line are unit distance apart.
These error chains will generate an undetectable logical error on any anti-commuting logical operators that crosses their path.
Hence, maintaining code distance requires careful placement of twist defects sufficiently far from the boundaries of a surface code patch, and from each other to avoid undetectable error chains.

\begin{figure}
    \centering
    \includegraphics[width=0.2\linewidth, valign=m]{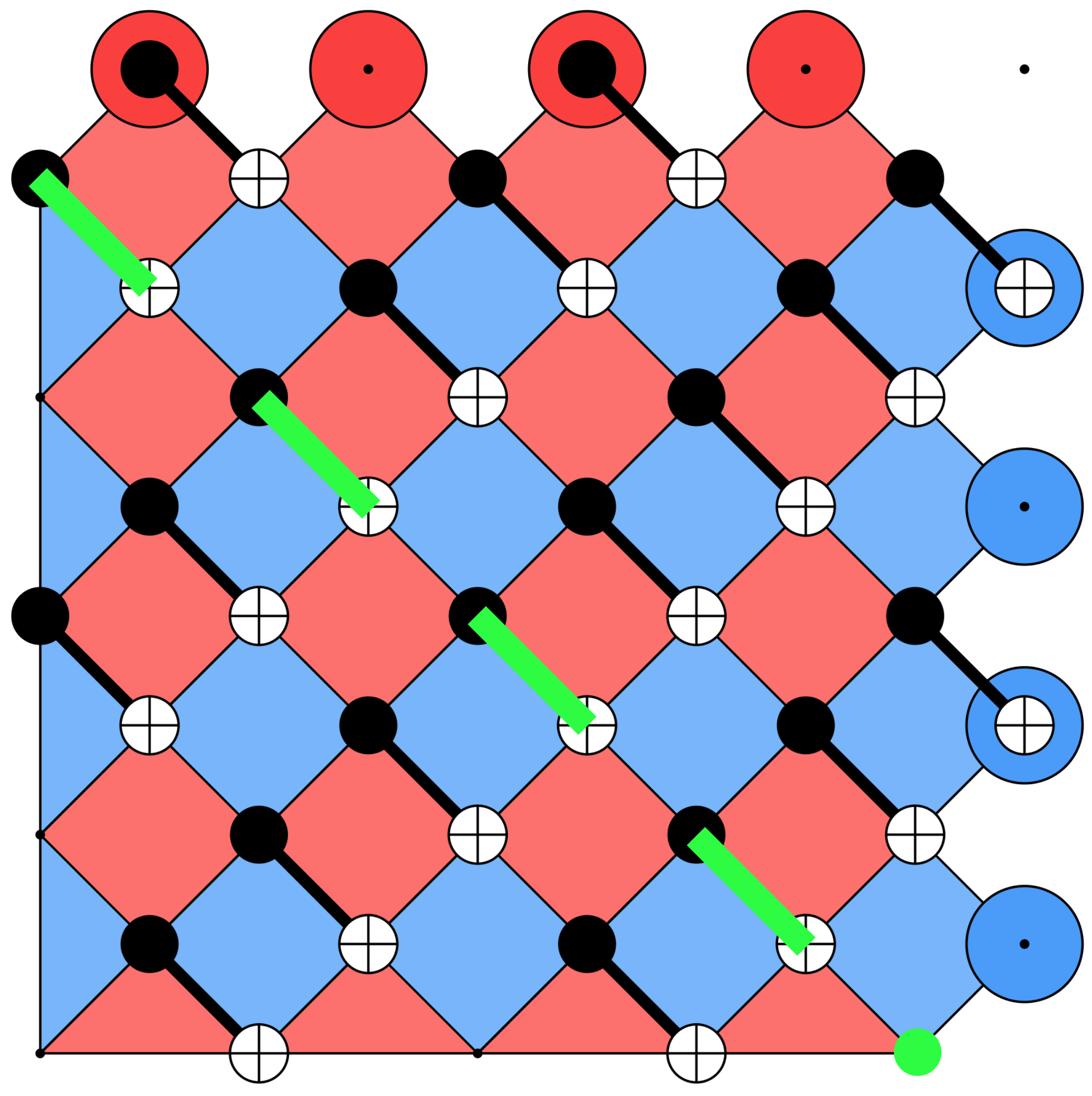}
    \hspace{2cm}
    \includegraphics[width=0.2\linewidth, valign=m]{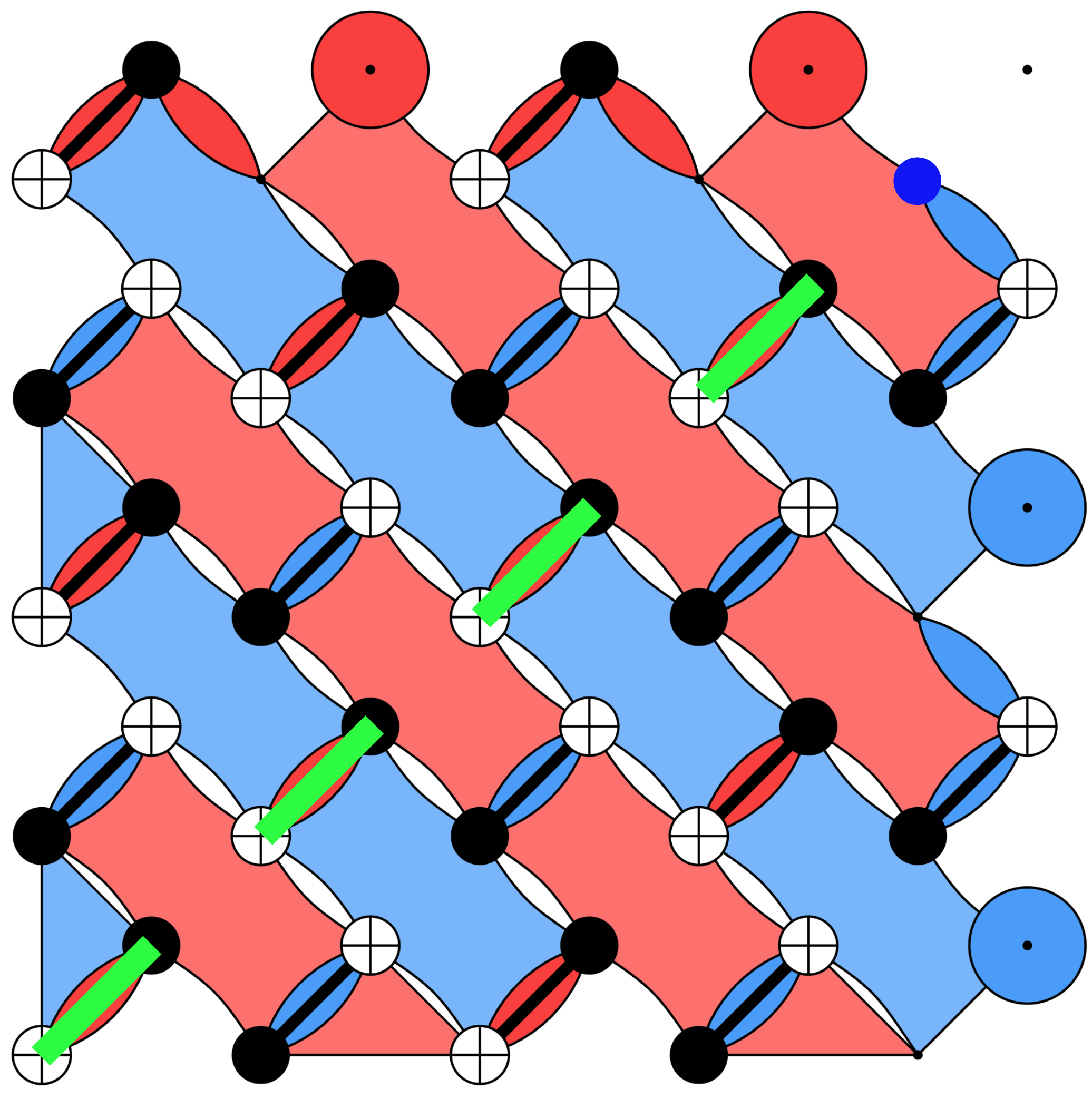}
    \caption{$YY$ errors (green lines) from the two-qubit depolarizing error channel that create a diagonally-propagating distance $d$ logical $Y$ error. However, the logical error probability of this minimum-distance error mechanism is suppressed compared to $X$ or $Z$ error chains in the benchmark~\crefpos{fig:square_patch_alt}{left} as it requires this very specific low-entropy configuration of errors. Moreover, in a physical noise model, each $YY$ error also has small probability of occurring, for instance, $p/15$ in a depolarizing channel.
    }
    \label{fig:y_errors}
\end{figure}

However, error chains in the end-cycle state only reflect possible single-qubit errors, and do not account for other error paths due to hook errors from the noisy gate sequence measuring the stabilizer.
For instance, a computation of the graphlike distance without decomposing hyperedges reveals the presence of diagonally propagating $Y$ hook errors, with distance scaling like the $L_\infty$ norm.
As shown in~\cref{fig:y_errors}, these errors occur from $YY$ hook errors from the two-qubit depolarizing error channel in specific layers of the gate sequence.
Such errors halve the circuit-level distance compared to the $L_1$ norm, which would require the placement of twist defects so far apart that any benefit in spatial packing is eliminated.
However, graphlike distances represent neither an upper bound nor a lower bound on the actual performance of a code.
As shown in~\cref{fig:square_patch_alt} and in prior work~\cite{Gidney2025Yoked}, the probability of a logical $Y$ error occurring is suppressed to almost that of both a logical $X$ and $Z$ error occurring. 
The reduced impact of these $YY$ hook errors can be explained by the combinatorial suppression of requiring a very specific distribution that each also occurs with a lower probability, like $p/15$ in the uniform depolarizing model, than all other single-qubit errors.
We exploit the larger effective distance, scaling close to the $L_1$ norm, of diagonally propagating $Y$ errors to optimize the placement of twist defects.
We also exploit the asymmetry in logical error rate between static ($X$ top boundary) and dynamically updated ($Z$ top boundary) logical operators to optimize the placement of boundary $Y$ defects.

\subsubsection{3-qubit rectangular patch}\label{sec:3_qubit_patch}
\begin{figure}
    \centering
    \includegraphics[width=0.5\linewidth, valign=m]{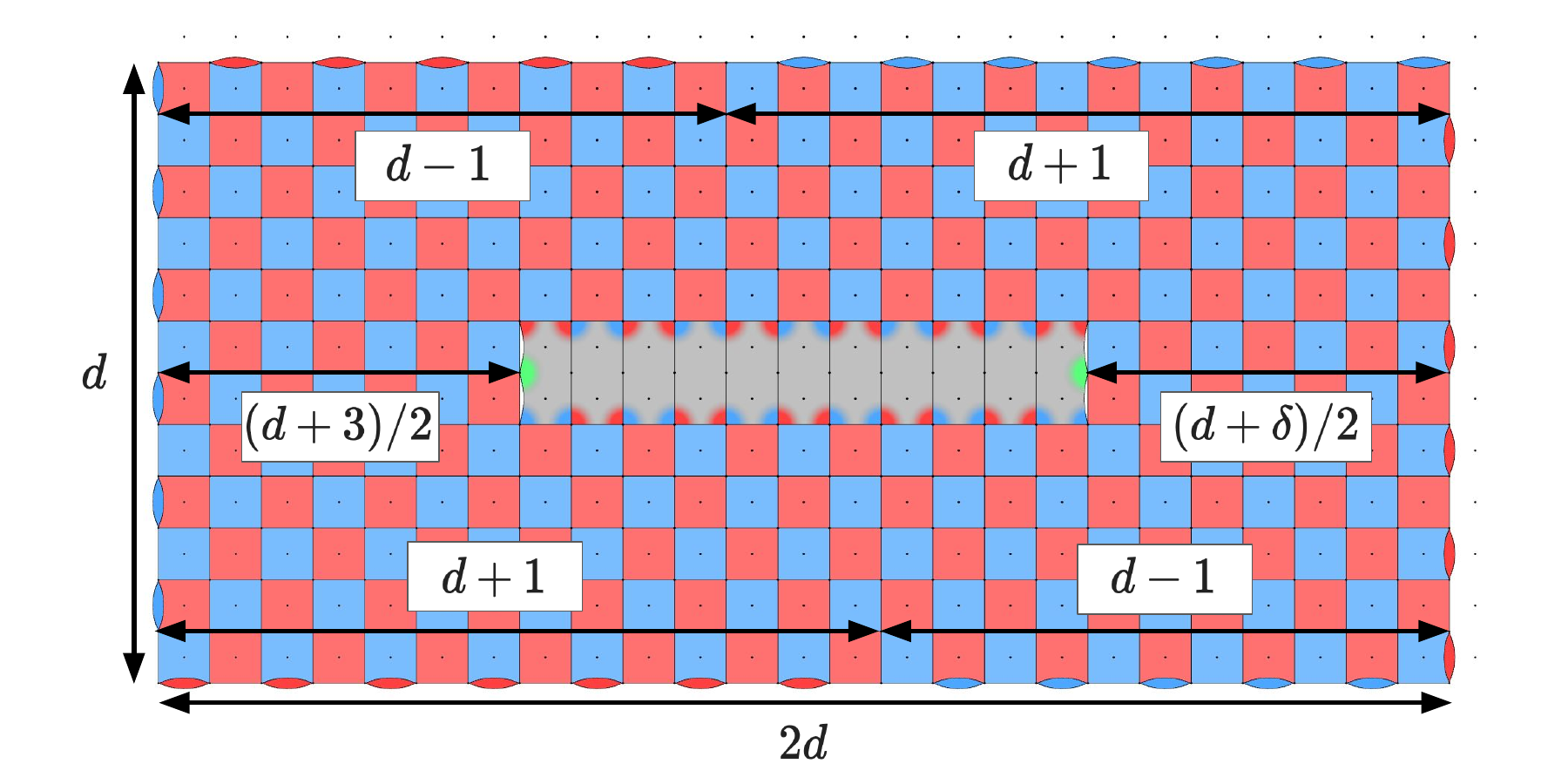}
    \hspace{1cm}
    \includegraphics[scale=0.8, valign=m]{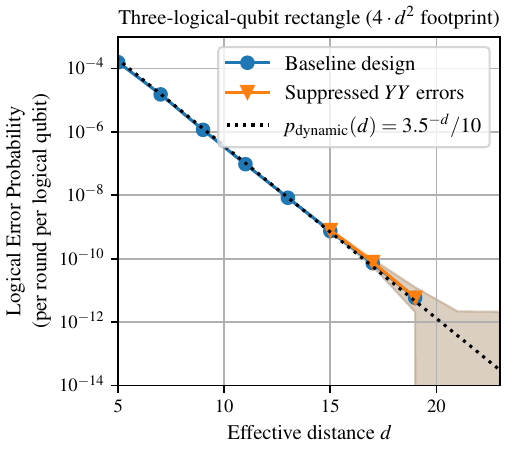}
    \caption{(Left) The dimensions of the three-qubit patch. $\delta=3$ in the `baseline' design and $\delta=5$ in `suppressed $YY$ errors' design. See~\cref{fig:straight_y_defects} for gate sequences implementing the horizontal boundary $Y$ defects. (Right) Simulated logical error rate of these designs in a $10^{-3}$ one- and two- qubit uniform depolarizing noise model.}
    \label{fig:3_qubit_error}
\end{figure}

\begin{figure}
    \centering
    \includegraphics[width=0.9\linewidth]{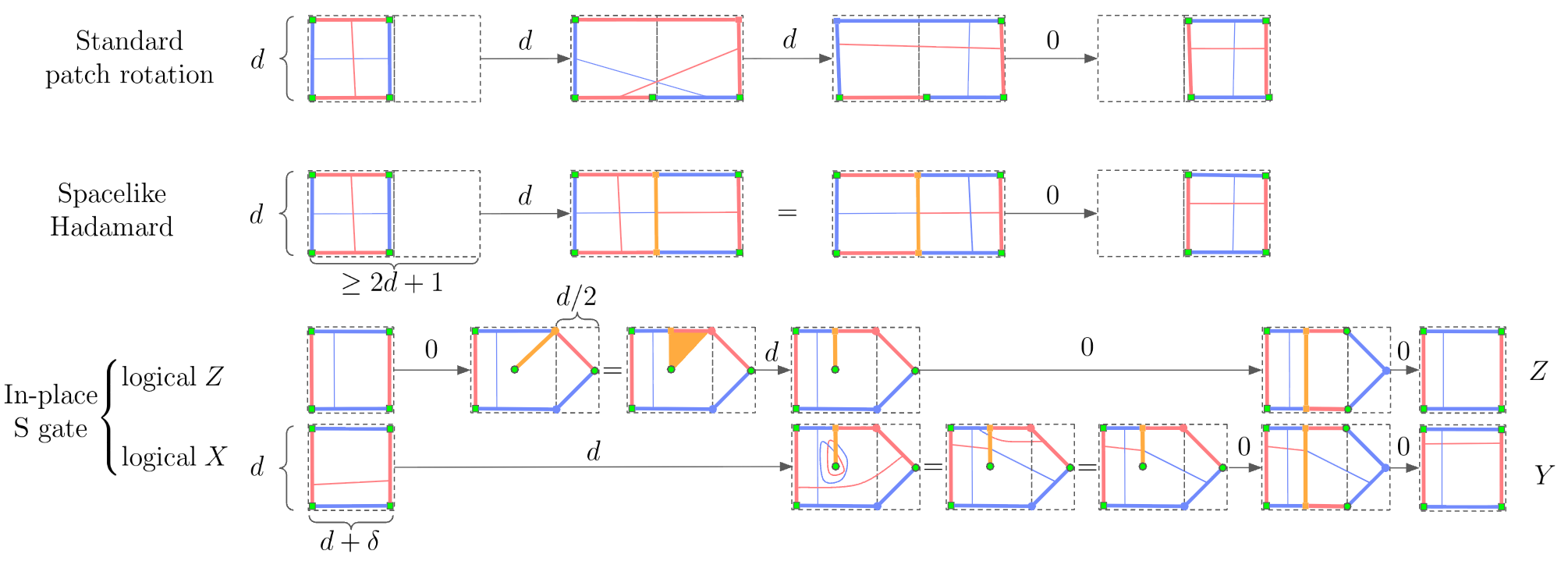}
    \caption{Defect diagrams for (top) a standard patch rotation like in prior art, (middle) a spatial Hadamard,
    and (bottom) an almost in-place $S$ gate, where the first step involves moving a corner $Y$ defect diagonally to the middle, which is equivalent to moving it horizontally from the right combined with single triangle-shaped time-like domain wall (orange) implemented by layer of transversal Hadamard gates, followed by holding for $d$ cycles to make measurement fault-tolerant. 
    Note that whenever $Y$ defects are moved, they create a spacelike line of $Y$ defects within a single cycle.
    Whenever a horizontal or vertical domain wall is merged with a boundary, such as in the last step, it is implicit that there is a similar triangle-shaped time-like domain wall to ensure that the spacelike $Y$ defect line is diagonal.
    The presented patch dimensions ensure that all possible $Y$ error chains have a $L_1$ distance of at least $d$.
    We leave a detailed benchmark of these primitives to future work, but the patch dimensions $d\times (d+\delta)$, where $\delta=1$ follow from taking the left half of the rectangle in~\crefpos{fig:3_qubit_error}{left}. Larger distances $[19,25]$ may require increasing $\delta$ to $2$ or $3$.
    Note that spatial Hadamard requires padding to width $2d+1$ as our domain wall implementation in~\cref{fig:twist_alternating_gate_sequence} is $2$ units wide. In our later resource estimates, we assume that wherever they are used with no padding, there is an equivalent compilation using timelike transversal Hadamards.}
    \label{fig:in-place-gates}
\end{figure}

We now demonstrate the design of a rectangular patch that encodes $3$ logical qubits to effective distance $d$ in a footprint containing $4d^2$ data and measurement qubits.
Our procedure highlights the importance of precise $Y$-defect placement to control the distance of different mechanisms, leading to a highly efficient packing.
We consider two designs: A baseline design with a logical error rate that matches that of the $Z$-top surface code patch for code distances up to $19$, and a modified design that may be relevant for larger code distances.
As different error mechanisms exhibit different scaling constants with $d$: $p_\text{static}(d)=\frac{1}{10}4^{-d}$ for static $X/Z$ logical operators, $p_\text{dynamic}(d)=\frac{1}{10}3.5^{-d}$ for dynamically updated $X/Z$ logical operators, and $\frac{1}{70}8.5^{-d}$ for $Y$ errors, we express our results in terms of an effective distance $d$, which is defined to simply be the height of the rectangular patch, identical to that of a distance-$d$ one-qubit rotated surface code patch.
The dimensions of the baseline design are chosen as a function of $d$ following heuristics based on the empirical logical error rates in~\cref{fig:square_patch_alt}.
\begin{enumerate}
    \item The distance of $Y$ logical operators propagating diagonally in the `$\backslash$' direction scales with some constant times the $L_1$ norm. This constant is slightly less than one, so we choose the $L_1$ distance between $Y$ defects in the `$\backslash$' direction to be $d+1$.
    \item The distance of $Y$ logical operators propagating diagonally in the `$/$' direction scales with some constant times the $L_1$ norm. This constant is closer to one, so we choose the $L_1$ distance between $Y$ defects in the `$/$' direction to be $d-2$.
    \item The vertical $X$-component or horizontal $Z$-component of logical operators scale better, like $4^{-d}$.
    Hence distances associated with these logical operators could be made slightly shorter, like $d-1$.
    \item The $Z$-component or horizontal $X$ component of logical operators scale worse like $3.5^{-d}$. 
    Hence distances associated with these logical operators could be made slightly longer, like $d+1$.
\end{enumerate}
We emphasize that these are heuristics, and they will lead to different logical qubits within the patch experiencing different rates of correlated logical errors.
These heuristics lead to the patch dimensions indicated in~\crefpos{fig:3_qubit_error}{left}, and the average logical error rate per logical qubit benchmarked in~\cref{fig:3_qubit_error}  is in very good agreement with the $Z$-top orientation of the one-qubit surface code patch.
We have only optimized the rectangular patch for odd $d$, and later, where applicable, we will assume that there exists a design for even $d$ with the same overall dimensions of $2d\times d$ and the same logical error rate.
By taking only the left half of each rectangular patch, we also obtain in~\cref{fig:in-place-gates} the patch dimensions to $S$ gates almost in-place as shown in~\cref{fig:in-place-gates}.

\begin{figure}
    \centering
    \includegraphics[width=0.3\linewidth]{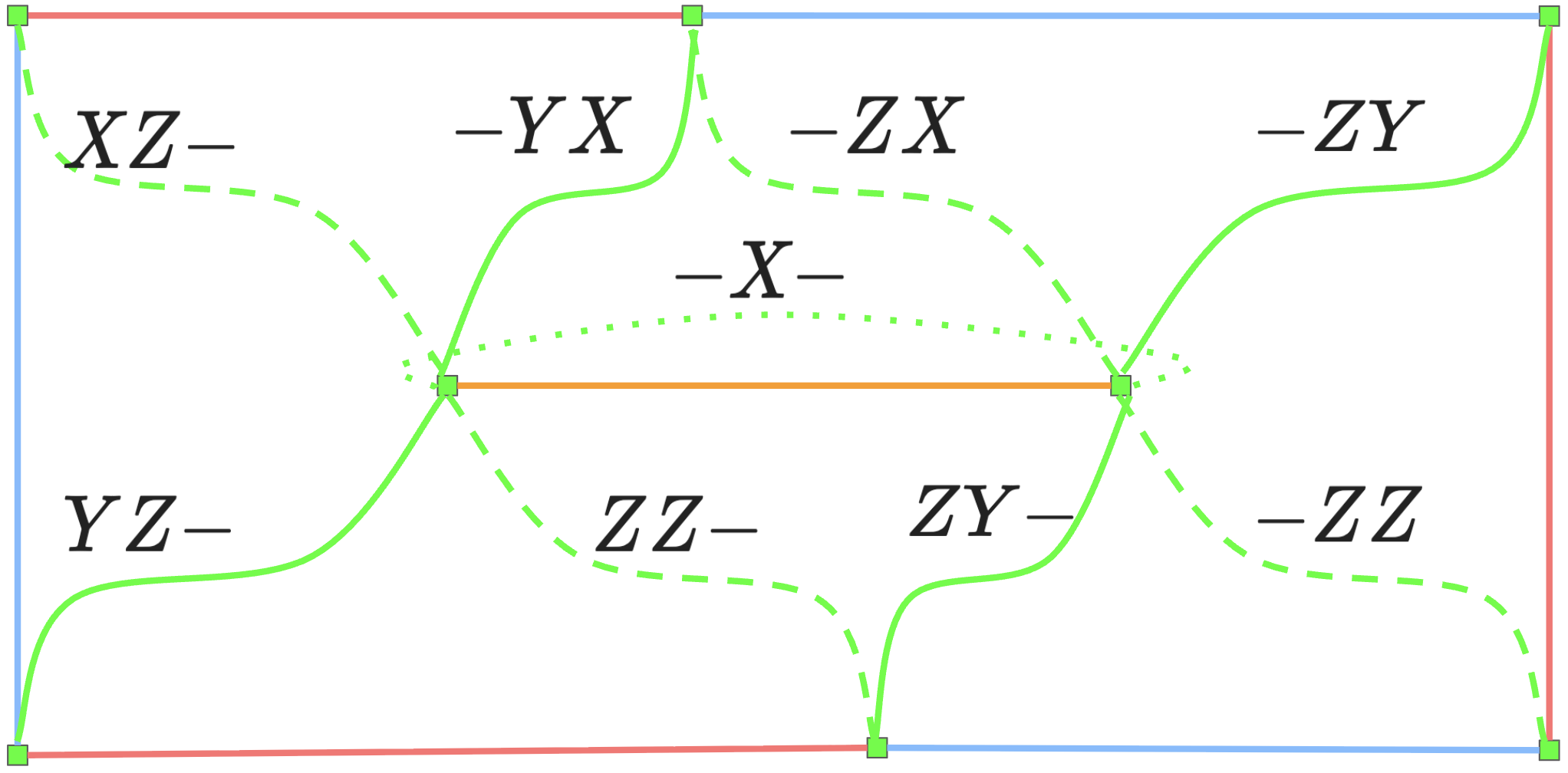}
    \hspace{2cm}
     \includegraphics[width=0.3\linewidth]{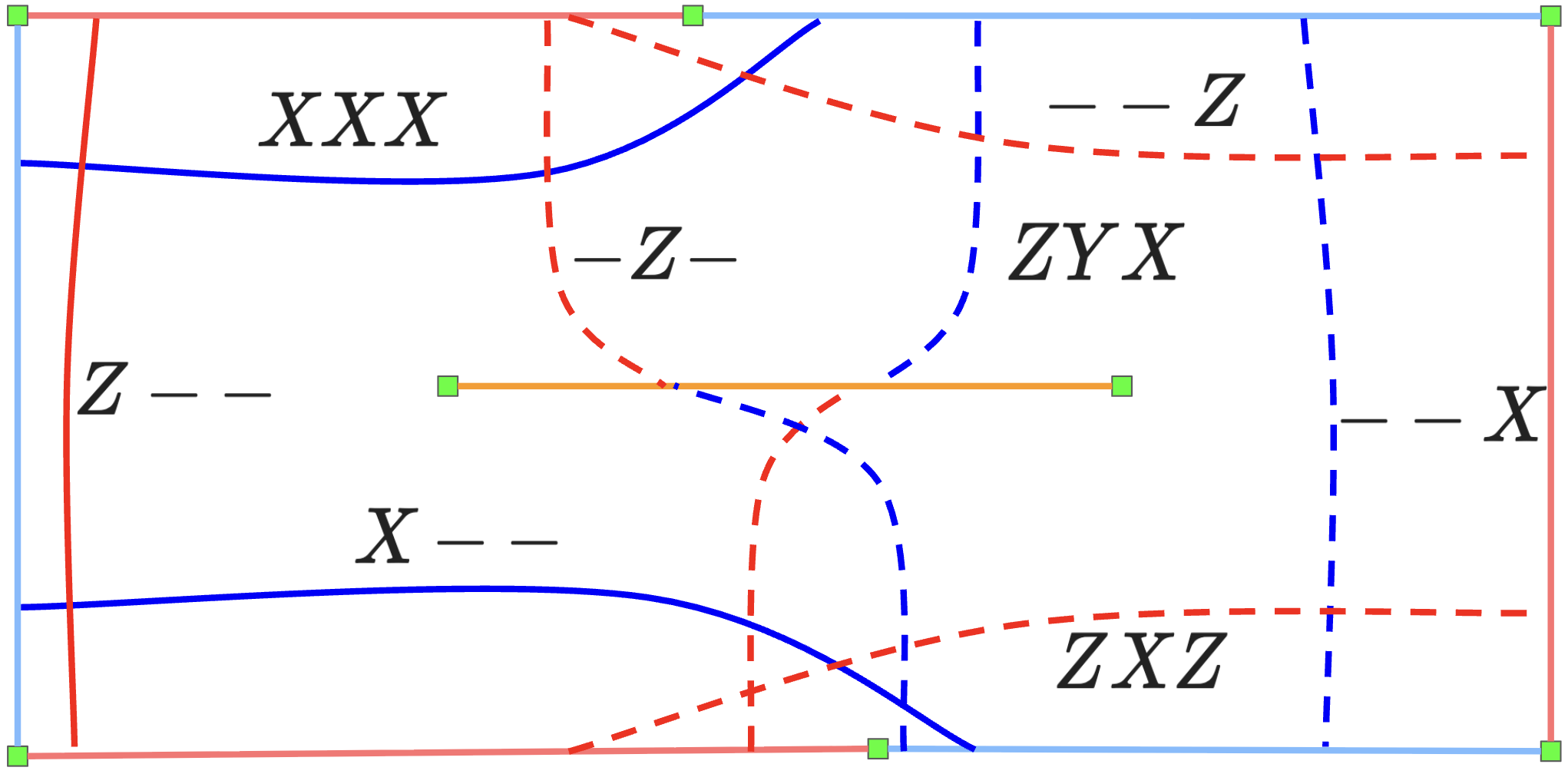}
    \vspace{0.5cm}
    \includegraphics[scale=0.8]{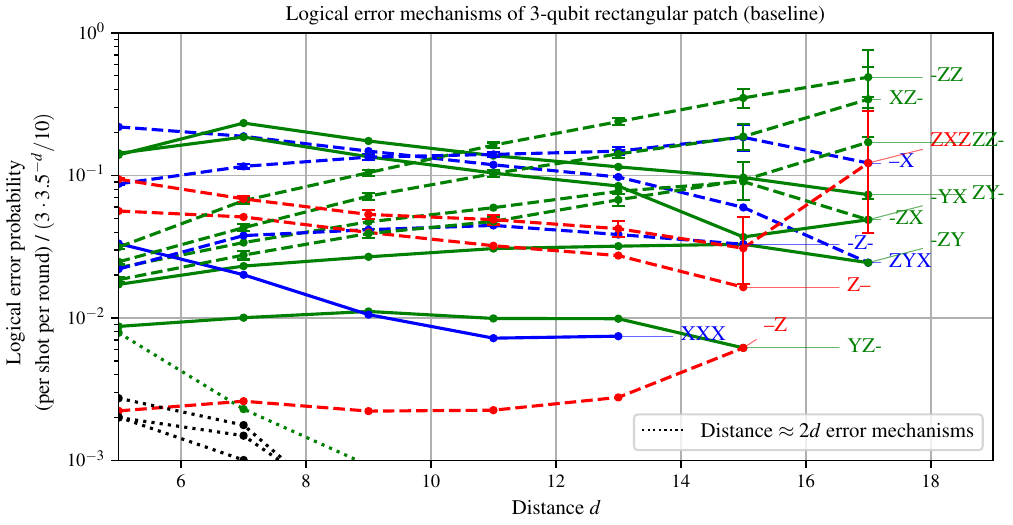}
     \includegraphics[scale=0.8]{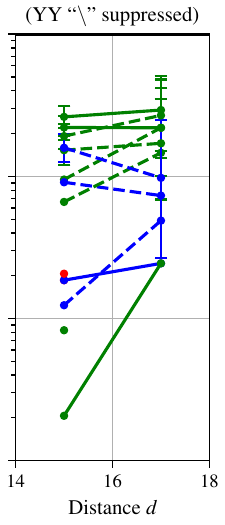}
    \vspace{-.5cm}
    \caption{(Top) All distance $\approx d$ logical operators, or error mechanisms with some degeneracies, labeled by the logical Pauli error they cause on logical qubits defined in~\cref{fig:twist_patch}. Solid lines represent error chain orientations with better error scaling than the orientations of the same color represented by dashed lines. (Bottom) Probability of the different error mechanisms occurring, normalized by the fitted probability $\frac{3}{10}3.5^{-d}$ from~\cref{fig:3_qubit_error} of any logical error occurring in the $3$-qubit rectangle. Error bars of dominant mechanisms at large $d$ are plotted.}
    \label{fig:3_qubit_error_mechanism}
\end{figure}

We may further optimize the location of the point $Y$ defects by evaluating the probability of these types of error mechanisms.
In~\crefpos{fig:3_qubit_error_mechanism}{top}, we enumerate all distinct undetectable distance $\approx d$ error mechanisms, and the corresponding logical Pauli errors they cause.
In~\crefpos{fig:3_qubit_error_mechanism}{bottom}, we evaluate the probability of these error mechanisms, which is consistent with our patch dimensions heuristics: The error scaling of dynamically updated logical operators (dashed lines) is noticeably worse than the static logical operators (solid lines), and $Y$ errors propagating from the topleft to bottom indeed have distance a small fraction less than the $L_1$ distance between $Y$ defects.
For distances above $17$ up to $25$, it should be sufficient to increase the overall patch width by one or two additional units.
However, considering that some other error mechanisms decay in relevance with distance, it may be possible to instead move $Y$ defects in a way that suppresses the bad $Y$ errors and amplifies all other errors without changing patch dimensions. 
For example, the `$YY$ suppressed' design increases $\delta\rightarrow \delta+2$, which leads to a more balanced distribution of $Y$ errors whilst maintaining at least the same error rate as the baseline design.
A key difficulty in such a study at large $d$ is the exponentially growing large number of Monte-Carlo samples required to gather actionable statistics.

\subsubsection{A dense packing of twist defects}
\begin{figure}
    \centering
    \begin{tabular}{ccc}
        \begin{tabular}[t]{@{}c@{}}
            \includegraphics[height=2cm, valign=t]{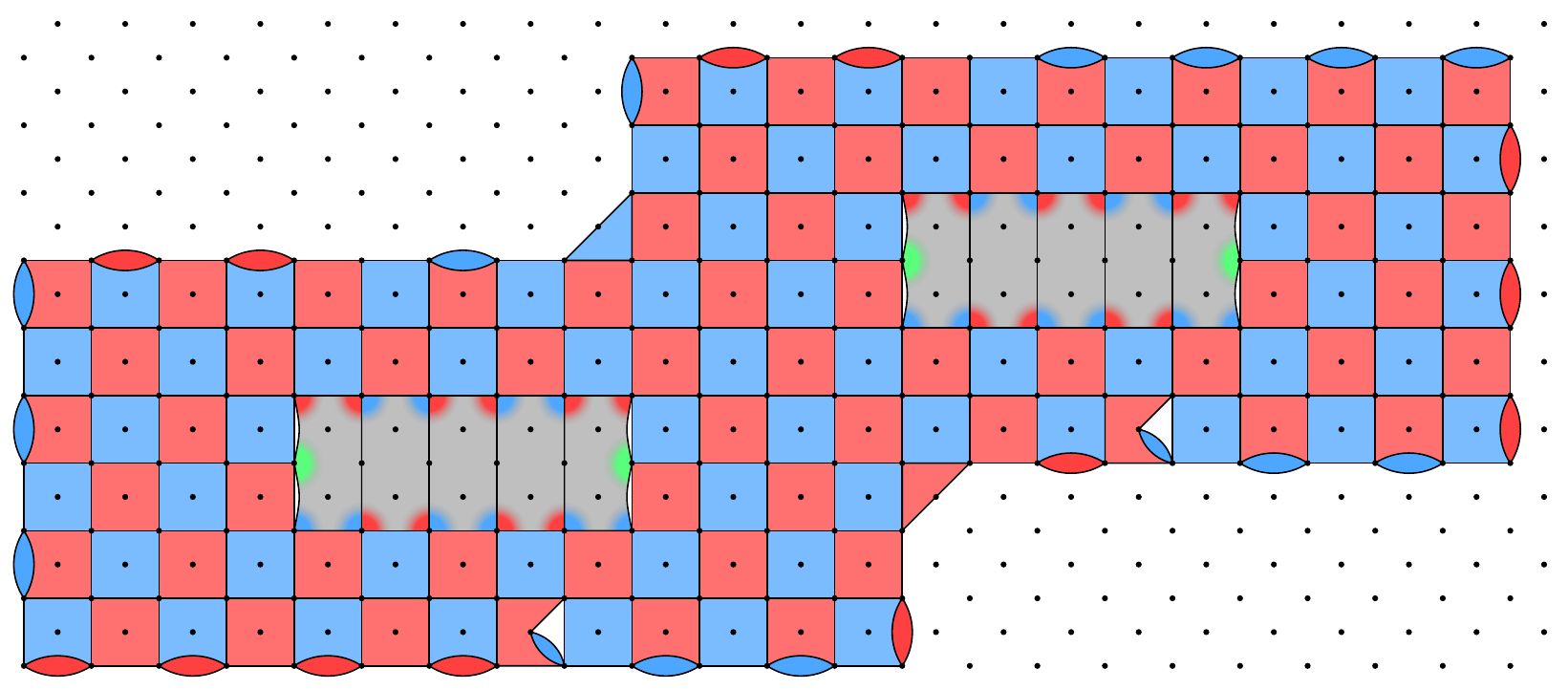} \\
            \\[-0.5em] 
            \includegraphics[height=2cm, valign=t]{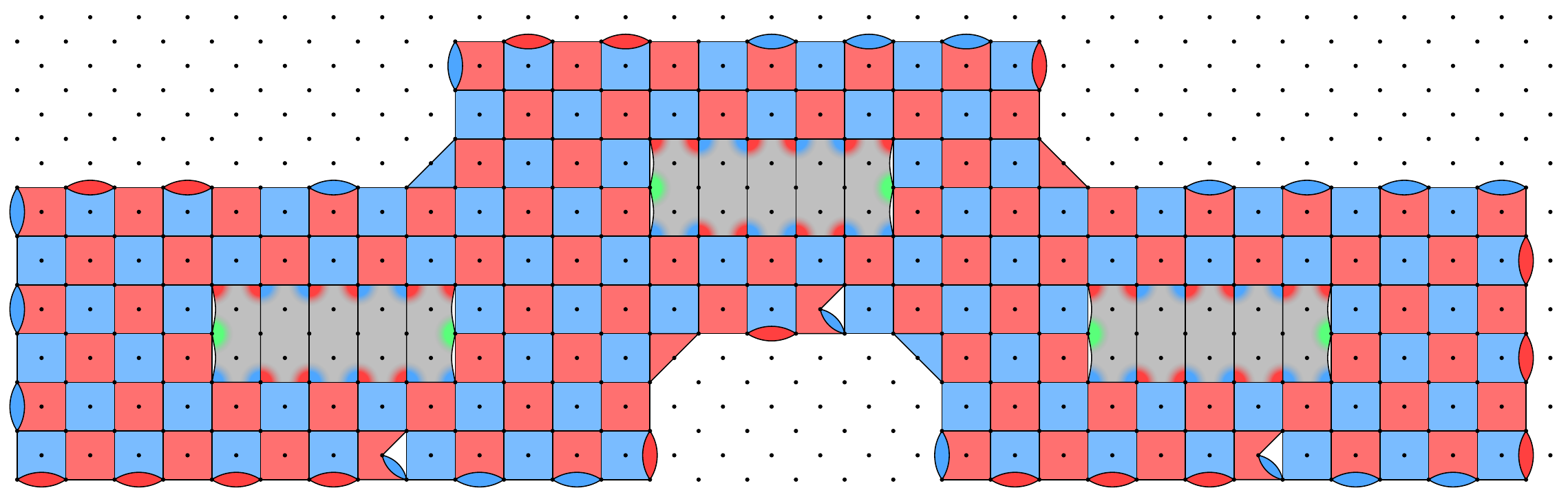}
        \end{tabular} 
        & 
        \includegraphics[height=3.333cm, valign=t]{2_denser_surface_code/figs/end_cycles/dense_d6_r2_c2_show_end_cycle.pdf} 
        & 
        \includegraphics[height=3.333cm, valign=t]{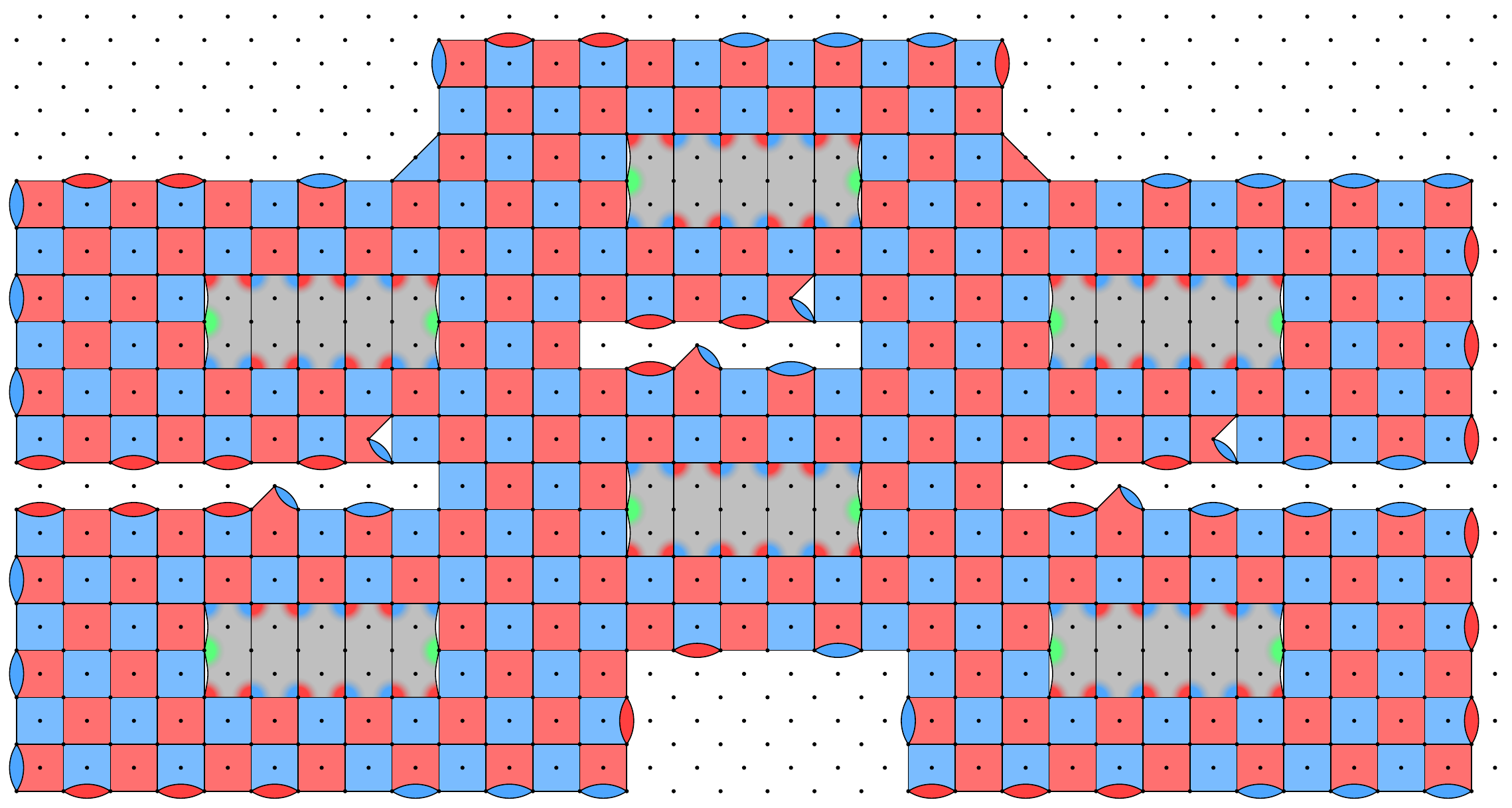} 
    \end{tabular}
    
    \vspace{-.5em} 
    
    \includegraphics[width=0.4\linewidth, valign=m]{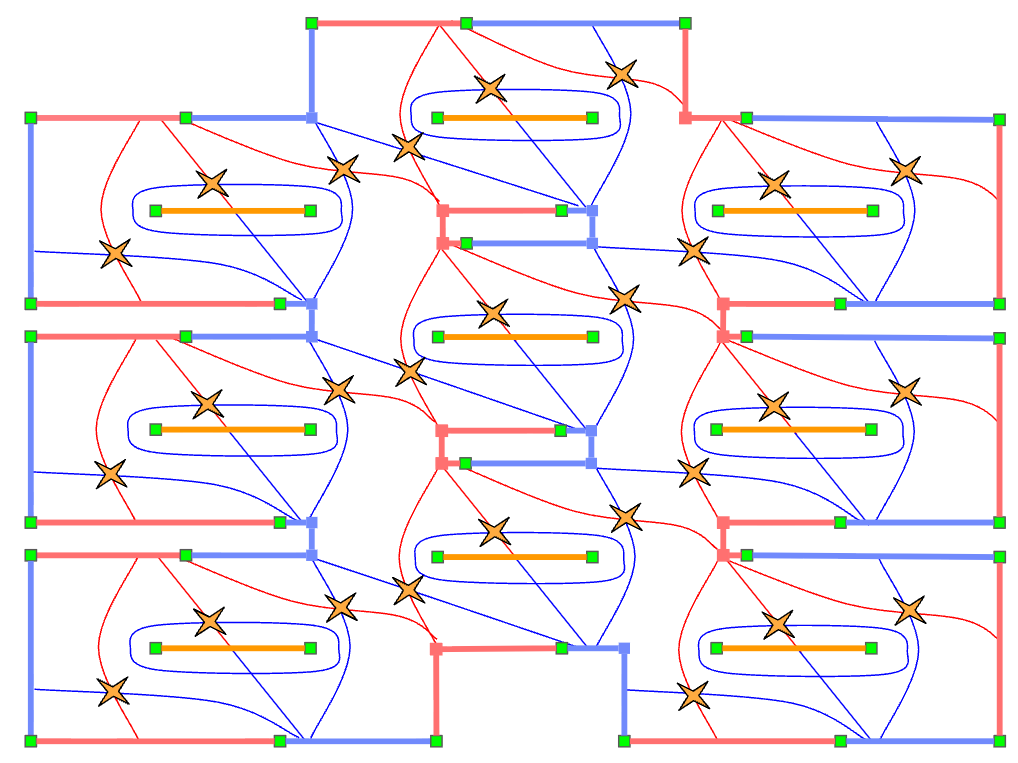}
    \hspace{1cm}
    \includegraphics[scale=0.8, valign=m]{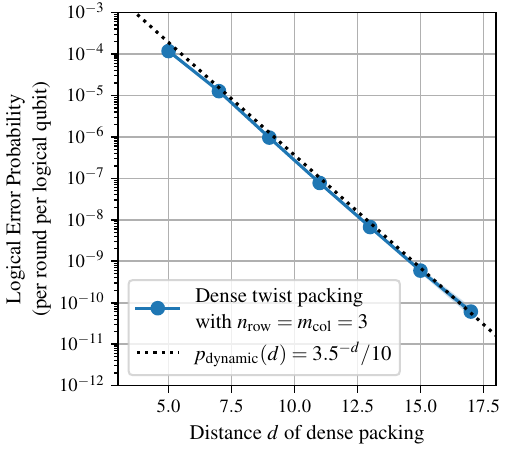}
    
    \caption{(Top) $r$ rows and $c$ cols of $3$-qubit rectangles may be merged in horizontally and vertically to form a single patch of densely packed twist defects with interior boundaries encoding $3rc$ logical qubits. 
    Shown are examples for $(r,c)\in\{(1,2),(1,3),(2,2),(2,3)\}$.
    (Bottom left) The logical operators defining the $27$ logical qubits of an $(r,c)=(3,3)$ patch and (bottom right) the logical error rate under a $10^{-3}$ uniform depolarizing noise model.}
    \label{fig:dense_packing}
\end{figure}

We may create an even denser packing of logical qubits by merging the boundaries of a $n_\text{row}\times m_{\text{col}}$ grid of 3-qubit rectangular patches encoding $3n_\text{row}m_{\text{col}}$ logical qubits, as shown in~\cref{fig:dense_packing}.
Using the same dimensions as the baseline design, the bounding box of this dense packing
\begin{align}\label{eq:footprint}
\text{width} = \frac{1}{2}\left(d+3\right)+m_\text{col}\frac{3}{2}\left(d-1\right),
\quad
\text{height} = \frac{1}{2}\left(d-1\right)+n_\text{row}d,
\end{align}
encloses $2\cdot\text{width}\cdot\text{height}$ physical qubits.
In the limit of large $n_\text{row}$ and $m_\text{col}$, the asymptotic footprint of physical qubits per logical qubit approaches
\begin{align}
\text{density}=\lim_{n_\text{row},m_\text{col}\rightarrow\infty}\frac{2\cdot\text{width}\cdot\text{height}}{3n_\text{row}m_{\text{col}}}=d(d-1),
\end{align}
which is less than half that of a $1$-qubit rotated surface code patch.
We benchmark this design in a $10^{-3}$ uniform depolarizing model on a $n_\text{row}=m_\text{col}=3$ grid.
These dimensions ensure our results are representative of larger grids as more than half of the logical operators are subject to error mechanisms between interior boundaries and defects.

\subsubsection{Computing on dense packing}\label{sec:compute_dense_packing}
We perform computations on logical qubits in either the $3$-qubit rectangle or the dense twist packing by lattice surgery.
Lattice surgery may be applied directly to perform, in $d$ rounds, two-logical-qubit joint Pauli measurements involving any logical operator lying on the boundary.
However, not all logical operators of these dense packings are so easily accessible.
Hence, we require a procedure to either move the logical operators of interest to the boundary, or to unload selected logical qubits of the dense representation into $1$-qubit surface patches for computation. 
Once computation is complete, the $1$-qubit patch may be loaded by the reverse procedure back into dense storage.
The steps for loading and unloading involve the elementary steps of moving boundaries, merging, and splitting, which are shared with patch rotations and other lattice surgery operations.

\begin{figure}
    \centering
\begin{tabular}{c}
    \includegraphics[width=0.9\linewidth]{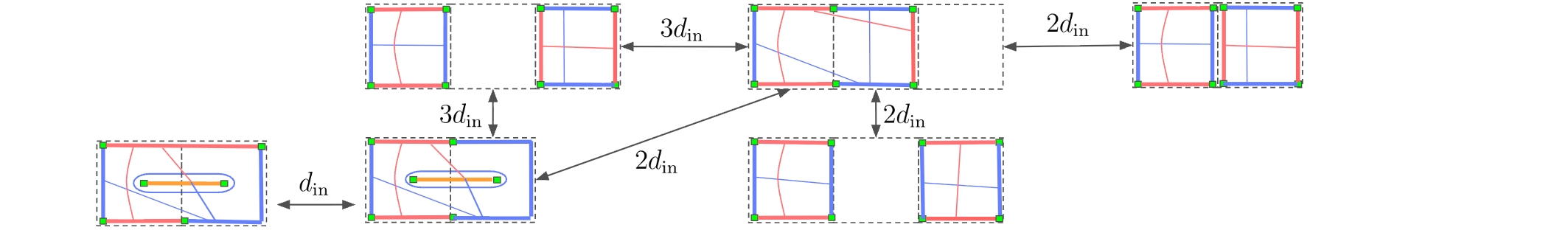}\\
    \hline
    \includegraphics[width=0.9\linewidth]{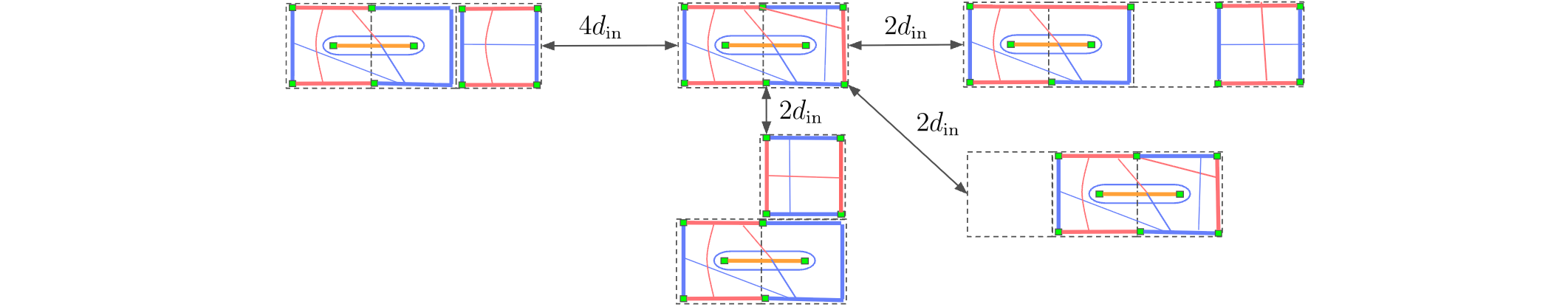}
\end{tabular}
    \caption{Non-exhaustive sequence of high-level operations for loading and unloading of $1$-qubit patches into a (top) $2$-qubit rectangle and (middle) $3$-qubit rectangle. See~\cref{fig:loading_unloading_proof} for the elementary steps that realize some of these operations.}
    \label{fig:loading_unloading}
\end{figure}

We present a web~\cref{fig:loading_unloading} of possible  loading/unloading steps and their execution time in units of rounds.
The elementary rules for generating these steps arise from the constraints of lattice surgery and fault-tolerant measurement as follows.
\begin{enumerate}
    \item Any final measurement or initial preparation of the surface code takes $0$ rounds.
    \item However, due to faulty measurement and initial preparation, any movement of boundaries or growth of the surface code must be held for $d$ rounds.
    \item Due to lattice surgery boundary constraints, we may only merge or split into parallel boundaries of the same type -- we do not assume use of spacelike Hadamards for these operations. 
    \item Logical operators may be topologically deformed freely. For instance, endpoints may be  along boundaries of the same Pauli type, and may be moved across twist defects using the identity double-loop~\cite{Brown2017PokingHoles}.
\end{enumerate}
It should be clear that there is a menagerie of possibilities, depending on the orientation of input and output patches, and also the intermediate states of the dense packing.
The steps we show for the $3$-qubit rectangle also apply readily to dense twist storage.
We do not perform benchmarks of any of these operations and simply assume that the logical error rate of these operations is approximated by the worst error rate across all our benchmarks of $p_\text{dynamic}(d)$ per unit of additional spacetime volume per round.
Moreover, it is quite likely that the time taken for these steps can be further optimized.
We leave the exploration and detailed design of such possibilities to future work.

\subsection{1D yoked twist-defect dense packing}\label{sec:dense_packing_yoking}
\begin{figure}
    \centering
    \includegraphics[width=0.9\linewidth]{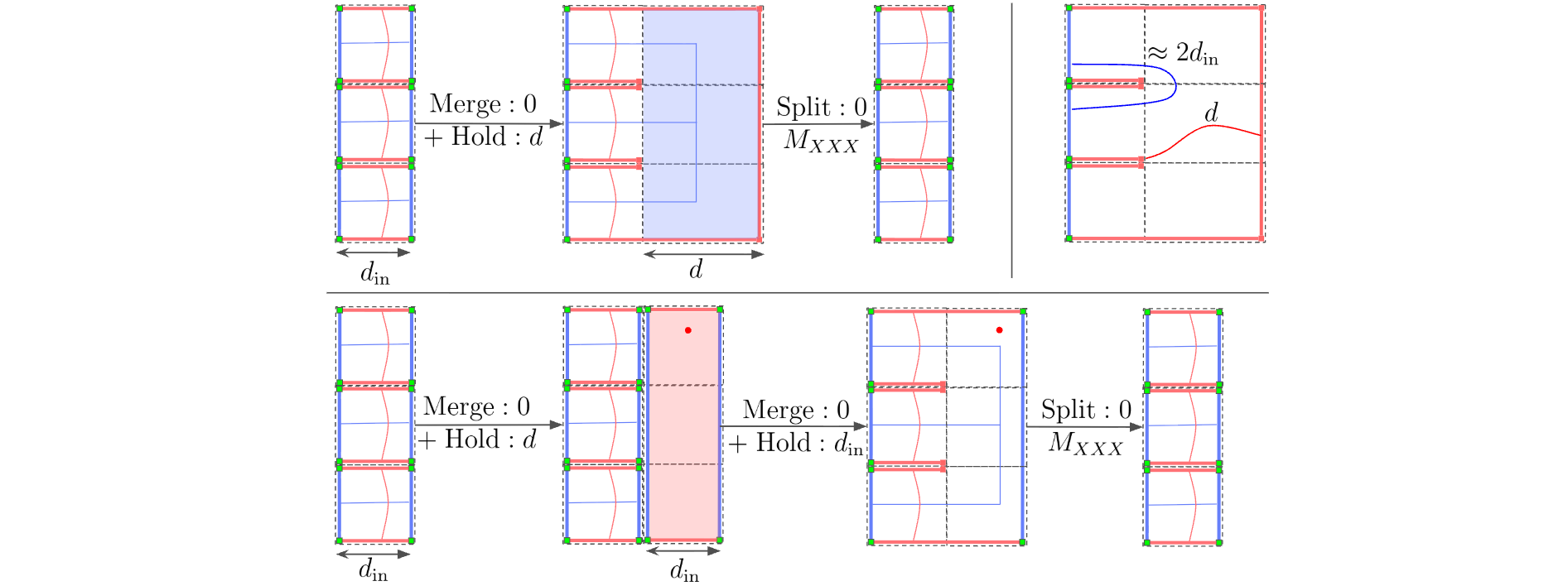}
    \caption{Lattice surgery operations that measure the logical $X^{\otimes3}$ stabilizer on three $1$-qubit patches. (Top) Measurement errors are suppressed by holding for $d$ rounds and space-like hook errors that flip the signs of logical $Z$ are suppressed to distance $d$ by choosing the measurement patch to be of width $d$. (Bottom) By bending the measurement patch downwards (or upwards) in a timelike direction, we may trade space for time. The red dot represents a timelike $X$ error string.}
    \label{fig:measure_X_parity}
\end{figure}

We now increase the density of logical qubits by concatenating each column of distance $d_\text{in}$ dense twist storage with a $[[n,n-2,2]]$ quantum parity check code to achieve a targeted effective distance of $d<2d_\text{in}$.
This builds upon the prior art of 1D yoked surface codes~\cite{Gidney2025Yoked} which instead concatenates each column of $1$-qubit distance $d_\text{in}$ patches with the parity check, as we will now review.
The stabilizers of the parity check code can be chosen to be $\{X^{\otimes n},Z^{\otimes n}\}$.
There are $n-2$ encoded logical qubits and the remaining two are called yoke qubits, assigned to rows $n-2$ and $n-1$.
One may then choose a basis for the $j^\text{th}$ logical qubit, such as $\{X_jX_{n-1},Z_jZ_{n-2}\}$. 
Although the inner patches have a finite error rate, the lattice surgery allows measurement of the parity check stabilizers with arbitrarily small error.
As shown in~\cref{fig:measure_X_parity}, hook errors in a spacelike direction that flip the unmeasured logical operation may be suppressed by measuring with a larger patch.
In a uniform depolarizing model, it should be sufficient for the measurement patch to be of distance $d_m\approx d$, and we approximate this error contribution using the lattice surgery error rate of $3.5^{-d_m}/30$.
The topological nature of the surface code also allows us to trade space for time by deforming the patch in a timelike direction, thus converting spacelike hook errors into timelike hooks.
In this latter case, we make a very conservative choice of $d_m=2d_\text{in}$, similar to~\cite{Gidney2025Yoked}, for the purposes of stabilizer measurement, and approximate the resulting error contribution as $0$.

The parity check is able to detect up to one distance $d_\text{in}$ logical $Z$ and/or $X$ error in each column, which would otherwise remain undetected and cause a catastrophic logical failure in the quantum algorithm.
The specific patch that experienced a logical failure may be determined with high confidence by an outer round of decoding based on computing the complementary gap.
The key empirical benchmark from~\cite{Gidney2025Yoked} under the $10^{-3}$ uniform depolarizing noise model is a fit for the logical error probability per round $p_{\text{yokeSC}}$ of this concatenated code as function of $n$, $d_\text{in}$, and the number of rounds $T_{\text{yoke}}$ between completion of all stabilizer measurements.
From~\cite{Gidney2025Factoring}, this fit is
\begin{align}\label{eq:1d_gidney_yoked_error}
p_{\text{total},\text{yokeSC}}(d_\text{in})=n^2T_{\text{yoke}}15^{-d_\text{in}}/100.
\end{align}
The $n$ and $T_\text{yoke}$ exponents are consistent with path-counting approximations: if $p\approx cT_\text{yoke}$ is the independent probability of a logical error on a single patch, then the probability of at least two logical errors on any of the $n$ patches is to leading order $1-(1-cT_\text{yoke})^n-\binom{n}{1}cT_\text{yoke}(1-cT_\text{yoke})^{n-1}= \binom{n}{2}(cT_\text{yoke})^2+\mathcal{O}((ncT_\text{yoke})^3)$. Dividing this probability by $T_{\text{yoke}}$ rounds yields the overall logical error rate scaling of $\mathcal{O}(n^2T_{\text{yoke}})$ per round.
The logical error rate of the single-qubit patch used in~\cite{Gidney2025Yoked,Gidney2025Cultivation} was reported to be between $p_{\text{SC}}(d)\in[4.0^{-d}/10,3.5^{-d}/40]$.
Then the fit $p_{\text{total},\text{yokeSC}}$ is approximated by choosing $p=nT_\text{yoke}p_{\text{SC}}(d_\text{in})$, corresponding to simply assuming an ideal outer decoder that translates single-qubit $X$ or $Z$ error detection into faultless determination of which logical patch failed. 

\begin{figure}
\centering
\begin{tabular}{l}
     \includegraphics[height = 4cm]{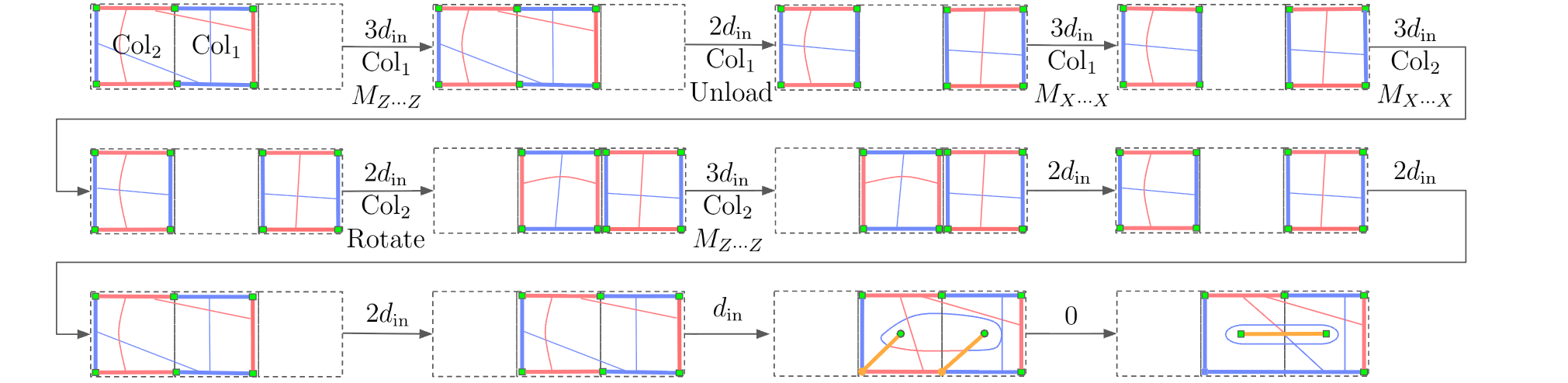}
     \\\hline\\
    \includegraphics[height = 4cm]{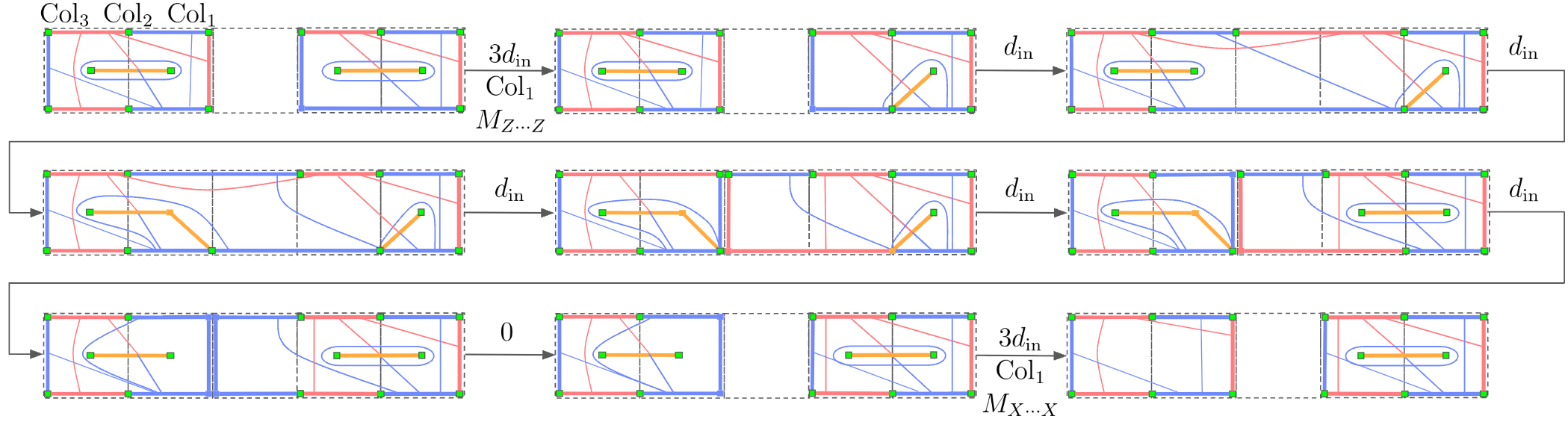}
     \\\hline\\
    \includegraphics[height = 5.5cm]{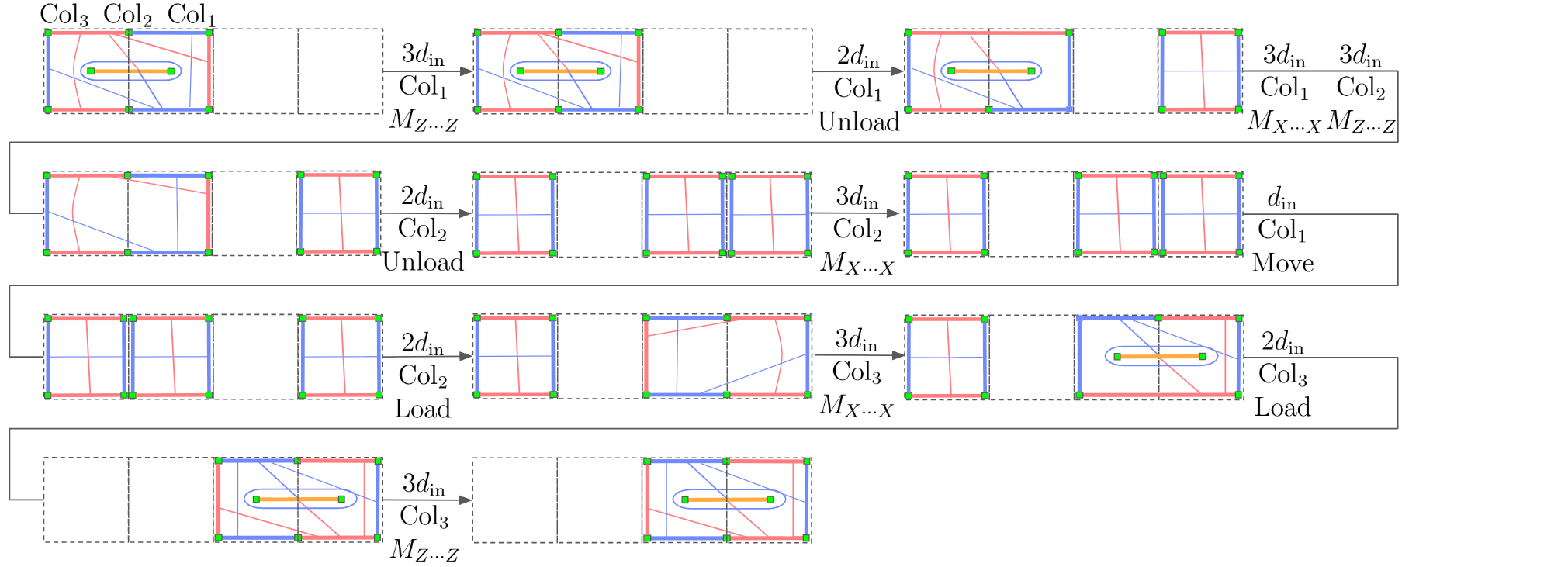}
    \end{tabular}
    \caption{Sequence of operations, with some drawn from~\cref{fig:loading_unloading,fig:measure_X_parity}, to measure all parity check stabilizers across $m_\text{col}$ columns of $n_\text{row}$ rows of logical qubits using a workspace column $d_\text{in}$ units wide
    (Top) The right-most column starts as a two-qubit rectangle. After $23d_\text{in}$ rounds, all four parity checks are measured, and the two qubits are shifted right to a partially packed state of the three-qubit rectangle.
    (Middle) With another column of three-qubit rectangles incident to the left, two parity check stabilizers of the left rectangle are measured in $11d_\text{in}$ rounds, after which a column of three-qubit rectangles and two-qubit rectangles is obtained to the right and left respectively, and the next set of parity checks can be measured using the top figure.
    Overall, the rectangles encode $(n_\text{row}-2)(3m_\text{col}-1)$ logical qubits, and all parity checks are measured in $T_\text{right}=(23m_\text{col}+11(m_\text{col}-1))d_\text{in}$ rounds.
    The entire sequence is then reversed much faster without measuring parity checks in at most $T_\text{left}=(3m_\text{col}+8(m_\text{col}-1))d_\text{in}$ to return to the original configuration in $T_\text{yoke,min}$ rounds.
    When yoking dense twist storage, the left and right ends of the rectangles are offset vertically by $d_\text{in}/2$. 
    If walking is permitted, we unload into a column of square patches, and walk vertically for $d_\text{in}$ cycles.
    Otherwise, we alternate unloading and loading even/odd rows.
    Without parity checks, this takes $8d_\text{in}$ cyceles, which increases $T_\text{right}$ to $(23m_\text{col}+19(m_\text{col}-1))d_\text{in}$ and $T_\text{left}$ to $(3m_\text{col}+16(m_\text{col}-1))d_\text{in}$ to return to the original configuration in $T_\text{yoke,min}$.
    (Bottom) A double-width workspace $2d_\text{in}$ units wide enables a shorter yoking sequence of $27d_\text{in}$ rounds per column of 3-qubit rectangles, but this ends up being overall less dense.
    }
    \label{fig:yoke_unload_load}
\end{figure}

We now evaluate the performance of concatenating the $n_{\text{col}}=3m_{\text{col}}$ columns of logical qubits in either multiple rows of $m_{\text{col}}$ columns of $3$-qubit patches ($3$ columns) or dense twist storage with a $[[n,n-2,2]]$ code.
The sequence of operations to measure all stabilizers is shown in~\cref{fig:yoke_unload_load}, averaging $T_\text{right}/3\approx 14 d_\text{in}$ rounds per column of logical qubits using a single-column workspace, or $T_\text{right}/3=9d_\text{in}$ rounds using a two-column workspace.
For comparison, it takes $8d_\text{in}$ rounds to yoke each column of one-qubit patches~\cite{Gidney2025Yoked}.
After measuring all stabilizers, the entire grid of qubits is shifted left in $T_\text{left}$ cycles, and the entire sequence is repeated.
With hex grid connectivity, shifting is performed by lattice surgery move operations as we cannot walk patches. 
With the two-column workspace, we use move operations, so $T_\text{left}$ costs $d_\text{in}$ cycles per column or $T_\text{left}=3m_{\text{col}}d_\text{in}$ cycles for all columns.
With the one-column workspace, we reverse the sequence of operations without measuring parity checks.

\begin{figure}
    \centering
    \includegraphics[width=0.33\linewidth, valign=m]{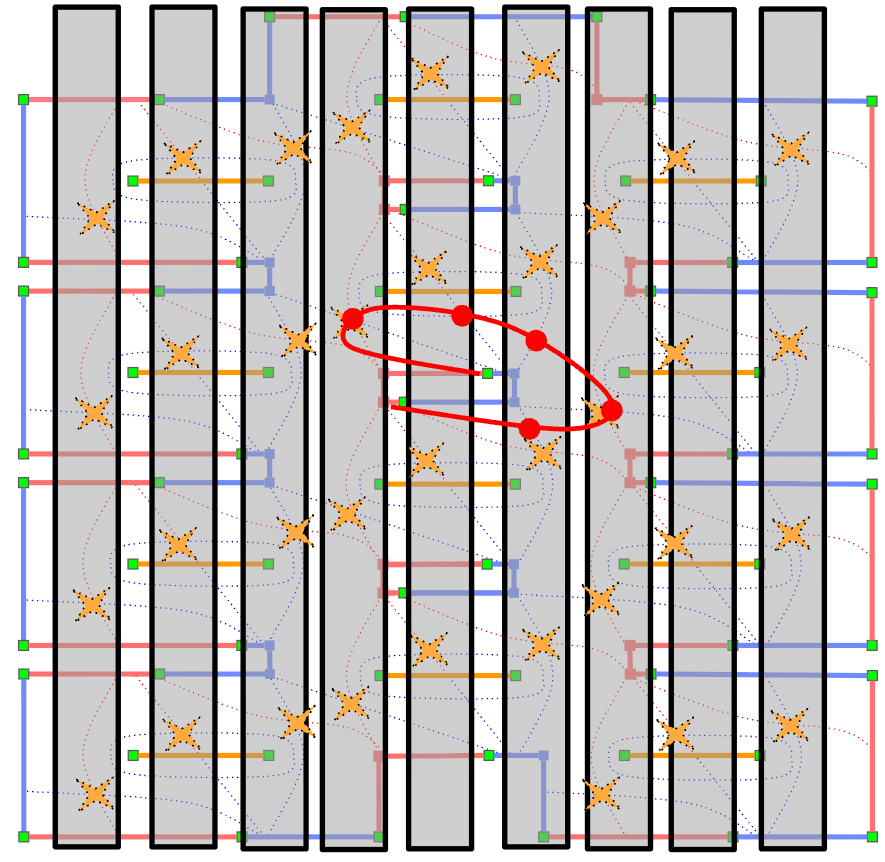}
    \hfill
    \includegraphics[width=0.66\linewidth, valign=m]{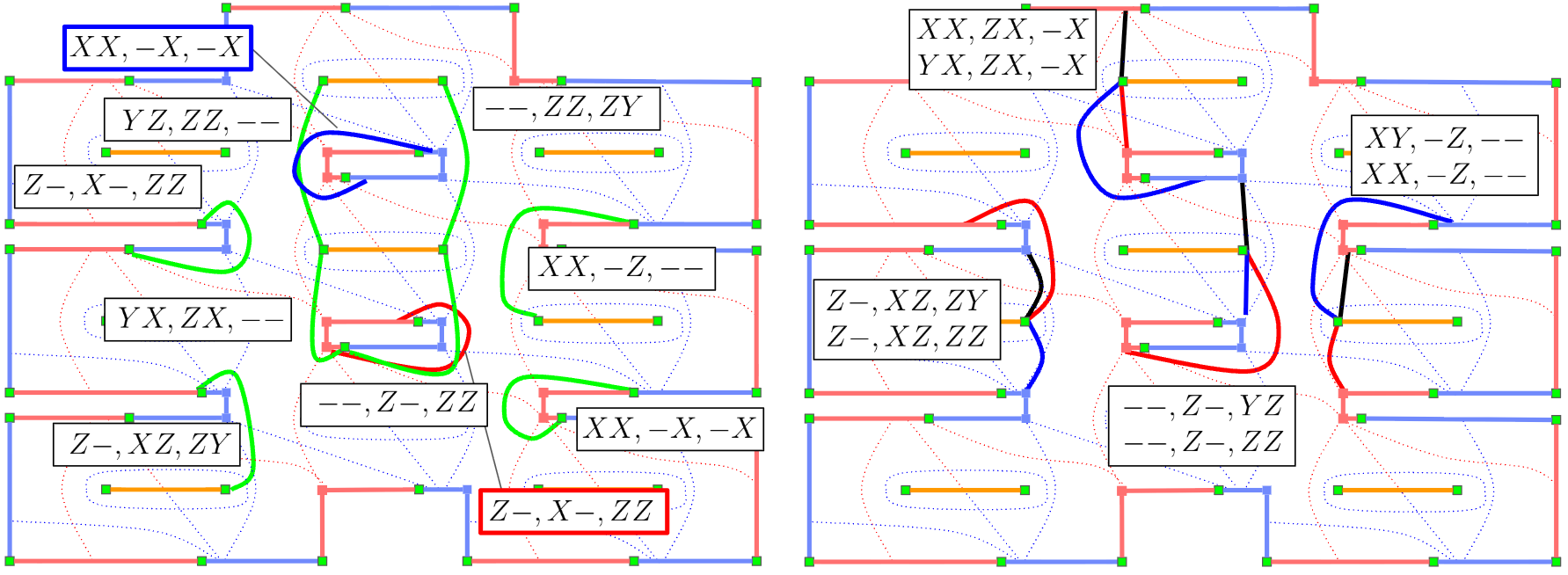}
    \caption{(Left) We group dense twist storage logical qubits into columns with an even number of qubits (gray boxes) and perform a parity check with $\{X\cdots X, Z\cdots Z\}$ stabilizers by sequentially unloading qubits to the right, and reloading into a new instance of dense twist storage. The middle column demonstrates a single distance $d$ error mechanism that flips two adjacent logical $Z$ operators in a single column, which is undetectable by a single-column parity check. (Middle) All distance $d$ error mechanisms, with some degeneracies included. We annotate logical errors by the tuple $(AB,CD,EF)$ where Pauli operators $A,C,E$ act on logical qubits $1,2,3$ of the row above and $B,D,F$ act on logical qubits $1,2,3$ of the row below. 
    We do not annotate logical errors in the next left and right columns. (Right) All inequivalent distance $\frac{3}{2}d_\text{in}$ error mechanisms. Note that every such error has two configuration: one where the $\frac{d_\text{in}}{2}$ error chain points down, and one where it points up, drawn in black for clarity.}
    \label{fig:correlated_failure}
\end{figure}

A new key difficulty is the presence of correlated logical errors.
Whereas all distance $d_\text{in}$ errors in a column of single-qubit patches cause distinct and independent logical errors, the dense twist storage features many distance $d_\text{in}$ and higher distance $\frac{3}{2}d_\text{in}$ error mechanisms that can cause multiple logical failures within a single column, which would be undetected by a distance $2$ parity check, such as those shown in~\cref{fig:correlated_failure}.
As intra-column correlated logical errors are always adjacent to each other, a simple solution is to perform separate parity checks on even and odd rows. 
However, this is wasteful as it doubles the number of yoke qubits and extends the syndrome extraction rounds $T_\text{yoke}$.
Note that such issues do not arise when concatenating the $3$-qubit rectangular patch as correlated errors across columns may be decoded column-wise independently and pose no issue.   
A key caveat of results in this section is that, like previous work, we do not simulate the lattice surgery of stabilizer measurement.
In our case, this would require an implementation of the loading and unloading steps of~\cref{fig:yoke_unload_load}, which is extremely involved and left to future work.
Additionally, a significant simplifying assumption we make is that we do not specify the outer code decoder -- we assume that the parity check will return the correct outer error syndrome, which should be validated in future work by a complementary gap simulation as done by~\cite{Gidney2025Yoked}.

\begin{figure}
    \centering
    \includegraphics[scale=0.8]{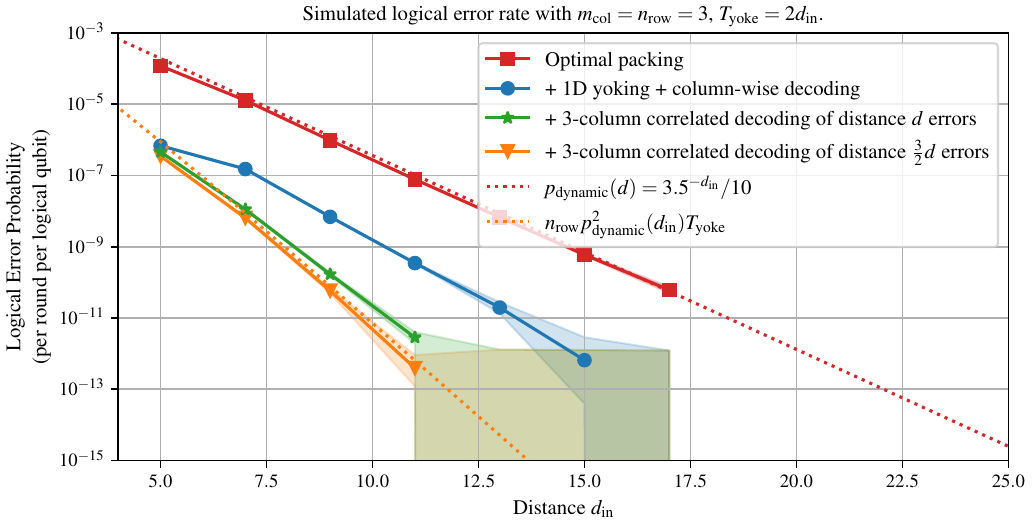}
    \caption{(Top) Logical error rate of 1D yoked twist dense memory assuming an ideal outer decoder. By correlating information on detectable single-qubit faults across three columns, we may uniquely decode all errors of distance $d_\text{in}$ and $\frac{3}{2}d_\text{in}$ that would be undetectable using only single-column-wise information.}
    \label{fig:dense_packing_error_rate}
\end{figure}

We now demonstrate that all distance $d_\text{in}$ and $\frac{3}{2}d_\text{in}$ correlated logical errors can be decoded by correlated decoding: combining parity check syndromes across 3 neighboring columns.
In~\cref{fig:dense_twist_error_mechanisms}(top), we enumerate all possible such error mechanisms and in~\cref{fig:dense_twist_error_mechanisms}(bottom), we show that every error mechanism causing the same pattern of logical failure across two rows modulo undetectable $ZZ$ or $XX$ logical errors has a unique 3-column syndrome.
Therefore, given that no distance $2d_\text{in}$ errors occur, we may determine all $d_\text{in}$ and $\frac{3}{2}d_\text{in}$ faults assuming that the outer code decoder returns the locations of all detectable single-error faults in each column.
We evaluate the logical error rate of this best-case outer decoder in~\cref{fig:dense_packing_error_rate}.
First, we exclude errors that are detected by the parity check on each column, without correlating parity check information across columns. 
Although a significant improvement in error rate is obtained, the error scaling with $d_\mathrm{in}$ is limited by correlated distance $d_\mathrm{in}$ two-qubit intracolumn error mechanisms.
Second, we exclude distance $d_\mathrm{in}$ errors that are detected by correlating parity check information across $3$ columns, which improves distance scaling to $\frac{3}{2}d_{\mathrm{in}}$.
Finally, we exclude all detectable distance $d$ and $\frac{3}{2}d_\mathrm{in}$ errors.
This leads to a final error per round per logical qubit scaling with effective distance $2d_\text{in}$, which is fit well by $n_\mathrm{row}T_\text{yoke}p_{\text{dynamic}}^2(d_\mathrm{in})$, where our simulations are equivalent to choosing  $T_\text{yoke}=2d_\mathrm{in}$.
Hence, we report the probability per round of any logical error of 1D yoked dense twist storage as
\begin{align}\label{eq:1d_yoke_twist_timesteps}
p_{\text{total},\text{yokeTwist}} \approx 3m_\mathrm{col}n_\mathrm{row}^2T_\text{yoke}p_{\text{dynamic}}^2(d_\mathrm{in}),\quad T_\mathrm{yoke}=
\begin{cases}
(26m_\text{col}+20(m_\text{col}-1))d_\text{in},&\text{with walking},\\
(26m_\text{col}+35(m_\text{col}-1))d_\text{in},&\text{without walking}.
\end{cases}
\end{align}
Excluding the overhead of two yoke qubits per column, the expected logical error rate per round per logical qubit is
\begin{align}\label{eq:1d_yoke_twist_error_rate}
p_{\text{yokeTwist}}\approx\frac{p_{\text{total,yokeTwist}}}{3m_\mathrm{col}(n_\mathrm{row}-2)} =\frac{n^2_\mathrm{row}}{n_\mathrm{row}-2}T_\text{yoke}p_\text{dynamic}^2(d_\mathrm{in}).
\end{align}
The bounding box of physical qubits used is given by~\cref{eq:footprint} plus a workspace of the same height and $2$ patches of distance $d_\mathrm{in}$ wide.
It turns out to be more space-efficient to use a workspace of only one patch of width $d_\mathrm{in}$, which is possible if each rectangular patch in the rightmost column stores two instead of three densely packed logical qubits.

We use these to enumerate over a wide range of yoked dense twist storage parameters to obtain the plot of~\cref{fig:square_patch_alt} along with effective logical qubit densities and encoding rates for finite-sized instances.
In principle, one could consider $2$D yoking of rows and columns with a higher distance-$4$ outer code, like in previous work~\cite{Gidney2025Yoked}.
However, we leave this to future work as there are some challenges: First, it is not clear whether the many more types of correlated errors can be decoded using only two yoke qubits per row or column. Second, measuring the row stabilizer is less straightforward as each row of densely packed qubits cannot be all unloaded into a single row of regular patches. Third, $T_\text{yoke}$ of our construction is already quite long, and a distance-$4$ outer code would have an even longer $T_\text{yoke}$ with error scaling like $T_\text{yoke}^3$, which makes the concatenated code harder to use in a practical setting.

\begin{figure}
    \centering
    \begin{tabular}{c|c|c}
    \hline\hline
        Error classification & Logical error & Decoded parity check syndrome\\
        \hline
         Distance $d_\text{in}$ $Y$ errors&  $YX,ZZ,--$ & $Z-,--,--$\\
& $YX,ZX,–$ & $Z-,ZX,--$ \\
& $Z-,XZ,ZY$ & $Z-,XZ,-X$\\
& $--,Z-,ZZ$ & $--,Z-,--$\\
& $--,ZZ,ZY$ & $--,--,-X$\\
& $XX,-Z,--$ & $--,-Z,--$\\
& $XX,-X,-X$ & $--,-X,-X$\\
& $Z-,X-,ZZ$& $Z-,X-,--$ \\
\hline
Distance $d_\text{in}$ $X,Z$ errors &  $XX,-X,-X$ & $--,-X,-X$\\
&  $Z-,X-,ZZ$ & $Z-,X-,--$\\
\hline
Distance $\frac{3}{2}d_\text{in}$ mixed $X,Z$ errors 
&  $XX,ZX,-X$ & $--,ZX,-X$\\
&  $YX,ZX,-X$ & $Z-,ZX,-X$\\
&  $Z-,XZ,ZY$ & $Z-,XZ,-X$\\
&  $Z-,XZ,ZZ$ & $Z-,XZ,--$\\
&  $XY,-Z,--$ & $-Z,-Z,--$\\
&  $XX,-Z,--$ & $--,-Z,--$\\
&  $--,Z-,YZ$ & $--,Z-,X-$\\
&  $--,Z-,ZZ$ & $--,Z-,--$\\
\hline\hline
    \end{tabular}
    \caption{Assuming that no distance $2d_\text{in}$ errors occur, every distance $d$ or $\frac{3}{2}d$ error mechanism in~\cref{fig:correlated_failure} causes a pattern of logical errors that is uniquely determined by combining parity check information across at most $3$ columns.}
    \label{fig:dense_twist_error_mechanisms}
\end{figure}

%% file: 3_architecture/architecture.tex

\ifSubfilesClassLoaded{\section{Architecture}}{}
In this section, we describe how the compilation of a quantum algorithm expressed in terms of logical quantum gates is organized into lattice surgery operations.
We assume that lattice surgery operations are distributed across specialized regions comprised of surface code patches laid out on a $2$D plane as shown in~\cref{fig:storage_layouts}.
In a practical modular architecture, these regions as well as individual patches within each region may be connected by noisy links~\cite{Marton2025NoisyLink,Haug2025NoisyLinks2}.
For the purposes of obtaining resource estimates though, we essentially assume a large monolithic grid of physical qubits with planar hex-grid connectivity, and count physical qubits within rectangular bounding boxes around each region.
This is similar to previous hybrid architectures~\cite{Gidney2025Factoring}.
The patch distances of these regions are chosen to bound the overall probability of any logical error occurring across the execution of the entire algorithm.
\begin{figure}
    \centering
    \includegraphics[width=.9\linewidth]{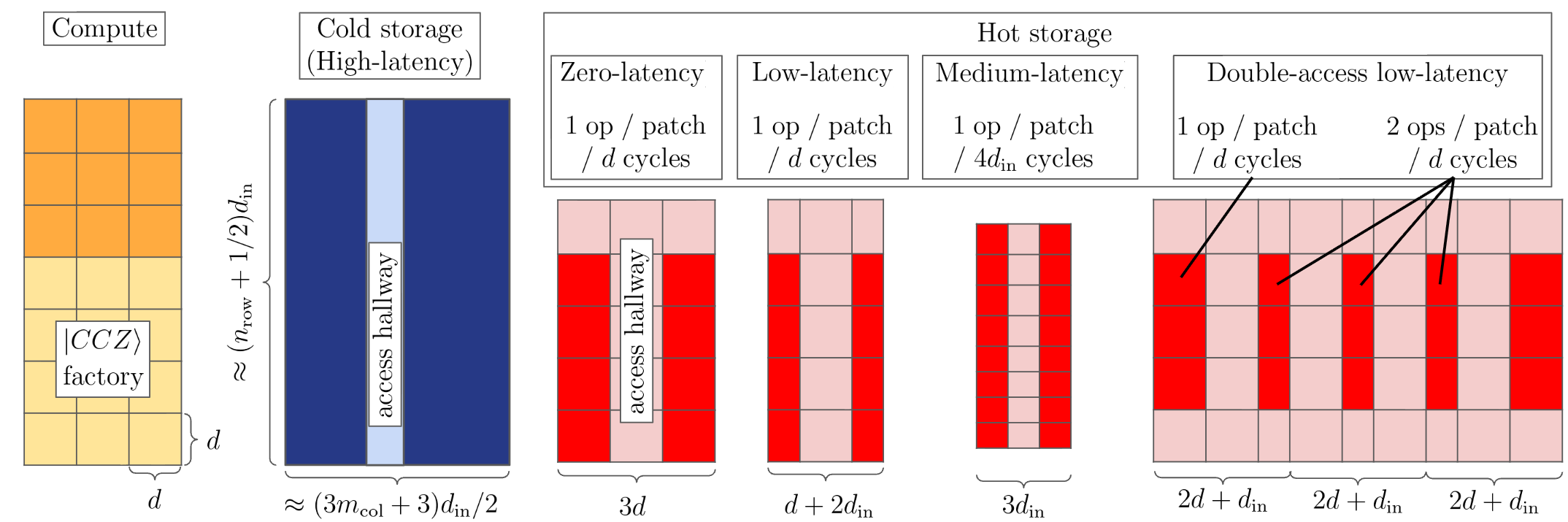}
    \caption{We organize lattice surgery of an algorithm into specialized regions: compute, cold storage, and hot storage, drawn to scale for $d=25, d_\text{in}=15$. Magic state production and distillation occurs in the compute region. We also present hot storage variants with different tradeoffs between latency, space, and throughput (ops / cycle). Zero-latency hot storage has two columns (red) of full-distance $d$ regular surface code patches adjacent to each access hallway (pink). Low-latency concatenates either each column or both columns together with a single parity check to reduce patch width to $d_\text{in}$. Medium-latency concatenates each column with two parity checks at the cost of a longer $d_\text{in}$ access time to perform a \textsc{PauliPM} targeting both columns. Double-access surrounds each column of patches with an access hallway, which permits two independent \textsc{PauliPM}s per $d$ cycles. To improve space efficiency, the sides are lined with regular surface code patches.}
    \label{fig:storage_layouts}
\end{figure}

At the high level, we define three types of regions, that may each have different dimensions and store patches of different distances.
\begin{enumerate}
    \item \textbf{Compute}: This is a general purpose region that can store a maximum of $n_C$ distance $d$ patches where arbitrary lattice surgery operations may be performed.
    Magic states may also be cultivated and distilled in this region.
    The physical qubit footprint is $n_C\cdot2\cdot d^2$.
    In a computation that takes a total of $T$ logical timesteps, we model the total error contribution from this region with the lattice surgery logical error rate $n_C\cdot T\cdot d\cdot p_L(d)$, where $p_L(d)\doteq\frac{1}{30}3.5^{-d}$.
    This is a conservative worst-case estimate as it does not account for how patches may be idling with the lower memory logical error rate $p_\text{static}(d)\doteq \frac{1}{10}4.0^{-d}$, or that there could be fewer than $n_C$ active patches, or that one could bias lattice surgery towards the lower-error $X$-top configurations using Hadamard gates.
    \begin{itemize}
        \item 
\textbf{Cultivation-based $\ket{CCZ}$ factories}: 
Magic state cultivation dramatically reduces the cost of preparing the magic $\ket{T}$ state with a logical error $p_{\ket{T}}$ as low as $10^{-8}$ in a $10^{-3}$ depolarizing noise model, using a volume $V_{\ket{T}}$ in units of (physical-qubit$\cdot$rounds) that can be read off Figure 1 of~\cite{Gidney2025Cultivation} and is comparable to a logical $\textsc{CX}$ gate.
However, magic state distillation remains necessary as we consider utility-scale simulations that require $\gg10^{8}$ such states.
We provide a simple estimate for the overall spacetime volume of distillation following~\cite{Gidney2025Cultivation,Gidney2025Factoring} for the $8\ket{T}\rightarrow\ket{CCZ}$ factory that distills a less noisy $\ket{CCZ}$ state with error $p_{\ket{CCZ}}= 28\cdot p_{\ket{T}}^2+p_{\text{factory}}$ with a logical qubit footprint of $n_{\text{CCZ}}=3\times 4$ and $6$ layers of lattice surgery in a distance $d_2$ surface code each executed in $2/3d_2$ cycles. As $d_2$ is finite, topological errors in the distillation protocol add an additional error $p_{\text{factory}}=n_{\text{CCZ}}\cdot 4\cdot d_{2}\cdot p_L(d_{2})$.
The overall physical volume of the factory is $V_{\text{factory}}=n_{\text{CCZ}}\cdot4\cdot d_2\cdot 2\cdot d_2^2$. 
By producing eight noisy $\ket{T}$ states with magic state cultivation, we tabulate in~\cref{table:ccz_factory} factory parameters that produce $\ket{CCZ}$ states every $d_{\ket{CCZ}}$ cycles for some target errors of interests. 
To accommodate the non-deterministic production of $\ket{CCZ}$ states, we add a slack buffer of $d_2$ cycles to $d_{\ket{CCZ}}$ and the $V_\text{factory}$ calculation.
    \end{itemize}
    \item \textbf{Cold storage}: This memory region stores $k_D$ logical qubits in a high-rate code comprised of the equivalent of $n_D$ inner distance $d_\text{in}$ surface code patches.
    It also includes any additional workspace needed to access quantum data in cold storage logical qubits.
    Access is associated with high latency, meaning that it may take many cycles to access the desired logical operator, and high error rates that must be carefully accounted for, or mitigated with a sufficiently large buffer.
    We implement cold storage using yoked dense twist storage in~\cref{sec:dense_memory}, where the $n_D$ patches are merged into a $n_D=n_\text{row}\times 3m_\text{col}$ dense packing with a footprint given by~\cref{eq:footprint}.
    With a workspace column one inner patch wide,
    the dense packing encodes $(n_\text{row}-2)\times (3m_\text{col}-1)$ logical qubits.
    All outer code stabilizers are measured every $T_\text{yoke}$ rounds, which can be variable, but at least $T_\text{yoke,min}$.
    From~\cref{eq:1d_yoke_twist_timesteps}, the error contribution across the entire storage per round is $p_{\text{yokeTwist}}=3m_\text{col} n_\text{row}^2T_\text{yoke} p_{\text{dynamic}}^2(d_\text{in})$, plus the topological error for measuring outer code stabilizers, which is approximated as zero using hook-error suppressed measurements.
    Using lattice surgery, we access individual logical qubits by Pauli Product Measurements  (\textsc{PauliPM}s), such as $ZZ$ or $XX$, between an exposed logical operator of a target encoded qubit in memory and a logical qubit in the $\textbf{Compute}$ region.
    Each cold storage access could have latency as high as $\gtrsim100d_\text{in}$ cycles for accessing a particular desired logical operator.
    This increases $T_\text{yoke}$ and thus contributes an excess error over usual memory operation that is detailed in~\cref{sec:latency}. 
    There, we also discuss variants that trade density for latency, such as by yoking copies of smaller dense twist packings or $3$-qubit rectangles or one-qubit rotated surface code patches like in the original work~\cite{Gidney2025Yoked}.
    Concatenating with a larger distance outer code, like 2D yoking could even further increase encoding rates, but only for $k\ge194$, which is too large for some applications.
    Using only distance-$2$ for the outer code provides latency advantages as well.
    As $p_\text{yoke}$ scales only linearly with $T_\text{yoke}$, we may pause syndrome extraction on demand for substantial durations with linear error accumulation, unlike $2$D yoking, with $T^3_\text{yoke}$ scaling.
    This places less demand on carefully synchronizing the cold storage yoking schedule with \textsc{PauliPM}s.
    
    \item \textbf{Hot storage}: In contrast to the high latency of \textbf{cold storage}, we define hot storage to have much lower latencies.
    Zero-latency is the simple case of a column of full-distance $d$ one-qubit patches with an access hallway on one side (single-access) or two sides (double-access)~\cite{Gidney2019CCZ}, so any desired lattice surgery operation can be performed without delay.
    Low-latency is defined to have an average delay of $\ll d$ cycles and to complete lattice surgery operations in $d$ cycles.
    Medium-latency has a maximum delay on the order of $6d$ cycles, but this could be much lower on average.
    The original version of hot storage~\cite{Gidney2025Yoked} is high-latency.
    It is implemented by two columns of $n_\text{row}$ one-qubit inner distance $d_\text{in}$ surface code patches with an access hallway of the same distance in between.
    Like cold storage, each column is concatenated with a quantum parity check code, leading to $2(n_\text{row}-2)$ total logical qubits.
    We assume the footprints of our compact surface code and lattice surgery operations in~\cref{sec:compact_patch} apply so the footprint of hot storage is $6n_\text{row}d_\text{in}^2$ physical qubits encoding $2(n_\text{row}-2)$ logical qubits with an error contribution every $T_\text{yoke}$ rounds of $2p_{\text{total},\text{yokeSC}}=2 n_\text{row}^2 T_\text{yoke}\frac{1}{100}15^{-d_\text{in}}$~\cite{Gidney2025Factoring}.
    Multi-patch lattice surgery for measuring outer code stabilizers and writing data to logical qubits may be performed using hook-error suppressed measurements, in which case the topological error is approximated as $0$.
    This takes $3d_\text{in}$ cycles for single-side access, or $4d_\text{in}$ cycles if both columns are addressed.
    Using patch rotations, all stabilizers are measured in $20d_\text{in}$ consecutive cycles.
    We present variations of this model in~\cref{sec:hot_storage} with tradeoffs between low-latency based on yoking with a single parity check with an encoding rate $\lesssim 1.4$ and medium latency with an encoding rate $\lesssim 2.5$ relative to zero-latency single-access hot storage.
\end{enumerate}

For our resources estimates, we choose timescales that characterize superconducting qubits.
The fundamental unit of time is the surface code cycle time of $\tau_C=1\mu$s, identical to previous work~\cite{berry2019qubitization,PRXQuantum.2.030305,low2025fast}.
The logical timestep $\tau_l=d\tau_C$ is the time to execute $d$ rounds ($6d$ layers) of stabilizer measurements, or $d$ surface code cycles.
The latency associated with decoding all stabilizer measurements to detect errors is the reaction time, assumed to be $\tau_{\text{r}}=10\mu$s as in previous work.
Reaction time imposes a minimum delay for performing subsequent lattice surgery operations that are classically controlled by the measurement results of the previous round of lattice surgery operations, such as in the fix-up step of magic state teleportation.
In some cases, we can reduce the duration of the fix-up step to equal the reaction time $\tau_\text{r}$ with an \emph{auto}-magic state~\cite{Gidney2019AutoCCZ} that uses additional space to precompute lattice surgery operations for all possible fixups -- this is reaction-limited computation. 
Importantly, reaction time impose a type of latency that is distinct from the latencies defined for zero-, medium-, and high-latency storage above.

\begin{table}
\begin{tabular}{c|c||c|c|c|c|c|c|c|c|c|c}
\hline\hline
&
\multicolumn{1}{c||}{Target} & 
\multicolumn{4}{c|}{CCZ factory} & \multicolumn{3}{c|}{Level 2 distillation} &
\multicolumn{3}{c}{Level 1 cultivation}
\\
Footprint&$p_{\ket{CCZ}}$ & $p_{\ket{CCZ}}$ & Qubits &$V_\text{factory}$  & $d_{\ket{CCZ}}$& $d_2$ & $p_{d_2}$ & Cycles & $p_{\ket{T}}$ & $V_{\ket{T}}$ & Cycles 
\\
\hline
\multirow{5}{*}{$3\times4$~\cite{Gidney2025Cultivation}}  & $10^{-9}$ & $7.8 \cdot 10^{-10}$ & $9.6 \cdot 10^{3}$ & $9.7 \cdot 10^{5}$ & 100.8 & 20 & $4.4 \cdot 10^{-13}$ & 100 & $3.0 \cdot 10^{-6}$ & $7.0 \cdot 10^{3}$ & 0.8 \\
 & $10^{-10}$ & $7.5 \cdot 10^{-11}$ & $1.2 \cdot 10^{4}$ & $1.3 \cdot 10^{6}$ & 114.8 & 22 & $3.6 \cdot 10^{-14}$ & 110 & $1.0 \cdot 10^{-6}$ & $1.5 \cdot 10^{4}$ & 4.8 \\
 & $10^{-11}$ & $4.5 \cdot 10^{-12}$ & $1.4 \cdot 10^{4}$ & $1.8 \cdot 10^{6}$ & 127.3 & 24 & $2.9 \cdot 10^{-15}$ & 120 & $1.0 \cdot 10^{-7}$ & $2.3 \cdot 10^{4}$ & 7.3 \\
 & $10^{-12}$ & $6.5 \cdot 10^{-13}$ & $1.6 \cdot 10^{4}$ & $2.2 \cdot 10^{6}$ & 134.8 & 26 & $2.4 \cdot 10^{-16}$ & 130 & $1.0 \cdot 10^{-7}$ & $2.3 \cdot 10^{4}$ & 4.8 \\
 & $10^{-13}$ & $3.5 \cdot 10^{-14}$ & $1.9 \cdot 10^{4}$ & $2.8 \cdot 10^{6}$ & 150.0 & 28 & $1.9 \cdot 10^{-17}$ & 140 & $1.0 \cdot 10^{-8}$ & $4.0 \cdot 10^{4}$ & 10.0 \\
\hline\hline
\end{tabular}
\caption{\label{table:ccz_factory}
Cost of CCZ factory with a footprint of $3\times 4$ surface code patches.
This CCZ factory produces one $\ket{CCZ}$ every $4d_2$ cycles plus a few more cycles for magic state cultivation.
$\ket{CCZ}$ production rate $f_{\ket{CCZ}}$ assumes a surface code cycle time of $1\mu$s. 
Temporally encoded lattice surgery~\cite{Chamberland2021Temporally} is used. To account for the non-determinism, we add a slack parameter of $d_2$ cycles to the average cycles $d_{\ket{CCZ}}$ taken to produce each magic state.}
\end{table}

\subsection{Latency and storage}\label{sec:latency}
In this work, \textsc{PauliPM} between compute and storage is implemented using lattice surgery between surface code patches and has a cost counted by surface code cycles. In general, though, this could be abstracted to a cost in units of noisy Bell pairs. 
As no compute operations can occur on our concatenated codes while its outer stabilizers are being measured, this introduces latency, where \textsc{PauliPM} operations must wait before they can be executed.
Unless managed carefully, this can add substantial runtime to the overall quantum algorithm.
Importantly, latency here in applying \textsc{PauliPM} is distinct and not to be confused with the reaction time, which is the latency of the decoder.
For instance, the yoked dense twist defect packing only permits sequential access and our typical instances will have $T_\text{yoke,min}\gtrsim1000\;\text{cycles}=1$ms.
\textsc{PauliPM} on a random logical qubit
takes $\frac{1}{8} T_\text{yoke,min}$ cycles on average as columns of logical qubits can only be loaded or unloaded from the left or right ends, and increases $T_\text{yoke}$ by twice that amount. 
Latency can be reduced substantially at the cost of density by instead yoking columns of distance $d_\text{in}$ single-qubit patches, or three-qubit rectangles, or using a two-column workspace.
This enables random access at a medium latency that grows linearly with the number of columns with a small constant, unless square-grid walking is permitted, in which case constant-time access is possible.

To perform such an access, outer code stabilizer measurements are paused and columns of patches are walked or moved using the gate sequences in~\cref{fig:walking}.
Any patch may be exposed to an access hallway in $2d_\text{in}$ cycles with walking or $\propto(\#\;\text{of columns})\times d_\text{in}$ cycles otherwise for \textsc{PauliPM} or other lattice surgery operations, such as when a \textsc{PauliPM} on many patches in the column is desired.
After which, the columns are walked or moved again to either resume stabilizer measurement, or to access some other patch.
Similarly, three-qubit rectangles can be moved or walked in the same time using the gate sequences in~\cref{fig:twist_walk_cycles}, though there will be additional latency for unloading and loading patches in the middle.
In general, latency is minimized by choosing a good indexing of logical qubits in patches to minimize the number of moves and load/unload steps across the entire algorithm.
We find it useful to reason about the latency in terms of the following parameters, which we also tabulate in~\cref{tab:dense_memory_characteristics} for our quantum codes.
\begin{itemize}
    \item $T_{\text{yoke},\text{min}}$: Minimum number of cycles for measuring all stabilizers.
    \item $T_{\text{yoke},{\text{max}}}$: Lattice surgery operations on storage delay the completion of a measuring all stabilizers and adds an excess error. We bound this by always choosing $T_{\text{yoke}}\le T_{\text{yoke},{\text{max}}}$, such as by pausing \textsc{PauliPM} operations.
    \item $p_\text{yoke}(T_\text{yoke})$: Error per encoded logical qubit per cycle, where $T_\text{yoke}\in[T_{\text{yoke},\text{min}},T_{\text{yoke},\text{max}}]$.
    \item $T_{\text{latency}}$: The number of cycles that must elapse before a scheduled lattice surgery operation can execute ($T_\text{yoke}<T_{\text{yoke},\text{max}}$).
    This latency exists as we do not interrupt an in-progress stabilizer measurement.
    We consider both maximum $T_{\text{latency,max}}$ and average cases $T_{\text{latency,avg}}$.
    \item $T_{\textsc{PauliPM}}$: Minimum number of cycles taken to perform a joint measurement between compute and a logical operator in memory.
\end{itemize}

\begin{table}[t]
    \centering
    \begin{tabular}{l|c|c|ccc|cc|c|c}
    \hline\hline
         \multicolumn{1}{c|}{\multirow{2}{*}{Storage type}}&\multirow{2}{*}{Distance}&Encoding&\multicolumn{3}{c|}{$T_{\text{yoke}}$}& \multicolumn{2}{c|}{$T_\text{latency}$}& \multirow{2}{*}{$
         T_{\textsc{PauliPM}}$} & $p_\text{yoke}$ \\
         \cline{5-9}
         &&rate& min& max& avg&max&avg  &&
         \\
         \hline
         \textbf{Compute} &$d\times d$&$1$&0&0&0&0& $0$&$d$&\cref{eq:compact_patch_error_formula}
         \\
         \textbf{Hot}: low-latency &$d\times d_\text{in}$ & $\lesssim1.4$ & $d$ & $\approx30d$& $\approx T_{\text{yoke,max}}$ &$d$&$\ll d$&$d$&\cref{eq:fast_hot_storage_error}
         \\
         \textbf{Hot}: medium-latency &$d_\text{in}\times d_\text{in}$ & $\lesssim2.5$ & $20d_\text{in}$ & $\approx20d_\text{in}+15d$& $\approx T_{\text{yoke,max}}$ &$7d_\text{in}$&-&$\{3d_\text{in},4d_\text{in}\}$&\cref{eq:1d_gidney_yoked_error}
         \\
         \textbf{Hot}: high-latency~\cite{Gidney2025Yoked}  &$d_\text{in}\times d_\text{in}$ & $\lesssim2.5$ & $20d_\text{in}$ & $\approx50d_\text{in}$& $\approx T_{\text{yoke,max}}$ &$20d_\text{in}$&-&$\{3d_\text{in},4d_\text{in}\}$&\cref{eq:1d_gidney_yoked_error}
         \\
         \textbf{Cold}: yoked twists
         &$d_\text{in}\times d_\text{in}$ & $\lesssim4.5$ & $\gg60d_\text{in}$ & $\approx4T_{\text{yoke,min}}$& $\approx T_{\text{yoke,min}}$ &$7d_\text{in}$&-&$\{3d_\text{in},4d_\text{in}\}$&\cref{eq:1d_gidney_yoked_error}
         \\
         \hline\hline
    \end{tabular}
    \caption{Latency parameters of our quantum codes. Encoding rates for hot storage include the access column and are calculated relative to zero-latency hot storage.}
    \label{tab:dense_memory_characteristics}
\end{table}

Suppose that the entire quantum algorithm requires $N_{\textsc{PauliPM}}$ \textsc{PauliPM}s on a memory encoding $k$ logical qubits. 
Each \textsc{PauliPM} at step $j$ of the algorithm delays completing the $j^\text{th}$ syndrome measurement of all yokes by $\Delta_j=T_{\textsc{PauliPM}}N_{\textsc{PauliPM},j}$ cycles.
This delay introduces an excess memory error across the entire memory that is
\begin{align}
p_{\text{yoke},\text{excess}}(T_\text{yoke})\doteq
k\cdot T_\text{yoke}\cdot[p_\text{yoke}(T_\text{yoke})-p_\text{yoke}(T_{\text{yoke},\text{min}})],\quad
T_\text{yoke}=T_{\text{yoke},\text{min}}+\Delta_j.
\end{align}
Across the quantum algorithm, the total excess memory error from all delays is then
\begin{align}
p_{\text{yoke},\text{excess}}=\sum_{j}p_{\text{yoke},\text{excess}}(T_{\text{yoke},\text{min}}+\Delta_j).
\end{align}
Depending on the algorithm, it is possible that \textsc{PauliPM}s are scheduled intermittently such that $\Delta_j\ll T_{\text{yoke},\text{min}}$, leading to a small excess error $p_{\text{yoke},j}$ and a small additional quantum algorithm runtime only due to memory latency.
However, the worst-case excess error occurs when all \textsc{PauliPM}s are scheduled in a row as $p_\text{yoke}(T_\text{yoke})$ scales at least linearly in $T_\text{yoke}$. 
The parameter $T_{\text{yoke},{\text{max}}}$ controls this worst-case error, leading to the upper bound on excess error
\begin{align}\label{eq:mem_excess_error}
p_{\text{yoke},\text{excess}}\le \left\lceil\frac{N_\textsc{PauliPM}}{\left\lfloor\frac{T_{\text{yoke},{\text{max}}}-T_{\text{yoke},{\text{min}}}}{T_\textsc{PauliPM}}\right\rfloor}\right\rceil\cdot p_{\text{yoke},\text{excess}}(T_{\text{yoke},{\text{max}}}),
\end{align}
and the upper bound on additional quantum algorithm cycles 
\begin{align}\label{eq:mem_excess_cycles}
T_{\text{yoke},\text{excess}}\le N_\textsc{PauliPM}\cdot (T_\text{latency}+T_\textsc{PauliPM})+\left\lceil\frac{N_\textsc{PauliPM}}{\left\lfloor\frac{T_{\text{yoke},{\text{max}}}-T_{\text{yoke},{\text{min}}}}{T_\textsc{PauliPM}}\right\rfloor}\right\rceil \cdot T_{\text{yoke},{\text{min}}}.
\end{align}
This runtime bound is very conservative as it also assumes that \textsc{PauliPM}s block operations outside the memory.
It is possible to trade-off excess errors for excess cycles. For instance, $p_{\text{yoke},\text{excess}}$ is minimized if we only permit one instruction per complete syndrome extraction, corresponding to $T_{\text{yoke},{\text{max}}}=T_{\text{yoke},{\text{min}}}+T_\textsc{PauliPM}$, but this also maximizes excess cycles. 
In our later resource estimates, we avoid these worst-case bounds by scheduling \textsc{PauliPM}s appropriately, using buffer logical qubits, or simply using less cold storage.

\subsubsection{Hot storage}\label{sec:hot_storage}
Hot storage defined in prior art~\cite{Gidney2025Yoked} consists of two columns of even-$n$ rows of distance $d_\text{in}$ surface code patches encoding $2(n-2)$ logical qubits.
Like cold storage, each column is yoked with a parity check code and all outer code $Z^{\otimes n}$ and $X^{\otimes n}$ stabilizers are measured in $T_{\text{yoke},\text{min}}=20d_\text{in}$ using one additional column and a sequence of patch rotations.
Subsequently, one performs hook-error-suppressed lattice surgery operations on encoded qubits with a $60\%$ uptime corresponding to $T_{\text{yoke},\text{max}}=50d_\text{in}$, which bounds the error of hot storage to $p_\text{mem}(T_\text{yoke,max})$ and is high-latency with $T_\text{latency,max}=20d_\text{in}$.
In general, any lattice surgery operation that is classically controlled by the measurement outcome of logical qubits in storage should also be performed only immediately after an additional measurement of the relevant outer code stabilizer.

We modify the parity check and syndrome extraction cycles to obtain medium-latency hot storage with an encoding rate $\lesssim2.5\times$ higher than single-access zero-latency storage and low-latency hot storage with an encoding rate $\lesssim1.4\times$, and double-access low-latency hot-storage with a per-operation encoding rate $\lesssim1.9\times$.
The medium-latency hot storage is a simple modification of the original.
We use the syndrome extraction cycle illustrated in~\cref{fig:hot_storage_medium_latency}, where a set of stabilizers is measured in $6d_\text{in}$ or $7d_\text{in}$ cycles.
Syndrome extraction of medium-latency hot storage is intended to be synchronized with $\ket{CCZ}$ production in parallel, after which there is uptime $T_\text{up}$, say $T_\text{up}\approx5d$ cycles to consume the $\ket{CCZ}$ state with lattice surgery operations on encoded data.
Overall, this leads to $T_\text{latency,max}\le7d_\text{in}$ and $T_\text{yoke,max}= 20d_\text{in}+3T_\text{up}$.
In practice, one enumerates over various configurations to find the minimum physical qubit footprint that meets a target error rate, such as plotted in~\crefpos{fig:hot_storage_footprint}{left} for a typical choice $d_\text{in}=\lceil0.6d\rceil d,\;T_\text{up}=5d$.
\begin{figure}
    \centering
    \includegraphics[width=\linewidth]{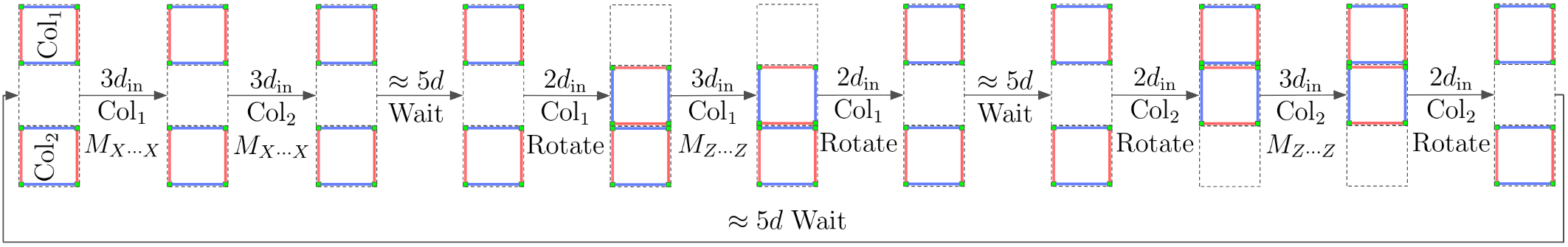}
    \caption{Hot storage syndrome extraction optimized for medium latency. In the first $6d_\text{in}$ cycles, $X\cdots X$ stabilizers are measured for the upper and lower columns. In the next $6d_\text{in}$ cycles, the upper $Z\cdots Z$ stabilizer is measured in between two patch rotations, and similarly for the lower $Z\cdots Z$ stabilizer in the last $6d_\text{in}$ cycles.
    In between the three syndrome extraction steps, we leave $5d$ cycles for other lattice surgery operations on the patches.}
    \label{fig:hot_storage_medium_latency}
\end{figure}

We may even further reduce latency by only yoking one type of logical operator.
A significant portion of compute in our later resource estimates is dedicated to implementing quantum lookup tables with a large number of output bits.
Data in output bits is written by classically-controlled multi-target $\textsc{CX}$ gates, implemented by a joint $X\cdots X$ measurement across logical qubits.
As these $X\cdots X$ measurements dominate the spacetime volume of the lookup table, we optimize logical error rates by storing quantum data in $X$-top orientation, which has lower error for both memory and lattice surgery.
Let us choose a one-qubit rectangular patch with dimensions $d\times d_\text{in}$, where $d$ ($d_\text{in}$) is the linear dimension of the vertical $Z$ (horizontal $X$) boundary, and $d>d_\text{in}$.
From~\cref{fig:square_patch_alt}, the logical error rate of an $Z$ error, an $X$ error, or any error occurring is then $\frac{1}{2}p_{\text{static}}(d),\frac{1}{2}p_{\text{static}}(d_\text{in}),\frac{1}{2}(p_{\text{static}}(d)+p_{\text{static}}(d_\text{in}))$ respectively.
Now consider hot storage with two columns of $n_\text{row}$ rectangular patches with an access hallway $d$ units wide -- the overall dimensions are $n_\text{row}d \times (2d_\text{in}+d)$.
Let us concatenate all patches in these two columns with a parity check with only a single $X_0\cdots X_{2n_\text{row}-1}$ stabilizer that is measured every $T_\text{yoke}$ cycles.
The $j^\text{th}$ logical qubit is then defined by $X$ ($\bar{Z}$) logical operator $X_j$ ($Z_jZ_{2n_\text{row}-1}$) for $j\in\{0,\cdots,2n_\text{row}-2\}$.
From~\cref{fig:h_bridge}, $X$-top lattice surgery also has lower error $\frac{2}{3}p_{\text{static}}$.
Hence, we model the topological error of measuring this stabilizer in $d$ cycles as $n_\text{row}\cdot d\cdot \frac{2}{3}p_{\text{static}}(d)$.
In between stabilizer measurements, we assume that the access hallway is fully utilized with the same lattice surgery error rate.
Assuming an ideal outer decoder, the error per round of this version of hot storage storing $2n_\text{row}-1$ logical qubits is then
\begin{align}\label{eq:fast_hot_storage_error}
p_{\text{yokeFast}}&=
\underbrace{n_\text{row}\cdot \frac{2}{3}p_{\text{static}}(d)}_{\text{access hallway lattice surgery}}
+
\underbrace{n_\text{row}\cdot p_{\text{static}}(d)}_{\text{logical $Z$ error}}
+
\underbrace{T_\text{yoke}\cdot n_\text{row}^2\cdot p^2_{\text{static}}(d_\text{in})}_{\text{yoked logical $X$ error}}.
\end{align}
Unlike medium-latency hot storage, all lattice surgery operations on encoded logical qubits take $d$ rather than $3d_\text{in}$ cycles.
We enumerate over these parameters to find the minimum physical qubit footprint of low-latency hot storage in~\crefpos{fig:hot_storage_footprint}{middle}, plotted for the choice of fixed $T_\mathrm{yoke}=\{4d,30d\}$.
Note that we may alternatively yoke with the $Z\cdots Z$ stabilizer, also with $p_\text{static}^2$ error scaling, by rotating the hot storage layout so that all patches remain in the $X$-top orientation, but with the logical-$Z$ operator exposed to the hallway. 

Another metric for hot storage performance is the average spacetime volume per lattice surgery operation.
Whereas single-access low-latency hot storage permits up to one operation per logical qubit per $d$, double-access hot storage permits two operations.
Each column of patches is concatenated with a single parity check and is surrounded on both sides by an access hallway, which permits independent lattice surgery operations on either side of a patch.
Although double-access low-latency storage is less space efficient, it is extremely spacetime-efficient, with up to twice the spacetime-operation encoding rate in~\crefpos{fig:hot_storage_footprint}{right} over conventional single-sided zero-latency storage.

\begin{figure}
    \centering
    \includegraphics[scale=0.7]{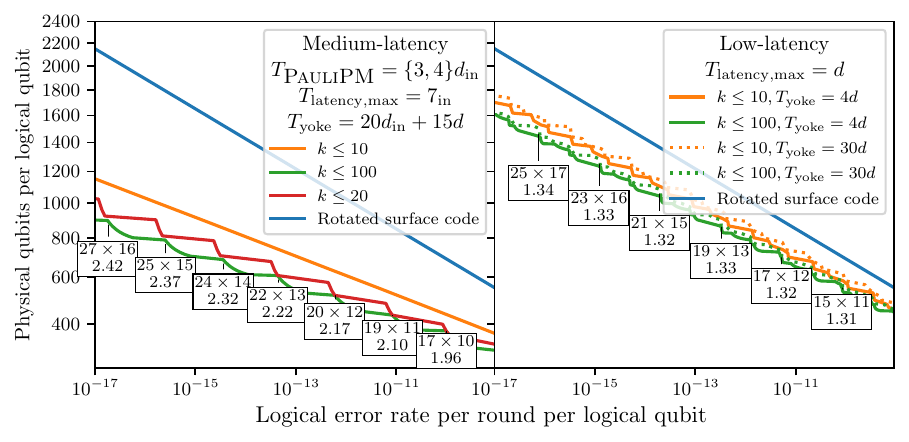}
    \hfill
    \includegraphics[scale=0.7]{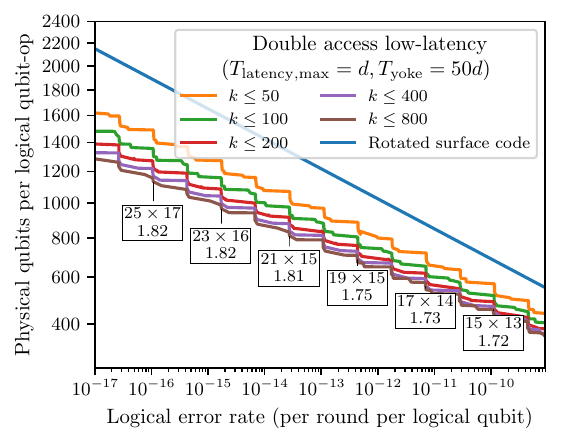}
    \caption{Density of logical qubit hot storage based on the layouts in~\cref{fig:storage_layouts}. Selected points are annotated with the patch dimensions $d\times d_\text{inner}$ and the encoding rate over zero-latency hot storage. Access hallway utilization is assumed to be $100\%$ for the purposes of computing lattice surgery error. Double-access hot storage has a worse density per logical qubit, but a smaller spacetime volume when normalized for up to $2$ concurrent lattice surgery operations on each logical qubit per $d$ cycles.}
    \label{fig:hot_storage_footprint}
\end{figure}

Our second version of low-latency hot storage is based on qubit hotels~\cite{Gidney2025Factoring,Fujiu2025DensePacking}.
As seen in~\cref{fig:yoked_hotels}, we have columns of single-access qubit hotels.
The idea is to yoke just the logical operators along the exposed side, similar to the first version of low-latency hot storage.
However, a key issue is the presence of intra-row correlated errors.
The distance $2d_{\text{in,d}}$ and $2d_{\text{in,s}}$ error chains that form around the interior boundaries cause three logical operators to fail at once.
As there are two failures in a column, this is undetected by a parity check.
As the logical operators that fail are adjacent to each other, we resolve this by yoking even and odd rows separately, with a total of two yoke qubits per column.
This increases $T_\text{yoke,min}$ to $2d$.
Unlike twist-defect dense storage, we cannot detect this by correlating syndromes across columns as the third logical operator that fails is not yoked.
Nevertheless, the two independent even/odd parity check provide enough information to resolve this from other full-distance errors.
Note that both $X$-top and $Z$-top configurations occur, unlike the previous case where we could bias all patch orientations towards the lower error $X$-top configuration.
Assuming an ideal outer decoder, we estimate the overall error per round per distance $d\times n_\text{row}d$ access hallway, averaged across orientations, to be
\begin{align}
p_{\text{yokeHotel}}=
\underbrace{n_\text{row}p_{L}(d)}_{\substack{\text{access hallway}\\\text{lattice surgery}}}
+
\underbrace{n_\text{row}\frac{p_{\text{static}}(d)+p_{\text{dynamic}}(d)}{2}}_{\text{unyoked logical error}}
+
\underbrace{T_\text{yoke}n^2_\text{row}\frac{p^2_{\text{static}}(2d_{\text{in}})+p^2_{\text{dynamic}}(2d_{\text{in}}+2\delta_{\text{in}})}{2}}_{\text{yoked logical error}}.
\end{align}
However, due to the use of additional yoke qubits, and scaling with the worse $p_\text{dynamic}$ error rate, yoked qubits hotels have only comparable density.
Nevertheless, this construction could be useful with surface code gate schedules without the $p_{\text{static}}, p_{\text{dynamic}}$ error asymmetry.
\begin{figure}
    \centering
    \includegraphics[width=0.5\linewidth]{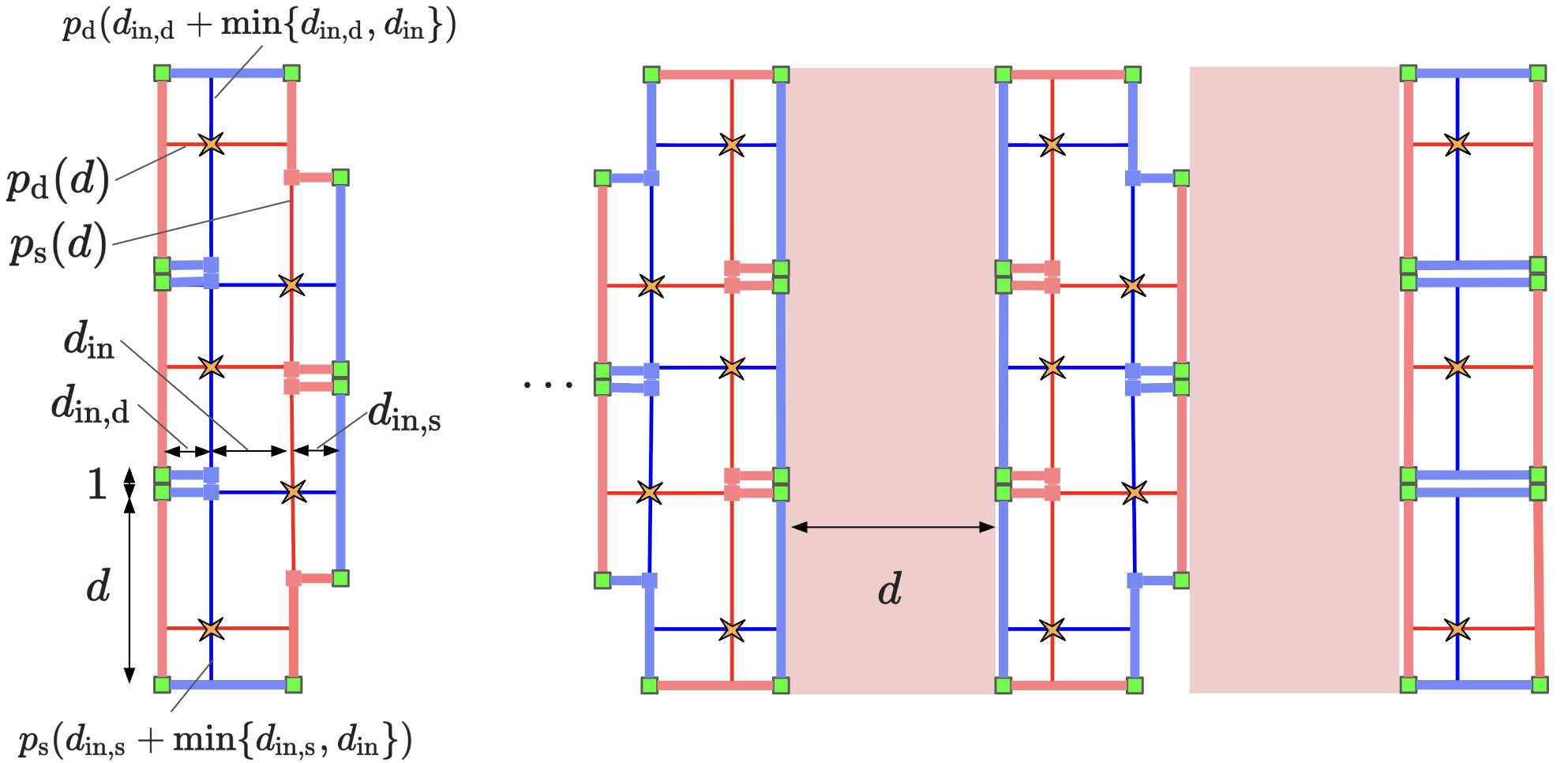}
    \caption{Probability of failure for each logical operator in a qubit hotel, where $p_{\text{s}}(d)=\frac{1}{2}p_\text{dynamic}(d)$ and $p_{\text{d}}(d)\approx\frac{1}{2}p_\text{static}(d)$ are based on fits from~\cref{fig:square_patch_alt}. We parameterize widths with $d_\text{in}=d_\text{in,s}=d_\text{in,d}-\delta_\text{in}$. (Right) Layout of qubit-hotel low-latency hot storage where each distance $d$ access hallway is adjacent to the sides of a qubit hotel.
    The boundaries of qubit hotels about each hallway is of the same type to enable yoking of all vertically oriented logical operators in the even or odd rows in $d$ cycles.
    %
    }
    \label{fig:yoked_hotels}
\end{figure}

\subsection{Layouts}
In this section, we describe how compute, hot storage, and cold storage are arranged to realize minimum-space, minimum-time, minimum-spacetime layouts, as well as interpolations between them.
Our later resource estimates optimize these layouts to execute the dominant-costs components of quantum simulation algorithms, specifically block-encoding operations comprised of large quantum lookup tables, and multiplexed quantum rotations.
Between minimum-space and minimum-spacetime, runtime will be dominated by Toffoli gate counts. 
Layouts for this regime are designed to realize a desired rate of $\ket{CCZ}$ state consumption.
Achieving this rate sets the compute region size through the number of magic state factories, space to route magic states to where they are needed, and space for the lattice surgery operations that consume magic states sufficiently quickly.
By optimizing lattice surgery compilation of these primitives, we significantly reduce space and time overhead compared to more general-purposes architectures, such as generic Clifford+$T$ quantum circuits~\cite{Litinski2019Game,Beverland2022Assessing}.

\subsubsection{Minimum-space layout}\label{sec:minimum_space}
This layout uses the smallest possible compute region.
With a $3\times 4$ logical qubit $\ket{CCZ}$ factory, this is a grid of $3\times 4$ distance $d$ rotated surface code patches.
In our later resource estimates, $d=25$ is typical.
Compute is attached to hot storage and cold storage with distance $d_\text{in}$ inner patches, such as depicted in~\crefpos{fig:example_layout}{left}.
When data in cold storage is required, the cold storage access hallway is connected to compute via the hot storage access hallway by moving at most two single hot storage patches out of the way and then back in a total of $\le 2d_\text{in}$ cycles.
We surround the remaining perimeter of compute with full-distance patches to buffer hot storage output and serve as storage that can be accessed simultaneously with hot storage.

The compute region alternates between a `factory' phase and a `work' phase.
In the factory phase, 
compute is occupied by a magic state factory for $d_{\ket{CCZ}}\approx5d$ cycles to  produce a $\ket{CCZ}$ state to a target error, typically $p_{\ket{CCZ}}=10^{-11}$, such as given by~\cref{table:ccz_factory}.
In the work phase, the $\ket{CCZ}$ state is consumed in the compute region, together with joint measurements on quantum data in hot storage, and occasionally, cold storage.
Our quantum circuits in this case are compiled to lattice surgery operations using only a $3\times 3$ sub-grid of the compute region -- the additional space is slack for the cases where do not provide an explicit compilation.
We choose medium-latency hot storage where $d_\text{in}=15$ is typical.
Medium-latency is particularly suitable as the $\{6d_\text{in},7d_\text{in}\}\approx\{90,105\}$ cycles to measure each set of outer code stabilizers is performed concurrently with the factory phase of duration $\approx125$ cycles and thus fully amortized.
In practice, we have found that $5d$ cycles is sufficient to compute the vast majority of $\ket{CCZ}$ states, including the fact that hook-error-suppressed lattice surgery on hot storage qubits is roughly twice as slow at $3d_\text{in}\approx 45$ instead of $d$ cycles, so $T_\text{yoke,max}=20d_\text{in}+3\cdot5d$ cycles.
On occasions where more cycles are needed, we idle the compute region to complete the next set of outer code stabilizer measurements.

In our later resource estimates (such as~\cref{fig:femoco_improvements_breakdown}) using this layout, a minimum-space dirty-ancilla variant~\cite{khattar2024riseconditionallycleanancillae} of the quantum lookup table accounts for $80\%-96\%$ of $\ket{CCZ}$ consumption.
The minimum dimensions of $3\times 3$ is the smallest amount of space with which we were able to compile this operation into lattice surgery operations using $\le5d$ cycles per $\ket{CCZ}$ state.
As the dirty-ancilla quantum lookup table involves a non-trivial sequence of measurements between compute and hot storage, this should be representative for consuming $\ket{CCZ}$ states in other quantum circuit components -- we later work out in some detail the lattice surgery overhead of other subdominant components.
\begin{figure}
    \centering
    \includegraphics[width=0.9\linewidth]{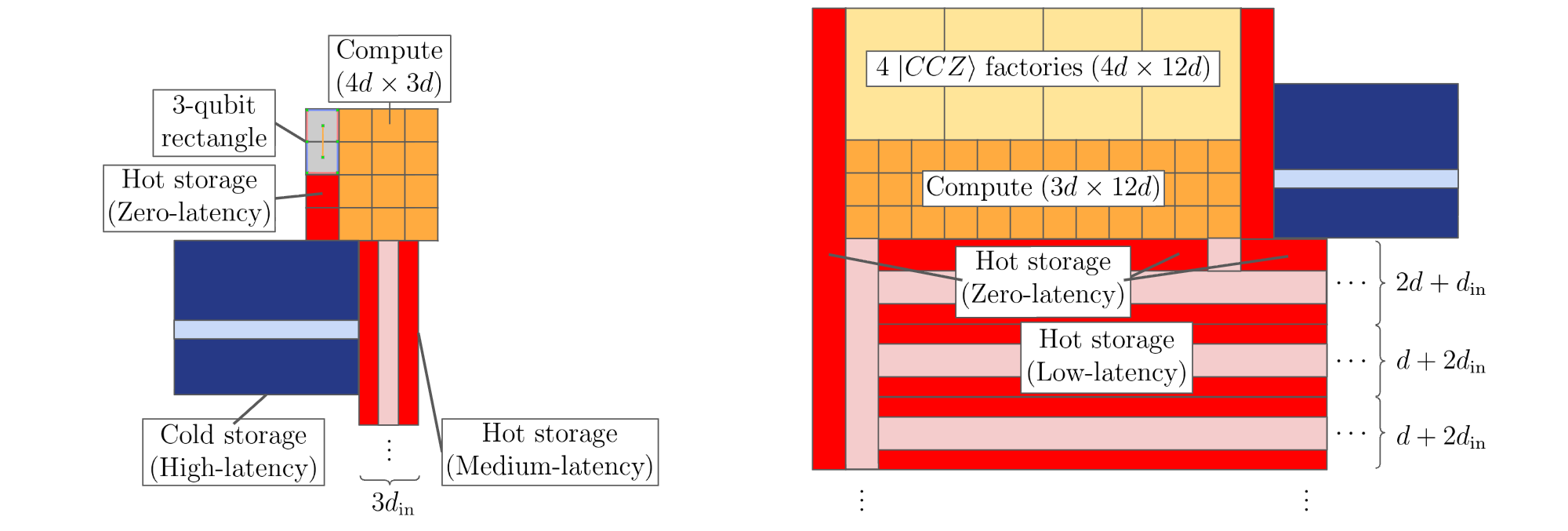}
    \\
    \vspace{0.5cm}
    \includegraphics[width=0.9\linewidth]{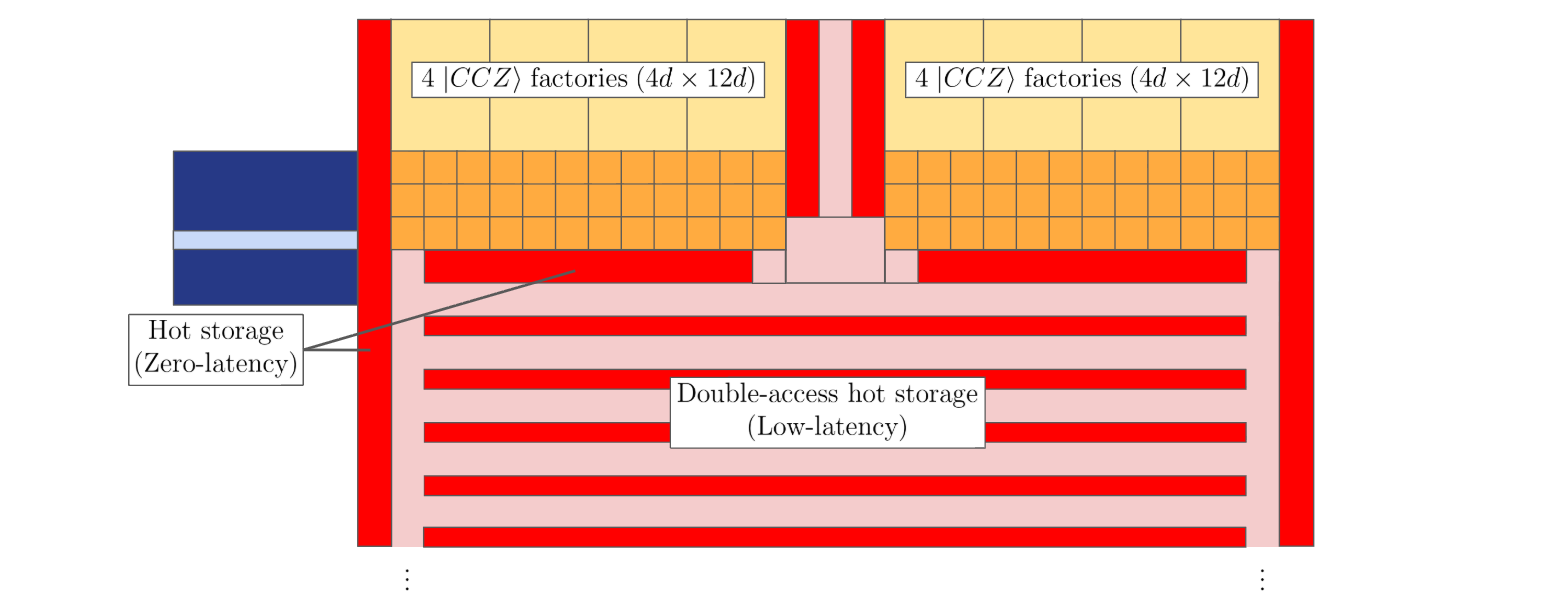}
    \caption{Example layouts drawn to scale with $d=25, d_\text{in}=15$. (Left) Minimum-space layout. 
    The compute (orange) region consists of at least $3\times 3$ distance $d$ rotated surface code patches.
    Compute alternates between a `factory' phase for producing a $\ket{CCZ}$ state and a `work' phase for consuming it.
    The factory phase occurs concurrently with hot storage (red) outer code stabilizers measurements using the access hallway (light red).
    In the work phase, each hot storage access with hook error suppression costs $3d_{\text{in}}$ or $4d_{\text{in}}$ cycles. 
    In the rare cases where cold storage interacts with compute, a passage linking the cold storage access hallway (light blue) and hot storage access hallway is made by moving at most two qubits out of the way.
    (Right) Minimum-spacetime layout with $4$ $\ket{CCZ}$ factories, with different variations of hot storage, and cold storage. We maximize space efficiency by lining access hallways with regular patches that serve as zero-latency hot storage.
    (Bottom) Minimum time layout featuring double-access hot storage and $\ge8$ $\ket{CCZ}$ factories.}
    \label{fig:example_layout}
\end{figure}

\subsubsection{Minimum-spacetime layout}\label{sec:minimum_spacetime}
This layout dedicates sufficient compute to operate $\ket{CCZ}$ factories continuously, and allocates enough additional space to consume $\ket{CCZ}$ at least as fast as they are produced, up to a maximum of $1$ per logical timestep of $d$ cycles.
Runtime is then limited by the rate of $\ket{CCZ}$ production.
The number of $\ket{CCZ}$ factories is variable and is at least one.
In our later resource estimates, $d\approx 23$ is typical.
Hence maximum throughput is achieved with around $6$ factories with the $3\times 4$ footprint.
We pair this layout with low-latency hot storage with distance $d\times d_\text{in}$ inner patches, where $d_\text{in}=15$ is typical, such as in~\crefpos{fig:example_layout}{right}.
There will be many opportunities to measure outer code stabilizers.
For instance, when $\ket{CCZ}$ production is low, e.g. $d_{\ket{CCZ}}\ge\frac{4}{3}d$, we may choose $T_\text{yoke}=4d$ as there will be one out of four timesteps where no $\ket{CCZ}$ is available for consumption so hot storage idles, and we may measure yokes.
With more factories, we insert an artificial delay to measure yokes and find that $T_\text{yoke}=50d$ is a convenient choice, though this could be further optimized.
We surround the compute region with regular surface code patches that serve as zero-latency hot storage to support cases where parallel $\textsc{PauliPM}$s are required.

In larger instances of this layout, cumulative cold storage latency can become a significant fraction of runtime.
We discuss in~\cref{sec:latency} how this can be mitigated by adding a modest number of buffer qubits.

\subsubsection{Minimum-time layout}\label{sec:minimum_time_layout}
The time-optimal layout uses a large amount of space to produce and consume $\ket{CCZ}$ states at a rate of one every $d/2$ cycles when implementing a quantum lookup table.
This uses roughly 10-12 CCZ factories and double-access low-latency hot storage.
Despite the large space requirements, this layout can actually be the most spacetime efficient in some regimes due to the high per-operation efficiency of double-access low-latency hot storage~\crefpos{fig:hot_storage_footprint}{right}.
As the number of logical qubits grows, other quantum circuit components can dominate Toffoli cost, in particular phase gradient adders that implement multiplexed rotation.

\subsection{Block-encoding with high-latency memories}\label{sec:block_encoding_communication}
Our architecture with dense but high-latency memory is particularly suited to the very general class of quantum simulation algorithms based on block-encodings~\cite{Low2016HamSim,low2019hamiltonian,GSLW2019QSVT} that incorporate a large amount of classical data.
A finite-dimensional $n$-qubit operator $H\in\mathbb{C}^{N\times N}$ is block-encoded with a positive normalization factor $\lambda\ge\|H\|$ if there is a unitary quantum circuit 
\begin{align}
\Be\left[{H}/{\lambda}\right]=
\left(
\begin{array}{cc}
H/\lambda&\cdots\\
\vdots&\ddots
\end{array}
\right)
\quad\implies\quad
(\bra{0}_{\text{l}}\otimes \mathcal{I})\Be\left[{H}/{\lambda}\right](\ket{0}_{\text{r}}\otimes \mathcal{I}) = H/\lambda,
\end{align}
where $H/\lambda$ is obtained by the projection on all-zero strings $\ket{0}_\text{l}$ and $\ket{0}_\text{r}$.
A typical block-encoding of $H=\sum_{j}|\alpha_j| U_j$ by linear combination of unitaries involves two quantum circuits:
\begin{align}\label{eq:rectangular_BE_primitives}
\textsc{Be}\left[\frac{H}{|\alpha|_1}\right]=\textsc{Prep}^\dagger\cdot\textsc{Sel}\cdot\textsc{Prep},\quad
\textsc{Sel} \coloneqq \sum_{j=0}^{L-1}\ket{j}\bra{j}_{\mathrm{a}}\otimes U_j,
\quad
\textsc{Prep}\ket{0}_\mathrm{a}=\frac 1 {\sqrt{|\alpha|_1}} \sum_{j=0}^{L-1}{\sqrt{|\alpha_j|}}\ket{j}_\mathrm{a}.
\end{align}
This creates a natural partition of the quantum circuits~\cref{eq:rectangular_BE_primitives} into components where $\textsc{Prep}$ is implemented entirely in the low-latency compute region, and we choose $H$ to act on a quantum state residing in dense, high-latency memory.
Hence, the cost of $\textsc{Sel}$ is characterized by the quantum communication complexity of moving logical qubits between dense memory and compute: This is precisely the number of $\textsc{PauliPM}_k$ instructions in~\cref{sec:latency}, which is more traditionally expressed in terms of Bell pairs shared between the two regions.

The key idea is that operators $H$ comprised of many $L$ terms can be block-encoded with communication complexity that scales with $\log L$.
This implies that only a few $\textsc{PauliPM}_k$ instructions need to be executed, which overall minimizes the excess runtime and error contribution of latency.
Although block-encodings can require a considerable amount of ancilla qubits, the ability to store large quantum states in dense memory can overall lead to a much smaller physical qubit footprint despite having many more logical qubits than algorithms such as Trotterization with $\Omega(L)$ communication complexity.

Our following result uses even fewer $\mathcal{O}(ln)$ units of quantum communication to block-encode any Hamiltonian that is any sum of $l$-local Pauli terms.
Crucially, the number of terms $L=\mathcal{O}(3^l\binom{n}{l})$ can be exponentially larger than $l$, yet the method requires a compute region of only $\mathcal{O}(l\log n)$ qubits.
The same result holds also for sums of products of up to $l$ Majorana operators.
We note that some results are known on the quantum communication complexity of block-encoding, such as a trivial $\mathcal{O}(n)$ upper bound~\cite{Montanaro2024BEcommunication} using an $\mathcal{O}(n)$-sized workspace, and these are typically presented in the context of distributed quantum computation.

\begin{theorem}[Communication complexity of block-encoding $l$-local Hamiltonians\label{thm:communication_BE}]
Let $\mathcal{S}$ be an $n$-qubit storage register, and let $\mathcal{C}$ be a $\mathcal{O}(l\log{n})$-sized compute register.
Let $H=\sum_{j}\alpha_jP_j$ be any linear combination of $l$-local Paulis, or products of up to $l$ Majorana operators acting on $\mathcal{S}$. 
Then the block-encoding $\Be[H/\alpha]$ can be constructed with $\mathcal{O}(ln)$ quantum communication complexity between $\mathcal{S}$ and $\mathcal{C}$.
\end{theorem}
\begin{proof}
Let us block-encode by linear combination of unitaries~\cref{eq:rectangular_BE_primitives} with 
\begin{align}
H=\sum_{\vec{q},\vec{j}}|\alpha_{\vec{q},\vec{j}}|U_{q_0,j_0}U_{q_1,j_1}\cdots U_{q_{l-1},j_{l-1}},
\end{align}
where $q_i\in[n]$ indexes qubits in $\mathcal{S}$ and $j_i$ indexes the type of term.
For Paulis, $U_{q,i}\in\{I_q,X_q,Y_q,Z_q\}$, and for Majorana operators, $U_{q,i}$ is either $\gamma_{q,0}$ or $\gamma_{q,1}$, where the anticommutator $\{\gamma_{p,x},\gamma_{q,y}\}=2\delta_{pq}\delta_{xy}$.
For this proof, we assume a Jordan-Wigner representation where $\gamma_{q,0}=X_qZ_{q+1}\cdots Z_{n-1}$ and $\gamma_{q,1}=Y_qZ_{q+1}\cdots Z_{n-1}$.
Let us define $\textsc{Sel}_{i}=\sum_{q,j}\ket{q,j}\bra{q,j}_{i}\otimes U_{q,j}$ to be controlled by the $i^\text{th}$ index register $\ket{q,j}_i$ with $\mathcal{O}(\log n)$ qubits.
Then we define $\textsc{Sel}=\prod_{i=0}^{l-1}\textsc{Sel}_{i}$, which has a control register with $\mathcal{O}(l\log n)$ qubits.
Note that any control register state $\otimes_{i=0}^{l-1}\ket{q,j}_i$, which has dimension $\mathcal{O}(n^l4^l)$, can be prepared by a quantum circuit $\textsc{Prep}$ using $\mathcal{O}(l\log n)$ space~\cite{shende2006synthesis}.

Hence, the proof reduces to showing that $\textsc{Sel}_i$ can be implemented in $\mathcal{O}(\log n)$ space, and $\mathcal{O}(n)$ units of quantum communication.
We provide a quantum circuit with sub-optimal gate complexity for the purposes of simplifying the proof -- the unary iteration gadget~\cite{BabbushPRX18} provides an optimal implementation with $\mathcal{O}(n)$ gates.
This is accomplished by iterating over all indices $\ket{q,j}_i\equiv\ket{q}_{i,\text{qubit}}\ket{j}_{i,\text{type}}$.
Let us start with the case of $U_{q,j}\in\{I,X,Y,Z\}$. 
\begin{enumerate}
    \item For $q=0,1,\cdots,n-1$:
    \begin{enumerate}
        \item Apply a multi-controlled Toffoli gate that checks whether $\ket{\cdots}_{i,\text{qubit}}=\ket{q}$. 
        If true, apply $X$ on an output register that starts in $\ket{0}_\text{out}$.
        \item Teleport qubit $q$ from $\mathcal{S}$ to a one-qubit register $\ket{\cdots}_{\text{tmp}}$ in $\mathcal{C}$ using $\Theta(1)$ units of quantum communication.
        \item Controlled by $\ket{1}_\text{out}$, then controlled by $\ket{j}_{i,\text{type}}$, apply $U_{q,j}$ to $\ket{\cdots}_{\text{tmp}}$.
        This can be implemented such as using another multi-controlled Toffoli gate.
        \item Uncompute the multi-controlled Toffoli gate in step (a).
        \item Teleport $\ket{\cdots}_{\text{tmp}}$ back to $\mathcal{S}$.
    \end{enumerate}
\end{enumerate}
The case of Majorana operators replaces the multi-controlled Toffoli in step 1a) with a comparator that checks whether the index in $\ket{\cdots}_{i,\text{qubit}}$ is equal to $q$, in which case a controlled-$X_q$ or $Y_q$ is applied depending on $\ket{\cdots}_{i,\text{type}}$, or whether the index is $<q$, in which case a controlled-$Z_q$ is applied. If neither condition is met, the identity operation is applied.
If signs or phases are required in the $U_{q,j}$, this can be implemented by having \textsc{Prep} additionally output a binary representation $\ket{\theta_{q,j}}$ of the phase, and using, for instance, a phase gradient adder~\cite{Gidney2018halvingcostof} to convert $\ket{\theta_{q,j}}\rightarrow e^{i2\pi\theta_{q,j}}\ket{\theta_{q,j}}$.

The proof is completed by noting that a multi-controlled Toffoli and the comparator can be implemented with at most a constant number of ancillae~\cite{khattar2024riseconditionallycleanancillae}.
\end{proof}
Overall,~\cref{thm:communication_BE} implies that the number of quantum gates spent implementing $\textsc{Prep}$ can scale like $\mathcal{O}(n^l)$, whereas $\textsc{Sel}$ scales like $\mathcal{O}(ln)$, but needs $\mathcal{O}(ln)$ units of quantum communication. 
In other words, the $\mathcal{O}(ln)$ cycles spent on memory latency, which may come with a large constant factor, can easily be dominated by $\mathcal{O}(n^l/(l\log n))$ logical cycles, with a smaller constant factor, for $\textsc{Prep}$ even assuming perfect parallelization.
Moreover, the amortized logical cycles spent on memory latency can be close to zero as memory access operations can occur in parallel with a careful sequencing of long compute operations. 

%% file: 4_compilation/quantum_circuits.tex
\begin{figure}
    \centering
    \includegraphics[width=0.9\linewidth]{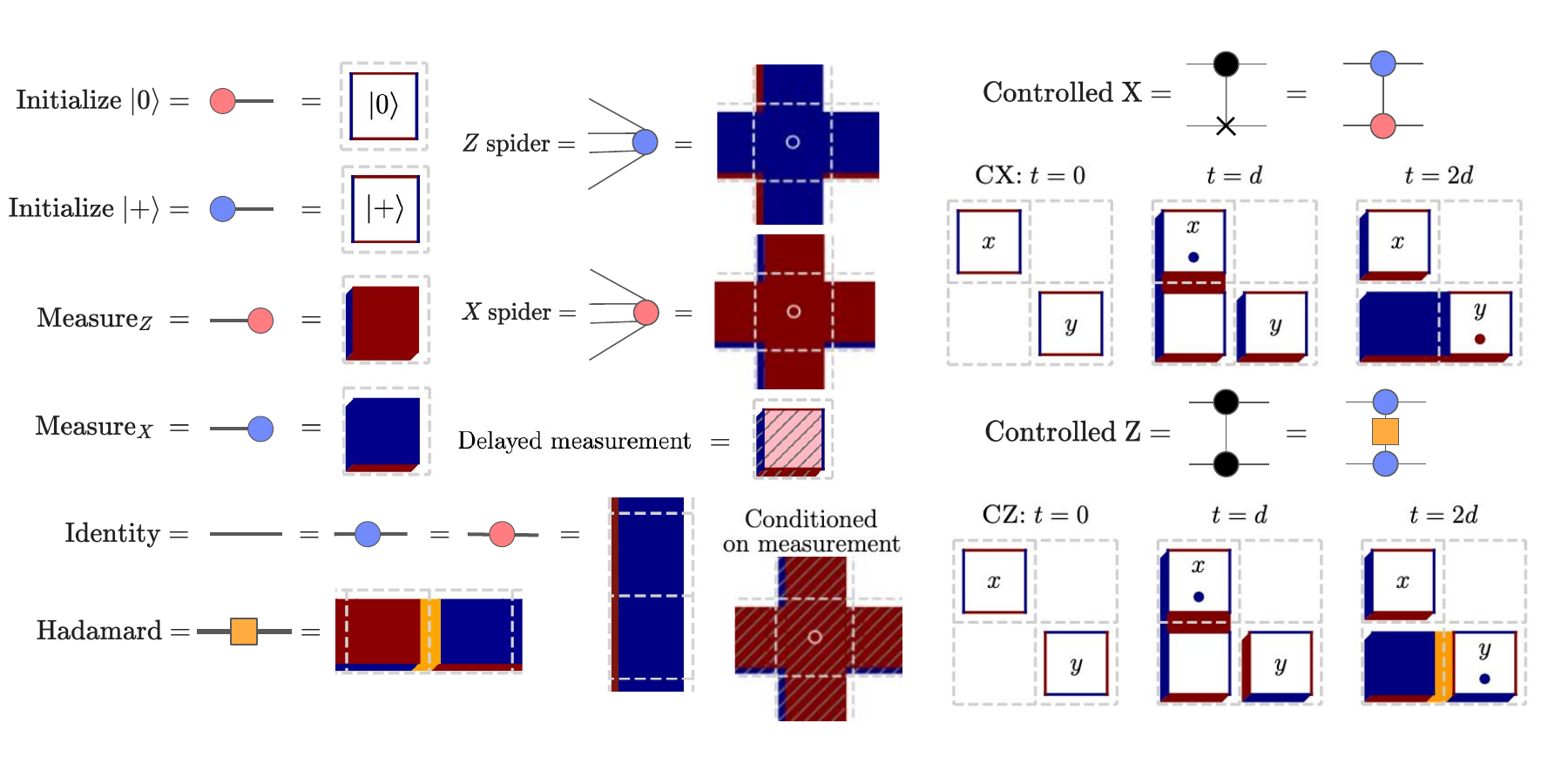}
    \caption{Map between quantum circuit, $ZX$ diagram, and lattice surgery primitive operations assuming $d$ surface code cycles per logical timestep. 
    Pipe diagram vertices represent a $ZX$ diagram node with a color representing the type of joint measurement and is determined by the unbent side. 
    Pipes with an open white face travel out of the page, corresponding to the positive time direction.
    Compare the pipe diagram of a logical $\textsc{CX}$ gate $\ket{x}_0\ket{y}_1\rightarrow\ket{x}_0\ket{y\oplus 1}_1$ with the physical gate-level implementation in~\cref{fig:CNOT_X_Z}. 
    At cycle $t=d$, measure $Z_0Z_a$ of $\ket{x}_0\ket{+}_a$. At $t=2d$, measure $X_aX_1$ of $\ket{+}_a\ket{y}_1$. Any required Pauli fixup is commuted through all Clifford operators and to the very end of the quantum algorithm by Pauli frame tracking. 
    Note that the space-like Hadamard in the logical $\textsc{CZ}$ gate $\ket{x}_0\ket{y}_1\rightarrow(-1)^{x\wedge y}\ket{x}_0\ket{y}_1$ may be replaced by two transversal Hadamards before and after the $y$ block at $t=2d$.
    (Bottom middle) We use gray diagonal lines to indicate a lattice surgery operation that is classically controlled by a  measurement outcome that occurred at least $d_\text{reaction}$ cycles ago.}
    \label{fig:lattice_surgery_primitives}
\end{figure}

Quantum circuits expressed in terms of logical quantum gates such as $\textsc{CZ}$, Toffoli and arbitrary-angle rotations are compiled to the surface code using lattice surgery for the Clifford components of the circuit, and gate teleportation with magic states for the Non-Clifford components. 
As lattice surgery instantiates $ZX$ calculus~\cite{Wetering2020ZXCalculus}, in addition to some specific rules for routing, the key intermediate step is to represent quantum circuits as $ZX$ diagrams.
At this stage, we abstract away details of surface code patches and lattice surgery operations, and represent them with blocks with a unit $d\times d$ area, and height up to $d+d_\text{reaction}$ as shown in~\cref{fig:lattice_surgery_primitives}.
The sides of a block are colored by the type of patch boundary, and the top and bottom are colored by the logical state of patch initialization ($\ket{0}$,$\ket{+}$) and measurement ($Z$,$X$).
Connecting these blocks with lattice surgery operations produces a network of pipe diagrams~\cite{Gidney2019AutoCCZ,Fowler2018LowOverhead}.

In this section, we provide some simple examples of how a logical $\textsc{CCZ}$ gate is implemented by consuming a $\ket{CCZ}$ magic state in~\cref{sec:consume_ccz}, then we
provide highly optimized lattice surgery implementations of quantum circuits that dominate the spacetime volume of our later resource estimates, specifically the quantum lookup table in~\cref{sec:lattice_surgery_lookup} and multiplexed quantum rotations in~\cref{sec:multiplex_rotations}.

\subsection{Consuming a \textsc{CCZ} magic state}\label{sec:consume_ccz}
A CCZ gate is implemented by consuming a CCZ magic state $\ket{CCZ}=\sum_{a,b,c\in\{0,1\}}(-1)^{abc}\ket{abc}$ by magic state teleportation. 
After measuring the magic state qubits, a Clifford fix-up operation of up to three \textsc{CZ} gates conditioned on the measurement outcomes is required on the input state -- see references for details~\cite{Paetznick2013Universal}. 
Overall, a $\ket{CCZ}$ gate can be implemented with various tradeoffs between space and time.
For instance,~\crefpos{tab:ccz_fixup}{top-left} shows a $2d+d_\text{r}$ implementation by waiting for the reaction time $d_\text{r}$ and applying the 3 \textsc{CZ} fixups in-place.
Alternatively~\crefpos{tab:ccz_fixup}{top-right} shows an auto-$\ket{CCZ}$~\cite{Gidney2019CCZ} state that pre-computes all possible fixups and postpones the combinations of fixups with a delay-choice measurement.
This uses more space, but allows the $\ket{CCZ}$ state to be consumed $d_\text{r}$ or $d+d_\text{r}$ cycles after it is introduced from a spacelike or timelike direction, respectively.
A common quantum circuit pattern is a ladder of Toffoli (\text{CCX}) gates, such as in a lookup-table. 
In this case, one of the fixups can be postponed and at most two need to be performed immediately.
This optimization allows fixups to be performed in fewer cycles and less time for both in-place and delayed-choice variants in~\crefpos{tab:ccz_fixup}{middle}.
The more complicated controlled-swap (\textsc{CSwap}) is shown in~\crefpos{tab:ccz_fixup}{bottom} using either an auto-$\ket{CCZ}$ state or a regular $\ket{CCZ}$ state:
both cases block an access hallway for $2d$ cycles.
\begin{table}[h]
    \centering
    \begin{tabular}{c|cc}
    \hline\hline
         $\textsc{CCZ}$ gate &In-place fixup on all qubits& Delayed-choice fixup on all qubits
         \\
         \hline
         Cycles &$2d+d_\text{reaction}$& $d+d_\text{reaction}$
         \\
         Footprint &$3\times 3$& $3\times 4$
         \\
         \shortstack{Pipe\\diagram} &
         {\includegraphics[scale=0.3,valign=m]{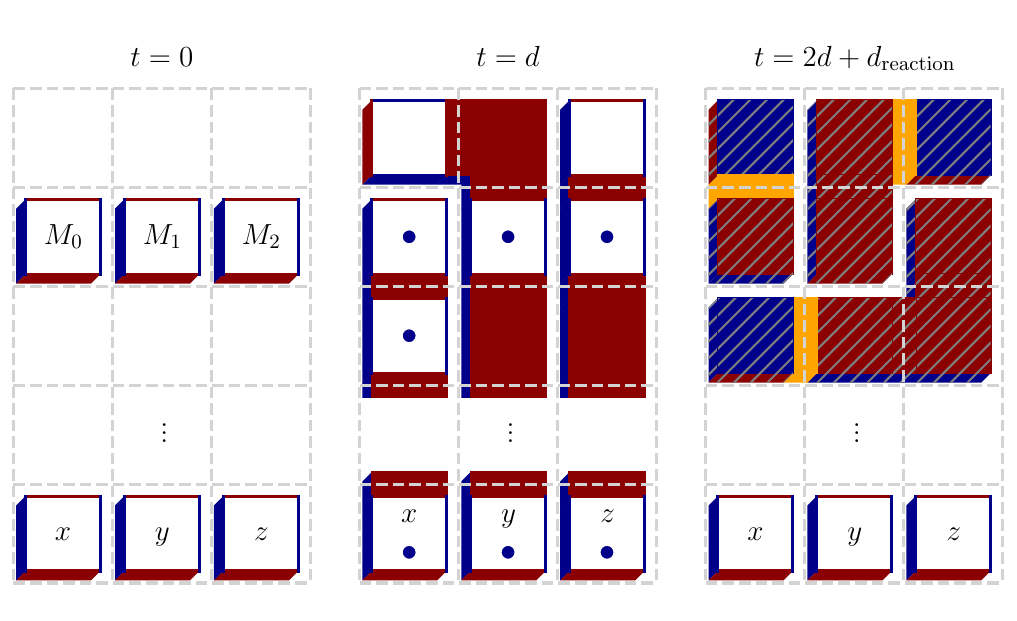}} 
         & See reference~\cite{Gidney2019CCZ}
         \\
         \hline\hline
         $\textsc{CCX}$ gate & In-place fixup only on target qubit $\ket{y}$& Delayed-choice fixup only on target qubit $\ket{y}$
         \\
         \hline
         Cycles  &$d+d_\text{reaction}$& $d_\text{reaction}$ 
         \\
         Footprint &$2\times 3$& $3\times 3$
         \\
         \shortstack{Pipe\\diagram} &{\includegraphics[scale=0.35,valign=m]{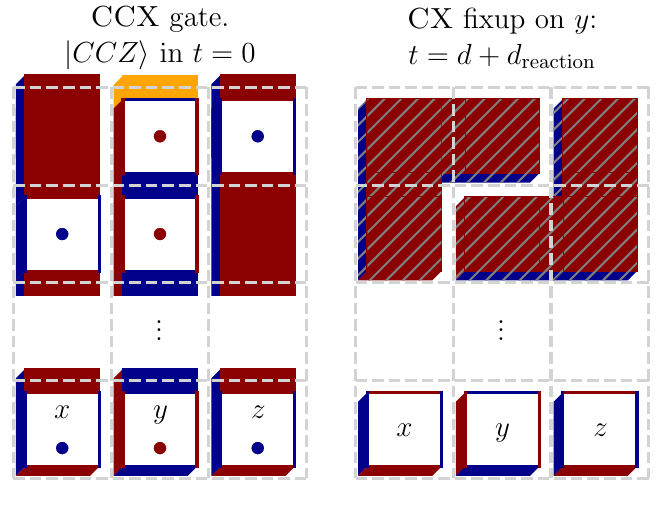}}& {\includegraphics[scale=0.35,valign=m]{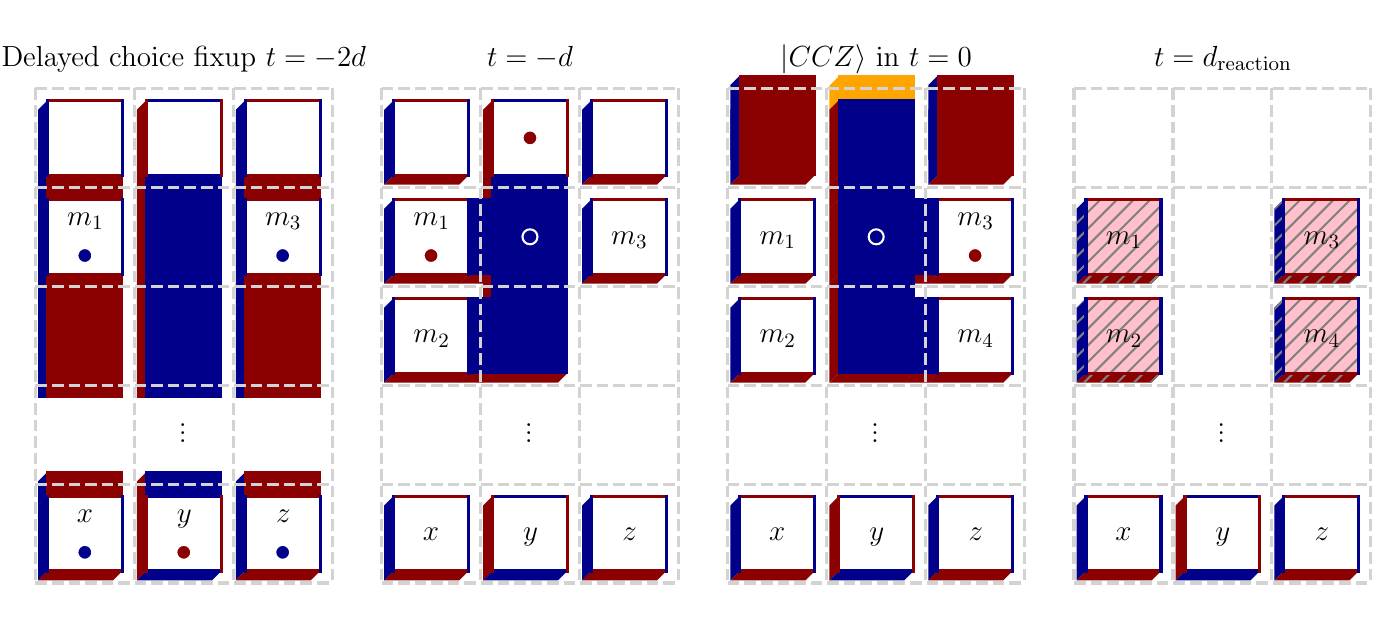}}
         \\
         \hline\hline
         $\textsc{CSwap}$ & $ZX$-diagram (top left) & With (top right) and without (bottom) auto-$\ket{CCZ}$
         \\
         \hline
         Cycles  & & Access hallway blocked for $2d$ cycles
         \\
         Footprint && $2\times 3$
         \\
         \shortstack{Pipe\\diagram} &\multicolumn{2}{c}{{\includegraphics[width=0.3\linewidth,valign=m]{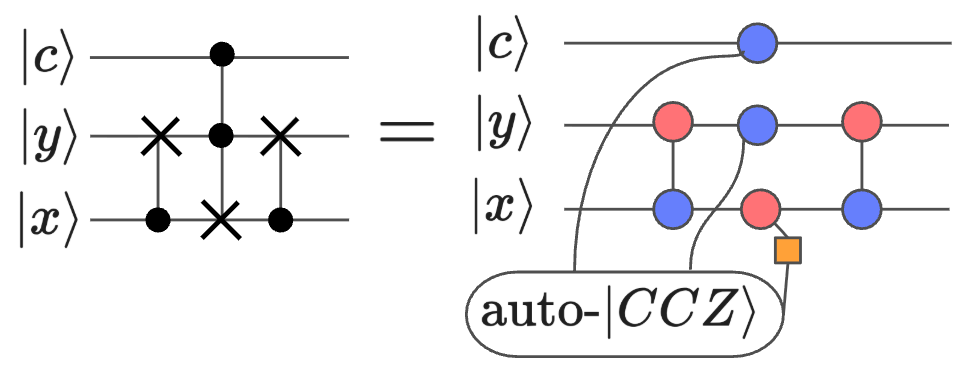} \includegraphics[scale=0.35,valign=m]{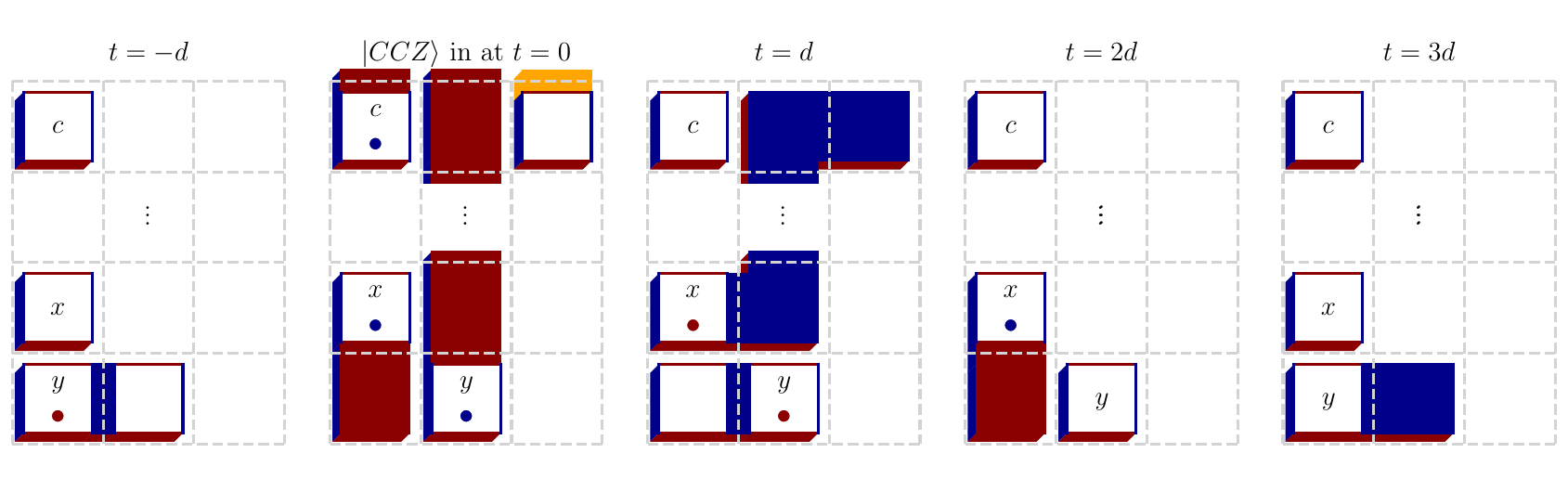}}}
         \\
         &\multicolumn{2}{c}{\includegraphics[width=0.88\linewidth,valign=m]{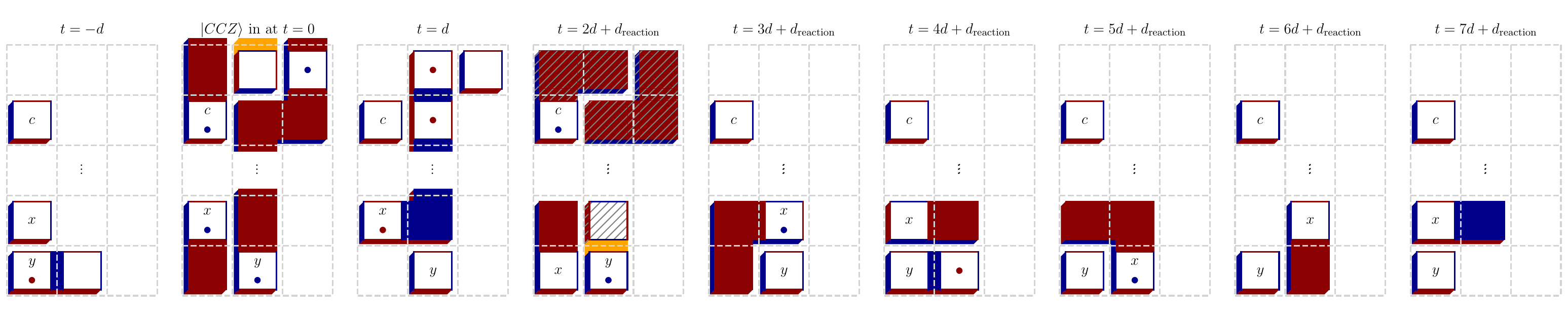}}
         \\
     \hline\hline
    \end{tabular}
    \caption{Examples of (top) $\textsc{CCZ}$ gates $\ket{x}\ket{y}\ket{z}=(-1)^{x\wedge y\wedge z}\ket{x}\ket{y}\ket{z}$ by consuming a $\ket{CCZ}$ state coming from a timelike direction on patches $M_{0},M_1,M_2$ and (middle) $\textsc{CCX}$ gates $\ket{x}\ket{y}\ket{z}=\ket{x}\ket{y\oplus(x\wedge z)}\ket{z}$ by measurement based computation using different fixup options with the $\ket{CCZ}$ state coming from a spacelike direction. 
    (Bottom) Controlled-\textsc{Swap} gate with or without an auto-$\ket{CCZ}$ that is consumed in two amortized timesteps with one access hallway.
    Any subsequent controlled-swap can be performed on any pair of qubits above $\ket{x},\ket{y}$ without being blocked by in-progress move operations.
    }
    \label{tab:ccz_fixup}
\end{table}

\subsection{Space-efficient quantum lookup table}\label{sec:lattice_surgery_lookup}
A quantum lookup table is a unitary quantum circuit defined to have the properties of~\cref{def:lookup}.
Some lookup table quantum circuits optimized for Toffoli count~\cite{BabbushPRX18,Low2024Trading} can be interpreted as traversing a balanced binary tree.
In this work, we focus on compiling to lattice surgery operations a variant that instead traverses a skew-tree~\cite{khattar2024riseconditionallycleanancillae}, which has several advantages.
First, dirty-qubit skew-trees overall use the fewest number of ancilla qubits. We use the variant with $\approx \frac{5}{4}X$ Toffoli gates. 
Second, half of the $\ket{CCZ}$ fixups can be parallelized with the other half, in contrast to the balanced binary approach, where all fixups are serial. 
This allows us to perform all fixups in-place without being reaction-limited for any $d\ge2d_\text{reaction}$, thus avoiding the large qubit footprint of the auto$\ket{CCZ}$ state.
Third, all skew-tree traversal steps that consume Toffoli gates occur with fixed periodicity, which simplifies compilation and tiling of pipe diagrams.
Overall, these factors allow us to compile the skew-tree table lookup to significantly smaller footprints summarized in~\cref{table:lattice_surgery_lookup} than prior work based on a balanced binary tree.
\begin{definition}[Quantum lookup table\label{def:lookup}] For any integers $b,X>0$, and any $\mathfrak{data}\subseteq[X]\times\{0,1\}^{b}$, the table-lookup unitary $\textsc{Lookup}_{\mathfrak{data}}$ implements
\begin{align}
&\forall z\in\{0,1\}^{b},\quad\textsc{Lookup}_{\mathfrak{data}}\ket{x}\ket{z}
=
\begin{cases}
\ket{x}\ket{z\oplus \mathfrak{data}_{x}}, & x\in[0,X),\\
\ket{x}\ket{z}, & x\in[X,2^{n_X}),
\end{cases}
\quad
n_X\doteq\lceil\log_2X\rceil.
\end{align}
\end{definition}

For example, we are able to compile clean-ancilla skew-tree lookup with a $\ket{CCZ}$ state consumption rate of one per $d/2$ cycles using a compute workspace with the dimensions of $3\times 24=72$ surface code patches, excluding the $\ket{CCZ}$ factories.
In contrast, an explicit layout in prior art~\cite{Gidney2019AutoCCZ,PRXQuantum.2.030305} with the same throughput requires at least $9\times 72=648$ patches.
We are also able to compile clean- and dirty- ancilla skew-tree lookup to a compute region as small as $3\times2$ and $3\times 3$ patches respectively.
Based on these explicit examples, our later resource estimates assume that it is possible to compile clean-ancilla skew-tree lookup with workspace and $\ket{CCZ}$ consumption rate parameters that interpolate linearly between a $3\times 3$ compute workspace consuming $1$ $\ket{CCZ}$ every $4d$ cycles, and the $3\times 24$ compute workspace consuming $1$ $\ket{CCZ}$ every $d/2$ cycles.

\begin{figure}
    \centering
    \includegraphics[width=0.3\linewidth,valign=m]{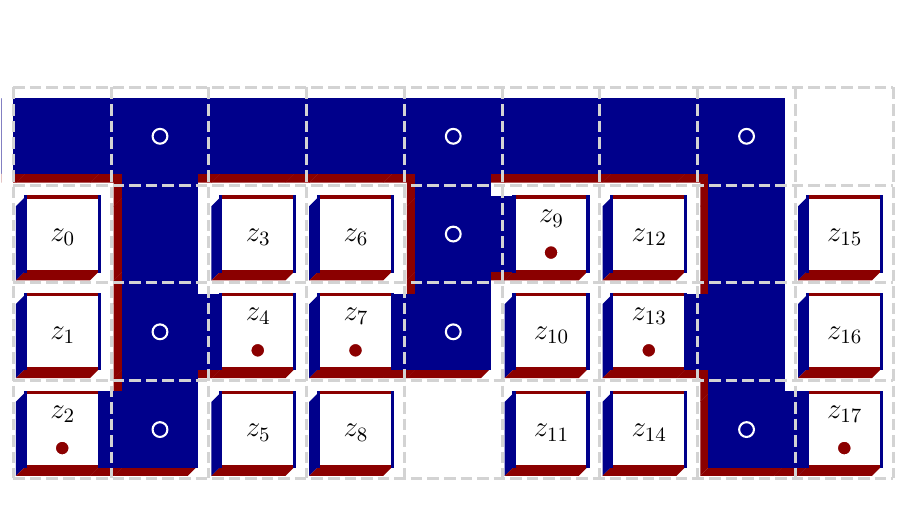}
    \hspace{0.5cm}
    \includegraphics[width=0.3\linewidth,valign=m]{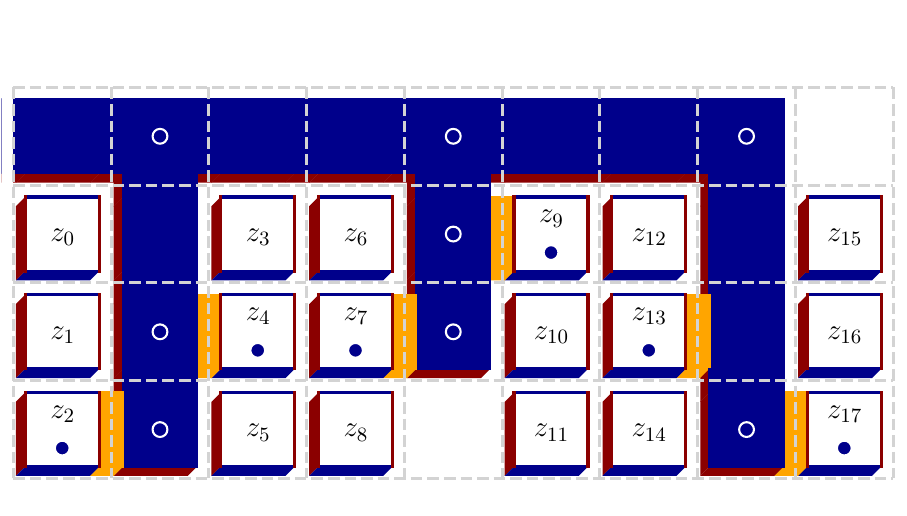}
    \hspace{0.5cm}
    \includegraphics[width=0.3\linewidth,valign=m]{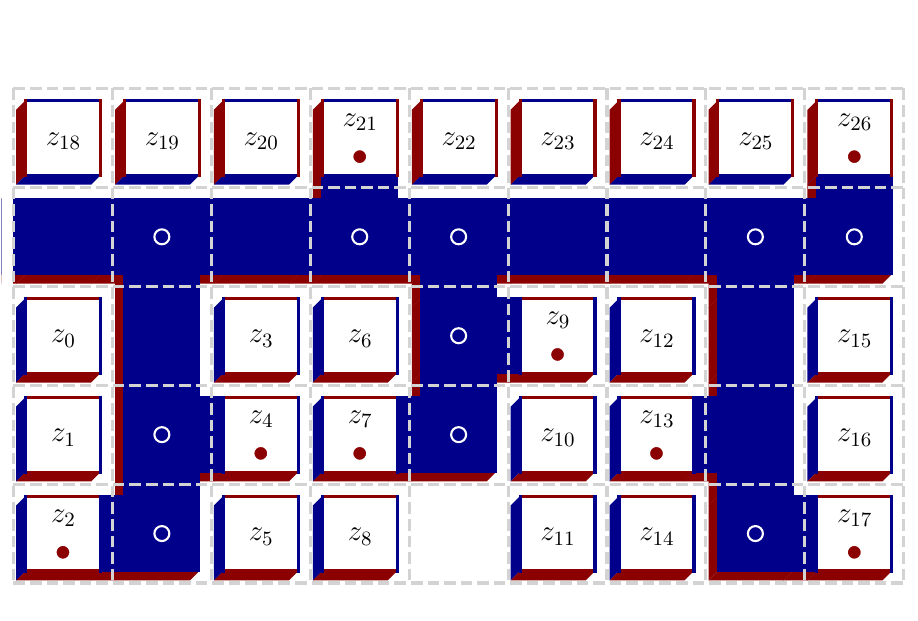}
    \caption{Pipe diagram showing how lookup table output bits are written to hot storage. The example shown is $6$ columns of single-access zero-latency hot storage, or just columns of unconcatenated surface code patches with data $\mathfrak{d}_x=(\mathfrak{d}_{x,j})_{j=0}^{b-1}$, where $b=18$ and $\mathfrak{d}_{x,j}=1$ if $j\in\{2,4,7,9,13,17\}$ and zero otherwise.
    Using a single access-hallway, all $b$ bits for each address $x$ can be written every $d$ cycles.
    With low-latency hot storage, the square patches $z_j$ become narrower rectangles of the same height, and lattice surgery targets logical operators that may be distributed across multiple patches in each column.
    With medium-latency hot storage, all patches are smaller distance $d_\text{in}$ square, and hook-error suppressed lattice surgery writes all $b$ bits using $4d_\text{in}$ cycles if targeting both columns next to a hallway and $3d_\text{in}$ cycles if targeting only a single side.
    Data can be written with either (left) space-like \textsc{CX} gates when $\ket{z}\in\{\ket{0},\ket{1}\}$ or (right) space-like \textsc{CZ} gates when $\ket{z}\in\{\ket{+},\ket{-}\}$.
    (Right) Where applicable, we increase density of storage with additional patches on exposed sides of any shared access hallway within a rectangular bounding box.
 }
 \label{fig:output_bits}
\end{figure}

The actual average cycles $d_\text{lookup}\ge d_\text{lookup,min}$ to consume each $\ket{CCZ}$ state will be larger than the theoretical minimum.
First, it is limited by the average production rate $d_\text{lookup}\ge d_{\ket{CCZ}}$ across some number $F$ of factories.
Second, $d_\text{lookup}\ge d$ if output bits are written to hot storage with a single access hallway for every two columns of patches like in~\cref{fig:output_bits}.
Third, the access hallway is blocked during hot storage outer code stabilizer measurements.
With low- and medium- latency hot storage, these blocks occurs in segments of $\mathfrak{Blocks}=\{d\}$ and $\mathfrak{Blocks}=\{6,7,7\}d_\text{in}$ cycles every $T_\text{yoke}$ cycles respectively. 
Zero-latency hot storage is trivially two columns of full distance patches with $\mathfrak{Blocks}=\emptyset, \;T_\text{yoke}=\infty$.
We model this as $d_\text{lookup}\ge d/\left(1-\frac{\sum_j\mathfrak{Blocks}_j}{T_\text{yoke}}\right)$.
Overall,
\begin{align}\label{eq:lookup_actual_cycles_ccz_consumed}
\text{Cycles per $\ket{CCZ}$ consumed with single-access}: \quad d_{\text{lookup}}=\max\left\{d_{\text{lookup,min}},d_{\ket{CCZ}},d/\left(1-\frac{\sum_j\mathfrak{Blocks}_j}{T_\text{yoke}}\right)\right\}.
\end{align}
One may also write bits directly to the logical operators on exposed sides of denser storage, such as the three-qubit rectangle.

\begin{table}
\begin{tabular}{c|c|c|c|c|c|c|c}
\hline\hline
\multicolumn{3}{c|}{Lookup table} & \multirow{2}{*}{$F$}& Compute & \multicolumn{2}{c|}{Toffolis} & Cycles to apply $A$ Toffolis 
\\
Method&Output storage&Proof&&patches&($A$) &Total& $(=A \times d_{\text{lookup,min}})$ 
\\
\hline
\multirow{2}{*}{Dirty}&Medium-latency&\multirow{2}{*}{\cref{fig:zx_skew_tree_dirty}}& \multirow{3}{*}{1}& \multirow{2}{*}{$\max\{3\times 3,n_\text{f}\}$}  & \multirow{2}{*}{5} &\multirow{2}{*}{$\frac{5}{4}X$}& $\approx 5(6d_\text{in}+\max\{10d_\text{in},d_\text{ccz}+d\}$
\\
\cline{2-2}
\cline{8-8}
 &Single access && & &&& $5d_\text{ccz}+17d+5d_\text{reaction}$
\\
\cline{1-1}\cline{3-3}\cline{5-8}
\multirow{6}{*}{Clean} &low-latency&\cref{fig:skew_tree_clean_ZX}& & $2\times3+n_\text{f}$  & 3 &\multirow{6}{*}{$X$}& $3\max\{d_\text{ccz},3d+d_\text{reaction}\}+\max\{d_\text{ccz},3d\}$
\\
\cline{3-8}
&or zero-latency&Interpolation& $2,3,5,6$&& $F$ &&\multirow{5}{*}{$\begin{gathered}\max\{\min(\{F,8\})d_\text{ccz},
 \\4d,3d+2d_\text{reaction}\}\end{gathered}$}  
\\
\cline{3-4}
\cline{6-6}
 &&\multirow{2}{*}{\cref{fig:zx_skew_tree_dirty_Clifford_rate}}& $4$&  & $4$ &&
\\
\cline{2-2}\cline{4-4}\cline{6-6}
 &Double access&& $8$&$(3\times 3F)$& $8$ && 
\\
\cline{3-4}\cline{6-6}
&zero-latency&Interpolation& $6,7$&$+Fn_\text{f}$& $F$ &&  
\\
\cline{3-4}\cline{6-6}
&&Extrapolation& $>8$&& $8$ &&  
\\
\hline\hline
\end{tabular}
\caption{High-level overview of our compilation of skew-tree quantum lookup tables to lattice surgery operations as a function of number of $\ket{CCZ}$ factories $F$.
Except for the first entry which assumes a minimum-space layout~\crefpos{fig:example_layout}{left}, all other compilations assume a  minimum-spacetime layout~\crefpos{fig:example_layout}{right} with three simultaneously accessible low-latency hot storages, one each for address, ancillae, and output qubits.
Each compilation uses $F$ $\ket{CCZ}$ factories with assumed footprint of $3\times\text{(factory height)}=n_\text{f}$ that are placed on one long side of the compute region.
The minimum number of cycles to apply $A$ Toffoli is $A \times d_{\text{lookup,min}}$, but achieving this rate will usually be limited by the cycles $d_{\mathrm{CCZ}}$ per $\ket{CCZ}$.
$d_{\text{lookup,min}}$ is a conservative upper bound as it includes volume and reaction time to implement an explicit \textsc{CCX} fixup that is eliminated entirely following~\cref{sec:low_reaction_depth_lookup}.
From these results, we linearly interpolate or extrapolate to other values of $F$.
Whenever $\ket{CCZ}$ states are to be consumed faster than one every $d$ cycles, output bits require double-access: each column of patches has access hallways on both sides.
}
\label{table:lattice_surgery_lookup}
\end{table}

\begin{figure}
    \centering
     \includegraphics[width=.3\linewidth]{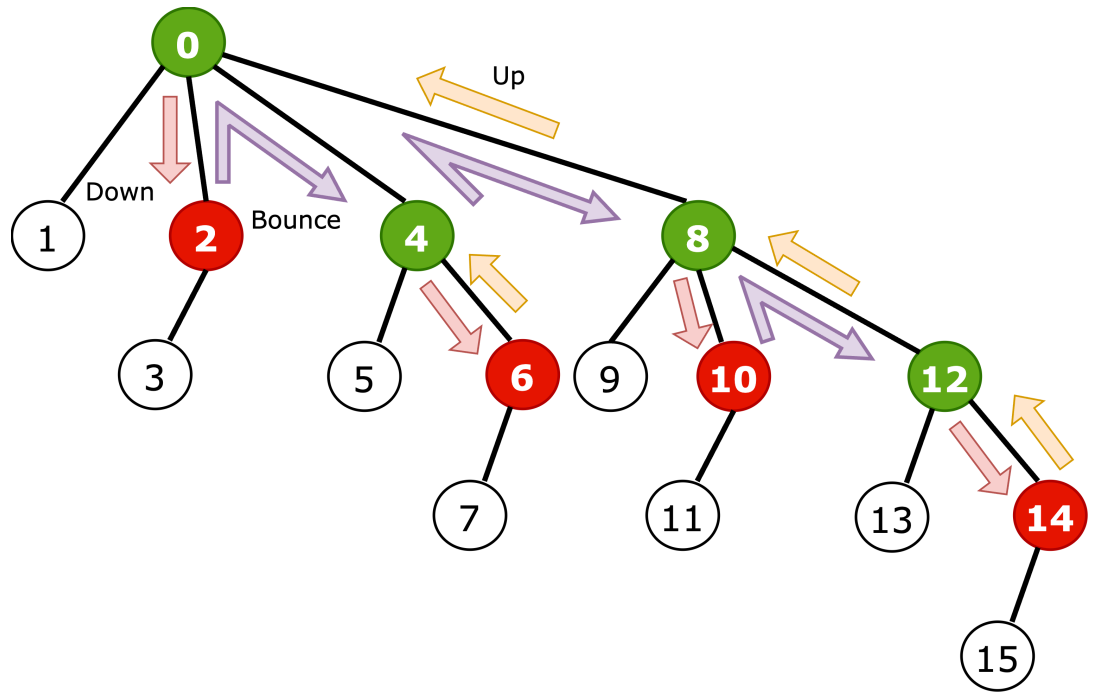}\includegraphics[width=.34\linewidth]{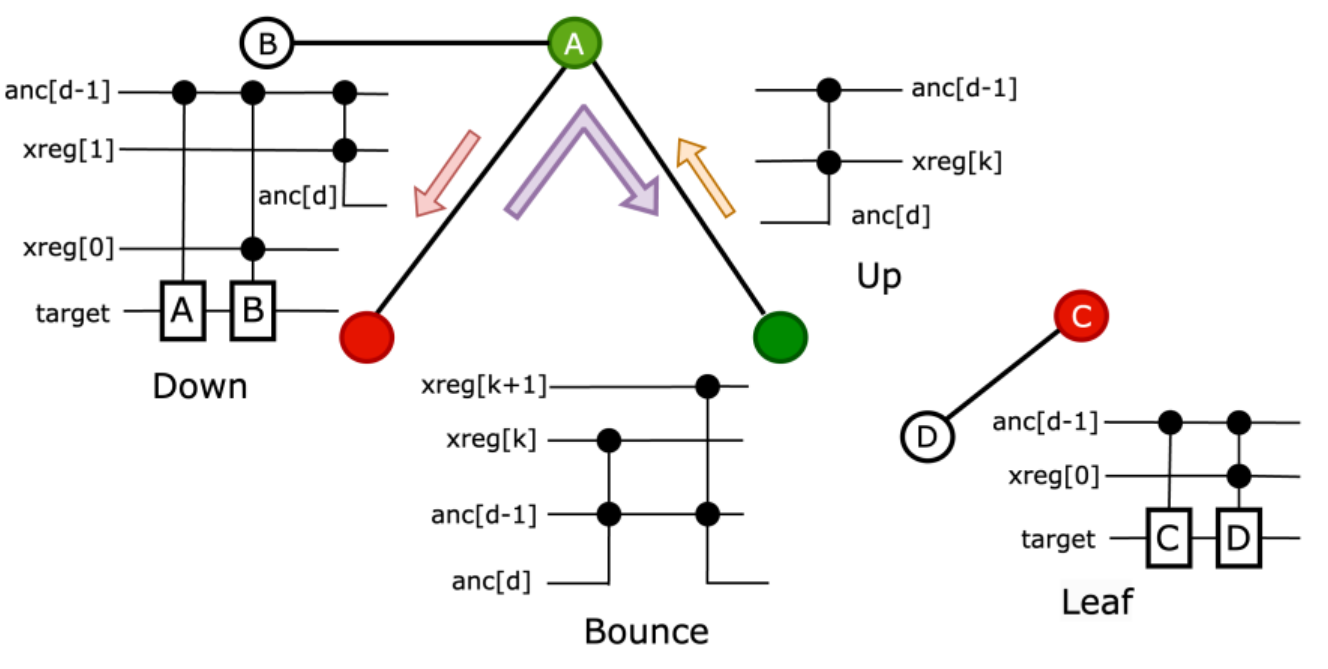}
\includegraphics[width=0.34\linewidth]{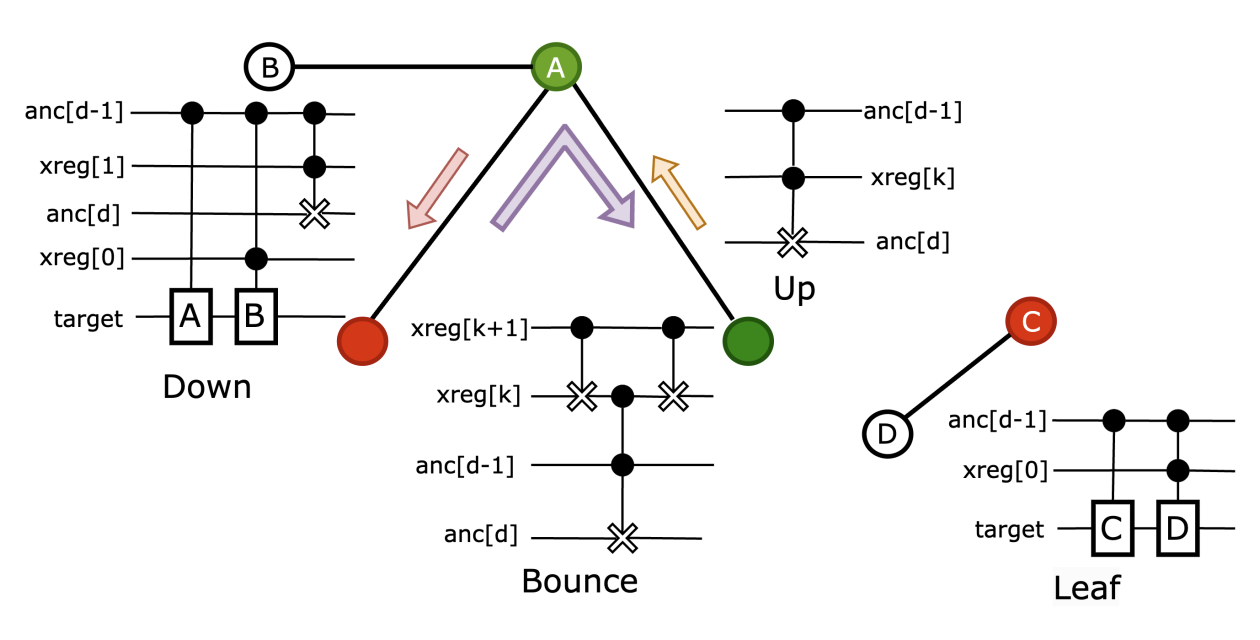}
    \caption{Reproduced verbatim from Figure. 10 of~\cite{khattar2024riseconditionallycleanancillae}: An $X$-address quantum lookup table~\cref{def:lookup} with address register $\text{xreg}[0]\cdots\text{xreg}[n_X-1]$ is implemented by (left) traversing a skew-tree, shown for $X=16$ on the left, where each move $\{$``Down'',``Leaf',``Bounce'',``Up''$\}$ is (middle) implemented with the aid of clean ancilla qubits by the corresponding  quantum circuit, or (right) with dirty ancilla.}
    \label{fig:skew_tree_clean_quantum_circuit}
\end{figure}

We now demonstrate the compilation of the simplest case: the clean-ancilla skew-tree lookup with a $3\times2$ workspace -- see~\cref{sec:lookup_appendix} for the other cases.
This begins with the definition of the skew-tree lookup circuit, which we reproduce verbatim in~\cref{fig:skew_tree_clean_quantum_circuit} from Figure.10 of~\cite{khattar2024riseconditionallycleanancillae}.
Observe that all steps that consume Toffoli gates occur with fixed periodicity (``Down'',``Leaf'',``Bounce'') as the ``Up'' step is performed using measurement. 
This greatly simplifies lattice surgery compilation as we may independently compile the quantum circuit for each step and tile them in the time-direction.
In~\cref{fig:skew_tree_clean_ZX}, we present $ZX$ diagrams for these steps, and their subsequent lattice surgery compilation.
We annotate the $ZX$ diagrams to provide information on what the lattice surgery pipe diagrams should resemble, such as reaction-time dependencies and logical qubit locations.
Although $ZX$ diagrams provide a guide for lattice surgery compilation, the final pipe diagrams may bear little resemblance to the original suggestions for how operations are ordered and connected. 
Observe that in this compilation, the $\ket{CCZ}$ factory operates continuously and injects a $\ket{CCZ}$ state from the top of the $2\times 3$ compute workspace every $d_{\ket{CCZ}}\approx 5d$ cycles.
Using three low-latency hot storages surrounding the workspace, each $\ket{CCZ}$ is then rapidly consumed in at most $3d+d_\text{reaction}$, indicating that the bottleneck with a single factory is the $\ket{CCZ}$ production rate.
Note that fixup is only required with probability $\frac{3}{4}$, which provides opportunities to further improve $\ket{CCZ}$ consumption rate, but the overall circuit will remain limited by $\ket{CCZ}$ production rates.
As a result, one may also set hot storage $T_\text{yoke}=5d$.
\begin{table}
    \centering
    \begin{tabular}{c|c|c}
    \hline\hline
         & ZX diagram & Pipe diagram \\
         \hline
         \rotatebox{90}{Down (Toffoli \#1)}&{\includegraphics[width=0.2\linewidth,valign=b]{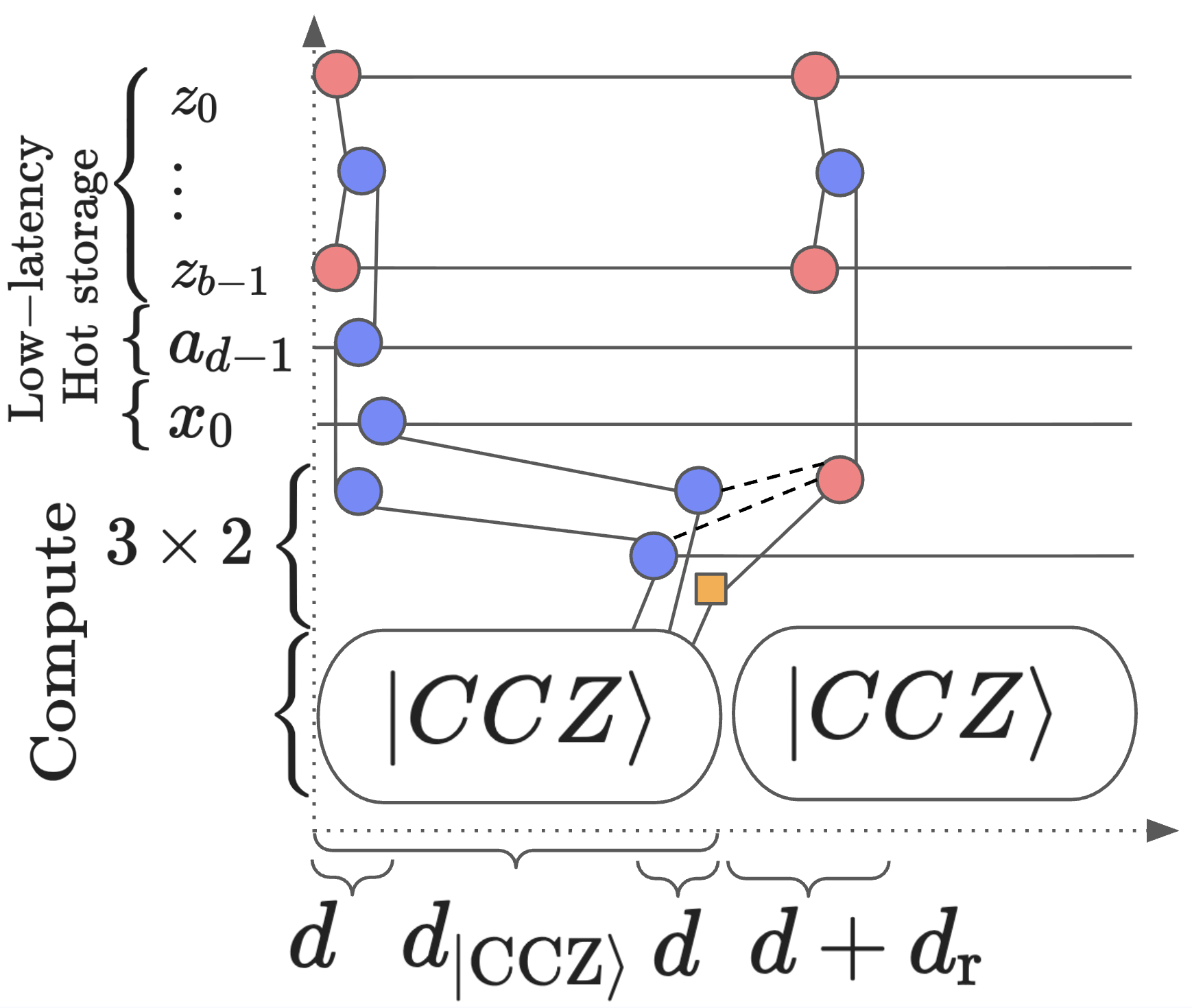}} & \includegraphics[width=0.75\linewidth,valign=b]{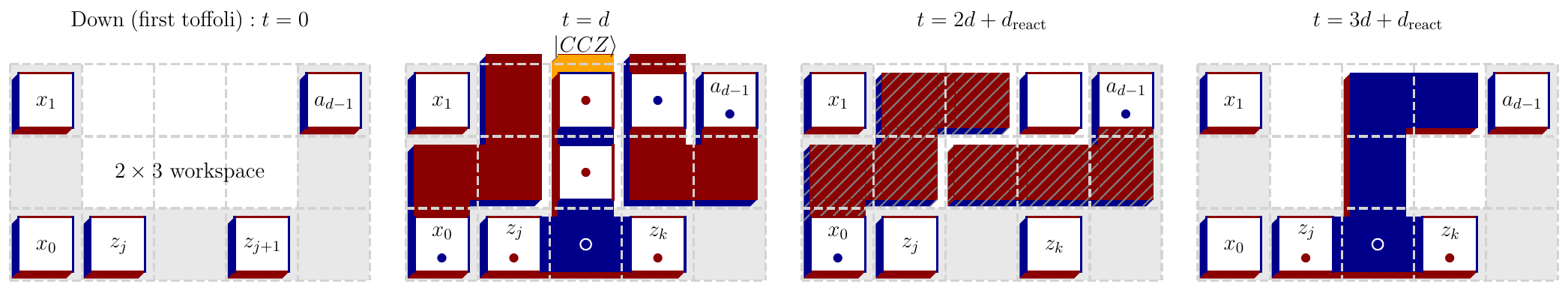}
         \\\hline
         \rotatebox{90}{Down (Toffoli \#2)}&{\includegraphics[width=0.2\linewidth,valign=b]{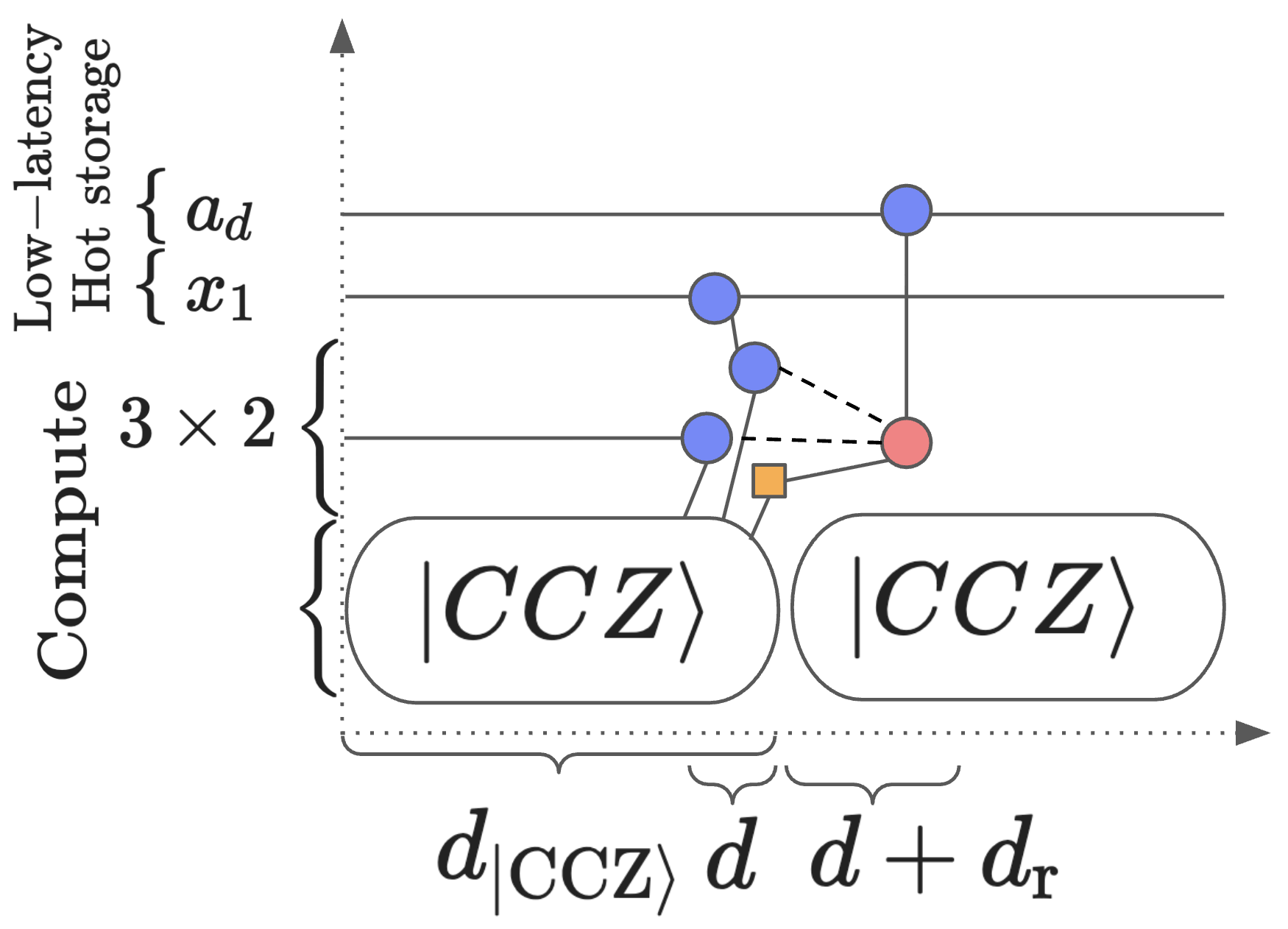}} & \includegraphics[width=0.75\linewidth,valign=b]{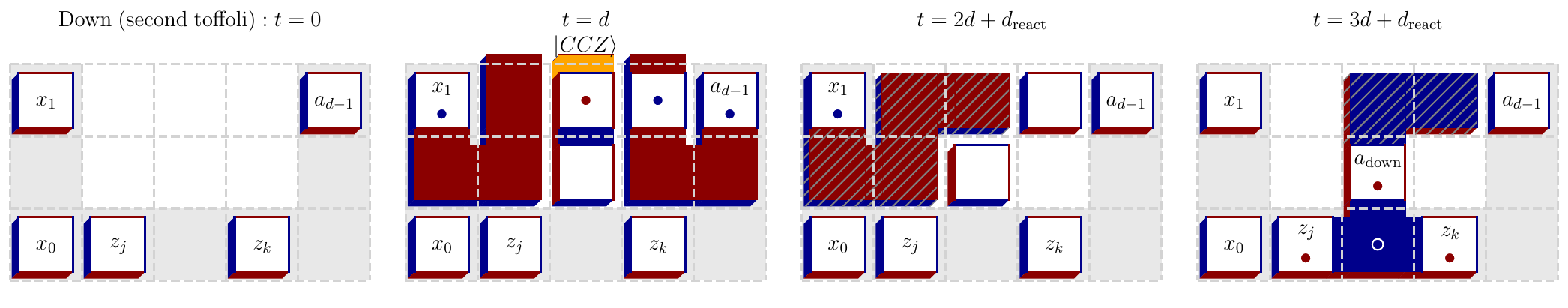}
         \\\hline
         \rotatebox{90}{Leaf}&{\includegraphics[width=0.2\linewidth,valign=m]{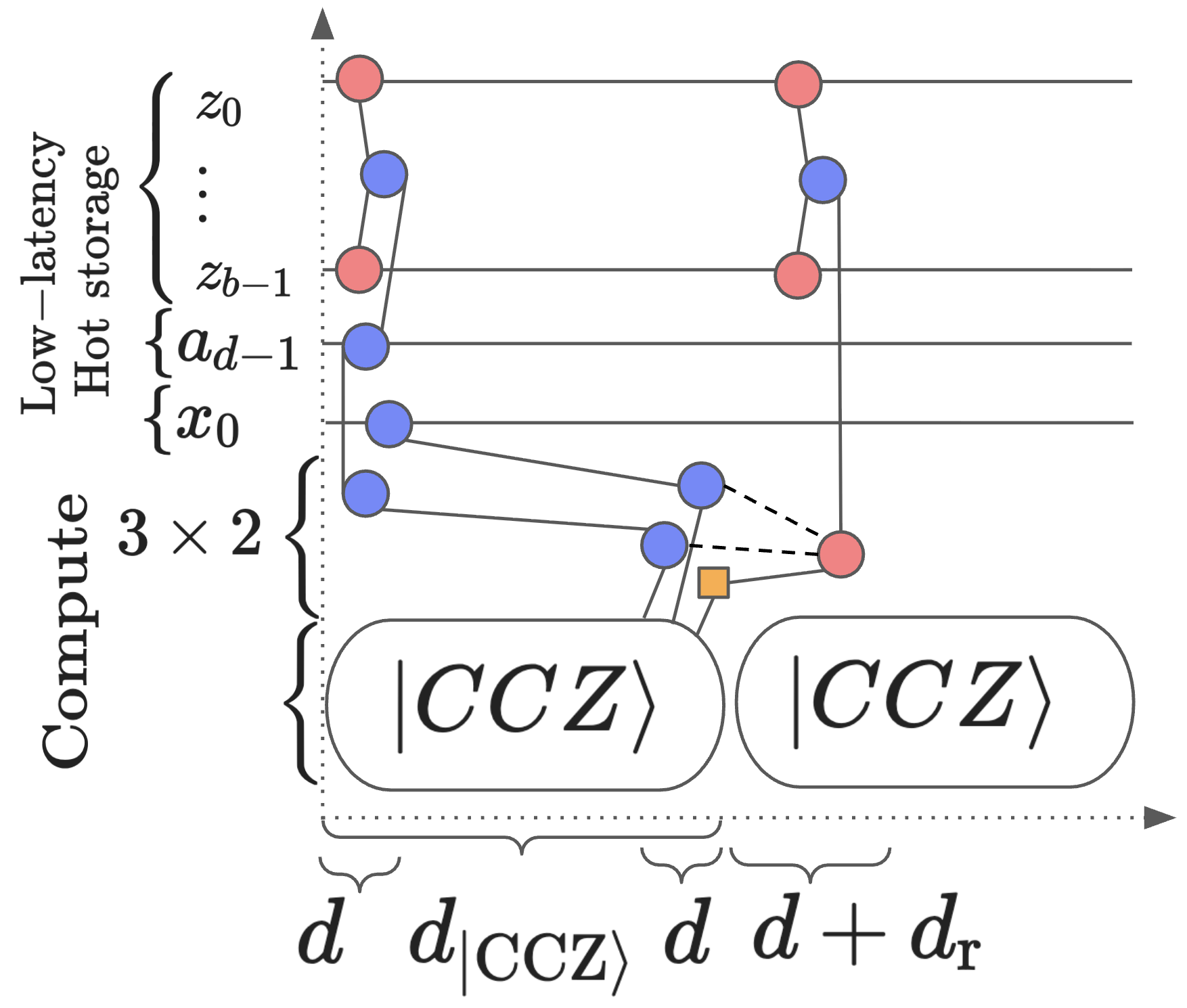}} & \includegraphics[width=0.75\linewidth,valign=m]{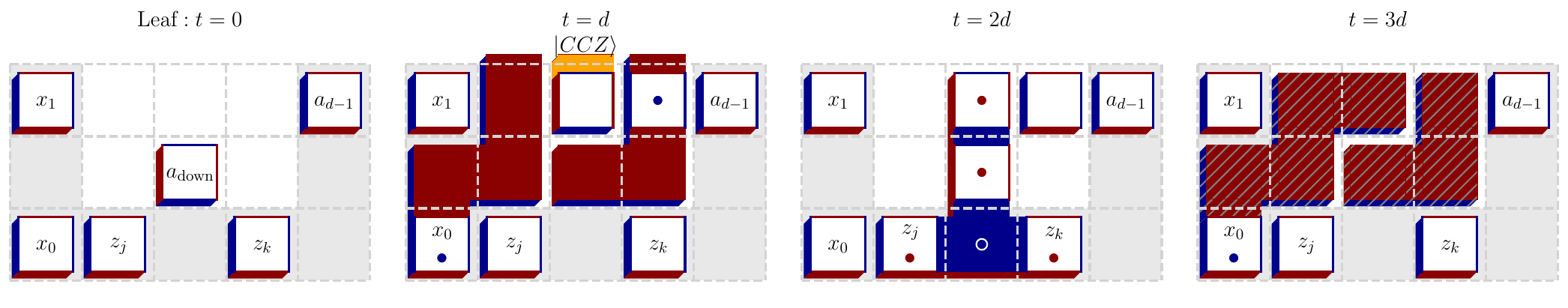}
         \\\hline
         \rotatebox{90}{Bounce}&{\includegraphics[width=0.2\linewidth,valign=m]{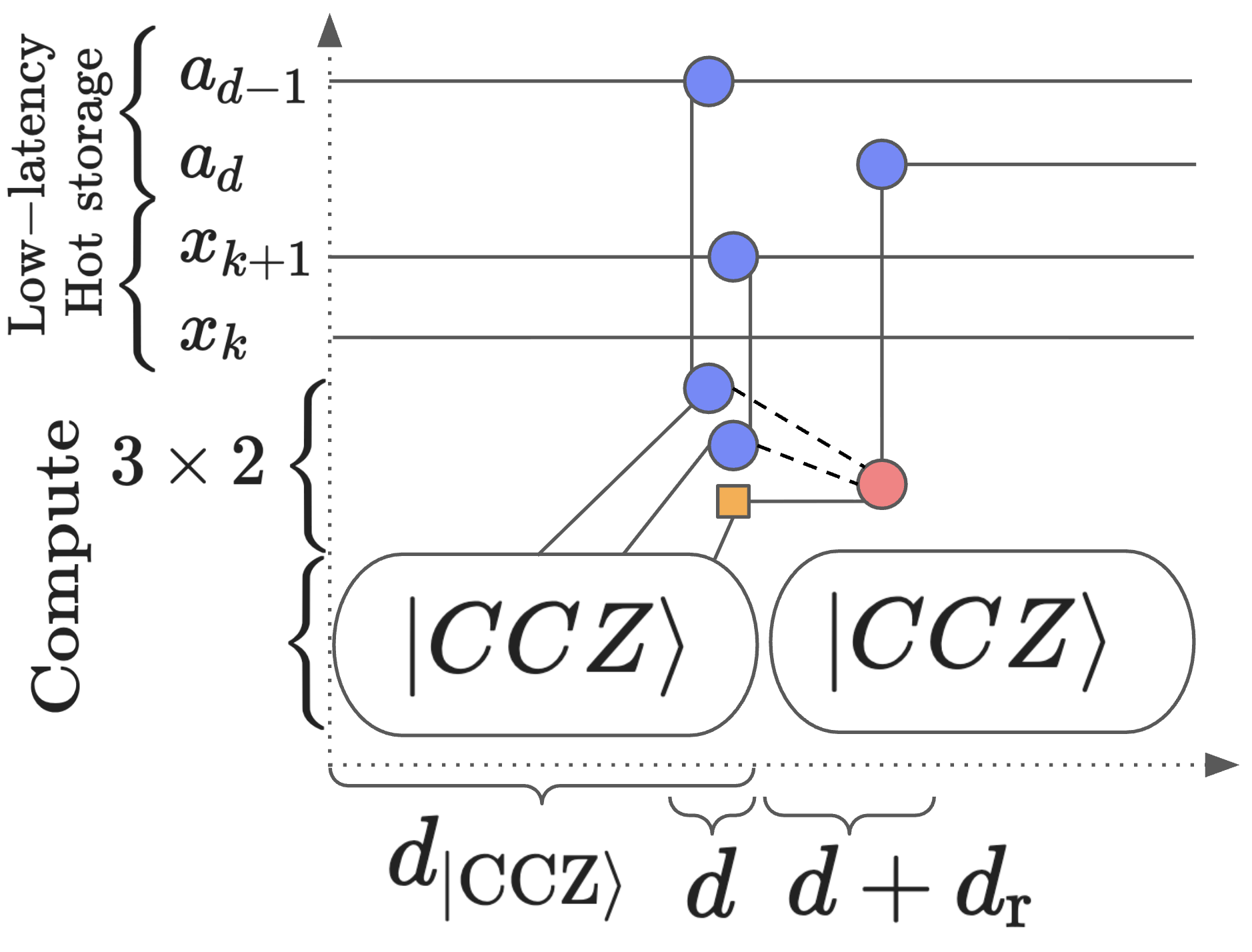}} & \includegraphics[width=0.75\linewidth,valign=m]{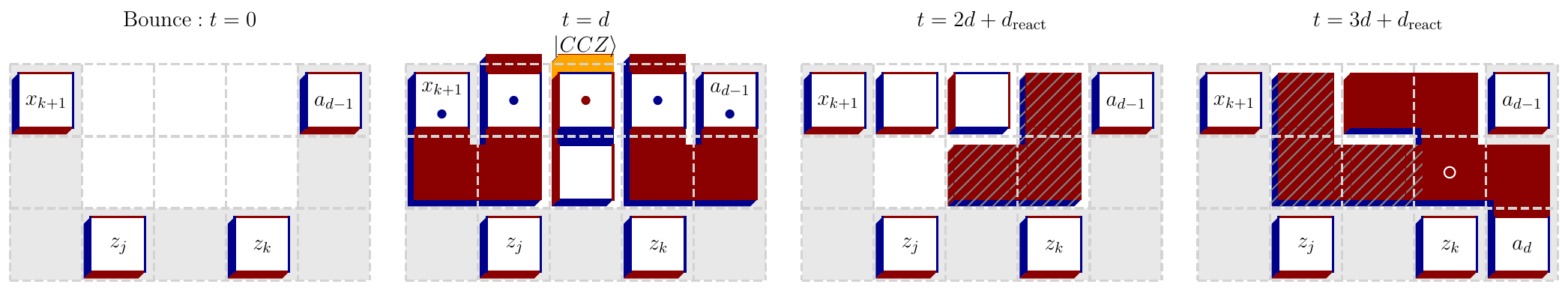}
         \\
         \hline\hline
    \end{tabular}
    \caption{ZX diagrams for clean skew-tree lookup with space going up and time going right.
    All ``up'' steps in~\cref{fig:skew_tree_clean_quantum_circuit} contribute only a $CZ$ fixup that can be deferred to the end of the lookup. 
    Annotated cycle times provide lower bounds the sequence of lattice surgery operations.
    When compiling to pipe diagrams, space and time constraints may push some operations to the preceding or succeeding $ZX$ diagram, or even reverse the order of some operations. All pipe diagrams above assume that the $\ket{CCZ}$ state comes in from a spacelike direction between cycles $t\in(0,d]$, such as when output by an adjacent magic state factory.}
    \label{fig:skew_tree_clean_ZX}
\end{table}

\subsubsection{Trading Toffoli gates for clean or dirty qubits}\label{sec:lookup_table_trading}
We may also implement a lookup table with sublinear scaling in $X$, to as low as $\Omega(\sqrt{Xb})$.
In this section, we present a quantum lookup table that uses skew-tree lookup as a subroutine to minimize Toffoli count under the constraint of using $n_\text{clean}$ and $n_\text{dirty}$ clean and dirty logical ancillae respectively.
This quantum lookup table generalizes the space-time tradeoffs discussed in~\cite{Low2024tradingtgatesdirty,khattar2024riseconditionallycleanancillae} and outputs
\begin{align}
\textsc{Lookup}_\mathfrak{data}\ket{x}\ket{0}^{\otimes b}=\ket{x}\ket{\mathfrak{data}_x}.
\end{align}
Let $\mathfrak{d}=\mathfrak{data}$ and split $\mathfrak{d}_x$ into $b_\text{sub}$-bit subsets.
For $j\in[J]$, where $J=\lceil b/b_\text{sub}\rceil$, let $\mathfrak{d}_{x,j,b_\text{sub}}=\mathfrak{d}_{x,jb_\text{sub}}\cdots \mathfrak{d}_{x,(j+1)b_\text{sub}-1}$, where we pad $\mathfrak{d}_{x,J-1}$ with zeros if $Jb_\text{sub}>b$.
Then $\mathfrak{d}_x=\mathfrak{d}_{x,0,b_\text{sub}}\cdots \mathfrak{d}_{x,J-1}$, and we define
\begin{align}\label{eq:lookup_subset_unitary}
\textsc{Lookup}_{j,b_\text{sub}}\ket{x}\ket{0}^{\otimes b_\text{sub}}=\ket{x}\ket{\mathfrak{d}_{x,j,b_\text{sub}}}.
\end{align}
We evaluate the cost of performing all $j\in[J]$ lookup tables $\textsc{Lookup}_{j,b_\text{sub}}$.
Split the address register $\ket{x}=\ket{x_{\text{h}}}\ket{x_{\text{m}}}\ket{x_{\text{l}}}$ into high, middle, and low components, comprising $n_\text{h},n_\text{m}$, and $n_\text{l}$ qubits respectively. Then $x=x_\text{h}x_\text{m}x_\text{l}=2^{n_\text{l}+n_\text{m}}x_{\text{h}}+2^{n_\text{l}}x_{\text{m}}+x_{\text{l}}$.
We will need the multi-controlled-$\textsc{Not}$ gate (\textsc{MCX}), and a controlled swap network (\textsc{CSwapN}) which have the following costs.
\begin{lemma}[Multi-controlled-\textsc{Not} gate\label{lem:circuit_MCX}]
Let $x$ be an $n$-bit integer.
There is a quantum circuit $\textsc{MCX}_n$ that computes
$\ket{x}\ket{z}\rightarrow \ket{x}\ket{z \oplus (x=2^{n}-1)}$ using the following resources
\begin{center}
\begin{tabular}{c|c|c}
\hline\hline
    Method & Minimum Toffoli~\cite{khattar2024riseconditionallycleanancillae} & Dirty ancilla~\cite{khattar2024riseconditionallycleanancillae} \\
    \hline
    Clean Ancilla & $1$ & $0$ \\    
    Dirty Ancilla & $0$ & $1$\\
    Toffoli gates & $2n-3$ & $4n-8$\\ 
    \hline\hline
\end{tabular}
\end{center}
\end{lemma}
\begin{lemma}[Controlled swap network \textsc{CSwapN}~\cite{Low2024Trading}\label{lem:circuit_swap_network}]
Let $x$ be an $n$-bit integer.
There is a quantum circuit $\textsc{CSwapN}_n$ where controlled on $\ket{x}$, moves the $b$-qubit register $\ket{\cdots}_0$ to register $\ket{\cdots}_x$ for any $x\le X< 2^n$, and applies some arbitrary permutation on the other registers.
This circuit uses $b(X-1)$ Controlled-$\textsc{Swap}$ gates~\crefpos{tab:ccz_fixup}{bottom} and no ancilla qubits.
\end{lemma}

We implement each $\textsc{Lookup}_{j,b_\text{sub}}$ as follows.
\begin{enumerate}
    \item For each address $x_\text{h}\in[X_\text{h}]$, $X_\text{h}=\lceil X/2^{n_\text{m}+n_\text{l}}\rceil$ apply~\cref{lem:circuit_MCX} to implement a $\textsc{MCX}_n$ gate controlled by $\ket{x_\text{h}}$ and targeting a single-qubit register $\ket{\cdot}_{\textsc{MCX}}$. We may choose one of the following two options.
    \begin{itemize}
        \item Clean $\textsc{MCX}$ output: Choose a newly initialized qubit $\ket{\cdot}_{\textsc{MCX}}=\ket{0}_{\textsc{MCX}}$ at $j=x_\text{h}=0$.
        \item Dirty $\textsc{MCX}$ output: Allocate an arbitrary pre-existing qubit to $\ket{\cdot}_{\textsc{MCX}}$ at $j=x_\text{h}=0$.
    \end{itemize}
    \item For each address $x_\text{h}\in[X_\text{h}]$, implement the skew-tree lookup-table $\textsc{Lookup}_{j,x_\text{h},b_\text{sub}}$ controlled by $\ket{\cdot}_{\textsc{MCX}}$.
    Each lookup-table outputs $2^{n_\text{l}}b_\text{sub}$ bits 
    \begin{align}
    \textsc{Lookup}_{j,x_\text{h},b_\text{sub}}\ket{x_\text{m}}\bigotimes_{x_\text{l}\in[2^{n_\text{l}}]}\ket{z_{x_\text{l}}}_{x_\text{l}}=\ket{x_\text{m}}\bigotimes_{{x_\text{l}}\in[2^{n_\text{l}}]}\ket{z_{x_\text{l}}\oplus(\mathfrak{d}_{x_\text{h}x_\text{m}x_\text{l},{j-1},b_\text{sub}}\oplus \mathfrak{d}_{x_\text{h}x_\text{m}x_\text{l},j,b_\text{sub}})}_{x_\text{l}}.
    \end{align}
    Note that the last lookup-table at $x_\text{h}=X_\text{h}-1$ is controlled by only $x_\text{m}\in[X_\text{m}]$ addresses, where $X_\text{m}=\lceil \frac{X-(X_\text{h}-1)2^{n_\text{m}+n_\text{l}}}{2^{n_\text{l}}}\rceil$.
    We may choose one of the following two options.
    \begin{itemize}
        \item Clean MCX output and clean lookup: The ancillae used by unary iteration in $\textsc{Lookup}_{j,x_\text{h},b_\text{sub}}$ are in the $\ket{0}$ state at $j=x_\text{h}=0$.
        \item Dirty MCX output or dirty lookup: The ancillae used by unary iteration in $\textsc{Lookup}_{j,x_\text{h},b_\text{sub}}$ may start in an arbitrary state at $j=x_\text{h}=0$.
    \end{itemize}
    \item Finally, the output bits of the lookup table may be written to one of the following two options, where $\textsc{CSwapN}_{n_\text{l}}$ is controlled by the low bits $\ket{x_\text{l}}$. 
    \begin{itemize}
        \item Clean output: Choose $z_{x_\text{l}}= 0$ at $j=x_\text{h}=0$ for all $x_\text{l}\in[2^{n_\text{l}}]$.
        After each $\textsc{Lookup}_{j,x_\text{h},b_\text{sub}}$, apply $\textsc{CSwapN}_{n_\text{l}}$, and measure all registers $\bigotimes_{x_\text{l}>0}\ket{\cdot}_{x_\text{l}}$ in the $X$ basis and reset them to $\ket{0}^{\otimes (x_\text{l}-1)b_\text{sub}}$.
        This introduces an $\ket{x}$-dependent $\pm$ phase.
        At $j=J-1$, the register $\ket{\cdot}_{x_\text{l}=0}$ can also be measured in the $X$ basis and reset to $\ket{0}^{\otimes b_\text{sub}}$.
        The overall $\ket{x}$-dependent phase can be canceled by a final lookup table.

        \item Dirty output: Choose $z_{x_\text{l}}\neq 0$ at $j=x_\text{h}=0$ for any $x_\text{l}>0$. Every instance of $\textsc{Lookup}_{j,x_\text{h},b_\text{sub}}$ is replaced by $\textsc{CSwapN}_{n_\text{l}}\cdot\textsc{Lookup}_{j,x_\text{h},b_\text{sub}}\cdot\textsc{CSwapN}_{n_\text{l}}^\dagger$.
    \end{itemize}
\end{enumerate}
The overall costs of these three steps are summarized in~\cref{table:qubit_constrained_lookup_cost}.
In practice, we set a maximum number of clean and dirty qubits that the lookup table has available, and enumerate over all possible options to select the option with the minimum Toffoli count.
For the resource estimates of this paper, most Toffoli gates across all instances of $\textsc{Lookup}$ will be spent implementing the skew-tree lookup $\textsc{Lookup}_{j,x_\text{h},b_\text{sub}}$. 
As seen from~\cref{tab:ccz_fixup}, \textsc{CSwap} gates consume each $\ket{CCZ}$ state in $2d$ cycles given one access hallway, but are more easily parallelized.
Although swap networks can account for up to half the Toffoli count of a lookup table, this occurs only when compiling to a very large number of ancilla logical qubits for lookups with $X\gg b$.
Considering that the large-qubit regime also opens many more parallelization options at zero additional space, such as moving columns of patches to instantiate additional magic state factories, it is quite likely that on average, each $\textsc{CSwap}$ gate takes substantially less than $d$ cycles.
Hence, we assume that the footprint and cycles needed to execute all Toffoli gates are upper bounded by the lattice surgery compilation parameters of~\cref{table:lattice_surgery_lookup}.
Only pathological scenarios related to having large $n_\text{h}$, such as providing only one clean qubit, can significantly violate this assumption; these are automatically excluded as they also have the worst Toffoli counts.
\begin{table}
\begin{tabular}{c|c|c|c|c}
\hline\hline
    Step &Method& Toffoli & Clean & Dirty\\
    \hline
    \multirow{2}{*}{1} & Clean & $\lceil \frac{b}{b_\text{sub}}\rceil X_\text{h}(2n_\text{h}-3)$ & $1$  &$0$
    \\ 
     & Dirty & $\lceil \frac{b}{b_\text{sub}}\rceil X_\text{h}(4n_\text{h}-8)$ & $0$ &$1$ (May use $\ket{x_\text{m}x_\text{l}}$)\\ 
    \hline
    \multirow{2}{*}{2} & Clean $\ket{\cdot}_{\textsc{MCX}}$ and lookup-ancillae & $\lceil \frac{b}{b_\text{sub}}\rceil ((X_\text{h}-1)2^{n_\text{m}}+X_\text{m})$ & $n_\text{m}+1$ & 0
    \\
    & Dirty $\ket{\cdot}_{\textsc{MCX}}$ or lookup-ancillae & $\frac{5}{4}\left(\lceil \frac{b}{b_\text{sub}}\rceil ((X_\text{h}-1)2^{n_\text{m}}+X_\text{m})+2^{n_\text{m}}\right)$ & $0$ & $n_\text{m}+1$
    \\
    \hline
    \multirow{2}{*}{3} & Clean output & $\lceil \frac{b}{b_\text{sub}}\rceil(2^{n_\text{l}}-1)b_\text{sub}$ & $2^{n_\text{l}}b_\text{sub}$ & 0
    \\
    & Dirty output & $2\lceil \frac{b}{b_\text{sub}}\rceil(2^{n_\text{l}}-1)b_\text{sub}$ & $b_\text{sub}$ & $(2^{n_\text{l}}-1)b_\text{sub}$
    \\
    \hline\hline
\end{tabular}
\begin{tabular}{c|c|c|c|c}
\hline\hline
    Step &Method& Toffoli & Clean & Dirty\\
    \hline
    \multirow{2}{*}{2} & Clean lookup-ancillae & $\lceil \frac{b}{b_\text{sub}}\rceil X_\text{m}$ & $n_\text{m}-1$ & 0
    \\
    & Dirty lookup-ancillae & $\frac{5}{2}\lceil \frac{b}{b_\text{sub}}\rceil X_\text{m}$ & $0$ & $n_\text{m}-1$
    \\
    \hline
    \multirow{2}{*}{3} & Clean output & $\lceil \frac{b}{b_\text{sub}}\rceil(2^{n_\text{l}}-1)b_\text{sub}$ & $2^{n_\text{l}}b_\text{sub}$ & 0
    \\
    & Dirty output & $2\lceil \frac{b}{b_\text{sub}}\rceil(2^{n_\text{l}}-1)b_\text{sub}$ & $b_\text{sub}$ & $(2^{n_\text{l}}-1)b_\text{sub}$
    \\
    \hline\hline
\end{tabular}
\caption{Overall cost of implementing a quantum lookup table from~\cref{sec:lookup_table_trading} that can be realized with the given budget of clean and dirty qubits. (Top) Costs when the number of address high bits $n_\text{h}>2$. (Bottom) Costs when $n_\text{h} = 0$, with the number of middle bits $n_\text{m}\doteq n_X-n_\text{l}$ with $n_\text{l}$ as a free parameter. 
In all cases, there is no space advantage when $n_\text{h}\in\{1,2\}$.}
\label{table:qubit_constrained_lookup_cost}
\end{table}

\subsubsection{Constant reaction depth via classical absorption of CCZ corrections}\label{sec:low_reaction_depth_lookup}

In the skew-tree lookup compilation described above, each AND gate is implemented by consuming a $\ket{CCZ}$ magic state via state injection.
The measurement outcomes $(m_1, m_2)$ from the injection produce up to three feed-forward corrections ($\textsc{CX}$ and $\textsc{X}$ gates) on the AND-gate output, incurring a $d_\text{reaction}$ wait.
When the $\ket{CCZ}$ production rate is sufficiently fast (e.g., with multiple factories), these serial reaction-time waits can dominate the overall lookup runtime.
We show that these corrections can be eliminated from the quantum circuit by absorbing them into a classical update of the skew data, reducing the reaction depth of a lookup table over $X = 2^{n_X}$ entries from $\mathcal{O}(X)$ to $\mathcal{O}(1)$.

Recall from~\cref{fig:skew_tree_clean_quantum_circuit} that each skew-tree node $p$ represents a subset of selection bits $[n_X]$ and is activated when $p \subseteq x$. The AND gate at skew-tree node $p > 0$ computes $a_p = x_{b(p)} \cdot a_{\mathrm{par}(p)}$, where $b(p)$ is the index of the lowest set bit and $\mathrm{par}(p) = p \;\&\; (p-1)$ is the parent node.
When compiled via $\ket{CCZ}$ state injection, the output before feed-forward corrections is $\tilde{a}_p = (x_{b(p)} \oplus m_1[p]) \cdot (a_{\mathrm{par}(p)} \oplus m_2[p])$, requiring up to three corrections.
A node $p$ is \emph{level-1} if $\mathrm{par}(p) = 0$ (i.e., $p$ is a power of two); there are exactly $n_X$ such nodes.

\begin{theorem}[Classical absorption of CCZ corrections; proof in~\cref{sec:deferred_corrections_appendix}]\label{thm:classical_absorption}
Let $\mathcal{T}$ be a skew-tree lookup table over $X = 2^{n_X}$ entries compiled via $\ket{CCZ}$ state injections, traversed in reverse DFS order. For any pattern of measurement outcomes $\{(m_1[p], m_2[p])\}$, there exists a classical update to the skew data $s \mapsto s'$ such that:
\begin{enumerate}
\item The circuit with modified data $s'$ and \textbf{no feed-forward corrections} loads the correct output $\mathrm{data}[x]$ for all $x$.
\item The update can be computed \textbf{on the fly}: when the measurement outcomes for node $p$ become available, the classical controller updates not-yet-loaded data entries in $O(|\mathrm{sub}(p)|)$ work. The total classical work is $O(X \log X)$.
\item Provided the lattice surgery duration satisfies $d_{LS} \ge d_\text{reaction}$, the classical computation is never on the critical path.
\end{enumerate}
Consequently, the lookup can be compiled with a constant reaction depth compared to $X \cdot d_\text{reaction}$ for the naive approach of applying all corrections in the circuit. 
A controlled lookup of $X$ entries can be reduced to an uncontrolled lookup of $2X$ entries, maintaining constant reaction depth.
\end{theorem}

The key observation is that a $\ket{CCZ}$ injection error at node $p$ redirects the accumulation of $s[d]$ (for each descendant $d$) from $d$ to a strict subset $d' \subsetneq d$ — obtained by clearing some bits of $p$ from $d$ — which is itself a valid skew-tree node with $d' < p$; the error is thus purely classical and corrected by $s'[d'] \mathrel{\oplus}= s[d]$ with no quantum feed-forward.
Since $d' < p$, the reverse DFS traversal guarantees that every correction target is always a future data load, so the classical update can be applied in-place before it is consumed.
The data load at node $p$ itself provides at least a $d_\mathrm{LS}$ buffer between the $\ket{CCZ}$ injection and the loading of any correction target.
Under the practical assumption $d_\mathrm{LS} \ge d_\text{reaction}$, measurements are always resolved before corrected data is consumed, and the circuit never stalls.
See~\cref{sec:deferred_corrections_appendix} for the full proof, and~\cref{fig:dc_skew_tree_qrom_circuit} for a worked example.


%% file: 4_compilation/multiplexed_rotations.tex

\begin{definition}[Multiplexed Z rotation\label{def:multiplexed_Z_rotation}] For any integer $X>1$, and any set of rotation angles $\vec{\theta}\in[0,2\pi)^{X}$, the unitary applied on an $n_X$-qubit address register (where $n_X\doteq\lceil\log_2X\rceil$) and a $1$-qubit target register is
	\begin{align}
		M R_Z (\vec{\theta})\ket{x}\ket{\psi}=\ket{x}e^{i\theta_{x} Z}\ket{\psi}.
	\end{align}
\end{definition}

Using the phase gradient technique, this can be implemented to $b$ bits of precision using an adder with the $b$-qubit phase gradient state $\ket{\mathcal{P}}$.
Let $\tilde{\theta}$ satisfy $|\tilde{\theta}-\vec{\theta}|_{\infty}\le\pi/2^{b}$, and $\tilde{\theta}2^b/(2\pi)\in\mathbb{Z}$.
Consider the sequence
\begin{enumerate}
	\item Compute $\ket{x}\ket{0}\ket{P} \rightarrow_{\textsc{Lookup}_{\tilde{\theta}}}\ket{x}\ket{\tilde{\theta}_x}\ket{P} \rightarrow_{\textsc{Add}_b}e^{i\tilde{\theta}_x}\ket{x}\ket{\tilde{\theta}_x}\ket{P} \rightarrow_{\textsc{Lookup}_{\tilde{\theta}}^\dagger}e^{i\tilde{\theta}_x}\ket{x}\ket{0}\ket{P}$.
\end{enumerate}
Let $\ket{\psi}$ be the control qubit to a $\textsc{CX}$ gate targeting each qubit of $\ket{P}$ before and after the $\textsc{Add}_b$ step of the sequence.
Then we obtain $M R_Z(\vec{\theta})$.
However, this uses a significant number of ancilla qubits: $b$ to store the phase gradient state, and the $b$ bits output by the lookup table.

\subsubsection{Matsumoto Amano normal form for rotations}\label{sec:Matsumoto_Amano_normal form}

Alternatively, the lookup table could output bits specifying rotation synthesis using the normal form of Matsumoto and Amano (MA).
One perspective on this normal form is that it converts a unitary \(U\) expressed as a product of \(k'\) T gates and single-qubit Clifford gates into a circuit of the form
\begin{equation}
	W_k T W_{k-1} T, \cdots, T W_1 T W_0,
\end{equation}
where \(k \leq k'\), \(W_k \in \left\{ I, H, SH \right\}\), \(W_i \in \left\{ H, SH \right\}\) for all \(0 < i < k\), and \(W_0 \in \mathcal{C_1}\) (the single-qubit Clifford group).

For compilation as a multiplexed controlled rotation gate, it is convenient to write the same normal form in an alternative way.
Let \(G_k \in \left\{ T, TS \right\}\) for odd \(k\) and \(G_k \in \left\{ HTH, HTSH \right\}\) for even \(k\).
Let \(V_1 \in \left\{ I, H, SH \right\}\) when \(k\) is odd and \(V_1 \in \left\{ H, I, S \right\}\) when \(k\) is even.
Let \(V_0\) denote a gate of the form \(H^b S^c X^d Z^e\) for \(b, c, d, e \in \left\{ 0, 1 \right\}\).
Then we claim that the following proposition holds:
\begin{proposition}
	Any MA normal form gate can be expressed as a product
	\begin{equation}
		V_1 G_k G_{k-1} \cdots G_1 V_0,
	\end{equation}
	with \(V_0\), \(V_1\), and \(G_i\) as defined above.
	The choice of gates for \(V_0\), \(V_1\), and each of the \(G_k\) can be determined straightforwardly from the MA normal form representation.
\end{proposition}

\begin{figure}
	\centering
	\includegraphics[]{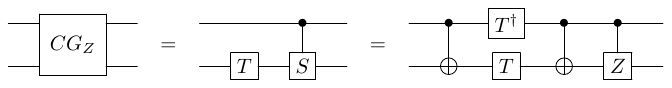}
	\caption{Implementation of the $CG_Z$ gate used to control the \(G_k\) gates for odd \(k\).
		This gate applies $T$ to the second qubit if the first qubit is $\ket{0}$ and $T^3$ (ST) if the first qubit is $\ket{1}$.
		It uses Clifford gates, one \(T\) gate, and one \(T^\dagger\) gate.
	}
	\label{fig:G_Z_gate_implementation}
\end{figure}

\begin{figure}
	\centering
	\includegraphics[]{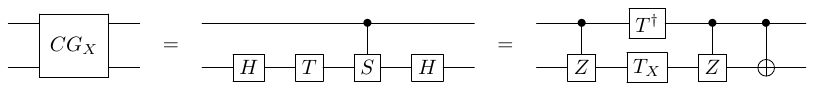}
	\caption{Implementation of the $CG_X$ gate used to control the \(G_k\) gates for even \(k\).
		It applies $T_X$ if the first qubit is $\ket{0}$ and $T_X^3$ if the first qubit is $\ket{1}$ (where \(T_X = H T H\)).
		It uses Clifford gates, along with one \(T_X\) gate and one \(T^\dagger\) gate.
	}
	\label{fig:G_X_gate_implementation}
\end{figure}

\begin{figure}
	\centering
	\includegraphics[]{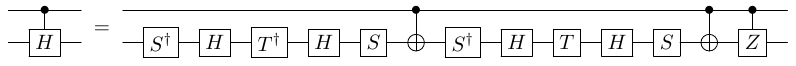}
	\caption{Implementation of the controlled Hadamard gate using Clifford gates, along with one \(T\) gate and one \(T^\dagger\) gate.
	}
	\label{fig:C_H_gate_implementation}
\end{figure}

\begin{figure}
	\centering
	\includegraphics[]{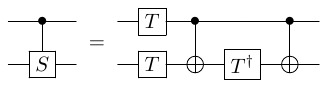}
	\caption{Implementation of the controlled phase gate using Clifford gates, along with two \(T\) gates and one \(T^\dagger\) gate.
	}
	\label{fig:C_S_gate_implementation}
\end{figure}

Using this modified normal form, we can implement a multiplexed rotation circuit using \(k + 6\) bits of data lookup and \(2k + 10\) \(T\) gates (or \(T^\dagger\) gates, which we treat as equivalent).
This cost is accounted for as follows: In order to implement the controlled \(G\) gates, we can use the constructions presented below in \Cref{fig:G_Z_gate_implementation,fig:G_X_gate_implementation}.
Each one requires a single control bit and \(2\) T gates.
All four choices for \(V_1\) can be implemented by taking a product of a controlled \(H\) gate and a controlled \(S\) gate.
All sixteen choices for \(V_0\) can be implemented as a product of four controlled gates: \(CZ\), \(CX\), \(CS\), and \(CH\).
The \(CH\) gate requires \(2\) \(T\) gates and the \(CS\) gate requires \(3\) \(T\) gates, as shown below in \Cref{fig:C_H_gate_implementation,fig:C_S_gate_implementation}.
Overall then, \(6\) control qubits and \(10\) \(T\) gates are required to implement \(V_0\) and \(V_1\).

Alternatively, we can implement the same multiplexed rotation using \(k + 9\) bits of data lookup and \(k + 9\) T gates by making the following observation: Each of the \(k\) controlled G gates as well as the two controlled S gates involves a T or T\(^\dagger\) gate acting on the input bit.
For any branch of the superposition, we therefore accumulate a phase equal to some eighth root of unity.
We can remove these \(k + 2\) T gates and instead implement this phase using a sequence of three data lookups, which we use to apply an arbitrary phase of the form \(\frac{p 2 \pi}{8}\) for some integer \(p\) by acting on the three looked-up qubits with a \(T\) gate, an \(S\) gate, and a \(Z\) gate.

Implementing multiplexed rotations using this modified normal form removes the need for storing the phase gradient state.
Moreover, it makes additional spacetime tradeoffs possible, e.g., performing $b$ table-lookups that each output $1$ bit rather than one table-lookup that outputs $b$ bits.
However, there are several issues that we must address in order to use this approach.

\subsubsection{Distilling \textsc{T} states}\label{sec:catalyzed_T}

\begin{figure}
	\centering
	\includegraphics[width=0.4\textwidth]{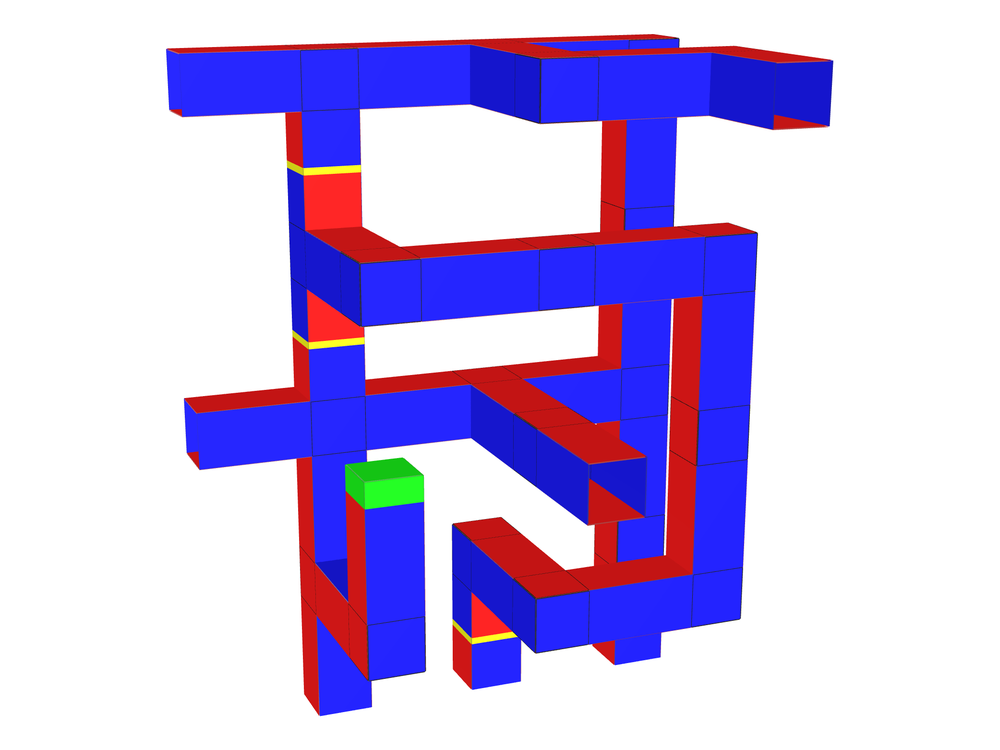}
	\caption{A lattice surgery pipe diagram that shows the Clifford component of our optimized catalyzed CCZ to 2T circuit.
		The pipe diagram fits into a \(2 \times 3\) spatial footprint and takes \(4\) units of time to execute.
		The ports on the bottom take in a CCZ state and the two ports on the wide face (angled slightly right) emit two T states.
		The ports on the narrower face (angled left) are used to consume a catalytic T state and emit its replacement.
	}
	\label{fig:ccz_2t_pipe}
\end{figure}

Using the MA normal form approach to multiplexed rotation synthesis requires a large number of high-fidelity T states.
To distill these states, we assume the use of CCZ factories to produce high-fidelity CCZ states, which we convert into T states using an optimized version of the catalyzed CCZ to 2T approach of~\cite{Gidney2019CCZ}.
In \Cref{fig:ccz_2t_pipe}, we present an optimized compilation of the CCZ to 2T catalysis circuit.
\footnote{We use 3D pipes here instead of 2D cross sections in the previous sections as in-place $Y$ measurements and other operations that have cycles that are not integer multiples of $d$.
	Additionally, this optimized compilation was developed using the LaSsynth software package, which programmatically generates such diagrams.
}
This compilation is designed to minimize the spacetime volume required to catalytically convert a CCZ state into two T states.
It occupies a \(2 \times 3\) spatial footprint and an additional \(2\) logical qubits worth of space are used to store and consume the catalytic T state, resulting in an overall footprint of \(2 \times 4\).
In \Cref{app:space_limited_MA_lattice_surgery}, we present a variation that instead occupies a \(3 \times 3\) footprint.
Using this construction, we can produce \(2\) high-fidelity T states every \(F + 4\) timesteps, where \(F\) is the number of timesteps it takes the CCZ factory to produce a CCZ state.

\subsubsection{MA normal form with a fixed bit length}

The implementation of the multiplexed rotation based on the MA normal form we described in the previous section applies an arbitrary multiplexed rotation for some fixed choice of \(k\).
Most prior work on rotation synthesis has focused on understanding the precision required to implement rotations using any sequence with \(k\) or fewer \(T\) gates.
A simple counting argument shows that the restriction to a fixed \(k\) removes roughly half of the possible unitaries.

\begin{figure}
	\centering
	\includegraphics[width=.7\textwidth]{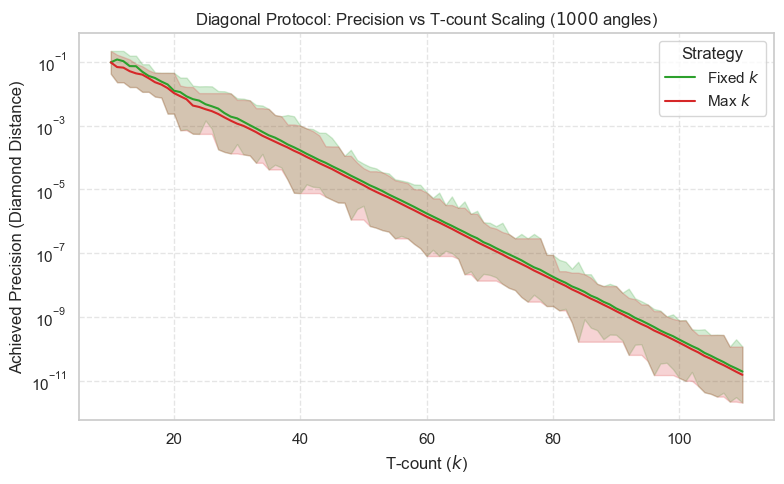}
	\caption{
		A plot showing the error achievable as a function of the number of T gates for diagonal rotation synthesis.
		We compare a strategy that uses up to \(k\) T gates (``Max k'') with a strategy that uses exactly \(k\) T gates (``Fixed k'').
		The line represents the median behavior over a set of \(1000\) randomly chosen angles and we shade the region between the minimum and maximum precision observed over the same set of angles.
		Fixing the number of T gates allowed to be exactly \(k\) has an almost negligible impact on the achievable precision.
	}
	\label{fig:diagonal_scaling}
\end{figure}

In \Cref{fig:diagonal_scaling}, we compare the standard Ross-Selinger rotation synthesis method of \cite{Ross2016-tm} (which allows gate sequences with up to \(k\) T gates), with the same approach restricted to solutions which have exactly \(k\) T gates when expressed in the MA normal form.
We do so using a slightly modified version of the rotation synthesis code available in Qualtran~\cite{Harrigan2024-rj}, which closely follows the presentation of \cite{Kliuchnikov2023shorterquantum}.
We generate \(1000\) \(R_Z\) rotations by sampling the rotation angle from a uniform distribution and plot the precision achievable by both methods as a function of \(k\).
We find that the restriction to a fixed MA normal form length only slightly decreases the achievable precision over a wide range of parameters.
Specifically, performing a linear regression on the asymptotic data (\(k \geq 40\)) reveals that the required T-count scales as \(k \approx (3.010 \pm 0.002) \log_2(1/\epsilon) + (1.33 \pm 0.05)\) when allowing up to \(k\) gates, and \(k \approx (3.009 \pm 0.002) \log_2(1/\epsilon) + (2.40 \pm 0.05)\) when restricted to exactly \(k\) gates.
Thus, the restriction imposes a small constant additive overhead of roughly one T-gate, without affecting the asymptotic slope (the difference in the computed slopes is within the error bars of the calculation).

\subsubsection{Fallback synthesis}

A direct application of the MA normal form to \(R_Z\) synthesis corresponds to the ``diagonal'' synthesis method of \cite{Kliuchnikov2023shorterquantum}.
This is expected to require \(\approx 3.0 \log_2 \left( 1/\epsilon \right) + 1.3\) T gates\footnote{The specific constants in the scaling laws presented in this section are derived from our numerical simulations.
}, where \(\epsilon\) is some allowable error tolerance in the diamond norm.
Using the diagonal approach, we deterministically obtain the same error each time we implement a specific rotation.
This is convenient because it can enable us to interpret errors as errors in the Hamiltonian we are simulating, which might allow a looser error bound than if we strictly bound the error over the whole algorithm in the diamond norm.

\begin{figure}
	\centering \includegraphics[]{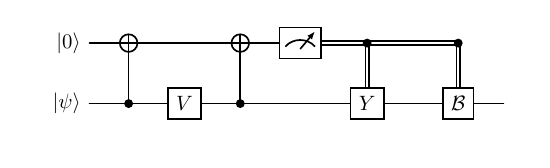} \caption{ A circuit diagram for the fallback protocol, reproduced from \cite{Kliuchnikov2023shorterquantum}.
		The single-qubit operations \(V\) and \(\mathcal{B}\) are both instances of the diagonal unitary approximation, and \(Y\) is a Pauli \(Y\) gate.
		We assume \(\mathcal{B}\) is implemented at most \(1\%\) of the time.
	}
	\label{fig:fallback_protocol}
\end{figure}

To reduce the cost of rotation synthesis, we can instead use the fallback synthesis approach of~\cite{Kliuchnikov2023shorterquantum}.
We sketch the circuit diagram for this approach in \Cref{fig:fallback_protocol}.
This allows us to implement a quantum channel using \(\approx 1.0 \log_2 \left( 1/\epsilon \right) + 6.1\) T gates in expectation.
Let \(\epsilon_{proj}\) be the error of the successful projection measurement relative to the ideal target, \(p_{fail}\) be the probability of failure, and \(\epsilon_{fixup}\) be the precision of the correction step relative to the projection.
By the triangle inequality, the overall single-rotation error \(\epsilon\) is bounded by:
\begin{equation}
	\epsilon_{proj} + p_{fail} \epsilon_{fixup} \le \epsilon.
\end{equation}
It is standard to demand \(p_{fail} \le 0.01\).

\begin{table}[ht]
	\centering
	\begin{tabular}{llcc}
		\toprule
		\textbf{Protocol} & \textbf{Strategy} & \textbf{Slope ($m$)} & \textbf{Intercept ($b$)}
		\\
		\midrule
		Fallback          & Fixed $k$         & $0.9861 \pm 0.0024$  & $7.1596 \pm 0.0710$
		\\
		Fallback          & Max $k$           & $0.9877 \pm 0.0024$  & $6.1297 \pm 0.0744$
		\\
		\midrule
		Diagonal          & Fixed $k$         & $3.0088 \pm 0.0021$  & $2.3962 \pm 0.0533$
		\\
		Diagonal          & Max $k$           & $3.0103 \pm 0.0022$  & $1.3344 \pm 0.0548$
		\\
		\bottomrule
	\end{tabular}
	\caption{
	Empirically determined scaling behavior for the number of T gates required (\(k\)) to achieve a fixed precision \(\epsilon\) in the synthesis of an \(R_Z\) rotation, modeled as $k \approx m \log_2(1/\epsilon) + b$.
	We generated the data by performing synthesis on 1,000 uniformly sampled random angles $\theta \in [0, 2\pi)$.
	For the fallback protocol, the target success probability of the projective step was set to $p_{th} = 0.99$ and the error represents the error assuming the initial projection step is successful.
	We performed linear regression on the data in the large \(k\) regime ($k \in [20, 50]$ for fallback and $k \in [40, 110]$ for diagonal).
	Uncertainties represent the $2\sigma$ ($95\%$) confidence intervals.
	The results show that fallback synthesis is approximately $3\times$ more efficient asymptotically, and that restricting the sequence to exactly $k$ T-gates incurs a small additive overhead without a noticeable impact on the asymptotic scaling.
	}
	\label{tab:synthesis_scaling}
\end{table}

The fallback approach to rotation synthesis requires fewer resources, but its error is not deterministic.
This prevents us from interpreting its error as coming from simulating a perturbed Hamiltonian (which would allow us to relax the precision requirements).
However, we can make the channel perform nearly deterministically by forcing the rarely executed fixup rotation to approximate the outcome of a successful projection step with extremely high precision.

Let $\mathcal{U}$ be the target unitary channel.
When the fallback measurement succeeds, the protocol applies a unitary channel $\mathcal{V}_{proj}$ such that $||\mathcal{U} - \mathcal{V}_{proj}||_\diamond \le \epsilon_{proj}$.
We interpret $\mathcal{V}_{proj}$ as simulating a static perturbed Hamiltonian $H' \approx H$.
In this section, we treat $\epsilon_{proj}$ as an independent parameter.
In practice, we estimate a value of \(\epsilon_{proj}\) based on the sensitivity of classical ground state electronic structure calculation to errors in the Hamiltonian.

During execution, the initial fallback measurement simultaneously succeeds or fails across all branches of the superposition with probability $p_{fail}$.
In the case of failure, we perform a fixup correction using a higher precision multiplexed fallback rotation.
If necessary, we execute this fallback protocol recursively until we achieve a successful projection.
The overall implemented stochastic channel $\mathcal{E}$ is therefore an infinite sum over the number of failures $k$ before a success: $\mathcal{E} = \sum_{k=0}^{\infty} p_{fail}^k(1-p_{fail}) \mathcal{V}^{(k)}$, where $\mathcal{V}^{(0)} = \mathcal{V}_{proj}$.

Assuming each fixup instance targets a uniform precision $||\mathcal{V}^{(k)} - \mathcal{V}_{proj}||_\diamond \le \epsilon_{fixup}$ for $k \ge 1$, we can strictly bound the single-channel error using the triangle inequality.
If we invoke the fallback channel $N_{rot}$ times across the entire algorithm, we can bound the total stochastic error from all \(N_{rot}\) invocations by $\epsilon_{stoc}$ by requiring that the error in each invocation is bounded by $\epsilon_{stoc} / N_{rot}$:
\begin{equation}
	||\mathcal{E} - \mathcal{V}_{proj}||_\diamond \le \sum_{k=1}^{\infty} p_{fail}^k(1-p_{fail}) ||\mathcal{V}^{(k)} - \mathcal{V}_{proj}||_\diamond \le \epsilon_{fixup}(1-p_{fail})\frac{p_{fail}}{1-p_{fail}} = p_{fail} \cdot \epsilon_{fixup} \le \frac{\epsilon_{stoc}}{N_{rot}}.
\end{equation}
To achieve this, we demand that the error in each fixup step is bounded by
\begin{equation}
	\epsilon_{fixup} \le \frac{\epsilon_{stoc}}{N_{rot} \cdot p_{fail}}.
\end{equation}
Using the scaling estimates from \Cref{tab:synthesis_scaling}, we determine the required sequence lengths for both multiplexed operations using the fallback rotation synthesis method.
The primary sequence requires length $k_{1} \approx 1.0 \log_2(1/\epsilon_{proj}) + 7.2$. The conditional high-precision fixup sequence requires length $k_{2} \approx 1.0 \log_2(1/\epsilon_{fixup}) + 7.2$.

\begin{figure}
	\centering \includegraphics[width=.7\textwidth]{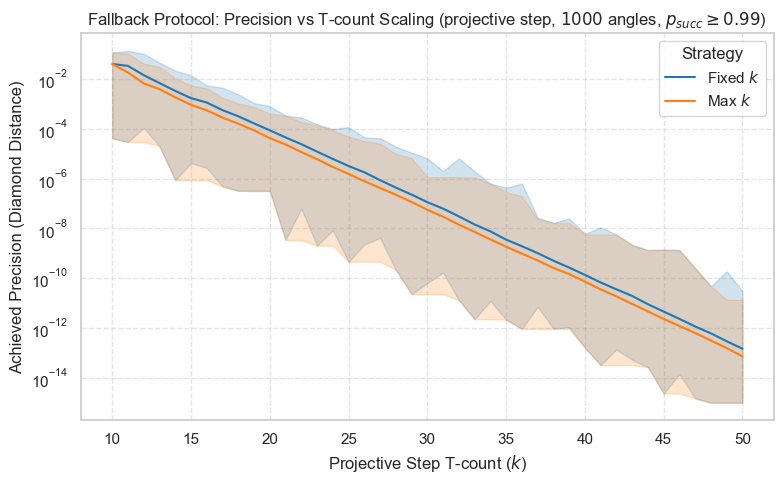} \caption{ A plot showing the error achievable as a function of the number of T gates for fallback rotation synthesis.
		This plot shows the error assuming that the projection is successful and no fixup is necessary.
		We compare a strategy that uses up to \(k\) T gates (``Max k'') with a strategy that uses exactly \(k\) T gates (``Fixed k'').
		The line represents the median behavior over a set of \(1000\) randomly chosen angles and we shade the region between the minimum and maximum precision observed over the same set of angles.
		Fixing the number of T gates allowed to be exactly \(k\) has an almost negligible impact on the achievable precision.
	}
	\label{fig:fallback_scaling}
\end{figure}

\subsubsection{Implementing a Givens rotation}
\begin{figure}[ht]
	\centering
	\includegraphics[]{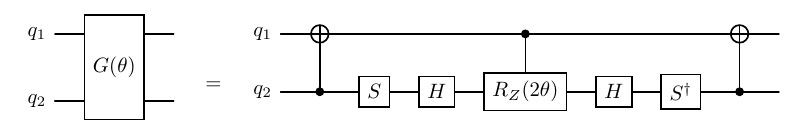}
	\caption{Implementation of a Givens rotation using a controlled \(R_Z\) gate.
	}
	\label{fig:givens_circuit}
\end{figure}

\begin{figure}[ht]
	\centering
	\includegraphics[]{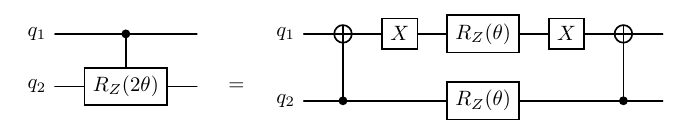}
	\caption{Implementation of a controlled \(R_Z(2 \theta)\) gate using two parallel \(R_Z(\theta)\) gates without controls.
	}
	\label{fig:C_RZ_circuit}
\end{figure}

A Givens rotation $G(\theta)$ can be exactly mapped to a $CR_Z(2\theta)$ by conjugating with Clifford gates~\cite{caesura2025faster}.
We show this decomposition in \Cref{fig:givens_circuit}.
As shown in \Cref{fig:C_RZ_circuit}, we can implement the controlled rotation using two parallel calls to uncontrolled rotations by the same angle.
We use this decomposition to implement a multiplexed Givens rotation.
\begin{definition}[Multiplexed controlled-Z rotation\label{def:multiplexed_CZ_rotation}] For any integer $X>1$, and any set of rotation angles $\vec{\theta}\in[0,2\pi)^{X}$, the unitary applied on an $n_X$-qubit address register (where $n_X\doteq\lceil\log_2X\rceil$), a $1$-qubit control register, and a $1$-qubit target register is
	\begin{align}
		M CR_Z(\vec{\theta})\ket{x}\ket{y}\ket{\psi}=\ket{x}\ket{y}
		                                                            \begin{cases}
			                                                            \ket{\psi},                  & y=0,
			                                                            \\
			                                                            e^{i\theta_{x} Z}\ket{\psi}, & y = 1.
		                                                            \end{cases}
	\end{align}
\end{definition}

\begin{figure}
	\centering
	\includegraphics[]{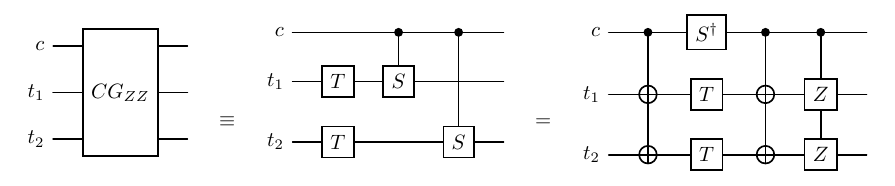}
	\caption{Implementation of the $CG_{ZZ}$ gate used when implementing two parallel multiplexed \(R_Z\) rotations (with the same angle).
		This gate applies $T\otimes T$ to the second and third qubits if the control qubit is $\ket{0}$ and $T^3 \otimes T^3$ if the control qubit is $\ket{1}$.
	}
	\label{fig:G_Z_Z_gate_implementation}
\end{figure}

This implementation has several advantages.
We can implement two parallel multiplexed \(R_Z\) gates using the same QROM data lookups.
Furthermore, we can take advantage of the parallelism to slightly reduce the cost.
In our implementation of a single multiplexed \(R_Z\) rotation, the cost is driven primarily by the alternating \(CG_Z\) and \(CG_X\) gates.
When implementing two such gates in parallel (with the same control), we can combine two \(T^\dagger\) gates on the control line into an \(S^\dagger\) gate, reducing the number of \(T\) gates to one per \(G_k\) without requiring the additional lookup described above.
See \Cref{fig:G_Z_Z_gate_implementation} for an example of this.
We can also slightly reduce the cost of implementing two \(CS\) gates in parallel for the same reason (provided they share a control).

We implement the two parallel \(R_Z\) gates using the fallback approach.
We first synthesize two identical diagonal rotations.
Taking \(k_{diag}\) to be the number of T gates in a non-multiplexed version of the diagonal rotation, we need to look up \(k_{diag} + 6\) bits.
These bits control the application of \(k_{diag}\) total calls to the \(CG_{ZZ}\) or \(CG_{XX}\) gates, along with two controlled \(S \otimes S\) gates, two controlled \(H \otimes H\) gates, a controlled \(X \otimes X\) gate, and a controlled \(Z \otimes Z\) gate.
The overall \(T\) gate requirement for these two identical multiplexed diagonal rotations is \(2 k_{diag} + 16\).

During execution, the fallback measurement succeeds or fails independently for each of the two parallel $R_Z$ rotations.
We dynamically route the execution based on the specific failure mode:
\begin{itemize}
	\item \textbf{Both rotations fail (probability $\le p_{fail}^2$):} We perform a single QROM lookup to fetch $k_2 + 6$ classical bits.
	      We then implement the fixup sequence on both target qubits simultaneously using the parallel architecture (e.g., $CG_{ZZ}$).
	      As derived above, this requires $2k_2 + 16$ T-gates.
	\item \textbf{One rotation fails (probability $\le 2p_{fail}(1-p_{fail})$):}
	      We perform a single QROM lookup to fetch $k_2 + 9$ classical bits.
	      We implement the fixup sequence only on the specific target qubit that failed, using the standard single-target $CG_{Z}$ architecture.
	      Using the optimization described previously, this requires $k_2 + 9$ T-gates.
\end{itemize}
Because the two $R_Z$ gates encode the same rotation angle, we only have to perform the QROM data loading once for the fixup sequence even if both rotations fail.


%% file: 4_compilation/adder_multiplexed_rotations.tex
The sequential architecture minimizes spatial footprint at the expense of execution time and overall space-time volume.
When sufficient space is available, we instead apply controlled multiplexed rotations using the phase gradient technique.
To optimize the space-time volume of this approach, we compile a variant of Gidney's controlled adder~\cite{Gidney2018halvingcostof} into a compact arrangement of logical qubits on a square grid.
We assume these logical qubits are implemented using a standard two-dimensional surface code.
Given the \(b\)-qubit phase gradient state
\begin{equation}
  \ket{\mathcal{P}_b} = \frac{1}{\sqrt{2^b}} \sum_{j=0}^{2^b - 1} \omega_b^{-j} \ket{j}, \qquad \omega_b = \exp\left(\frac{2\pi i}{2^b}\right),
\end{equation}
a controlled addition of $a = \lfloor \theta \cdot 2^b / 2\pi \rceil$ to the phase gradient register applies a $Z$-rotation by angle $\theta$ via phase kickback.

\begin{figure}[htbp]
  \centering
  \subfloat[The temporary AND operation computes the AND of the control qubit \(q\) and the \(k\)-th bit of the input \(i\), storing the result in \(i'_k\).\label{fig:and_circuit_logical}]{
    \begin{minipage}[b]{0.31\textwidth}
      \centering
      \includegraphics[scale=.75]{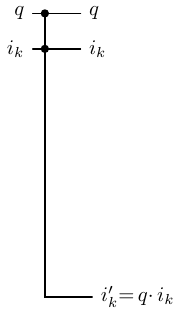}
    \end{minipage}
  }
  \hfill
  \subfloat[The MAJ operation takes the \(k\)-th carry bit along with the \(k\)-th bit of each input and outputs the \(k+1\)-th carry bit.\label{fig:maj_circuit_logical}]{
    \begin{minipage}[b]{0.31\textwidth}
      \centering
      \includegraphics[scale=.75]{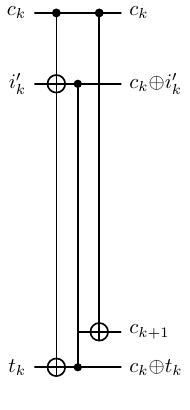}
    \end{minipage}
  }
  \hfill
  \subfloat[The UMA operation outputs the \(k\)-th bit of the sum while also uncomputing the \(k\)-th carry bit. In our construction we also uncompute the \(k\)-th bit of \(i'\).\label{fig:uma_circuit_logical}]{
    \begin{minipage}[b]{0.31\textwidth}
      \centering
      \includegraphics[scale=.75]{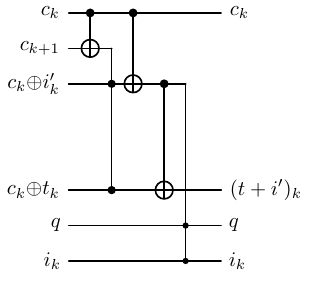}
    \end{minipage}
  }
  \caption{
    Diagrams of the logical circuits for the three main components we use to compile the controlled adder.
  }
  \label{fig:logical_circuits_controlled_adder}
\end{figure}

Gidney's ripple-carry adder~\cite{Gidney2018halvingcostof} decomposes the controlled addition into two building blocks: carry generation (MAJ) and carry uncomputation with sum extraction (UMA). 
We avoid modifying the standard UMA block. 
Instead, we compute the controlled addition of the \(b\)-bit input register \(i\) and target register \(t\) by substituting \(i\) with \(i'\), where \(i' = i\) if the control qubit \(q\) is in the \(1\) state and \(0\) otherwise.
This compilation strategy allows us to pack the components of the controlled adder tightly in time and space.
In particular, we avoid the need for a large number of long-lived routing chimneys (temporary ancilla qubits used to implement delayed-choice Clifford corrections).
In this approach, we separately compile three small circuit components: the initial delayed-choice AND, MAJ, and UMA.
We show the logical circuits for these components in \Cref{fig:logical_circuits_controlled_adder} (see \Cref{app:controlled_adder_compilation} for more details on their implementation as quantum circuits and ZX calculus diagrams).

We compile the controlled adder under standard device parameters.
Our descriptions use an $I$, $J$, $K$ coordinate system, where $I$ and $J$ label spatial dimensions (in units of logical qubits) and $K$ labels the temporal dimension (in units of \(d\)).
We assume that CCZ magic states are supplied every \(d_{factory} \approx 5.3d\) surface code cycles by a CCZ factory with a \(4 \times 3\) spatial footprint~\cite{Gidney2019CCZ}.
We assume a decoder reaction time $d_\text{reaction} \approx 10$ cycles, and code distance $d \ge 20$.

\begin{figure}[htbp]
  \centering
  \subfloat[Schematic that shows how the components of our controlled adder are arranged in time (\(K\), y axis) and one of two spatial dimensions (\(I\), x axis). We assume \(d_{\text{factory}} = 5\) in this figure.\label{fig:lattice_surgery_schematic}]{%
    \begin{minipage}[b][225pt][c]{0.35\textwidth}
      \centering
      \includegraphics[width=.7\textwidth]{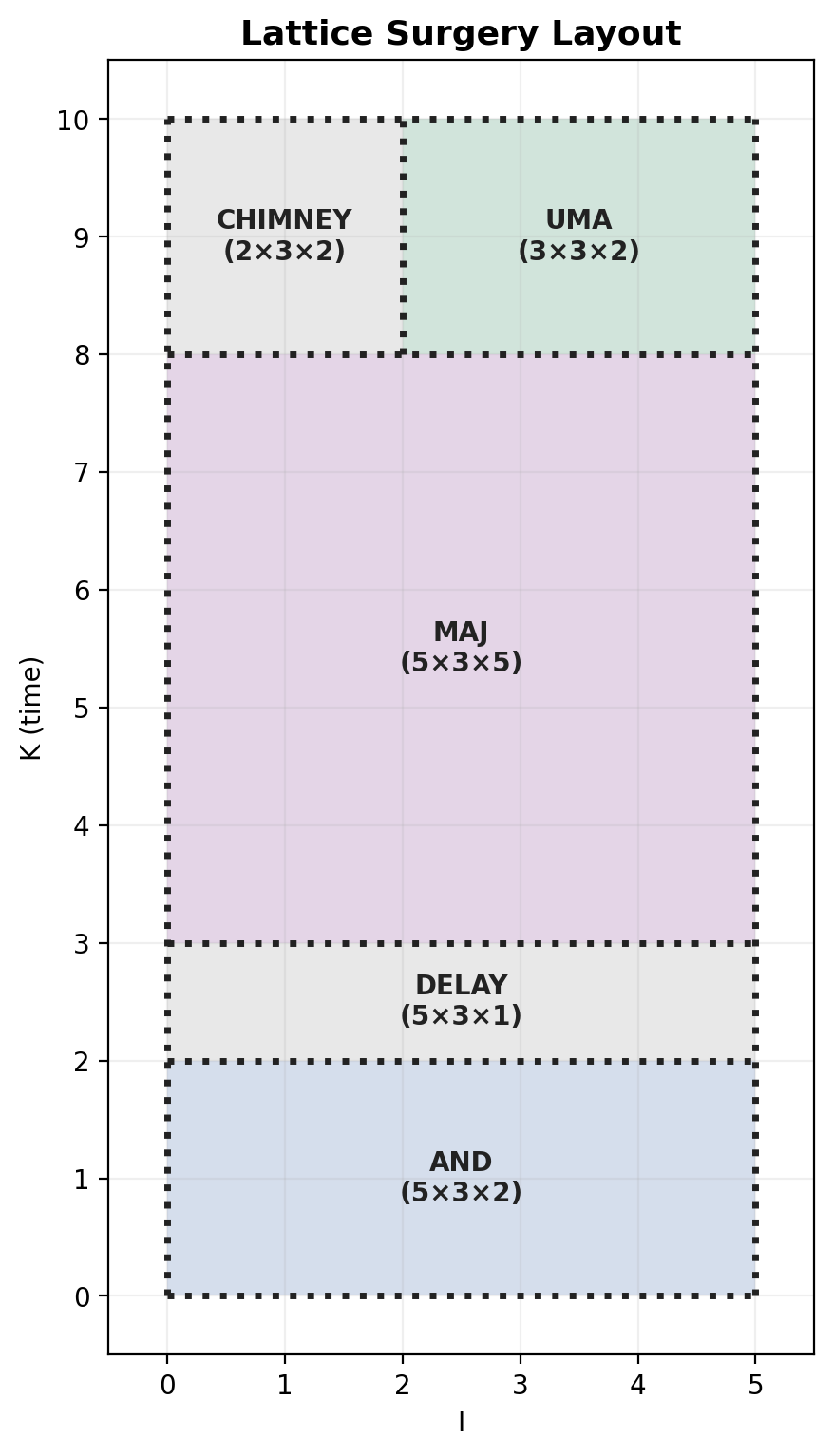}
    \end{minipage}%
  }
  \hfill
  \subfloat[A ``pipe diagram'' visualizing the lattice surgery implementation of MAJ.\label{fig:maj_pipe}]{%
    \begin{minipage}[b][225pt][c]{0.35\textwidth}
      \centering
      \includegraphics[width=\textwidth]{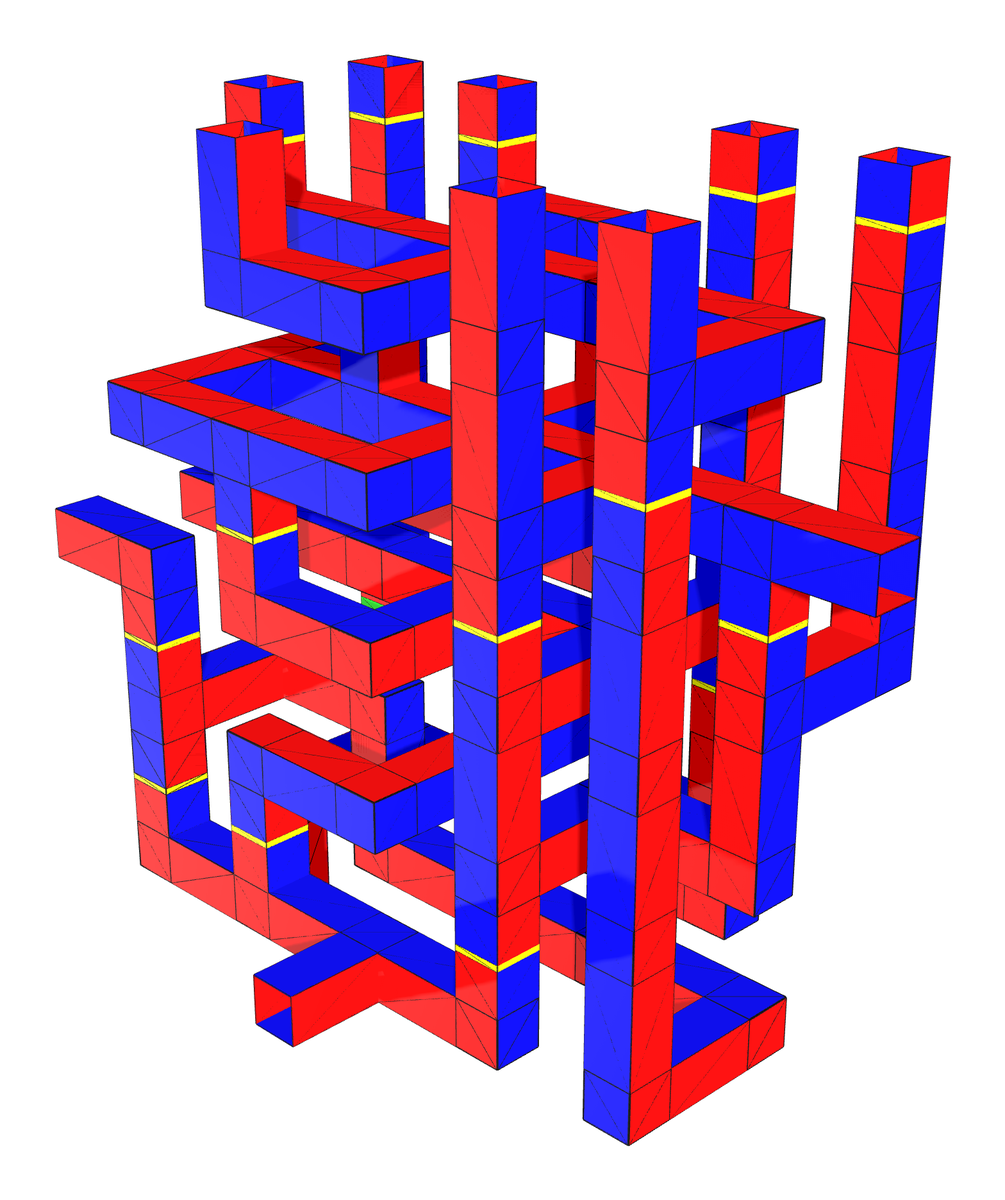}
    \end{minipage}%
  }
  \hfill
  \subfloat[Lattice surgery ``pipe diagrams'' showing the implementation of the temporary AND with delayed choice corrections (bottom) and UMA (top).\label{fig:uma_pipe}]{%
    \begin{minipage}[b][225pt][c]{0.28\textwidth}
      \centering
      \includegraphics[width=.6\textwidth]{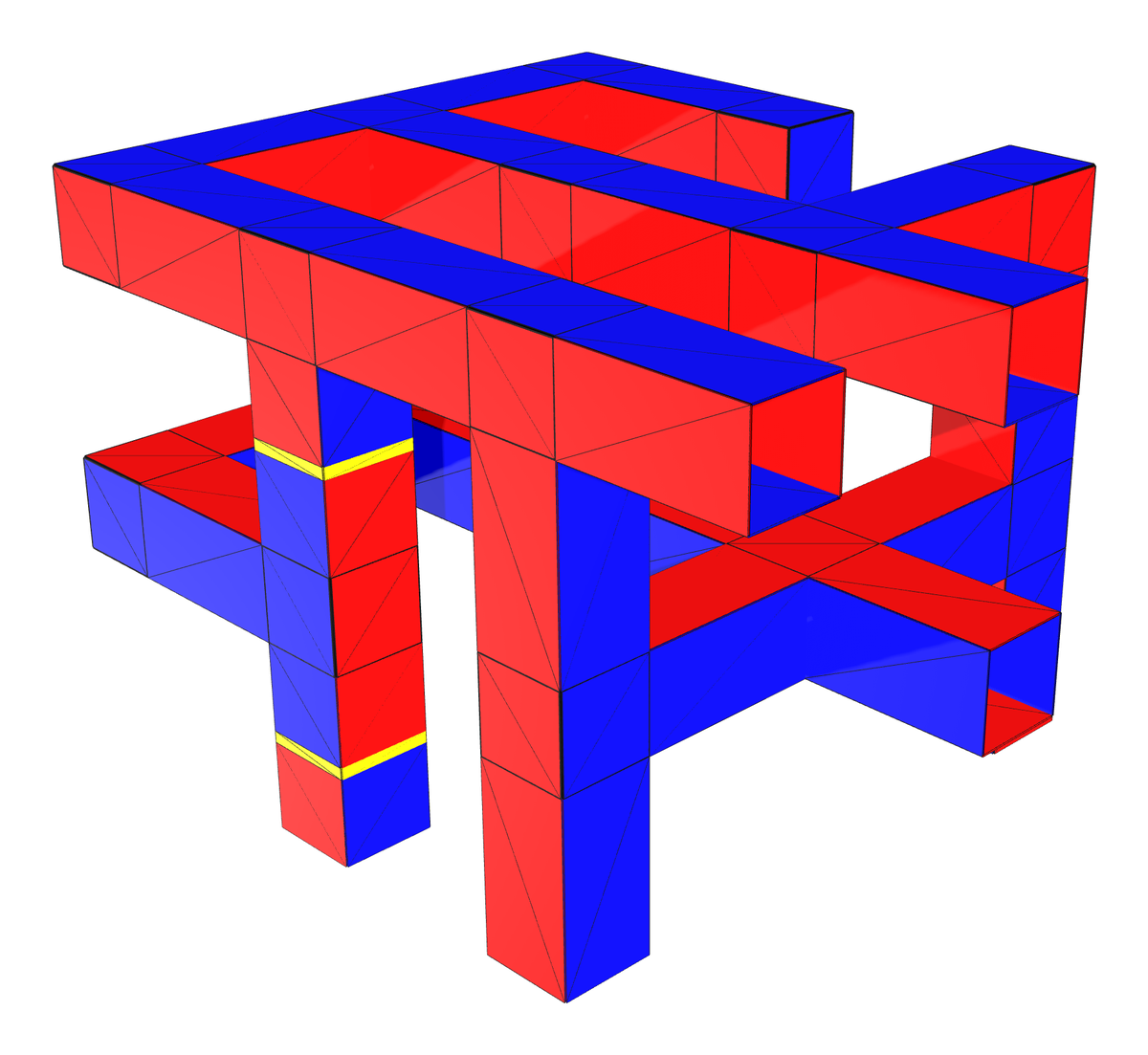}
      \par\vspace{0.2cm}
      \includegraphics[width=\textwidth]{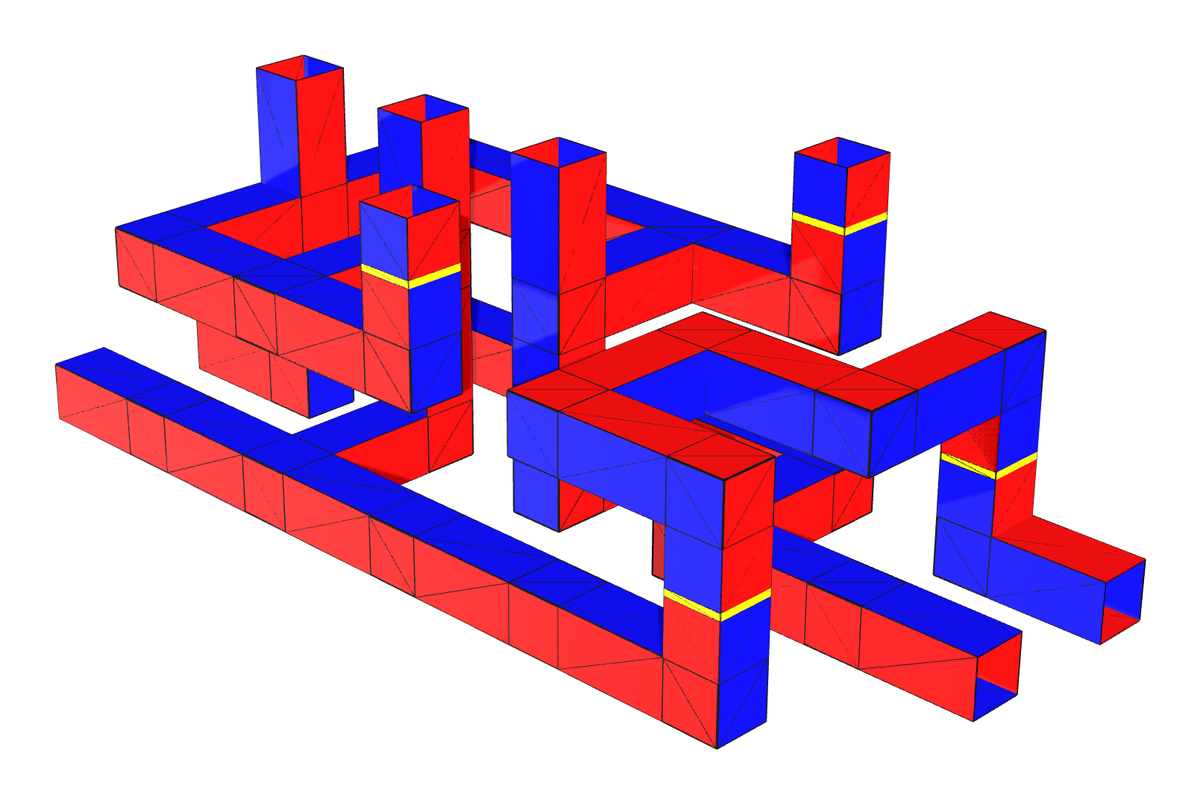}
    \end{minipage}%
  }
  \caption{
    Our controlled adder is implemented by combining the AND, MAJ, and UMA components in a \(5 \times 3 \times 10\) volume of spacetime.
    At the bottom, the AND operation is attached using a \(Z\)-port (see \Cref{app:controlled_adder_compilation} and~\cite{Gidney2025Factoring}) to the control qubit and the input bit \(i_k\).
    It consumes a CCZ state and outputs four delayed-choice routing chimneys and the bit \(i_k'\) upwards.
    In the delay region, the \(i_k'\) bit is passed upwards and the routing chimneys are measured in the \(X\) or \(Z\) bases in order to implement (or not) the Clifford corrections necessitated by the CCZ gate teleportation in the AND block.
    The MAJ operation takes the carry bit \(c_k\) as an input and passes the next carry bit \(c_{k+1}\) as an output along the \(J\) axis.
    It takes the \(i_k'\) bit as an input on the bottom and the CCZ state and \(t_k\) bit as inputs on opposite faces in the \(I, J\) plane.
    It outputs four delayed-choice routing chimneys into the CHIMNEY region and four other bits (the inputs to UMA) up into the UMA component.
    In UMA, the \(c_{k+1}\)-th carry bit is uncomputed and the sum qubit along with two delayed-choice routing chimneys are output in the \(+I\) direction.
  }
  \label{fig:pipe_diagrams}
\end{figure}

We design our compilation to fit in a \(5 \times 3\) region adjacent to a CCZ factory and to continuously consume CCZ states every \(d_{factory}\) cycles for any \(d_{factory} \geq 5\).
We present a two-dimensional schematic of our compilation in \Cref{fig:lattice_surgery_schematic}, showing one spatial dimension (\(I\)) on the x axis and the time dimension (\(K\)) on the y axis. 
The other spatial dimension (\(J\)) is not visible, but all components of our compilation are three units wide in this dimension.
In this schematic, the CCZ factory would sit to the left (from \(I=-4\) to \(I=0\)) and the inputs and outputs would come from the right-hand side.
We do not account for the space required to store the inputs or outputs in our explicit compilation here, nor the space required to temporarily store the two additional routing chimney qubits (temporary ancilla that must be stored for most of the duration of the adder's execution) generated while processing each bit.

We used a combination of hand-design and brute-force search to explore possible arrangements of these components.
For each component, given a bounding box size we randomly sampled \(\approx 400,000\) different arrangements of input and output ports.
We used the \texttt{LaSsynth} software package~\cite{Tan2024Scalpel} to search for a valid lattice surgery compilation for each bounding box size.
We show the arrangement of our optimized components and three-dimensional rendering of each of them in \Cref{fig:pipe_diagrams}.
Our compilation fits most of the components for processing the \(k\)-th bit of the controlled adder in a \(5 \times 3 \times 10\) bounding box (excluding the CCZ factories themselves and the storage area for input and output bits).

We tile the blocks in space along the \(J\) axis to process multiple bits in parallel.
This tiling is limited by the need to sequentially resolve the delayed-choice Clifford corrections.
The corrections that arise in the first temporary AND block do not lead to any sequential dependence and can be parallelized without issue.
When consuming the CCZ state to implement the temporary AND gate in the MAJ block, we have two delayed-choice CZ corrections that we implement by sending four routing chimneys out of the top of the MAJ block into the area labeled CHIMNEY on the diagram in \Cref{fig:lattice_surgery_schematic}.
These delayed-choice corrections must be resolved sequentially, leading to a maximum chimney height of \(F d_\text{reaction}\) when performing \(F\) MAJ blocks in parallel. 

The UMA block uncomputes two separate temporary AND gates, each of which may produce a CZ correction.
We implement one of these CZ corrections using a delayed-choice CZ and route the chimneys for this correction back into the data storage area. 
We must finish all MAJ operations before resolving these chimneys by measuring them in the X or Z bases.
As a result, we need space to store two additional routing chimney qubits per bit in the adder's arguments.
The final possible Clifford correction resulting from the UMA block's uncomputation is a CZ between the control qubit for the adder (\(q\)) and the input bit \(i_k\).
After performing the entire adder we can implement the necessary collection of CZs using a single-control multi-target Z gate in one lattice surgery time step.
In the case of the multiplexed rotations, we perform this fixup immediately before uncomputing \(i_k\) using measurement-based uncomputation.

Accounting for the sequential dependence of the delayed-choice corrections, we calculate the overall CCZ consumption rate when using \(F\) factories to process \(F\) bits in parallel.
Each bit uses two $\ket{CCZ}$ states from factories that output a state every $d_\text{factory}$ cycles. 
The MAJ block receives its $\ket{CCZ}$ state $d_\text{factory}$ cycles after the AND block. Counting $1d$ for the first round of the AND block, $d_\text{factory}$ to wait for the MAJ $\ket{CCZ}$ state, and $2d$ to finish the MAJ block, the routing chimneys exit the top of MAJ after $d_\text{factory} + 3d$ surface code cycles.  
Because we resolve the delayed-choice corrections sequentially, the $F$-th MAJ block produces a chimney of height $F \cdot d_\text{reaction}$. 
To avoid slowing down the factories, these chimneys must finish before the next batch starts at $2 d_\text{factory}$. 
The total time for the batch is therefore
\begin{align}\label{eq:controlled_adder_batch_cycles}
    T_{\text{batch}} = \max(2 d_{\text{factory}}, d_{\text{factory}} + 3d + F \cdot d_{\text{reaction}}).
\end{align}

For example, let us assume a code distance \(d=22\), a reaction time $d_\text{reaction} =10$, and a total time for CCZ production of $d_\text{factory} \approx 5.3d$.
Then we have $d_\text{factory} - 3d = 2.3d = 50.6$ cycles of room for the chimneys. 
For $F=5$, the largest chimney has a height of \(50\), fitting within this room without incurring additional delay.
The total time to process the \(5\) bits is therefore $10.6d,$ corresponding to a CCZ consumption rate of roughly $0.94$  per $d$ cycles ($10 / 10.6d$). 
For $F=6$, the chimneys delay the next batch slightly, pushing the total time to $11.03d$, but the rate still increases to $1.09$ CCZs per $d$ cycles ($12 / 11.03d$). 
Pushing this further to $F=10$ delays the next batch to $12.85d$ but yields an even higher rate of $1.56$ CCZs per $d$ cycles ($20 / 12.85d$).

%% file: 5_simulation/quantum_simulation.tex
Quantum computers promise efficient solutions to the broad class of electronic structure simulation problems: systems with more than $50$ active space molecular orbitals, where the goal is ground state energy estimation to chemical accuracy of $1\text{kcal/mol}\approx 1.6$mHa or to lower spectroscopic accuracy.
Representative electronic structure Hamiltonians include the FeMo cofactor~\cite{Reiher2017Elucidating} for Nitrogen fixation and a Ruthenium-based catalyst~\cite{PhysRevResearch.3.033055} for Carbon fixation.
Quantum methods are particularly well-suited to such problems as they attain chemical accuracy with polynomial cost in $N$, even when accounting for the cost of preparing highly-correlated trial wavefunctions that overlap strongly with the true quantum ground state~\cite{berry2024rapid}. Furthermore, quantum methods can provide accuracy guarantees that are missing when comparing heuristic classical methods and extrapolating.

At a high-level, we perform quantum phase estimation of the sum-of-squares spectrum-amplified~\cite{King2025SOSSA} quantum walk operator of a low Double-Factorized-Tensor-Hyper-Contracted (DFTHC) representation of electronic structure~\cite{low2025fast}, which is state-of-the-art.
We consider quantum systems represented by the spin-free Hamiltonian with $N$ orbitals
\begin{align}\label{eq:spinfree_Hamiltonian}
H&=\sum_{p,q=0}^{N-1}h^{(1)}_{pq}E_{pq}+\frac{1}{2}\sum_{p,q,r,s=0}^{N-1}h_{pqrs}^{(2)}E_{pq}E_{rs},
\quad
E_{pq}\coloneqq \sum_{\sigma\in\{0,1\}}a^\dagger_{p\sigma}a_{q\sigma},
\end{align}
where $h^{(1)}$ and $h^{(2)}$ are the one- and two-body interaction tensors respectively.
Ground state phase estimation is accomplished by making queries to the quantum walk formed by the qubitization~\cite{low2019hamiltonian} of a block-encoding of~\cref{eq:spinfree_Hamiltonian}.
The number of queries scales linearly with the block-encoding normalization factor $\Lambda$ for arbitrary states.
Overall, estimating the ground energy to standard deviation $\sigma_\text{PEA}$ requires $\lceil \frac{\pi \Lambda}{2\sigma_\text{PEA}}\rceil $ queries to this quantum walk.
For consistency with previous resource estimates, we choose $\sigma_\text{PEA}=1.0$mHa, assume perfect ground state overlap and ignore the cost of preparing a good trial state, which should be strongly subdominant~\cite{berry2024rapid} for the systems in~\cref{tab:dfthcblisssossa_costs}.
The overall phase estimation quantum circuit then requires only two additional ancillae~\cite{Najafi2023QPE2qubit}, and a negligible number of additional quantum gates beyond that needed to implement the quantum walks.

Spectrum amplification~\cite{King2025SOSSA} improves the quantum phase estimation accuracy of estimating low-lying ground state energies by reducing $\Lambda$ to a smaller effective normalization $\lambda_\text{eff}$.
Suppose that an $n$-qubit Hamiltonian $H$ with ground state energy $E_\text{gs}$ has a Sum-Of-Squares (SOS) representation \begin{align}\label{eq:gap_amplifiable_Hamiltonian}
H=H_{\text{SA}}+E_\text{SOS}\identity,
\quad
H_{\text{SA}} \coloneqq \sum_{\alpha=0}^{L-1}O_{\alpha}^{\dagger}O^{}_{\alpha}=H_\text{sqrt}^\dagger H_\text{sqrt},
\quad
H_\text{sqrt}={\sum_{\alpha=0}^{L-1}|\chi_\alpha\rangle \otimes
 O_{\alpha}},
\end{align}
where $\{|\chi_\alpha\rangle\}_\alpha$ is any set of mutually orthogonal quantum states.
Then, assuming there exists a block-encoding $\Be\left[\frac{H_\text{sqrt}}{\lambda_\text{sqrt}}\right]$, one may construct the quantum walk:
\begin{align}
W=\textsc{Ref}_{\text{c}}\cdot\Be^\dagger\left[\frac{H_\text{sqrt}}{\lambda_\text{sqrt}}\right]\cdot\textsc{Ref}_{\text{r}}\cdot\Be\left[\frac{H_\text{sqrt}}{\lambda_\text{sqrt}}\right]=\textsc{Ref}_{\text{c}}\cdot\Be\left[\frac{H_\text{SA}}{\Lambda}-\mathbb{I}\right],\quad
\Lambda=\frac{1}{2}\lambda^2_\text{sqrt},
\end{align}
where $\textsc{Ref}_{\text{r}}$ and $\textsc{Ref}_{\text{c}}$ are reflections about the subspace spanned by rows and columns of $H_\text{sqrt}$.
The eigenphases of $W$ are $\arccos{((E_k-E_\text{SOS})/\Lambda-1)}$, so they are amplified for small $E_k-E_\text{SOS}$.
Hence, using quantum phase estimation,
one may estimate the ground energy $E_\text{gs}$ to standard deviation $\sigma$ with the following cost:
\begin{align}
\left(\text{Queries to}\;W\right)=\left\lceil\frac{\pi\lambda_\text{eff}}{2\sigma_\text{PEA}}\right\rceil,\quad \lambda_\text{eff}=\sqrt{E_\text{gap}(2\Lambda - E_\text{gap})},\quad E_\text{gap}=E_\text{gs}-E_\text{SOS}.
\end{align}

The DFTHC Hamiltonian is an ansatz for the second quantized Hamiltonian of the form
\begin{align}\label{eq:DFTHC_representation}
H_{\text{DFTHC}}&=\sum_{pq}h^{(1)\prime}E_{pq} + \frac{1}{2}\sum_{r\in[R]}\sum_{c\in[C]}\left(w^{(rc)}_B\identity+\sum_{b=0}^{B-1}w^{(rc)}_{b}\sum_{pq}u^{(r)}_{b,p}u^{(r)}_{b,q}E_{pq}\right)^2-\frac{1}{2}\sum_{r\in[R]}\sum_{c\in[C]}\left|w^{(rc)}_B\right|^2 \identity ,\\
h^{(2)\prime}_{pqmn}&=\sum_{r\in[R],c\in[C]}\left(\sum_{b\in[B]}w^{(rc)}_{b}u^{(r)}_{b,p}u^{(r)}_{b,q}\right)\left(\sum_{b'\in[B]}w^{(rc)}_{b'}u^{(r)}_{b',m}u^{(r)}_{b',n}\right),\label{eq:DFTHC_h2_tensor}
\end{align}
where the identity term $W^{(rc)}\identity$ provides variational freedom to enable optimizing for a small $E_\text{gap}$.
Although the two-body interaction tensor $h^{(2)}$ has large dimension $\mathcal{O}(N^4)$, it admits the low rank approximate factorization in \cref{eq:DFTHC_h2_tensor}.
Previously, we established that a chemically accurate representation is obtained with the parameters in~\cref{tab:dfthcblisssossa_costs} where $u^{(r)}_{b,:}$ are unit vectors representing single-particle basis rotations specified to $b_\text{rot}$ bits of precision, the coefficients $w^{(rc)}_b$ are specified to $b_\text{coeff}$ bits of precision, and the tensor factorization ranks $R=\mathcal{O}(1)$, $B=\mathcal{O}(N)$, and $C=\mathcal{O}(N)$.
Crucially,~\cref{eq:DFTHC_representation} has the SOS representation
\begin{align}
H\approx H_\text{DFTHC}&=\sum_{G\in\{\mathrm{D}_1,\mathrm{Q}_1\}}\sum_{r\in[N_G]}\sum_{\sigma\in\{0,1\}}{O^\dagger_{G^\sigma,r}}{O_{G^\sigma,r}}+\sum_{r\in[R],c\in[C]}O^\dagger_{\text{SF},rc}O_{\text{SF},rc}+E_\text{SOS},
\\
{O_{\mathrm{D}_1^\sigma,r}}&=\sqrt{w^{(r)}_{+}}\frac{\gamma_{\vec{u}^{(r)}_+\sigma 0}+i\gamma_{\vec{u}^{(r)}_+\sigma 1}}{2},
\qquad
{O_{\mathrm{Q}_1^\sigma,r}}={\sqrt{w^{(r)}_{-}}\frac{\gamma_{\vec{u}^{(r)}_-\sigma 0}-i\gamma_{\vec{u}^{(r)}_-\sigma 1}}{2}},\quad\gamma_{\vec{u}\sigma x}=\sum_pu_p\gamma_{p\sigma x},\label{eq:SOS_generator_D1Q1}
\\
O_{\text{SF},rc}&=\frac{1}{\sqrt{2}}\left(w^{(rc)}_B+\frac{i}{2}\sum_{b\in[B]}\sum_{\sigma\in\{0,1\}}w^{(rc)}_b\gamma_{\vec{u}^{(r)}_b\sigma 0}\gamma_{\vec{u}^{(r)}_b\sigma 1}\right),\label{eq:SOS_generator_SF}\\
E_\text{SOS}&=-2\sum_{r}w^{(r)}_--\frac{1}{2}\sum_{r\in[R]}\sum_{c\in[C]}\left|W^{(rc)}\right|^2,
\end{align}
where the one-body terms factorize into a SOS of $N=N_{\mathrm{D}_{1}}$rotated Majorana operators with positive coefficients $w_\pm^{(r)}$.
The $(G,r,c)$ indices may be flattened into a single integer 
\begin{equation}
    x_o(G,r,c) \doteq \left\{\begin{array}{lll}
        r&\in[0,N_{\mathrm{D}_1}), & G=G_{\mathrm{D}_1}=0,\\
        r &\in[N_{\mathrm{D}_1},N), & G=G_{\mathrm{Q}_1}=1,\\
        N+rC+c&\in[N,X_\text{o}), & G=G_\text{SF}=2,
    \end{array}\right.
    \quad x_\text{o}\in[X_\text{o}],\quad X_\text{o}\doteq N+RC.
\end{equation}
so that the Hamiltonian is expressed as a sum of squares as
\begin{equation}
    H_\text{DFTHC}  = \sum_{x_o} O^\dagger_{x_o} O_{x_o}.
\end{equation}
A space-optimized quantum circuit that implements the square-root block-encoding $\textsc{Ref}_{p} \cdot \Be\left[\frac{H_\text{sqrt}}{\lambda_\text{sqrt}}\right]$ is shown in~\cref{fig:dfthc_min_qubits}.

\begin{table}
    \centering 
    \begin{tabular}{|c|c|c|c|c|c|c|c|}
    \hline \hline
Molecule &  \multicolumn{2}{c|}{FeMoco}  &  Cpd1X (P450X)  & \multicolumn{3}{c|}{CO$_2$ [XVIII]}     \\
     \cline{2-3}\cline{5-7}
&[54e, 54o]&[113e, 76o]&[63e, 58o]& [64e, 56o]& [100e, 100o]& [150e, 150o]
     \\
     \hline \hline    
     $(R,B,C)$ &$(10, 27, 27)$ &$(15, 57, 19)$ & $(9, 29, 14)$ &$(5, 28, 28)$ & $(8, 75, 25)$ & $(9,112,37)$ \\
     $\Lambda$/Ha &58.3 &179.7  &97.4 &55.5 & 155.5 & 336 \\
     $E_\text{gap}$/Ha &4.05 &5.38 &5.68 &2.69 & 4.57& 6.45 \\
     $\lambda_{\text{eff}}$/Ha  &21.4&43.7&32.8&17.07& 37.4& 65.5 \\
     \hline
     $(b_\text{coeff},b_\text{rot})$ & (9,16) & (9,15) & (10,15) & (7,12) & (8,16) & (9,16)
     \\
      \hline \hline
\end{tabular}
    \caption{
    DFTHC factorization and spectrum amplification parameters used to simulate electronic structure to chemical accuracy. We take the number of block-encoding queries to be $\lceil \frac{\pi \lambda_\text{eff}}{2\sigma_\text{PEA}}\rceil $~\cite{BabbushPRX18,PRXQuantum.2.030305} where, similar to previous work, $\sigma_\text{PEA} = 1.0\text{mHa}$. }
    \label{tab:dfthcblisssossa_costs}
\end{table}

\begin{figure}
    \centering
    \includegraphics[scale = 0.8]{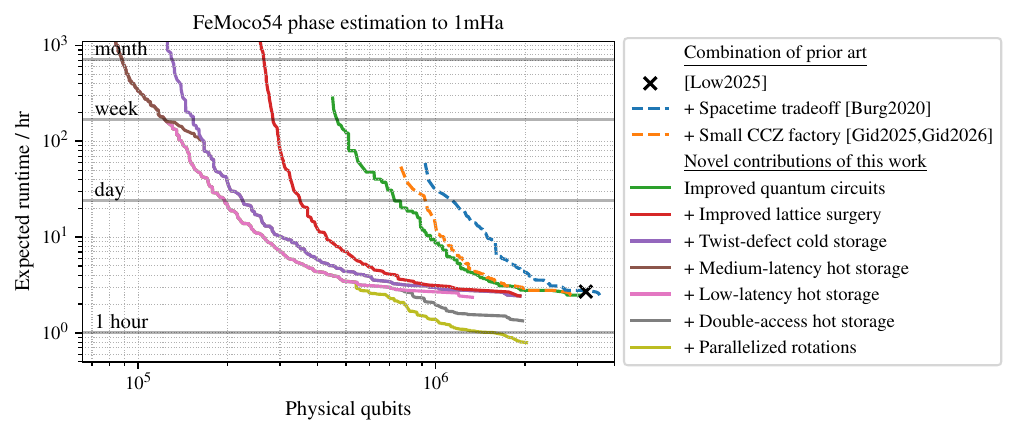}
    \caption{
    Our best physical-qubit-runtime tradeoff in~\cref{fig:intro_pareto} is obtained by a combination of methods, with their cumulative impact illustrated for the example of $108=2N$ spin-orbital FeMoco. 
    We distinguish between improvements arising from straightforward combinations of known techniques and the novel contributions of this work.
    Most of our new runtime improvements arise from highly optimized lattice surgery compilation and parallelization of the most costly logical circuit components, and most of our physical qubit improvements arise from using a careful combination of high-rate-high-latency and low-rate-low-latency error correcting codes.
   }
    \label{fig:femoco54_improvements}
\end{figure}

We introduce new logical quantum circuit optimizations specific to block-encoding~\cref{eq:spinfree_Hamiltonian} in~\cref{sec:sossa_logical_resources}, and optimize for physical-level resources across different layout options in~\cref{sec:sossa_physical_resources} to report the following key resource estimates:
\begin{enumerate}
    \item A minimum space compilation uses $89$k physical qubits for FeMoco-54, which is almost two orders of magnitude smaller than our previous state-of-the-art resource estimates for the same problems~\cite{low2025fast}, but with a long runtime on the order of a month.
    \item A Pareto frontier of tradeoffs between time and space in~\cref{fig:femoco54_improvements} that demonstrates an extremely rapid reduction in runtime as slightly more space becomes available, with a minimum physical spacetime volume of  $1.3$-mega-qubit-hours, which is almost an order of magnitude improvement over prior art~\cite{low2025fast}.
    \item We obtain similar tradeoffs in~\cref{fig:intro_pareto} for other benchmark molecules, which demonstrates that our methods are generically applicable. 
    The carbon fixation XVIII-56 complex uses an even smaller number of $79$k physical qubits, and another popular benchmark with significantly more orbitals FeMoco-76 is possible in under a month with $167$k physical qubits.
\end{enumerate}

The cumulative contribution of our optimizations is shown in~\cref{fig:femoco54_improvements} for FeMoco-54.
We begin by augmenting the original minimum-Toffoli compilation using $3.2\times 10^6$ physical qubits~\cite{low2025fast}, assuming the same layout as~\cite{PRXQuantum.2.030305}, with the `programmable gate array'~\cite{PhysRevResearch.3.033055} that streams output rotation angle bits using known lookup-table space-time trade-offs ~\cite{Low2024tradingtgatesdirty}.
We then use the much smaller cultivation-based~\cite{Gidney2025Cultivation} $\ket{CCZ}$ factories~\cite{Gidney2025Factoring}, reviewed earlier in~\cref{table:ccz_factory}.
This forms a space-time tradeoff representative of a `combination of prior art'.

\subsection{Minimizing logical qubits and Toffoli gate count}\label{sec:sossa_logical_resources}

\begin{figure}
    \centering
    \includegraphics[width=0.6\linewidth,valign = m]{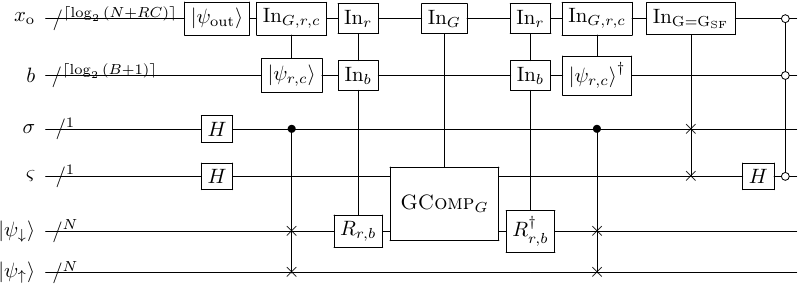}
    \caption{Quantum circuit for block-encoding the DFTHC square-root Hamiltonian after all logical qubit count optimizations. We define the following unitaries: $\textsc{Outer}$ prepares $\ket{\psi_\text{out}}$ with basis states $x_\text{o}\in[X_\text{o}]$ representing a flattened composite index $(G,r,c)\in\{(0,r,0)\}_{r=0}^{N_{\text{D}_1-1}}\cup \{(1,r,0)\}_{r=N_{\text{D}_1}}^{N-1}\cup\{(2,r,c)\}_{(r,c)\in[R]\times[C]}$. \textsc{Inner} prepares $\ket{\psi_{r,c}}$ controlled by $(2,r,c)$. \textsc{Givens} implements the Givens rotation unitary $R_{r,b}$ controlled by $(r,b)$, and is comprised of a lookup table $\textsc{RPrep}$ with $N+RB$ address and $(N-1)b_\text{rot}$ output bits, and \textsc{Rot}, which applies $N-1$ controlled phase rotations with each angle specified by $b_\text{rot}$ bits.}
    \label{fig:dfthc_min_qubits}
\end{figure}

Our novel contributions begin with further quantum-circuit-level optimizations that enable deeper space-time trade-offs down to a minimum-logical-qubit quantum circuit block-encoding that uses only $20$ to $30$ ancilla logical qubits and $2N$ data qubits for storing the fermion wavefunction, and is explained in detail in~\cref{sec:sec:appendix_tradeoffs}.
Almost all Toffolis in the block-encoding are accounted for by the following unitaries in~\cref{fig:dfthc_min_qubits}.
\begin{itemize}
    \item \textsc{Outer}: Prepares the dimension $X_\text{o}\doteq N+RC$ state $\ket{\psi_\text{out}}$ spanned by $\ket{x_\text{o}}$, where $x_\text{o}\in[X_\text{o}]$ represents a flattened composite index $(G,r,c)\in\{(0,r,0)\}_{r=0}^{N_{\text{D}_1}-1}\cup \{(1,r,0)\}_{r=N_{\text{D}_1}}^{N-1}\cup\{(2,r,c)\}_{(r,c)\in[R]\times[C]}$.
    \item \textsc{Inner}: Prepares the dimension $B'\doteq B+1$ state $\ket{\psi_{r,c}}$ controlled on indices $(2,r,c)$ of $\ket{x_\text{o}}$.
    \item \textsc{Givens}: Apply $N-1$ multiplexed-controlled-phase-rotations, where the rotation angle is controlled by the indices $(r,b)$ of $\ket{x_\text{o}}\ket{b}$.
    \begin{itemize}
        \item In a minimum-space compilation, the lookup-table \textsc{RPrep} that outputs rotation angle bits dominates the Toffoli count of the entire block-encoding, and these Toffolis are executed serially by the lattice surgery compilation of~\cref{table:lattice_surgery_lookup}.
        \item In a minimum-time compilation, the circuit \text{Rot} that converts rotation bits to phase rotations, e.g.\ by phase gradient adders, dominates the Toffoli count of the entire block-encoding, and the majority of these Toffolis gates can be executed by the lattice surgery compilation of~\cref{sec:multiplex_rotations} in parallel.
    \end{itemize}
\end{itemize}

\begin{figure}
    \centering
    \includegraphics[height=5cm]{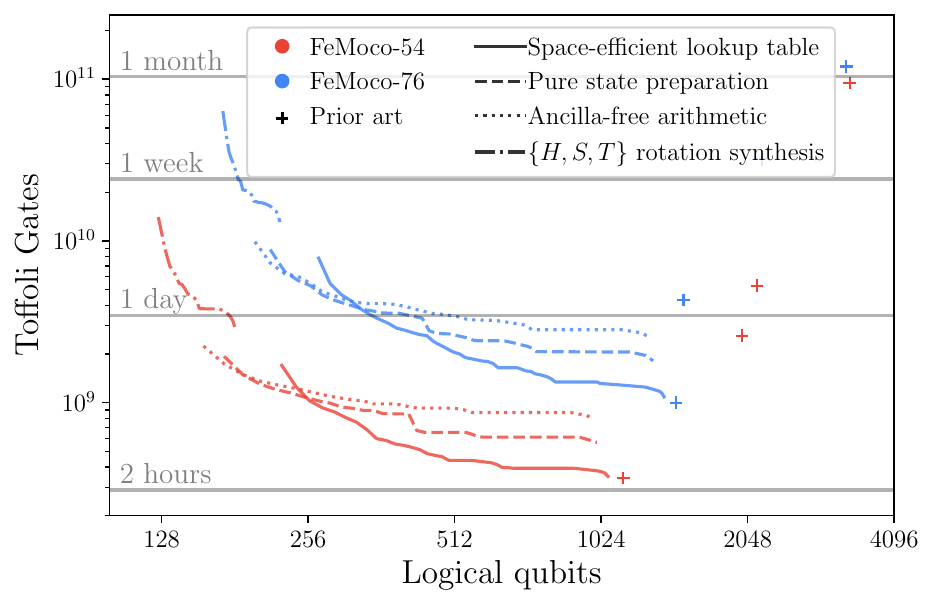}
    \includegraphics[height=5cm]{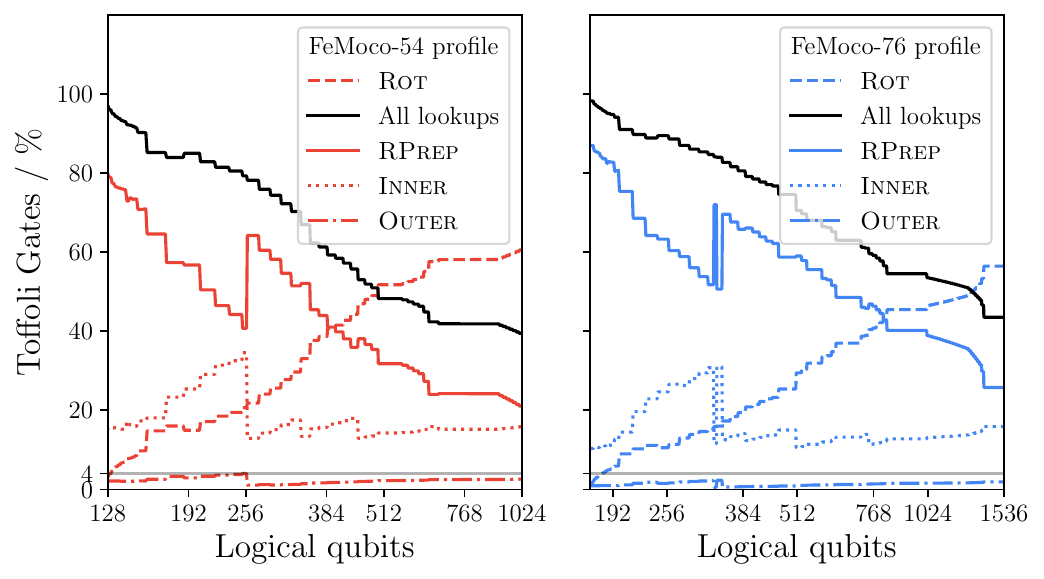}
    \caption{
    (Left) We block-encode the DFTHC Hamiltonian~\cref{sec:sec:appendix_tradeoffs} with different quantum circuit compilation options described in~\cref{sec:sossa_logical_resources} that span a broad logical-qubit-Toffoli gate trade-off down to as few as $2N+n_{B'}+n_{X_\text{o}}+\mathcal{O}(1)$ logical qubits.
    Plotted is the tradeoff curve with the maximum number of dirty qubits, which includes the $2N$ storing the fermion wave function.
    A conversion of $25\mu$s per Toffoli gate is portrayed.
    (Right) Across all regimes, $>96\%$ of  Toffolis gates for the best compilation are accounted for by: the \textsc{RPrep} lookup table for the angle bits of the $N-1$ Givens rotation angles, which dominates at minimum-space;
    the \textsc{Rot} rotations that convert angles bits to quantum rotations, which dominates at minimum-time; and the \textsc{Inner} lookup table  that prepares $\ket{\psi_{r,c}}$.
    The other components have negligible Toffoli cost.
    }
    \label{fig:femoco_improvements_breakdown}
\end{figure}

The four different space-time tradeoffs in~\crefpos{fig:femoco_improvements_breakdown}{left} each represent the cumulative qubit-Toffoli tradeoffs enabled by the following, where $n_X\doteq\lceil\log_2X\rceil$:
\begin{enumerate}
    \item Remove redundant registers~(\cref{sec:fewer_qubits_A}): The previous implementation of $\ket{\psi_\text{out}}$ was entangled with $\ket{r}\ket{g}$ and some registers indexing spin. 
    These can be computed on the fly when needed from $x_\text{o}$, which removes roughly $4+n_R$ qubits at a small additive Toffoli cost. 
    We always use this by default.
    \item Space-efficient quantum lookup tables~(\cref{sec:fewer_qubits_B}):
    We show that in cases where we stream the output bits $b$ in $J$ batches of $b_\text{sub}=\lceil b/J\rceil$ bits, dirty outputs qubits can be used with the same amortized Toffoli cost as using clean output qubits, which is a $\approx2\times$ improvement in the Toffoli count for implementing all swap networks.
    We also implement all lookup tables with the skew-tree variant in~\cref{sec:lattice_surgery_lookup} that has the same Toffoli count as \textsc{Select}-\textsc{Swap}~\cite{Low2024Trading} but can be implemented with even fewer ancilla.
    \item Pure state preparation~(\cref{sec:fewer_qubits_C}): $\ket{\psi_\text{out}}$ and $\ket{\psi_{r,c}}$ are $n_{X_\text{o}}$- and $n_{B'}$-qubit states respectively.
    Preparing these as pure states saves $n_{X_\text{o}}+2b_\text{coeff}$ garbage qubits over the previous alias sampling technique~\cite{BabbushPRX18} but at least doubles Toffoli count.
    \item Ancilla-free arithmetic~(\cref{sec:fewer_qubits_D}): Each multiplexed rotations to $b$ bits of precision using the phase gradient approach uses a $b$-bit adder. 
    Previously, the optimal-Toffoli adder using $b-1$ clean ancilla qubits was used. Here, we use the ancilla-free version with twice the Toffoli count.
    Ancilla-free arithmetic is also used in other places, such as to compute $r$ from $x_\text{o}$ on the fly, to flatten the $\ket{x_\text{o}}\ket{b}$ register in the lookup table. 
    This frees up a few dirty qubits to reduce the Toffoli count of the lookup table by \textsc{Select}-\textsc{Swap}.
    \item Direct rotation synthesis~(\cref{sec:fewer_qubits_E}): Rotations by the phase gradient approach requires a $b$-qubit Fourier resource state.
    By instead implementing rotations directly using $\{H,S,T\}$ gates we save $b$ logical qubits by not storing this state. 
    However, more lookup table output bits are needed to specify the rotation.
    We choose to increase $b_\text{coeff}$ and $b_\text{rot}$ by $12$ bits, which is almost double.
    This choice follows from the phase gradient bits of precision bound $b_{\text{rot}}\le\lceil\log_2\frac{\pi}{\epsilon}\rceil\approx\log_2\frac{1}{\epsilon}+1.65$ in~\cite{low2021halvingcostquantummultiplexed} compared to fallback synthesis of $k_1+6\approx\log_2\frac{1}{\epsilon}+13.16$ (see~\cref{sec:multiplex_rotations}),
    and is very conservative.
    The actual number needed could be possibly halved by computing the shift in CCSD(T) correlation energy, like in~\cite{low2025fast,PRXQuantum.2.030305}, combined with randomized rounding.
    As we stream output bits, this does not necessarily require a larger output register, but does increase the Toffoli count by a proportionate amount.
    We also require low-error $\ket{T}$ states that take additional cycles to synthesize using a catalyzed $\ket{CCZ}+\ket{T}\rightarrow3\ket{T}$ factory.
    Finally, we free a few more dirty qubit to assist the $\textsc{RPrep}$ lookup table by using reversible in-place arithmetic to compute a flattened index $\ket{x_\text{o}}\ket{b}\rightarrow \ket{x_{\textsc{RPrep}}\in[N+RB]}\ket{\cdots}$.
\end{enumerate}

However, we emphasize that once compiled to lattice surgery operations on a hybrid architecture~\cref{sec:architectures}, final physical-level resources map non-uniformly to logical quantum circuit-level resources of qubits and Toffolis.
For instance, the minimum-space compilation uses significantly fewer physical qubits than the logical qubit count might indicate as most reside in cold storage.
The minimum-space compilation may also use significantly more logical qubits than the least-possible, as the space penalty of having more logical qubits to improve runtime may be outweighed by the reduction in code distance.
The minimum-time compilation may also use significantly more Toffoli gates than expected as most in that regime implement single-particle basis rotations that may consume many $\ket{CCZ}$ statess in parallel.
In general, finding the optimal set of compilation options across all possible layouts is \textsc{NP}-hard.
This necessitates the use of heuristic optimization, and we report the Pareto frontier of our search space.
We believe that a more exhaustive enumeration would reveal options with slightly fewer qubits and slightly less runtime.

We generate candidate logical-level quantum circuits by the following heuristic.
\begin{enumerate}
    \item We enumerate over many pairs of non-zero integers $(n_\text{clean,max}, n_\text{dirty,max})$, which are the maximum number of clean and dirty logical ancilla qubits available, respectively, to implement the block-encoding.
    $n_\text{clean,max}$ excludes the absolute minimum number of qubits needed to run the algorithm, which is $n_{\text{min}}\doteq2N+n_{B'}+n_{X_\text{o}}+4$ (for all registers in~\cref{fig:dfthc_min_qubits} plus two for phase estimation).
    The largest possible choice of $n_\text{dirty,max}$ is $n_{\text{min}}$, which is typical of a monolithic architecture where all logical qubits are treated equally, and so may assist any quantum circuit except those that have it as explicit inputs.
    In this work, we choose $n_\text{dirty,max}\ge n_{B'}+n_{X_\text{o}}$, which represents that cold storage contains at most the $2N$ fermion wavefunction qubits, the two phase estimation ancilla, and the two spin indices $\sigma,\varsigma$, and cannot assist any computation. 
    These qubits are chosen as they are the least-often accessed qubits in the algorithm and so contribute the  least to cold storage latency. 
    \item  We find the minimum Toffoli count for each $(n_\text{clean,max}, n_\text{dirty,max})$ and for each set of compilation options, which is parameterized solely by the different ways to implement lookup tables. 
    Toffoli count is also monotonically decreasing with increasing $n_\text{clean,max}$ or $n_\text{dirty,max}$, and we plot in~\crefpos{fig:femoco_improvements_breakdown}{left} the case $n_\text{dirty,max}=n_{\text{min}}$.
    \item For each compilation option, we profile the number of Toffoli gates used by the main components \textsc{Inner}, \textsc{Outer}, \textsc{RPrep}, and \textsc{Rot}.
    An example of this profile as a function of total logical qubits is shown in~\crefpos{fig:femoco_improvements_breakdown}{right} where we only display the minimum-Toffoli option.
\end{enumerate}
Cost profiles like~\cref{fig:femoco_improvements_breakdown} are easily converted to physical qubit counts and runtimes using heuristic cost models.
A common model, for example, assumes a serial Toffoli execution model at one Toffoli per $d_{\text{lookup}}\approx 25$ cycles using a large ($\gtrsim 324$ patch) workspace on top of a  $50\%$ routing overhead~\cite{PRXQuantum.2.030305}, and adding the footprint of enough magic state factories.
Using our optimized lookup table lattice surgery compilation~\cref{table:lattice_surgery_lookup} with $F=1$ factory, which slows execution by $\approx 5\times$, the minimum logical qubit compilation of FeMoco54 requires roughly 
\begin{align}
\text{Surface code patches}\approx 130\times 1.5+(3\times 5F)=210\quad\Rightarrow\quad\text{Physical qubits}\approx210\times2\times25^2=262500.
\end{align}
Although already a respectable $12\times$ improvement over our previous work, this is still $>3\times$ larger than our final result.
Meanwhile, rigorous Pauli-based-computation models for completely general Clifford+$\textsc{T}$ circuits~\cite{Litinski2019Game} usually require one to two orders of magnitude more physical spacetime volume.
Although such simple models are easy to use, they tend to be very loose as they do not account for non-Clifford parallelization and non-uniform encoding rates, which we do in the next section.

\subsection{Minimizing physical qubits and runtime}\label{sec:sossa_physical_resources}
We obtain the Pareto physical qubits-runtime frontier in~\cref{fig:femoco54_improvements}, from enumerating candidate circuits over a range of layout options, lattice surgery compilation options, and code distances.
The compilation options for select points are tabulated in~\cref{sec:parto_frontier_parameters}.
We report expected runtimes that have already been divided by $(1-p)$ to fairly account for the need to repeat the algorithm in event of failure -- failure can be detected, for instance, by a post-hoc complementary gap analysis.
Importantly, as we assume a hex-grid architecture, our resource estimates do not assume the ability to walk surface code patches.
This forbids fast random access by walking in cold storage: only move operations are allowed, which increases $T_\text{yoke}$ by a modest factor.
We compute timing and error contributions from latency separately and find that it has limited impact on our reported resource estimates due to the communication complexity advantage of block-encodings~\cref{sec:block_encoding_communication}.
We now describe the layout options we enumerate over.
\subsubsection{Minimum-space layout}\label{sec:simulation_minimum_space}
This layout described in~\cref{sec:minimum_space} uses a $4\times 3$ compute region with at most $F=1$ $\ket{CCZ}$ factory~\cref{table:ccz_factory} to implement the dirty-ancilla lookup table in~\cref{table:lattice_surgery_lookup}.
Cold storage and medium-latency hot storage is used.
We modify the lookup table compilation of~\cref{fig:zx_skew_tree_dirty} that uses only a $3\times 3$ sub-grid of compute to use one storage instead of three: Observe that the address bits $x_k$ for $k>1$ are used only in the ``bounce'' and ``up'' step, where the output register access hallway is unoccupied.
Hence, all these address bits may share hot storage with the output register.
The two bits $x_0,x_1$ are used in ``down'' and ``leaf'', but at most only once per $\gtrsim8d$ cycles.
Hence, we place them to the left of the $4\times3$  compute region in a $4\times1$ column that store five logical qubits using two regular patches and one 3-qubit rectangular patch. 
We place a catalyst $\ket{T}$ in this column as well.
Now observe that each sequence of $5$ Toffoli gates queries only two ancilla bits $a_{d},a_{d-1}$.
We assume that the slack granted by a larger $4\times 3$ compute region is sufficient to these bits around and place them in the same column.
As $d_{\ket{CCZ}}\approx150$ cycles is longer than the $7d_\text{in}\approx105$ cycles to measure a hot storage stabilizer by about $3d_\text{in}$, we assume there is sufficient time to rotate in the next ancilla qubits in the sequence from the slow hot storage to one of these three patches.

As each candidate circuit has a known Toffoli count, we have from~\cref{table:ccz_factory} an initial guess for the error per magic state $p_{\ket{CCZ}}$, the minimum code distance $d_2$, and the number of cycles $d_{\ket{CCZ}}$ to produce each $\ket{CCZ}$ state.
After lookup tables, controlled-angle rotations \textsc{Rot} are the next dominant consumer of $\ket{CCZ}$ states, but still less than $4\%$ by~\cref{fig:femoco_improvements_breakdown} in a minimum-space compilation.
Using $\{H,S,T\}$ rotation synthesis,~\cref{sec:catalyzed_T} and~\cref{tab:space_limited_resources} indicates that $2\ket{T}$ states are produced and consumed in $\approx d_{\ket{CCZ}}+13d_2$ cycles.
Overall, this indicates that a reasonable simplification is to apply the average $\ket{CCZ}$ consumption rate given by~\cref{table:lattice_surgery_lookup} to all $\ket{CCZ}$ states and $\ket{T}$ states.
With the average $\ket{CCZ}$ and $\ket{T}$ consumption rates, we obtain the number of cycles required to execute the algorithm.
We repeat this for a few trial distances $d>d_2$ for the compute region and enumerate over cold and hot storage parameters to obtain the smallest physical qubit footprint and the corresponding runtime with overall error $\le p_0$.

\subsubsection{Minimum-spacetime layout}
\begin{figure}
    \centering
    \includegraphics[width=0.8\linewidth]{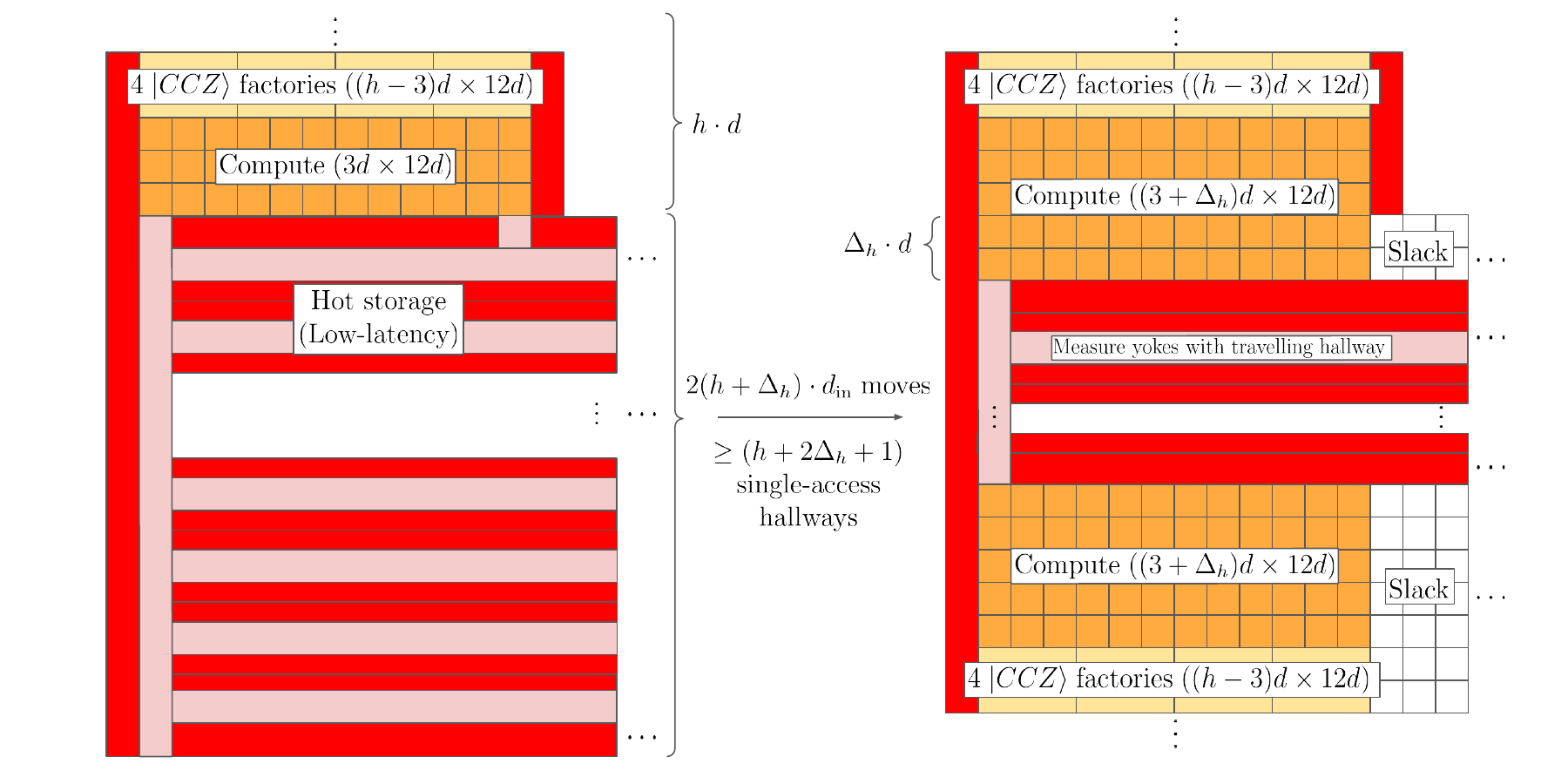}
    \caption{After implementing the quantum lookup table, rows of output bits are packed towards the top in order to instantiate more $\ket{CCZ}$ factories in a new compute region. This layout provides the resources to perform two Givens rotations in parallel, with unused slack.}
    \label{fig:multiplex_reflected_layout}
\end{figure}

This layout described in~\cref{sec:minimum_spacetime} uses a $3F\times 7$ compute region for any number of factories $F$ to implement the clean-ancilla lookup table in~\cref{table:lattice_surgery_lookup}.
Cold storage and low-latency hot storage is used.
We only consider single-access hallways where the cycles $d_{\text{lookup}}$ per $\ket{CCZ}$ consumed is given by~\cref{eq:lookup_actual_cycles_ccz_consumed}, and we enumerate over $F\in\{1,2,3,4,5,6\}$ for each candidate circuit.
The computation for initial guess of code distance and cycles for the all lookup table is identical to~\cref{sec:simulation_minimum_space}.
However, the Toffoli cost of lookup tables tends to be significantly lower as many output qubits are allocated.
As a result, controlled-angle rotations \textsc{Rot} is no longer have a negligible Toffoli cost.
Although a workspace $3d$ units high is more than sufficient to consume Toffolis as quickly as they are produced during the lookup table phase, the adders for phase gradient rotations appear to require more space.
Hence, we use at least three access hallways: During \textsc{Rot}, two of these will be moved to create a compute workspace that is effectively $5d$ units high and bordered by hot storage patches.
The remaining access hallway is used to measure outer code stabilizers.
As we choose $T_\text{yoke}=50d$, this provides a very large uptime to also access any other required data.
As shown in~\cref{sec:multiplex_rotations}, this allows us to also consume Toffolis almost as fast as they are produced.
As described in~\cref{sec:parallel_givens}, Givens rotations can also be parallelized to execute in logarithmic depth, where the $j^\text{th}$ layer applies $2^j$ rotations.
Parallelization requires more $\ket{CCZ}$ factories, and a larger workspace.
If there are a sufficient number of output bits in a sufficient number of columns,~\cref{sec:parallel_givens} shows how we may widen selected access hallways by moving patches to instantiate new $\ket{CCZ}$ factories in the created space, like in~\cref{fig:multiplex_reflected_layout}.
In this regime, rotations may also be implemented by $\ket{CCZ}$ states, and this larger workspace allows rotation $\ket{CCZ}$ states to be consumed at the rate given by~\cref{sec:adder_multiplex_rotations}.
Moreover, rotation bits are measured out after each rotation, which leaves even more workspace to implement the remaining rotations.
Further parallelization by instantiating more $\ket{CCZ}$ factories in the newly created workspace is possible, but we leave this optimization to future work.
Overall, we assume that the overall layout suffices to consume $\ket{CCZ}$ states at the above rates, and we also leave to future work the explicit indexing od logical qubits in memory.

\subsubsection{Minimum-time layout}
We use the minimum-time layout~\cref{sec:minimum_time_layout} in much the same way as above.
The main differences are the use of low-latency double-access hot-storage, which supports implementing lookup table as double the Clifford rate, or down to $d/2$ cycles per $\ket{CCZ}$ consumed.
Saturating this rate requires a large number of factories $F\approx 12$.
As Givens rotations can consume $\ket{CCZ}$ states even faster by~\cref{eq:controlled_adder_batch_cycles}, we enumerate over a range $F\in[4,16]$.
Like the minimum-spacetime layout, we move access hallways per~\cref{sec:parallel_givens} 
during the Givens rotation phase to create a workspace that implements them in parallel by consuming $\ket{CCZ}$ states as quickly as they are produced, down to $<d/4$ cycles.

\subsection{Cold storage access and latency}\label{sec:5_simulation_latency}
In this section, we show that the additional runtime due to accessing cold storage with \textsc{PauliPM}s and the associated latency is either negligible, or can be made negligible with a buffer of at most three low-latency logical qubits.
We estimate the excess cycles and errors from the cold storage \textsc{PauliPM}s, guided by the upper bounds~\cref{eq:mem_excess_error,eq:mem_excess_cycles}, for implementing the block-encoding~\cref{fig:fullcircuit}.
There are $4N$ \textsc{CSwap} gates on cold storage, targeting the $N$ spin-up and $N$ spin-down fermions, and $4(N-1)$ Givens rotations between pairs of qubits storing the spin-down fermions.
In the worst case, each of these are implemented by moving the two cold storage targets to \textbf{compute} and back in $4$ \textsc{PauliPM}s taking $12d_\text{in}$ cycles.
Observe the linear scaling of \textsc{PauliPM}s with $N$ even though the Hamiltonian has $\mathcal{O}(N^4)$ terms when fully expanded, consistent with our block-encoding communication complexity results~\cref{thm:communication_BE}.

Let us give additional runtime estimates for typical electronic structure parameters with $2N\approx 100,r=\frac{\pi}{2 \epsilon}\lambda_\text{eff}\approx 5\cdot 10^4$ using typical code parameters from~\cref{sec:parto_frontier_parameters}.
In the minimum-space layout, we have $d\approx25,d_\text{in}\approx15,T_\text{yoke,min}\approx200d_\text{in},T_\text{yoke,max}\approx4T_\text{yoke,min}$.
Then \textsc{PauliPM}s for all \textsc{CSwap}s take $48rNd_\text{in}<24rT_\text{yoke,min}$ cycles, or $0.5$ hours.
In cold storage, latency is minimized by placing wavefunction qubits with the same orbital index next to each other.
This allows most \textsc{CSwap} gates to be performed without moving columns. So to first order, we ignore this contribution.
As we chose $T_\text{yoke,max}/T_\text{yoke,min}=4$,~\cref{eq:mem_excess_cycles} indicates that an excess time contribution of $\frac{48rNd_\text{in}}{T_\text{yoke,max}-T_\text{yoke,min}}T_\text{yoke,min}=4rT_\text{yoke,min}$, or $10$ minutes.
As we use a distance $2$ outer code, the excess error contribution to cold storage is the equivalent of running the algorithm for $\frac{T_\text{yoke,max}}{T_\text{yoke,min}}4rT_\text{yoke,min}$ cycles longer, or $40$ minutes.
The Givens rotations can be analyzed the same way, giving a total additional runtime of 1 hour, which is negligible when runtimes are on the order of days to months.

In the minimum-spacetime layout, the parameters for $d,d_\text{in}$, and $T_\text{yoke}$ are not too different from that of the minimum-space layout.
In this case, the additional runtime of 1.0 hours appears to be a significant fraction of total runtime, especially at the minimum-spacetime runtime of $\approx 1$ hours.
However, there are some mitigating factors. 
As $\ket{CCZ}$ factories are operating continuously, the $12d_\text{in}$ cycles for all $\textsc{PauliPM}$ in each $\textsc{CSwap}$ is also amortized by $d_{\ket{CCZ}}\ge d$ cycles for an effective duration of $12d_\text{in}-d_{\ket{CCZ}}$, which is a significant reduction when few factories are operating.
When more space is available, a simple solution that completely eliminates latency considerations is to add at least three logical qubits in low-latency hot storage as a buffer.
This buffer stores the next three fermion wavefunction qubits scheduled to interact with quantum gates.
There will be a very large number of cycles $\gg T_{\text{yoke,min}}$ spent producing Toffoli gates for lookup tables between each Givens rotation.
Moreover, each Givens rotation uses at least $2b_\text{rot}-1\approx 30$ Toffoli gates.
Even when parallelizing Givens rotations by a factor of four, wavefunction qubits only need to be cycled in every $(2b_\text{rot}-1)d/4\gtrsim150$ cycles, which is comparable to that for $12d_\text{in}\approx 180$ for swapping two logical qubits in and out of cold storage with $\textsc{PauliPM}$s that are executed concurrently.
The \textsc{CSwap} gates may appear to be more challenging as they occur in sequence and only cost one Toffoli gate each.
However, we see in~\cref{fig:fullcircuit} that they are applied concurrently with the large \textsc{Inner} lookup table, and do not need to be scheduled consecutively.
Hence, we may use the buffer qubits in a similar manner.

In general, we find that the cold storage is always advantageous in space-optimized block-encodings where the additional runtime is negligible, but more care is required when used with time-optimized block-encodings.
As Toffoli count scales like $\mathcal{O}(N^2)$ for these utility-scale electronic structure problems, the relative importance of cold storage access and latency in any case can be expected to diminish even further for larger $N$.
In our resource estimates, we store significantly fewer qubits in cold storage than the theoretical maximum.
By moving access hallways, we parallelize the Givens rotation to obtain a $\ket{CCZ}$ consumption rate that is a factor of $G\in\{1,2,3,4\}$ higher.
For slack, when $\ket{CCZ}$ states are produced faster than one per $d$ cycles, we place $3G$ buffer qubits in hot storage and assume that this is sufficient buffer to ignore cold storage latency, though this should be validated in future work by an explicit lattice surgery compiler and scheduler.

\subsection{Summary of key assumptions}

Our resource estimates for the physical costs of simulating electronic structure are qualitatively accurate.
However, as they are mostly compiled by hand without automatic large-scale tooling, and also rely on extrapolation and interpolation, we do not expect them to be quantitatively exact, and also expect that they can be further improved.
For instance, our benchmarks are based on $d$ cycles per lattice surgery operation, which could be improved to below $\frac{2}{3}d$ with more sophisticated compilation based on temporally encoded lattice surgery~\cite{Chamberland2021Temporally}.
Furthermore, our compilation has significant slack as it does not yet make full use of the additional $50\%$ logical qubit density offered by three-qubit rectangles.
Our lattice surgery complication of lookup tables also has significant slack as all $\textsc{CX}$ fixup operations are explicit -- this choice was made to support our approximation that all other circuit components, which are subdominant in Toffoli gate count, can be implemented at the same rate of $\ket{CCZ}$ consumption, possibly with some implicit movement of access hallways to create a larger workspace whenever required.
There is also significant opportunity to parallelize  Toffoli gates in lookup table \textsc{CSwap} operations.

There are also a few key caveats mentioned or implied throughout the manuscript to keep in mind when interpreting our resource estimates.
We now summarize them for convenience.
\begin{enumerate}
    \item We benchmark our quantum codes up to distance $18$, but our resource estimates require distances $20$ to $27$ depending on compilation parameters and the size of the molecule.
    \item We estimate logical error rates of our yoked codes assuming an ideal outer decoder.
    This is motivated by the observation that the empirical performance of yoked codes in prior art~\cite{Gidney2025Yoked} is well-approximated by making this assumption.
    A future study should instantiate and benchmark this outer decoder for our dense twist storage, such as using complementary gaps.
    \item We do not benchmark the logical error rate of lattice surgery operations involving twists, which are present when loading/unloading densely encoded qubits. Instead we implicitly assume the dense-packing memory error rate of $3.5^{-d}/10$, which is in any case more conservative than that of regular lattice surgery in~\cref{fig:h_bridge}.
    \item When a lattice surgery operation is classically conditioned on the outcome of a measurement involving a yoked logical qubit, the probability of a catastrophic error occurring is higher than expected, depending on how far in the past the outer code stabilizer was previously measured. We neither account for overhead of measuring the yoke stabilizer an additional time before each such operation nor evaluate the probability of these errors occurring.
    \item When the number of $\ket{CCZ}$ factories is not 1, 4, or 8, we assume interpolated or extrapolated $\ket{CCZ}$ consumption rates for quantum lookup tables given by~\cref{table:lattice_surgery_lookup}. Similarly, we assume simplified average magic state consumption rates for other parts for which we have not provided pipe diagrams.
    However, this assumption is quite reasonable as the marginal cost of adding more access hallways is quite low, and we may easily move access hallways to create even larger compute workspaces when required.
    For example, we move access hallways to create a $67\%$ larger workspace for the multiplexed rotations.
    \item We use a simplified cold storage access and latency model in~\cref{sec:latency} using a few logical buffer qubits and move up to $xb_\text{rot}$ wavefunction qubits from cold storage to hot storage, where $x\in\{0,1,2,4\}$ depends on how much we parallelize the multiplexed rotations.
\end{enumerate}


%% file: 1_compact_patch/noise_model.tex

For convenience,~\cref{tab:noise_gates} and~\cref{tab:noise_model} reproduce the definition of noise operations (with $p=10^{-3}$) used in this work, as established by Refs.~\cite{McEwen2023RelaxingHardware,Gidney2022Pentagons}.

\begin{table}[ht]
    \centering
    \begin{tabular}{c|c}
    \hline\hline
    Noise channel & Probability distribution of effects
    \\
    \hline
    $\text{Classical measurement error}$ & $\begin{aligned}
        1-p &\rightarrow \text{(report previous measurement correctly)}
        \\
        p &\rightarrow \text{(report previous measurement incorrectly; flip its result)}
    \end{aligned}$
    \\
    \hline
    $\text{Bitflip channel}$ & $\begin{aligned}
        1-p &\rightarrow I
        &\;\;
        p &\rightarrow X
    \end{aligned}$
    \\
    \hline
    $\text{Phaseflip channel}$ & $\begin{aligned}
        1-p &\rightarrow I
        &\;\;
        p &\rightarrow Z
    \end{aligned}$
    \\
    \hline
    $\text{One-qubit depolarizing channel}$ & $\begin{aligned}
        1-p &\rightarrow I
        &\;\;
        p/3 &\rightarrow X
        &\;\;
        p/3 &\rightarrow Y
        &\;\;
        p/3 &\rightarrow Z
    \end{aligned}$
    \\
    \hline
    $\text{Two-qubit depolarizing channel}$ & $\begin{aligned}
        1-p &\rightarrow I \otimes I
        &\;\;
        p/15 &\rightarrow I \otimes X
        &\;\;
        p/15 &\rightarrow I \otimes Y
        &\;\;
        p/15 &\rightarrow I \otimes Z
        \\
        p/15 &\rightarrow X \otimes I
        &\;\;
        p/15 &\rightarrow X \otimes X
        &\;\;
        p/15 &\rightarrow X \otimes Y
        &\;\;
        p/15 &\rightarrow X \otimes Z
        \\
        p/15 &\rightarrow Y \otimes I
        &\;\;
        p/15 &\rightarrow Y \otimes X
        &\;\;
        p/15 &\rightarrow Y \otimes Y
        &\;\;
        p/15 &\rightarrow Y \otimes Z
        \\
        p/15 &\rightarrow Z \otimes I
        &\;\;
        p/15 &\rightarrow Z \otimes X
        &\;\;
        p/15 &\rightarrow Z \otimes Y
        &\;\;
        p/15 &\rightarrow Z \otimes Z
    \end{aligned}$
    \\
    \hline\hline
    \end{tabular}
    \caption{
        Noise channel definitions used by \cref{tab:noise_gates}.}
    \label{tbl:noise_channels}
\end{table}

\begin{table}[ht]
    \centering
    \begin{tabular}{c|c}
         \hline\hline
         {Noisy Gate} & {Definition} \\
         \hline
         $\text{AnyClifford}_2(p)$ & \text{Any two-qubit Clifford gate, followed by a two-qubit depolarizing channel of strength $p$.} \\
         \hline
         $\text{AnyClifford}_1(p)$ & Any one-qubit Clifford gate, followed by a one-qubit depolarizing channel of strength $p$. \\
         \hline
         $\text{R}_{Z}(p)$ & Initialize the qubit to $\ket{0}$, followed by a bitflip channel of strength $p$. \\
         \hline
         $\text{R}_{X}(p)$ & Initialize the qubit to $\ket{+}$, followed by a phaseflip channel of strength $p$. \\
         \hline
         $M_Z(p, q)$ & Measure the qubit in the $Z$-basis, followed by a one-qubit depolarizing channel of strength $p$, \\
         & and flip the value of the classical measurement result with probability $q$.\\
         \hline
         $M_X(p, q)$ & Measure the qubit in the $X$-basis, followed by a one-qubit depolarizing channel of strength $p$, \\
         & and flip the value of the classical measurement result with probability $q$. \\
         \hline
         $\text{Idle}(p)$ & If the qubit is not used in this time step, apply a one-qubit depolarizing channel of strength $p$. \\
         \hline\hline
    \end{tabular}
    \caption{
    Noise channels and the rules used to apply them. Noisy rules stack with each other.
    }
    \label{tab:noise_gates}
\end{table}

\begin{table}[ht]
    \centering
    \begin{tabular}{c|c}
        \hline\hline
         {Name}
             & Uniform depolarizing
        \\\hline
        {Noisy Gateset}
            &\noindent\begin{tabular}{c|c}
                $\text{CX}(p)$ & \multirow{3}{*}{$\text{AnyClifford}_2(p)$}\\
                $\text{CZ}(p)$ &\\
                $\text{CXSWAP}(p)$ &\\
                \multicolumn{2}{c}{$\text{AnyClifford}_1(p)$}\\
                \multicolumn{2}{c}{$\text{R}_{Z/X}(p)$}\\
                \multicolumn{2}{c}{$M_{Z/X}(p, p)$}\\
            \multicolumn{2}{c}{$\text{Idle}(p)$}\\
            \end{tabular}
        \\\hline\hline
    \end{tabular}
    \caption{
        Details of the error models for gates used in this paper.
    }
    \label{tab:noise_model}
\end{table}


%% file: 1_compact_patch/lattice_surgery_appendix.tex
In this section, we present the full gate sequences that implement various surface code operations. We also present additional benchmarks that may be of interest.

\begin{figure}[htbp]
    \centering
    \includegraphics[width=0.9\linewidth]{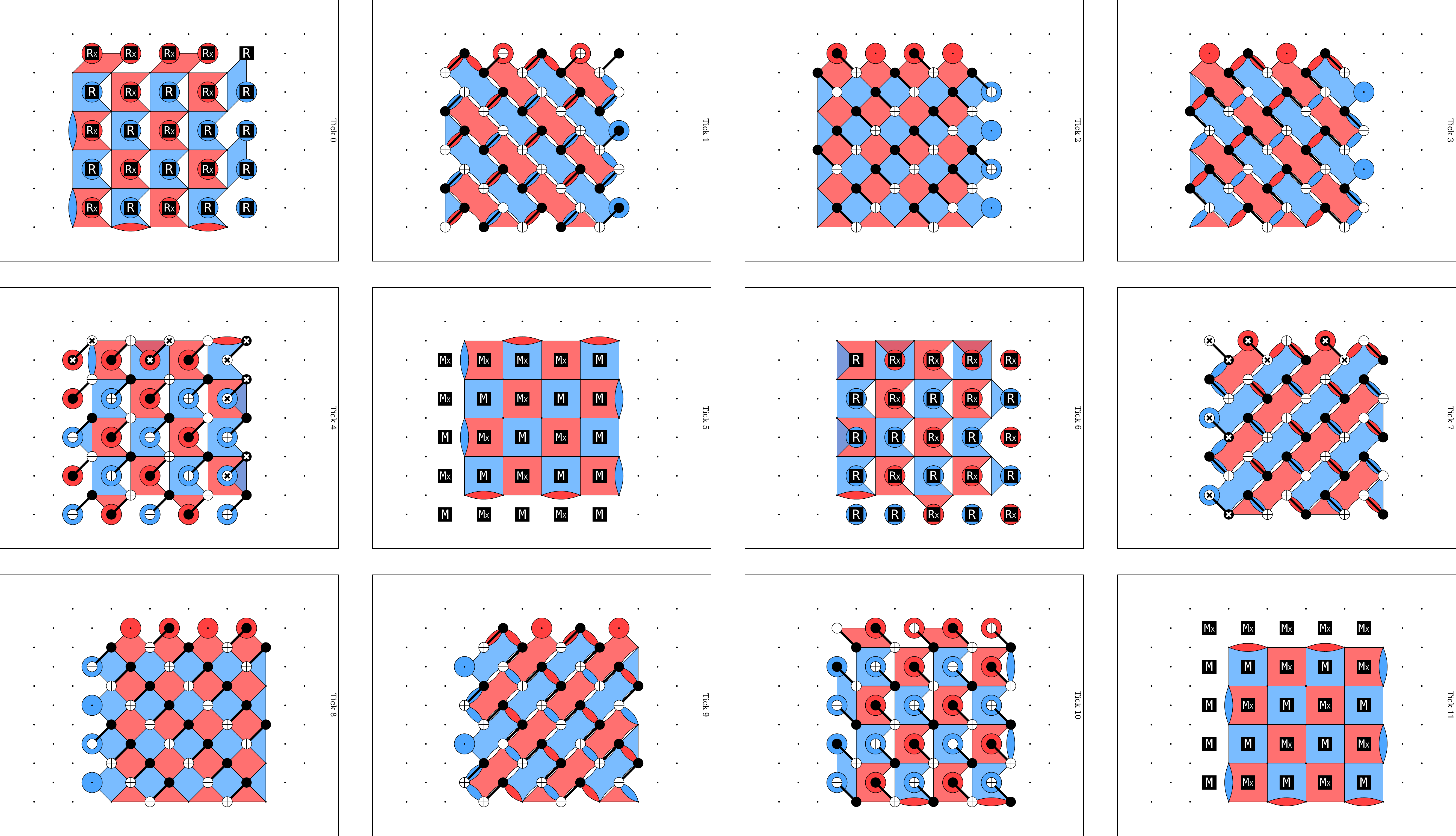}
    \\
    \vspace{0.5cm}
    \includegraphics[width=0.9\linewidth]{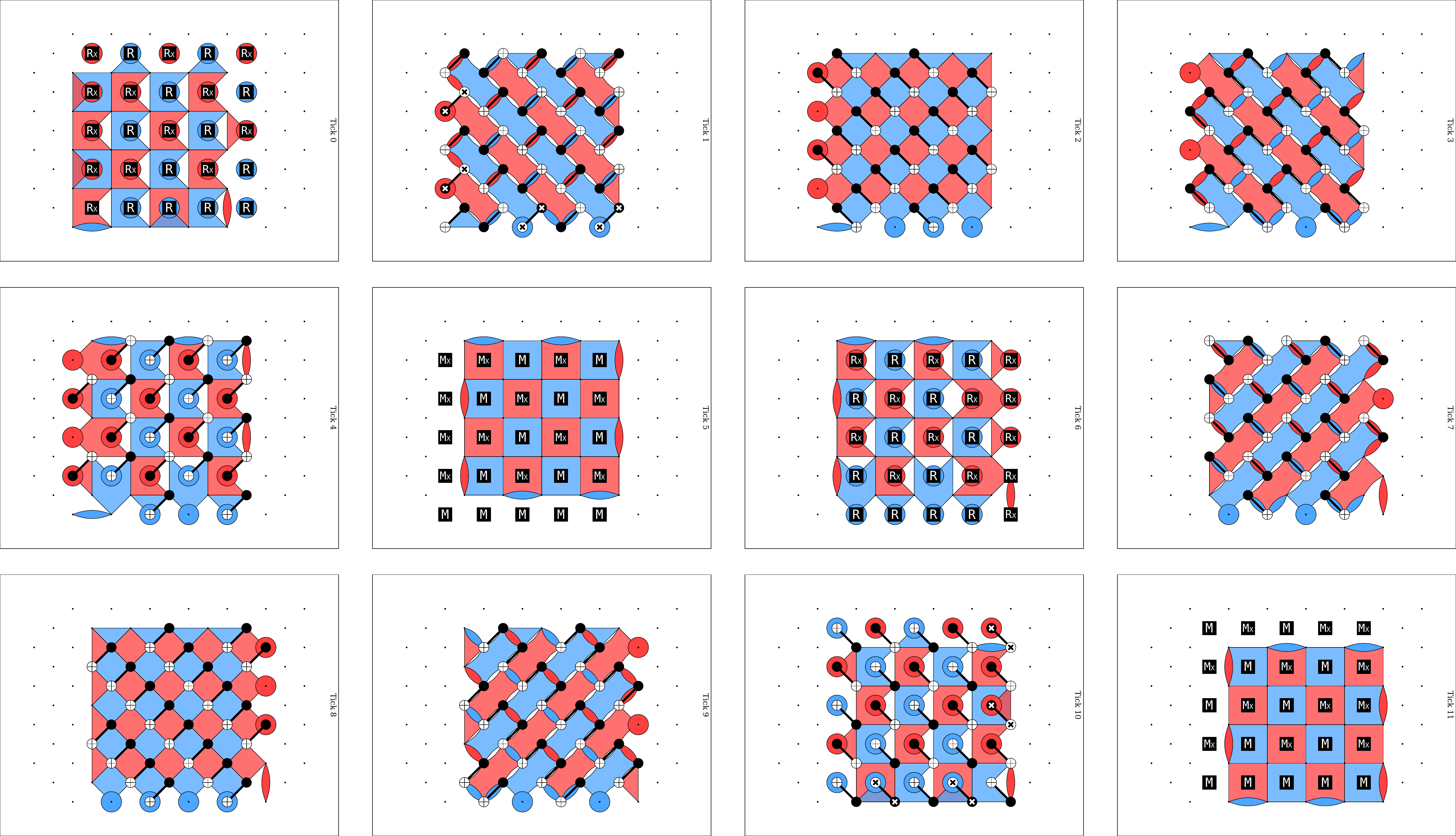}
    
    \caption{Gate sequences implementing walking of a surface code patch for the cases of (top) an $X$-top configuration, and (bottom) a $Z$-top configuration.}
    \label{fig:walking}
\end{figure}

\begin{figure}[htbp]
    \centering
    \includegraphics[width=\textwidth]{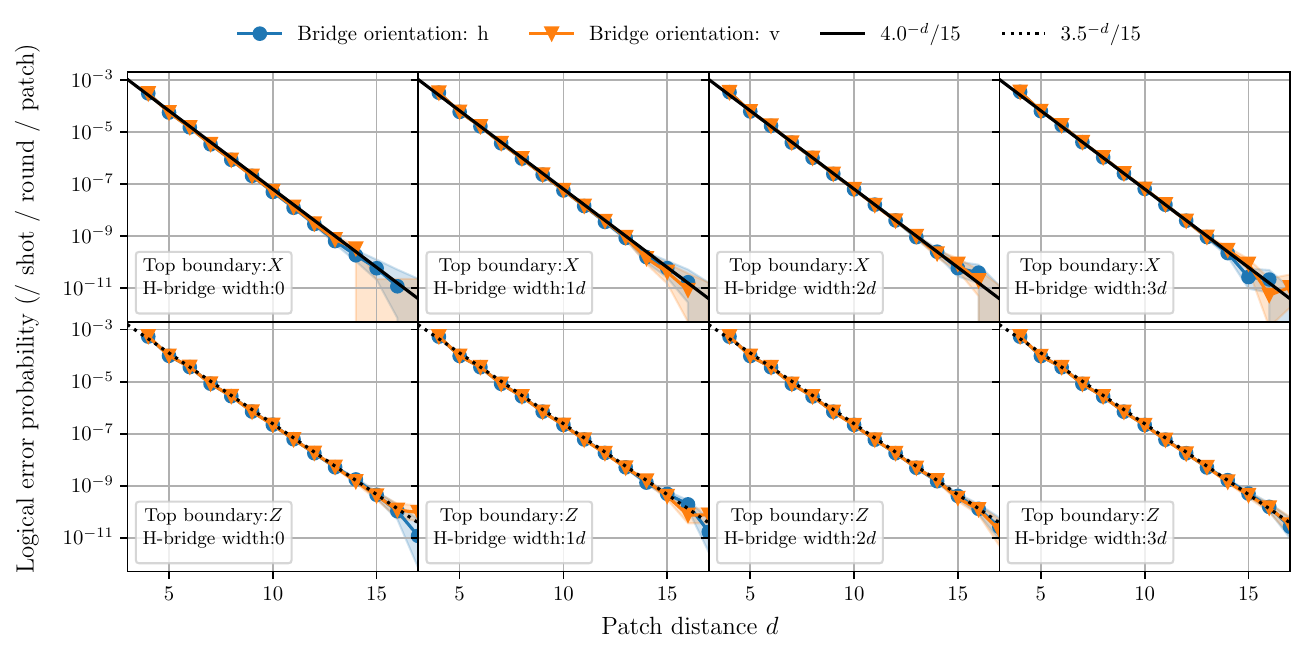}
    \caption{The logical error rate per round per patch of H-bridge lattice surgery between distance $d$ patches in a $10^{-3}$ uniform depolarizing noise model for each configuration, enumerated over all combinations of a $\{Z,X\}$-top orientation, a horizontal or vertical bridge orientation, and a bridge distance of $\{0,d,2d,3d\}$.}
    \label{fig:h_bridge_breakdown}
\end{figure}

\begin{figure}[htbp]
    \centering
    \begin{minipage}{0.9\linewidth}
        \centering
        \includegraphics[width=\linewidth]{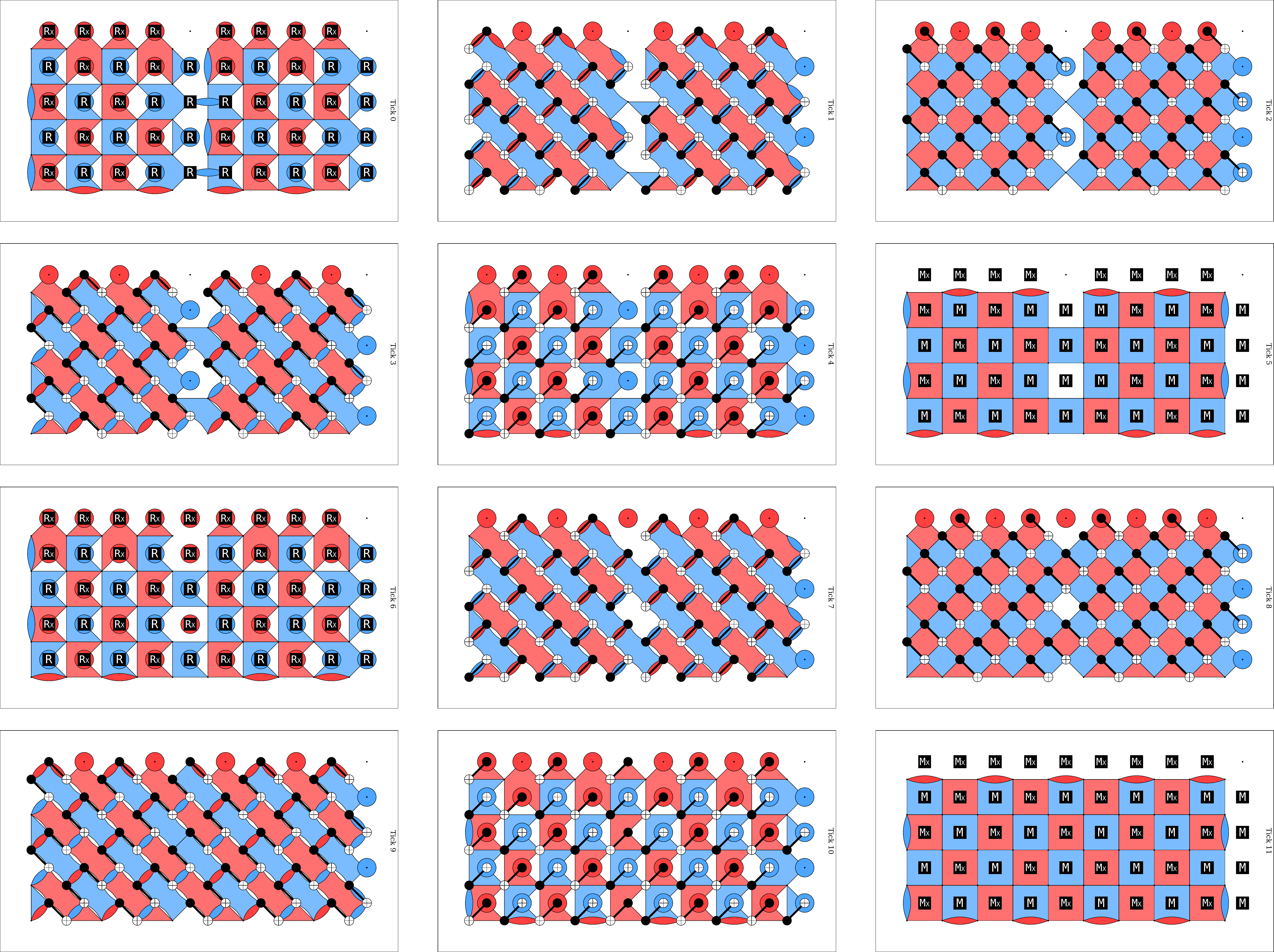}
        
        \vspace{0.2cm}
        (a) $X$-top configuration, horizontal merge.
    \end{minipage}
    
    \vspace{2em} 
    
    \begin{minipage}{0.9\linewidth}
        \centering
        \includegraphics[width=\linewidth]{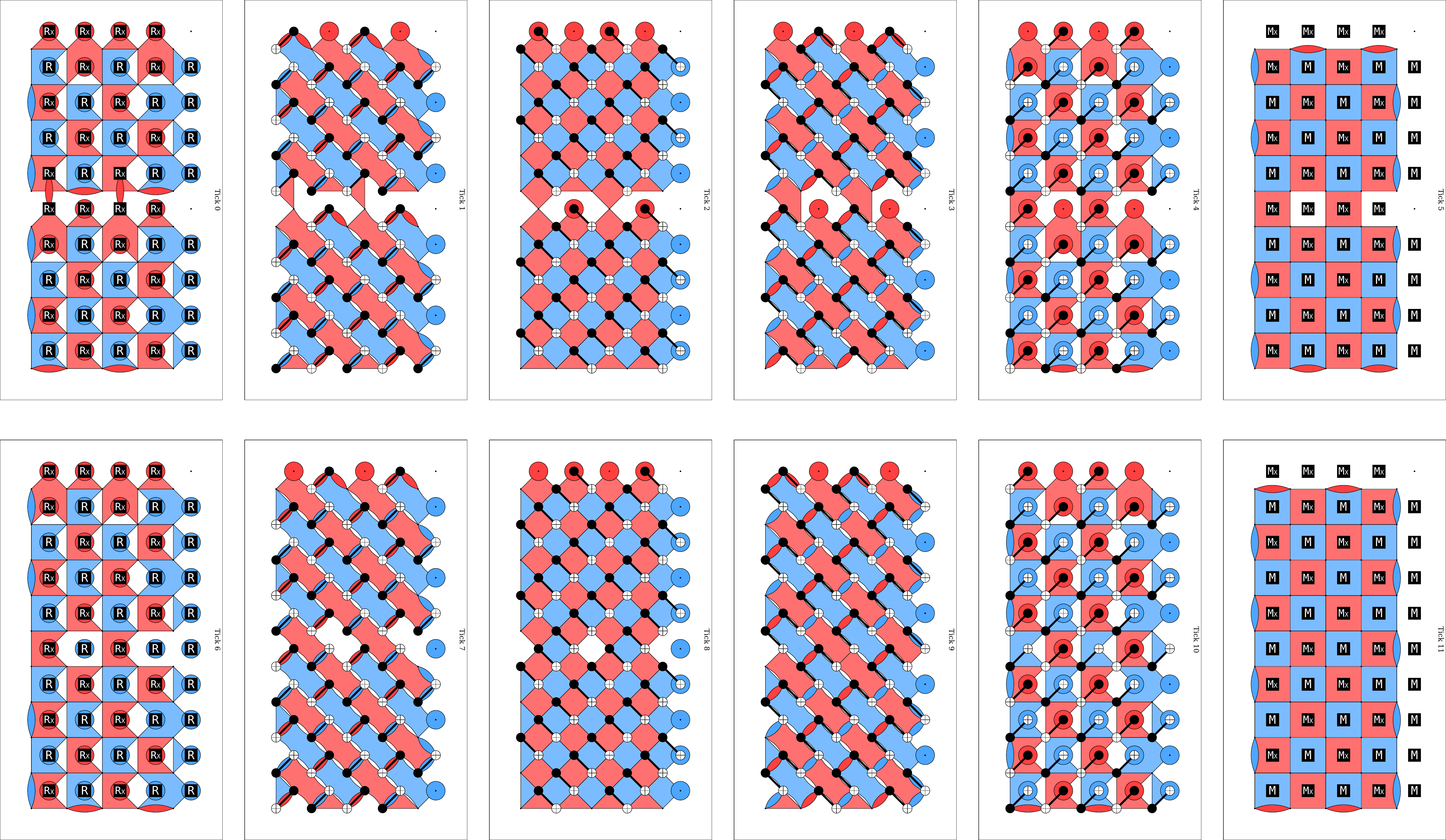}
        
        \vspace{0.2cm}
        (b) $X$-top configuration, vertical merge.
    \end{minipage}
    
    \caption{Lattice surgery merge operation for all possible patch orientations.}
    \label{fig:lattice_surgery_merge}
\end{figure}


\begin{figure}[htbp]
    \addtocounter{figure}{-1} 
    
    \centering
    \begin{minipage}{0.9\linewidth}
        \centering
        \includegraphics[width=\linewidth]{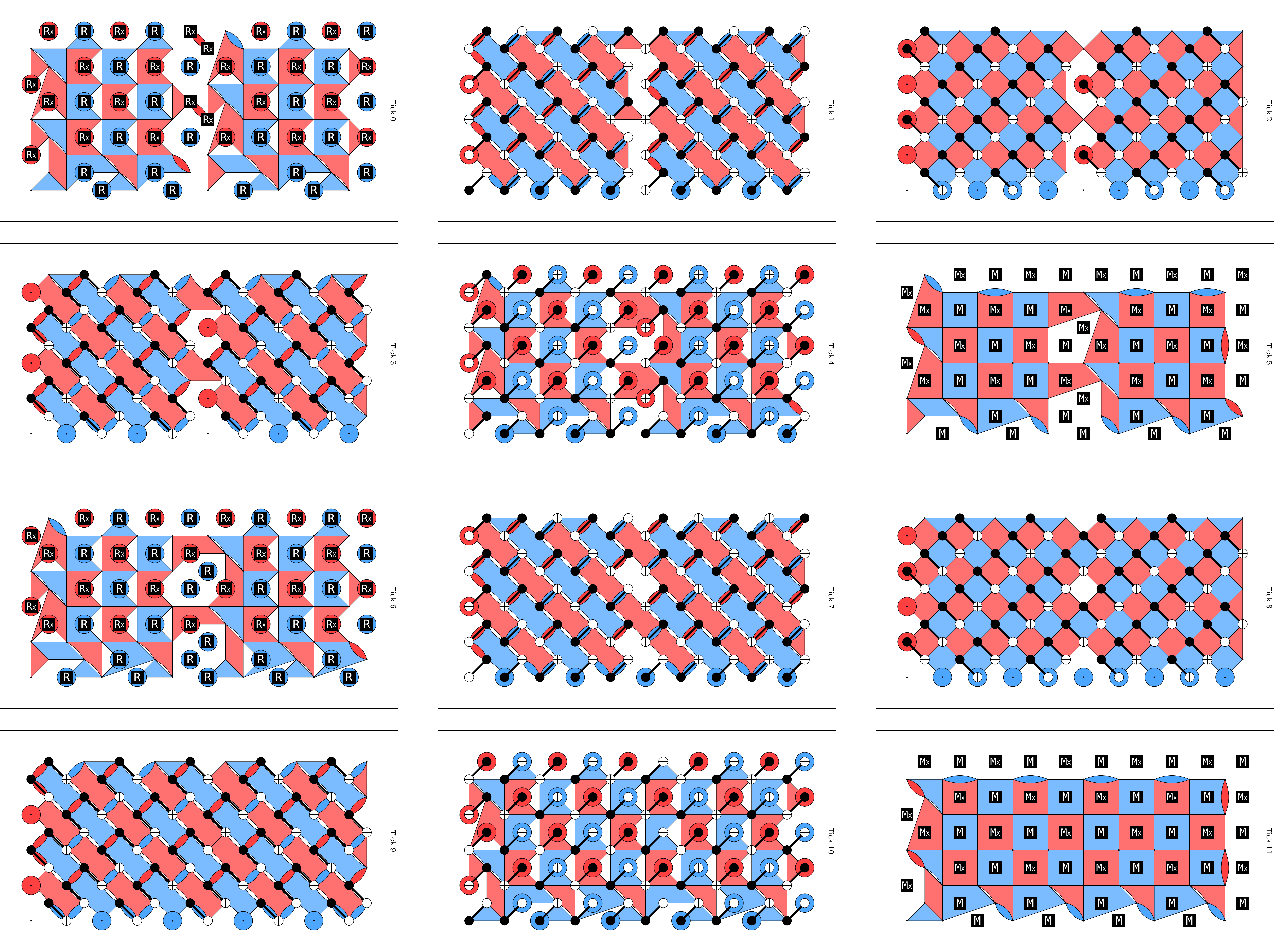}
        
        \vspace{0.2cm}
        (c) $Z$-top configuration, horizontal merge.
    \end{minipage}
    
    \vspace{2em} 
    
    \begin{minipage}{0.9\linewidth}
        \centering
        \includegraphics[width=\linewidth]{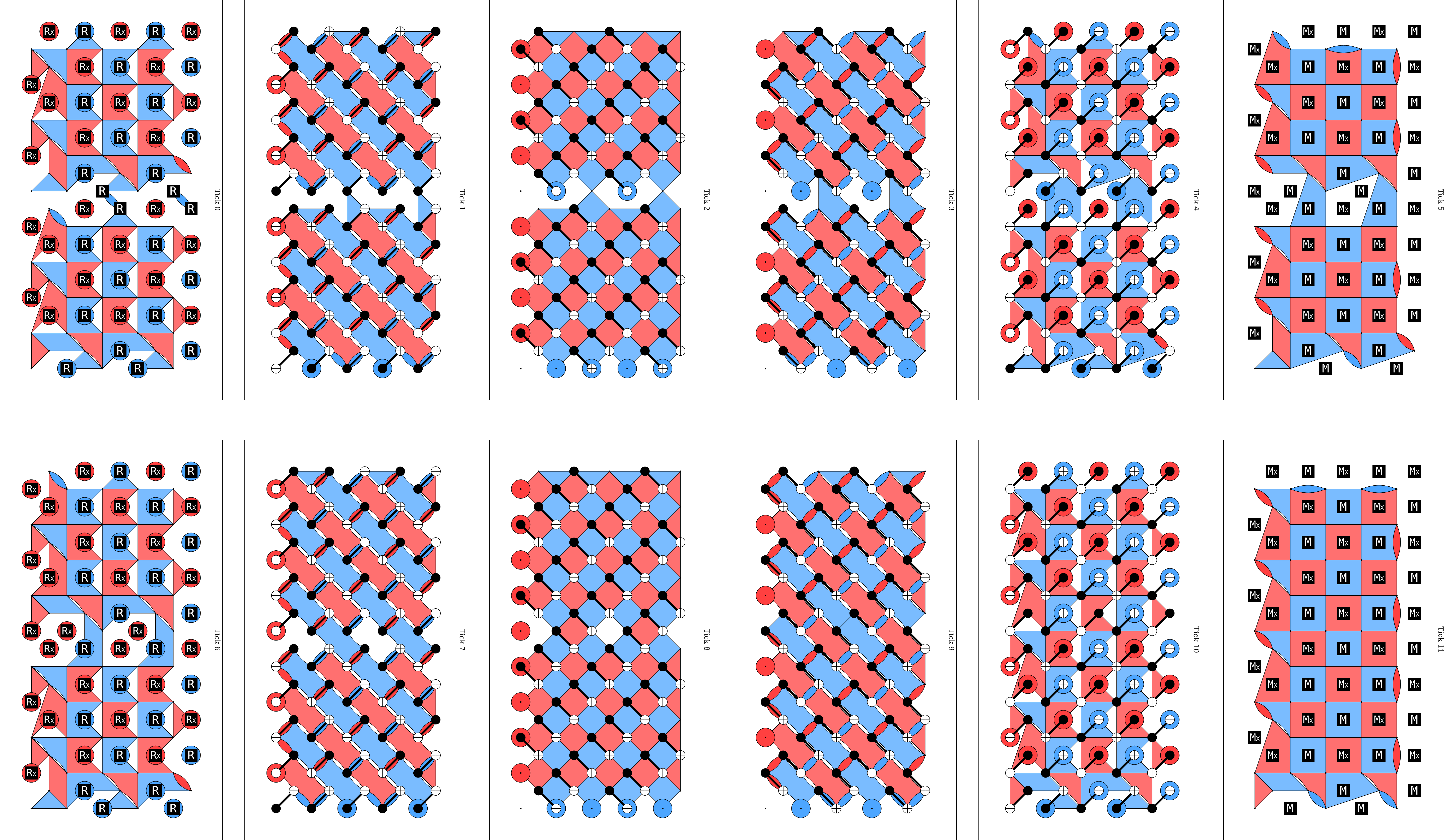}
        
        \vspace{0.2cm}
        (d) $Z$-top configuration, vertical merge.
    \end{minipage}
    
    \caption{(Continued) Lattice surgery merge operation for all possible patch orientations.}
    \label{fig:lattice_surgery_merge_continued}
\end{figure}

\begin{figure}[htbp]
    \centering
    \begin{minipage}{0.9\linewidth}
        \centering
        \includegraphics[width=\linewidth]{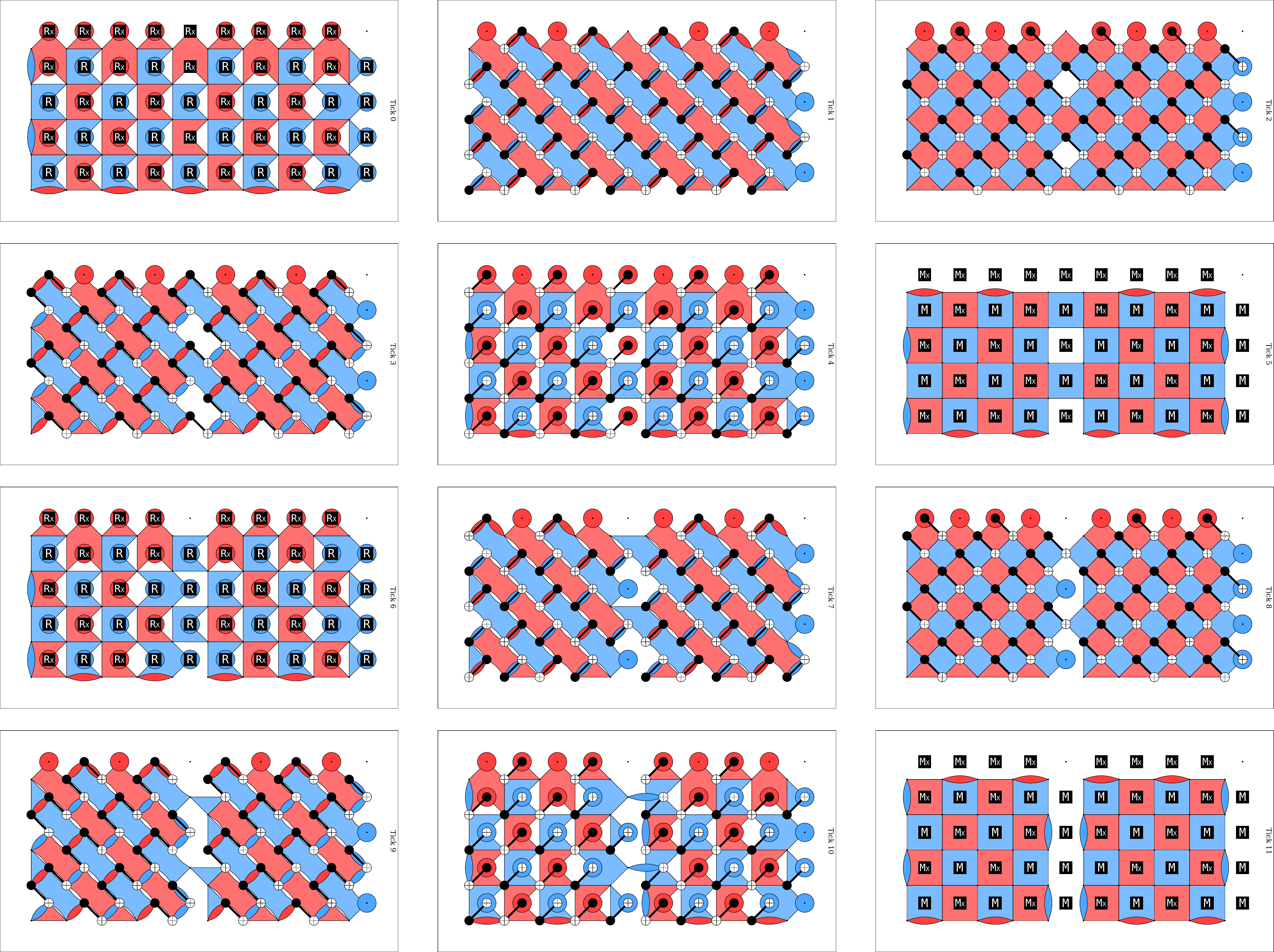}
        
        \vspace{0.2cm}
        (a) $X$-top configuration, horizontal split.
    \end{minipage}
    
    \vspace{2em} 
    
    \begin{minipage}{0.9\linewidth}
        \centering
        \includegraphics[width=\linewidth]{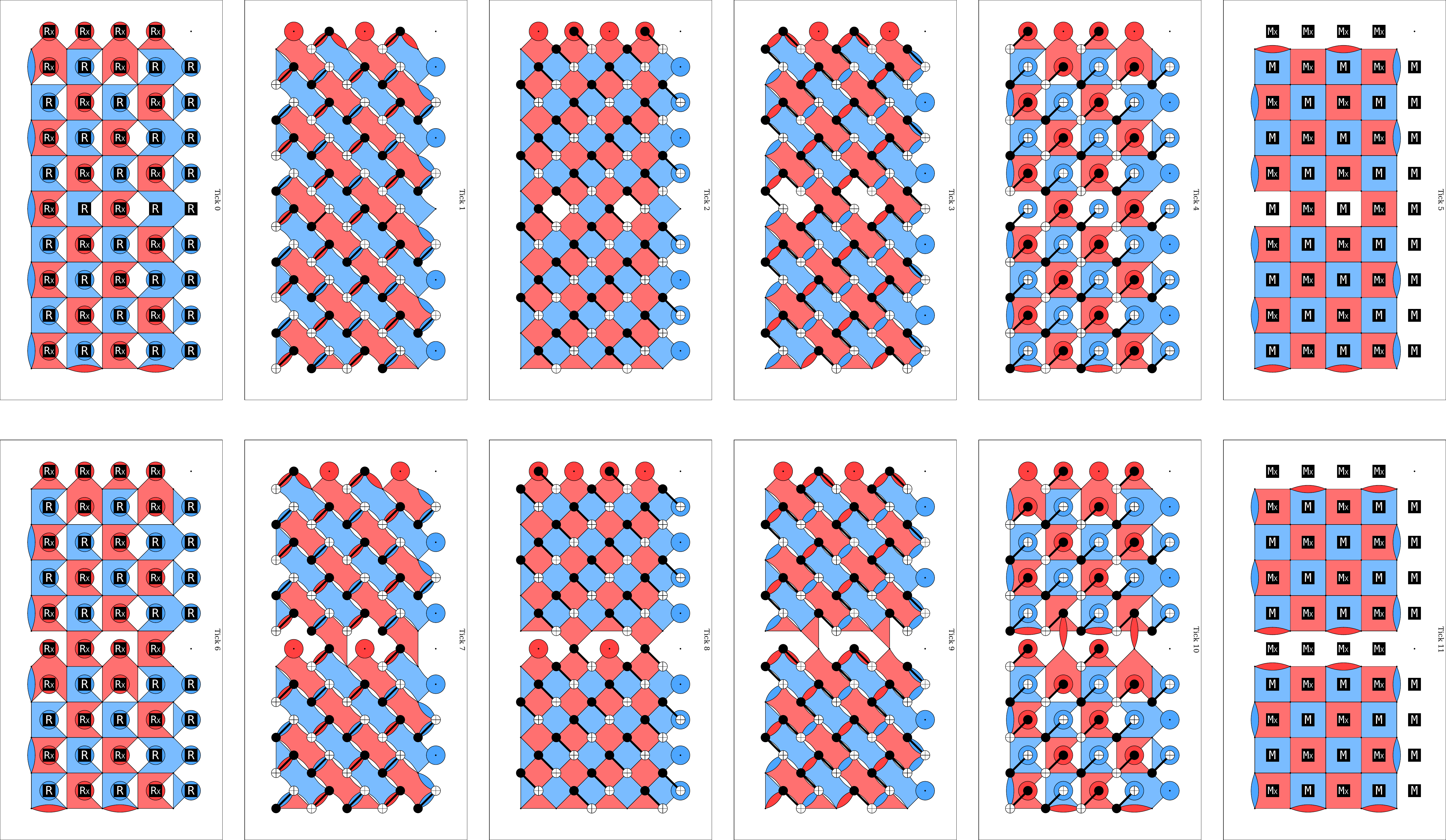}
        
        \vspace{0.2cm}
        (b) $X$-top configuration, vertical split.
    \end{minipage}
    
    \caption{Lattice surgery split operation for all possible patch orientations.}
    \label{fig:lattice_surgery_split}
\end{figure}


\begin{figure}[htbp]
    \addtocounter{figure}{-1} 
    
    \centering
    \begin{minipage}{0.9\linewidth}
        \centering
        \includegraphics[width=\linewidth]{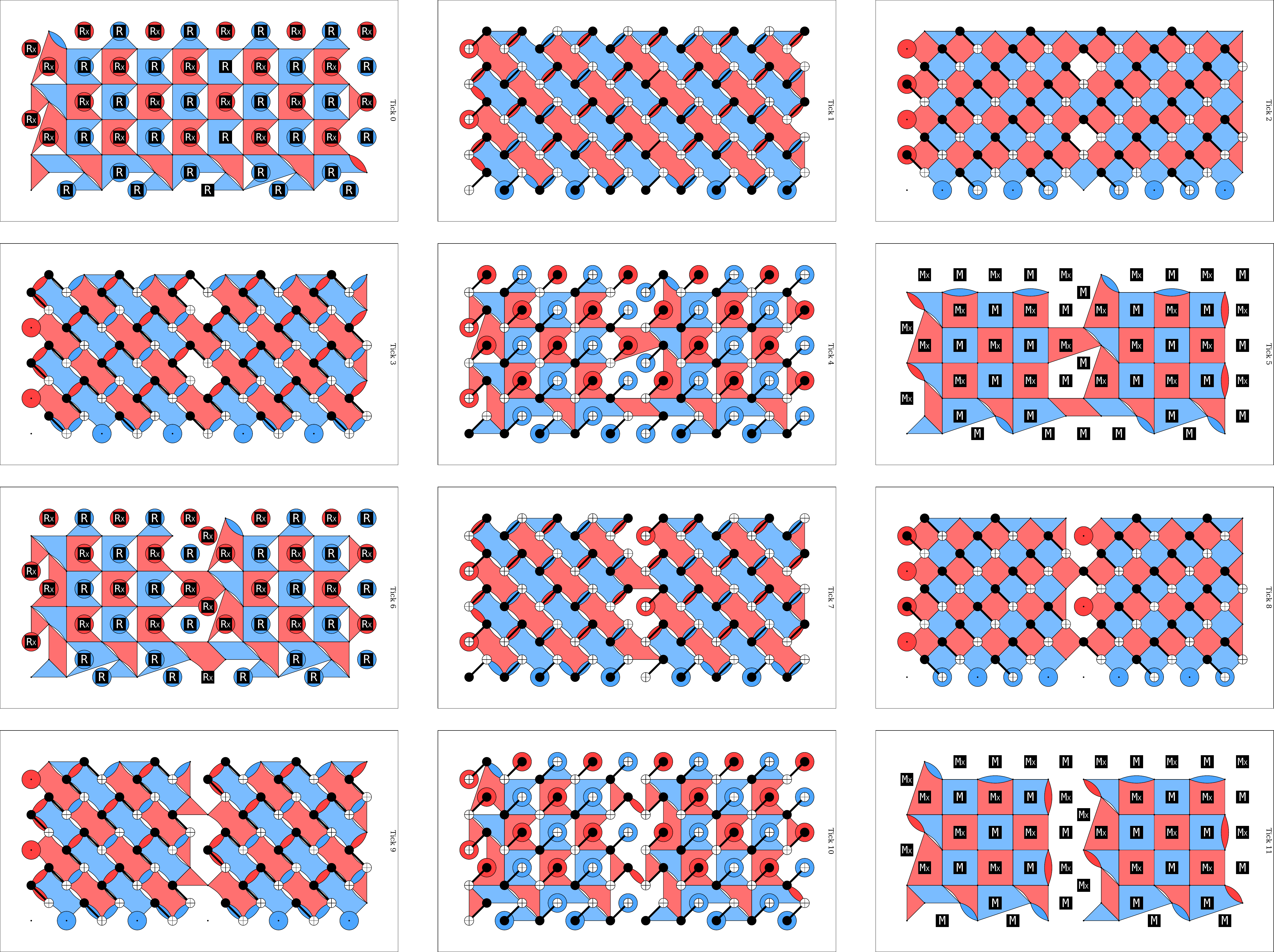}
        
        \vspace{0.2cm}
        (c) $Z$-top configuration, horizontal split.
    \end{minipage}
    
    \vspace{2em} 
    
    \begin{minipage}{0.9\linewidth}
        \centering
        \includegraphics[width=\linewidth]{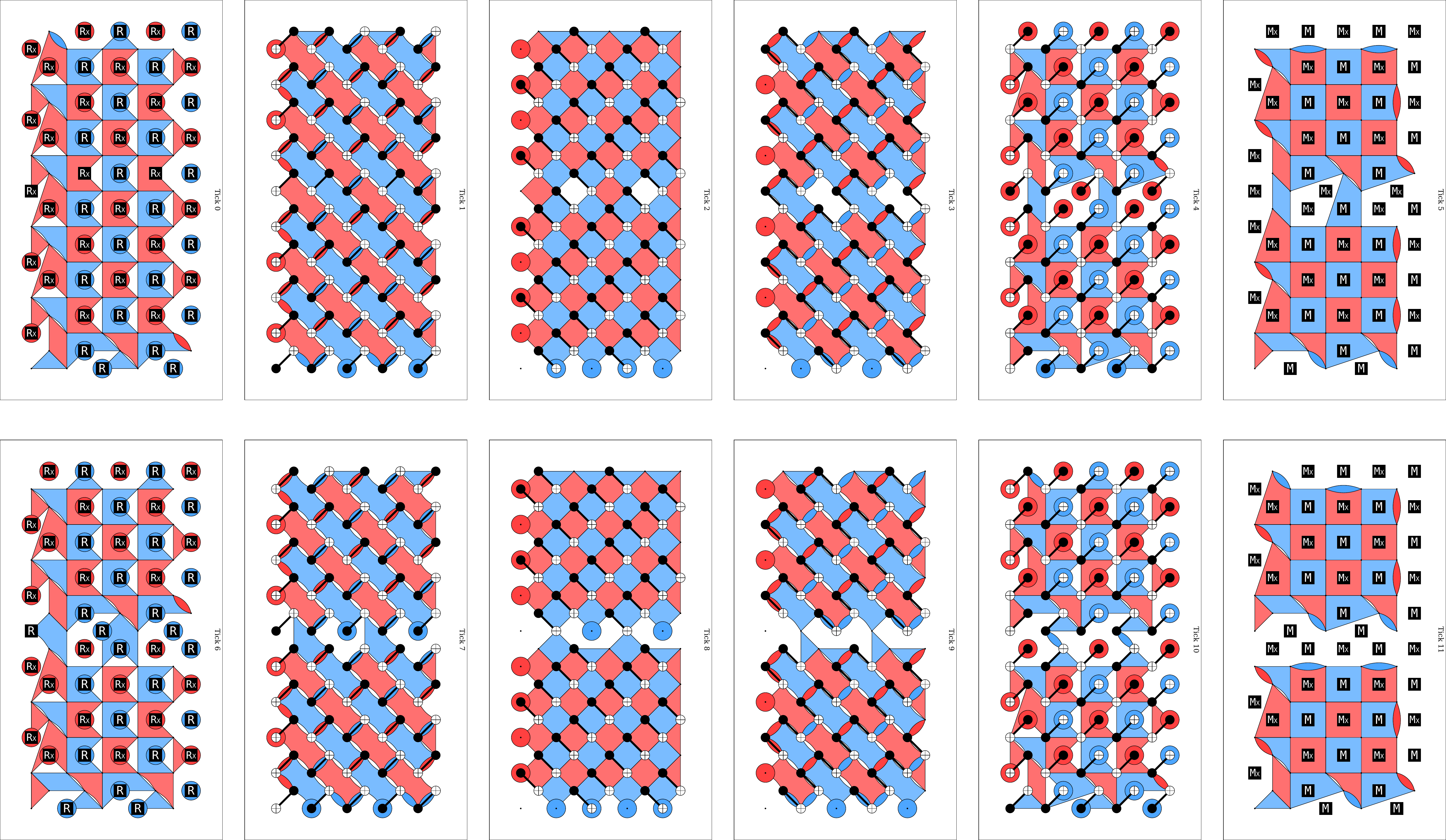}
        
        \vspace{0.2cm}
        (d) $Z$-top configuration, vertical split.
    \end{minipage}
    
    \caption{(Continued) Lattice surgery split operation for all possible patch orientations.}
    \label{fig:lattice_surgery_split_continued}
\end{figure}

\begin{figure}[htbp]
    \centering
    \begin{minipage}{\linewidth}
    \includegraphics[width=\linewidth]{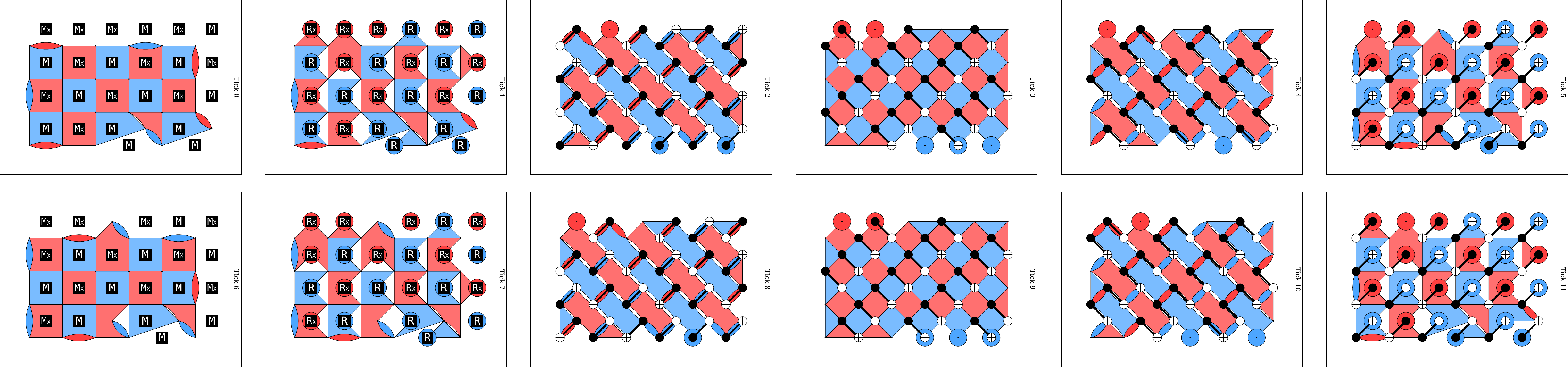}
    \end{minipage}
    \begin{minipage}{\linewidth}
    \includegraphics[width=\linewidth]{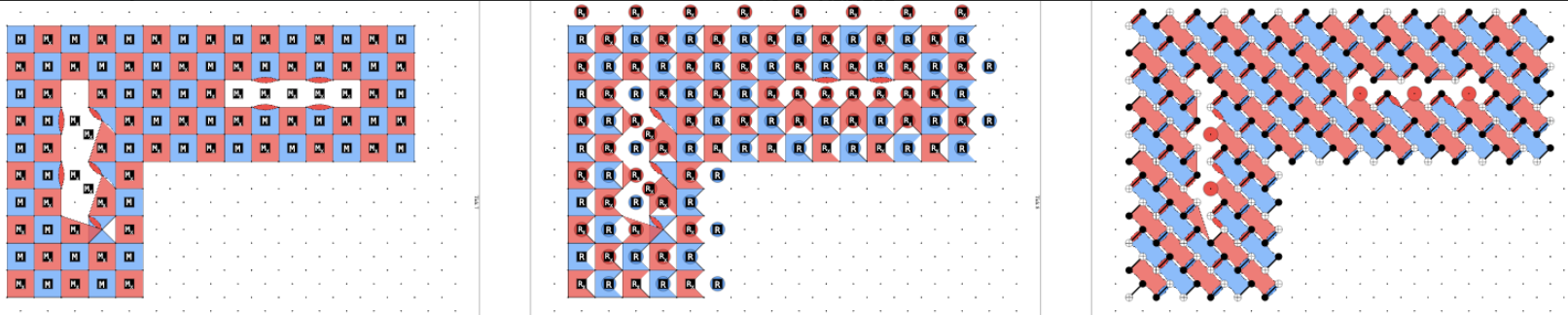}
    \end{minipage}
    \hfill 
    \begin{minipage}{\linewidth}
    \includegraphics[width=\linewidth]{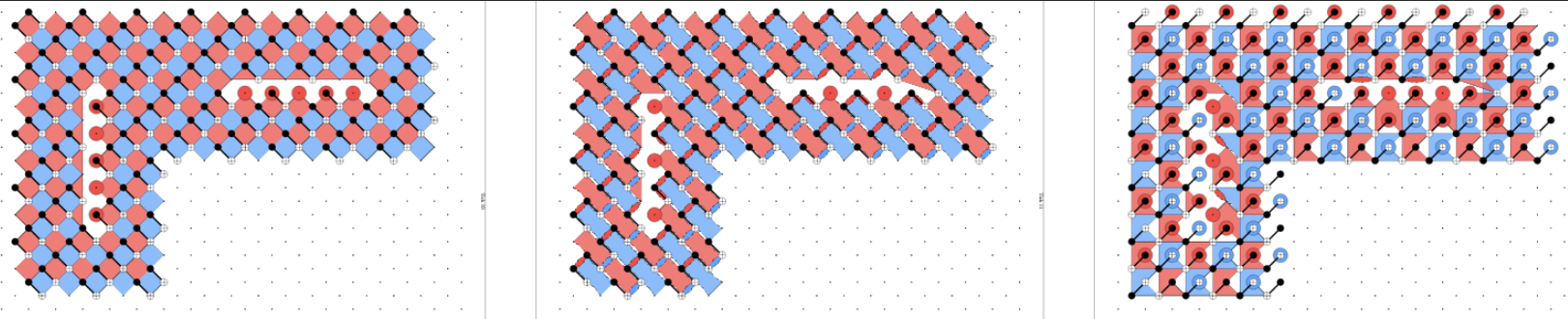}
    \end{minipage}
    \caption{(Top) Gate sequence implementing a rectangular patch encoding two logical qubits. Note the point $Y$ defect on the horizontal boundary. (Bottom) Gate sequence implementing the most challenging case: simultaneous horizontal and vertical interior boundaries, achieved without requiring additional padding.}
    \label{fig:straight_y_defects}
\end{figure}

\begin{figure}[htbp]
    \centering
    \includegraphics[width=\linewidth]{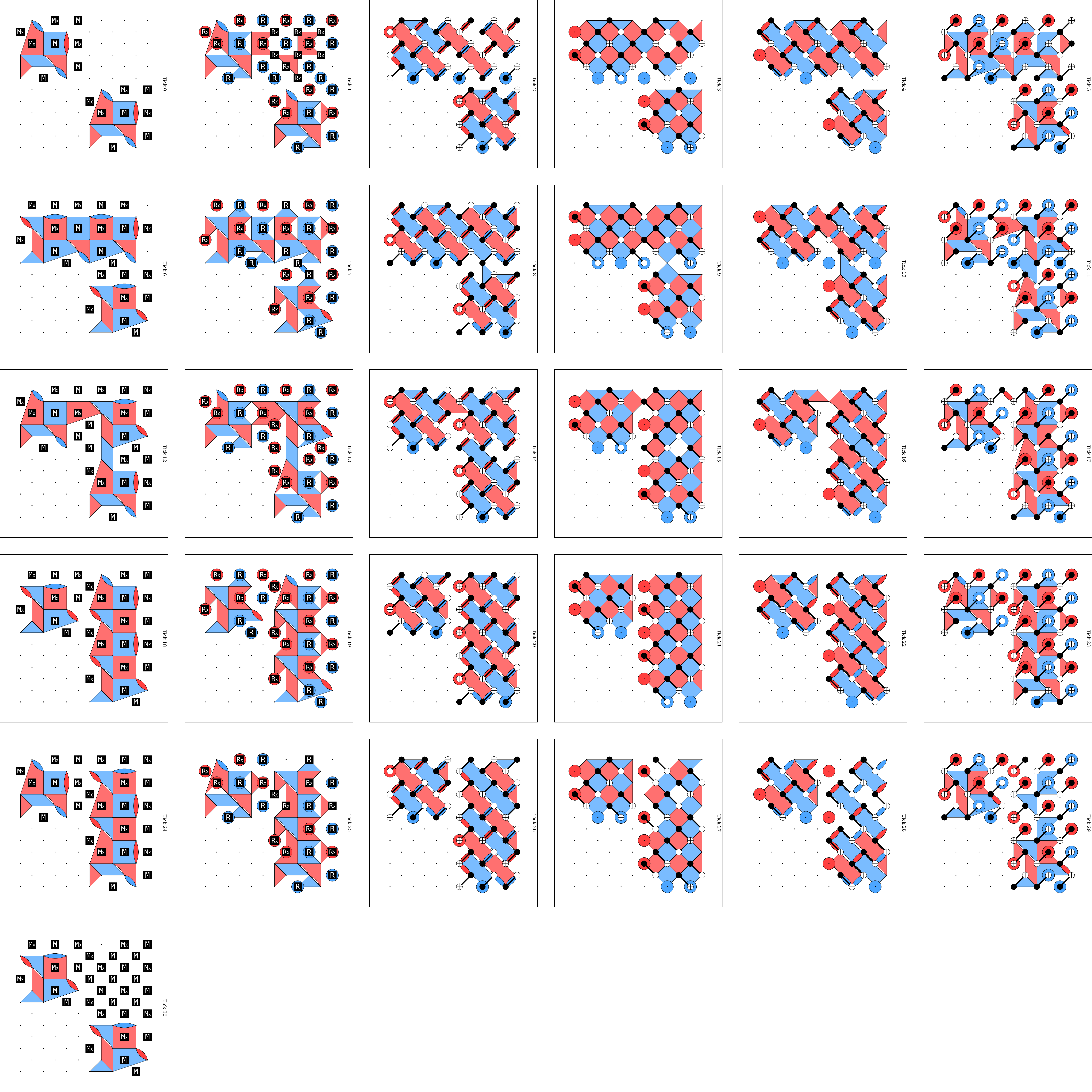}
    \caption{Lattice surgery implementing a logical $\textsc{CX}$ with a $2d^2$ bounding box per surface code patch in the $Z$-top orientation, which is more difficult than the $X$-top orientation. The $d$ rounds between merge and split for fault-tolerant measurement is omitted for brevity.}
    \label{fig:CNOT_X_Z}
\end{figure}

\begin{figure}[htbp]
    \centering
    \includegraphics[width=0.48\linewidth]{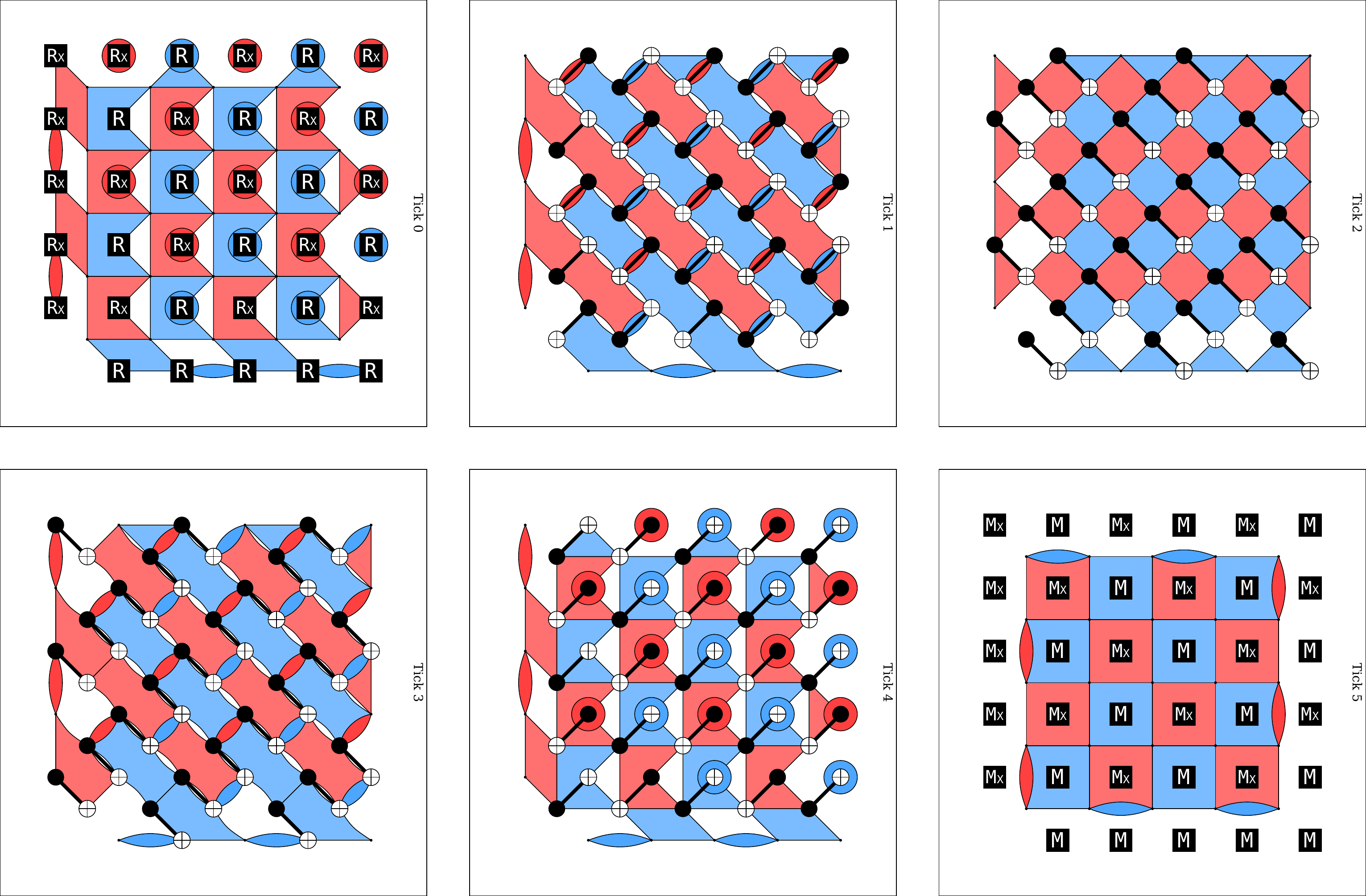}
        \includegraphics[width=0.48\linewidth]{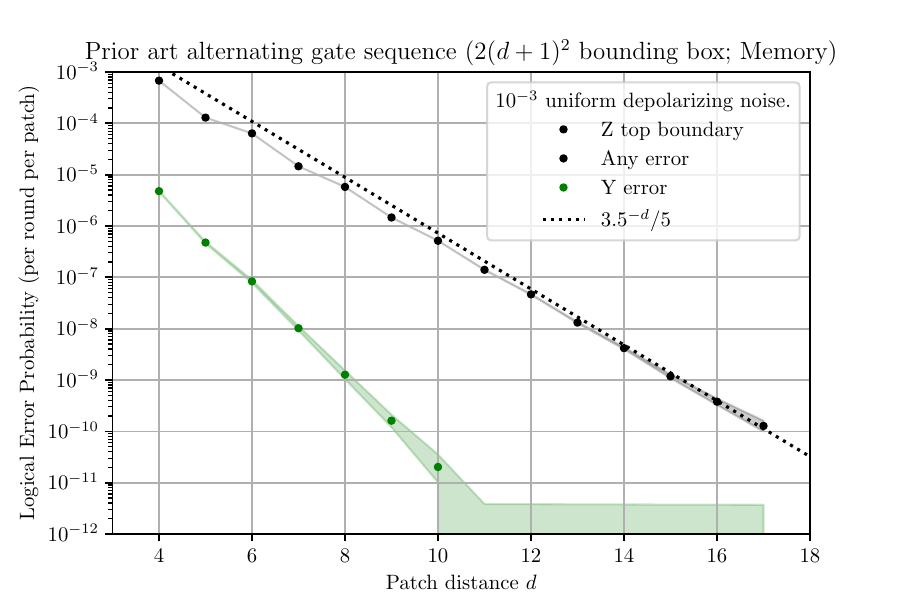}

    \caption{Benchmark of alternating gate sequence using one additional column and row~\cite{Yuga2026NoHook} to implement a $Z$-top configuration.}
    \label{fig:square_patch_big}
\end{figure}

\begin{figure}[htbp]
\centering
\includegraphics[width=\linewidth]{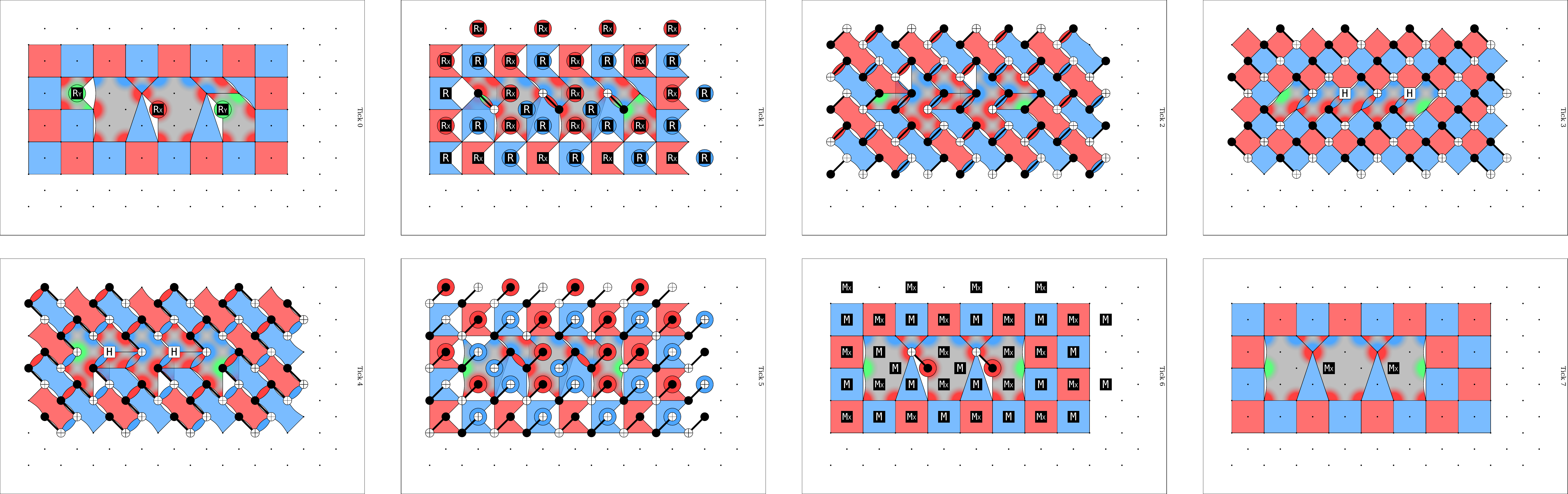}
    \caption{An $8$-cycle implementation of a twist defect with no hook errors. Note that all resets may be moved to execute in the same layer, and similarly for measurements.}
    \label{fig:twist_8_cycles}
\end{figure}

\begin{table}[htbp]
\centering
\begin{tabular}{r|c|}
\hline\hline
&Detecting region and gate sequence\\
\hline
     \rotatebox{90}{Walking twist defect}& \adjustbox{valign=m}{\includegraphics[width=0.9\linewidth]{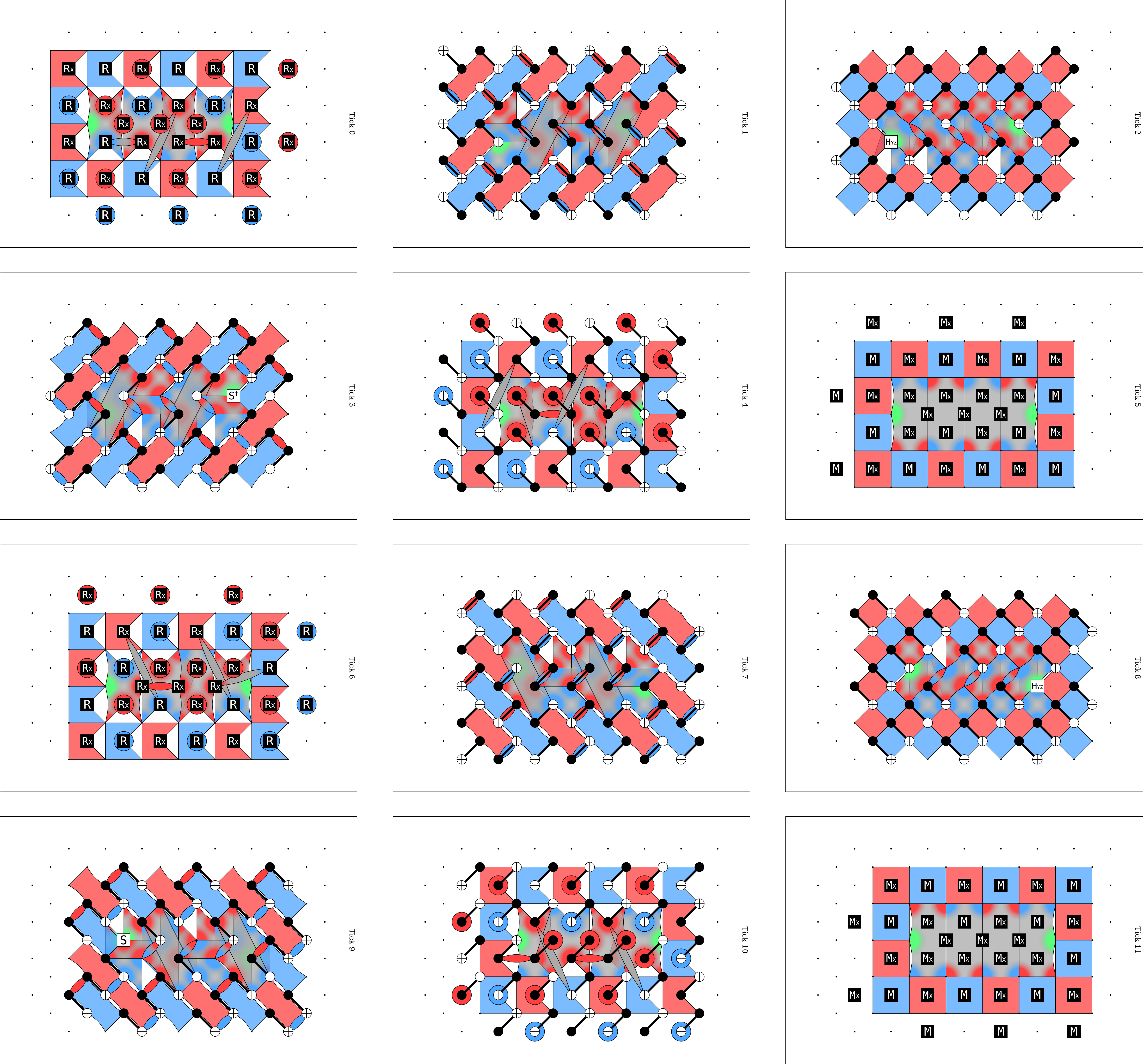}} \\
     \hline\hline
\end{tabular}
    \caption{A gate sequence that walks the twist defects and domain walls one unit to the right every $2$ rounds (12 layers).}
    \label{fig:twist_walk_cycles}
\end{table}

\begin{table}[htbp]
\centering
\begin{tabular}{r|c|}
\hline\hline
&Detecting region and gate sequence\\
\hline
     \rotatebox{90}{Domain wall with right-angle bend}& \adjustbox{valign=b}{\includegraphics[width=0.9\linewidth]{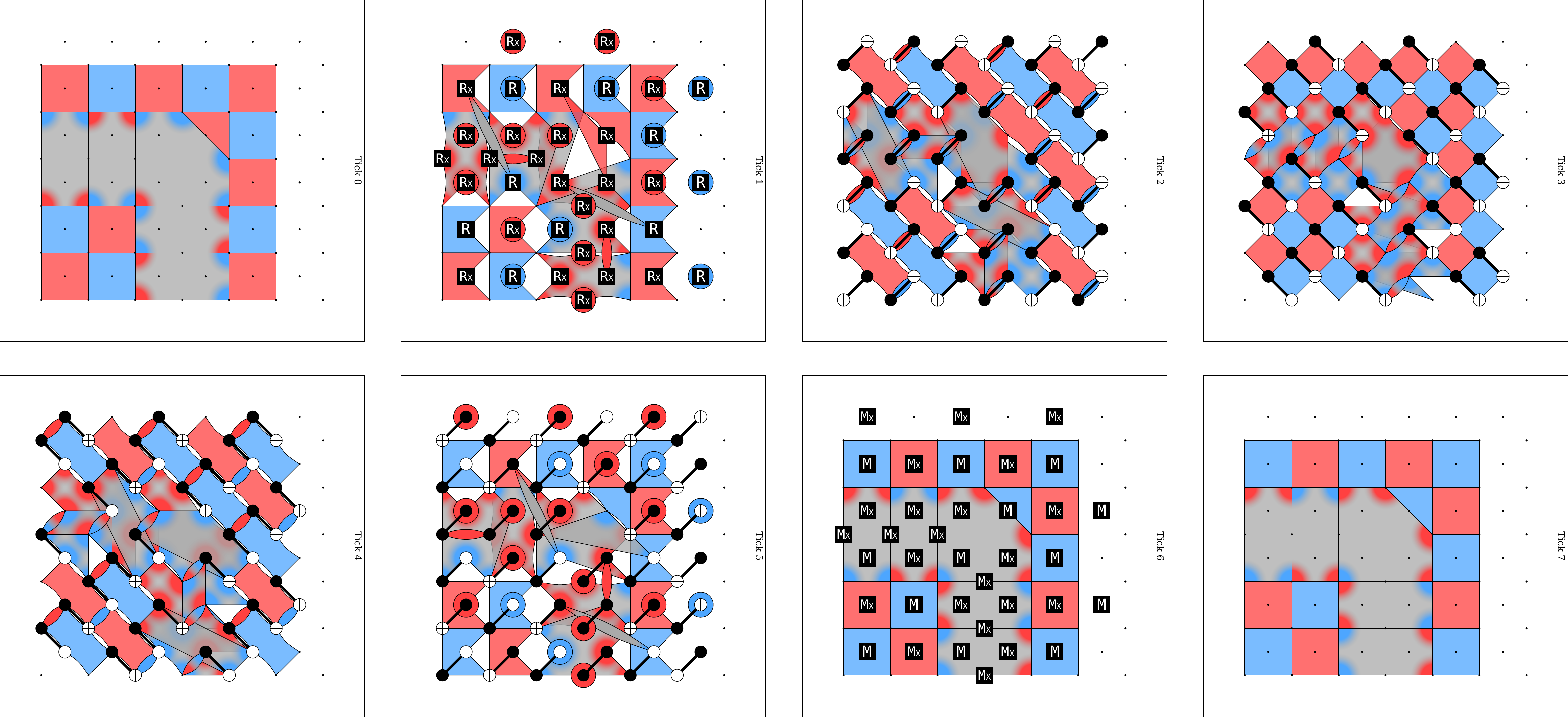}} \\
     \hline
          \rotatebox{90}{Domain wall with four right-angle bends}& \adjustbox{valign=b}{\includegraphics[width=0.9\linewidth]{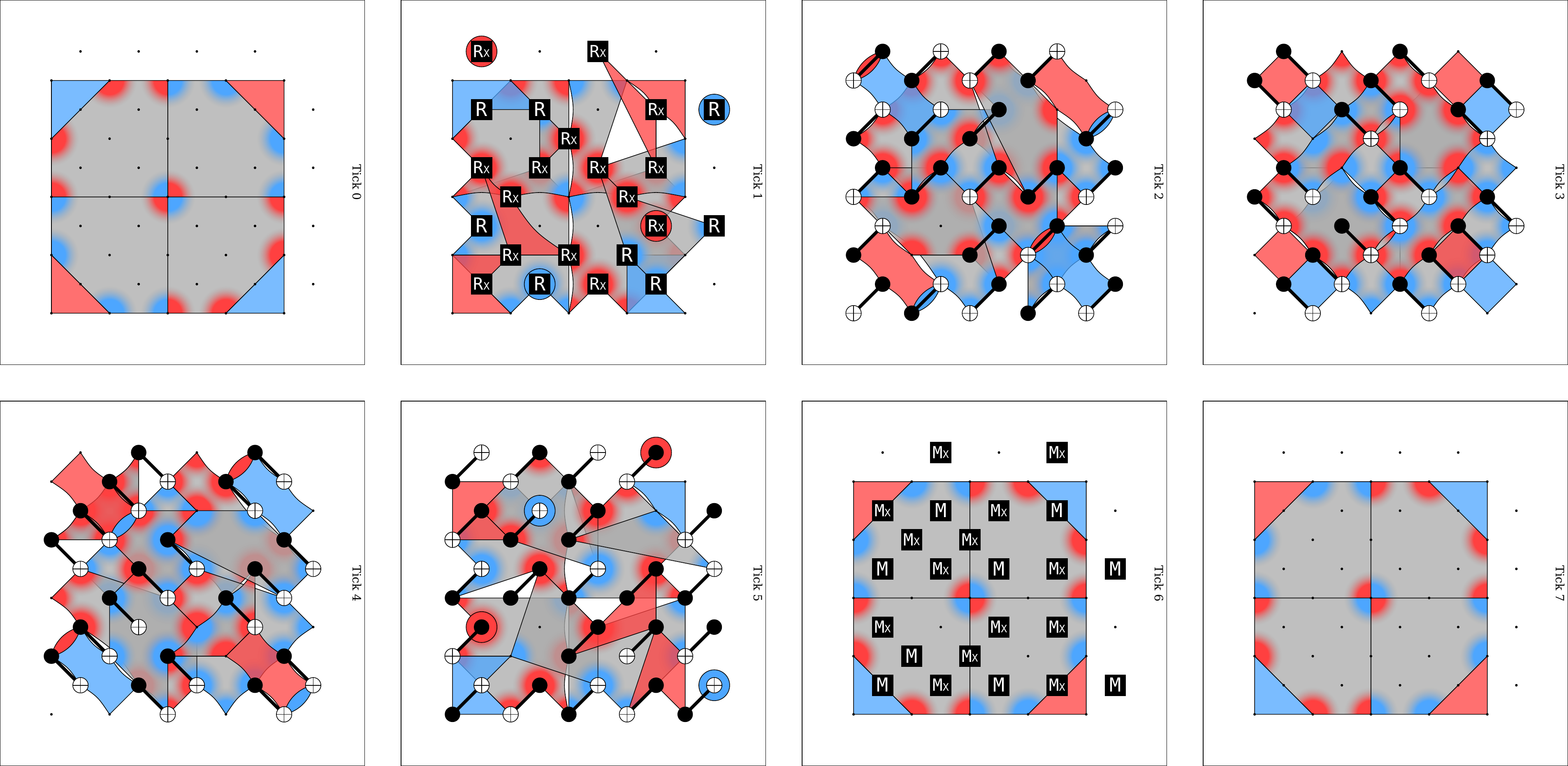}} \\
     \hline\hline
\end{tabular}
    \caption{Gate sequences that (top) implement a horizontal, a vertical, and a right-angled domain wall, and 
    (Bottom) that implement all possible right-angled domain wall bends.
    Note that every data qubit in the bulk supports exactly two $X$ and two $Z$ Paulis of four detectors, though this may be difficult to see as detecting regions overlap.
    From these examples, any orientation of twist defects and domain walls can be implemented on a hex grid in $6$ layers.}
    \label{fig:domain_wall_bend_gate_sequence}
\end{table}

\begin{figure}[htbp]
    \centering
\begin{tabular}{c}
    \includegraphics[width=0.8\linewidth]{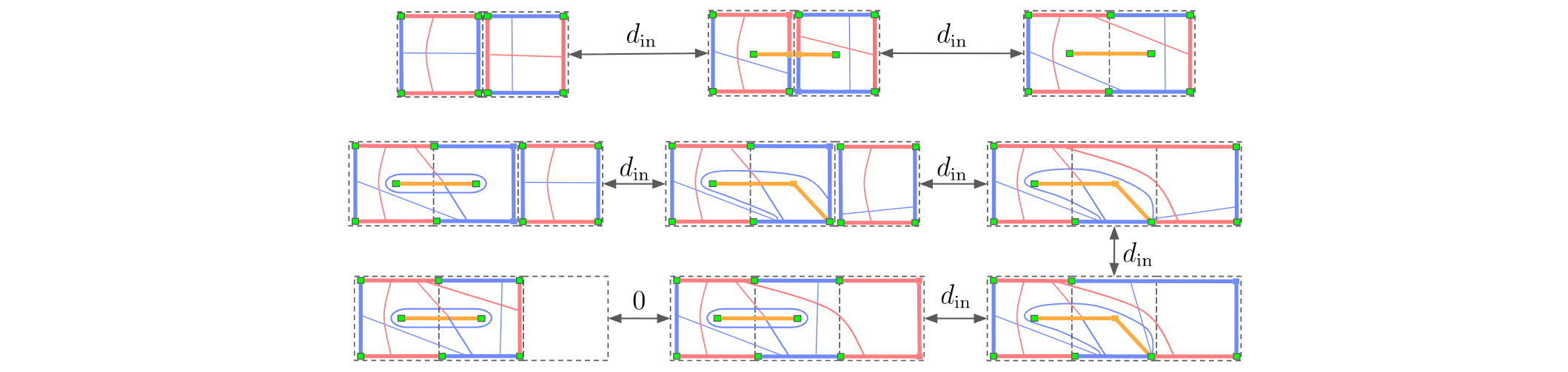}
\end{tabular}
    \caption{Example of elementary steps that realize some of the high-level operations in~\cref{fig:loading_unloading}.}
    \label{fig:loading_unloading_proof}
\end{figure}


%% file: 4_compilation/lookup_table_appendix.tex

\begin{figure}
\centering
    \begin{tabular}{ccc}
         \begin{tabular}{c}
              \includegraphics[width=0.33\linewidth,valign=m]{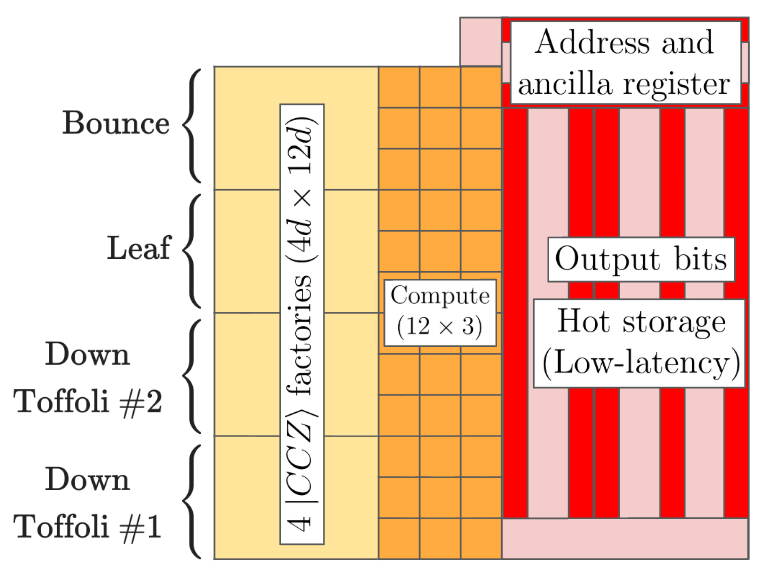}  
              \\
              \\
              \\\includegraphics[width=0.33\linewidth,valign=m]{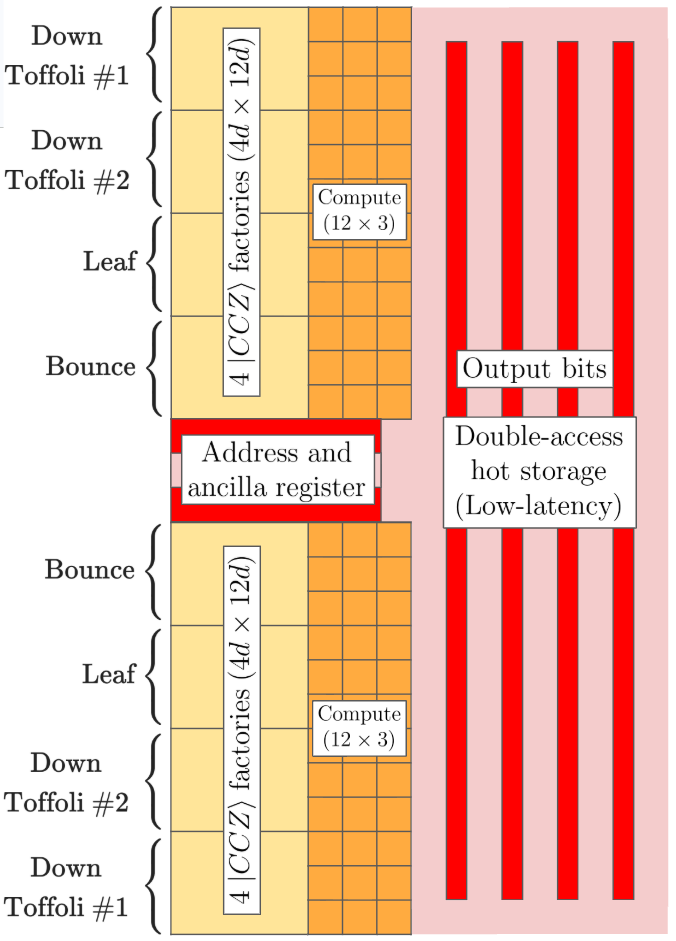}
         \end{tabular}
         & &{\includegraphics[width=0.45\linewidth,valign=m]{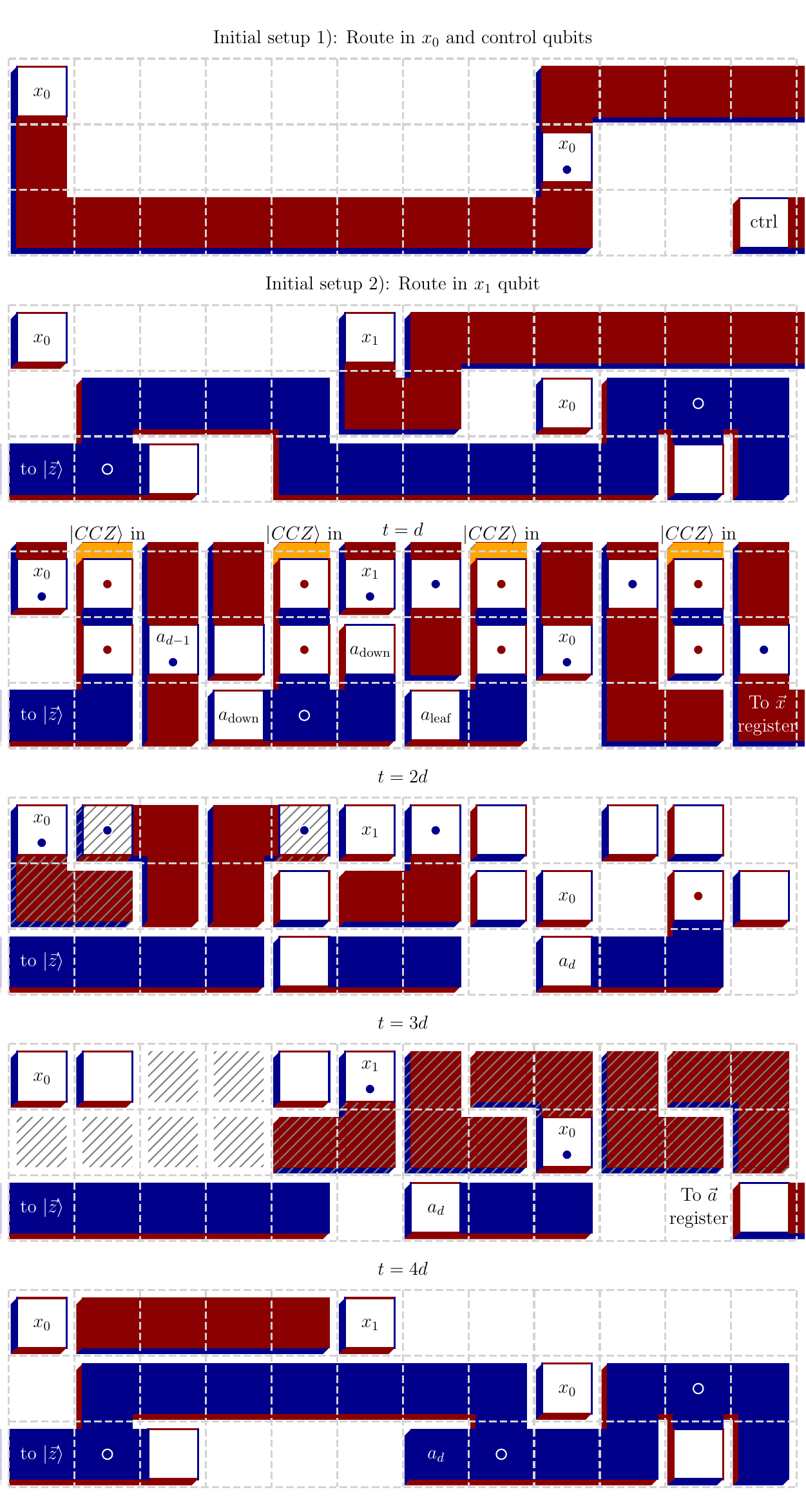}}
    \end{tabular} 
    \caption{(Top left) Example minimum-spacetime layout of skew-tree lookup with clean ancilla~\cref{fig:skew_tree_clean_quantum_circuit} with $3\times12$ surface code patches that consumes up to four $\ket{CCZ}$ states every $\max\{4d,3d+2d_\text{react}\}$ cycles. 
    (Right) Lattice surgery compilation based on the $ZX$ diagram~\cref{fig:skew_tree_clean_ZX}.
    The $x_0$ and $x_1$ address qubits are routed into position just once before the start of the lookup table. 
    Note that all fixup operations are completed by $t=3d$, which incurs no delay so long as the reaction time $d_\text{react}\le d/2$. 
    (Bottom left) Example minimum-time layout by mirroring the minimum-spacetime layout and executing the mirrored copy offset by $d/2$ cycles. This implements a double-access lookup table that consumes 8 $\ket{CCZ}$ states every $4d$ cycles on average.
    Our construction uses a compute region $3$ patches wide, which is half that of the previous double-access lookup~\cite{Gidney2019AutoCCZ}.}
    \label{fig:zx_skew_tree_dirty_Clifford_rate}
\end{figure}

In this section, we provide $ZX$ diagrams and lattice surgery compilation of the other skew-tree lookup table instances mentioned in~\cref{table:lattice_surgery_lookup}.
First, the dirty-ancilla lookup table with a $3\times 3$ workspace has $ZX$ and pipe diagrams shown in~\cref{fig:zx_skew_tree_dirty}.
Second, the clean-ancilla lookup table that consumes $1$ Toffoli gate every $d$ or $d/2$ cycles is compiled in~\cref{fig:zx_skew_tree_dirty_Clifford_rate}.
We also provide the full proof of~\cref{thm:classical_absorption}, which shows that $\ket{CCZ}$ injection corrections in a skew-tree lookup table can be classically absorbed into the lookup data, reducing the reaction depth from $\Theta(X)$ to $\mathcal{O}(1)$.

\begin{table}
    \centering
    \begin{tabular}{c|c|c}
    \hline\hline
         & ZX diagram & Pipe diagram \\
         \hline
         \rotatebox{90}{Down}&{\includegraphics[width=0.2\linewidth,valign=m]{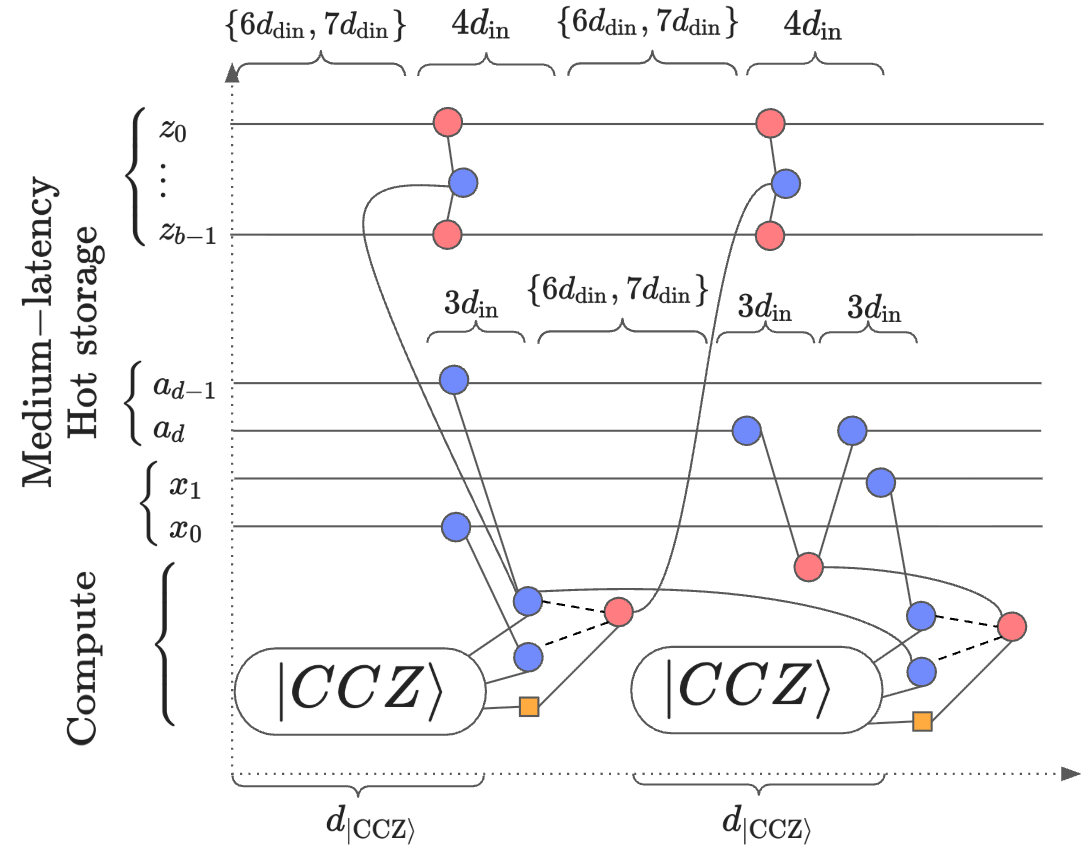}} & \includegraphics[width=0.75\linewidth,valign=m]{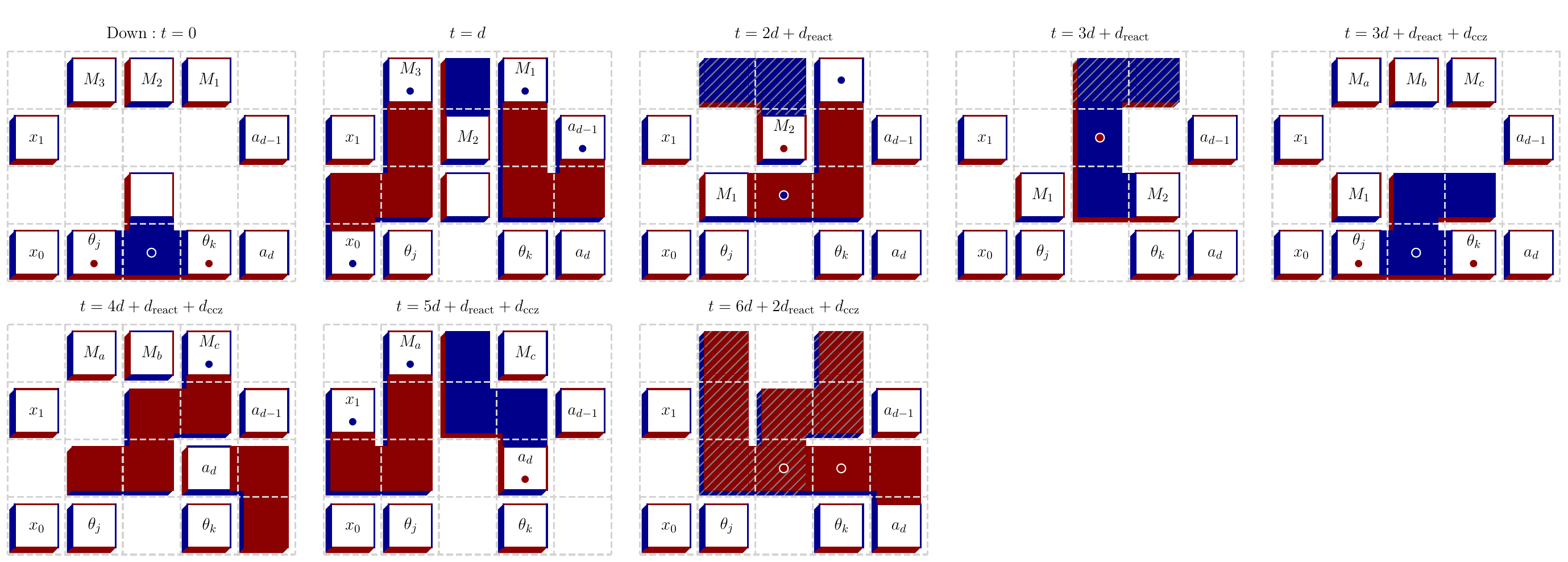}
         \\\hline
         \rotatebox{90}{Leaf}&{\includegraphics[width=0.15\linewidth,valign=m]{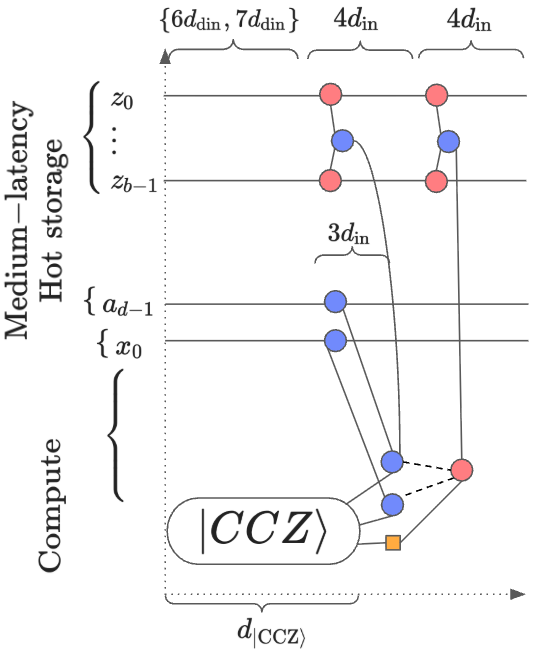}} & \includegraphics[width=0.75\linewidth,valign=m]{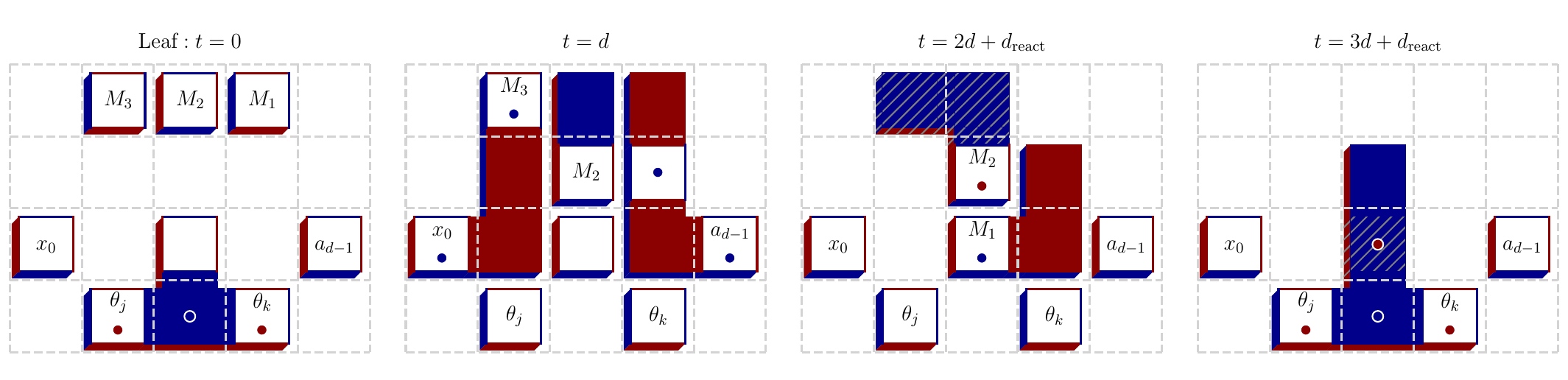}
         \\\hline
         \rotatebox{90}{Bounce}&{\includegraphics[width=0.15\linewidth,valign=m]{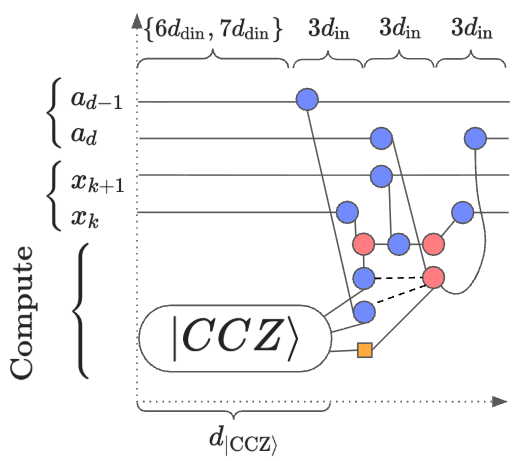}} & \includegraphics[width=0.75\linewidth,valign=m]{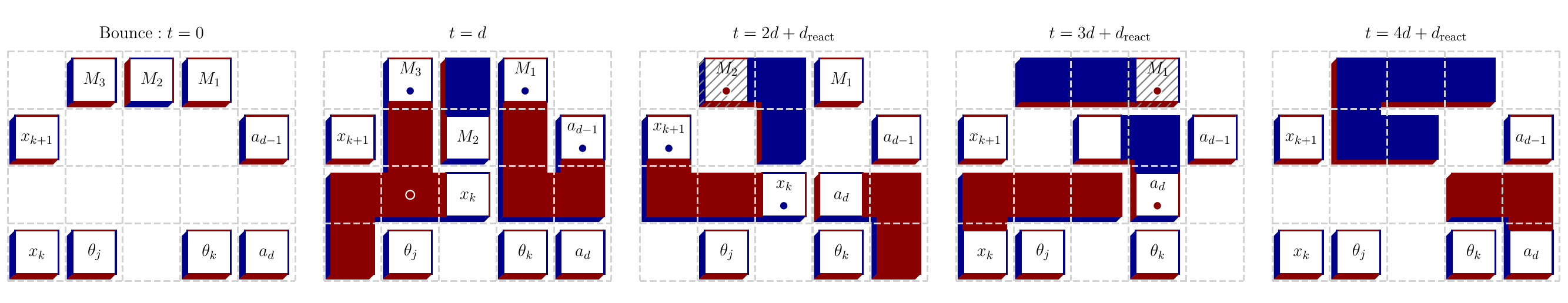}
         \\\hline
         \rotatebox{90}{Up}&{\includegraphics[width=0.15\linewidth,valign=m]{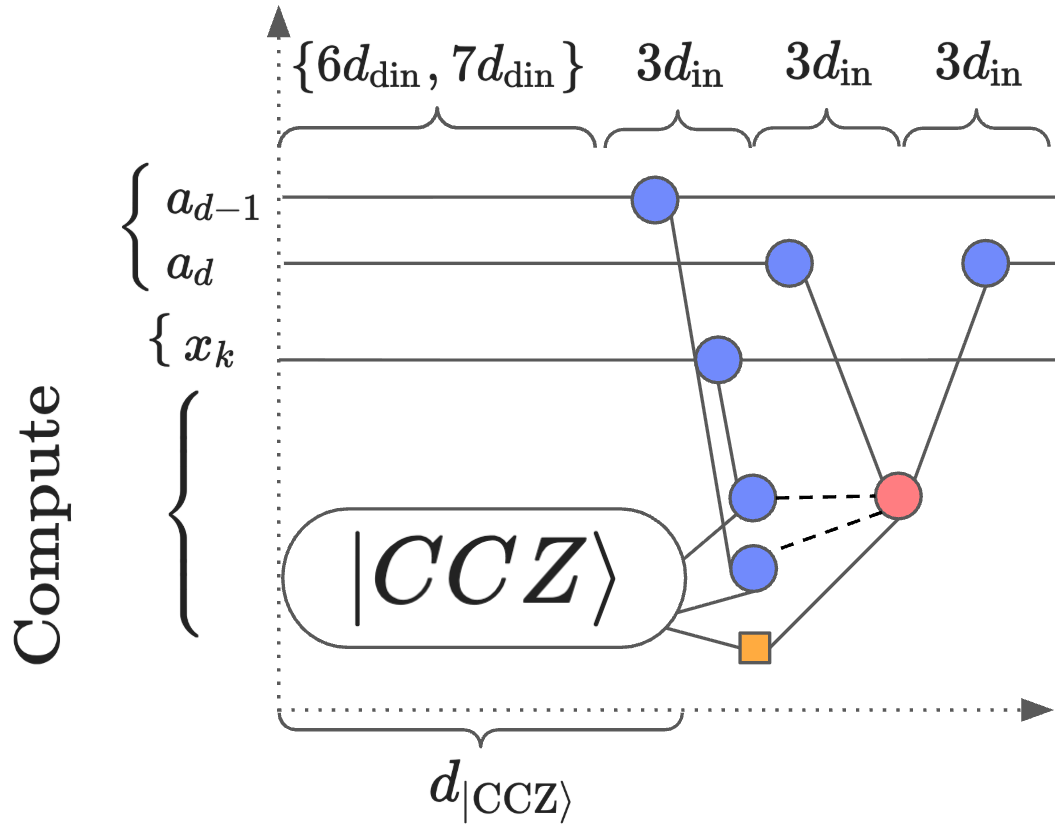}} & \includegraphics[width=0.75\linewidth,valign=m]{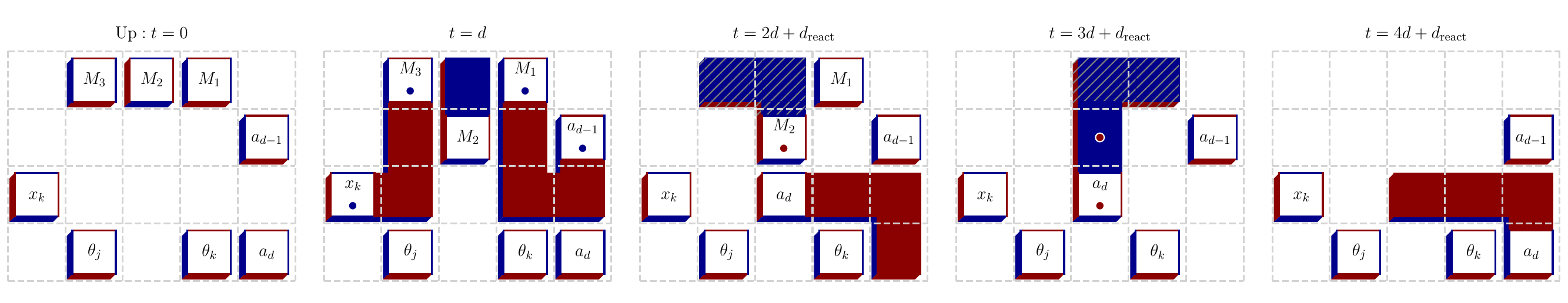}
         \\
         \hline\hline
    \end{tabular}
    \begin{tabular}{c|c|c}
\hline\hline
Step & \multicolumn{2}{c}{With medium-latency hot storage ($d_\text{syn}\in\{6,7\}d_\text{in},\;d_\text{acc}=3d_\text{in}$)}
\\
 & Cycles & Assuming $d_\text{in}=14$, $d_\text{ccz}=125$, $d=22$
\\
\hline
Down & $\max\{10d_\text{in},d_\text{ccz}+3d+d_\text{react}\}+\max\{7d_\text{in},d_\text{ccz}\}+6d_\text{in}$ & 410
\\
Bounce &$\max\{10d_\text{in},d_\text{ccz}+d\}+6d_\text{in}$ & 231
\\
Up &$\max\{10d_\text{in},d_\text{ccz}+d\}+6d_\text{in}$ & 231 
\\
Leaf  &$\max\{10d_\text{in},d_\text{ccz}+3d+d_\text{react}\}$ & 201 
\\
\hline\hline
Step & \multicolumn{2}{c}{With low-latency hot storage ($d_\text{syn}=0,\;d_\text{acc}=1$) and a $\ket{CCZ}$ factory to the top.}
\\
 & Cycles & Assuming $d_\text{ccz}=125$, $d=22$
\\
\hline
Down & $6d+2d_\text{react}+2d_\text{ccz}$ & 402 
\\
Bounce &$d_\text{ccz}+4d+d_\text{react}$ & 223 
\\
Up &$d_\text{ccz}+4d+d_\text{react}$ & 223 
\\
Leaf  &$d_\text{ccz}+3d+d_\text{react}$ & 201 
\\
\hline\hline
\end{tabular}
    \caption{(Top) ZX and pipe diagrams of the dirty skew-tree lookup quantum circuit~\cref{fig:skew_tree_clean_quantum_circuit} with a $3\times 3$ workspace and medium-latency hot storage, annotated with suggested timing information.
    Accessing one side of hot storage takes $d_\text{acc}=3d_\text{in}$ cycles and accessing two takes $d_\text{acc,2}=4d_\text{in}$ cycles
    Each set of hot storage stabilizers is measured in $\{6d_\text{in},7d_\text{in}\}$ cycles.
    As a hot storage patch rotation and access takes $8d_\text{in}$, it is faster to route the logical qubit out and back using two single-side accesses in $6d_\text{in}$ cycles.
    Dashed edges are classically conditioned on measurement results.
    In some cases, the classical condition only applies to the node indicating a $Z$- or $X$- spider, and not the pipe.
    It is only necessary to apply a fix-up operation on the target qubit of each Toffoli gate as the last $\textsc{CZ}$ fixup on the $x,a$ registers can be deferred to the end of the lookup-table. 
    Note that the second target access in ``Leaf'' can be delayed until after one syndrome cycle and occurs concurrent with next CCZ factory.
    (Bottom) Number of cycles needed to implement each tree-traversal step, with an example of typical parameters. 
    Note that the cycles used to implement each CCZ factory can be amortized with the last $3d_\text{in}$ cycles in all steps but ``leaf''.
    }
    \label{fig:zx_skew_tree_dirty}
\end{table}


%% file: 4_compilation/low_reaction_depth_lookup_appendix.tex

\subsubsection{Setup and notation}

We use the skew-tree QROM construction of~\cite{khattar2024riseconditionallycleanancillae}, reviewed in~\cref{sec:lattice_surgery_lookup} and \cref{fig:skew_tree_clean_quantum_circuit}.

\paragraph{Skew tree.} A skew tree over $n = n_X$ selection bits has $N = X = 2^n$ nodes labeled $0, 1, \ldots, N-1$, where labels are identified with subsets of $[n] = \{0, \ldots, n-1\}$ via their binary representation. The root is node~$0$. For each non-root node $p > 0$, define
\begin{align}
\mathrm{pbit}(p) &= p \;\&\; (-p) && \text{(lowest set bit of $p$)} \nonumber\\
b(p) &= \log_2 \left(\mathrm{pbit}(p)\right) && \text{(index of lowest set bit)} \nonumber\\
\mathrm{par}(p) &= p \;\&\; (p-1) && \text{(parent: $p$ with lowest bit cleared)}\label{dc:eq:par}
\end{align}
We define $\mathrm{level}(p) = \mathrm{popcount}(p)$; there are exactly $n$ level-1 nodes: $\{1, 2, 4, \ldots, 2^{n-1}\}$. 
We denote by $\mathrm{sub}(p)$ the set of tree-descendants of $p$ (all nodes $d$ such that $p$ lies on the root-to-$d$ path), including $p$ itself. Concretely, since the root-to-$d$ path repeatedly clears the lowest set bit of $d$, node $p$ is an ancestor of $d$ iff clearing the lower $b(p)$ bits of $d$ recovers $p$ exactly, i.e.,
\begin{equation}
  d \in \mathrm{sub}(p) \iff d \;\&\; (-\mathrm{pbit}(p)) = p. \label{dc:eq:sub}
\end{equation}
This simultaneously requires $p \subseteq d$ and that every set bit of $d \setminus p$ lies strictly below $b(p)$.

\paragraph{Branches affected by each node.}
Each node $p$ affects exactly those branches $x \in \{0,1\}^n$ for which $p \subseteq x$ (every bit set in $p$ is also set in $x$):
\begin{equation}\label{dc:eq:branches}
\mathrm{Branches}(p) = \{x \in \{0,1\}^n : p \subseteq x\}, \qquad |\mathrm{Branches}(p)| = 2^{n - \mathrm{popcount}(p)}
\end{equation}

\paragraph{Skew tree QROM circuit.}
An \emph{uncontrolled} skew tree QROM loads an $N$-entry table $\mathrm{data}[0], \ldots, \mathrm{data}[N-1]$ into a target register, indexed by a selection register $x \in \{0,1\}^n$. The circuit has two components at each node:
\begin{itemize}
\item \textbf{Control qubits.} Each node $p$ has a control qubit $a_p$ satisfying $a_p = 1$ iff $p \subseteq x$. For non-root nodes:
\begin{equation}\label{dc:eq:and}
a_p = x_{b(p)} \cdot a_{\mathrm{par}(p)}
\end{equation}
The root activation is $a_0 = 1$ (always active), so level-1 nodes satisfy $a_{2^b} = x_b \cdot 1 = x_b$: the control qubit is the selection bit itself. No magic state injection is needed and no measurement errors arise. Only non-level-1 nodes ($\mathrm{par}(p) \neq 0$) require a Toffoli gate via $\ket{CCZ}$ magic state injection.

\item \textbf{Data loading.} At each node $p$, the circuit applies a controlled-XOR of the skew data $s[p]$ onto the target register, controlled on $a_p$. The target accumulates:
\begin{equation}\label{dc:eq:qrom}
\mathrm{output}(x) = \bigoplus_{p:\, a_p = 1} s[p] = \bigoplus_{p \subseteq x} s[p]
\end{equation}
To ensure $\mathrm{output}(x) = \mathrm{data}[x]$, the skew data is the M\"obius transform: $s[p] = \bigoplus_{q \subseteq p} \mathrm{data}[q]$.
\end{itemize}

\paragraph{CCZ injection errors.} Each non-level-1 AND gate is implemented via $\ket{CCZ}$ magic state injection. This produces, \emph{before} applying feed-forward corrections, the erroneous output
\begin{equation}\label{dc:eq:ccz}
\tilde{a}_p = (x_{b(p)} \oplus m_1[p]) \cdot (a_{\mathrm{par}(p)} \oplus m_2[p])
\end{equation}
where $m_1[p], m_2[p] \in \{0,1\}$ are measurement outcomes. Recovering $a_p$ requires up to three corrections: $\mathrm{CX}(a_{\mathrm{par}(p)}, a_p)$, $\mathrm{CX}(x_{b(p)}, a_p)$, and $\mathrm{X}(a_p)$, incurring a $d_\text{reaction}$ wait.

\subsubsection{Classical absorption of Clifford corrections at non-level-1 nodes}

\begin{lemma}\label{dc:lem:absorb}
Let $p$ be a non-level-1 node ($\mathrm{par}(p) \ne 0$) whose AND gate is implemented via $\ket{CCZ}$ state injection with measurement outcomes $(m_1[p], m_2[p]) \in \{0,1\}^2$, and suppose all other nodes have $m_1 = m_2 = 0$. 
Let $s'$ be the skew data initialized to $s' = s$ with correction targets $d'$ updated in-place by
\begin{equation}\label{dc:eq:correction}
\begin{cases}
s'[d \oplus \mathrm{pbit}(p)] \mathrel{\oplus}= s[d] & \text{if } m_1[p] = 1 \\
s'[d \oplus \mathrm{par}(p)] \mathrel{\oplus}= s[d] & \text{if } m_2[p] = 1 \\
s'[d \oplus p] \mathrel{\oplus}= s[d] & \text{if } m_1[p] = m_2[p] = 1
\end{cases}
\quad \text{for each } d \in \mathrm{sub}(p)
\end{equation}
Then the circuit with skew data $s'$ and no feed-forward corrections at node $p$ produces the correct output $\mathrm{data}[x]$ for all $x$. Moreover, every correction target $d'$ satisfies $d' < p$, so $d' \notin \mathrm{sub}(p)$ (the update is in-place) and $d'$ is loaded \emph{after} $\mathrm{sub}(p)$ in the decreasing-order traversal (the update is timely).
\end{lemma}

\begin{proof}
The error at node $p$ propagates to every descendant $d \in \mathrm{sub}(p)$, since $a_p$ appears on the root-to-$d$ path. Nodes outside $\mathrm{sub}(p)$ are unaffected.

\textbf{Affected branch analysis.}
For each $d \in \mathrm{sub}(p)$, since $p \subseteq d$, we have the disjoint bit decomposition
\begin{equation}\label{dc:eq:factored}
d = \mathrm{par}(p) \cup \mathrm{pbit}(p) \cup \rho, \qquad \rho := d \setminus p,
\end{equation}
where $\rho$ consists of the (possibly empty) bits of $d$ strictly below $b(p)$. Since the three parts are disjoint, the control qubit $a_d$ factors as their AND:
\begin{equation}\label{dc:eq:factored_a}
a_d = a_{\mathrm{par}(p)} \cdot a_{\mathrm{pbit}(p)} \cdot a_\rho = a_{\mathrm{par}(p)} \cdot x_{b(p)} \cdot a_\rho
\end{equation}
using $a_{\mathrm{pbit}(p)} = x_{b(p)}$ since $\mathrm{pbit}(p)$ is a level-1 node.
Substituting the erroneous control $\tilde{a}_d = (a_{\mathrm{par}(p)} \oplus m_2)\cdot (x_{b(p)} \oplus m_1) \cdot a_\rho$ from~\cref{dc:eq:ccz} gives the error term:
\begin{equation}\label{dc:eq:delta}
\Delta_d := \tilde{a}_d \oplus a_d = (m_2\,a_{\mathrm{par}(p)} \oplus m_1\,x_{b(p)} \oplus m_1 m_2) \cdot a_\rho
\end{equation}

\textbf{Correctness.} We show that the update in~\cref{dc:eq:correction} cancels every $\Delta_d \cdot s[d]$ error term. Since node $d$ contributes $\tilde{a}_d \cdot s[d]$ to the output instead of the correct $a_d \cdot s[d]$, the erroneous branch introduces an output error of $(\tilde{a}_d \oplus a_d)\cdot s[d] = \Delta_d \cdot s[d]$.
To see why, consider $m_2 = 1, m_1 = 0$: the erroneous control is $\tilde{a}_d = \overline{a_{\mathrm{par}(p)}} \cdot x_{b(p)} \cdot a_\rho$, so $s[d]$ is XOR'd onto the branches where $\overline{a_{\mathrm{par}(p)}} \cdot x_{b(p)} \cdot a_\rho = 1$ instead of where $a_{\mathrm{par}(p)} \cdot x_{b(p)} \cdot a_\rho = 1$. XORing $s[d]$ onto \emph{all} branches where $x_{b(p)} \cdot a_\rho = 1$---i.e., updating $s'[d \oplus \mathrm{par}(p)]$---both cancels the erroneous load (on the $\overline{a_{\mathrm{par}(p)}}$ branches) and adds the missing correct load (on the $a_{\mathrm{par}(p)}$ branches), since XOR is its own inverse. The same logic applies to each summand of $\Delta_d$: each equals the control qubit of a node $d' < p$:

\begin{align}
a_{\mathrm{par}(p)} \cdot a_\rho &= a_{d \,\oplus\, \mathrm{pbit}(p)} & &\text{($b(p)$ cleared)} \label{dc:eq:term1} \\
x_{b(p)} \cdot a_\rho           &= a_{d \,\oplus\, \mathrm{par}(p)} & &\text{($\mathrm{par}(p)$'s bits cleared)} \label{dc:eq:term2} \\
a_\rho                          &= a_{d \,\oplus\, p}               & &\text{(all of $p$'s bits cleared)} \label{dc:eq:term3}
\end{align}

\textbf{Causal ordering.} Write $d = p \mid \rho$ with $\rho < \mathrm{pbit}(p)$. Each correction target satisfies $d' < p$:
\begin{itemize}
  \item $d \oplus \mathrm{pbit}(p) = \mathrm{par}(p) \mid \rho < \mathrm{par}(p) \mid \mathrm{pbit}(p) = p$, \quad since $\rho < \mathrm{pbit}(p)$.
  \item $d \oplus \mathrm{par}(p) = \mathrm{pbit}(p) \mid \rho < 2\,\mathrm{pbit}(p) < p$, \quad since $p$ is non-level-1 so $p > 2\,\mathrm{pbit}(p)$.
  \item $d \oplus p = \rho < \mathrm{pbit}(p) < p$.
\end{itemize}
Since every $d' < p$ lies outside $\mathrm{sub}(p)$ and the circuit loads data in decreasing order, $d'$ is loaded after the entire subtree $\mathrm{sub}(p)$.
\end{proof}

\begin{figure*}
    \centering
    \begin{minipage}{\linewidth}
        \centering
        \includegraphics[width=\linewidth]{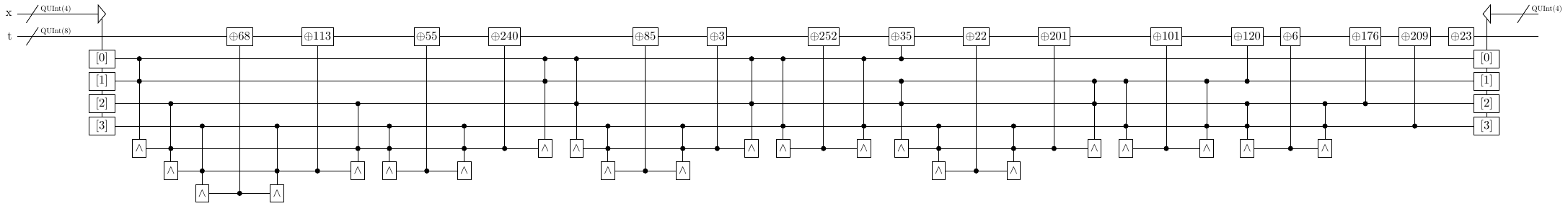}
        \vspace{0.5em}\\
        \textbf{(a)} A Skew Tree QROM circuit to load $N=16$ elements - [23, 198, 167, 112, 111, 219,  22, 178, 52, 25, 135, 249, 188, 195, 183, 201], using an $n=4$ bit selection register $x$. Traverses the skew tree from~\cref{fig:skew_tree_clean_quantum_circuit} in reverse DFS order.
    \end{minipage}
    \vspace{1.5em}
    \begin{minipage}{\linewidth}
        \centering
        \includegraphics[width=\linewidth]{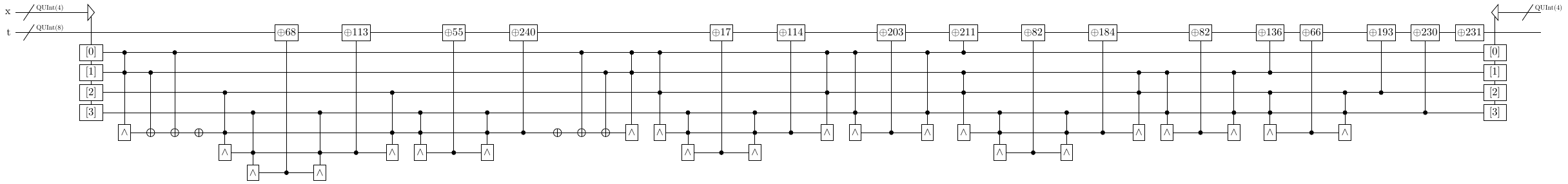}
        \vspace{0.5em}\\
        \textbf{(b)} Same circuit as (a) where the Toffoli gate at node $12$ is erroneous due to both measurements $m_1[12]=1$ and $m_2[12]= 1$.
        The data to be loaded is updated as per~\cref{dc:eq:correction} to absorb the CCZ injection corrections.
        The circuit still loads the correct data on all inputs.
    \end{minipage}
    \caption{Skew Tree QROM circuit (a) without errors and (b) with classical absorption of CCZ injection errors at node $12$.}
    \label{fig:dc_skew_tree_qrom_circuit}
\end{figure*}

See~\cref{fig:dc_skew_tree_qrom_circuit} for the skew-tree QROM circuit and for an example showing how the data is updated when the AND gate at node $12$ has errors $m_1[12] = m_2[12] = 1$.

\subsubsection{Proof of~\texorpdfstring{\cref{thm:classical_absorption}}{the classical absorption theorem}}

\emph{Construction.} Level-1 nodes use the selection bit $x_{b(p)}$ directly as their control qubit and require no $\ket{CCZ}$ injection. Each non-level-1 node computes $a_p = x_{b(p)} \cdot a_{\mathrm{par}(p)}$ via $\ket{CCZ}$ injection with no feed-forward corrections; instead, as each AND gate completes, the classical controller applies~\cref{dc:eq:correction} to update the not-yet-loaded skew data in-place.

\emph{Correctness.} Level-1 control qubits are exact. For non-level-1 nodes, \cref{dc:lem:absorb} guarantees that~\cref{dc:eq:correction} exactly cancels the CCZ injection error in the output. When multiple nodes have errors, corrections are processed in DFS traversal order; causal ordering from~\cref{dc:lem:absorb} ensures each correction target $d' < p$ is not yet loaded when its update is applied.

\emph{Reaction depth.} The classical computation of~\cref{dc:eq:correction} need only complete before the correction target's data is loaded — not at the time of the $\ket{CCZ}$ injection. Since the circuit loads at least one data word (at node $p$ itself) between the injection and any correction target, this provides a buffer of at least $d_\mathrm{LS}$. Under $d_\mathrm{LS} \ge d_\text{reaction}$, the circuit never stalls, giving constant reaction depth independent of $N$.

\emph{Classical work.} Total work is $O(N \log N)$: $\sum_p |\mathrm{sub}(p)| = \sum_d \mathrm{popcount}(d) = n \cdot 2^{n-1}$, since each node $d$ appears in the subtree of exactly $\mathrm{popcount}(d)$ ancestors.

\begin{remark}[Timing assumption]\label{dc:rem:timing}
The constant reaction depth claim rests on $d_\mathrm{LS} \ge d_\text{reaction}$. For fast-clock architectures such as superconducting qubits with surface codes, typical values are $d_\text{reaction} \approx 10\,\mu\text{s}$ and $d_\mathrm{LS} \ge 20\,\mu\text{s}$, comfortably satisfying this condition. Even if $d_\mathrm{LS}$ were reduced below $d_\text{reaction}$ (reaction-limited regime), each ancilla qubit responsible for loading the data at node $x$ need only remain alive for at most $d_\text{reaction}$ after being computed, before the corrected data value is consumed. Crucially, the Clifford feed-forward corrections are never explicitly applied to the quantum state---they are absorbed classically---which simplifies compilation and reduces the overall spacetime volume compared with approaches that use AutoCCZ states or explicit Clifford corrections.
\end{remark}

\begin{remark}[Controlled QROMs]\label{dc:rem:controlled}
If $a_0 = \mathrm{ctrl}$ is a quantum control qubit rather than the constant $1$, then level-1 nodes require genuine Toffoli gates ($a_{2^b} = x_b \cdot \mathrm{ctrl}$) and their $m_2$ corrections involve $\mathrm{CX}(\mathrm{ctrl}, \tilde{a}_p)$, which cannot be classically absorbed. The reason is immediate from the branch perspective: with $\mathrm{par}(2^b) = 0$, an $m_2$ error gives $\tilde{a}_{2^b} = x_b \cdot (\mathrm{ctrl} \oplus 1)$, which is active on the complement of the correct branches with respect to the $\mathrm{ctrl}$ qubit. The correction target $d \oplus \mathrm{par}(p) = d \oplus 0 = d$ is a self-reference, so no data update can compensate. This forces $n$ sequential $d_\text{reaction}$ waits, yielding $\mathcal{O}(\log X)$ reaction depth for controlled QROM with $N-1$ Toffoli gates.
A controlled QROM can also be reduced to an uncontrolled one by treating the control bit $c$ as an additional selection bit: define a $2N$-entry table $\mathrm{data}'[c \cdot N + x] = c \cdot \mathrm{data}[x]$ (entries are zero when $c = 0$). The controlled lookup of $N$ entries becomes an uncontrolled lookup of $2N$ entries over an $(n+1)$-bit selection register $(c, x)$, restoring constant reaction depth at the expense of doubling the number of Toffoli gates.
\end{remark} \qed

%% file: 4_compilation/multiplexed_rotations_appendix.tex

\subsection{Space-limited lattice surgery compilation}\label{app:space_limited_MA_lattice_surgery}

In this section, we compile the multiplexed MA normal form implementation of the multiplexed Givens rotations down to surface code lattice surgery pipe diagrams.
We focus on the space-limited regime, where the goal is to minimize the footprint used for rotation synthesis.
We compile these multiplexed rotations by breaking the circuit down into small components that we can implement as lattice surgery pipe diagrams using the LaSsynth software package~\cite{Tan2024Scalpel}.
To synthesize an individual circuit component, we begin by converting it to a ZX diagram.
We rewrite the diagram in a canonical form where the non-Clifford T gates can be implemented by performing a joint parity measurement with a magic state and then measuring that magic state in either the X or Y basis.
This reduces the compilation to the task of compiling a small ZX diagram that implements Clifford operations only.

Given a bounding box in space and time and a fixed configuration of input and output ports, the LaSsynth software package uses a SAT solver to translate such a ZX diagram into a valid lattice surgery pipe diagram.
We perform a non-exhaustive search over possible inputs in the following way: For each circuit component, we enumerate a series of candidate bounding box sizes.
Given a particular bounding box size, we randomly sample several hundred possible configurations of input and output ports.
We specifically sample input and output specifications to allow us to tile them in space and time in a straightforward way.
We allow LaSsynth a fixed budget of time to find a valid compilation for each set of inputs.
This process allows us to build a library of valid compilations in different shapes and sizes for each circuit component.

In order to minimize the number of physical qubits, we compile the multiplexed rotation pipeline into a minimal footprint: a fixed $3 \times 3$ grid of distance $d$ patches.
To execute the MA normal form within this 9-patch footprint, we stack operations entirely in time (the $K$ axis).
We do not allocate a dedicated, persistent region for a magic state factory alongside the execution region.
Instead, we embed the $3 \times 2$ CCZ-to-2T factory directly beneath each circuit component.
To connect these stacked blocks without widening the footprint, we route a single catalytic T-state vertically through the entire pipeline.

We lock one column of the $3 \times 3$ grid (the edge column) to carry the two target qubits ($t_1, t_2$) and this catalytic passthrough wire.
The SAT solver randomizes the remaining six interior patches to route the control qubit and the required magic states.
Because the edge column positions remain invariant across all components, the output ports of one block align perfectly with the input ports of the next block stacked above it.

\begin{figure*}[htbp]
	\centering
	\begin{tabular}{ccc}
		\begin{tabular}{c}
			\includegraphics[width=0.32\textwidth]{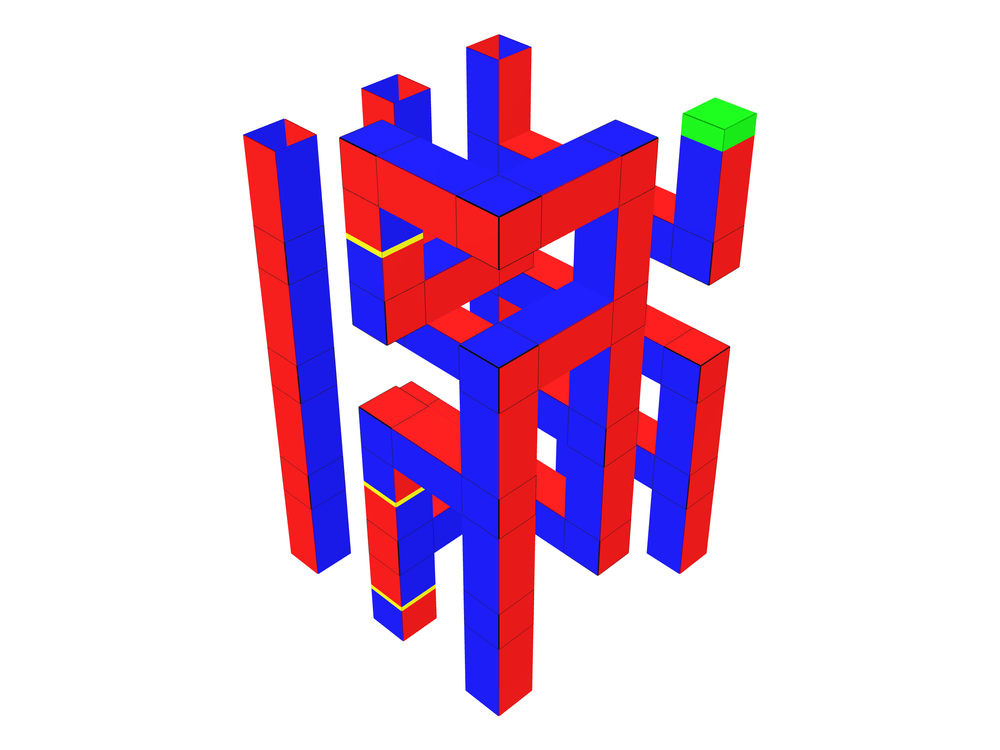}
			\\
			{$G_{ZZ}$ Passthrough}
		\end{tabular}
		 &
		\begin{tabular}{c}
			\includegraphics[width=0.32\textwidth]{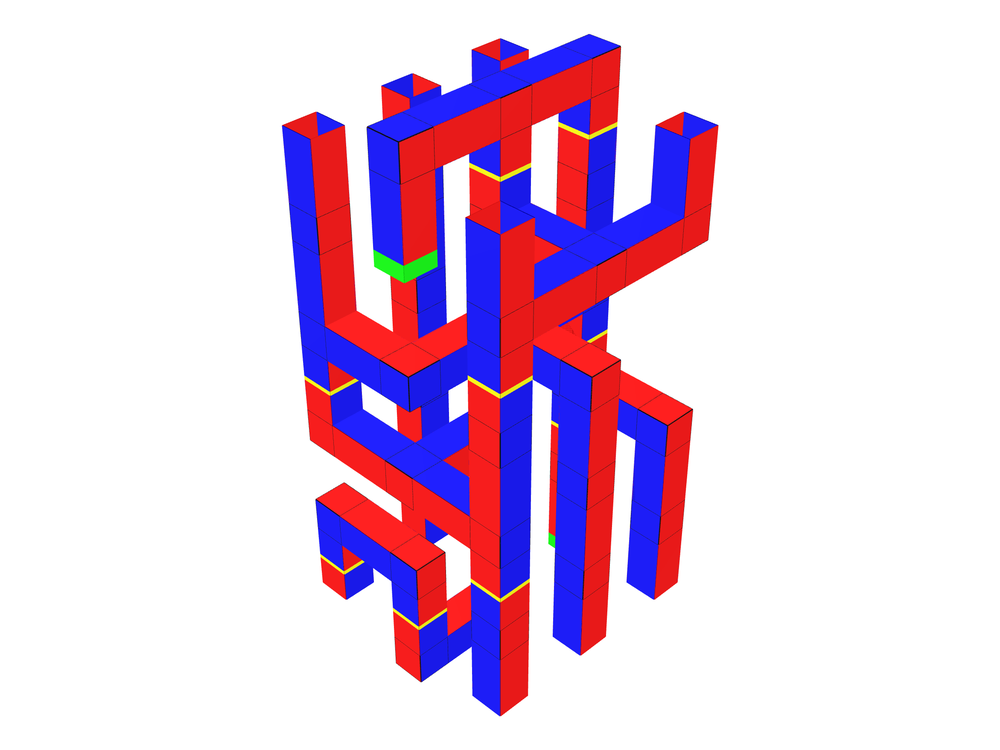}
			\\
			{CCZ to 2T catalysis (ZZ branch)}
		\end{tabular}
		 &
		\begin{tabular}{c}
			\includegraphics[width=0.32\textwidth]{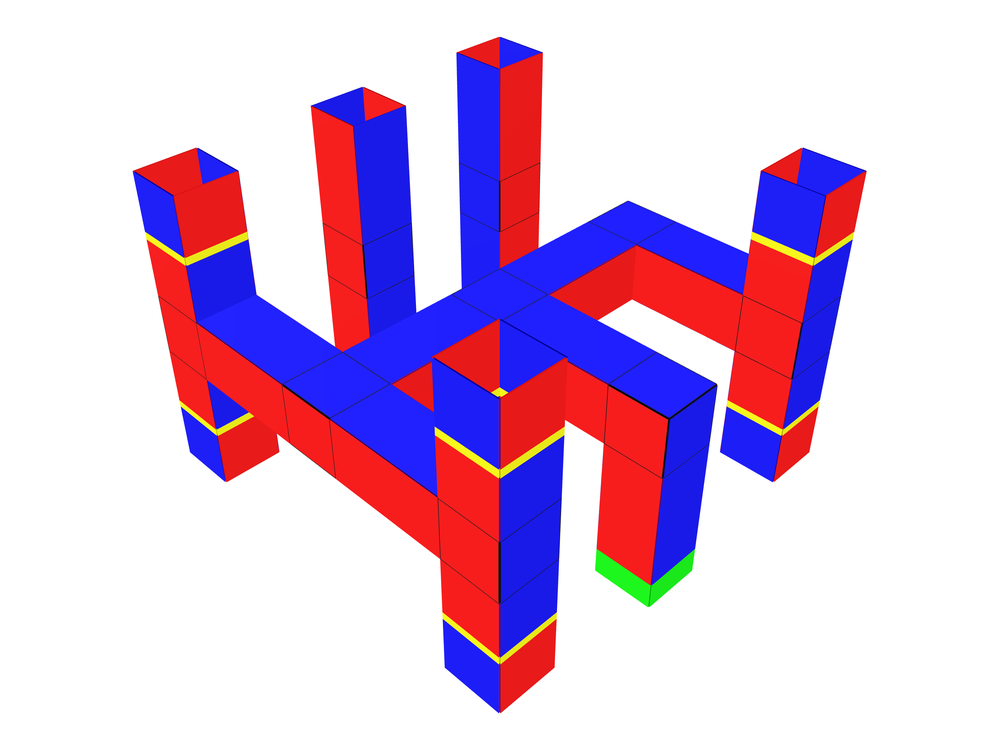}
			\\
			{S-Correction (ZZ branch)}
		\end{tabular}
		\\
		\begin{tabular}{c}
			\includegraphics[width=0.32\textwidth]{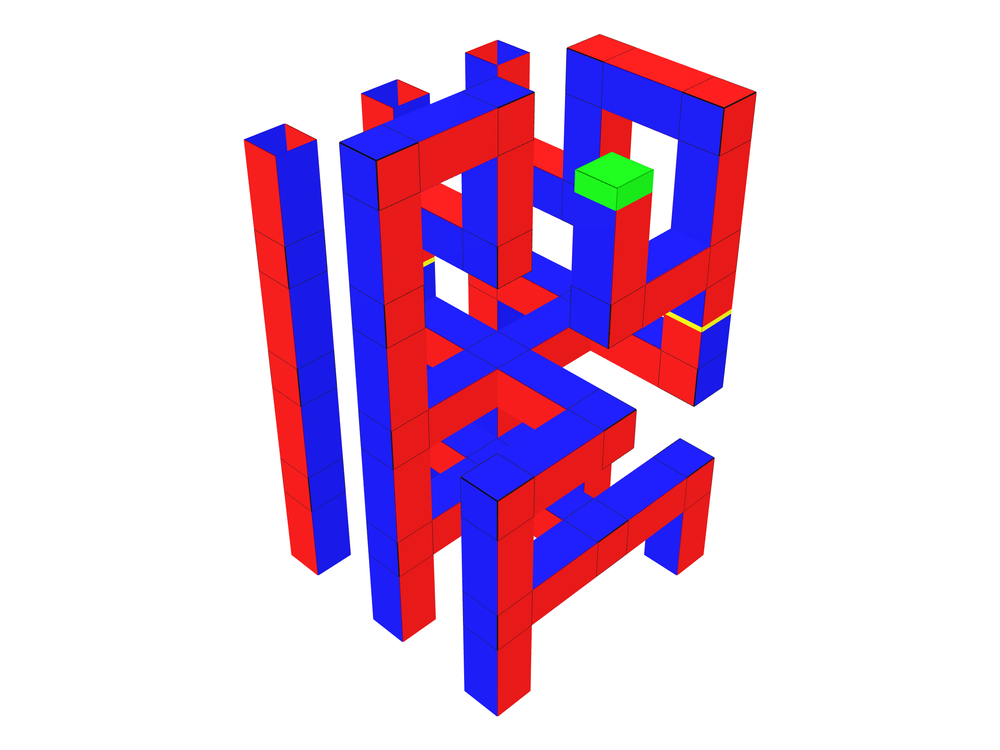}
			\\
			{$G_{XX}$ Passthrough}
		\end{tabular}
		 &
		\begin{tabular}{c}
			\includegraphics[width=0.32\textwidth]{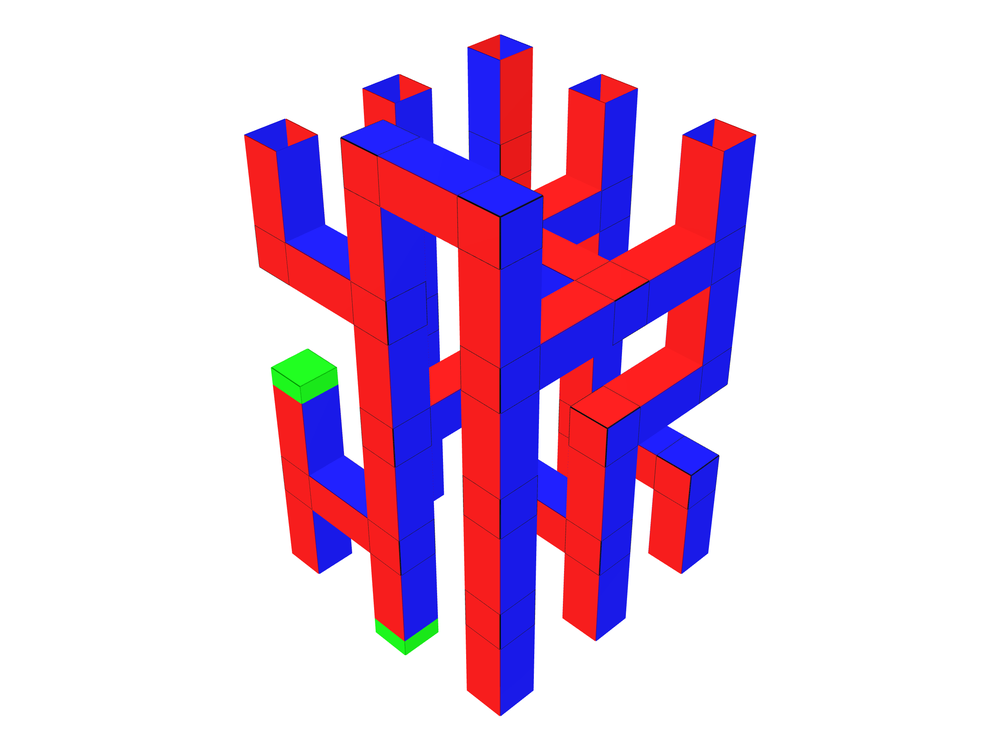}
			\\
			{CCZ to 2T catalysis (XX branch)}
		\end{tabular}
		 &
		\begin{tabular}{c}
			\includegraphics[width=0.32\textwidth]{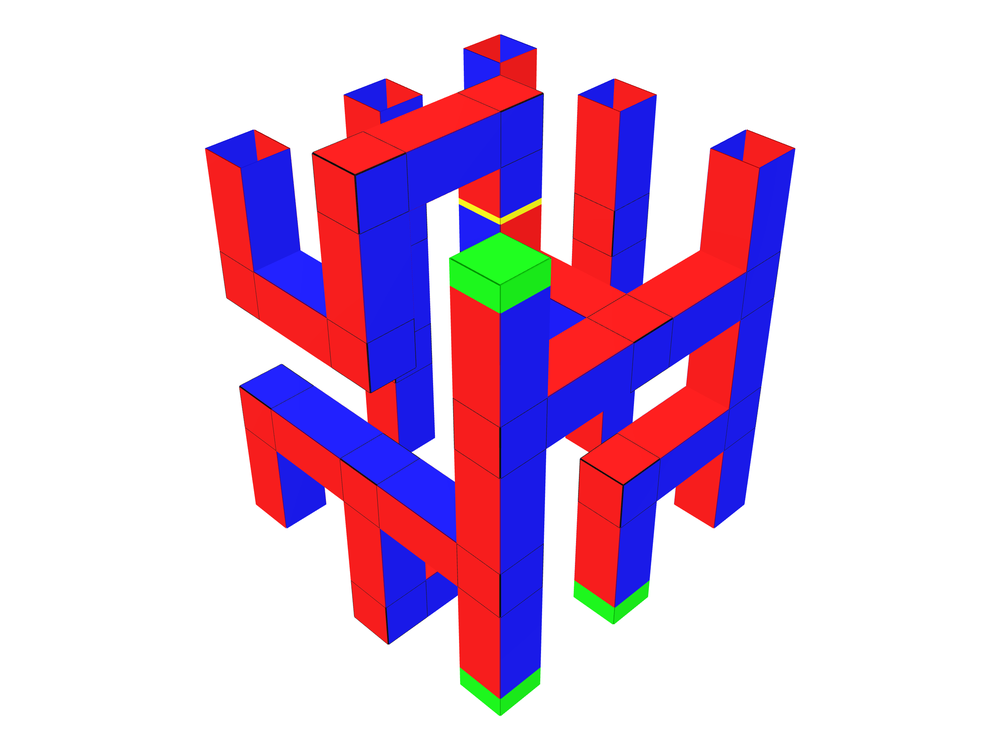}
			\\
			{S-Correction (XX branch)}
		\end{tabular}
	\end{tabular}
	\caption{The core $3 \times 3$ lattice surgery blocks for the space-limited bulk steps.
		Each Givens rotation step requires stacking three components vertically: a CCZ-to-2T catalysis block converts an input CCZ state (produced by a factory not shown here, as discussed in the main text) into two T-states, a probabilistic S-correction fixes the catalyst phase $50\%$ of the time, and the $G_{ZZ}$ or $G_{XX}$ block consumes the T-states to rotate the target qubits.
		The catalytic passthrough wire threads through the invariant edge column on the left face.
	}
	\label{fig:space_limited_bulk}
\end{figure*}

We compile the bulk $G_{ZZ}$ and $G_{XX}$ steps by stacking three blocks: a CCZ factory, a probabilistic S-correction block, and the Givens passthrough block (\Cref{fig:space_limited_bulk}).
We apply the correction block $50\%$ of the time based on the catalyst measurement.

The MA framing operations ($V_0, V_1$) require double-controlled operations, such as a Controlled-$(S \otimes S)$.
The $3 \times 3$ footprint cannot support the parallel $T$-state delivery required for a simultaneous double-control; therefore, we serialize these operations.
We apply a single-target $CS$ gate carrying a phase shift ($CS_{\text{phase}}$) to the first target, followed by a bare $CS$ gate ($CS_{\text{bare}}$) to the second target.
The phases introduced by the sequential components cancel exactly, reproducing the required single-control two-target operation.

We maintain the control state across sequential blocks without an extra memory patch.
We copy the control qubit into the first framing block with a \textsc{CX} gate.
We route the original control qubit into the second framing block and measure it for uncomputation.

We used LaSsynth to find a single edge port configuration that aligns all ten components (two bulk gates, four framing components, and four factory/correction variants).
Each framing component minimizes routing overhead by selecting the CCZ factory variant (ZZ or XX) that delivers magic states closest to the active target qubit.

\begin{table}[ht]
	\centering
	\begin{tabular}{l c c}
		\toprule
		\textbf{Component Stack}                                                & \textbf{Effective Height ($K$)} & \textbf{T-gate Cost}
		\\
		\midrule
		CCZ Factory                                                             & $F$                             & ---
		\\
		Routing Buffer\textsuperscript{b}                                       & 1.0                             & 0
		\\
		$G_{ZZ}$                                                                & 5.0                             & 2
		\\
		$G_{XX}$                                                                & 5.0                             & 2
		\\
		$C(S \otimes S)$\textsuperscript{a}                                     & 7.0                             & 4
		\\
		$C(H \otimes H)$\textsuperscript{a}                                     & 8.0                             & 4
		\\
		$C(Z \otimes Z)$\textsuperscript{b}                                     & 2.0                             & 0
		\\
		$C(X \otimes X)$\textsuperscript{b}                                     & 2.0                             & 0
		\\
		CCZ-to-2T + Correction (ZZ branch)                                      & $6.0 + 0.5(3.0) = 7.5$          & ---
		\\
		CCZ-to-2T + Correction (XX branch)                                      & $5.0 + 0.5(4.0) = 7.0$          & ---
		\\
		\midrule
		Total per $G_{ZZ}$ Step ($K_{ZZ}$)                                      & $13.5 + F$                      & 2
		\\
		Total per $G_{XX}$ Step ($K_{XX}$)                                      & $13.0 + F$                      & 2
		\\
		Total for $V_1$ Framing Op ($C(S \otimes S)$, $C(H \otimes H)$)         & $48.0 + 4F$                     & 8
		\\
		Total for $V_0$ Framing Op ($V_1$ + $C(Z \otimes Z)$, $C(X \otimes X)$) & $54.0 + 4F$                     & 8
		\\
		\midrule
		\textbf{Total ($k=24$ sequence)}                                        & \textbf{420.0 + 32F}            & \textbf{64}
		\\
		\bottomrule
	\end{tabular}
	\caption{Resource breakdown for the $3 \times 3$ space-limited regime.
		The effective height includes the probabilistic S-correction step and is parameterized by the height of the chosen CCZ factory, $F$.
		A 1-unit routing buffer is added before each logical gate operation to allow for control qubit loading.
		The total physical depth for a rotation of length $k$ is $K_{\text{phys}} = \lceil k/2 \rceil K_{ZZ} + \lfloor k/2 \rfloor K_{XX} + K_{V_0} + K_{V_1}$.
		For $k=24$, the sequence requires $420 + 32F$ logical timesteps in expectation, consuming $T$-states at a rate of $64 / (420 + 32F)$ per timestep with a T-depth rate of $32 / (420 + 32F)$.
		\textsuperscript{a}
		Each single-control two-target operation requires one ZZ-branch CCZ-to-2T unit and one XX-branch CCZ-to-2T unit.
		\textsuperscript{b}
		The temporal heights for the buffer and Clifford components are estimates.
	}
	\label{tab:space_limited_resources}
\end{table}

\Cref{tab:space_limited_resources} summarizes the temporal height and T-gate costs of this sequential architecture.
For a multiplexed rotation of length $k=24$, the total execution requires $420 + 32F$ logical timesteps in expectation.
This compact layout operates entirely within $9$ physical patches, optimizing for a minimal spatial footprint at the cost of execution speed.

\subsection{Lattice surgery compilation for multiplexed rotations using a controlled adder}\label{app:controlled_adder_compilation}

The sequential architecture minimizes spatial footprint at the expense of the overall execution time and space-time volume.
When sufficient space is available, we instead implement multiplexed rotations using the standard phase gradient addition technique.
To minimize the space-time volume of this approach, we compile the primary building blocks of the controlled adder to be compatible with our space-time limited layout.
We compile each sub-block of this adder through the following pipeline: starting from a logical circuit, we expand it into a Clifford circuit that implements the temporary non-Clifford operations by consuming a CCZ state (if necessary) and includes the necessary delayed-choice corrections.
We then convert this circuit into a ZX-calculus graph, and compile the graph into a 3D lattice surgery geometry using \texttt{LaSsynth}. 
As we discussed in the main text, we perform a randomized search over candidate geometries, including the size of the bounding box and the configuration of the input and output ports.
This randomized search is designed to sample from solutions that can be tiled together in space and time.
Given a specific configuration, \texttt{LaSsynth}'s SAT-based compiler generates a valid solution if one exists.
In all cases considered here, a few minutes of runtime per solution was sufficient to resolve if a particular configuration was SAT or UNSAT.

\paragraph{Delayed-choice corrections and Z-port calling convention.}
Each CCZ gate teleportation produces classically-controlled CZ corrections whose application can be deferred~\cite{Gidney2019AutoCCZ}. 
We route each correction as a vertical chimney, corresponding to a Hadamard-conjugated CZ bubble in the ZX graph, that exits on the top face of the block and resolves during the chimney gap or in a subsequent pipeline stage. 
This decouples the physical routing from the decoder reaction time.

Control qubits attach to sub-blocks via Z-ports~\cite{Gidney2025Factoring} rather than passing through as input-output pairs. 
To connect a qubit to a Z-port, we split it in the $Z$ basis into two qubits and merge one copy into the port.
This removes the need to return many of the inputs as outputs, which helps compress the overall spatial footprint.

\begin{figure}[htbp]
  \centering
  \subfloat[Clifford skeleton circuit.\label{fig:and_circuit}]{%
    \begin{minipage}[b]{0.64\textwidth}
      \centering
      \includegraphics[scale=.68]{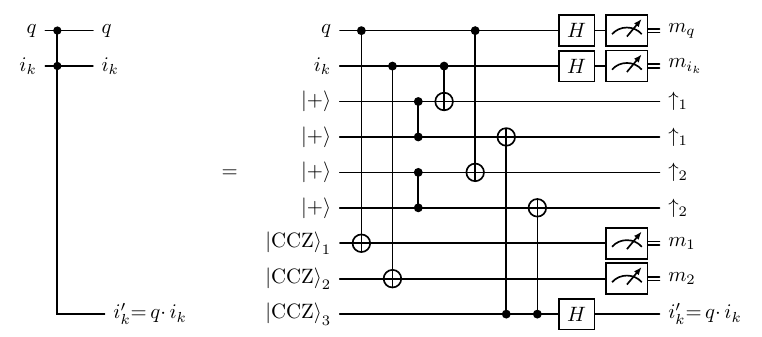}
    \end{minipage}%
  }
  \hfill
  \subfloat[ZX graph (capped \& tapped).\label{fig:and_zx}]{%
    \begin{minipage}[b]{0.35\textwidth}
      \centering
      \includegraphics[scale=.055]{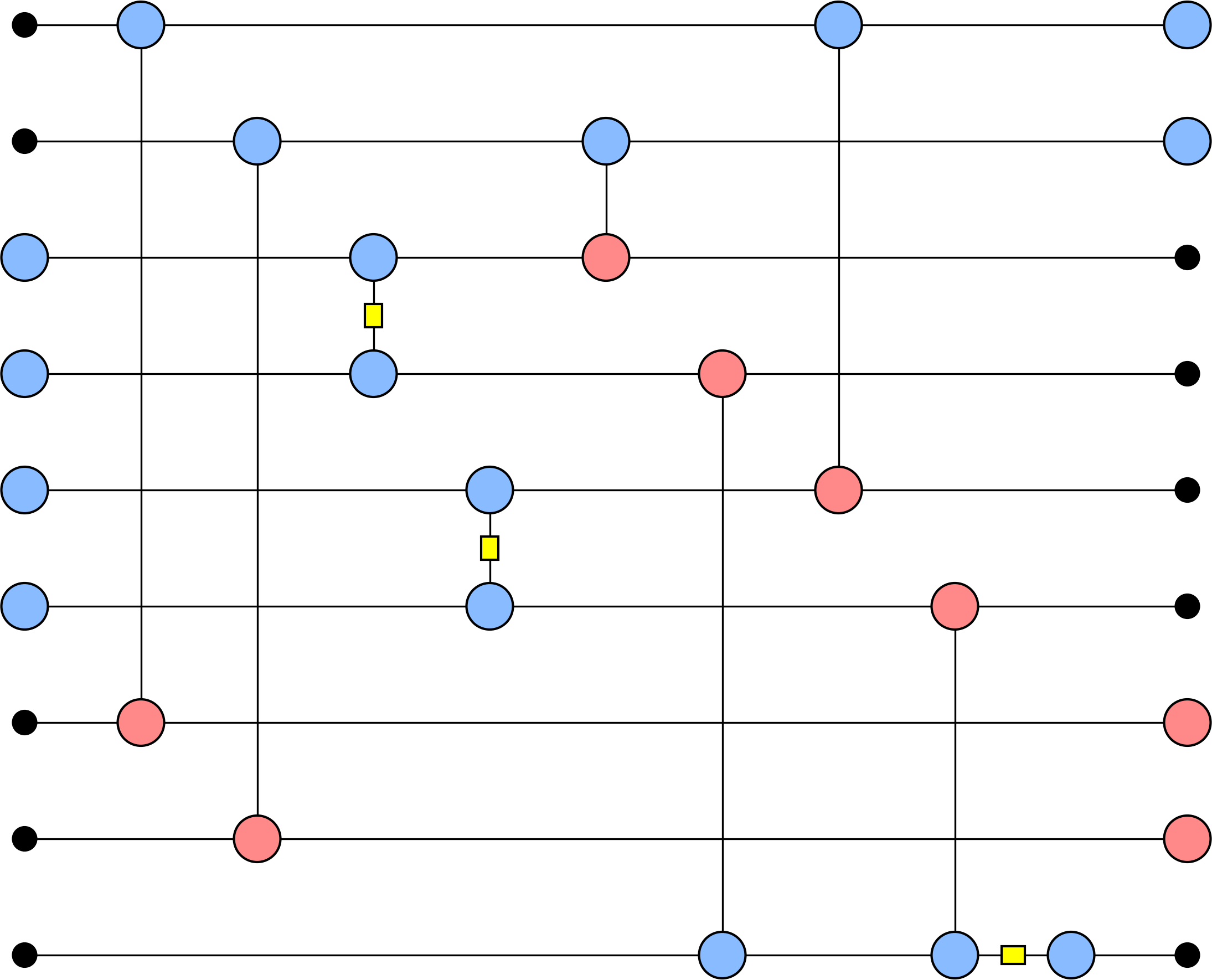}
      \par\vspace{6.55pt}
    \end{minipage}%
  }
  \caption{%
    \textbf{AND sub-block}.
    This block implements a temporary logical-AND~\cite{Gidney2018halvingcostof} via gate teleportation through a CCZ state. 
    Because the target of the CCZ is an ancilla starting in a known $\ket{+}$ state, only the two control qubits ($q$, $i_k$) require teleportation via $X$-basis measurement.
    There are two possible CZ corrections resulting from this teleportation, which we implement using the delayed-choice CZ of \cite{Gidney2019AutoCCZ}.
    Each delayed-choice CZ leads to a pair of routing chimneys (labeled $\uparrow_1$ and $\uparrow_2$). 
  }
  \label{fig:and_block}
\end{figure}

\begin{figure}[htbp]
  \centering
  \subfloat[Clifford skeleton circuit.\label{fig:maj_circuit}]{%
    \begin{minipage}[b]{0.64\textwidth}
      \centering
      \includegraphics[scale=.68]{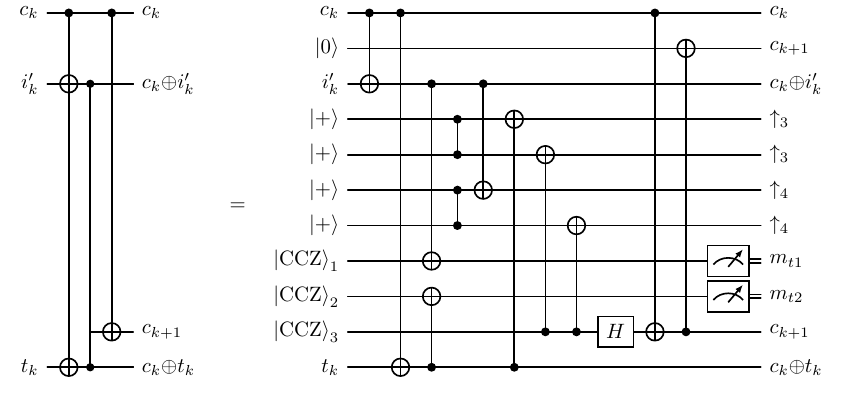}
    \end{minipage}%
  }
  \hfill
  \subfloat[ZX graph (capped \& tapped).\label{fig:maj_zx}]{%
    \begin{minipage}[b]{0.35\textwidth}
      \centering
      \includegraphics[scale=.055]{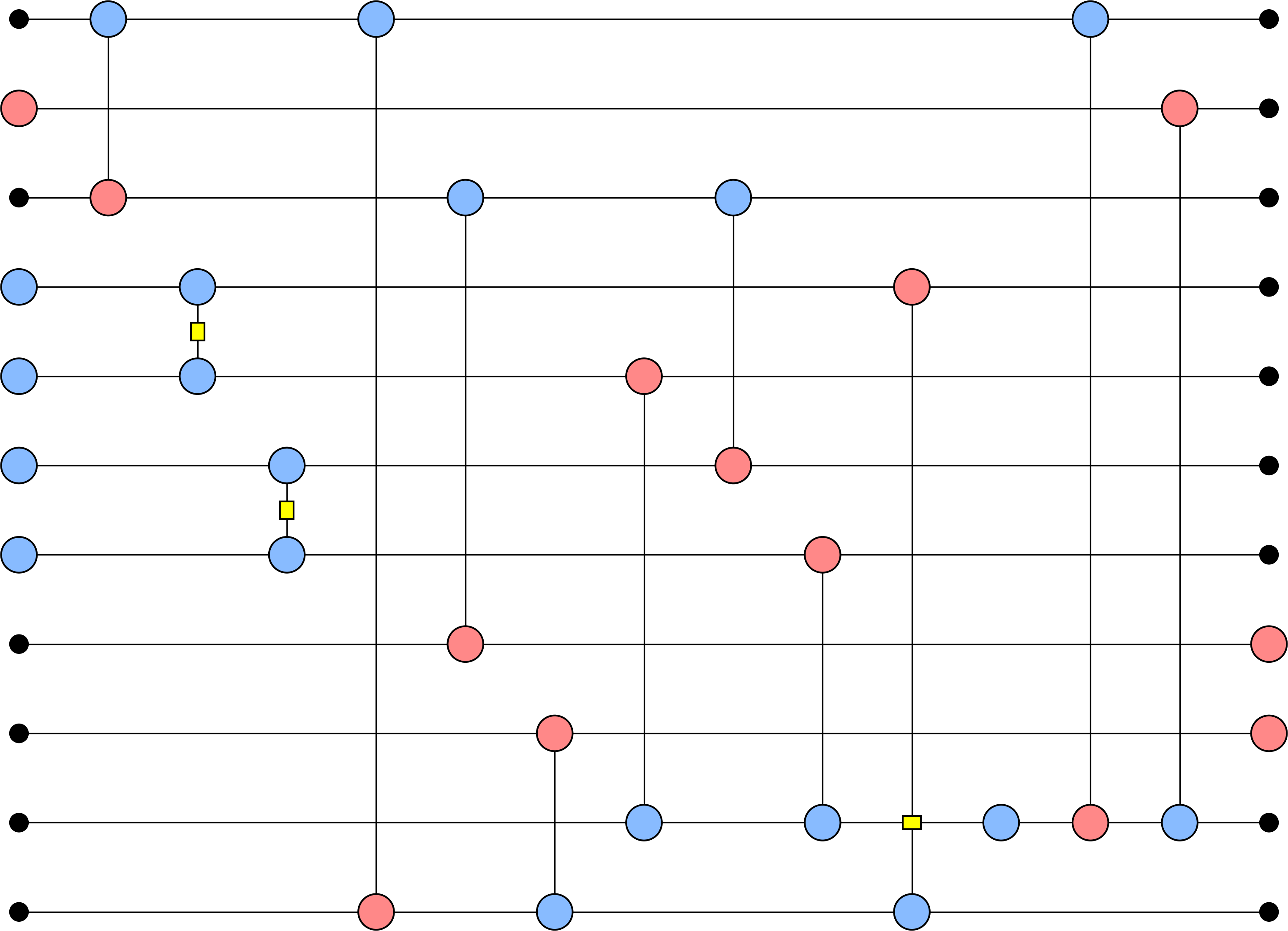}
      \par\vspace{10.98pt}
    \end{minipage}%
  }
  \caption{%
    \textbf{MAJ sub-block}.
    This block computes carry-out $c_{k+1} = \text{MAJ}(c_k, i'_k, t_k)$.
    We proposed candidate geometries for the SAT solver that satisfy several boundary conditions: the \(i'_k\) input on the bottom face aligns with the output from the top of the delayed-choice AND block.
    The various outputs on the top face align with the inputs ports for the UMA block placed directly above.
    The four routing chimneys (labeled by $\uparrow_3$, $\uparrow_4$ to denote which delayed-choice CZ they control) exit on the exposed region $I = 0$ to $2$ of the top face.
  }
  \label{fig:maj_block}
\end{figure}

\begin{figure}[htbp]
  \centering
  \subfloat[Clifford skeleton circuit.\label{fig:uma_circuit}]{%
    \begin{minipage}[b]{0.64\textwidth}
      \centering
      \includegraphics[scale=.68]{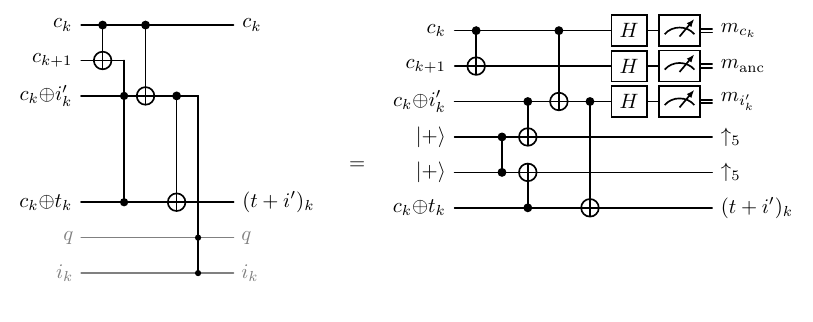}
    \end{minipage}%
  }
  \hfill
  \subfloat[ZX graph (capped \& tapped).\label{fig:uma_zx}]{%
    \begin{minipage}[b]{0.35\textwidth}
      \centering
      \includegraphics[scale=.055]{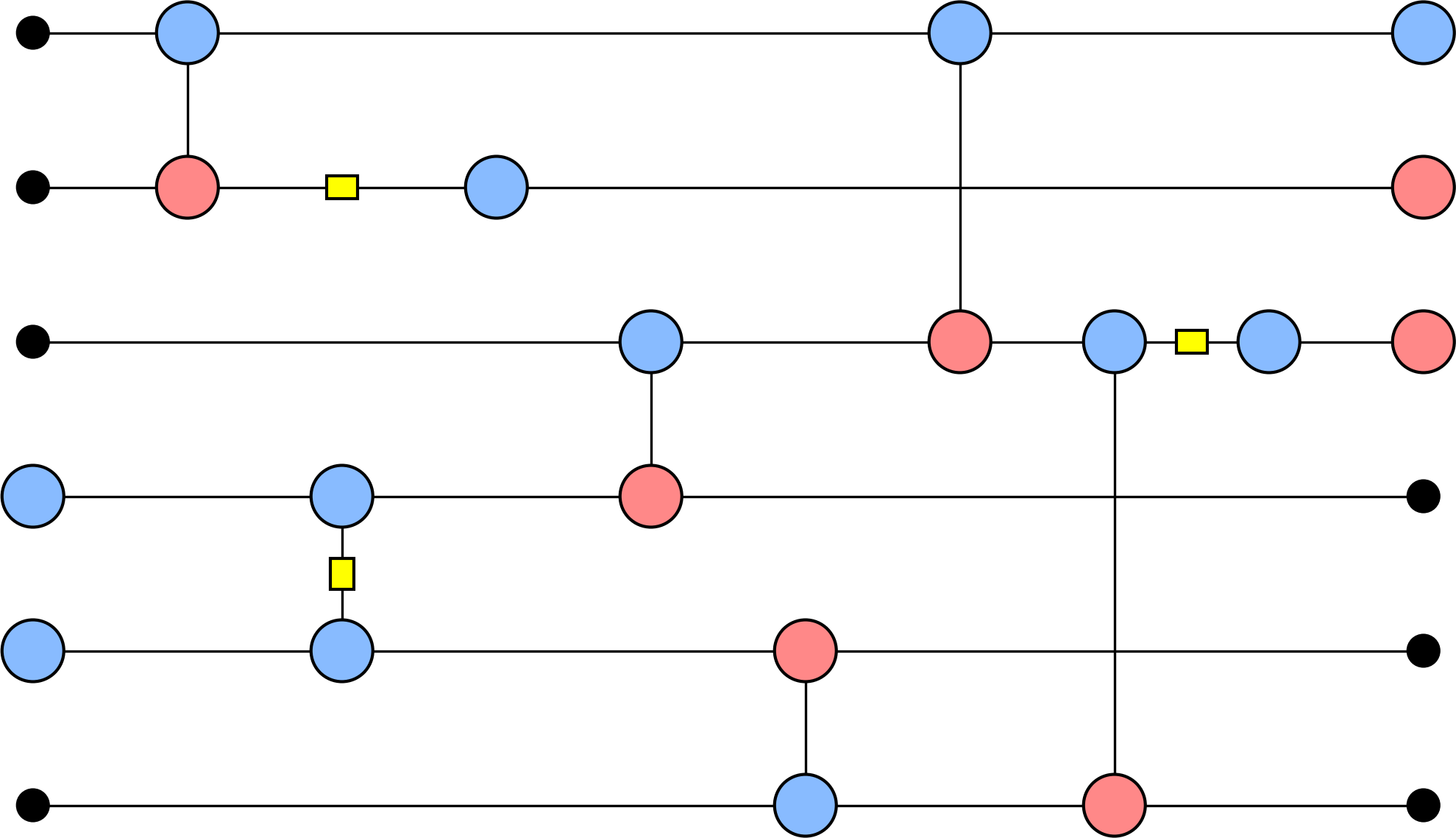}
      \par\vspace{36.64pt}
    \end{minipage}%
  }
  \caption{%
    \textbf{UMA sub-block}.
    This block uncomputes the carry $c_{k+1}$, restores the carry-in $c_k$, and extracts the sum bit $(t \oplus i')_k$. 
    The temporary AND gates from the initial AND block and from the MAJ block are uncomputed by $X$-basis measurement of $c_{k+1}$ and $i'_k$~\cite{Gidney2018halvingcostof}, possibly necessitating CZ corrections on the original control qubits. 
    We defer these corrections, one using the two routing chimneys (labeled by \(\uparrow_5\)), and one by waiting until after the full adder completes and a single-control multi-target Z gate can be applied from the control qubit \(q\) to various input qubits \(i_k\).
  }
  \label{fig:uma_block}
\end{figure}

Figures~\ref{fig:and_block}--\ref{fig:uma_block} show the Clifford circuits and related ZX graphs for each sub-block. 
The ZX graphs generate the stabilizer flows that serve as constraints for the \texttt{LaSsynth} SAT solver.

\begin{figure}[htbp]
  \centering
  \includegraphics[scale=0.60]{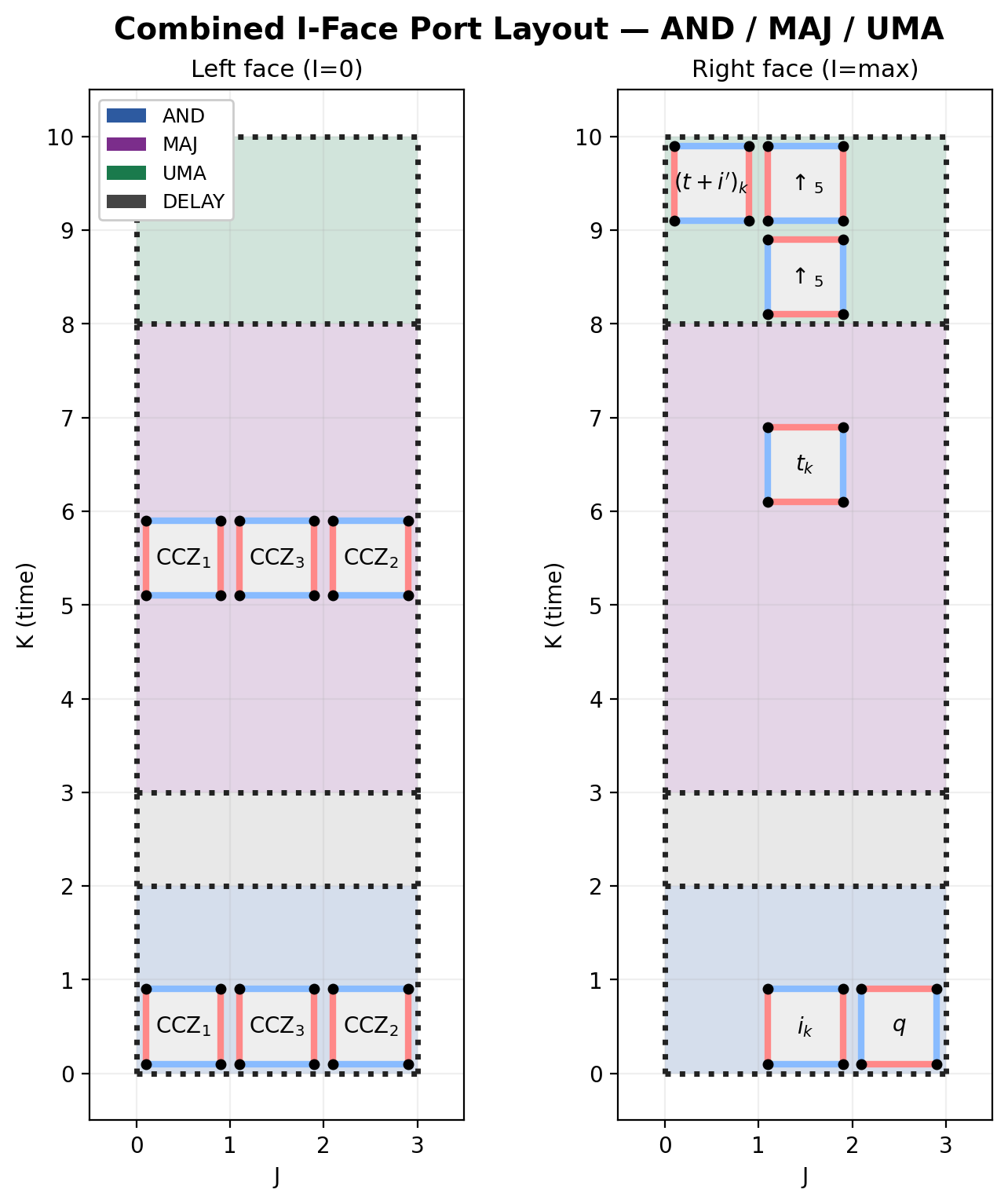}
  \caption{%
    \textbf{$I$-face port layout} for the composed AND/MAJ/UMA triple.
    This layout maps the $J$-axis against the temporal $K$-axis ($0 \le K \le 10$) on the left ($I=0$) and right ($I=5$) boundary faces. The vertical tiling order is visible, including the $1d$ chimney gap between the AND and MAJ stages.
  }
  \label{fig:port_layout_I}
\end{figure}

\begin{figure}[htbp]
  \centering
  \includegraphics[scale=0.60]{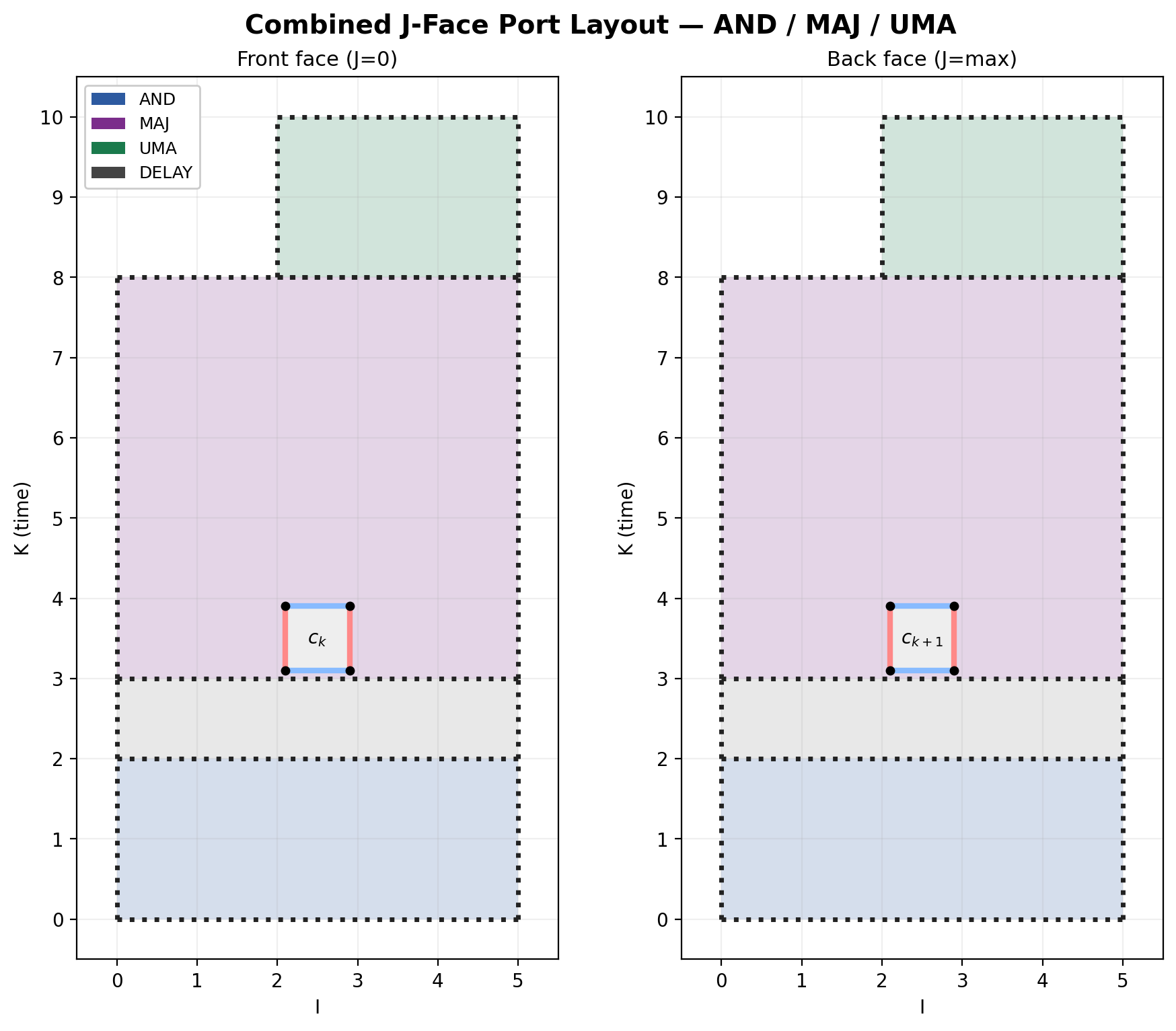}
  \caption{%
    \textbf{$J$-face port layout} for the composed AND/MAJ/UMA triple.
    This layout maps the $I$-axis against the temporal $K$-axis ($0 \le K \le 10$) on the front ($J=0$) and back ($J=2$) boundary faces. The $+2$ spatial offset of the UMA block ($K = 8$ to $10$, spanning $I = 2$ to $5$) is visible, showing the chimney exit window at $I = 0$ to $2$.
  }
  \label{fig:port_layout_J}
\end{figure}

\begin{figure}[htbp]
  \centering
  \includegraphics[scale=0.60]{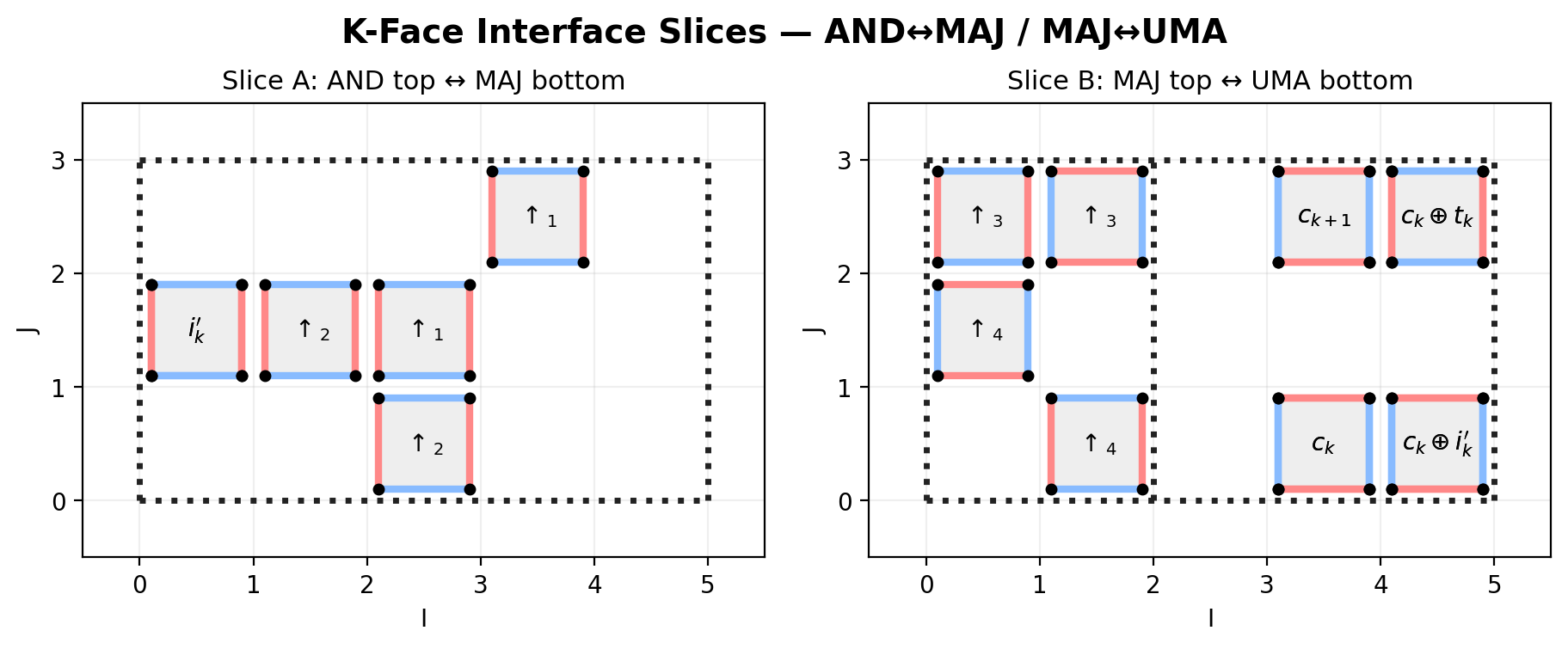}
  \caption{%
    \textbf{$K$-face interface slices} for the composed AND/MAJ/UMA triple.
    (Left) Slice at $K = 2$--$3$ showing the AND-to-MAJ interface, with the $i'_k$ port at $(I=0, J=1)$. (Right) Slice at $K = 8$ showing the MAJ-to-UMA interface: UMA bottom ports occupy $I = 2$ to $5$, adjacent to the MAJ chimney exits at $I = 0$ to $2$.
  }
  \label{fig:port_layout_K}
\end{figure}

The three blocks tile vertically along the $K$-axis. 
We insert a gap of at least \(1d\) between the AND and MAJ stages to provide time for the AND block's correction chimneys to resolve. 
The UMA block is translated by $+2$ along the $I$-axis to create an exit window on the top face of MAJ ($I = 0$ to $2$) for the MAJ correction chimneys.
The design allows for subsequent blocks to be stacked tightly along the \(K\) axis (time).
Data and CCZ state input happens along the \(I\) axis, while the carry bits are passed along the \(J\) axis.
In \Cref{fig:port_layout_I}, \Cref{fig:port_layout_J}, and~\Cref{fig:port_layout_K}, we present detailed cross-sections that show how the input and output ports are configured.

\begin{figure}[htbp]
  \centering
  \subfloat[AND block.\label{fig:and_pipe_bottom}]{%
    \begin{minipage}[b]{0.32\textwidth}
      \centering
      \includegraphics[width=\textwidth]{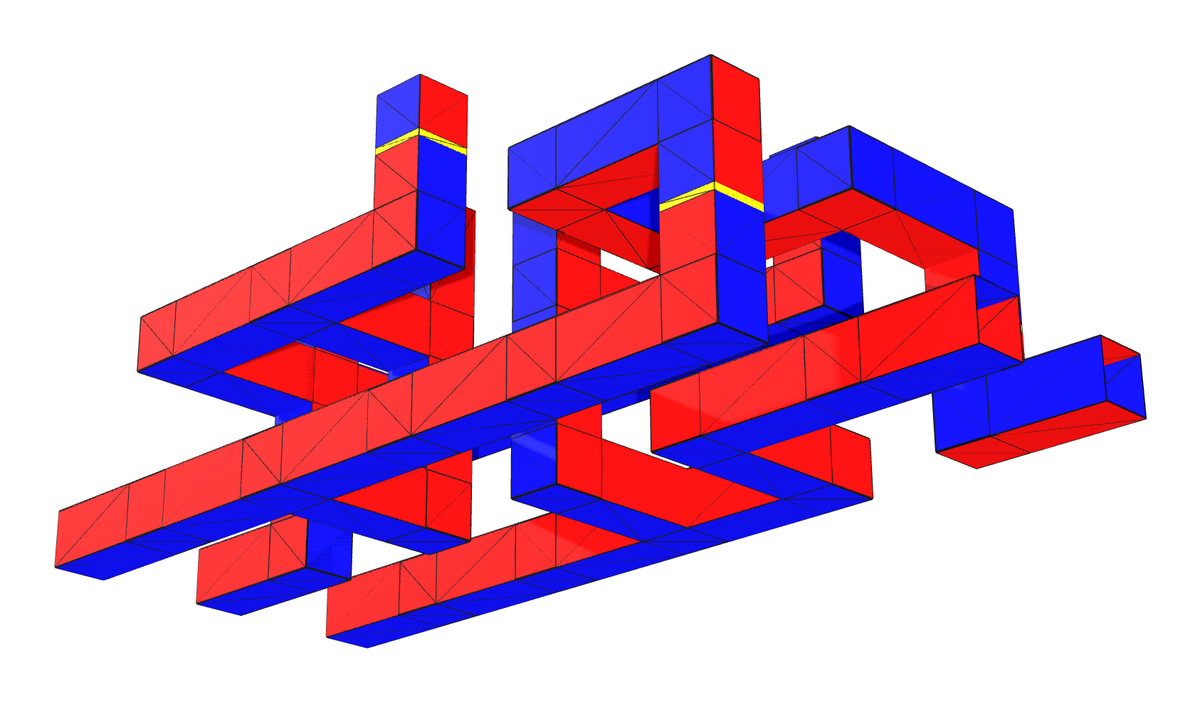}
    \end{minipage}%
  }
  \hfill
  \subfloat[MAJ block.\label{fig:maj_pipe_bottom}]{%
    \begin{minipage}[b]{0.32\textwidth}
      \centering
      \includegraphics[width=.95\textwidth]{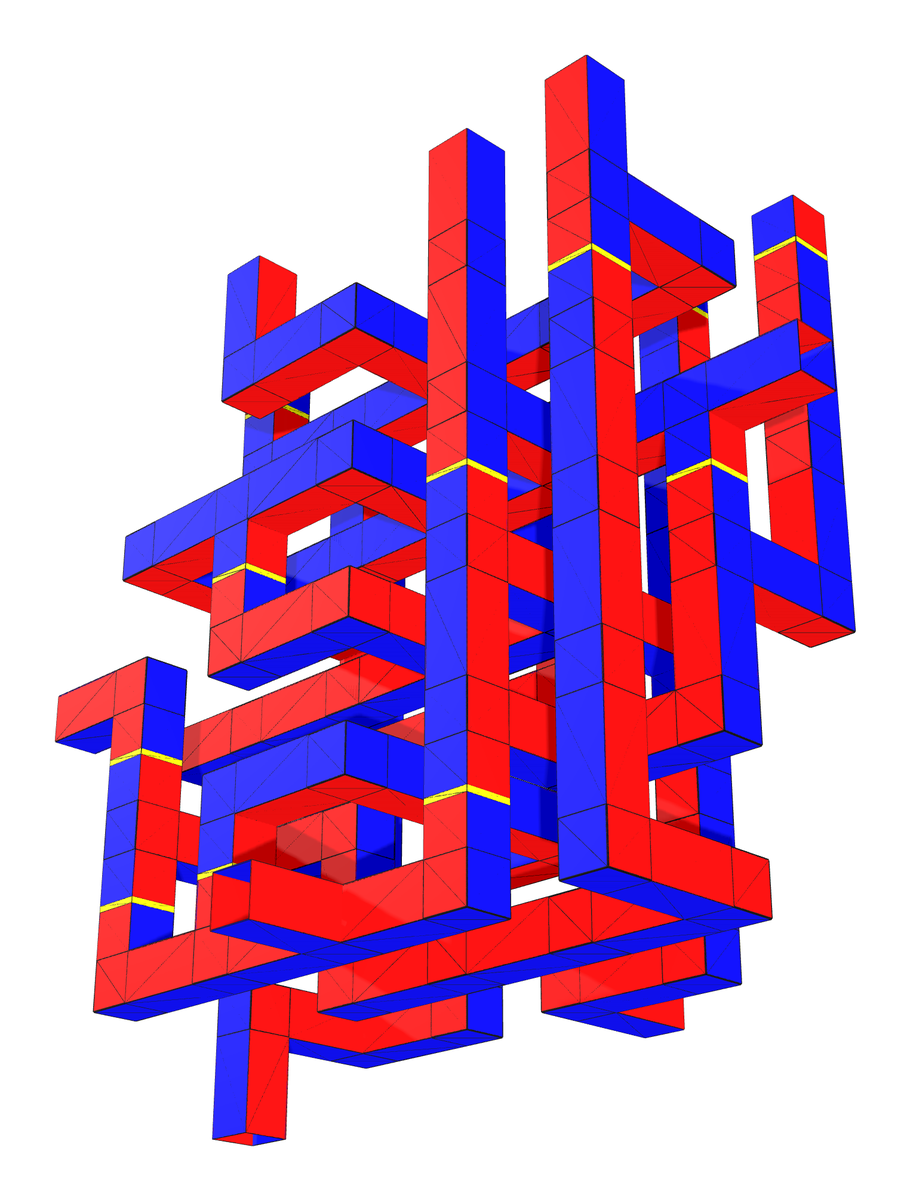}
    \end{minipage}%
  }
  \hfill
  \subfloat[UMA block.\label{fig:uma_pipe_bottom}]{%
    \begin{minipage}[b]{0.32\textwidth}
      \centering
      \includegraphics[width=.7\textwidth]{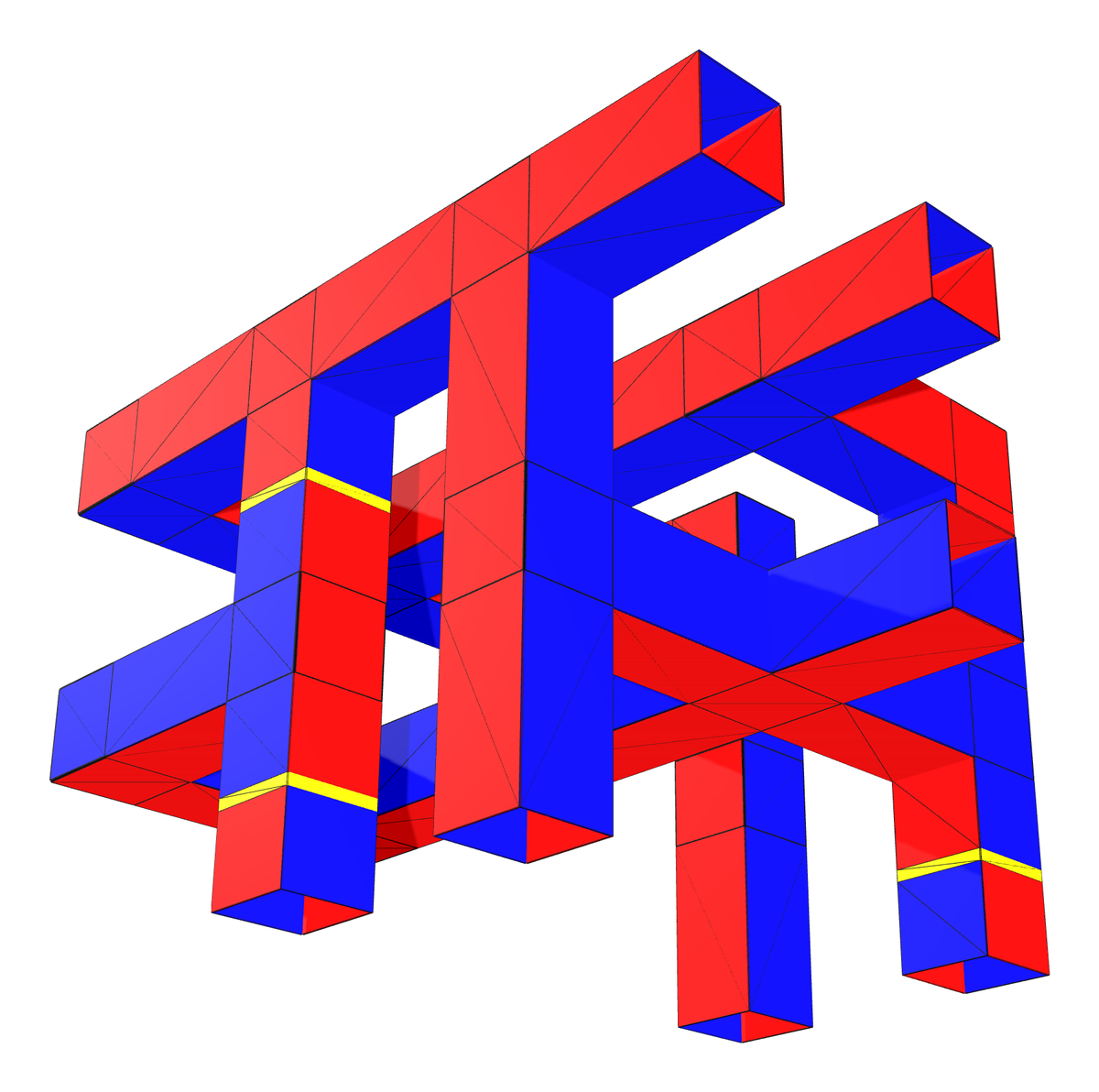}
    \end{minipage}%
  }
  \caption{%
    Alternative bottom-view camera angle of the compiled spacetime geometries in Figure~\ref{fig:pipe_diagrams}, showing the $I$-$J$ routing footprint.
  }
  \label{fig:pipe_diagrams_bottom}
\end{figure}

\begin{figure}[htbp]
  \centering
  \subfloat[AND block.\label{fig:and_pipe_back}]{%
    \begin{minipage}[b]{0.32\textwidth}
      \centering
      \includegraphics[width=\textwidth]{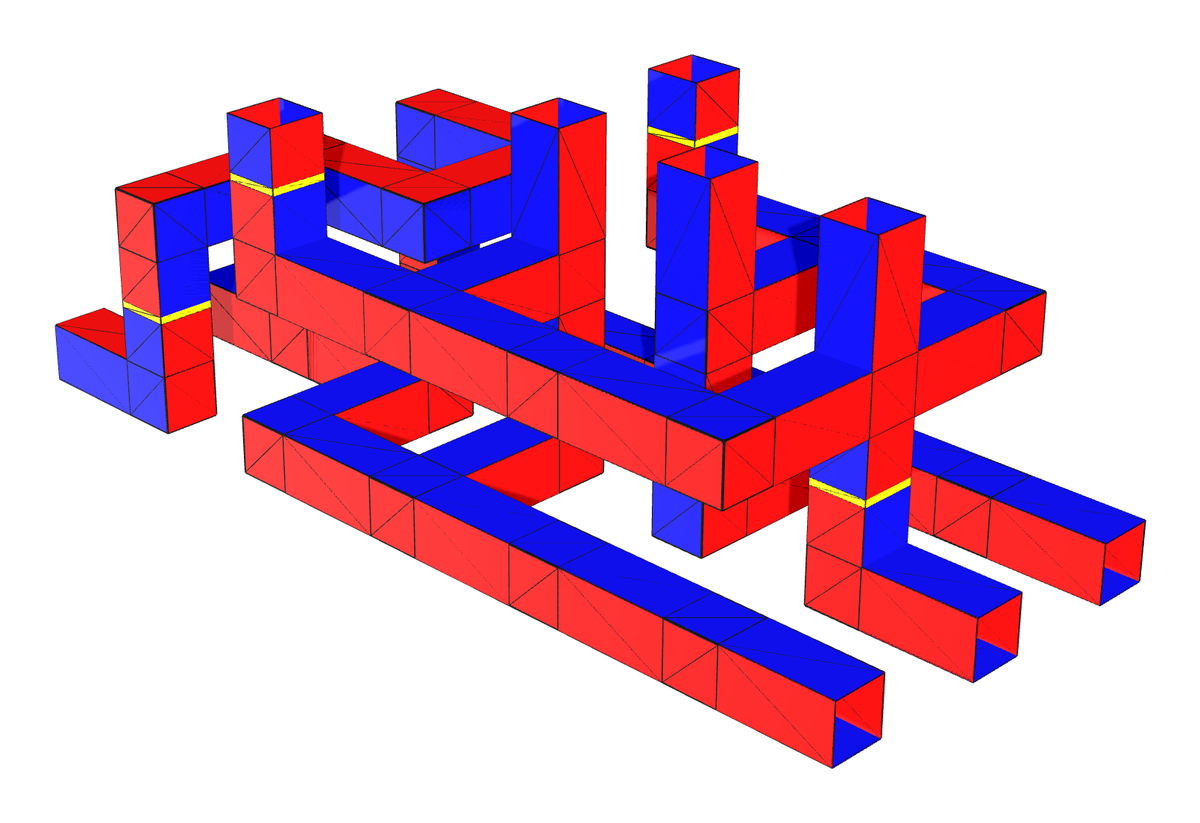}
    \end{minipage}%
  }
  \hfill
  \subfloat[MAJ block.\label{fig:maj_pipe_back}]{%
    \begin{minipage}[b]{0.32\textwidth}
      \centering
      \includegraphics[width=.95\textwidth]{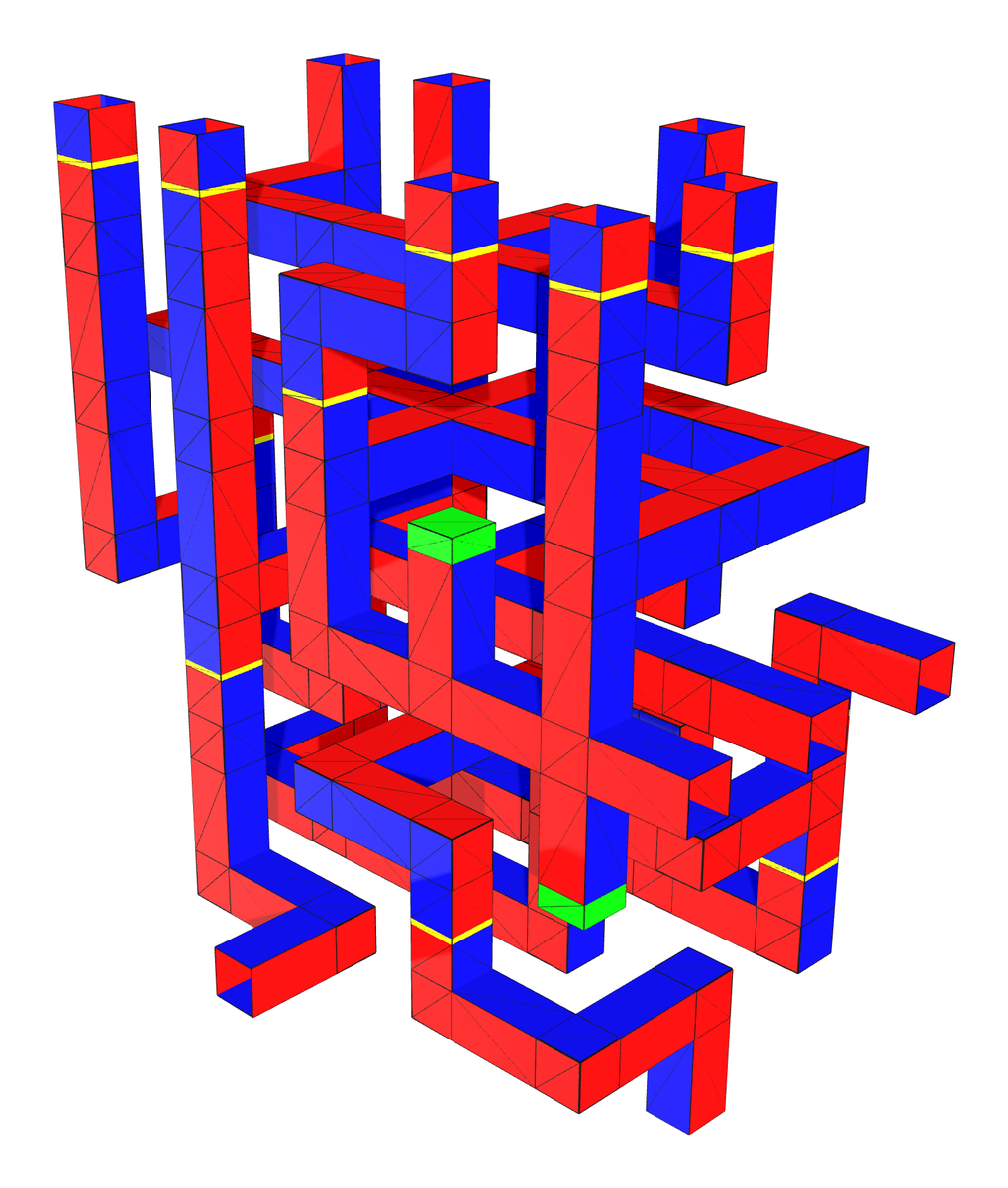}
    \end{minipage}%
  }
  \hfill
  \subfloat[UMA block.\label{fig:uma_pipe_back}]{%
    \begin{minipage}[b]{0.32\textwidth}
      \centering
      \includegraphics[width=.7\textwidth]{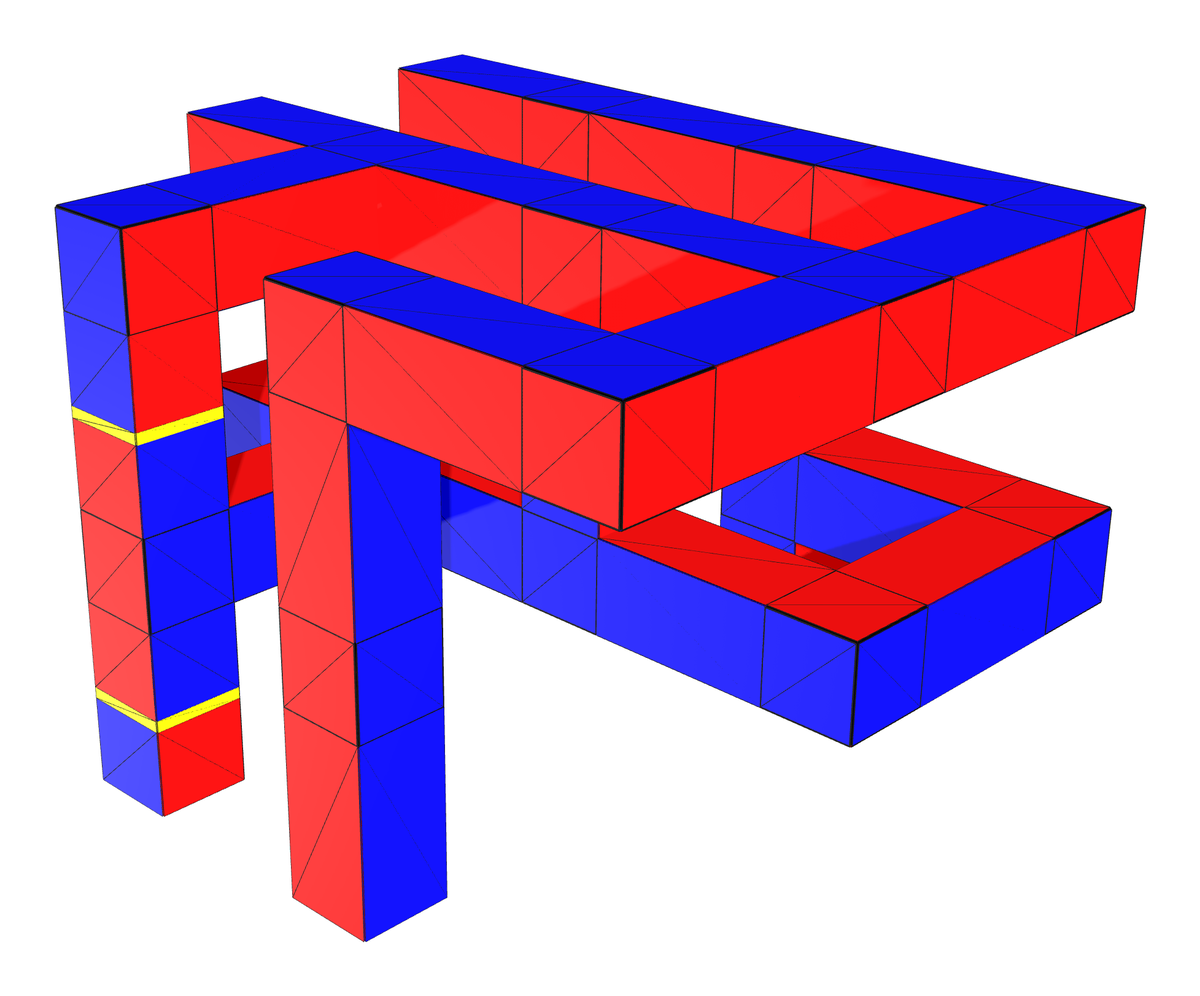}
    \end{minipage}%
  }
  \caption{%
    Alternative back-view camera angle of the compiled spacetime geometries in Figure~\ref{fig:pipe_diagrams}, showing the chimney exit surfaces.
  }
  \label{fig:pipe_diagrams_back}
\end{figure}

We compiled the blocks modularly: the UMA ($3 \times 3 \times 2$) and AND ($5 \times 3 \times 2$) blocks were compiled independently first. We then locked their interface port coordinates and ran a constrained search for the MAJ block ($5 \times 3 \times 5$) to weld the stack together. This modular approach reduces solver complexity and guarantees interface compatibility. The front-view pipe diagrams appear in Figure~\ref{fig:pipe_diagrams}; alternative bottom and back camera angles are shown in Figures~\ref{fig:pipe_diagrams_bottom} and~\ref{fig:pipe_diagrams_back}.

\subsection{Logarithmic-depth tree-structured Givens rotations}\label{sec:parallel_givens}
The rotated Majorana operator is specified by a unit vector $\vec{u}\in\mathbb{R}^{N}$
\begin{align}
	\gamma_{\vec{u}}\doteq \sum_{j=0}^{N-1}u_j\gamma_j,
\end{align}
where the Majorana operators satisfy the anti-commutation rule $\{\gamma_x,\gamma_y\}=2\delta_{xy}\mathbb{I}$.
The quantum circuit that rotates the product of Majorana operators $\gamma_{00}\gamma_{01}\rightarrow\gamma_{\vec{u}0}\gamma_{\vec{u}1}$ can be written as the product of Givens rotations
\begin{align}
	\forall p\neq q,\quad U_{(p,q),x}(\theta) & \doteq e^{\theta \gamma_{px}\gamma_{qx}}=\sum_{j\;\text{even}\ge0}\frac{(\theta \gamma_{px}\gamma_{qx})^j}{j!
	}+\sum_{j\;\text{odd}>0}\frac{(\theta \gamma_{px}\gamma_{qx})^j}{j!}
	=\cos(\theta)\mathbb{I}+\sin(\theta)\gamma_{px}\gamma_{qx}.
\end{align}
Conjugating a Majorana operator by $U_{(p,q),x}(\theta)$ rotates it to
\begin{align}
	U_{(p,q),x}(\theta)^\dagger\gamma_{px}U_{(p,q),x}(\theta)=\cos(2\theta)\gamma_{px}+\sin(2\theta)\gamma_{qx}.
\end{align}
Due to the factor of two, it suffices to consider only $\theta\in[0,\pi)$~\cite{caesura2025faster}.
Moreover, as $[U_{(p,q),0},U_{(p,q),1}]=0$, both $\gamma_{p0}$ and $\gamma_{p1}$ may be simultaneously rotated by the combined operator
\begin{align}
	U_{(p,q)}(\theta) & \doteq U_{(p,q),0}(\theta)U_{(p,q),1}(\theta)=C_{pq}^\dagger\cdot\left[
		                                                                                           \begin{array}{cccc} 1 & 0 & 0            & 0
             \\
             0        & 1 & 0            & 0
             \\
             0        & 0 & e^{2i\theta} & 0
             \\
             0        & 0 & 0            & e^{-2i\theta}
             \\
		                                                                                           \end{array}
		\right]\cdot C_{pq}.
\end{align}
By choosing an appropriate fermion-to-Pauli mapping of $\gamma_{px}$, all Clifford gates $C_{pq}$ only act on two wavefunction qubits.
If $2\theta$ is given to $b$ bits of precision, then $2\theta=2\pi\sum_{j=0}^{b-1}x_j2^{j-b}$ and the multiplexed unitary is defined as
\begin{align}
	\textsc{MU}_{(p,q),a}\ket{x}_a\otimes \mathbb{I}=\ket{x}_a\otimes U_{(p,q)}(\pi x/2^{b})
\end{align}

The sequence of Givens rotations transforming the product $\gamma_{00}\gamma_{01}\rightarrow \gamma_{\vec{u}0}\gamma_{\vec{u}1}$ is
\begin{align}
	\left(\prod_{j=0}^{N-2}U_{\mathcal{E}_{j}}(\theta_j)\right)^\dagger \gamma_{00}\gamma_{01}\left(\prod_{j=0}^{N-2}U_{\mathcal{E}_{j}}(\theta_j)\right)=\gamma_{\vec{u}0}\gamma_{\vec{u}1},
\end{align}
where $\mathcal{E}$ is the edge list of a binary tree defined as follows: At depth $l=0$, the root node of the binary tree is labeled by $0$.
At any depth $l\ge0$, the node with label $j$ has left and right child with label $j$ and $j+2^l$ respectively.
Let us traverse this tree breadth-first, and let the list of edges seen be labeled by the node values.
Let $\mathcal{E}$ be this list excluding self-loops, e.g. for a complete binary tree, the first $7$ elements of $\mathcal{E}$ are $\{(0,1),(0,2),(1,3),(0,4),(2,6),(1,5),(3,7)\}$.

\begin{figure}
	\centering
	\includegraphics[width=0.9\linewidth]{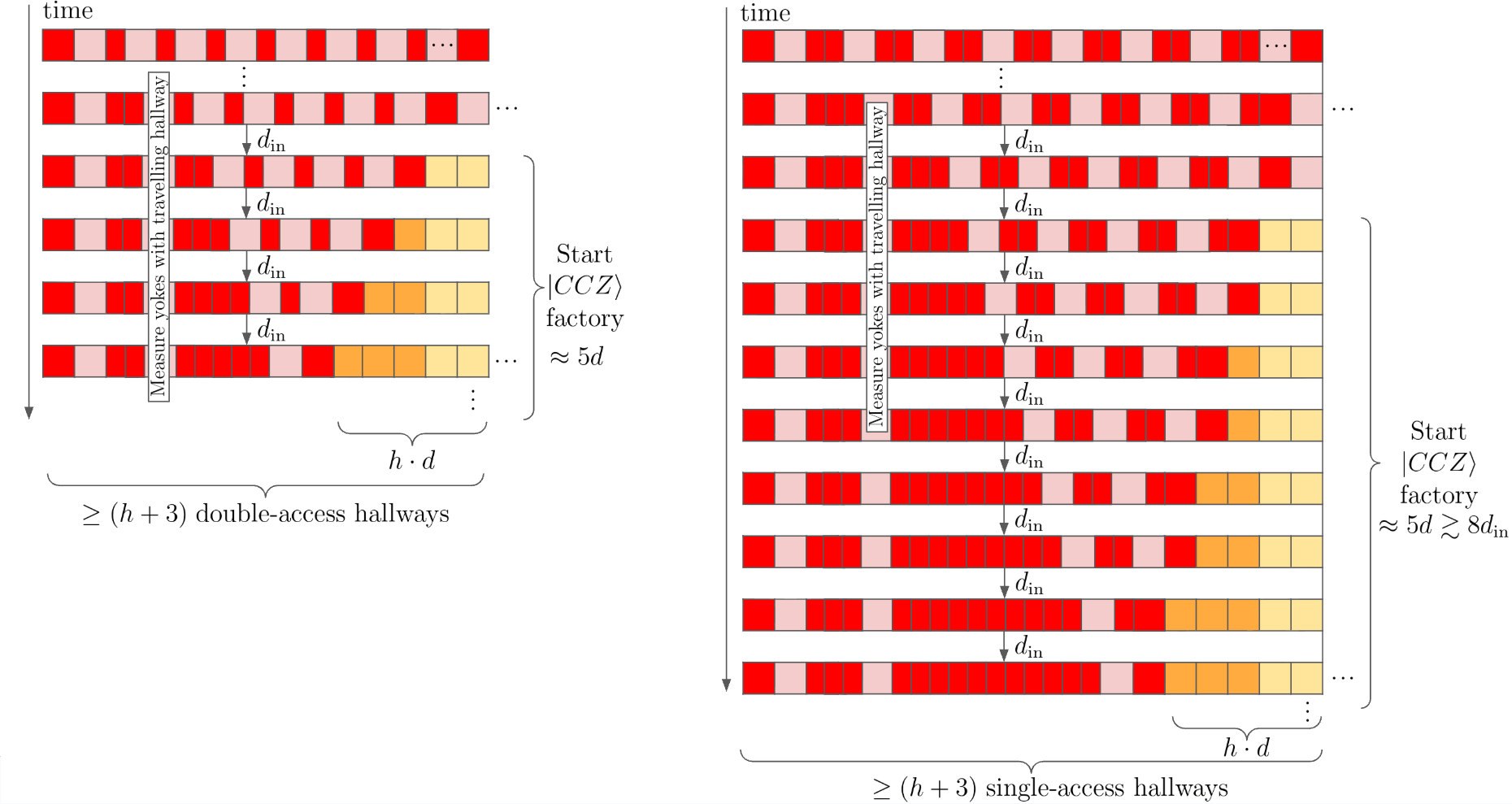}
	\caption{Sequence of moves showing how columns of single-access hallways may be moved to open up new space for another compute region in at most $2(3+h)d_\text{in}$ cycles or $(3+h)d_\text{in}$ cycles with double-access hallways, of which all but $3d_\text{in}$ are amortized with $\ket{CCZ}$ production.
		Note that the right square patch can be moved in $d_\text{in}$ cycles if it is yoked, but otherwise its moves should be performed in larger steps taking $d$ cycles. Instead of leaving three access hallways, one access hallway can also be sufficient if the zero-latency hot storage patches store the logical qubits that the compute region needs to access -- this is the case for the Givens rotations, where the patches that store output bits-of-precision are measured out after each rotation, which provides access to other patches.
        Low-latency hot storage logical qubits are easily expanded into zero-latency hot storage logical qubits by measuring the yoke qubit, and optionally increasing their footprint from $d\times d_\text{in}$ to $d\times d$ if prolonged storage is required.
	},
	\label{fig:faster_givens}
\end{figure}

Whenever we have enough single- or double-access distance $d$ hallways,~\cref{fig:faster_givens} shows that hallways can be moved to create a new compute region that is the mirror image of the original minimum-spacetime or minimum-time layout in at most $20d_\text{in}$ cycles or $10d_\text{in}$ cycles respectively, much of which is amortized with the cycles taken to produce a $\ket{CCZ}$ state.
Hence, we may double the rate at which we consume $\ket{CCZ}$ states to implement the multiplexed rotation, such as with the phase gradient adder.
After each phase gradient adder, the set of bits specifying one set of rotation angle may be measured out to reveal the next set of bits.
This doubles the number of qubits needed to store the phase gradient state.
The time-optimal layout provides up to quadruple the throughput, using $4$ phase gradient states.


%% file: 5_simulation/circuit_tradeoffs.tex
\begin{figure}
    \centering
    \includegraphics[width=0.48\linewidth,valign = m]{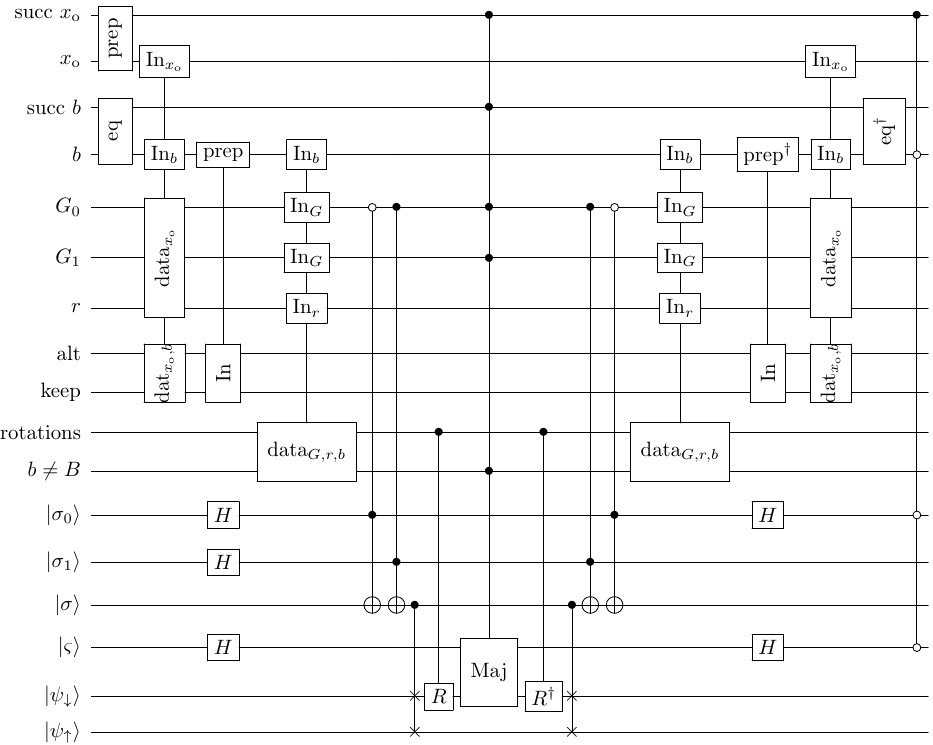}
\includegraphics[width=0.48\linewidth,valign = m]{circuits/dfthc_be3.pdf}
\caption{\label{fig:fullcircuit}
(Left) The circuit for the sum of squares from Ref.~\cite{low2025fast}, with the qubit used for the superposition of Majorana operators shown explicitly as $\ket{\varsigma}$. (Right) The simplified form of the block encoding after qubit count optimizations, where registers containing $G,r,\sigma_0,\sigma_1$ are eliminated.}
\end{figure}

In previous work, the DFTHC Hamiltonian is block-encoded using the circuit shown in~\crefpos{fig:fullcircuit}{left}, which was designed in part for readability, simplicity, and minimum Toffoli count. However, this leads to using a few more registers than is strictly necessary. 
In this section, we discuss optimizations that reduce the qubits needed to block-encode DFTHC to obtain the DFTHC block-encoding circuit~\crefpos{fig:fullcircuit}{right} using as few as $2N+n_{N+RC}+n_{B+1}+2$ qubits, where we use the notation $n_X\doteq{\bits{X}}$.
Beyond the block-encoding circuit, the overall algorithm for phase estimation requires only two additional qubits~\cite{Najafi2023QPE2qubit}.

A particular feature to note in this circuit is that the value of $x_\text{o}$ is used to output $G_0,G_1,r$.
The values of $G_0,G_1$ are bits used to select between $\mathrm{D}_1$, $\mathrm{Q}_1$, and $\text{SF}$ in the outer sum.
In turn, those are used to control the data output for the rotations as well as the choice of spin qubit $\sigma_0$ or $\sigma_1$.
To reduce qubit usage in the implementation here, these are controlled directly by $x_\text{o}$, and $G_0,G_1,r$ are not output on separate registers.

A further feature of that implementation is that it uses three spin qubits.
That was a choice made for clarity, but only two are needed.
The reason for using two is that there is an outer sum over $\sigma$ for $O^\dagger_{G^\sigma,r}$, but an inner sum over $\sigma$ within $O_{\text{SF},rc}$.
The $\sigma$ used for the inner sum is $\sigma_0$ in Fig.~\ref{fig:fullcircuit}, which is why a reflection is applied to it.
The value of $G_0$ is used to select between these spin qubits, and copies them into $\sigma$ which is then used to control the swap between the spin up and spin down components of the system state.
Rather than using this procedure, we could simply use $\sigma$ and apply a controlled Hadamard gate, but this is not what we ultimately end up doing.

A further qubit $\ket{\varsigma}$ is used for the block encoding of the sum of Majorana operators for $\mathrm{D}_1$ and $\mathrm{Q}_1$.
This is explicitly shown in Fig.~\ref{fig:fullcircuit}, though it was omitted for simplicity in Ref.~\cite{low2025fast}.
The Hadamard on this qubit is not needed for $O_{\text{SF},rc}$, so rather than making the Hadamard controlled we perform a controlled swap between the qubits with $\sigma$ and $\varsigma$ to ensure the Hadamard is performed on $\sigma$ for $O_{\text{SF},rc}$ but $\varsigma$ for $\mathrm{D}_1,\mathrm{Q}_1$.
Moreover, the central reflection should only be on $\sigma$ for $O_{\text{SF},rc}$, not $\mathrm{D}_1,\mathrm{Q}_1$.
This controlled swap ensures that this is the case.
The resulting simplification is illustrated in Fig.~\ref{fig:fullcircuit} up to the central reflection.
See~\cref{sec:fewer_qubits_A} for details.

The Givens rotation angles are output in the register labeled `rotations'.
The optimizations given in~\cref{sec:fewer_qubits_B} show how to reduce the qubit requirement here.
The registers labeled `alt' and `keep' are those used for alias sampling in Ref.~\cite{low2025fast}.
\cref{sec:fewer_qubits_C} shows how the need for those registers may be eliminated by instead using pure state preparation.

First we explain the procedure from Ref.~\cite{low2025fast} to explain where the qubits were originally used.
The overall principle of the block encoding for the sum of squares is that there is a variable $x_\text{o}$ that is used for the outer sum, and a variable $b$ that is used for the inner sum within the square.
Initially there is a state preparation for $x_\text{o}$, which is used to control the state preparation for $b$.
The register with $b$ is used for the block encoding of the operator to be squared, which is implemented with a single step of oblivious amplitude amplification.
The original procedure from Ref.~\cite{low2025fast} is illustrated in Fig.~\ref{fig:fullcircuit}.

More specifically, $x_\text{o}$ was used to select each value of $r$ for $\mathrm{D}_1$, each value of $r$ for $\mathrm{Q}_1$, then each value of $r$ and $c$ for $\text{SF}$.
This variable was therefore chosen to be
\begin{equation}\label{eq:xo_indexing}
    x_\text{o}(G,r,c) = \begin{cases}
        r, & G=G_{\mathrm{D}_1},\\
        N_{\mathrm{D}_1}+r, & G=G_{\mathrm{Q}_1},\\
        N+rC+c, & G=G_\text{SF}.
    \end{cases}
\end{equation}
The Hamiltonian was then expressed as a sum of squares as
\begin{equation}
    H_{SOS}  = \sum_{x_\text{o}} O^\dagger_{x_\text{o}} O_{x_\text{o}}.
\end{equation}
The square is implemented via a single step of oblivious amplitude amplification.

To prepare the superposition for the inner sum over $b$, an equal superposition over $b$ was prepared, and QROM over $x_\text{o}$ and $b$ was used to output alt and keep values to use with coherent alias sampling on $b$.
Because QROM was used on $x_\text{o}$ at this stage, it was also used to output registers recording $G$ (with two qubits $G_0$ and $G_1$ to encode the three possibilities) and $r$.

Within each operator $O_{x_\text{o}}$, the Majorana operators depend on $r$, but do not depend on $c$.
Therefore, QROM was used on $G$, $b$, and $r$ to output rotations used to encode the Majorana operators.
The data output was used to apply a sequence of Givens rotations, so that the Majorana operators may be applied on just the one qubit.
The rotations are then inverted, as is the QROM and the state preparation on $b$.
For each $x_\text{o}$, this gives the block encoding of $O_{x_\text{o}}$.
There is a reflection on the qubits used for the block encoding, then $O^\dagger_{x_\text{o}}$ is block encoded, and then the preparation of $x_\text{o}$ for the outer sum is inverted.

Up to this point, we have omitted the discussion of the spin.
For the spin there are two separate parts: for $\mathrm{D}_1$ and $\mathrm{Q}_1$ there is only an outer sum over the spin, whereas for $\text{SF}$ the sum over spin is within $O_{x_\text{o}}$.
As with block encodings of this type, the spin qubit is used to control the swap between the spin up and spin down components of the system register.
To account for the spin, two qubits were used, one for the outer sum (for $\mathrm{D}_1$ and $\mathrm{Q}_1$) and the other for the inner sum (for $\text{SF}$).
That was then copied in a controlled way to a temporary register used to control the swap of the components of the system register.

As a result, the block encoding used the following registers.
\begin{itemize}
    \item The register used for preparing $x_\text{o}$, along with ancillary registers for the coherent alias sampling preparation.
    These include the alt and keep registers, temporary registers for the QROM, an ancilla in an equal superposition for the inequality test with keep, and a flag register for success of preparing the initial equal superposition (not needed for a power of 2).
    \item That storing $b$, as well as its ancillary registers for coherent alias sampling.
    \item The output registers for $G$ and $r$ that are later used to output the rotation angles.
    \item The output register for the angles for the Givens rotations.
    \item A qubit flagging that $b\ne B$, because $b=B$ is used to indicate the identity term in the sum.
    \item Three qubits for the two spin registers (inner and outer sums), and a temporary register.
    \item An extra qubit for the sum over the two Majorana operators in $\mathrm{D}_1$ and $\mathrm{Q}_1$.
\end{itemize}
The overall Toffoli and data qubit count is the special case of $\lambda=N-1$ summarized in~\cref{tab:fewer_qubits_A}.

By using more ancilla-free arithmetic throughout, such as for multi-controlled $X$ gates, lookup tables, and adders in~\cref{sec:fewer_qubits_D}, we further reduce qubit use by around the bits of precision $b_\text{rot}$ used for each rotation.
Finally, we compile rotations using $\{H,S,T\}$ gates in~\cref{sec:fewer_qubits_E} instead of phase gradient adders to save another $b_\text{rot}$ qubits.

\subsection{Primitive quantum circuits}
Earlier, we have already defined the following quantum circuits and discussed their cost when compiled to lattice surgery operations.
\begin{enumerate}
    \item Quantum lookup table that combines skew-tree unary iteration with \textsc{Select}-\textsc{Swap} \textsc{QROAM}:~\cref{def:lookup}
    \item Multi-$n$-qubit-controlled $\textsc{Not}$ ($\textsc{MCX}_n$):~\cref{lem:circuit_MCX}
    \item $n$-qubit controlled swap network ($\textsc{CSwapN}_n$):~\cref{lem:circuit_swap_network}
    \item Multiplexed-\textsc{Z} rotation in Matsumoto-Amano normal form without control~\cref{def:multiplexed_Z_rotation} and with control~\cref{def:multiplexed_CZ_rotation}
\end{enumerate}
We additionally require the following quantum circuit operations.
\begin{lemma}[Phase gradient addition and subtraction\label{def:phase_gradient_addition}] For any integers $b>0$ and $x\in[0,2^b)$, the $b$-bit phase gradient adder $\textsc{Grad}_{a,\mathcal{F}}$ on the Fourier resource state $\ket{\mathcal{F}_b}=\frac{1}{2^{b/2}}\sum_{j=0}^{2^b-1}e^{-2\pi ij/2^{b}}\ket{j}$ implements
\begin{align}\label{eq:phase_gradient_addition}
\textsc{Grad}_{a,\mathcal{F}}\ket{x}_{a}\ket{\mathcal{F}_b}_\mathcal{F}&=\ket{x}_a\frac{1}{2^{b/2}}\sum_{j=0}^{2^b-1}e^{-2\pi ij/2^{b}}\ket{j+x\mod 2^b}_\mathcal{F}
=
e^{i2\pi x/2^{b}}\ket{x}_a\ket{\mathcal{F}_b}_\mathcal{F},
\end{align}
where $\textsc{Grad}_{a,\mathcal{F}}$ can be implemented by any $b$-bit quantum-quantum adder, such as those in~\cref{eq:circuit_QQadder}.
\end{lemma}

\begin{lemma}[Multiplexed $R_Z$ by phase gradient adder~\cite{Gidney2018halvingcostof}]\label{eq:circuit_MRZ} For any integers $b>0$ and $x\in[0,2^b)$, the 
operation
\begin{align}
M {R_Z}_{a,c}\ket{x}_a\ket{\mathcal{F}_b}_\mathcal{F}\otimes \mathbb{I}_c&=
\ket{x}_a\ket{\mathcal{F}_b}_\mathcal{F}\otimes\exp\left(i\frac{2\pi x}{2^{b}}Z_c\right),
\end{align}
can be implemented using a single $b$-bit adder from~\cref{eq:circuit_QQadder}.
\end{lemma}
\begin{proof}
By the identity $a-b=\neg(\neg a+b)$ where $\neg$ is the bit-wise $\textsc{Not}$ operator, 
\begin{align}\label{eq:phase_gradient_subtraction}
(\mathbb{I}\otimes X^{\otimes b})\cdot\textsc{Grad}_{a,\mathcal{F}}\cdot(\mathbb{I}\otimes X^{\otimes b})\ket{x}_a\ket{\mathcal{F}_b}_\mathcal{F}=e^{-i2\pi x/2^{b}}\ket{x}_a\ket{\mathcal{F}_b}_\mathcal{F}.
\end{align}
Using $b$ controlled $\textsc{Not}$ gates $\textsc{CNot}_{c,\mathcal{F}_j}$ controlled by qubit $c$ and targeting qubits $\mathcal{F}_j$, in~\cref{eq:phase_gradient_subtraction}, the circuit
\begin{align}
\left(\bigotimes_{j=0}^{b-1}\textsc{CNot}_{c,\mathcal{F}_j}\right)\cdot\textsc{Grad}_{a,_\mathcal{F}}\cdot\left(\bigotimes_{j=0}^{b-1}\textsc{CNot}_{c,\mathcal{F}_j}\right)\ket{x}_a\ket{\mathcal{F}_b}_{\mathcal{F}}\ket{y}_c=e^{(-1)^yi2\pi x/2^{b}}\ket{x}_a\ket{\mathcal{F}_b}_{\mathcal{F}}\ket{y}_c,
\end{align}
implements the required operation.
\end{proof}

\begin{lemma}[Multiplexed controlled-$R_Z$ by phase gradient adder]\label{eq:circuit_MCRZ} For any integers $b>0$ and $x\in[0,2^b)$, the unitary $M C{R_Z}_{a,c}$ that implements
\begin{align}
M C{R_Z}_{a,c}\ket{x}_a\otimes \mathbb{I}_c&=\ket{x}_a\otimes\left[
\begin{array}{cccc}
1 &0&0&0\\
0&1&0&0\\
0 &0&e^{i2\pi x/2^{b}}&0\\
0 &0&0&e^{-i2\pi x/2^{b}}\\
\end{array}
\right]_c,
\quad
\mathbb{I}_c=\left[\begin{array}{cccc}
1 &0&0&0\\
0&1&0&0\\
0 &0&1&0\\
0 &0&0&1\\
\end{array}
\right]_c,
\end{align}
can be implemented using a single $b$-bit controlled adder from~\cref{eq:circuit_QQadder}.
\end{lemma}
\begin{proof}
Replace the unitary $\textsc{Grad}$ in~\cref{eq:circuit_MRZ} with its controlled version.
\end{proof}

\begin{lemma}[Quantum-quantum integer addition]\label{eq:circuit_QQadder}
Let $x$ and $y$ be non-zero $n$-bit integers.
There is a quantum circuit $\textsc{Add}_n$ that computes
$\ket{x}\ket{y}\rightarrow \ket{x}\ket{(x + y)\mod 2^{n}}$ and a quantum circuit $\textsc{CAdd}_n=\proj{0}\otimes\mathcal{I}+\proj{1}\otimes \textsc{Add}_n$ using the following resources.

\begin{center}
\centering
\begin{tabular}{c|c|c|c}
\hline\hline
    Gate & Method & Minimum Toffoli~\cite{Gidney2018halvingcostof}  & Minimum ancilla \\
    \hline
    \multirow{2}{*}{$\textsc{Add}_n$} & Clean Ancilla & $n-1$ & $0$ \\    
    & Toffoli gates & $n-1$ & $2n-2^\text{a}$~\cite{Takahashi2010Addition}\\ 
    \hline
    \multirow{2}{*}{$\textsc{CAdd}_n$} & Clean Ancilla & $n-1$ &$0$\\    
    & Toffoli gates & $2n-1$ &$3n-2$~\cite{Munoz2019ControlledAdder}\\ 
    \hline\hline
    \multicolumn{4}{l}{\small $^{\text{a}}$$2n-1$ Toffoli gates if a carry bit is output.}
    \\
\end{tabular}
\end{center}
The minimum ancilla controlled adder with $3n+\mathcal{O}(1)$ Toffoli gates is obtained from the adder building block Figure. 4 of~\cite{Gidney2018halvingcostof}, where all logical-\textsc{And}s and uncomputes are replaced with Toffoli gates. Equivalently, the two ancilla qubits required by~\cite{Munoz2019ControlledAdder} can be removed for modular addition without carry bits.
\end{lemma}

\begin{lemma}[Quantum-classical integer addition]\label{lem:qc_addition}
Let $x$ and $y$ be non-zero $n$-bit integers.
There is a quantum circuit $\textsc{AddC}_{n}$that computes
$\ket{x}\rightarrow \ket{(x+y)\mod 2^{n}}$ and its controlled version $\textsc{CAddC}_{n}$ using the following resources.

\begin{center}
\begin{tabular}{c|c|c|c|c|c}
\hline\hline
    Gate&Method & Minimum Toffoli & \multicolumn{2}{c|}{Fewer ancilla~\cite{Gidney2025QCAdder}} & Minimum ancilla~\cite{Haner2017Adder}\\
    \hline
    \multirow{3}{*}{$\textsc{AddC}_n$}&Clean Ancilla & $n-3$ & $3$& $2 $& $0$\\    
    &Dirty Ancilla & $0$ &$0$& $n-2$& $1$ \\
    &Toffoli gates & $n-2$~\cite{Fedoriaka2025ConstantAdder} &$4n\pm\mathcal{O}(1)$&$3n\pm\mathcal{O}(1)$&$8n(\log_2 n-2) + \mathcal{O}(1)$\\ 
    \hline
    \multirow{3}{*}{$\textsc{CAddC}_n$}&Clean Ancilla & $n - 1$ &$3$&$2$& $0$\\    
    &Dirty Ancilla & $0$ &$0$&$n-2$& $1$ \\
    &Toffoli gates & $n-1$ (new) &$4n\pm\mathcal{O}(1)$&$3n\pm\mathcal{O}(1)$&$8n(\log_2 n-2) + \mathcal{O}(1)$\\ 
    \hline\hline
\end{tabular}
\end{center}
\end{lemma}
\begin{proof}{Minimum Toffoli $\textsc{CAdd}_n$:}
The controlled adder computes $\ket{x}|\text{ctrl}\rangle\rightarrow \ket{(x+y\cdot\text{ctrl})\mod 2^{n}}$. Let the classical constant $y=\sum_{j=0}^{n-1}y_j2^j$, and let the $n-1$ clean ancilla qubits start in $\ket{0}_{a_0}\cdots\ket{0}_{a_{n-2}}$. Let $c_i$ be the $i^\text{th}$ carry bit. For $i>0$, $c_i$ is in ancilla $a_{i-1}$. The algorithm proceeds in three phases:
\begin{enumerate}
    \item Compute the carry bits iteratively, starting from the least significant bit. When $\text{ctrl}=0$, all carries $c_i=0$, and when $\text{ctrl}=1$, this reduces to the correct carry output $c_i=\textsc{Majority}(x_i,c_i,1)$ if $y_i=1$.
    \begin{itemize}
        \item Base case $i=0$: If $y_0=1$, compute $\ket{\text{ctrl}}\ket{x_0}\ket{0}_{a_0}\rightarrow\ket{\text{ctrl}}\ket{x_0}\ket{x_0\wedge\text{ctrl}}_{a_0}$ with a $\textsc{CCX}$.
        \item For $i=1,2,\cdots, n-2$: If $y_i=0$, compute $\ket{0}_{a_i}\ket{x_i}\ket{c_i}_{a_{i-1}}\rightarrow\ket{x_i\wedge c_i}_{a_i}\ket{x_i}\ket{c_i}_{a_{i-1}}$ with a $\textsc{CCX}$. Otherwise if $y_i=1$, the carry bit is computed using the identity for $\textsc{Majority}(x,c,\text{ctrl})=x\oplus ((x\oplus \text{ctrl})\wedge (x\oplus c))$. First, we compute the two inputs of $\textsc{Majority}$ by $\ket{\text{ctrl}}\ket{c_i}_{a_{i-1}}\ket{x_i}\rightarrow\ket{\text{ctrl}}\ket{c_i\oplus x_i}_{a_{i-1}}\ket{x_i}\rightarrow\ket{\text{ctrl}}\ket{c_i\oplus x_i}_{a_{i-1}}\ket{x_i\oplus\text{ctrl}}$ using $\textsc{CX}$ gates. Second, we compute in the ancilla $\ket{0}_{a_i}\rightarrow\ket{(x_i\oplus \text{ctrl})\wedge (x_i\oplus c)}_{a_i}$ using a $\textsc{CCX}$ gate. Third, we uncompute with $\textsc{CX}$ gates to recover $\ket{c_i}_{a_{i-1}}\ket{x_i}$. Finally, we complete the computation of $\textsc{Majority}$ with $\ket{x_i}\ket{(x_i\oplus \text{ctrl})\wedge (x_i\oplus c_i)}_{a_i}\rightarrow\ket{x_i}\ket{\textsc{Majority}(x_i,c_i,\text{ctrl})}_{a_{i}}$ with a $\textsc{CX}$ gate.
    \end{itemize}
    \item Compute the most significant bit, which in modular addition is $s_{n-1}=x_{n-1}\oplus(y_{n-1}\cdot \text{ctrl})\oplus c_{n-1}$. This is performed inplace using $\textsc{CX}$ gates. If $y_{n-1}=1$, we obtain $\ket{\text{ctrl}}\ket{x_{n-1}}\ket{c_{n-1}}_{a_{n-2}}\rightarrow\ket{\text{ctrl}}\ket{s_{n-1}}\ket{c_{n-1}}_{a_{n-2}}$. 
    \item We now iteratively uncompute the ancilla qubits and compute the sum bits.
    \begin{itemize}
        \item For $i=n-2,\cdots,2,1$: In order to apply measurement-based uncomputation, we restore the qubit states to just after the application of the $\textsc{CCX}$ gate. So, if $y_i=1$, we first use $\textsc{CX}$ gates to again compute $\ket{\text{ctrl}}\ket{c_i}_{a_{i-1}}\ket{x_i}\ket{\textsc{Majority}(x_i,c_i,\text{ctrl})}_{a_{i}}\rightarrow\ket{\text{ctrl}}\ket{c_i\oplus x_i}_{a_{i-1}}\ket{x_i\oplus\text{ctrl}}\ket{(x_i\oplus \text{ctrl})\wedge (x_i\oplus c)}_{a_{i}}$. Second, measure $\ket{\cdot}_{a_i}$ in the $X$ basis and apply the conditional $\textsc{CZ}$ fixup on $\ket{c_i\oplus x_i}_{a_{i-1}}\ket{x_i\oplus\text{ctrl}}$. Third, we use $\textsc{CX}$ gates to compute the sum bit in-place like $\ket{\text{ctrl}}\ket{c_i\oplus x_i}_{a_{i-1}}\ket{x_i\oplus\text{ctrl}}\rightarrow \ket{\text{ctrl}}\ket{c_i}_{a_{i-1}}\ket{x_i}\rightarrow\ket{\text{ctrl}}\ket{c_i}_{a_{i-1}}\ket{x_i\oplus c_i\oplus\text{ctrl}}$. If $y_i=0$, the state is $\ket{x_i\wedge c_i}_{a_i}\ket{x_i}\ket{c_i}_{a_{i-1}}$ and we uncompute by measuring $\ket{\cdot}_{a_i}$ in the $X$ basis, perform a conditional $\textsc{CZ}$ fixup, and compute the sum bit $\ket{x_i}\ket{c_i}_{a_{i-1}}\rightarrow\ket{x_i\oplus c_i}\ket{c_i}_{a_{i-1}}$.
        \item Base case $i=0$: If $y_0=1$, the state is $\ket{\text{ctrl}}\ket{x_0}\ket{x_0\wedge\text{ctrl}}_{a_0}$ so we uncompute by measuring $\ket{\cdot}_{a_0}$ in the $X$ basis, perform a conditional $\textsc{CZ}$ fixup, and compute $\ket{\text{ctrl}}\ket{x_0}\rightarrow\ket{\text{ctrl}}\ket{x_0\oplus\text{ctrl}}$.
    \end{itemize}
\end{enumerate}
Any qubit after $X$ basis measurement can be reset to $\ket{0}$ and made clean, and the overall procedure uses $n-1$ clean ancilla qubits and Toffoli gates. Note that this can be modified to simultaneously add a different constant controlled on $\ket{\text{ctrl}}=\ket{0}$ with the same cost.
\end{proof}

\begin{lemma}[Quantum-classical comparison]\label{lem:quantum_classical_comparison}
Let $x$ and $c$ be non-zero $n$-bit integers and let $z\in\{0,1\}$.
There is a quantum circuit that computes
$\ket{x}\ket{z}\rightarrow \ket{x}\ket{z\oplus(x<c)}$ using the following resources.

\begin{center}
\begin{tabular}{c|c|c|c|c}
\hline\hline
    Method & Minimum Toffoli~\cite{khattar2024riseconditionallycleanancillae} & Few ancilla~\cite{khattar2024riseconditionallycleanancillae} & Fewer ancilla (new) & Minimum ancilla (new) \\
    \hline
    Clean Ancilla & $n - 1$  & $\log_{2}^{*}n$ & 3 & 0 \\
    Dirty Ancilla & $0$ & $0$ & $0$ & $1$ \\
    Toffoli gates & $n - 1$ & $3n$ & $8n\pm\mathcal{O}(1)$ & $16n(\log_2n-2)+\mathcal{O}(1)$\\ 
    \hline\hline
\end{tabular}
\end{center}
\end{lemma}
\begin{proof}
Pad the $n$-qubit register $\ket{x}$ with one qubit $\ket{z}_\text{out}$. Using the quantum-classical integer subtraction~\cref{lem:qc_addition}, subtract $c$ from $\ket{x}\ket{z}_\text{out}\equiv\ket{2^nz+x}$. There are four cases:
\begin{align}
\ket{(2^{n}z+x-c) \;\mathrm{mod}\;2^{n+1}}
=
\begin{cases}
\ket{x-c}_{\text{in}}\ket{0}_\text{out},& z=0,\; c\le x < 2^{n},\\
\ket{2^n-(c-x)}_{\text{in}}\ket{1}_\text{out},& z=0,\; 0\le x<c, \\
\ket{x-c}_{\text{in}}\ket{1}_\text{out},& z=1,\; c\le x < 2^{n},\\
\ket{2^n-(c-x)}_{\text{in}}\ket{0}_\text{out},& z=1,\; 0\le x<c.
\end{cases}
\end{align}
Now perform quantum-classical integer addition to add $c$ to the $n$-qubit register $\ket{\cdot}_{\text{in}}$ to obtain $\ket{x}\ket{z\oplus(x<c)}$.
\end{proof}

\begin{lemma}[Integer Multiply--add]\label{lem:multiply-add}

Let $y,Y,z,Z$ be non-zero integers satisfying $y<Y,z<Z$. For any $\ket{y}\ket{z}$, there is a quantum circuit that computes $\ket{y}\ket{z}\rightarrow\ket{Zy+z}$ where $\ket{y}$ and $\ket{z}$ are $n_{Y}$ and $n_Z$ qubit registers respectively. This circuit uses the following resources.

\begin{center}
\begin{tabular}{c|c|c|c}
\hline\hline
    Method & Minimum Toffoli & Fewer ancilla &Minimum ancilla \\
    \hline
    Clean Ancilla & $2(n_Z+n_Y)-1$ & $n_Z+n_Y$ & 1\\    
    Dirty Ancilla & $0$ &$0$ & 0 \\
    Toffoli gates & $4n_Y(n_Z + 3)$ & $5n_Y(n_Z + 4)$ & $24 n_Y n_Z (\log_{2}n_Z-2) + \mathcal{O}(n_Y)$ \\ 
    \hline\hline
\end{tabular}
\end{center}
\end{lemma}
\begin{proof}
As $Z$ is a classical constant, we compute $\ket{y}\ket{z}\rightarrow\ket{Zy+z}$ using a shift-and-add algorithm. 
As $Y$ is also a classical constant, this remains reversible.
Let us use one clean ancilla qubit, and let $\ket{y}$ in binary be
$\ket{y}\ket{0}\ket{z}_\text{in}=\ket{y_{n_Y-1}}\cdots \ket{y_0}\ket{0}\ket{z}_\text{in}$. For loop index $j=0,\cdots,n_Y-1$, let $\ket{z_j}$ be the value of $\ket{z}_\text{in}$ at iteration $j$.
\begin{enumerate}
    \item Quantum-classical addition: Controlled on the $j^\text{th}$ bit $\ket{y_j}$, perform the addition $\ket{y_j}\ket{0}\ket{z}_\text{in}\rightarrow\ket{y_j}\ket{z+y_j2^j Z}_\text{in}=\ket{y_j}\ket{z_{j+1}}$, where 
    $\ket{z_{j+1}}\doteq\ket{z+y_j2^j Z}$ and has $n_Z+1+j$ qubits.
    \item Quantum-classical comparison: Compute the comparison $\ket{y_j}\ket{z_{j+1}}\rightarrow \ket{y_j\oplus(z_{j+1}\ge 2^jZ)}\ket{z_{j+1}}=\ket{0}\ket{z_{j+1}}$. This zeros out the $j^\text{th}$ bit $\ket{y_j}$ and allows it to be used as a clean qubit for the next iteration. Set $\ket{z}_\text{in}=\ket{z_{j+1}}$. 
\end{enumerate}
The final state is $\ket{0}\ket{Zy+z}$. Note that we always add or compare a number that is a multiple of $2^j$. Hence, the adders and comparisons do not affect the lowest $j$ bits, and are of fixed size $n_Z+1$, acting only on the highest bits. 
\end{proof}
Running~\cref{lem:multiply-add} in reverse furnishes in-place integer divider $\ket{x}\rightarrow\ket{\lfloor x / Z\rfloor}\ket{x\mod{Z}}$, which is reversible with the promise $\lfloor x / Z\rfloor<Y$ for classical constants $Y>0$ and $Z>0$.

\subsection{Removing registers storing redundant information}\label{sec:fewer_qubits_A}

In this section, we describe how to replace the registers $\ket{G}\ket{r}\ket{c}\ket{b=B}\ket{\sigma_0}\ket{\sigma_1}\ket{\sigma}\ket{\varsigma}$ consisting of $7+n_R+n_C$ qubits with just two qubits $\ket{\sigma}\ket{\varsigma}$, leading to the overall cost summarized in~\cref{tab:fewer_qubits_A}.
This replacement comes at a small additive Toffoli cost. 
Let us pad $R$ to a power of $2$ and slightly modify the indexing scheme
\begin{align}\label{eq:xo_indexing_scheme}
x_\text{o}(G,r,c)=
\begin{cases}
2^{n_R}c+r,&G=G_\text{SF},\\
2^{n_R}C+r,&G=G_{\text{D}_1},\\
2^{n_R}C+N_{\text{D}_1}+r,&G=G_{\text{Q}_1},
\end{cases}
\in\begin{cases}
[0,2^{n_R}C),&G=G_\text{SF},\\
[2^{n_R}C,2^{n_R}C+N_{\text{D}_1}),&G=G_{\text{D}_1},\\
[2^{n_R}C+N_{\text{D}_1},2^{n_R}C+N),&G=G_{\text{Q}_1}.
\end{cases}
\end{align}
Padding $R$ simplifies iterating over lookup tables address in some cases, but increases the number of addresses to $X_\text{o}=2^{n_R}C+N$.
When $G=G_\text{SF}$, this indexing scheme factors the $n_{X_\text{o}}$-qubit register $\ket{x_\text{o}}=\ket{0}^{\otimes (n_{X_\text{o}}-n_R-n_C)}\ket{c}\ket{r}$ into qubit registers storing the binary representation of the integers $r$ and $c$.
Later in~\cref{sec:fewer_qubits_E}, we will not pad $R$ to save some qubits.
\begin{table}[h]
    \centering
    \begin{tabular}{|c|c|c|c|c|c|}
    \hline \hline
    & Definitions & \multicolumn{4}{c|}{$X_\text{o}=N+2^{n_R}C,\quad  b_1\doteq n_{X_\text{o}}+b_\text{coeff}, \quad B'\doteq B+1,\quad b_2 = n_{B'}+b_\text{coeff}$}\\
    \hline
    & Subroutine & Cost & Persistent Ancilla & Temporary Ancilla & \# \\
    \hline
    \multirow{4}{*}{\rotatebox[origin=c]{0}{Outer \textsc{Prep}}}
    & Uniform State prep  & $4n_{X_\text{o}}$  & $n_{X_\text{o}} + 2$ &  $n_{X_\text{o}} - 2$ & 1\\
    & AliasSampling & $\left\lceil\frac{X_\text{o}}{2^{k_1}}\right\rceil + 2^{k_1} b_1$ & $b_1+ b_\text{coeff}$ & $n_{{X_\text{o}}/{2^{k_{1}}}}+2^{k_{1}}b_{1}$ & 1 \\
    & AliasSampling$^{\dagger}$ & $b_1 + \left\lceil \frac{X_\text{o}}{2^{k_5}}\right\rceil + 2^{k_5}$ &  & & 1 \\
    & Uniform State prep$^{\dagger}$ & $ 4n_{X_\text{o}}$ & & &  1 \\
    \hline
    \multirow{4}{*}{\rotatebox[origin=c]{0}{Inner \textsc{Prep}}}    & uniform($b$) & $4n_{B'}$ & $n_{B'} +2$ & $n_{B'}- 2 $ & 2 \\
    & for $x_{o}$ do: QROAM$(b)$  & $RC \left\lceil \frac{B'}{2^{k_2}}\right\rceil + 2^{k_2} b_2$ & $b_2 + b_\text{coeff}$ & $n_{RC \left\lceil {B'}/{2^{k_2}}\right\rceil}  + (2^{k_2}-1)b_2$ & 2 \\
    & (for $x_{o}$ do: QROAM$(b)$)$^{\dagger}$ & $b_2 + N + \left\lceil \frac {RC}{2^{k_4}}\right\rceil + 2^{k_4}B'$ & & & 2 \\
    & uniform($b$)$^{\dagger}$ & $4n_{B'}$ & & &  2 \\
    \hline 
    \multirow{2}{*}{\rotatebox[origin=c]{0}{\textsc{RPrep}}}
    & QROM($\vec{u}$) & $N+RB$ & $(N-1)b_{\text{rot}}$ &$n_{X_\text{o}}+  n_{RB}$& 2 \\
    & QROM($\vec{u}$)$^{\dagger}$ &  $R + B$ &  &  & 2 \\
    \hline
    \multirow{4}{*}{\rotatebox[origin=c]{0}{\textsc{Sel}}}
    & \textsc{Rot} & $2 (2b_{\text{rot}}-1) (N - 1)$ & $b_{\text{rot}}$ & $b_{\text{rot}}-1$ & 2 \\
    & $\textsc{GComp}$ 1 and 3& $12n_{X_\text{o}}$ & $2$ & $1+\log^*_2{n_{X_\text{o}}}$ & 2 \\
    & CSWAP & $2N$ &  &  & 2\\
    & Maj-control & $12n_{X_\text{o}}+n_{B'}$ &  & $\max(1+\log^*_2{n_{X_\text{o}}},n_{B'})$ & 2 \\
    \hline
     \multirow{2}{*}{\rotatebox[origin=c]{0}{\textsc{Ref}}}
    & $T_{2}$ & $n_{B'} + b_{k2} + 1$ & $ $ &  & 1 \\
    & Walk & $n_{X_\text{o}} +n_{B'}+ 2b_\text{coeff} + 2$ &  &  & 1 \\
    \hline \hline
    \end{tabular}
    \caption{\label{tab:fewer_qubits_A}
    Cost breakdown using resource estimates from Table III of \cite{low2025fast} (where we set $b_{k_1}=b_{k_2}=b_\text{coeff}$) combined with optimizations described in~\cref{sec:fewer_qubits_A}. 
    Note that the temporary ancilla of Outer and Inner $\textsc{Prep}$ can use all of the $(N-1)b_\text{rot}+n_{X_\text{o}}+n_{B'}$ persistent and temporary ancillae of $\textsc{RPrep}$, and we ensure that these temporary ancillae do not exceed this number.
    }
\end{table}
\begin{itemize}
    
    \item \textbf{Quantum-classical comparator with fewer qubits.} We will require the following circuit primitive in this section. For any $n$-bit integers $x,y$, and any bit $z$, there is a quantum circuit $\textsc{Comp}_{<y}$ that computes $\textsc{Comp}_{<y}\ket{x}\ket{z}=\ket{x}\ket{z\oplus(x<y)}$ using $3n$ Toffoli gates and $\log_2^*n$ clean qubits~\cite{khattar2024riseconditionallycleanancillae}. Note that the iterated logarithm has values 
    \begin{align}
        \log_2^*n=\begin{cases}
        0,&n\le 1,\\
        1,&n\in(1, 2],\\
        2,&n\in(2, 4],\\
        3,&n\in(4,16],\\
        4,&n\in(16, 65536],\\
        5,&n\in(65536, 2^{65536}].
        \end{cases}
    \end{align}

\item \textbf{Remove $(G,r,c)$ computation from $x_0$}: The $\ket{G},\ket{r},\ket{c}$ registers can be computed from the index $\ket{x_0}$, as is done in inner $\textsc{Prep}$ using unary iteration.
    The $\ket{G}$ register is used later in four steps:
    \begin{enumerate}
        \item With the spin registers $\ket{\sigma_0}\ket{\sigma_1}$ in the uniform superposition state over two qubits to control a multi-controlled-$\text{Not}$ gate on $\ket{\sigma}$.
        \item As one of the controls of the $\textsc{Maj}$ unitary.
        \item Again with the spin registers $\ket{\sigma_0}\ket{\sigma_1}$ to control a multi-controlled-$\text{Not}$ gate on $\ket{\sigma}$.
        \item As a control for the type of rotation is output by $\textsc{PRep}$ $\textsc{QROM}$.
    \end{enumerate}
    The $\ket{r}$ register is used to control the $\textsc{RPrep}$ QROM.
    We now describe how the registers $\ket{x_\text{o}}\ket{G}\ket{r}\ket{c}\ket{\sigma_0}\ket{\sigma_1}\ket{\sigma}\ket{\varsigma}$ can be replaced with just $\ket{x_\text{o}}\ket{\sigma}\ket{\varsigma}$.
    Consider the following computation on $\ket{x_\text{o}}$ in~\cref{fig:fullcircuit}.
    \begin{align}
    \ket{x_\text{o}}\mapsto \ket{x_\text{o}}\ket{G}\mapsto \ket{x_\text{o}}\ket{G}\otimes U_G,
    \end{align}
    where $U_G$ is some unitary that is controlled by $\ket{G}$.
    We can always replace such a sequence without explicitly computing $\ket{G}$ by using a sequence of arithmetic comparisons on $\ket{x_\text{o}}$ to realize the equivalent controlled unitary
    \begin{align}\nonumber
    &\textsc{GComp}\ket{x_\text{o}}\otimes I=\ket{x_\text{o}}\otimes \begin{cases}
    U_{G_{\mathrm{SF}}}, &x_\text{o}<2^{n_R}C,\\
    U_{G_{\mathrm{D}_1}}, &x_\text{o}\in[2^{n_R}C,2^{n_R}C+ N_{\mathrm{D}_1}), \\
    U_{G_{\mathrm{Q}_1}}, &x_\text{o}\ge2^{n_R}C+N_{\mathrm{D}_1},
    \end{cases}
    \\
    &\equiv\ket{x_\text{o}}\otimes \left[\left(
        \begin{cases}
    U_{G_{\mathrm{SF}}}U_{G_{\mathrm{D}_1}}^\dagger, &x_\text{o}< 2^{n_R}C,\\
    \one, & \text{otherwise}.
    \end{cases}
    \right)
    \cdot\left(
        \begin{cases}
    U_{G_{\mathrm{D}_1}}U_{G_{\mathrm{Q}_1}}^\dagger, &x_\text{o}< 2^{n_R}C+N_{\mathrm{D}_1}, \\
    \one, & \text{otherwise},
    \end{cases}\right)\cdot U_{G_{\mathrm{Q}_1}}\right],
    \end{align}
    where the last line lets us implement these comparisons sequentially using the same output qubit, with each computation or uncomputation costing at most $3\lceil\log_2 X_\text{o}\rceil$ Toffoli gates, and $\log^*_2n_{X_\text{o}}\le3$ clean qubits.
   Then the unitaries $U_G$ in steps 1 to 3 are defined as in Table \ref{tab:steps}, where we have replaced six qubits $\ket{G\sigma_0\sigma_1\sigma\varsigma}$ with two qubits $\ket{\sigma}\ket{\varsigma}$.

\begin{table}
\begin{tabular}{c|c|c|c|c}
\hline\hline
Step &Registers & $U_{G_{\mathrm{D}_1}}$ & $U_{G_{\mathrm{Q}_1}}$& $U_{G_{\mathrm{SF}}}$ \\
\hline
1 & $\ket{\sigma}\ket{\varsigma}$ & $\textsc{H}\otimes\textsc{H}$ &$\textsc{H}\otimes\textsc{H}$&$\textsc{H}\otimes\textsc{H}$ \\\hline
1-2 & \multicolumn{4}{c}{Perform system state spin-up-spin-down controlled swaps \& Givens rotations}
\\\hline
2 &$\ket{\varsigma}\ket{\psi_\downarrow}_0$& $(\one\otimes X)\cdot(\proj{0}\otimes \one-\proj{1}\otimes Z)$& $(\one\otimes X)\cdot(\proj{0}\otimes \one+\proj{1}\otimes Z)$&  $-\one\otimes Z$\\\hline
2-3 & \multicolumn{4}{c}{Perform system state spin-up-spin-down controlled swaps \& Givens rotations}
\\\hline
3 &$\ket{\sigma}\ket{\varsigma}$ & $\one\otimes\textsc{H}$ &$\one\otimes\textsc{H}$&$(\one\otimes\textsc{H})\cdot\textsc{Swap}$\\
\hline\hline
\end{tabular}
\caption{The main steps used for the block encoding of the Majorana operators.
Step 2 corresponds to $\textsc{GComp}_g$, and its implementation is illustrated in Fig.~\ref{fig:GComp}. The subscript $\ket{\psi_\downarrow}_0$ indexes the zeroth qubit of the fermion wavefunction register.
In between steps 1 and 2, and in between steps 2 and 3, the controlled swaps of the spin up and spin down components of the system state are performed, as well as the Givens rotations; see Fig.~\ref{fig:fullcircuit}.
\label{tab:steps}
}
\end{table}

\begin{figure}
    \centering
\includegraphics[width=0.5\linewidth]{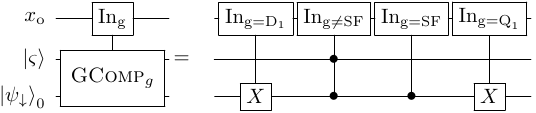}
\caption{The circuit diagram for the procedure shown in step 2 of Table \ref{tab:steps}.
This is the central step implementing the Majorana operators in~\crefpos{fig:fullcircuit}{right}, where the dependence of \textsc{RPrep} on $G$ is implicit on $r$ through $x_\text{o}$ in~\cref{eq:xo_indexing}.
The choice of applying the controlled $X$ at the beginning or end is used to apply the sign flip between ${\mathrm{D}_1}$ and ${\mathrm{Q}_1}$.
\label{fig:GComp}}
\end{figure}
In steps 1 and 3, $U_{G_{\mathrm{Q}_1}}=U_{G_{\mathrm{D}_1}}$ so we only need to compute and uncompute the comparison $\ket{x_\text{o}<2^{n_R}C}$. 
Hence, these steps use a total of $12\lceil\log_2 X_\text{o}\rceil$ Toffolis for the comparisons, and a negligible additional equivalent number for any controlled Hadamards. In step 2, computing the two comparisons and uncomputing them each sequentially costs $12\lceil\log_2 X_\text{o}\rceil$. We ignore the negligible number of gates needed to perform the controlled $U_G$ gates.

The $\ket{G}\ket{r}\ket{b}$ registers are used as the control over unary iteration to apply QROM to output the $N-1$ rotation angles $\vec{\theta}^{(G,r,b)}$.
The index set $(G,r,b)$ has $N+RB$ elements: The first $N$ correspond to the case $G\in\{G_{\text{D}_1},G_{\text{Q}_1}\}$ and $b=0$ and the remaining correspond to the case $G=G_\text{SF}$, where the QROMs are additionally controlled by $\ket{b}$.
Then, instead of computing $\ket{G}\ket{r}$ from $\ket{x_\text{o}}$, we do the following.
\begin{enumerate}
    \item From $\ket{x_\text{o}}$, compute in a single ancillary qubit the comparison $\ket{x_{\text{o}}<2^{n_R}C}$.
    \item Controlled on $\ket{x_{\text{o}}<2^{n_R}C}=\ket{1}$, perform unary iteration over the first $R$ addresses of the $n_R$ qubits.
    These correspond to the $G_\text{SF}$ components.
    \item Subtract $2^{n_R}C$ from $\ket{x_\text{o}}$ to obtain $\ket{x_\text{o}-2^{n_R}C}$. The high bit can be used as a control to perform unary iteration over the first $N$ addresses of the $n_R+n_C$ qubits. These correspond to the $G_{\text{D}_1}$ and $G_{\text{Q}_1}$ components.
\end{enumerate}
    \item \textbf{Removing the single-qubit $\ket{b=B}$ register} QROAM in $\textsc{RPrep}$ outputs a single qubit state flagging the condition $\ket{b=B}$. This is then used to control $\textsc{Maj}$: If $\ket{b=B}=\ket{1}$, then $\textsc{Maj}$ is not applied. 
    Note that $b=B$ is only possible when in the case $G=G_\text{SF}$. Hence after completing $\textsc{GCOMP}$ step 2, it suffices to use a single multi-controlled-$Z$ targeting $\ket{\psi_\downarrow}_0$, controlled on the condition $\ket{b}=\ket{B}$ to cancel $U_{G_\text{SF}}$. This uses at most $\lceil\log( B+1)\rceil$ Toffoli gates and the same number of temporary ancillary qubits. 

    \item \textbf{Reimplementing $\textsc{Maj}$-control} At this point, $\textsc{Maj}$-control is implemented using $\textsc{GComp}$ step 2 and multi-controlled $Z$ gate by $\ket{b}$. The application of these unitaries are further controlled by $\ket{\text{succ}\;x_\text{o}}\ket{\text{succ}\;b}$.
    The controls can be added to $U_{G_{\text{D}_1}}, U_{G_{\text{Q}_1}}, U_{G_\text{SF}}$, and the multi-controlled $Z$ gate using a negligible number of Toffoli gates that will be ignored, and a number of clean ancillary qubits that is subsumed by other steps.
\end{itemize}
\subsection{Reducing \textsc{RPrep} bits and using dirty qubits}\label{sec:fewer_qubits_B}
The qubit overhead is dominated by the $(N-1)b_{\text{rot}}$ qubits output by the $\textsc{QROM}$ in $\textsc{RPrep}$. This large overhead was chosen to minimize the Toffoli cost of $\textsc{RPrep}$. In this section, we show that for any integer $\lambda\in[1,(N-1)]$, we can reduce this qubit overhead to just $\lambda b_\text{rot}$. We also use QROAM with either clean or dirty qubits for inner and outer $\text{Prep}$, whichever gives the lowest Toffoli count. This leads to the overall cost summarized in~\cref{tab:fewer_qubits_B}. As $\textsc{Sel}$ dominates the cost of $\textsc{RPrep}$, much of this qubit reduction only increases the overall Toffoli cost by a small fraction. 
In the interests of minimizing qubit usage, we no longer uncompute the dirty qubit $\textsc{QRoam}$ by binary-to-unary conversion of the $\ket{b}$ register.
\begin{table}[h]
    \centering
    \resizebox{\textwidth}{!}{
    \begin{tabular}{|c|c|c|c|c|c|}
    \hline \hline
    & Subroutine & Cost & Persistent & Temporary & Dirty \\
    \hline
    \multirow{2}{*}{\rotatebox[origin=c]{90}{\parbox[c]{1cm}{\centering Outer \textsc{Prep}}}}  
        & Dirty AliasSampling & $2\big(\lceil\frac{N}{2^{k_1}}\rceil+\lceil\frac{R}{2^{k_1}}\rceil C\big) + (4(2^{k_1}-1)+1)b_1$ & $b_1+ b_\text{coeff}$ & $\max(n_{{X_\text{o}}/{2^{k_{1}}}},b_\text{coeff})$ &$(2^{k_{1}}-1)b_{1}$ \\
    & Dirty AliasSampling$^{\dagger}$ & $b_1 + 2\big(\lceil\frac{N}{2^{k_5}}\rceil+\lceil\frac{R}{2^{k_5}}\rceil C\big) + 2^{k_5+2}$ & & $\max(1+n_{X_\text{o}/2^{k_5}},b_{\text{coeff}})$ & $2^{k_{5}}-1$ \\
    \hline
    \multirow{2}{*}{\rotatebox[origin=c]{90}{\parbox[c]{1cm}{\centering Inner \textsc{Prep}}}} & Dirty QROAM$(b)$  & $2\lceil \frac{R}{2^{k_r}}\rceil\lceil \frac{C}{2^{k_c}}\rceil\lceil \frac{B'}{2^{k_b}}\rceil  + (4(2^{{k_r+k_c+k_b}}-1)+1) b_2$ & $b_2 + b_\text{coeff}$ & $\max(n_{X_\text{o}}+n_{B'}-(k_r+k_c+k_b),b_\text{coeff})$ & $(2^{k_r+k_c+k_b}-1)b_2$  \\
    & Dirty QROAM$(b)^{\dagger}$ & $b_2 + 2\lceil\frac{R}{2^{k^\prime_r}}\rceil\lceil \frac{C}{2^{k^\prime_c}}\rceil\lceil \frac{B'}{2^{k^\prime_b}}\rceil + 2^{k^\prime_r+k^\prime_b+k^\prime_c+2}$ &&$\max(1+n_{X_\text{o}}+n_{B'}-(k^\prime_r+k^\prime_b+k^\prime_c),b_\text{coeff})$  & $2^{k^\prime_r+k^\prime_b+k^\prime_c}-1$  \\
    \hline 
    \multirow{2}{*}{\textsc{RPrep}}
    & QROM($\vec{u}$) & $2\big(\big\lceil\frac{N-1}{\lambda}\big\rceil (N+RB)+R+B\big)$ & $\lambda b_{\text{rot}}$ &$n_{X_\text{o}}+  n_{RB}$& \\
    \cline{2-6}
    & \begin{tabular}{c}Dirty\\ QROAM $(\vec{u})$\end{tabular} & $2(\big\lceil\frac{N-1}{\lambda}\big\rceil+1) \big(\lceil\frac{N}{2^{k_6}}\rceil+\lceil\frac{R}{2^{k_6}}\rceil B+2(2^{k_6}-1)\lambda b_\text{rot}\big)$ & $\lambda b_{\text{rot}}$ &$n_{X_\text{o}}+  n_{B'}-k_6$ & $(2^{k_6}-1)\lambda b_\text{rot}$\\
    \hline \hline
    \end{tabular}
    }
    \caption{\label{tab:fewer_qubits_B}
    Cost breakdown using the optimizations described in~\cref{sec:fewer_qubits_B} to reduce qubit costs as compared to~\cref{tab:fewer_qubits_A}, where the integer $\lambda\in[1,N-1]$ is a free parameter. 
    Note that the cost of $\textsc{QROM}(\vec{u})$ uncomputation is included in the modified $\textsc{QROM}(\vec{u})$ calls.
    The number of dirty qubits available for each step is the total number of qubits minus the temporary and persistent ancilla.
    For the DFTHC factorizations in this work, total temporary ancillae is $\max\{b_\text{rot}, n_{X_\text{o}}+n_{B'}-k_6\}$, which is rarely more than $b_\text{rot}$.}
\end{table}
\begin{itemize}

    \item \textbf{QROAM with dirty qubits}: When $\lambda b_\text{rot}$ is smaller than the number of system qubits $2N$, it can become advantageous to use dirty qubits to reduce Toffoli counts in QROAM~\cite{Low2024tradingtgatesdirty,berry2019qubitization}. 
    The dirty qubit version of QROAM has higher constant factors: Given $d$ addresses $b$ output bits and any $K=2^k$ that is a power of $2$, computing an arbitrary look-up table costs $2\lceil d/K\rceil+4b(K-1)$ Toffoli gates, $b(K-1)$ dirty ancilla qubits, and $\lceil\log_2(d/K)\rceil$ clean ancilla qubits. 
    Uncomputing costs $2\lceil d/K\rceil+4K$ Toffoli gates, $K-1$ dirty ancilla qubits, and $\lceil\log_2(d/K)\rceil+1$ clean ancilla qubits.
    However the number of dirty qubits available can be significant as it includes all other registers, including the $2N$ qubits used to store the fermionic wavefunction.

    Moreover, further optimizations are possible when combined with the programmable gate array.
    Here we describe an implementation of dirty qubit QROAM with an amortized complexity almost that of clean qubit QROAM.
    Let us define a lookup table that maps the address $x$ to the $b$-bit string $\theta_x$, and review the dirty QROAM implementation of this lookup table.
    Let $\ket{\theta_x}_0$ be the register storing output bit-string $\theta_x$.
    The address $\ket{x}=\ket{h}\ket{l}$ is split into a high and low component.
    For some integer $\lambda$, there are dirty qubit registers $\ket{r_1}_1\cdots\ket{r_{\lambda-1}}_{\lambda-1}$, where $\ket{r_j}$ are computational basis states.
    Let $\textsc{S}$ be a swapping operation that, controlled by $\ket{l}$, moves $\ket{\cdots}_0$ to register $\ket{\cdots}_l$ and applies some arbitrary permutation on the other registers.
    Let $\textsc{T}$ be an operator which controlled by $\ket{h}$, XORs the bit-string $\theta_{h\lambda+l}$ into register $\ket{\cdot}_l$ for all $l=0,\cdots,\lambda-1$.
    Then dirty QROAM is implemented as the sequence
    \begin{align}\nonumber
   & \ket{h}\ket{l}\ket{0}^{\otimes b}_0\ket{r_1}_1\cdots\ket{r_l}_l\cdots\ket{r_{\lambda-1}}_{\lambda-1}
    \\\nonumber
    &\rightarrow_{\text{Hadamard}} \ket{h}\ket{l}\ket{+}^{\otimes b}_0\ket{r_1}_1\cdots
    &\rightarrow _\textsc{S}&\ket{h}\ket{l}\cdots\ket{+}^{\otimes  b}_l\cdots
    \\\nonumber
    &\rightarrow _\textsc{T}\ket{h}\ket{l}\cdots\ket{+}^{\otimes  b}_l\cdots       &\rightarrow _{\textsc{S}^\dagger}&\ket{h}\ket{l}\ket{+}^{\otimes b}_0\ket{r_1\oplus\cdots}_1\cdots
    \\\nonumber
    &\rightarrow_{\text{Hadamard}} \ket{h}\ket{l}\ket{0}^{\otimes b}_0\ket{r_1\oplus\cdots}_1\cdots
    &\rightarrow _\textsc{S}&\ket{h}\ket{l}\cdots\ket{0}^{\otimes  b}_l\cdots\\   
    &\rightarrow _\textsc{T}\ket{h}\ket{l}\cdots\ket{\theta_{h\lambda+l}}_l\cdots   
    &\rightarrow _{\textsc{S}^\dagger}&\ket{h}\ket{l}\ket{\theta_{h\lambda+l}}_0\ket{r_1}_1\cdots \ket{r_{\lambda-1}}_{\lambda-1}.
    \end{align}
    We see that this costs $2$ blocks of the form $\textsc{S}^\dagger\textsc{T}\textsc{S}$.

    Let us instead consider a sequence of QROM applications where for $j=0,1,\cdots,J$ the $j^\text{th}$ QROM is controlled by $\ket{x}$ for $x\in[d]$ to output $b$ bits $\theta_{j,x}$ that in turn control a $j^\text{th}$ unitary $\textsc{CU}_j=\sum_{y}\proj{y}_0\otimes U_y$, where $U_y$ is applied to a register $\ket{\psi}$ that does not overlap any of those used by $\textsc{S}^\dagger\textsc{T}_j\textsc{S}$.
    In this case, the total number of $\textsc{S}\textsc{T}\textsc{S}^\dagger$ blocks can be reduced to just $J+1$. Let us define $\vec{\theta}_j=\bigotimes_{l=0}^{\lambda-1}\theta_{j,h\lambda+l}$ and let $\textsc{T}_j$ controlled on $\ket{h}$ output $\vec{\theta}_{j}\oplus\vec{\theta}_{j-1}$, where we define $\vec{\theta}_{-1}=\vec{\theta}_J\doteq (0,\cdots,0)$ and $\textsc{CU}_{J}=\one$.
    Then consider the sequence where we apply $\prod_{j=0}^{J}\textsc{CU}_j\textsc{S}^\dagger\textsc{T}_j\textsc{S}$.
    \begin{align}
    \nonumber
    &\ket{h}\ket{l}\ket{0}^{\otimes b}_0\ket{r_1}_1\cdots\ket{r_l}_l\cdots\ket{r_{\lambda-1}}_{\lambda-1}\ket{\psi}
    \\\nonumber
    &\rightarrow _\textsc{S}\ket{h}\ket{l}\cdots\ket{0}^{\otimes  b}_l\cdots \ket{\psi}
    &\rightarrow _{\textsc{T}_0}&\ket{h}\ket{l}\cdots\ket{\theta_{0,h\lambda+l}}_l\cdots \ket{\psi}
    \\\nonumber
    &\rightarrow _{\textsc{S}^\dagger}\ket{h}\ket{l}\ket{\theta_{0,h\lambda+l}}_0\ket{r_1\oplus\cdots}_1\cdots\ket{\psi}
    &\rightarrow _{\textsc{CU}_0}&\ket{h}\ket{l}\ket{\theta_{0,h\lambda+l}}_0\ket{r_1\oplus\cdots}_1\cdots U_{\theta_{0,h\lambda+l}}\ket{\psi}
    \\\nonumber
    &\rightarrow _\textsc{S}\ket{h}\ket{l}\cdots\ket{\theta_{0,h\lambda+l}}_l\cdots U_{\theta_{0,h\lambda+l}}\ket{\psi}
    &\rightarrow _{\textsc{T}_1}&\ket{h}\ket{l}\cdots\ket{\theta_{1,h\lambda+l}}_l\cdots U_{\theta_{0,h\lambda+l}}\ket{\psi}
    \\\nonumber
    &\rightarrow _{\textsc{S}^\dagger}\ket{h}\ket{l}\ket{\theta_{1,h\lambda+l}}_0\ket{r_1\oplus\cdots}_1\cdots U_{\theta_{0,h\lambda+l}}\ket{\psi}
    &\rightarrow _{\textsc{CU}_1}&\ket{h}\ket{l}\ket{\theta_{1,h\lambda+l}}_0\ket{r_1\oplus\cdots}_1\cdots U_{\theta_{1,h\lambda+l}}U_{\theta_{0,h\lambda+l}}\ket{\psi}
    \\\nonumber
    \vdots
    \\
    &\rightarrow \ket{h}\ket{l}\ket{0}^{\otimes b}_0\ket{r_1}_1\cdots \ket{r_{\lambda-1}}_{\lambda-1} \prod_{j=0}^{J-1}U_{\theta_{j,h\lambda+l}}\ket{\psi}.
    \end{align}
    In this situation, the cost of dirty qubit QROAM is $(J+1)(\lceil d/K\rceil+2b(K-1))$ Toffoli gates, and no uncomputation is required by definition as the $0^\text{th}$ register is already reset to $\ket{0}_0$ after the final step.
    We note concurrent work~\cite{Motlagh2026QROM} that shows a similar optimization, and further improves upon the $J+1$ case.
    Later, when $\textsc{CU}_j$ is a Givens rotation, it is useful to perform measurement-based uncomputation of the $0^\text{th}$ register in the $X$ basis to free up clean ancilla qubits as $\textsc{CU}_j$ progresses.
    In this case, the $\text{T}_j$ outputs $\vec{\theta}_j$ instead of $\vec{\theta}_j\oplus\vec{\theta}_{j-1}$, and the measurement only introduces a phase flip dependent on $\ket{h}\ket{l}$ that can be fixed by one more much cheaper $b$-independent lookup at the end of the entire sequence.

    When using the $2N$ system qubits as dirty qubits in the programmable gate array for $\textsc{RPrep}$, we use the observation that each $\ket{\vec{\theta}_j}$ only controls rotations applied to at most $\lambda+1$ system qubits at a time. The system qubits targeted change for each $\textsc{CU}_j$. Hence, we can further subdivide $J$ into subblocks of $J'<J$, where in each subblock, we perform the QROAM that costs at most $(J'+1)(\lceil d/K\rceil+2b(K-1))$ Toffoli gates and restores the dirty qubits to their original state. This QROAM excludes from the dirty qubit register the $J'\lambda+1$ system qubits affected by the sequence of the $J'$ successive $\textsc{CU}_j$s. However, this correction to Toffoli cost is small when $\lambda$ is small, which is precisely the regime where we use dirty qubit QROAM. 
    In our physical resource estimates, this correction is not needed due to either 1) Many wavefunction qubits are in cold storage and so may not be used as dirty qubits, or 2) Clean qubit QROAM provides better resource estimates.
    
    To adjust for this, we require the number of dirty qubits $b(K-1)$ used to be no more than $2N-J'\lambda-1$ in each block of $J'$ $\textsc{CU}_j$s, and also need the sum over $J'\lambda$ for the steps to be equal to $N-1$.    
    As an example, if we aim for $K=4$ to provide a significant Toffoli reduction from dirty QROAM, then for $N=76$, $(R,B,C)=(15,17,19)$, $b_{\rm rot}=15$, then $d = N+RB = 931$, and choosing $\lambda=3$ gives $b(K-1)< 2N$.
    If we require $b(K-1)\le 2N-J'\lambda-1$, then the maximum $J'$ value is only 5.
    We would normally have $J=25$, so choosing $J'=5$ means we have a factor in the cost of $5(J'+1)=5(5+1)=30$ rather than $J+1=26$, increasing the cost by $15\%$.
    In this example, it is advantageous to take $K=2$ and $J'=25$, which reverts to the original method.
    
    More generally, can choose a sequence of values for block $l$ of $J'$ QROAMs such that
    \begin{align}
        \sum_l J'_l\lambda_l &\ge N-1 , \\
        b_l(K_l-1)&\le 2N-J'_l\lambda_l-1 , \\
        b_l &= \lambda_l b_{\rm rot} ,
    \end{align}
    with Toffoli cost for the QROAMs
    \begin{equation}
        \sum_l (J'_l+1)\left(\left\lceil \frac d K_l \right\rceil+2b_l(K_l-1)\right).
    \end{equation}
    Here $\lambda_l\le \lambda$, since $\lambda$ is a parameter chosen to reduce the qubit usage.
    In cases where $\lambda_l<\lambda$, we could alternatively allow
    \begin{align}
        b_l(K_l-1)&\le 2N-J'_l\lambda_l-1+(\lambda-\lambda_l)b_{\rm rot},
    \end{align}
    if we are considering an upper bound on clean qubit usage.
    The dependence on $l$ is most useful for $J'$, as the last block may not need $J'$ as large.

    \item \textbf{QROAM controlled by non-contiguous registers:} Given an $n$-bit address register $\ket{x}$, QROAM splits $\ket{x}=\ket{h}\ket{l}$, where choosing the low register indices $l$ to be base $2$ allows this split to be performed without using integer arithmetic. The high part $\ket{h}$, which is the most significant bits of the integer $x$, controls unary iteration, and the low part $\ket{l}$ controls a swapping operation.

    In the situation where $\ket{x}=\ket{y}\ket{z}$ for some integers $y\in[0,Y), z\in[0,Z)$, and unary iteration is performed only over indices $y,z$ that are not a power of $2$, the most convenient way to split $\ket{x}$ into a high or low part is to let the low part to be any $k_2\le n_Z$ qubits of $\ket{z}$. For example, unary iteration is controlled by the entirety of $\ket{y}$, and the swapping operation is never controlled by any part of $\ket{y}$.
    However, this limits the maximum amount of qubit-Toffoli tradeoff to be at most $Z$ copies as the number of addresses that unary iteration is performed over is then $Y\lceil Z/2^{k_2}\rceil$.
    To overcome this limitation, let non-zero integers $k_1\le n_Y$ and $k_2<n_Y$. Then split $\ket{y}=\ket{y_{h}}\ket{y_{l}}$ and $\ket{z}=\ket{z_{h}}\ket{z_{l}}$, where $\ket{y_{l}}$ the $k_1$-bit low part of $\ket{y}$ and similarly for $\ket{z}$ with $k_2$. 
    Then we choose the low part of $\ket{x}$ to be $\ket{l}=\ket{y_{l}}\ket{z_{l}}$, and the high part to be $\ket{h}=\ket{y_{h}}\ket{z_{h}}$. The number of addresses that unary iteration is performed over is then $\lceil Y/2^{k_1}\rceil\lceil Z/2^{k_2}\rceil$.
    This generalizes to unary iteration any composite integer $\ket{x}=\bigotimes_j\ket{x_j}$, where $x_j\in[0,X_j)$. For any choice $k_j\in[0,n_{X_j}]$, the number of unary iteration addresses is $\prod_j\lceil X_j/2^{k_j}\rceil$.
We apply this to Inner \textsc{Prep} QROAM, which performs a outer loop of unary iteration over $\ket{x_\text{o}}=\ket{c}\ket{r}$ for $c\in[0,C),r\in[0,R)$, or a total of $RC$ addresses. It then performs an inner loop over $\ket{b}$ for $b\in[0,B')$.  
    For any non-negative $k_1,k_2,k_3$ QROAM iterates over only $\lceil R/{2^{k_1}}\rceil\lceil C/{2^{k_2}}\rceil\lceil B'/2^{k_3}\rceil$ addresses.

\item \textbf{\textsc{RPrep} QROAM}: 
The lookup table $\textsc{RPrep}$ performs unary iteration over a total of $N+RB$ addresses in $\ket{x_\text{o}}\ket{b}$. There are two cases: First, when $x_\text{o}<2^{n_R}C$, unary iteration is performed over $\ket{x_\text{o}}\ket{b}$ for $x_\text{o}=2^{n_R}c+r$ with $c=0$, $r\in[0,R)$, and $b\in[0,B)$. Second, when $x_\text{o}\ge2^{n_R}C$, unary iteration is performed over $\ket{x_\text{o}}\ket{b}=\ket{x_\text{o}}\ket{0}^{\otimes n_{B'}}$ for $x_\text{o}\in[2^{n_R}C,2^{n_R}C+N)$.
This is straightforward in the case of QROM and costs only $N+RB+\mathcal{O}(1)$ Toffoli gates, but making the most efficient use of dirty-qubit QROAM requires a simple split between high and low address bits.
    The simplest approach to realize QROAM in $\textsc{RPrep}$ is to compute a contiguous address $f_{x_\text{o},b} \in[0, N+RB)$ in an ancilla register and perform unary iteration over it, but this increases the qubit overhead and requires an expensive multiply-add. Instead, we can compute the flattened index in-place using the fact that $R=2^{n_R}$ is padded to a power of $2$, allowing us to cleanly separate $\ket{x_\text{o}}$ into high bits $\ket{x_{\text{o},h}}$ (representing $c$) and low bits $\ket{x_{\text{o},l}}$ (representing $r$, taking exactly $n_R$ qubits).
    Although this makes QROAM slightly less efficient as we may be iterating over padded values of $R\le r< 2^{n_R}$ that index no rotations, this provides a simple formula for the \textsc{Select}-\textsc{Swap} Toffoli-qubit trade-off.
    In~\cref{sec:fewer_qubits_E}, we evaluate the case where $R$ is not padded.
    
    Let us compute in a single ancilla qubit the comparison $\ket{a}=\ket{x_\text{o}\ge 2^{n_R}C}$ to obtain
    \begin{align}
    \begin{cases}
    \ket{0}_a\ket{x_{\text{o},h}}\ket{x_{\text{o},l}}\ket{b},&x_\text{o}< 2^{n_R}C,
    \\
    \ket{1}_a\ket{x_\text{o}}\ket{0}^{\otimes n_{B'}},&x_\text{o}\ge 2^{n_R}C.
    \end{cases}
    \end{align}
    Controlled on $\ket{1}_a$, we add the classical constant $2^{n_R}(B-C)$ to $\ket{x_\text{o}}$. This effectively shifts the index for the second case to start immediately after the first case:
    \begin{align}
    \begin{cases}
    \ket{0}_a\ket{x_{\text{o},h}}\ket{x_{\text{o},l}}\ket{b},&x_\text{o}< 2^{n_R}C,
    \\
    \ket{1}_a\ket{x_\text{o}+2^{n_R}(B-C)}\ket{0}^{\otimes n_{B'}},&x_\text{o}\ge 2^{n_R}C.
    \end{cases}
    \end{align}
    Then, controlled on $\ket{0}_a$, we swap the registers $\ket{x_{\text{o},h}}$ and $\ket{b}$, padding with leading zeros if they are of different sizes. Because $x_{\text{o},h}=c$, this moves $c$ into the $\ket{b}$ register and $b$ into the high bits of $\ket{x_\text{o}}$:
    \begin{align}
    \begin{cases}
    \ket{0}_a\ket{b}\ket{x_{\text{o},l}}\ket{c},&x_\text{o}< 2^{n_R}C,
    \\
    \ket{1}_a\ket{x_\text{o}+2^{n_R}(B-C)}\ket{0}^{\otimes n_{B'}},&x_\text{o}\ge 2^{n_R}C.
    \end{cases}
    \end{align}
    At this point, the register $\ket{x_\text{o}}$ completely contains the flattened, contiguous address. 
    When $a=0$, its value is $b \cdot 2^{n_R} + r$, which uniquely spans $[0, 2^{n_R}B)$ since $b \in[0, B)$ and $r \in[0, R)$. 
    When $a=1$, the original $x_\text{o} \in[2^{n_R}C, 2^{n_R}C+N)$, so its new value spans $[2^{n_R}B, 2^{n_R}B+N)$. The total range is perfectly contiguous over $[0, 2^{n_R}B+N)$. Unary iteration can then be performed directly on this single, flattened $\ket{x_\text{o}}$ register.
    Overall, this uses one quantum-classical comparison and controlled quantum-classical addition costing $\mathcal{O}(n_{X_\text{o}})$ Toffolis, and one controlled swap using no more than $n_{B'}$ Toffolis.
    As we have at least $b_\text{rot}> n_{X_\text{o}},n_{B'}$ clean ancilla qubits available, this overall costs at most $2(n_{X_\text{o}}-1)+n_{B'}$ Toffoli gates.

\item \textbf{Arithmetic-free QROAM}: 
The approach for the QROAM over the $x_\text{o}$ and $B$ registers can be modified to avoid the need for arithmetic, and also to ensure the iteration is over a range of only $N+RB$ values instead of $N+2^{n_R}B$. The tradeoff that the minimum Toffoli count from \textsc{Select}-\textsc{Swap} is limited to a fewer number of address qubits.
First we explain the method for QROM, then the modifications for dirty QROAM.

The unary iteration for QROM can start with the range $x_\text{o}\in[2^{n_R}C,2^{n_R}C+N)$.
This is also used to flip an ancilla qubit in the case where $x_\text{o}\ge 2^{n_R}C$, similar to the qubit containing the result of the inequality test above.
The QROM starting from $2^{n_R}C$ has a cost slightly larger than $N-2$ Toffolis, but the extra cost is less than for the inequality test.
Controlled by that ancilla qubit, now perform the unary iteration over the registers containing $r$ (the low bits of $x_\text{o}$) and $b$.
That will have a cost of $RB-1$ Toffolis, for a total of $N+RB$ (omitting single Toffoli costs).

For the dirty QROAM there are two modifications.
First, for the QROAM some of the least-significant bits of $x_\text{o}$ are omitted in the unary iteration.
For $k_6$ qubits omitted this results in the unary iteration cost being
\begin{equation}
    \lceil N/2^{k_6} \rceil + \lceil R/2^{k_6} \rceil B .
\end{equation}
The cost of the controlled swap in the QROAM is unchanged.

The second modification is that there are $J+1$ rounds of the QROAM.
In the first round the ancilla qubit needs to be set, but in rounds after that it need not be set, and can be used as a control.
In the last round the ancilla qubit should be reset.
That may be achieved by performing the controlled iteration over the registers containing $r$ and $b$ first, then perform the QROAM over the range $x_\text{o}\in[2^{n_R}C,2^{n_R}C+N)$ using the result to reset the ancilla qubit.

A similar approach may also be used to reduce the cost of the alias sampling for the outer state preparation.
The unary iteration need not be over all values of $x_\text{o}$ up to $2^{n_R}C+N$.
It can be over the $c$ and $r$ subregisters, with $c\in[0,C)$ and $r\in[0,R)$, then from $2^{n_R}C$ to $N$.
When $k_1$ least-significant qubits of $r$ are omitted, then the unary iteration cost is
\begin{equation}
    \lceil N/2^{k_1} \rceil + \lceil R/2^{k_1} \rceil C .
\end{equation}
Again the controlled swap cost is unchanged.
The cost of the initial preparation of an equal superposition state for the alias sampling is slightly increased, because the inequality test needs to eliminate cases where $c<C$ and $r\ge R$.

\end{itemize}

\subsection{Pure state preparation instead of alias sampling}\label{sec:fewer_qubits_C}
Storing a pure state $\sum_{j=0}^{X-1}c_j\ket{j}$ requires only $n_X$ qubits.
In contrast, alias sampling of the same state to $b_\text{coeff}$ bits of precision uses a large number of $2(n_X+b_\text{coeff})$ persistent bits.
In the limit of very few ancillae, it is beneficial to replace alias sampling in both Inner and Outer $\textsc{Prep}$ with alternate methods that prepare the pure state.
The dominant cost in these methods is from QROAM, and our resource estimates select the most efficient of either the clean or dirty versions. As we are focused on minimizing qubit counts, the overall cost summarized in~\cref{tab:fewer_qubits_C} reports that of dirty QROAM.
\begin{table}[h]
    \centering
    \resizebox{\textwidth}{!}{
    \begin{tabular}{|cc|c|c|c|c|c|c|}
    \hline \hline
    && Subroutine & Cost & Persistent & Temporary & Dirty & \# \\
    \hline
    \multirow{3}{*}{\rotatebox[origin=c]{0}{Outer}}
        &\multirow{3}{*}{\rotatebox[origin=c]{0}{\textsc{Prep}}}
        & QROAM & $\sum_{l=0}^{n_{X_\text{o}}-1}\left\lceil\frac{f(X_\text{o},l)}{2^{k_{l,1}}}\right\rceil + 2(2^{k_{l,1}}-1)b_\text{rot}$ & $n_{X_\text{o}}$ & $n_{X_\text{o}}-k_1-1$ &$(2^{k_{1}}-1)b_\text{rot}$  & 2\\
        && QROAM$^\dagger$ & $\left\lceil\frac{f(X_\text{o})}{2^{k_{5}}}\right\rceil + 2(2^{k_{5}}-1)b_\text{rot}$ &  &   &$(2^{k_{5}}-1)b_\text{rot}$ & 2\\
    && Adders & $n_{X_\text{o}}(b_\text{rot}-1)$ & & $b_{\text{rot}}-1$ & & 2 \\
    \hline
    \multirow{3}{*}{\rotatebox[origin=c]{0}{Inner}}
        &\multirow{3}{*}{\rotatebox[origin=c]{0}{\textsc{Prep}}}
        & QROAM & $\sum_{l=0}^{n_{B'}-1}\lceil\frac{R}{2^{k_r}}\rceil\lceil \frac{C}{2^{k_c}}\rceil\left\lceil\frac{f(B',l)}{2^{k_{l,b}}}\right\rceil + 2(2^{k_r+k_c+k_{l,b}}-1)b_\text{rot}$ & $n_{B'}$ & $n_{X_\text{o}}+n_{B'}-(k_r+k_c+k_{l,b})-1$ &$(2^{k_r+k_c+k_{l,b}}-1)b_\text{rot}$ & 4\\
        && QROAM$^\dagger$ & $\lceil\frac{R}{2^{k^\prime_r}}\rceil\lceil \frac{C}{2^{k^\prime_c}}\rceil\left\lceil\frac{f(B')}{2^{k_4}}\right\rceil + 2(2^{k^\prime_r+k^\prime_c+k_4}-1)b_\text{rot}$ &  &  &$(2^{k^\prime_r+k^\prime_c+k_{4}}-1)b_\text{rot}$ & 4 \\
    && Adders & $n_{B'}(b_\text{rot}-1)$ & & $b_{\text{rot}}-1$ &  & 4 \\
    \hline \hline
    \end{tabular}
    }
    \caption{\label{tab:fewer_qubits_C}
    Cost breakdown using the optimizations described in~\cref{sec:fewer_qubits_C} compared to~\cref{tab:fewer_qubits_B}, where the integer $\lambda\in[1,N-1]$ is a free parameter, $k_1=\max_l(k_{l,1},k_5)$, and $k_2=\max_l(k_{l,2},k_4)$.
    The number of dirty qubits available for each step is the total number of qubits minus the temporary and persistent ancilla.
    For the DFTHC factorizations in this work, total temporary ancillae is $\max\{b_\text{rot}, n_{X_\text{o}}+n_{B'}-k_6\}$. Note that the $b_\text{rot}$ output bits of QROAM are already contained in the $\lambda b_{\text{Rot}}$ persistent qubits of  $\textsc{QROM}(\vec{u})$ in~\cref{tab:fewer_qubits_B}, and the $b_\text{rot}$ bits storing the phase gradient state that go into the adder are already accounted for by $\textsc{Rot}$ in~\cref{tab:fewer_qubits_A}.
    }
\end{table}

\begin{itemize}
    \item \textbf{Outer $\textsc{Prep}$ without alias sampling}: Instead of preparing the arbitrary superposition
    \begin{align}
\ket{\psi_\text{alias}}=\sum_{j=0}^{X_\text{o}-1}|c_j|\ket{j}\ket{\mathrm{sign}[c_j]}_s\ket{\mathrm{garb}_j},    
    \end{align} 
    using alias sampling, for some real coefficients $c_j$, and some arbitrary garbage state we can use a variant~\cite{Low2024tradingtgatesdirty} of Shende-Bullock-Markov state preparation to prepare $\ket{\psi}=\sum_{j=0}^{X_\text{o}-1}c_j\ket{j}$. Any such quantum state of dimension $2^{n_{X_\text{o}}}$, with only real amplitudes can be synthesized in $n_{X_\text{o}}$ steps over a loop $l=0,\cdots,n_{X_\text{o}}-1$, where at step $l$, an arbitrary single-qubit rotation is applied on the $l^{\text{th}}$ qubit and is controlled by the $0$-th to $(l-1)$-th qubits. This prepares
    \begin{align}
\sum_{j=0}^{X_\text{o}-1}|c_j|\ket{j}=\prod_{l=0}^{n_{X_{\text{o}}}-1}U_{\vec{\theta}_l}\ket{0},\quad U_{\vec{\theta}_l}\doteq\sum_{j=0}^{2^{l}-1}\proj{j}\otimes e^{-i\theta_{l,j}Y},
    \end{align}
    for some angles $\theta_{l,j}\in[0,\pi/2]$. When $X_\text{o}$ is not a power of $2$, the final controlled-rotation only needs to iterate over the first $j\in[0,X_\text{o}-2^{n_{X_\text{o}}-1})$ addresses and all other rotations only iterate over the first $\lceil X_\text{o}/2^{n_{X_\text{o}}-l}\rceil$ addresses.
    Let the number of addresses to iterate over at iteration $l$ be
    \begin{align}\label{eq:sbm_addresses}
    f(X_\text{o},l)=
    \begin{cases}
        f(X_\text{o}),& l=n_{X_\text{o}}-1\\
        \lceil X_\text{o}/2^{n_{X_\text{o}}-l}\rceil,& l=0,\cdots, n_{X_\text{o}}-2,
    \end{cases}
    \quad f(X_\text{o})=X_\text{o}-2^{n_{X_\text{o}}-1}.
    \end{align}
    If any $c_j$ are negative, we can either widen the domain $\theta_{l,j}\in[0,\pi]$ or apply after this loop a diagonal unitary to introduce signs and obtain the desired pure state
   \begin{align}
\sum_{j=0}^{X_\text{o}-1}c_j\ket{j}=\left(\sum_{j=0}^{X_\text{o}-1} \mathrm{sign}[c_j]\proj{j}\right)\cdot \sum_{j=0}^{X_\text{o}-1}|c_j|\ket{j}.
    \end{align}

    Overall, this circuit is similar to that used in $\textsc{RPrep}$ programmable gate array for $\textsc{RPrep}$, except that the number of control qubits changes as a function of the loop index, and that we use Clifford gates to convert a $Z$ rotation by the phase gradient method to a $Y$ rotation.

    For simplicity in our resource estimates, we assume that the number of rotation bits needed is also $b_\text{rot}$, rather than the smaller $b_\text{coeff}$ used in alias sampling. We also choose $\theta_{l,j}\in[0,\pi]$ so that the last diagonal unitary is not required.
    The non-Clifford gate cost of this procedure is then the sum of 1) $n_{X_\text{o}}$ adders; 2) QROAM over $\min(2^{l}, X_\text{o}-2^{n_{X_\text{o}}-l}-1)$ addresses to output $b_\text{coeff}$ for $l=0,\cdots,n_{X_\text{o}}-1$; 3) Uncompute QROAM over at most $2^{n_{X_\text{o}}-1}$ addresses -- when doing the forward preparation, the QROAM uncomputes can all be bundled into the diagonal unitary for the signs, but this trick does not  work in reverse.

    Pure state preparation removes the need for the register $\ket{\text{succ}\;x_\text{o}}$ and preparing an initial dimension $X_\text{o}$ uniform superposition.
    When used in the context of block-encoding a linear combination of unitaries then instead of the pattern $\bra{\psi_\text{alias}}\textsc{Select}\otimes Z_s\ket{\psi_\text{alias}}$, we account for the removed sign qubit $\ket{\mathrm{sign}[c_j]}$ by defining two state preparation unitaries: $U\ket{0}=\ket{\psi}=\sum_{j=0}^{X_\text{o}-1}|c_j|\ket{j}$ and $U'\ket{0}=\ket{\psi'}=\sum_{j=0}^{X_\text{o}-1}c_j\ket{j}$. Then the same linear combination is block-encoded by $\textsc{Be}=U^\dagger\cdot\textsc{Select}\cdot U^\prime$. Note that $\textsc{Be}$ is still self-inverse as it is equivalent to having a separate diagonal operation performing the signs, and that commutes through SELECT.

    \item \textbf{Inner $\textsc{Prep}$ without alias sampling}: With the above optimizations, QROAM for Inner $\textsc{Prep}$ only needs to prepare an arbitrary dimension $B+1$ superposition over the $\ket{b}$ register consisting of $n=\lceil\log_2(B+1)\rceil$ qubits, and this state is controlled by the index $\ket{x_\text{o}}$ and non-trivial only for $x_\text{o}\in[N,N+RC)$. Using optimization in~\cref{sec:fewer_qubits_A}, there is also no longer any need for QROAM to additionally output the $\ket{G}\ket{r}\ket{c}$ registers. If we use the variant of Shende-Bullock-Markov state preparation as described above for Outer $\textsc{Prep}$ instead of alias sampling, there is also no need for the $\ket{\text{alt}}\ket{\text{keep}}$.
    This method of controlled-state preparation is identical to Outer $\textsc{Prep}$ above, except that the QROAM queries are additionally controlled by $\ket{x_\text{o}}$. 
    It may be possible to further reduce cost using more efficient pure state preparation techniques~\cite{gosset2024quantumstatepreparationoptimal,berry2024rapid} than considered here.
    However, this will be dominated by \textsc{RPrep}.
\end{itemize}

\subsection{Clean-ancilla-free arithmetic}\label{sec:fewer_qubits_D}
The arithmetic operations (multiple-controlled $X$, quantum-classical comparators, adders, unary iteration) used in our block-encoding previously required temporary clean ancilla qubits. However, there exist alternate implementations that cost a constant factor more in Toffoli gates but only use dirty qubits. By using only dirty qubits, the block-encoding only needs a constant number of temporary ancilla qubits with the overall cost summarized in~\cref{tab:fewer_qubits_D}.

\begin{table}[h]
    \centering
    \resizebox{\textwidth}{!}{
    \begin{tabular}{|c|c|c|c|c|c|c|}
    \hline \hline
    & Subroutine & Cost & Per. & T. & Dirty& \# \\
    \hline
    \multirow{3}{*}{\rotatebox[origin=c]{0}{\parbox[c]{1cm}{\centering Outer \textsc{Prep}}}}
        & QROAM & $\sum_{l=0}^{n_{X_\text{o}}-1}\frac{5}{4}\left\lceil\frac{f(X_\text{o},l)}{2^{k_{l,1}}}\right\rceil + 2(2^{k_{l,1}}-1)b_\text{rot}$ & $n_{X_\text{o}}$ & &$(2^{k_{l,1}}-1)b_\text{rot}+n_{X_\text{o}}-k_{l,1}-1$  & 2\\
        & QROAM$^\dagger$ & $\frac{5}{4}\left\lceil\frac{f(X_\text{o})}{2^{k_5}}\right\rceil + 2(2^{k_5}-1)b_\text{rot}$ &  &  &$(2^{k_{5}}-1)b_\text{rot}+n_{X_\text{o}}-k_{5}-1$ & 2\\
    & Adders & $n_{X_\text{o}}(2b_\text{rot}-2)$ & & & $b_{\text{rot}}$ & 2 \\
    \hline
    \multirow{3}{*}{\rotatebox[origin=c]{0}{\parbox[c]{1cm}{\centering Inner \textsc{Prep}}}}
        & QROAM & $\sum_{l=0}^{n_{B'}-1}\frac{5}{4}\lceil\frac{R}{2^{k_r}}\rceil\lceil \frac{C}{2^{k_c}}\rceil\left\lceil\frac{f(B',l)}{2^{k_{l,2}}}\right\rceil + 2(2^{k_r+k_c+k_{l,2}}-1)b_\text{rot}$ & $n_{B'}$ &  &$(2^{k_r+k_c+k_{l,2}}-1)b_\text{rot}+n_{X_\text{o}}+n_{B'}-(k_r+k_c+k_{l,2})-1$ & 4\\
        & QROAM$^\dagger$ & $\frac{5}{4}\lceil\frac{R}{2^{k^\prime_r}}\rceil\lceil \frac{C}{2^{k^\prime_c}}\rceil\left\lceil\frac{f(B')}{2^{k_4}}\right\rceil  + 2(2^{k^\prime_r+k^\prime_c+k_4}-1)b_\text{rot}$ &  &  & $(2^{k_r+k_c+k_{4}}-1)b_\text{rot}+n_{X_\text{o}}+n_{B'}-(k_r+k_c+k_{4})-1$& 4 \\
    & Adders & $n_{B'}(2b_\text{rot}-2)$ & &  & $b_{\text{rot}}$ & 4 \\
    \hline 
    \rotatebox{0}{\textsc{RPrep}}
    & \begin{tabular}{c}Dirty\\ QROAM $(\vec{u})$\end{tabular} & $2(\big\lceil\frac{N-1}{\lambda}\big\rceil+1) \times\big(\frac{5}{4}\big(\lceil\frac{N}{2^{k_6}}\rceil+\lceil\frac{R}{2^{k_6}}\rceil B\rceil\big)+2(2^{k_6}-1)\lambda b_\text{rot}\big)$
    & $\lambda b_{\text{rot}}$ && $(2^{k_6}-1)\lambda b_\text{rot}+n_{X_\text{o}}+n_{B'}-k_6$& 2\\
    \hline
    \multirow{4}{*}{{\textsc{Sel}}}
    & \textsc{Rot} & $2\big((2b_{\text{rot}}-1)(N-1-\lceil\frac{N-1}{\lambda}\big\rceil) + \lceil\frac{N-1}{\lambda}\big\rceil(3b_{\text{rot}}-2)\big)$ & $b_{\text{rot}}$ & & & 2 \\
    & $\textsc{GComp}$ 1 and 3& $64n_{X_\text{o}}(\log_2n_{X_\text{o}}-2) +\mathcal{O}(1)$& $2$ & $1$ & $1$&  2 \\
    & CSWAP & $2N$ & & && 2\\
    & Maj-control & $64n_{X_\text{o}}(\log_2n_{X_\text{o}}-2)+4n_{B'}+\mathcal{O}(1)$ &  & $1$ & $1$& 2 \\
    \hline
     \multirow{2}{*}{{\textsc{Ref}}}
    & $T_{2}$ & $4n_{B'}$ & & &$1$& $1$ \\
    & Walk & $4(n_{X_\text{o}} +n_{B'})$ & &  &$1$& $1$ \\
    \hline \hline
    \end{tabular}
    }
    \caption{\label{tab:fewer_qubits_D}
    Cost breakdown using the optimizations described in~\cref{sec:fewer_qubits_D}. Persistent (Per.) and Temporary (Tmp.) qubits are abbreviated. The temporary ancilla of all steps can use all the $\lambda b_\text{rot}$ persistent ancillae of $\textsc{RPrep}$. Since $\lambda,b_\text{rot} \ge 1$, this implies a total qubit count of $2N+(\lambda+1)b_\text{rot}+n_{X_\text{o}}+n_{B'}+2$. Note that presented Toffoli and qubit counts for QROAM are leading order representations of the exact cost in~\cref{table:qubit_constrained_lookup_cost}.}
\end{table}
\begin{itemize}
\item \textbf{Multiple-controlled-$X$}: The $X$ gate with multiple $n$ controls is used to implement the reflections, select on the condition $\ket{b}=\ket{B}$, and implement the controlled unitaries $U_{G}$ in $\textsc{GComp}$. Previously this was implemented using $n$ Toffoli gates and $n$ clean ancillae. However, this can instead be implemented using $4n-8$ Toffoli gates and one dirty qubits~\cite{khattar2024riseconditionallycleanancillae}. 
\item \textbf{Quantum-classical comparator:} The quantum-classical comparator is used in implementing $\textsc{GComp}$. Previously, this was implemented using $3n$ Toffoli gates and $\log_2^*n$ clean qubits. By~\cref{lem:quantum_classical_comparison}, this can instead be performed using $16n(\log_2n-2)+\mathcal{O}(1)$ Toffoli gates and $1$ dirty qubits~\cite{khattar2024riseconditionallycleanancillae}. 

\item \textbf{Adders}: We perform arbitrary single-qubit $Z$ rotations by the phase gradient technique where the rotation angle is controlled by $b_{\text{rot}}$ qubits using an adder and a $b_{\text{rot}}$-qubit Fourier resource state. In total, this method of rotation uses $2b_{\text{rot}}$ persistent qubits. Previously, this was implemented using at most $b_\text{rot}$ Toffoli gates~\cite{Gidney2018halvingcostof} and $b_\text{rot}$ clean qubits. This can instead be performed using at most $2b_\text{rot}$ Toffoli gates and $0$ qubits~\cite{Takahashi2010Addition}, or the equivalent of $\frac{3}{2}b_\text{rot}$ Toffoli gates each for the special case of Givens rotations, which is equivalent to a controlled-$R_Z$ and applying~\cref{eq:circuit_MCRZ}.
Note that only the first adder in each sequence of $\lambda$ Givens rotations needs to be ancilla-free.
After the first rotation is performed, the first set of rotation bits may be measured out as part of the measurement-based QROM uncomputation~\cite{berry2019qubitization}.
This frees up clean ancilla qubits to perform all subsequent clean-ancilla adders using $b_\text{rot}$ Toffoli gates each.

\item \textbf{Unary iteration}: We use unary iteration extensively in QRAOM. Previously, on $d$ addresses with $k$ output copies, this was implemented using at most $d$ Toffoli gates and  $\lceil\log_2(d/k)\rceil$ clean qubits.
This can instead be performed using skew-tree construction~\cite{khattar2024riseconditionallycleanancillae} with $5d/4 +\mathcal{O}(\sqrt{d}\log d)$  Toffoli gates and $\lceil\log_2(d/k)\rceil$ dirty qubits. 
We compute the Toffoli count exactly using~\cref{table:qubit_constrained_lookup_cost}.
In the case of \textsc{Inner} with three address registers, we approximate the skew-tree Toffoli count, which is subdominant, by as a single register with $RCf(B',l)$ addresses.
\end{itemize}
\subsection{Rotation synthesis with fewer qubits}\label{sec:fewer_qubits_E}
Up to this point, we have been synthesizing $b$-bit $Z$ rotations using the phase gradient technique. 
This method always uses at least $2b_\text{rot}$ qubits: Half for storing a $b_\text{rot}$-qubit Fourier resource state, and half for the $b_\text{rot}$-qubit bit-string specifying the rotation angle.
However, there are other methods of rotation synthesis that may use fewer output qubits, though at the cost of more non-Clifford gates, and a longer effective bits of rotation $b_{\text{rot},T}\lesssim b_\text{rot}+12$.
Moreover, we may choose a parameter $\gamma\in[1,b_\text{rot}]$ and have QROAM output $\gamma$ of the $b_\text{rot}$ rotation bits at a time.
Unlike earlier sections, we no longer pad $R$ to a power of $2$, so $X_\text{o}=N+RC$.
In the $\lambda = 1$ case, this leads to a block-encoding with overall cost summarized in~\cref{tab:fewer_qubits_E}.

\begin{table}[h]
    \centering
    \resizebox{\textwidth}{!}{
    \begin{tabular}{|c|c|c|c|c|c|c|c|}
    \hline \hline
    & Subroutine & Cost & Pers. & T.& Dirty& \# \\
    \hline
    \multirow{2}{*}{\rotatebox[origin=c]{90}{\parbox[c]{1cm}{\centering Outer \textsc{Prep}}}}
        & QROAM & $\sum_{l=0}^{n_{X_\text{o}}-1}\lceil\frac{b_{\text{rot},T}}{\gamma}\rceil\left(\frac{5}{4}\left\lceil\frac{f(X_\text{o},l)}{2^{k_{l,1}}}\right\rceil + 2(2^{k_{l,1}}-1)\gamma\right)$ & $n_{X_\text{o}}$ & &$(2^{k_{l,1}}-1)\gamma+n_{X_\text{o}}-k_{l,1}-1$  & 2\\
        & QROAM$^\dagger$ & $\frac{5}{4}\left\lceil\frac{f(X_\text{o})}{2^{k_5}}\right\rceil + 2(2^{k_5}-1)\gamma$ &  &  &$(2^{k_{5}}-1)\gamma+n_{X_\text{o}}-k_{5}-1$ & 2\\
    & Rotations & $n_{X_\text{o}}(2b_{\text{rot},T}+4)$ & & 1 & $\gamma$ & 2 \\
    \hline
    \multirow{2}{*}{\rotatebox[origin=c]{90}{\parbox[c]{1cm}{\centering Inner \textsc{Prep}}}}
        & QROAM & $\sum_{l=0}^{n_{B'}-1}\lceil\frac{b_{\text{rot},T}}{\gamma}\rceil\left(\frac{5}{4}\lceil\frac{RC}{2^{k_{rc}}}\rceil\left\lceil\frac{f(B',l)}{2^{k_{l,2}}}\right\rceil  + 2(2^{k_{rc}+k_{l,2}}-1)\gamma\right)$ & $n_{B'}$ &  &$(2^{k_{rc}+k_{l,2}}-1)\gamma+n_{X_\text{o}}+n_{B'}-(k_{rc}+k_{l,2})-1$ & $4$\\
        & QROAM$^\dagger$ & $\frac{5}{4}\lceil\frac{RC}{2^{k^\prime_{rc}}}\rceil\left\lceil\frac{f(B',l)}{2^{k_{4}}}\right\rceil + 2(2^{k^\prime_r+k^\prime_c+k_4}-1)\gamma$ &  &  &$(2^{k_{rc}+k_{4}}-1)\gamma+n_{X_\text{o}}+n_{B'}-(k_{rc}+k_{4})-1$ & $4$ \\
    & Rotations & $n_{B'}(2b_{\text{rot},T}+4)$ & & $1$ & $\gamma$ & 4 \\
    \hline 
    \rotatebox{0}{\textsc{RPrep}}
    & \begin{tabular}{c}Dirty\\ QROAM $(\vec{u})$\end{tabular} & $2(\lceil\frac{b_\text{rot,T}}{\gamma}\rceil+1)(N-1)\big(\frac{5}{4}\lceil\frac{N+RB}{2^{k_6}}\rceil+2(2^{k_6}-1)\gamma\big)$ & $\gamma$ & & $(2^{k_6}-1)\gamma+n_{N+RB}-k_6$& 2\\
    \hline
    \multirow{4}{*}{{\textsc{Sel}}}
    & \textsc{Rot} & $4(N - 1)(2b_{\text{rot},T}+4)$ & $\begin{cases}\gamma,&\gamma>1\\
    0,&\gamma=1\end{cases}$ & $1$ && 2 \\
    & $\textsc{GComp}$ 1 and 3& $64n_{X_\text{o}}(\log_2n_{X_\text{o}}-2) +\mathcal{O}(1)$ & $2$ & $1$ & $1$&  2 \\
    & CSWAP & $2N$ & & && 2\\
    & Maj-control & $64n_{X_\text{o}}(\log_2n_{X_\text{o}}-2) +4n_{B'}+\mathcal{O}(1)$ &  & $1$ & $1$& 2 \\
    \hline
     \multirow{2}{*}{{\textsc{Ref}}}
    & $T_{2}$ & $4n_{B'}$ & & &$1$& $1$ \\
    & Walk & $4(n_{X_\text{o}} +n_{B'})$ & &  &$1$& $1$ \\
    \hline \hline
    \end{tabular}
    }
    \caption{\label{tab:fewer_qubits_E}
    Cost breakdown using the optimizations described in~\cref{sec:fewer_qubits_E}.Persistent (Per.) and Temporary (Tmp.) qubits are abbreviated. The temporary ancilla of all steps except $\textsc{Rot}$ can use all the $\gamma$ persistent ancillae of $\textsc{RPrep}$. 
    Note that presented Toffoli and qubit counts for QROAM are leading order representations of the exact cost in~\cref{table:qubit_constrained_lookup_cost}.
    Total qubit count if $\gamma>1$ is $2N+2\gamma+n_{X_\text{o}}+n_{B'}+3$, else if $\gamma=1$ is $2N+4+n_{X_\text{o}}+n_{B'}$. Here, $X_\text{o}=N+RC$}
\end{table}
\begin{itemize}
    \item \textbf{Arbitrary controlled phase rotations}: Given the $b_\text{rot}$-qubit state $\ket{\theta_0\theta_1\cdots\theta_{b_\text{rot}-1}}$ storing the binary representation of some angle $\theta$, one may instead perform $b_\text{rot}$ separate controlled-$Z$ rotations $\proj{0}\otimes \one+\proj{1}\otimes e^{i\pi Z/2^{j+1}}$, where the $j^{\text{th}}$ rotation is controlled by $\ket{\theta_j}$. 
    In principle, each controlled rotation can be approximated with two arbitrary single-qubit $Z$ rotations synthesized to diamond distance $\epsilon$ using a sequence of gates from $\{H,S,T\}$, where the number of $T$ gates required is $\approx 0.56\log_2(1/\epsilon
    )+5.3$ with one clean ancilla or $1.52 \log_2
    (1/\epsilon)-0.01$ with no ancilla by ~\cite{Kliuchnikov2023shorterquantum}, or a total of $\mathcal{O}(b_\text{rot}\log\frac{1}{\epsilon})$ T states. 
    We instead implement arbitrary controlled phase rotations using the method of~\cref{sec:multiplex_rotations}, where output bits specify rotations by the Matsumoto-Amano normal form.
    Rotations specified this way require more rotation bits, at most $b_{\text{rot},T}\lesssim b_{\text{rot}}+12$, leading to a larger lookup.
    However, the number of $T$ states needed is only linear like $2b_{\text{rot},T}+4$.
    In this work, we synthesize $\ket{T}$ states from $\ket{CCZ}$ states and found that the combined $\ket{CCZ}$ factory and the catalyzed $\ket{CCZ}\rightarrow 2\ket{T}$ pipe diagram is close to double the spacetime volume of the small $2\times 3$ footprint $\ket{CCZ}$. 
    Hence, we count the cost of each $\ket{T}$ as being equal to a $\ket{CCZ}$.
    This is a conservative estimate compared to another common convention of $4\ket{T}$ states being the equivalent of one Toffoli (or $\ket{CCZ}$ state).
    The difference $b_{\text{rot},T}-b_{\text{rot}}\approx12$ is based on the diamond distance and can be considered a rigorous estimate.
    In future work, a smaller estimate for $b_{\text{rot},T}$ can be determined as that achieving a chemically accurate CCSD(T) correlation energy as performed in~\cite{low2025fast} for $b_{\text{rot}}$

\item \textbf{QROAM on $\gamma$ rotation bits at a time}: Previously, we implemented controlled rotations following the pattern of 1) Perform QROAM on some number of $d$ addresses to output $b_\text{rot,T}$ Matsumoto Amano normal form rotation bits from ~\cref{sec:Matsumoto_Amano_normal form} in the state $\ket{\theta}=\ket{\theta_0\theta_1\cdots\theta_{b_\text{rot}-1}}$; 2) Perform controlled rotations $\ket{\theta}\rightarrow \ket{\theta}\otimes e^{i\pi Z\theta}$; 3) Uncompute QROAM. 
For any integer $\gamma\in[1,b_\text{rot}]$,
one may instead define a loop over $l=0,1,l_\text{max}-1$, where $l_{\max}=\lceil b_\text{rot}/\gamma\rceil$.
Then at iteration $l$: 1) Perform QROAM on some number of $d$ addresses to output $\gamma$ rotation bits $\ket{\vec{\theta}_l}$ where $\vec{\theta}_l\doteq(\theta_{0+l\gamma},\theta_{1+l\gamma},\cdots,\theta_{\gamma-1+l\gamma})$.
Similar to the programmable gate array, the $(l+1)^\text{th}$ iteration can XOR in $\vec{\theta}_l\oplus\vec{\theta}_{l+1}$ to realize $\ket{\vec{\theta}_l}\rightarrow \ket{\vec{\theta}_{l+1}}$ without needing to uncompute; 2) Apply the sub-circuit of Matsumoto Amano normal rotations corresponding to the output bits.
\item \textbf{Flatten address register}: QROAM for rotation bits is performed over $N+RB$ non-contiguous addresses in the joint address register $\ket{x_\text{o}}\ket{b}$ with $n_\text{total}=n_{X_\text{o}}+n_{B'}$ qubits. We reversibly flatten these addresses into a single contiguous integer index $x_{\textsc{RPrep}}\in[N+RB]$ stored in $n_{N+RB}$ qubits.
As seen in~\cref{tab:address_flattening}, the Toffoli count of address flattening with minimal ancilla is extremely high.
Nevertheless, when extremely few qubits are available, this is more than offset by the freed $n_\text{total}-n_{N+RB}$ dirty qubits, which are used to reduce the QROAM Toffoli count.
In an extreme case with $4$ clean qubits and $0$ dirty qubits, address flattening can reduce the Toffoli count of \textsc{RPrep} by $90\%$.
When more ancilla qubits are available, address flattening is significantly cheaper and the freed qubits can still be worthwhile, but is no longer effective with $\gtrsim40$ ancilla qubits.
The arithmetic operations to flatten the addresses requires three clean ancilla and at most one dirty ancilla qubit as follows.

\begin{enumerate}
    \item Compute the flags $\ket{f_{\text{SF}}}=\ket{x_\text{o}< RC}$ and $\ket{f_{B'}}=\ket{b'=B}$. This uses as most one dirty ancilla for the comparison by~\cref{lem:quantum_classical_comparison}, which also dominates the Toffoli count.
    \item Controlled by $\ket{f_{B'}}=\ket{1}$, perform a quantum-classical subtraction $\ket{b}\rightarrow\ket{b-B}$ using one dirty ancilla. This guarantees that $\ket{b}$ stores an integer $<B$.
    \item Controlled by $\ket{f_\text{SF}}$, apply a multiply-add to compute $\ket{x_\text{o}}\ket{b}\rightarrow\ket{x_\text{o} B+b}$, where $x_\text{o} B+b=cRB+rB+b$. The multiply-add uses one clean qubit and one dirty qubit.
    \item Controlled by $\ket{f_\text{SF}}$, apply a multiply-add in reverse to compute $\ket{x_\text{o} B+b}\rightarrow \ket{rB+b}\ket{c}$.
    \item Controlled by $\ket{f_\text{SF}}=\ket{1}$, add the constant $N$ with a quantum-classical adder to obtain $\ket{rB+b}\rightarrow\ket{rB+b+N}$, and controlled by $\ket{f_\text{SF}}=\ket{0}$, subtract the constant $RC$ from $\ket{x_\text{o}}$ to index the $G_{\text{D}_1},G_{\text{Q}_1}$ terms.
    \item Uncompute the flag $\ket{f_\text{SF}}$ using a comparison.
    \item At this point, the lowest $n_{N+RB}$ qubits hold the contiguous address, and the remaining addresses may be used as dirty qubits for QROAM.
\end{enumerate}

One may also flatten the $RBC$ addresses for \textsc{Inner}, but this does not appear worth doing as it only frees one or two qubits since $RBC\gg N+RB$, and \textsc{Inner} is a much smaller lookup than \textsc{RPrep} anyway. 
Address flattening also changes some of the comparisons in \textsc{GComp} from a $n_{X_\text{o}}$-bit register to a $n_{N+RB}$-bit register, with a negligible difference in Toffoli cost.
\begin{table}
    \centering
    \begin{tabular}{c|c|c|c|c|ccc|ccc}
    \hline\hline
         &\multirow{2}{*}{Molecule}& \multirow{2}{*}{$(R,B,C)$} & \multirow{2}{*}{$\lceil\frac{\pi \lambda_\text{eff}}{2\sigma_{\text{PEA}}}\rceil$} & Toffolis&  \multicolumn{3}{c|}{Qubits before}  &  \multicolumn{3}{c}{Qubits after}  \\
         &&&& / $10^9$ &$n_{X_\text{o}}+n_{B'}$ & $n_\text{clean,in}$ & $n_\text{dirty,in}$ &$n_{N+RB}$ & $n_\text{clean,out}$ & $n_\text{dirty,out}$
         \\\hline
\multirow{6}{*}{\rotatebox{90}{\begin{tabular}{c}Minimum\\ancilla\end{tabular}}} & XVIII56o64e & (5, 28, 28) & 26816 & 0.63 & 13 & 3 & 1 & 8 & 2 & 7 \\
 & XVIII100o100e & (8, 75, 25) & 59188 & 2.08 & 16 & 3 & 1 & 10 & 2 & 8 \\
 & XVIII150o150e & (9, 112, 37) & 103469 & 3.95 & 16 & 3 & 1 & 11 & 2 & 7 \\
 & FeMoCo54o54e & (10, 27, 27) & 33564 & 0.91 & 14 & 3 & 1 & 9 & 2 & 7 \\
 & FeMoCo76o113e & (15, 57, 19) & 68572 & 2.20 & 15 & 3 & 1 & 10 & 2 & 7 \\
 & cpd1X58o63e & (9, 29, 14) & 51510 & 1.18 & 13 & 3 & 1 & 9 & 2 & 6 \\
\hline
\multirow{6}{*}{\rotatebox{90}{\begin{tabular}{c}Minimum\\toffoli\end{tabular}}} & XVIII56o64e & (5, 28, 28) & 26816 & 0.03 & 13 & 10 & 0 & 8 & 9 & 6 \\
 & XVIII100o100e & (8, 75, 25) & 59188 & 0.10 & 16 & 11 & 0 & 10 & 10 & 7 \\
 & XVIII150o150e & (9, 112, 37) & 103469 & 0.19 & 16 & 11 & 0 & 11 & 10 & 6 \\
 & FeMoCo54o54e & (10, 27, 27) & 33564 & 0.05 & 14 & 11 & 0 & 9 & 10 & 6 \\
 & FeMoCo76o113e & (15, 57, 19) & 68572 & 0.11 & 15 & 11 & 0 & 10 & 10 & 6 \\
 & cpd1X58o63e & (9, 29, 14) & 51510 & 0.06 & 13 & 10 & 0 & 9 & 9 & 5 \\
\hline
\multirow{6}{*}{\rotatebox{90}{\begin{tabular}{c}Fewer\\ancilla\end{tabular}}} & XVIII56o64e & (5, 28, 28) & 26816 & 0.15 & 13 & 6 & 0 & 8 & 5 & 6 \\
 & XVIII100o100e & (8, 75, 25) & 59188 & 0.44 & 16 & 6 & 0 & 10 & 5 & 7 \\
 & XVIII150o150e & (9, 112, 37) & 103469 & 0.84 & 16 & 6 & 0 & 11 & 5 & 6 \\
 & FeMoCo54o54e & (10, 27, 27) & 33564 & 0.21 & 14 & 6 & 0 & 9 & 5 & 6 \\
 & FeMoCo76o113e & (15, 57, 19) & 68572 & 0.48 & 15 & 6 & 0 & 10 & 5 & 6 \\
 & cpd1X58o63e & (9, 29, 14) & 51510 & 0.28 & 13 & 6 & 0 & 9 & 5 & 5 \\
\hline
\hline\hline
    \end{tabular}
    \caption{Cumulative Toffoli count across entire phase estimation of flattening and unflattening an initial address register $\ket{x_\text{o}}\ket{b}\ket{\cdot}^{\otimes n_\text{dirty,in}}\ket{0}^{\otimes n_\text{clean,in}}$, where $x_\text{o}\in[N+RC]$ and $b\in[B+1]$ to $\ket{x_\textsc{RPrep}}\ket{\cdot}^{\otimes n_\text{dirty,out}}\ket{0}^{\otimes n_\text{clean,out}}$, where $x_\textsc{RPrep}\in[N+RB]$. ``Qubits after'' shows the number of qubits freed to assist \textsc{RPrep}. The different Toffoli and qubit counts correspond to using different choices of quantum-classical adders from~\cref{lem:qc_addition}}
    \label{tab:address_flattening}
\end{table}
\end{itemize}

\section{Electronic structure compilation parameters}\label{sec:parto_frontier_parameters}
In this section, we tabulate for each molecule in~\cref{fig:intro_pareto} the compilation parameters for select points on the physical space-time Pareto frontier. 

\input{5_simulation/tables_3x4}


%% file: 5_simulation/tables_3x4.tex
\begin{table}
\tiny
\begingroup
\setlength{\tabcolsep}{1pt}
\begin{tabular}{c|c|cc|c|ccc|cccc|ccccc|cccc|cccc|cc}
\hline
\multicolumn{26}{c}{FeMoco-54o54e}
\\
\hline
Qubits&Volume&\multicolumn{2}{c|}{Runtime/hr} & \multirow{2}{*}{$p_\text{fail}$}& \multicolumn{3}{c|}{Factories}&\multicolumn{4}{c|}{Compute}&\multicolumn{5}{c|}{Hot storage}&\multicolumn{4}{c|}{Cold storage}&\multicolumn{4}{c|}{Toffolis/$10^9$}&Rotation&\\
$\mathfrak{n}$/$10^3$&$/10^6\mathfrak{n}\cdot\text{hr}$&$\langle.\rangle$& $1$ shot & &$F$&$d_{\textsc{CCZ}}$&$p_\text{fail}$& $k$&$d$&$\mathfrak{n}$/$10^3$&$p_\text{fail}$ & Latency&$k$&$d_\text{in}$&$\mathfrak{n}$/$10^3$&$p_\text{fail}$ & $k$&$d_\text{in}$&$\mathfrak{n}$/$10^3$&$p_\text{fail}$&Total&\textsc{RPrep}&\textsc{IPrep}&\textsc{Rot}& synthesis&\\
\hline
77.27 & 873 & 11300 & 1290 & 0.885 & 1& 123.7& 0.238 & 16& 23& 16.9& 0.391 & M($12,1$) &20& 14& 14.1& 0.201 & 112& 16& 46.2& 0.691 & 18.9 & 9.48 & 8.52 & 0.91 & HST\\
81.83 & 136 & 1660 & 597 & 0.640 & 1& 123.7& 0.121 & 16& 23& 16.9& 0.201 & M($16,1$) &28& 14& 18.8& 0.168 & 110& 16& 46.1& 0.383 & 8.98 & 5.06 & 2.99 & 0.91 & HST\\
86.53 & 72.6 & 839 & 397 & 0.526 & 1& 123.7& 0.084 & 16& 23& 16.9& 0.137 & M($20,1$) &36& 14& 23.5& 0.174 & 110& 16& 46.1& 0.275 & 6.13 & 3.14 & 2.05 & 0.91 & HST\\
91.19 & 49.6 & 544 & 302 & 0.446 & 1& 123.7& 0.066 & 16& 23& 16.9& 0.104 & M($24,1$) &44& 14& 28.2& 0.188 & 104& 16& 46.0& 0.183 & 4.77 & 2.47 & 1.36 & 0.91 & HST\\
97.06 & 38.7 & 399 & 288 & 0.278 & 1& 127.3& 0.020 & 16& 24& 18.4& 0.029 & M($24,1$) &44& 15& 32.4& 0.014 & 112& 16& 46.2& 0.230 & 4.42 & 2.47 & 1.01 & 0.91 & HST\\
102.3 & 31.9 & 312 & 242 & 0.224 & 1& 127.3& 0.017 & 16& 24& 18.4& 0.024 & M($28,1$) &52& 15& 37.8& 0.016 & 110& 16& 46.1& 0.178 & 3.78 & 2.08 & 0.76 & 0.91 & HST\\
107.7 & 29.2 & 271 & 221 & 0.186 & 1& 127.3& 0.016 & 16& 24& 18.4& 0.022 & M($32,1$) &60& 15& 43.2& 0.019 & 104& 16& 46.0& 0.138 & 3.49 & 1.80 & 0.76 & 0.91 & HST\\
113.1 & 26.4 & 234 & 194 & 0.169 & 1& 127.3& 0.014 & 16& 24& 18.4& 0.019 & M($36,1$) &68& 15& 48.6& 0.021 & 104& 16& 46.0& 0.122 & 3.13 & 1.60 & 0.59 & 0.91 & HST\\
121.2 & 21.8 & 180 & 151 & 0.157 & 1& 127.3& 0.011 & 16& 24& 18.4& 0.015 & M($42,1$) &80& 15& 56.7& 0.022 & 110& 16& 46.1& 0.115 & 2.54 & 1.22 & 0.38 & 0.91 & HST\\
128.3 & 19.1 & 149 & 113 & 0.242 & 1& 123.7& 0.046 & 35& 23& 37.0& 0.094 & L($8,2$) &34& 13& 45.1& 0.028 & 112& 16& 46.2& 0.097 & 3.29 & 2.06 & 0.71 & 0.50 & dirty\\
134.9 & 15.7 & 117 & 73.3 & 0.372 & 1& 120.3& 0.099 & 35& 22& 33.9& 0.203 & L($11,2$) &46& 13& 54.9& 0.072 & 110& 16& 46.1& 0.058 & 2.19 & 1.43 & 0.39 & 0.36 & dirty\\
141.7 & 12.9 & 91.0 & 75.4 & 0.171 & 1& 123.7& 0.031 & 35& 23& 37.0& 0.064 & L($11,2$) &46& 13& 58.6& 0.029 & 110& 16& 46.1& 0.059 & 2.19 & 1.43 & 0.39 & 0.36 & dirty\\
149.0 & 9.47 & 63.5 & 40.5 & 0.363 & 2& 60.2& 0.109 & 56& 22& 54.2& 0.163 & L($10,2$) &42& 12& 48.6& 0.115 & 112& 16& 46.2& 0.036 & 2.42 & 1.60 & 0.42 & 0.37 & dirty\\
159.6 & 7.62 & 47.8 & 41.6 & 0.129 & 2& 61.9& 0.034 & 56& 23& 59.2& 0.050 & L($10,2$) &42& 13& 54.1& 0.014 & 112& 16& 46.2& 0.037 & 2.42 & 1.60 & 0.42 & 0.37 & dirty\\
169.2 & 7.03 & 41.5 & 23.3 & 0.439 & 2& 60.2& 0.064 & 56& 22& 54.2& 0.097 & L($17,2$) &70& 12& 76.9& 0.165 & 102& 15& 38.1& 0.204 & 1.39 & 0.87 & 0.20 & 0.30 & dirty\\
177.8 & 5.79 & 32.6 & 24.4 & 0.250 & 3& 40.1& 0.099 & 77& 22& 74.5& 0.129 & L($7,3$) &45& 13& 57.0& 0.023 & 112& 16& 46.2& 0.022 & 2.19 & 1.43 & 0.38 & 0.36 & dirty\\
188.4 & 4.93 & 26.2 & 24.0 & 0.084 & 2& 61.9& 0.020 & 56& 23& 59.2& 0.029 & L($17,2$) &70& 13& 85.7& 0.018 & 102& 16& 43.5& 0.020 & 1.39 & 0.87 & 0.20 & 0.30 & dirty\\
199.5 & 4.46 & 22.3 & 17.0 & 0.238 & 3& 40.1& 0.070 & 77& 22& 74.5& 0.092 & L($11,3$) &69& 12& 78.9& 0.087 & 104& 16& 46.0& 0.011 & 1.53 & 1.00 & 0.20 & 0.30 & dirty\\
211.9 & 3.86 & 18.2 & 12.8 & 0.299 & 4& 30.1& 0.070 & 98& 22& 94.9& 0.085 & L($11,3$) &69& 12& 78.9& 0.066 & 102& 15& 38.1& 0.118 & 1.53 & 1.00 & 0.20 & 0.30 & dirty\\
225.4 & 3.13 & 13.9 & 11.7 & 0.161 & 4& 30.1& 0.064 & 98& 22& 94.9& 0.078 & L($12,3$) &75& 13& 88.7& 0.019 & 98& 16& 41.8& 0.009 & 1.39 & 0.87 & 0.20 & 0.30 & dirty\\
238.6 & 3.14 & 13.1 & 9.63 & 0.268 & 4& 30.1& 0.053 & 98& 22& 94.9& 0.065 & L($15,3$) &93& 12& 103& 0.085 & 112& 15& 40.5& 0.095 & 1.15 & 0.66 & 0.20 & 0.28 & dirty\\
253.6 & 2.79 & 11.0 & 9.32 & 0.152 & 5& 24.1& 0.064 & 119& 22& 115& 0.074 & L($12,3$) &75& 14& 92.4& 0.015 & 104& 16& 46.0& 0.006 & 1.39 & 0.87 & 0.20 & 0.30 & dirty\\
267.7 & 2.60 & 9.73 & 8.49 & 0.128 & 4& 30.1& 0.047 & 98& 22& 94.9& 0.057 & L($18,3$) &111& 13& 127& 0.022 & 110& 16& 46.1& 0.007 & 1.02 & 0.53 & 0.20 & 0.27 & dirty\\
281.1 & 2.41 & 8.56 & 7.10 & 0.170 & 4& 30.1& 0.040 & 98& 22& 94.9& 0.048 & L($21,3$) &129& 13& 146& 0.023 & 112& 15& 40.5& 0.071 & 0.850 & 0.44 & 0.12 & 0.27 & dirty\\
299.2 & 2.19 & 7.33 & 4.90 & 0.332 & 5& 22.2& 0.137 & 119& 21& 105& 0.129 & L($23,3$) &141& 13& 148& 0.108 & 112& 16& 46.2& 0.004 & 0.791 & 0.40 & 0.11 & 0.27 & dirty\\
319.8 & 1.92 & 6.00 & 5.40 & 0.100 & 5& 24.1& 0.038 & 119& 22& 115& 0.044 & L($22,3$) &135& 14& 158& 0.017 & 112& 16& 46.2& 0.005 & 0.808 & 0.40 & 0.12 & 0.27 & dirty\\
339.5 & 1.88 & 5.53 & 5.01 & 0.095 & 5& 24.1& 0.035 & 119& 22& 115& 0.041 & L($25,3$) &153& 14& 178& 0.019 & 110& 16& 46.1& 0.004 & 0.749 & 0.35 & 0.11 & 0.27 & dirty\\
359.4 & 1.86 & 5.18 & 4.70 & 0.092 & 5& 24.1& 0.033 & 119& 22& 115& 0.038 & L($28,3$) &171& 14& 198& 0.020 & 112& 16& 46.2& 0.004 & 0.703 & 0.31 & 0.11 & 0.26 & dirty\\
379.8 & 1.80 & 4.73 & 4.28 & 0.095 & 6& 20.1& 0.033 & 140& 22& 136& 0.043 & L($28,3$) &171& 14& 198& 0.018 & 112& 16& 46.2& 0.004 & 0.703 & 0.31 & 0.11 & 0.26 & dirty\\
412.8 & 1.78 & 4.32 & 3.93 & 0.091 & 6& 20.1& 0.030 & 140& 22& 136& 0.040 & L($33,3$) &201& 14& 231& 0.021 & 112& 16& 46.2& 0.004 & 0.646 & 0.27 & 0.11 & 0.25 & clean\\
445.8 & 1.79 & 4.01 & 3.65 & 0.089 & 6& 20.1& 0.028 & 140& 22& 136& 0.037 & L($38,3$) &231& 14& 264& 0.023 & 112& 16& 46.2& 0.003 & 0.602 & 0.22 & 0.11 & 0.25 & clean\\
479.6 & 1.73 & 3.62 & 3.22 & 0.110 & 6& 20.1& 0.025 & 140& 22& 136& 0.032 & L($44,3$) &267& 14& 304& 0.025 & 112& 15& 40.5& 0.033 & 0.534 & 0.18 & 0.08 & 0.26 & dirty\\
511.2 & 1.75 & 3.42 & 3.17 & 0.073 & 6& 20.1& 0.025 & 140& 22& 136& 0.032 & L($46,3$) &280& 15& 329& 0.016 & 112& 16& 46.2& 0.003 & 0.526 & 0.18 & 0.08 & 0.25 & clean\\
541.6 & 1.58 & 2.93 & 2.25 & 0.230 & 5& 22.2& 0.088 & 119& 21& 105& 0.042 & L($15,12$) &374& 13& 403& 0.106 & 88& 15& 33.9& 0.014 & 0.495 & 0.14 & 0.08 & 0.26 & Pdirty\\
570.9 & 1.57 & 2.76 & 2.25 & 0.182 & 5& 22.2& 0.088 & 119& 21& 105& 0.042 & L($15,12$) &376& 14& 420& 0.063 & 110& 16& 46.1& 0.002 & 0.495 & 0.14 & 0.08 & 0.26 & Pdirty\\
610.1 & 1.58 & 2.58 & 2.43 & 0.058 & 5& 24.1& 0.023 & 119& 22& 115& 0.014 & L($15,12$) &375& 14& 449& 0.020 & 110& 16& 46.1& 0.002 & 0.495 & 0.14 & 0.08 & 0.26 & Pdirty\\
685.0 & 1.68 & 2.45 & 1.84 & 0.248 & 6& 18.5& 0.080 & 140& 21& 123& 0.045 & L($20,12$) &494& 13& 521& 0.130 & 110& 15& 40.4& 0.017 & 0.447 & 0.09 & 0.08 & 0.26 & Pdirty\\
724.3 & 1.65 & 2.28 & 1.84 & 0.193 & 6& 18.5& 0.080 & 140& 21& 123& 0.045 & L($21,12$) &520& 14& 568& 0.071 & 84& 15& 32.8& 0.012 & 0.447 & 0.09 & 0.08 & 0.26 & Pdirty\\
775.5 & 1.61 & 2.08 & 1.93 & 0.069 & 6& 20.1& 0.021 & 140& 22& 136& 0.014 & L($21,12$) &519& 14& 607& 0.023 & 84& 15& 32.8& 0.013 & 0.447 & 0.09 & 0.08 & 0.26 & Pdirty\\
818.4 & 1.38 & 1.69 & 1.36 & 0.195 & 9& 12.3& 0.090 & 224& 21& 198& 0.055 & DL($27,12$) &380& 14& 580& 0.050 & 112& 15& 40.5& 0.014 & 0.507 & 0.14 & 0.08 & 0.27 & Pdirty\\
867.9 & 1.34 & 1.54 & 1.24 & 0.194 & 10& 11.1& 0.090 & 245& 21& 216& 0.055 & DL($30,12$) &417& 13& 621& 0.057 & 77& 15& 30.7& 0.006 & 0.507 & 0.14 & 0.08 & 0.27 & Pdirty\\
939.4 & 1.33 & 1.42 & 1.32 & 0.068 & 10& 12.0& 0.024 & 245& 22& 237& 0.018 & DL($30,12$) &417& 13& 672& 0.022 & 77& 15& 30.7& 0.007 & 0.507 & 0.14 & 0.08 & 0.27 & Pdirty\\
1019 & 1.34 & 1.32 & 1.22 & 0.069 & 11& 10.9& 0.024 & 266& 22& 257& 0.018 & DL($33,12$) &456& 13& 731& 0.023 & 77& 15& 30.7& 0.006 & 0.507 & 0.14 & 0.08 & 0.27 & Pdirty\\
1076 & 1.34 & 1.25 & 1.01 & 0.193 & 12& 9.3& 0.081 & 287& 21& 253& 0.057 & DL($39,12$) &534& 13& 786& 0.061 & 98& 15& 36.6& 0.008 & 0.455 & 0.09 & 0.08 & 0.27 & Pdirty\\
1144 & 1.32 & 1.16 & 1.08 & 0.069 & 11& 10.9& 0.021 & 266& 22& 257& 0.015 & DL($39,12$) &534& 13& 850& 0.025 & 98& 15& 36.6& 0.009 & 0.455 & 0.09 & 0.08 & 0.27 & Pdirty\\
1206 & 1.32 & 1.10 & 1.03 & 0.062 & 13& 9.3& 0.021 & 308& 22& 298& 0.019 & DL($39,12$) &536& 14& 871& 0.014 & 98& 15& 36.6& 0.009 & 0.455 & 0.09 & 0.08 & 0.27 & Pdirty\\
1364 & 1.41 & 1.03 & 0.968 & 0.063 & 15& 8.0& 0.021 & 350& 22& 339& 0.023 & DL($45,12$) &614& 14& 993& 0.015 & 84& 15& 32.8& 0.006 & 0.439 & 0.09 & 0.08 & 0.25 & Pclean\\
1683 & 1.62 & 0.964 & 0.772 & 0.200 & 13& 8.5& 0.068 & 308& 21& 272& 0.046 & DL($71,12$) &950& 13& 1370& 0.093 & 102& 15& 38.1& 0.008 & 0.379 & 0.02 & 0.07 & 0.26 & Pdirty\\
1774 & 1.59 & 0.895 & 0.727 & 0.188 & 16& 6.9& 0.068 & 371& 21& 327& 0.060 & DL($71,12$) &952& 14& 1410& 0.066 & 102& 15& 38.1& 0.007 & 0.379 & 0.02 & 0.07 & 0.26 & Pdirty\\
1864 & 1.57 & 0.841 & 0.771 & 0.083 & 15& 8.0& 0.018 & 350& 22& 339& 0.017 & DL($71,12$) &950& 13& 1490& 0.044 & 104& 15& 40.4& 0.006 & 0.379 & 0.02 & 0.07 & 0.26 & Pdirty\\
2002 & 1.59 & 0.794 & 0.745 & 0.061 & 16& 7.5& 0.017 & 371& 22& 359& 0.018 & DL($75,12$) &1004& 14& 1600& 0.020 & 112& 15& 40.5& 0.008 & 0.371 & 0.02 & 0.07 & 0.25 & Pclean\\
\hline
88.84 & 59.6 & 671 & 351 & 0.477 & 1& 123.7& 0.076 & 16& 23& 16.9& 0.121 & M($22,1$) &40& 14& 25.9& 0.185 & 104& 16& 46.0& 0.210 & 5.47 & 2.95 & 1.58 & 0.91 & HST\\
196.2 & 4.47 & 22.8 & 15.5 & 0.318 & 3& 40.1& 0.064 & 77& 22& 74.5& 0.084 & L($12,3$) &75& 12& 85.0& 0.092 & 98& 15& 36.6& 0.123 & 1.39 & 0.87 & 0.20 & 0.30 & dirty\\
1683 & 1.62 & 0.964 & 0.772 & 0.200 & 13& 8.5& 0.068 & 308& 21& 272& 0.046 & DL($71,12$) &950& 13& 1370& 0.093 & 102& 15& 38.1& 0.008 & 0.379 & 0.02 & 0.07 & 0.26 & Pdirty\\
1149 & 1.30 & 1.13 & 1.05 & 0.071 & 12& 10.0& 0.021 & 287& 22& 278& 0.017 & DL($38,12$) &521& 13& 830& 0.024 & 112& 15& 40.5& 0.011 & 0.455 & 0.09 & 0.08 & 0.27 & Pdirty\\
\hline\hline
\end{tabular}
\endgroup
\caption{\label{tab:pareto_optimal}Selected compilation parameters on space-time pareto frontier for performing ground state phase estimation of utility-scale electronic structure to chemical accuracy. Hot storage latencies `M',`L', and `DL' refer to `medium-latency', `low-latency', and `double-access-low-latency' respectively. The suffix in parenthesis is the number ($n_\text{row}\times n_\text{col}$) of access hallways. Rotation synthesis methods `HST', `clean', and `dirty' refer to $\{H,S,T\}$, phase gradient by clean-ancilla adders, and by dirty-ancilla adders respectively. The prefix `P' indicates whether parallelization of Givens rotations is performed. 
We highlight in the bottom rows estimates that run within a month, a day, an hour (if possible) and with minimum spacetime volume.}
\end{table}

\begin{table}
\addtocounter{table}{-1}
\tiny
\begingroup
\setlength{\tabcolsep}{1pt}
\begin{tabular}{c|c|cc|c|ccc|cccc|ccccc|cccc|cccc|cc}
\hline
\multicolumn{26}{c}{FeMoco-76o113e}
\\
\hline
Qubits&Volume&\multicolumn{2}{c|}{Runtime/hr} & \multirow{2}{*}{$p_\text{fail}$}& \multicolumn{3}{c|}{Factories}&\multicolumn{4}{c|}{Compute}&\multicolumn{5}{c|}{Hot storage}&\multicolumn{4}{c|}{Cold storage}&\multicolumn{4}{c|}{Toffolis/$10^9$}&Rotation&\\
$\mathfrak{n}$/$10^3$&$/10^6\mathfrak{n}\cdot\text{hr}$&$\langle.\rangle$& $1$ shot & &$F$&$d_{\textsc{CCZ}}$&$p_\text{fail}$& $k$&$d$&$\mathfrak{n}$/$10^3$&$p_\text{fail}$ & Latency&$k$&$d_\text{in}$&$\mathfrak{n}$/$10^3$&$p_\text{fail}$ & $k$&$d_\text{in}$&$\mathfrak{n}$/$10^3$&$p_\text{fail}$&Total&\textsc{RPrep}&\textsc{IPrep}&\textsc{Rot}& synthesis&\\
\hline
104.7 & 4420 & 42200 & 6040 & 0.857 & 1& 127.3& 0.314 & 16& 24& 18.4& 0.485 & M($14,1$) &24& 15& 18.9& 0.096 & 154& 17& 67.3& 0.552 & 84.0 & 55.65 & 25.73 & 2.52 & HST\\
110.1 & 967 & 8790 & 3120 & 0.645 & 1& 127.3& 0.179 & 16& 24& 18.4& 0.288 & M($18,1$) &32& 15& 24.3& 0.082 & 154& 17& 67.3& 0.339 & 43.9 & 29.92 & 11.44 & 2.52 & HST\\
116.4 & 442 & 3800 & 2250 & 0.409 & 1& 131.0& 0.047 & 16& 25& 20.0& 0.066 & M($22,1$) &40& 15& 29.7& 0.090 & 153& 17& 66.7& 0.270 & 31.3 & 21.12 & 7.65 & 2.52 & HST\\
122.4 & 309 & 2520 & 1680 & 0.335 & 1& 131.0& 0.036 & 16& 25& 20.0& 0.050 & M($26,1$) &48& 15& 35.1& 0.093 & 154& 17& 67.3& 0.200 & 23.7 & 16.53 & 4.60 & 2.52 & HST\\
129.9 & 241 & 1850 & 1290 & 0.304 & 1& 131.0& 0.028 & 16& 25& 20.0& 0.038 & M($32,1$) &60& 15& 43.2& 0.108 & 153& 17& 66.7& 0.165 & 18.6 & 12.58 & 3.40 & 2.52 & HST\\
138.0 & 208 & 1500 & 1070 & 0.288 & 1& 131.0& 0.024 & 16& 25& 20.0& 0.031 & M($38,1$) &72& 15& 51.3& 0.125 & 153& 17& 66.7& 0.139 & 15.6 & 10.79 & 2.26 & 2.52 & HST\\
146.3 & 192 & 1310 & 1070 & 0.183 & 1& 131.0& 0.024 & 16& 25& 20.0& 0.031 & M($38,1$) &72& 15& 51.3& 0.125 & 153& 18& 75.0& 0.013 & 15.6 & 10.79 & 2.26 & 2.52 & HST\\
156.4 & 176 & 1120 & 1040 & 0.074 & 1& 131.0& 0.023 & 16& 25& 20.0& 0.031 & M($40,1$) &76& 16& 61.4& 0.010 & 153& 18& 75.0& 0.013 & 14.9 & 10.61 & 1.73 & 2.52 & HST\\
164.6 & 127 & 773 & 529 & 0.316 & 1& 127.3& 0.065 & 35& 24& 40.3& 0.123 & L($10,2$) &42& 13& 57.6& 0.101 & 153& 17& 66.7& 0.071 & 14.9 & 10.94 & 2.76 & 1.20 & dirty\\
174.8 & 104 & 593 & 425 & 0.284 & 1& 127.3& 0.052 & 35& 24& 40.3& 0.100 & L($12,2$) &50& 13& 67.2& 0.111 & 154& 17& 67.3& 0.055 & 12.0 & 9.16 & 1.71 & 1.09 & dirty\\
183.9 & 80.4 & 437 & 334 & 0.237 & 2& 63.7& 0.081 & 56& 24& 64.5& 0.110 & L($8,2$) &34& 14& 49.9& 0.015 & 160& 17& 69.4& 0.052 & 18.9 & 13.93 & 3.52 & 1.38 & dirty\\
193.6 & 58.0 & 299 & 237 & 0.208 & 2& 63.7& 0.058 & 56& 24& 64.5& 0.080 & L($11,2$) &46& 13& 62.4& 0.055 & 153& 17& 66.7& 0.033 & 13.4 & 9.92 & 2.31 & 1.14 & dirty\\
203.6 & 48.0 & 236 & 192 & 0.188 & 2& 63.7& 0.047 & 56& 24& 64.5& 0.065 & L($14,2$) &58& 13& 76.8& 0.068 & 140& 17& 62.3& 0.021 & 10.8 & 7.95 & 1.82 & 1.02 & dirty\\
214.1 & 39.5 & 184 & 126 & 0.315 & 2& 61.9& 0.100 & 56& 23& 59.2& 0.145 & L($18,2$) &74& 13& 90.2& 0.097 & 143& 17& 64.7& 0.014 & 7.35 & 5.58 & 0.91 & 0.80 & dirty\\
224.9 & 37.3 & 166 & 112 & 0.323 & 3& 41.2& 0.131 & 77& 23& 81.5& 0.164 & L($10,3$) &63& 13& 81.1& 0.056 & 140& 17& 62.3& 0.013 & 9.79 & 7.13 & 1.64 & 0.97 & dirty\\
237.2 & 29.0 & 122 & 83.9 & 0.314 & 3& 41.2& 0.100 & 77& 23& 81.5& 0.126 & L($12,3$) &75& 14& 98.5& 0.030 & 143& 16& 57.2& 0.101 & 7.32 & 5.58 & 0.89 & 0.80 & dirty\\
250.5 & 25.9 & 104 & 69.2 & 0.332 & 4& 30.9& 0.109 & 98& 23& 104& 0.127 & L($11,3$) &69& 13& 87.9& 0.040 & 153& 16& 58.9& 0.106 & 8.05 & 5.58 & 1.61 & 0.80 & dirty\\
263.1 & 21.8 & 83.0 & 62.9 & 0.242 & 4& 30.9& 0.100 & 98& 23& 104& 0.116 & L($12,3$) &75& 13& 94.7& 0.041 & 143& 17& 64.7& 0.007 & 7.32 & 5.58 & 0.89 & 0.80 & dirty\\
276.8 & 21.0 & 76.0 & 64.7 & 0.148 & 4& 31.8& 0.032 & 98& 24& 113& 0.035 & L($12,3$) &75& 15& 109& 0.007 & 140& 16& 55.1& 0.081 & 7.32 & 5.58 & 0.89 & 0.80 & dirty\\
290.9 & 17.1 & 58.9 & 48.1 & 0.183 & 4& 30.9& 0.077 & 98& 23& 104& 0.090 & L($16,3$) &99& 14& 127& 0.023 & 132& 17& 60.5& 0.005 & 5.59 & 3.92 & 0.87 & 0.75 & dirty\\
306.1 & 15.1 & 49.2 & 38.5 & 0.218 & 5& 24.7& 0.077 & 119& 23& 126& 0.086 & L($16,3$) &99& 14& 127& 0.034 & 132& 16& 53.5& 0.041 & 5.59 & 3.92 & 0.87 & 0.75 & dirty\\
323.2 & 14.8 & 45.8 & 37.4 & 0.183 & 4& 30.9& 0.061 & 98& 23& 104& 0.071 & L($22,3$) &135& 14& 169& 0.025 & 126& 16& 50.6& 0.040 & 4.35 & 3.16 & 0.49 & 0.66 & clean\\
341.3 & 13.9 & 40.7 & 32.6 & 0.200 & 5& 24.7& 0.066 & 119& 23& 126& 0.073 & L($21,3$) &129& 14& 162& 0.043 & 132& 16& 53.5& 0.035 & 4.74 & 3.16 & 0.87 & 0.66 & clean\\
358.8 & 12.5 & 34.9 & 29.9 & 0.143 & 5& 24.7& 0.061 & 119& 23& 126& 0.067 & L($22,3$) &135& 15& 176& 0.019 & 126& 17& 57.3& 0.003 & 4.35 & 3.16 & 0.49 & 0.66 & clean\\
378.0 & 11.8 & 31.2 & 26.3 & 0.155 & 5& 24.7& 0.054 & 119& 23& 126& 0.059 & L($26,3$) &159& 14& 197& 0.048 & 119& 17& 55.1& 0.002 & 3.83 & 2.64 & 0.48 & 0.66 & clean\\
399.3 & 11.8 & 29.5 & 24.8 & 0.161 & 5& 24.7& 0.051 & 119& 23& 126& 0.056 & L($27,3$) &165& 15& 212& 0.020 & 160& 16& 61.3& 0.045 & 3.60 & 2.41 & 0.45 & 0.69 & dirty\\
419.4 & 11.3 & 26.9 & 23.0 & 0.146 & 5& 24.7& 0.047 & 119& 23& 126& 0.052 & L($30,3$) &183& 15& 234& 0.020 & 154& 16& 59.5& 0.035 & 3.35 & 2.15 & 0.45 & 0.69 & dirty\\
444.0 & 10.6 & 23.9 & 21.2 & 0.111 & 5& 24.7& 0.043 & 119& 23& 126& 0.048 & L($32,3$) &195& 15& 249& 0.020 & 160& 17& 69.4& 0.003 & 3.09 & 1.90 & 0.45 & 0.68 & dirty\\
466.2 & 10.5 & 22.5 & 19.9 & 0.118 & 6& 20.6& 0.043 & 140& 23& 148& 0.057 & L($32,3$) &195& 15& 249& 0.019 & 160& 17& 69.4& 0.003 & 3.09 & 1.90 & 0.45 & 0.68 & dirty\\
492.0 & 10.5 & 21.3 & 18.2 & 0.146 & 6& 20.6& 0.040 & 140& 23& 148& 0.052 & L($37,3$) &225& 14& 274& 0.059 & 160& 17& 69.4& 0.003 & 2.83 & 1.64 & 0.45 & 0.68 & dirty\\
517.1 & 9.62 & 18.6 & 16.8 & 0.094 & 5& 24.7& 0.035 & 119& 23& 126& 0.038 & L($42,3$) &255& 15& 322& 0.022 & 160& 17& 69.4& 0.003 & 2.45 & 1.39 & 0.34 & 0.68 & dirty\\
545.9 & 9.26 & 17.0 & 15.0 & 0.118 & 6& 20.6& 0.033 & 140& 23& 148& 0.043 & L($44,3$) &267& 15& 336& 0.020 & 160& 16& 61.3& 0.027 & 2.34 & 1.30 & 0.32 & 0.68 & dirty\\
575.9 & 9.27 & 16.1 & 14.6 & 0.095 & 6& 20.6& 0.032 & 140& 23& 148& 0.042 & L($47,3$) &285& 15& 358& 0.021 & 160& 17& 69.4& 0.002 & 2.28 & 1.30 & 0.27 & 0.66 & clean\\
612.5 & 9.14 & 14.9 & 13.6 & 0.091 & 6& 20.6& 0.030 & 140& 23& 148& 0.039 & L($52,3$) &315& 15& 395& 0.022 & 160& 17& 69.4& 0.002 & 2.13 & 1.13 & 0.27 & 0.67 & dirty\\
655.1 & 9.04 & 13.8 & 12.4 & 0.103 & 5& 24.7& 0.030 & 119& 23& 126& 0.024 & L($15,12$) &374& 14& 479& 0.039 & 126& 16& 50.6& 0.013 & 2.14 & 1.13 & 0.27 & 0.69 & Pdirty\\
699.7 & 8.93 & 12.8 & 12.1 & 0.052 & 5& 25.5& 0.009 & 119& 24& 137& 0.006 & L($15,12$) &374& 14& 509& 0.024 & 132& 16& 53.5& 0.013 & 2.05 & 1.04 & 0.27 & 0.69 & Pdirty\\
743.1 & 8.60 & 11.6 & 10.6 & 0.084 & 5& 24.7& 0.027 & 119& 23& 126& 0.020 & L($17,12$) &423& 15& 556& 0.021 & 160& 16& 61.3& 0.019 & 1.88 & 0.88 & 0.27 & 0.68 & Pdirty\\
790.0 & 8.16 & 10.3 & 9.33 & 0.097 & 6& 20.6& 0.025 & 140& 23& 148& 0.023 & L($19,12$) &470& 14& 591& 0.042 & 126& 16& 50.6& 0.010 & 1.79 & 0.79 & 0.27 & 0.68 & Pdirty\\
878.2 & 8.43 & 9.60 & 8.71 & 0.093 & 8& 15.5& 0.033 & 203& 23& 215& 0.031 & DL($24,12$) &339& 14& 610& 0.020 & 136& 16& 53.8& 0.012 & 2.36 & 1.30 & 0.27 & 0.73 & Pdirty\\
951.5 & 7.90 & 8.30 & 7.57 & 0.087 & 8& 15.5& 0.030 & 203& 23& 215& 0.027 & DL($27,12$) &378& 14& 675& 0.020 & 160& 16& 61.3& 0.014 & 2.09 & 1.04 & 0.27 & 0.72 & Pdirty\\
1010 & 7.45 & 7.38 & 6.79 & 0.080 & 9& 13.7& 0.030 & 224& 23& 237& 0.026 & DL($29,12$) &404& 14& 719& 0.019 & 132& 16& 53.5& 0.007 & 2.09 & 1.04 & 0.27 & 0.72 & Pdirty\\
1061 & 7.01 & 6.61 & 4.71 & 0.287 & 11& 10.9& 0.083 & 266& 22& 257& 0.073 & DL($34,12$) &469& 13& 751& 0.089 & 154& 15& 52.1& 0.080 & 1.81 & 0.79 & 0.27 & 0.70 & Pdirty\\
1128 & 6.65 & 5.90 & 5.25 & 0.110 & 10& 12.4& 0.026 & 245& 23& 259& 0.021 & DL($34,12$) &469& 13& 810& 0.059 & 154& 16& 59.5& 0.008 & 1.81 & 0.79 & 0.27 & 0.70 & Pdirty\\
1189 & 6.25 & 5.26 & 4.89 & 0.069 & 11& 11.2& 0.026 & 266& 23& 281& 0.022 & DL($35,12$) &482& 14& 851& 0.017 & 143& 16& 57.2& 0.006 & 1.81 & 0.79 & 0.27 & 0.70 & Pdirty\\
1365 & 6.70 & 4.91 & 3.78 & 0.230 & 11& 10.9& 0.069 & 266& 22& 257& 0.057 & DL($49,12$) &664& 13& 1050& 0.118 & 153& 16& 58.9& 0.006 & 1.50 & 0.53 & 0.23 & 0.69 & Pdirty\\
1452 & 6.53 & 4.50 & 3.61 & 0.198 & 14& 8.6& 0.069 & 329& 22& 318& 0.078 & DL($49,12$) &666& 14& 1070& 0.060 & 153& 16& 58.9& 0.006 & 1.50 & 0.53 & 0.23 & 0.69 & Pdirty\\
1536 & 6.32 & 4.12 & 3.85 & 0.064 & 12& 10.3& 0.021 & 287& 23& 304& 0.019 & DL($50,12$) &677& 14& 1180& 0.020 & 136& 16& 53.8& 0.005 & 1.50 & 0.53 & 0.23 & 0.69 & Pdirty\\
1718 & 6.70 & 3.90 & 3.61 & 0.074 & 16& 7.7& 0.021 & 371& 23& 393& 0.028 & DL($54,12$) &729& 14& 1270& 0.021 & 153& 16& 58.9& 0.006 & 1.47 & 0.53 & 0.23 & 0.66 & Pclean\\
2248 & 7.99 & 3.56 & 2.75 & 0.227 & 10& 12.0& 0.051 & 245& 22& 237& 0.034 & DL($47,24$) &1228& 13& 1950& 0.153 & 153& 16& 58.9& 0.004 & 1.10 & 0.14 & 0.23 & 0.67 & Pdirty\\
2366 & 6.59 & 2.79 & 2.30 & 0.174 & 16& 7.5& 0.051 & 371& 22& 359& 0.057 & DL($92,12$) &1225& 14& 1950& 0.074 & 153& 16& 58.9& 0.004 & 1.10 & 0.14 & 0.23 & 0.67 & Pdirty\\
2543 & 6.71 & 2.64 & 2.38 & 0.099 & 16& 7.7& 0.016 & 371& 23& 393& 0.017 & DL($92,12$) &1223& 14& 2100& 0.027 & 153& 15& 51.6& 0.043 & 1.10 & 0.14 & 0.23 & 0.67 & Pdirty\\
2692 & 6.73 & 2.50 & 2.36 & 0.058 & 16& 7.7& 0.015 & 371& 23& 393& 0.017 & DL($49,24$) &1280& 14& 2240& 0.023 & 154& 16& 59.5& 0.004 & 1.08 & 0.14 & 0.23 & 0.66 & Pclean\\
\hline
166.9 & 120 & 716 & 529 & 0.262 & 1& 127.3& 0.065 & 35& 24& 40.3& 0.123 & L($10,2$) &42& 14& 59.9& 0.030 & 153& 17& 66.7& 0.071 & 14.9 & 10.94 & 2.76 & 1.20 & dirty\\
444.0 & 10.6 & 23.9 & 21.2 & 0.111 & 5& 24.7& 0.043 & 119& 23& 126& 0.048 & L($32,3$) &195& 15& 249& 0.020 & 160& 17& 69.4& 0.003 & 3.09 & 1.90 & 0.45 & 0.68 & dirty\\
1170 & 6.16 & 5.26 & 4.89 & 0.070 & 11& 11.2& 0.026 & 266& 23& 281& 0.022 & DL($34,12$) &469& 14& 829& 0.017 & 154& 16& 59.5& 0.007 & 1.81 & 0.79 & 0.27 & 0.70 & Pdirty\\
\hline\hline
\end{tabular}
\endgroup
\caption{Selected compilation parameters on space-time pareto frontier for performing ground state phase estimation of utility-scale electronic structure to chemical accuracy (Continued).}
\end{table}

\begin{table}
\addtocounter{table}{-1}
\tiny
\begingroup
\setlength{\tabcolsep}{1pt}
\begin{tabular}{c|c|cc|c|ccc|cccc|ccccc|cccc|cccc|cc}
\hline
\multicolumn{26}{c}{Cpd1X-58o63e}
\\
\hline
Qubits&Volume&\multicolumn{2}{c|}{Runtime/hr} & \multirow{2}{*}{$p_\text{fail}$}& \multicolumn{3}{c|}{Factories}&\multicolumn{4}{c|}{Compute}&\multicolumn{5}{c|}{Hot storage}&\multicolumn{4}{c|}{Cold storage}&\multicolumn{4}{c|}{Toffolis/$10^9$}&Rotation&\\
$\mathfrak{n}$/$10^3$&$/10^6\mathfrak{n}\cdot\text{hr}$&$\langle.\rangle$& $1$ shot & &$F$&$d_{\textsc{CCZ}}$&$p_\text{fail}$& $k$&$d$&$\mathfrak{n}$/$10^3$&$p_\text{fail}$ & Latency&$k$&$d_\text{in}$&$\mathfrak{n}$/$10^3$&$p_\text{fail}$ & $k$&$d_\text{in}$&$\mathfrak{n}$/$10^3$&$p_\text{fail}$&Total&\textsc{RPrep}&\textsc{IPrep}&\textsc{Rot}& synthesis&\\
\hline
81.13 & 879 & 10800 & 1500 & 0.862 & 1& 127.3& 0.092 & 16& 24& 18.4& 0.149 & M($14,1$) &24& 14& 16.5& 0.302 & 112& 16& 46.2& 0.743 & 21.6 & 15.79 & 4.35 & 1.45 & HST\\
85.83 & 162 & 1890 & 686 & 0.637 & 1& 127.3& 0.045 & 16& 24& 18.4& 0.070 & M($18,1$) &32& 14& 21.2& 0.238 & 112& 16& 46.2& 0.463 & 10.3 & 6.98 & 1.84 & 1.45 & HST\\
90.53 & 86.5 & 955 & 450 & 0.529 & 1& 127.3& 0.031 & 16& 24& 18.4& 0.045 & M($22,1$) &40& 14& 25.9& 0.234 & 112& 16& 46.2& 0.335 & 7.01 & 4.30 & 1.22 & 1.45 & HST\\
95.13 & 72.5 & 762 & 440 & 0.423 & 1& 123.7& 0.096 & 16& 23& 16.9& 0.148 & M($22,1$) &40& 14& 25.9& 0.225 & 112& 17& 52.3& 0.034 & 7.01 & 4.30 & 1.22 & 1.45 & HST\\
100.5 & 52.5 & 523 & 457 & 0.125 & 1& 127.3& 0.031 & 16& 24& 18.4& 0.046 & M($22,1$) &40& 15& 29.7& 0.019 & 112& 17& 52.3& 0.035 & 7.01 & 4.30 & 1.22 & 1.45 & HST\\
105.9 & 43.9 & 415 & 370 & 0.108 & 1& 127.3& 0.026 & 16& 24& 18.4& 0.037 & M($26,1$) &48& 15& 35.1& 0.021 & 112& 17& 52.3& 0.029 & 5.81 & 3.52 & 0.81 & 1.45 & HST\\
111.3 & 38.9 & 349 & 315 & 0.098 & 1& 127.3& 0.022 & 16& 24& 18.4& 0.031 & M($30,1$) &56& 15& 40.5& 0.024 & 112& 17& 52.3& 0.024 & 5.06 & 2.99 & 0.58 & 1.45 & HST\\
118.2 & 37.8 & 320 & 299 & 0.067 & 1& 131.0& 0.007 & 16& 25& 20.0& 0.008 & M($34,1$) &64& 15& 45.9& 0.029 & 112& 17& 52.3& 0.023 & 4.74 & 2.67 & 0.58 & 1.45 & HST\\
124.4 & 34.3 & 275 & 229 & 0.170 & 1& 123.7& 0.055 & 16& 23& 16.9& 0.076 & M($38,1$) &72& 15& 51.3& 0.027 & 120& 17& 56.2& 0.023 & 3.94 & 2.15 & 0.31 & 1.45 & HST\\
130.7 & 30.9 & 236 & 157 & 0.338 & 1& 123.7& 0.063 & 35& 23& 37.0& 0.128 & L($8,2$) &34& 13& 45.1& 0.039 & 119& 16& 48.6& 0.156 & 4.56 & 3.13 & 0.66 & 0.74 & dirty\\
137.4 & 24.8 & 181 & 130 & 0.281 & 1& 123.7& 0.053 & 35& 23& 37.0& 0.108 & L($10,2$) &42& 13& 54.1& 0.043 & 112& 16& 46.2& 0.111 & 3.78 & 2.67 & 0.42 & 0.66 & dirty\\
146.2 & 20.9 & 143 & 117 & 0.183 & 1& 123.7& 0.048 & 35& 23& 37.0& 0.097 & L($10,2$) &42& 13& 54.1& 0.039 & 119& 17& 55.1& 0.011 & 3.40 & 2.45 & 0.36 & 0.56 & dirty\\
153.5 & 14.7 & 96.1 & 56.9 & 0.408 & 2& 60.2& 0.149 & 56& 22& 54.2& 0.221 & L($10,2$) &42& 13& 50.7& 0.051 & 119& 16& 48.6& 0.060 & 3.40 & 2.45 & 0.36 & 0.56 & dirty\\
162.0 & 11.6 & 71.7 & 58.5 & 0.185 & 2& 61.9& 0.048 & 56& 23& 59.2& 0.070 & L($10,2$) &42& 13& 54.1& 0.020 & 119& 16& 48.6& 0.061 & 3.40 & 2.45 & 0.36 & 0.56 & dirty\\
173.2 & 10.8 & 62.3 & 36.2 & 0.419 & 2& 60.2& 0.098 & 56& 22& 54.2& 0.147 & L($16,2$) &66& 12& 72.9& 0.223 & 110& 16& 46.1& 0.029 & 2.17 & 1.37 & 0.30 & 0.47 & dirty\\
182.2 & 8.66 & 47.5 & 34.0 & 0.285 & 2& 60.2& 0.092 & 56& 22& 54.2& 0.138 & L($18,2$) &74& 13& 84.5& 0.059 & 102& 16& 43.5& 0.029 & 2.03 & 1.37 & 0.16 & 0.47 & dirty\\
192.9 & 7.68 & 39.8 & 35.0 & 0.122 & 2& 61.9& 0.029 & 56& 23& 59.2& 0.042 & L($18,2$) &74& 13& 90.2& 0.028 & 102& 16& 43.5& 0.029 & 2.03 & 1.37 & 0.16 & 0.47 & dirty\\
203.0 & 6.26 & 30.8 & 23.2 & 0.248 & 3& 40.1& 0.094 & 77& 22& 74.5& 0.123 & L($11,3$) &69& 13& 82.4& 0.035 & 110& 16& 46.1& 0.019 & 2.08 & 1.37 & 0.21 & 0.47 & dirty\\
214.2 & 5.99 & 28.0 & 17.4 & 0.378 & 4& 30.1& 0.094 & 98& 22& 94.9& 0.114 & L($11,3$) &69& 12& 78.9& 0.089 & 110& 15& 40.4& 0.149 & 2.08 & 1.37 & 0.21 & 0.47 & dirty\\
226.7 & 5.10 & 22.5 & 17.4 & 0.226 & 4& 30.1& 0.094 & 98& 22& 94.9& 0.114 & L($11,3$) &69& 14& 85.8& 0.022 & 110& 16& 46.1& 0.014 & 2.08 & 1.37 & 0.21 & 0.47 & dirty\\
241.3 & 4.70 & 19.5 & 17.9 & 0.082 & 4& 30.9& 0.029 & 98& 23& 104& 0.035 & L($11,3$) &69& 14& 91.5& 0.006 & 110& 16& 46.1& 0.014 & 2.08 & 1.37 & 0.21 & 0.47 & dirty\\
255.0 & 4.26 & 16.7 & 13.5 & 0.192 & 4& 30.1& 0.074 & 98& 22& 94.9& 0.090 & L($16,3$) &99& 13& 114& 0.031 & 110& 16& 46.1& 0.011 & 1.62 & 1.04 & 0.16 & 0.38 & clean\\
270.4 & 3.69 & 13.6 & 11.1 & 0.185 & 5& 24.1& 0.076 & 119& 22& 115& 0.088 & L($14,3$) &87& 14& 106& 0.021 & 120& 16& 49.6& 0.013 & 1.66 & 1.05 & 0.16 & 0.43 & dirty\\
284.2 & 3.56 & 12.5 & 9.73 & 0.223 & 5& 24.1& 0.067 & 119& 22& 115& 0.077 & L($17,3$) &105& 13& 120& 0.088 & 119& 16& 48.6& 0.010 & 1.46 & 0.85 & 0.16 & 0.42 & dirty\\
303.4 & 3.17 & 10.4 & 8.81 & 0.156 & 5& 24.1& 0.061 & 119& 22& 115& 0.070 & L($19,3$) &117& 14& 139& 0.023 & 120& 16& 49.6& 0.010 & 1.32 & 0.72 & 0.16 & 0.41 & dirty\\
322.2 & 3.10 & 9.62 & 8.19 & 0.149 & 5& 24.1& 0.056 & 119& 22& 115& 0.065 & L($22,3$) &135& 14& 158& 0.026 & 119& 16& 48.6& 0.009 & 1.22 & 0.67 & 0.12 & 0.41 & dirty\\
346.1 & 3.01 & 8.71 & 7.49 & 0.140 & 5& 24.1& 0.052 & 119& 22& 115& 0.060 & L($26,3$) &159& 14& 185& 0.029 & 110& 16& 46.1& 0.006 & 1.12 & 0.59 & 0.09 & 0.41 & dirty\\
369.4 & 2.79 & 7.55 & 6.57 & 0.129 & 5& 24.1& 0.046 & 119& 22& 115& 0.053 & L($29,3$) &177& 14& 205& 0.029 & 120& 16& 49.6& 0.008 & 0.983 & 0.46 & 0.09 & 0.40 & dirty\\
389.7 & 2.68 & 6.88 & 5.97 & 0.132 & 6& 20.1& 0.046 & 140& 22& 136& 0.059 & L($29,3$) &177& 14& 205& 0.026 & 120& 16& 49.6& 0.007 & 0.983 & 0.46 & 0.09 & 0.40 & dirty\\
416.1 & 2.63 & 6.33 & 5.52 & 0.127 & 6& 20.1& 0.042 & 140& 22& 136& 0.055 & L($33,3$) &201& 14& 231& 0.029 & 120& 16& 49.6& 0.006 & 0.911 & 0.40 & 0.09 & 0.38 & clean\\
445.9 & 2.61 & 5.86 & 5.19 & 0.114 & 6& 20.1& 0.040 & 140& 22& 136& 0.052 & L($36,3$) &220& 15& 261& 0.020 & 120& 16& 49.6& 0.006 & 0.860 & 0.34 & 0.09 & 0.40 & dirty\\
482.1 & 2.61 & 5.41 & 4.76 & 0.120 & 6& 20.1& 0.037 & 140& 22& 136& 0.047 & L($43,3$) &261& 14& 297& 0.035 & 120& 16& 49.6& 0.006 & 0.791 & 0.27 & 0.09 & 0.39 & dirty\\
521.5 & 2.67 & 5.12 & 4.85 & 0.051 & 6& 20.6& 0.011 & 140& 23& 148& 0.014 & L($44,3$) &267& 14& 324& 0.022 & 120& 16& 49.6& 0.006 & 0.779 & 0.27 & 0.08 & 0.39 & dirty\\
548.0 & 2.57 & 4.69 & 3.22 & 0.314 & 5& 22.2& 0.125 & 119& 21& 105& 0.057 & L($15,12$) &374& 13& 403& 0.148 & 104& 15& 40.4& 0.025 & 0.719 & 0.21 & 0.08 & 0.40 & Pdirty\\
582.7 & 2.45 & 4.20 & 3.47 & 0.172 & 5& 24.1& 0.034 & 119& 22& 115& 0.019 & L($15,12$) &374& 13& 431& 0.101 & 98& 15& 36.6& 0.029 & 0.719 & 0.21 & 0.08 & 0.40 & Pdirty\\
613.6 & 2.28 & 3.72 & 3.42 & 0.081 & 5& 24.1& 0.033 & 119& 22& 115& 0.019 & L($15,12$) &375& 14& 449& 0.028 & 120& 16& 49.6& 0.004 & 0.702 & 0.21 & 0.08 & 0.38 & Pclean\\
696.3 & 2.42 & 3.47 & 3.11 & 0.103 & 6& 20.1& 0.033 & 140& 22& 136& 0.023 & L($18,12$) &447& 14& 528& 0.031 & 84& 15& 32.8& 0.021 & 0.702 & 0.21 & 0.08 & 0.38 & Pclean\\
745.6 & 2.45 & 3.28 & 3.02 & 0.081 & 5& 24.1& 0.030 & 119& 22& 115& 0.015 & L($20,12$) &495& 14& 581& 0.034 & 120& 16& 49.6& 0.004 & 0.648 & 0.14 & 0.08 & 0.39 & Pdirty\\
785.7 & 2.31 & 2.94 & 2.60 & 0.116 & 8& 15.0& 0.038 & 203& 22& 197& 0.029 & DL($24,12$) &339& 13& 553& 0.032 & 98& 15& 36.6& 0.022 & 0.814 & 0.27 & 0.08 & 0.43 & Pdirty\\
826.6 & 2.16 & 2.61 & 1.95 & 0.254 & 9& 12.3& 0.128 & 224& 21& 198& 0.077 & DL($27,12$) &380& 14& 580& 0.071 & 119& 16& 48.6& 0.002 & 0.737 & 0.21 & 0.08 & 0.42 & Pdirty\\
870.0 & 2.12 & 2.44 & 1.78 & 0.269 & 10& 11.1& 0.128 & 245& 21& 216& 0.077 & DL($30,12$) &417& 13& 621& 0.081 & 84& 15& 32.8& 0.012 & 0.737 & 0.21 & 0.08 & 0.42 & Pdirty\\
941.5 & 1.98 & 2.11 & 1.90 & 0.099 & 10& 12.0& 0.034 & 245& 22& 237& 0.025 & DL($30,12$) &417& 13& 672& 0.031 & 84& 15& 32.8& 0.013 & 0.737 & 0.21 & 0.08 & 0.42 & Pdirty\\
1021 & 1.99 & 1.95 & 1.76 & 0.100 & 11& 10.9& 0.034 & 266& 22& 257& 0.025 & DL($33,12$) &456& 13& 731& 0.033 & 84& 15& 32.8& 0.012 & 0.737 & 0.21 & 0.08 & 0.42 & Pdirty\\
1101 & 2.03 & 1.84 & 1.66 & 0.102 & 12& 10.0& 0.033 & 287& 22& 278& 0.028 & DL($36,12$) &495& 13& 791& 0.035 & 84& 15& 32.8& 0.011 & 0.702 & 0.21 & 0.08 & 0.38 & Pclean\\
1164 & 1.98 & 1.70 & 1.54 & 0.099 & 11& 10.9& 0.031 & 266& 22& 257& 0.021 & DL($40,12$) &547& 13& 870& 0.037 & 98& 15& 36.6& 0.013 & 0.660 & 0.14 & 0.08 & 0.41 & Pdirty\\
1226 & 1.96 & 1.60 & 1.46 & 0.087 & 13& 9.3& 0.031 & 308& 22& 298& 0.026 & DL($40,12$) &549& 14& 891& 0.020 & 98& 15& 36.6& 0.012 & 0.660 & 0.14 & 0.08 & 0.41 & Pdirty\\
1368 & 2.07 & 1.51 & 1.37 & 0.091 & 15& 8.0& 0.030 & 350& 22& 339& 0.031 & DL($45,12$) &614& 14& 993& 0.022 & 98& 15& 36.6& 0.012 & 0.636 & 0.14 & 0.08 & 0.38 & Pclean\\
1723 & 2.47 & 1.43 & 1.08 & 0.245 & 13& 8.5& 0.096 & 308& 21& 272& 0.062 & DL($71,12$) &952& 14& 1410& 0.097 & 119& 15& 42.6& 0.013 & 0.545 & 0.04 & 0.08 & 0.39 & Pdirty\\
1826 & 2.34 & 1.28 & 1.13 & 0.118 & 13& 9.3& 0.026 & 308& 22& 298& 0.019 & DL($71,12$) &950& 13& 1490& 0.064 & 119& 15& 42.6& 0.014 & 0.545 & 0.04 & 0.08 & 0.39 & Pdirty\\
1919 & 2.25 & 1.17 & 1.08 & 0.083 & 15& 8.0& 0.026 & 350& 22& 339& 0.023 & DL($72,12$) &965& 14& 1540& 0.027 & 102& 15& 38.1& 0.011 & 0.545 & 0.04 & 0.08 & 0.39 & Pdirty\\
2181 & 2.41 & 1.11 & 1.07 & 0.034 & 16& 7.7& 0.008 & 371& 23& 393& 0.007 & DL($76,12$) &1015& 14& 1750& 0.009 & 110& 15& 40.4& 0.010 & 0.534 & 0.04 & 0.08 & 0.38 & Pclean\\
\hline
96.64 & 63.6 & 658 & 450 & 0.316 & 1& 127.3& 0.031 & 16& 24& 18.4& 0.045 & M($22,1$) &40& 14& 25.9& 0.234 & 112& 17& 52.3& 0.035 & 7.01 & 4.30 & 1.22 & 1.45 & HST\\
223.3 & 5.04 & 22.6 & 17.4 & 0.229 & 4& 30.1& 0.094 & 98& 22& 94.9& 0.114 & L($11,3$) &69& 13& 82.4& 0.026 & 110& 16& 46.1& 0.014 & 2.08 & 1.37 & 0.21 & 0.47 & dirty\\
1029 & 1.93 & 1.88 & 1.69 & 0.098 & 11& 10.9& 0.033 & 266& 22& 257& 0.025 & DL($33,12$) &456& 13& 731& 0.032 & 104& 15& 40.4& 0.013 & 0.702 & 0.21 & 0.08 & 0.38 & Pclean\\
\hline\hline
\end{tabular}
\endgroup
\caption{Selected compilation parameters on space-time pareto frontier for performing ground state phase estimation of utility-scale electronic structure to chemical accuracy (Continued).}
\end{table}

\begin{table}
\addtocounter{table}{-1}
\tiny
\begingroup
\setlength{\tabcolsep}{1pt}
\begin{tabular}{c|c|cc|c|ccc|cccc|ccccc|cccc|cccc|cc}
\hline
\multicolumn{26}{c}{XVIII-56o64e}
\\
\hline
Qubits&Volume&\multicolumn{2}{c|}{Runtime/hr} & \multirow{2}{*}{$p_\text{fail}$}& \multicolumn{3}{c|}{Factories}&\multicolumn{4}{c|}{Compute}&\multicolumn{5}{c|}{Hot storage}&\multicolumn{4}{c|}{Cold storage}&\multicolumn{4}{c|}{Toffolis/$10^9$}&Rotation&\\
$\mathfrak{n}$/$10^3$&$/10^6\mathfrak{n}\cdot\text{hr}$&$\langle.\rangle$& $1$ shot & &$F$&$d_{\textsc{CCZ}}$&$p_\text{fail}$& $k$&$d$&$\mathfrak{n}$/$10^3$&$p_\text{fail}$ & Latency&$k$&$d_\text{in}$&$\mathfrak{n}$/$10^3$&$p_\text{fail}$ & $k$&$d_\text{in}$&$\mathfrak{n}$/$10^3$&$p_\text{fail}$&Total&\textsc{RPrep}&\textsc{IPrep}&\textsc{Rot}& synthesis&\\
\hline
75.91 & 243 & 3200 & 351 & 0.890 & 1& 120.3& 0.232 & 16& 22& 15.5& 0.369 & M($14,1$) &24& 13& 14.2& 0.688 & 112& 16& 46.2& 0.273 & 5.55 & 3.20 & 1.69 & 0.65 & HST\\
80.53 & 51.5 & 640 & 267 & 0.583 & 1& 120.3& 0.182 & 16& 22& 15.5& 0.296 & M($16,1$) &28& 14& 18.8& 0.077 & 112& 16& 46.2& 0.215 & 4.23 & 2.30 & 1.25 & 0.65 & HST\\
85.83 & 23.6 & 275 & 204 & 0.260 & 1& 127.3& 0.014 & 16& 24& 18.4& 0.021 & M($18,1$) &32& 14& 21.2& 0.078 & 112& 16& 46.2& 0.169 & 3.17 & 1.58 & 0.92 & 0.65 & HST\\
90.39 & 17.5 & 194 & 153 & 0.213 & 1& 127.3& 0.011 & 16& 24& 18.4& 0.015 & M($22,1$) &40& 14& 25.9& 0.086 & 110& 16& 46.1& 0.116 & 2.46 & 1.24 & 0.54 & 0.65 & HST\\
95.09 & 15.0 & 158 & 126 & 0.204 & 1& 127.3& 0.009 & 16& 24& 18.4& 0.012 & M($26,1$) &48& 14& 30.6& 0.099 & 110& 16& 46.1& 0.097 & 2.08 & 1.06 & 0.36 & 0.65 & HST\\
100.8 & 14.1 & 140 & 118 & 0.158 & 1& 123.7& 0.028 & 16& 23& 16.9& 0.040 & M($28,1$) &52& 15& 37.8& 0.008 & 110& 16& 46.1& 0.091 & 1.98 & 0.98 & 0.33 & 0.65 & HST\\
106.3 & 11.7 & 110 & 77.2 & 0.298 & 1& 120.3& 0.067 & 16& 22& 15.5& 0.086 & M($38,1$) &72& 14& 44.7& 0.123 & 110& 16& 46.1& 0.061 & 1.46 & 0.62 & 0.17 & 0.65 & HST\\
112.6 & 11.4 & 101 & 76.8 & 0.241 & 1& 123.7& 0.020 & 16& 23& 16.9& 0.025 & M($40,1$) &76& 14& 47.0& 0.137 & 119& 16& 48.6& 0.080 & 1.43 & 0.59 & 0.17 & 0.65 & HST\\
118.7 & 9.47 & 79.7 & 39.6 & 0.503 & 1& 120.3& 0.055 & 35& 22& 33.9& 0.115 & L($8,2$) &34& 13& 42.2& 0.029 & 119& 15& 42.6& 0.388 & 1.19 & 0.75 & 0.20 & 0.22 & dirty\\
124.8 & 6.35 & 50.9 & 39.6 & 0.222 & 1& 120.3& 0.055 & 35& 22& 33.9& 0.115 & L($8,2$) &34& 13& 42.2& 0.029 & 119& 16& 48.6& 0.042 & 1.19 & 0.75 & 0.20 & 0.22 & dirty\\
131.5 & 5.19 & 39.5 & 18.3 & 0.537 & 2& 55.5& 0.198 & 56& 21& 49.4& 0.240 & L($8,2$) &34& 13& 39.5& 0.047 & 119& 15& 42.6& 0.203 & 1.19 & 0.75 & 0.20 & 0.22 & dirty\\
139.0 & 4.12 & 29.7 & 19.8 & 0.332 & 2& 60.2& 0.055 & 56& 22& 54.2& 0.083 & L($8,2$) &34& 13& 42.2& 0.014 & 119& 15& 42.6& 0.218 & 1.19 & 0.75 & 0.20 & 0.22 & dirty\\
146.5 & 3.35 & 22.8 & 12.9 & 0.433 & 2& 55.5& 0.144 & 56& 21& 49.4& 0.177 & L($13,2$) &54& 12& 56.7& 0.093 & 110& 15& 40.4& 0.113 & 0.839 & 0.47 & 0.17 & 0.18 & dirty\\
155.2 & 2.92 & 18.8 & 11.4 & 0.394 & 2& 55.5& 0.128 & 56& 21& 49.4& 0.157 & L($14,2$) &58& 13& 63.2& 0.050 & 119& 15& 42.6& 0.132 & 0.738 & 0.47 & 0.10 & 0.15 & clean\\
165.0 & 2.53 & 15.3 & 12.9 & 0.158 & 2& 60.2& 0.036 & 56& 22& 54.2& 0.055 & L($14,2$) &58& 12& 64.8& 0.068 & 104& 16& 46.0& 0.009 & 0.771 & 0.47 & 0.10 & 0.18 & dirty\\
173.4 & 2.13 & 12.3 & 8.12 & 0.340 & 3& 37.0& 0.136 & 77& 21& 67.9& 0.145 & L($9,3$) &57& 13& 65.1& 0.035 & 110& 15& 40.4& 0.073 & 0.789 & 0.47 & 0.12 & 0.18 & dirty\\
182.2 & 1.88 & 10.3 & 7.34 & 0.288 & 3& 37.0& 0.124 & 77& 21& 67.9& 0.132 & L($10,3$) &63& 12& 68.0& 0.056 & 112& 16& 46.2& 0.007 & 0.713 & 0.43 & 0.09 & 0.17 & dirty\\
191.3 & 1.64 & 8.60 & 6.17 & 0.282 & 3& 37.0& 0.105 & 77& 21& 67.9& 0.113 & L($12,3$) &75& 13& 82.9& 0.035 & 112& 15& 40.5& 0.062 & 0.600 & 0.33 & 0.09 & 0.17 & dirty\\
203.2 & 1.50 & 7.37 & 5.47 & 0.258 & 3& 37.0& 0.094 & 77& 21& 67.9& 0.100 & L($14,3$) &87& 13& 94.8& 0.036 & 112& 15& 40.5& 0.055 & 0.532 & 0.27 & 0.09 & 0.16 & dirty\\
213.4 & 1.32 & 6.20 & 4.62 & 0.256 & 4& 27.8& 0.105 & 98& 21& 86.4& 0.104 & L($12,3$) &76& 14& 86.4& 0.026 & 112& 15& 40.5& 0.047 & 0.598 & 0.33 & 0.09 & 0.17 & dirty\\
225.7 & 1.21 & 5.35 & 4.10 & 0.234 & 4& 27.8& 0.094 & 98& 21& 86.4& 0.093 & L($14,3$) &88& 14& 98.8& 0.027 & 112& 15& 40.5& 0.042 & 0.532 & 0.27 & 0.09 & 0.16 & dirty\\
240.1 & 1.11 & 4.63 & 3.51 & 0.242 & 5& 22.2& 0.100 & 119& 21& 105& 0.094 & L($14,3$) &87& 13& 94.8& 0.039 & 110& 15& 40.4& 0.032 & 0.568 & 0.30 & 0.09 & 0.17 & dirty\\
252.5 & 0.997 & 3.95 & 3.10 & 0.214 & 5& 22.2& 0.089 & 119& 21& 105& 0.084 & L($15,3$) &94& 14& 105& 0.022 & 119& 15& 42.6& 0.038 & 0.501 & 0.23 & 0.09 & 0.16 & dirty\\
268.9 & 0.900 & 3.35 & 2.71 & 0.189 & 5& 22.2& 0.078 & 119& 21& 105& 0.074 & L($18,3$) &112& 14& 123& 0.023 & 112& 15& 40.5& 0.028 & 0.438 & 0.20 & 0.06 & 0.16 & dirty\\
282.5 & 0.855 & 3.03 & 2.49 & 0.178 & 5& 22.2& 0.072 & 119& 21& 105& 0.068 & L($19,3$) &118& 15& 135& 0.021 & 119& 15& 42.6& 0.030 & 0.402 & 0.17 & 0.05 & 0.16 & dirty\\
302.3 & 0.818 & 2.71 & 2.18 & 0.197 & 6& 18.5& 0.068 & 140& 21& 123& 0.075 & L($21,3$) &129& 13& 136& 0.043 & 119& 15& 42.6& 0.027 & 0.377 & 0.15 & 0.05 & 0.16 & dirty\\
319.3 & 0.809 & 2.53 & 2.06 & 0.187 & 5& 22.2& 0.060 & 119& 21& 105& 0.056 & L($27,3$) &165& 13& 172& 0.060 & 119& 15& 42.6& 0.025 & 0.332 & 0.11 & 0.05 & 0.16 & dirty\\
337.8 & 0.796 & 2.36 & 1.91 & 0.190 & 6& 18.5& 0.060 & 140& 21& 123& 0.066 & L($27,3$) &165& 13& 172& 0.056 & 119& 15& 42.6& 0.023 & 0.332 & 0.11 & 0.05 & 0.16 & dirty\\
361.4 & 0.806 & 2.23 & 1.88 & 0.158 & 6& 18.5& 0.059 & 140& 21& 123& 0.065 & L($30,3$) &184& 14& 198& 0.027 & 110& 15& 40.4& 0.017 & 0.326 & 0.11 & 0.05 & 0.15 & clean\\
382.2 & 0.805 & 2.11 & 1.78 & 0.157 & 6& 18.5& 0.056 & 140& 21& 123& 0.061 & L($33,3$) &202& 14& 216& 0.028 & 119& 15& 42.6& 0.022 & 0.310 & 0.09 & 0.05 & 0.16 & dirty\\
406.9 & 0.811 & 1.99 & 1.69 & 0.153 & 6& 18.5& 0.053 & 140& 21& 123& 0.058 & L($37,3$) &226& 14& 241& 0.030 & 119& 15& 42.6& 0.021 & 0.295 & 0.09 & 0.04 & 0.15 & clean\\
442.1 & 0.837 & 1.89 & 1.77 & 0.063 & 6& 20.1& 0.014 & 140& 22& 136& 0.018 & L($38,3$) &231& 14& 264& 0.011 & 119& 15& 42.6& 0.022 & 0.295 & 0.09 & 0.04 & 0.15 & clean\\
468.8 & 0.829 & 1.77 & 1.56 & 0.118 & 4& 27.8& 0.051 & 98& 21& 86.4& 0.024 & L($12,12$) &304& 14& 346& 0.036 & 98& 15& 36.6& 0.013 & 0.281 & 0.07 & 0.04 & 0.16 & Pdirty\\
533.0 & 0.842 & 1.58 & 1.04 & 0.340 & 5& 20.2& 0.182 & 119& 20& 95.2& 0.058 & L($16,12$) &399& 13& 397& 0.135 & 110& 15& 40.4& 0.010 & 0.258 & 0.05 & 0.04 & 0.16 & Pdirty\\
571.7 & 0.731 & 1.28 & 1.11 & 0.129 & 5& 22.2& 0.047 & 119& 21& 105& 0.019 & L($16,12$) &398& 13& 426& 0.059 & 110& 15& 40.4& 0.010 & 0.258 & 0.05 & 0.04 & 0.16 & Pdirty\\
629.8 & 0.755 & 1.20 & 1.04 & 0.135 & 6& 18.5& 0.047 & 140& 21& 123& 0.024 & L($18,12$) &446& 13& 474& 0.065 & 80& 15& 32.5& 0.005 & 0.258 & 0.05 & 0.04 & 0.16 & Pdirty\\
663.3 & 0.743 & 1.12 & 0.942 & 0.159 & 7& 15.9& 0.052 & 182& 21& 161& 0.031 & DL($22,12$) &313& 12& 462& 0.076 & 110& 15& 40.4& 0.009 & 0.288 & 0.07 & 0.04 & 0.17 & Pdirty\\
705.5 & 0.732 & 1.04 & 0.839 & 0.192 & 8& 13.9& 0.052 & 203& 21& 179& 0.031 & DL($24,12$) &339& 12& 498& 0.078 & 80& 14& 28.3& 0.046 & 0.288 & 0.07 & 0.04 & 0.17 & Pdirty\\
751.1 & 0.686 & 0.913 & 0.808 & 0.116 & 8& 13.9& 0.050 & 203& 21& 179& 0.030 & DL($25,12$) &352& 13& 529& 0.031 & 119& 15& 42.6& 0.010 & 0.274 & 0.07 & 0.04 & 0.15 & Pclean\\
789.8 & 0.684 & 0.867 & 0.730 & 0.158 & 9& 12.3& 0.050 & 224& 21& 198& 0.030 & DL($27,12$) &378& 12& 552& 0.082 & 104& 15& 40.4& 0.006 & 0.274 & 0.07 & 0.04 & 0.15 & Pclean\\
843.6 & 0.681 & 0.808 & 0.677 & 0.162 & 9& 12.3& 0.048 & 224& 21& 198& 0.027 & DL($30,12$) &417& 12& 606& 0.090 & 112& 15& 40.5& 0.007 & 0.263 & 0.05 & 0.04 & 0.16 & Pdirty\\
896.0 & 0.621 & 0.693 & 0.620 & 0.105 & 10& 11.1& 0.048 & 245& 21& 216& 0.027 & DL($31,12$) &430& 13& 640& 0.030 & 104& 15& 40.4& 0.005 & 0.263 & 0.05 & 0.04 & 0.16 & Pdirty\\
943.4 & 0.617 & 0.654 & 0.585 & 0.105 & 11& 10.1& 0.048 & 266& 21& 235& 0.028 & DL($33,12$) &456& 13& 676& 0.030 & 80& 15& 32.5& 0.003 & 0.263 & 0.05 & 0.04 & 0.16 & Pdirty\\
1049 & 0.647 & 0.617 & 0.594 & 0.037 & 11& 10.9& 0.012 & 266& 22& 257& 0.008 & DL($34,12$) &469& 13& 751& 0.012 & 110& 15& 40.4& 0.006 & 0.253 & 0.05 & 0.04 & 0.15 & Pclean\\
1261 & 0.730 & 0.579 & 0.556 & 0.040 & 14& 8.6& 0.012 & 329& 22& 318& 0.012 & DL($42,12$) &573& 13& 910& 0.015 & 80& 15& 32.5& 0.003 & 0.253 & 0.05 & 0.04 & 0.15 & Pclean\\
1355 & 0.728 & 0.537 & 0.480 & 0.107 & 11& 10.1& 0.041 & 266& 21& 235& 0.022 & DL($55,12$) &742& 13& 1080& 0.044 & 110& 15& 40.4& 0.004 & 0.225 & 0.01 & 0.04 & 0.16 & Pdirty\\
1429 & 0.698 & 0.489 & 0.434 & 0.112 & 15& 7.4& 0.041 & 350& 21& 309& 0.032 & DL($55,12$) &742& 13& 1080& 0.039 & 110& 15& 40.4& 0.004 & 0.225 & 0.01 & 0.04 & 0.16 & Pdirty\\
1567 & 0.726 & 0.463 & 0.444 & 0.042 & 16& 7.5& 0.011 & 371& 22& 359& 0.011 & DL($55,12$) &742& 13& 1170& 0.017 & 110& 15& 40.4& 0.004 & 0.225 & 0.01 & 0.04 & 0.16 & Pdirty\\
\hline
79.62 & 55.0 & 691 & 366 & 0.470 & 1& 123.7& 0.077 & 16& 23& 16.9& 0.128 & M($14,1$) &24& 14& 16.5& 0.083 & 112& 16& 46.2& 0.283 & 5.55 & 3.20 & 1.69 & 0.65 & HST\\
145.1 & 3.44 & 23.7 & 19.8 & 0.164 & 2& 60.2& 0.055 & 56& 22& 54.2& 0.083 & L($8,2$) &34& 13& 42.2& 0.014 & 119& 16& 48.6& 0.021 & 1.19 & 0.75 & 0.20 & 0.22 & dirty\\
709.8 & 0.706 & 0.994 & 0.839 & 0.156 & 8& 13.9& 0.052 & 203& 21& 179& 0.031 & DL($24,12$) &339& 12& 498& 0.078 & 80& 15& 32.5& 0.004 & 0.288 & 0.07 & 0.04 & 0.17 & Pdirty\\
877.8 & 0.609 & 0.693 & 0.620 & 0.106 & 10& 11.1& 0.048 & 245& 21& 216& 0.027 & DL($30,12$) &417& 13& 621& 0.029 & 112& 15& 40.5& 0.006 & 0.263 & 0.05 & 0.04 & 0.16 & Pdirty\\
\hline\hline
\end{tabular}
\endgroup
\caption{Selected compilation parameters on space-time pareto frontier for performing ground state phase estimation of utility-scale electronic structure to chemical accuracy (Continued).}
\end{table}

\begin{table}
\addtocounter{table}{-1}
\tiny
\begingroup
\setlength{\tabcolsep}{1pt}
\begin{tabular}{c|c|cc|c|ccc|cccc|ccccc|cccc|cccc|cc}
\hline
\multicolumn{26}{c}{XVIII-100o100e}
\\
\hline
Qubits&Volume&\multicolumn{2}{c|}{Runtime/hr} & \multirow{2}{*}{$p_\text{fail}$}& \multicolumn{3}{c|}{Factories}&\multicolumn{4}{c|}{Compute}&\multicolumn{5}{c|}{Hot storage}&\multicolumn{4}{c|}{Cold storage}&\multicolumn{4}{c|}{Toffolis/$10^9$}&Rotation&\\
$\mathfrak{n}$/$10^3$&$/10^6\mathfrak{n}\cdot\text{hr}$&$\langle.\rangle$& $1$ shot & &$F$&$d_{\textsc{CCZ}}$&$p_\text{fail}$& $k$&$d$&$\mathfrak{n}$/$10^3$&$p_\text{fail}$ & Latency&$k$&$d_\text{in}$&$\mathfrak{n}$/$10^3$&$p_\text{fail}$ & $k$&$d_\text{in}$&$\mathfrak{n}$/$10^3$&$p_\text{fail}$&Total&\textsc{RPrep}&\textsc{IPrep}&\textsc{Rot}& synthesis&\\
\hline
121.5 & 6160 & 50700 & 5390 & 0.894 & 1& 127.3& 0.287 & 16& 24& 18.4& 0.446 & M($14,1$) &24& 15& 18.9& 0.086 & 204& 17& 84.1& 0.706 & 75.3 & 50.54 & 21.77 & 2.93 & HST\\
127.8 & 953 & 7460 & 2590 & 0.652 & 1& 127.3& 0.153 & 16& 24& 18.4& 0.245 & M($20,1$) &36& 15& 27.0& 0.084 & 196& 17& 82.3& 0.407 & 37.0 & 27.04 & 6.97 & 2.93 & HST\\
134.7 & 427 & 3170 & 1820 & 0.424 & 1& 131.0& 0.039 & 16& 25& 20.0& 0.054 & M($24,1$) &44& 15& 32.4& 0.087 & 196& 17& 82.3& 0.307 & 25.9 & 17.82 & 5.07 & 2.93 & HST\\
141.6 & 305 & 2160 & 1210 & 0.439 & 1& 127.3& 0.078 & 16& 24& 18.4& 0.120 & M($30,1$) &56& 15& 40.5& 0.088 & 200& 17& 82.7& 0.243 & 18.0 & 12.35 & 2.70 & 2.93 & HST\\
150.9 & 235 & 1560 & 1060 & 0.320 & 1& 131.0& 0.024 & 16& 25& 20.0& 0.031 & M($36,1$) &68& 15& 48.6& 0.112 & 196& 17& 82.3& 0.192 & 15.6 & 9.95 & 2.70 & 2.93 & HST\\
159.3 & 205 & 1290 & 820 & 0.362 & 1& 127.3& 0.055 & 16& 24& 18.4& 0.081 & M($42,1$) &80& 15& 56.7& 0.115 & 204& 17& 84.1& 0.170 & 12.7 & 8.39 & 1.36 & 2.93 & HST\\
168.6 & 185 & 1100 & 909 & 0.174 & 1& 131.0& 0.021 & 16& 25& 20.0& 0.026 & M($40,1$) &76& 15& 54.0& 0.118 & 204& 18& 94.6& 0.018 & 13.7 & 9.29 & 1.38 & 2.93 & HST\\
177.3 & 150 & 847 & 553 & 0.347 & 1& 127.3& 0.068 & 35& 24& 40.3& 0.129 & L($9,2$) &38& 13& 52.8& 0.089 & 204& 17& 84.1& 0.118 & 15.6 & 11.76 & 2.23 & 1.59 & dirty\\
186.9 & 117 & 625 & 434 & 0.306 & 1& 127.3& 0.054 & 35& 24& 40.3& 0.102 & L($11,2$) &46& 13& 62.4& 0.098 & 204& 17& 84.1& 0.094 & 12.3 & 9.64 & 1.42 & 1.16 & dirty\\
197.5 & 91.2 & 462 & 243 & 0.474 & 2& 61.9& 0.184 & 56& 23& 59.2& 0.260 & L($10,2$) &42& 13& 54.1& 0.080 & 204& 17& 84.1& 0.054 & 14.1 & 10.56 & 2.05 & 1.48 & dirty\\
208.5 & 65.1 & 312 & 250 & 0.199 & 2& 63.7& 0.061 & 56& 24& 64.5& 0.084 & L($10,2$) &42& 14& 59.9& 0.014 & 204& 17& 84.1& 0.055 & 14.1 & 10.56 & 2.05 & 1.48 & dirty\\
219.2 & 53.5 & 244 & 195 & 0.202 & 2& 63.7& 0.048 & 56& 24& 64.5& 0.066 & L($13,2$) &54& 13& 72.0& 0.061 & 200& 17& 82.7& 0.044 & 11.0 & 8.32 & 1.36 & 1.29 & dirty\\
230.5 & 46.1 & 200 & 115 & 0.424 & 2& 61.9& 0.092 & 56& 23& 59.2& 0.133 & L($19,2$) &78& 14& 98.5& 0.043 & 196& 16& 72.7& 0.235 & 6.71 & 5.00 & 0.70 & 0.96 & dirty\\
242.8 & 34.3 & 141 & 120 & 0.153 & 2& 63.7& 0.030 & 56& 24& 64.5& 0.041 & L($18,2$) &74& 13& 96.0& 0.067 & 196& 17& 82.3& 0.024 & 6.76 & 5.00 & 0.75 & 0.96 & dirty\\
256.2 & 30.1 & 117 & 76.9 & 0.345 & 3& 41.2& 0.092 & 77& 23& 81.5& 0.116 & L($13,3$) &81& 14& 106& 0.030 & 187& 16& 69.2& 0.159 & 6.71 & 5.00 & 0.70 & 0.96 & dirty\\
271.1 & 23.0 & 84.7 & 57.8 & 0.318 & 4& 30.9& 0.092 & 98& 23& 104& 0.107 & L($12,3$) &75& 13& 94.7& 0.037 & 196& 16& 72.7& 0.126 & 6.73 & 5.00 & 0.71 & 0.96 & dirty\\
287.6 & 21.1 & 73.5 & 57.6 & 0.216 & 4& 30.9& 0.092 & 98& 23& 104& 0.107 & L($13,3$) &81& 14& 106& 0.022 & 187& 17& 78.3& 0.011 & 6.71 & 5.00 & 0.70 & 0.96 & dirty\\
304.6 & 18.2 & 59.9 & 45.0 & 0.249 & 4& 30.9& 0.072 & 98& 23& 104& 0.084 & L($17,3$) &105& 14& 134& 0.023 & 180& 16& 67.2& 0.095 & 5.23 & 3.59 & 0.70 & 0.90 & dirty\\
324.2 & 16.8 & 51.9 & 47.6 & 0.084 & 5& 25.5& 0.030 & 119& 24& 137& 0.031 & L($12,3$) &75& 14& 105& 0.016 & 196& 17& 82.3& 0.010 & 6.73 & 5.00 & 0.71 & 0.96 & dirty\\
340.9 & 14.7 & 43.2 & 36.0 & 0.168 & 5& 24.7& 0.072 & 119& 23& 126& 0.080 & L($17,3$) &105& 15& 139& 0.018 & 180& 17& 76.1& 0.007 & 5.23 & 3.59 & 0.70 & 0.90 & dirty\\
358.7 & 13.9 & 38.7 & 30.6 & 0.208 & 5& 24.7& 0.062 & 119& 23& 126& 0.069 & L($22,3$) &135& 14& 169& 0.043 & 168& 16& 63.9& 0.053 & 4.46 & 3.14 & 0.39 & 0.87 & dirty\\
377.9 & 12.8 & 34.0 & 28.5 & 0.161 & 5& 24.7& 0.058 & 119& 23& 126& 0.064 & L($23,3$) &141& 14& 176& 0.043 & 180& 17& 76.1& 0.006 & 4.15 & 2.93 & 0.39 & 0.78 & clean\\
397.3 & 12.3 & 30.9 & 24.8 & 0.198 & 5& 24.7& 0.051 & 119& 23& 126& 0.056 & L($26,3$) &159& 14& 197& 0.045 & 204& 16& 74.3& 0.063 & 3.61 & 2.37 & 0.36 & 0.83 & dirty\\
418.4 & 11.1 & 26.5 & 21.7 & 0.182 & 5& 24.7& 0.044 & 119& 23& 126& 0.049 & L($29,3$) &177& 14& 218& 0.047 & 204& 16& 74.3& 0.055 & 3.16 & 1.92 & 0.36 & 0.82 & dirty\\
446.5 & 10.9 & 24.3 & 20.0 & 0.178 & 5& 24.7& 0.041 & 119& 23& 126& 0.045 & L($33,3$) &201& 14& 246& 0.054 & 204& 16& 74.3& 0.051 & 2.91 & 1.68 & 0.36 & 0.82 & dirty\\
471.5 & 10.5 & 22.3 & 19.1 & 0.143 & 6& 20.6& 0.042 & 140& 23& 148& 0.055 & L($32,3$) &195& 14& 239& 0.049 & 204& 17& 84.1& 0.004 & 2.99 & 1.76 & 0.36 & 0.82 & dirty\\
495.6 & 10.3 & 20.9 & 18.5 & 0.113 & 6& 20.6& 0.041 & 140& 23& 148& 0.053 & L($34,3$) &207& 15& 263& 0.019 & 204& 17& 84.1& 0.004 & 2.90 & 1.67 & 0.36 & 0.81 & dirty\\
527.8 & 10.3 & 19.6 & 17.1 & 0.129 & 5& 24.7& 0.035 & 119& 23& 126& 0.039 & L($43,3$) &261& 15& 329& 0.023 & 196& 16& 72.7& 0.039 & 2.48 & 1.35 & 0.28 & 0.81 & dirty\\
563.0 & 9.72 & 17.3 & 14.9 & 0.139 & 6& 20.6& 0.033 & 140& 23& 148& 0.043 & L($45,3$) &273& 14& 331& 0.067 & 204& 17& 84.1& 0.003 & 2.34 & 1.26 & 0.22 & 0.81 & dirty\\
605.0 & 9.51 & 15.7 & 14.4 & 0.087 & 5& 24.7& 0.030 & 119& 23& 126& 0.033 & L($52,3$) &315& 15& 395& 0.024 & 204& 17& 84.1& 0.003 & 2.09 & 1.01 & 0.22 & 0.80 & dirty\\
638.3 & 9.12 & 14.3 & 10.8 & 0.246 & 5& 24.1& 0.091 & 119& 22& 115& 0.067 & L($15,12$) &375& 14& 449& 0.085 & 204& 16& 74.3& 0.028 & 2.02 & 0.93 & 0.22 & 0.82 & Pdirty\\
677.0 & 8.66 & 12.8 & 11.7 & 0.087 & 5& 24.7& 0.030 & 119& 23& 126& 0.021 & L($15,12$) &374& 14& 479& 0.037 & 170& 17& 72.5& 0.002 & 2.11 & 1.01 & 0.22 & 0.82 & Pdirty\\
715.0 & 8.19 & 11.5 & 10.5 & 0.082 & 5& 24.7& 0.027 & 119& 23& 126& 0.019 & L($16,12$) &398& 14& 507& 0.037 & 196& 17& 82.3& 0.002 & 1.94 & 0.85 & 0.22 & 0.82 & Pdirty\\
762.6 & 8.26 & 10.8 & 8.35 & 0.229 & 6& 20.1& 0.080 & 140& 22& 136& 0.065 & L($19,12$) &471& 14& 554& 0.086 & 196& 16& 72.7& 0.019 & 1.76 & 0.68 & 0.22 & 0.81 & Pdirty\\
812.0 & 7.83 & 9.64 & 8.68 & 0.100 & 6& 20.6& 0.025 & 140& 23& 148& 0.020 & L($19,12$) &470& 14& 591& 0.039 & 196& 16& 72.7& 0.020 & 1.76 & 0.68 & 0.22 & 0.81 & Pdirty\\
896.2 & 8.16 & 9.10 & 7.10 & 0.220 & 6& 20.1& 0.072 & 140& 22& 136& 0.053 & L($24,12$) &591& 14& 686& 0.096 & 204& 16& 74.3& 0.018 & 1.56 & 0.52 & 0.19 & 0.80 & Pdirty\\
942.9 & 7.99 & 8.47 & 6.28 & 0.258 & 9& 13.4& 0.093 & 224& 22& 217& 0.081 & DL($29,12$) &404& 13& 652& 0.095 & 204& 16& 74.3& 0.016 & 2.06 & 0.93 & 0.22 & 0.85 & Pdirty\\
998.2 & 6.94 & 6.95 & 5.44 & 0.217 & 10& 12.0& 0.089 & 245& 22& 237& 0.075 & DL($30,12$) &419& 14& 688& 0.057 & 200& 16& 73.0& 0.014 & 1.97 & 0.85 & 0.22 & 0.85 & Pdirty\\
1056 & 6.63 & 6.28 & 5.59 & 0.111 & 10& 12.4& 0.028 & 245& 23& 259& 0.022 & DL($30,12$) &417& 13& 724& 0.051 & 200& 16& 73.0& 0.015 & 1.97 & 0.85 & 0.22 & 0.85 & Pdirty\\
1122 & 6.54 & 5.83 & 4.49 & 0.230 & 11& 10.9& 0.082 & 266& 22& 257& 0.067 & DL($36,12$) &495& 13& 791& 0.091 & 204& 16& 74.3& 0.012 & 1.79 & 0.68 & 0.22 & 0.84 & Pdirty\\
1205 & 6.63 & 5.50 & 5.07 & 0.078 & 11& 11.2& 0.027 & 266& 23& 281& 0.023 & DL($35,12$) &482& 14& 851& 0.018 & 200& 16& 73.0& 0.013 & 1.90 & 0.85 & 0.22 & 0.78 & Pclean\\
1271 & 6.27 & 4.93 & 4.57 & 0.074 & 12& 10.3& 0.025 & 287& 23& 304& 0.023 & DL($37,12$) &508& 14& 894& 0.017 & 196& 16& 72.7& 0.011 & 1.79 & 0.68 & 0.22 & 0.84 & Pdirty\\
1364 & 6.39 & 4.68 & 3.79 & 0.191 & 12& 10.0& 0.073 & 287& 22& 278& 0.063 & DL($46,12$) &627& 14& 1010& 0.060 & 200& 16& 73.0& 0.010 & 1.59 & 0.52 & 0.19 & 0.82 & Pdirty\\
1447 & 6.22 & 4.29 & 4.01 & 0.067 & 11& 11.2& 0.023 & 266& 23& 281& 0.017 & DL($46,12$) &625& 14& 1090& 0.019 & 204& 16& 74.3& 0.010 & 1.59 & 0.52 & 0.19 & 0.82 & Pdirty\\
1536 & 6.20 & 4.04 & 3.74 & 0.073 & 15& 8.2& 0.023 & 350& 23& 370& 0.025 & DL($46,12$) &625& 14& 1090& 0.018 & 204& 16& 74.3& 0.010 & 1.59 & 0.52 & 0.19 & 0.82 & Pdirty\\
1872 & 7.12 & 3.80 & 3.04 & 0.201 & 16& 7.5& 0.065 & 371& 22& 359& 0.076 & DL($67,12$) &900& 14& 1440& 0.070 & 196& 16& 72.7& 0.007 & 1.41 & 0.35 & 0.19 & 0.81 & Pdirty\\
1968 & 6.97 & 3.54 & 3.30 & 0.068 & 13& 9.5& 0.020 & 308& 23& 326& 0.017 & DL($68,12$) &911& 14& 1570& 0.025 & 187& 16& 69.2& 0.007 & 1.41 & 0.35 & 0.19 & 0.81 & Pdirty\\
2088 & 7.02 & 3.36 & 3.14 & 0.065 & 16& 7.7& 0.020 & 371& 23& 393& 0.023 & DL($69,12$) &926& 15& 1630& 0.018 & 170& 16& 64.0& 0.006 & 1.41 & 0.35 & 0.19 & 0.81 & Pdirty\\
3076 & 9.51 & 3.09 & 2.52 & 0.186 & 11& 10.9& 0.052 & 266& 22& 257& 0.032 & DL($66,24$) &1705& 14& 2750& 0.107 & 182& 16& 68.3& 0.005 & 1.13 & 0.09 & 0.19 & 0.80 & Pdirty\\
3261 & 8.90 & 2.73 & 2.21 & 0.192 & 16& 7.5& 0.052 & 371& 22& 359& 0.051 & DL($68,24$) &1755& 14& 2830& 0.097 & 196& 16& 72.7& 0.005 & 1.12 & 0.09 & 0.19 & 0.78 & Pclean\\
3456 & 8.44 & 2.44 & 2.30 & 0.060 & 16& 7.7& 0.016 & 371& 23& 393& 0.015 & DL($65,24$) &1682& 15& 2990& 0.024 & 204& 16& 74.3& 0.006 & 1.13 & 0.09 & 0.19 & 0.80 & Pdirty\\
\hline
182.9 & 127 & 697 & 500 & 0.283 & 1& 127.3& 0.061 & 35& 24& 40.3& 0.117 & L($10,2$) &42& 14& 59.9& 0.029 & 200& 17& 82.7& 0.109 & 14.1 & 10.56 & 2.05 & 1.48 & dirty\\
449.3 & 10.7 & 23.9 & 20.5 & 0.139 & 5& 24.7& 0.042 & 119& 23& 126& 0.047 & L($32,3$) &195& 14& 239& 0.052 & 204& 17& 84.1& 0.005 & 2.99 & 1.76 & 0.36 & 0.82 & dirty\\
1228 & 6.16 & 5.02 & 4.65 & 0.072 & 11& 11.2& 0.025 & 266& 23& 281& 0.020 & DL($36,12$) &495& 14& 872& 0.017 & 204& 16& 74.3& 0.012 & 1.79 & 0.68 & 0.22 & 0.84 & Pdirty\\
\hline\hline
\end{tabular}
\endgroup
\caption{Selected compilation parameters on space-time pareto frontier for performing ground state phase estimation of utility-scale electronic structure to chemical accuracy (Continued).}
\end{table}

\begin{table}
\addtocounter{table}{-1}
\tiny
\begingroup
\setlength{\tabcolsep}{1pt}
\begin{tabular}{c|c|cc|c|ccc|cccc|ccccc|cccc|cccc|cc}
\hline
\multicolumn{26}{c}{XVIII-150o150e}
\\
\hline
Qubits&Volume&\multicolumn{2}{c|}{Runtime/hr} & \multirow{2}{*}{$p_\text{fail}$}& \multicolumn{3}{c|}{Factories}&\multicolumn{4}{c|}{Compute}&\multicolumn{5}{c|}{Hot storage}&\multicolumn{4}{c|}{Cold storage}&\multicolumn{4}{c|}{Toffolis/$10^9$}&Rotation&\\
$\mathfrak{n}$/$10^3$&$/10^6\mathfrak{n}\cdot\text{hr}$&$\langle.\rangle$& $1$ shot & &$F$&$d_{\textsc{CCZ}}$&$p_\text{fail}$& $k$&$d$&$\mathfrak{n}$/$10^3$&$p_\text{fail}$ & Latency&$k$&$d_\text{in}$&$\mathfrak{n}$/$10^3$&$p_\text{fail}$ & $k$&$d_\text{in}$&$\mathfrak{n}$/$10^3$&$p_\text{fail}$&Total&\textsc{RPrep}&\textsc{IPrep}&\textsc{Rot}& synthesis&\\
\hline
171.8 & 30700 & 178000 & 19500 & 0.891 & 1& 131.0& 0.333 & 16& 25& 20.0& 0.454 & M($16,1$) &28& 15& 21.6& 0.349 & 300& 18& 130& 0.540 & 264 & 194.20 & 62.18 & 7.62 & HST\\
180.8 & 4390 & 24300 & 9570 & 0.606 & 1& 143.2& 0.046 & 16& 26& 21.6& 0.082 & M($22,1$) &40& 15& 29.7& 0.334 & 299& 18& 130& 0.325 & 124 & 87.04 & 29.72 & 7.62 & HST\\
190.4 & 2340 & 12300 & 8290 & 0.325 & 1& 146.5& 0.012 & 16& 27& 23.3& 0.021 & M($24,1$) &44& 16& 36.9& 0.029 & 300& 18& 130& 0.282 & 105 & 75.06 & 21.92 & 7.62 & HST\\
200.3 & 1440 & 7180 & 5220 & 0.273 & 1& 143.2& 0.025 & 16& 26& 21.6& 0.045 & M($32,1$) &60& 16& 49.2& 0.032 & 299& 18& 130& 0.193 & 68.6 & 49.18 & 11.69 & 7.62 & HST\\
210.9 & 1190 & 5650 & 3820 & 0.324 & 1& 131.0& 0.079 & 16& 25& 20.0& 0.110 & M($40,1$) &76& 16& 61.4& 0.036 & 299& 18& 130& 0.145 & 53.9 & 39.35 & 6.79 & 7.62 & HST\\
221.6 & 1030 & 4640 & 3770 & 0.187 & 1& 146.5& 0.006 & 16& 27& 23.3& 0.009 & M($42,1$) &80& 16& 64.5& 0.040 & 306& 18& 134& 0.140 & 49.7 & 36.14 & 5.88 & 7.62 & HST\\
233.3 & 1010 & 4350 & 4040 & 0.071 & 1& 146.5& 0.006 & 16& 27& 23.3& 0.010 & M($42,1$) &80& 16& 64.5& 0.043 & 300& 19& 145& 0.014 & 53.0 & 39.35 & 5.88 & 7.62 & HST\\
245.2 & 568 & 2320 & 868 & 0.625 & 2& 63.7& 0.198 & 56& 24& 64.5& 0.262 & L($11,2$) &46& 14& 64.9& 0.054 & 300& 17& 116& 0.330 & 49.1 & 37.81 & 7.51 & 3.66 & dirty\\
257.9 & 413 & 1600 & 884 & 0.448 & 2& 65.5& 0.072 & 56& 25& 70.0& 0.084 & L($11,2$) &46& 14& 68.9& 0.022 & 306& 17& 119& 0.336 & 48.6 & 37.81 & 6.98 & 3.66 & dirty\\
271.0 & 300 & 1110 & 893 & 0.192 & 2& 65.5& 0.072 & 56& 25& 70.0& 0.085 & L($11,2$) &46& 15& 71.5& 0.012 & 299& 18& 130& 0.036 & 49.1 & 37.81 & 7.51 & 3.66 & dirty\\
285.2 & 228 & 798 & 461 & 0.422 & 2& 63.7& 0.110 & 56& 24& 64.5& 0.149 & L($19,2$) &78& 14& 105& 0.055 & 300& 17& 116& 0.192 & 26.1 & 20.15 & 3.35 & 2.48 & dirty\\
299.5 & 197 & 657 & 461 & 0.298 & 2& 63.7& 0.110 & 56& 24& 64.5& 0.149 & L($19,2$) &78& 14& 105& 0.055 & 300& 18& 130& 0.018 & 26.1 & 20.15 & 3.35 & 2.48 & dirty\\
320.0 & 145 & 453 & 298 & 0.343 & 3& 42.4& 0.107 & 77& 24& 88.7& 0.126 & L($13,3$) &81& 14& 112& 0.034 & 306& 17& 119& 0.129 & 25.3 & 20.15 & 2.98 & 2.01 & clean\\
337.1 & 128 & 378 & 338 & 0.108 & 3& 43.7& 0.042 & 77& 25& 96.3& 0.042 & L($12,3$) &75& 14& 111& 0.014 & 299& 18& 130& 0.014 & 27.8 & 20.15 & 5.09 & 2.48 & dirty\\
355.0 & 107 & 301 & 234 & 0.224 & 4& 32.8& 0.039 & 98& 25& 123& 0.036 & L($13,3$) &81& 14& 119& 0.076 & 289& 17& 113& 0.093 & 25.7 & 20.15 & 2.98 & 2.48 & dirty\\
375.8 & 89.1 & 237 & 178 & 0.248 & 4& 31.8& 0.086 & 98& 24& 113& 0.094 & L($18,3$) &111& 15& 156& 0.028 & 272& 17& 107& 0.065 & 20.2 & 14.78 & 2.96 & 2.32 & dirty\\
400.0 & 74.1 & 185 & 143 & 0.229 & 5& 25.5& 0.086 & 119& 24& 137& 0.089 & L($18,3$) &111& 15& 156& 0.023 & 272& 17& 107& 0.052 & 20.2 & 14.78 & 2.96 & 2.32 & dirty\\
420.7 & 69.6 & 165 & 145 & 0.126 & 5& 26.2& 0.030 & 119& 25& 149& 0.026 & L($17,3$) &105& 15& 157& 0.009 & 299& 17& 115& 0.066 & 19.9 & 14.78 & 2.96 & 2.01 & clean\\
446.6 & 58.8 & 132 & 108 & 0.182 & 5& 25.5& 0.066 & 119& 24& 137& 0.068 & L($24,3$) &147& 15& 202& 0.024 & 266& 17& 107& 0.037 & 15.2 & 11.31 & 1.55 & 2.25 & dirty\\
470.6 & 57.3 & 122 & 111 & 0.090 & 5& 26.2& 0.023 & 119& 25& 149& 0.020 & L($24,3$) &147& 15& 215& 0.012 & 266& 17& 107& 0.038 & 15.2 & 11.31 & 1.55 & 2.25 & dirty\\
496.4 & 53.0 & 107 & 92.3 & 0.137 & 5& 25.5& 0.057 & 119& 24& 137& 0.059 & L($27,3$) &165& 15& 226& 0.024 & 306& 18& 134& 0.004 & 13.0 & 9.30 & 1.51 & 2.13 & dirty\\
532.9 & 48.6 & 91.3 & 77.5 & 0.151 & 5& 25.5& 0.048 & 119& 24& 137& 0.050 & L($34,3$) &207& 15& 280& 0.027 & 300& 17& 116& 0.035 & 11.0 & 7.24 & 1.51 & 2.11 & dirty\\
570.6 & 47.2 & 82.8 & 71.7 & 0.134 & 5& 25.5& 0.044 & 119& 24& 137& 0.046 & L($37,3$) &226& 16& 314& 0.018 & 306& 17& 119& 0.033 & 10.1 & 6.42 & 1.51 & 2.09 & dirty\\
606.0 & 45.4 & 74.9 & 64.8 & 0.135 & 5& 25.5& 0.040 & 119& 24& 137& 0.042 & L($43,3$) &261& 15& 350& 0.031 & 306& 17& 119& 0.030 & 9.16 & 5.46 & 1.51 & 2.08 & dirty\\
636.3 & 40.5 & 63.7 & 57.3 & 0.101 & 5& 25.5& 0.036 & 119& 24& 137& 0.037 & L($45,3$) &273& 15& 365& 0.030 & 306& 18& 134& 0.002 & 8.10 & 4.98 & 0.93 & 2.08 & dirty\\
668.2 & 40.3 & 60.3 & 53.1 & 0.119 & 5& 25.5& 0.033 & 119& 24& 137& 0.034 & L($51,3$) &309& 15& 412& 0.033 & 306& 17& 119& 0.024 & 7.51 & 4.50 & 0.83 & 2.07 & dirty\\
714.9 & 39.0 & 54.6 & 49.0 & 0.102 & 6& 21.2& 0.033 & 140& 24& 161& 0.040 & L($52,3$) &315& 15& 420& 0.031 & 306& 18& 134& 0.002 & 7.37 & 4.36 & 0.83 & 2.07 & dirty\\
756.1 & 38.5 & 50.9 & 45.6 & 0.103 & 6& 21.2& 0.030 & 140& 24& 161& 0.037 & L($57,3$) &346& 16& 476& 0.018 & 306& 17& 119& 0.021 & 6.89 & 3.88 & 0.83 & 2.06 & dirty\\
804.5 & 36.6 & 45.5 & 41.6 & 0.087 & 5& 25.5& 0.031 & 119& 24& 137& 0.023 & L($15,12$) &376& 16& 548& 0.018 & 306& 17& 119& 0.019 & 6.93 & 3.88 & 0.83 & 2.11 & Pdirty\\
847.1 & 35.4 & 41.7 & 38.1 & 0.086 & 5& 25.5& 0.028 & 119& 24& 137& 0.020 & L($17,12$) &423& 15& 591& 0.023 & 306& 17& 119& 0.018 & 6.44 & 3.40 & 0.83 & 2.10 & Pdirty\\
902.4 & 35.7 & 39.6 & 36.0 & 0.090 & 6& 21.2& 0.028 & 140& 24& 161& 0.026 & L($18,12$) &447& 15& 622& 0.023 & 306& 17& 119& 0.017 & 6.36 & 3.40 & 0.83 & 2.01 & Pclean\\
953.0 & 34.2 & 35.9 & 32.9 & 0.083 & 6& 21.2& 0.026 & 140& 24& 161& 0.023 & L($20,12$) &495& 15& 684& 0.024 & 272& 17& 107& 0.012 & 5.95 & 2.93 & 0.83 & 2.09 & Pdirty\\
1003 & 33.3 & 33.2 & 30.7 & 0.076 & 5& 25.5& 0.024 & 119& 24& 137& 0.016 & L($22,12$) &543& 15& 746& 0.025 & 306& 17& 119& 0.014 & 5.37 & 2.45 & 0.74 & 2.08 & Pdirty\\
1055 & 31.8 & 30.1 & 27.8 & 0.076 & 6& 21.2& 0.023 & 140& 24& 161& 0.019 & L($23,12$) &567& 15& 778& 0.024 & 300& 17& 116& 0.013 & 5.21 & 2.45 & 0.58 & 2.08 & Pdirty\\
1109 & 31.3 & 28.2 & 21.6 & 0.236 & 10& 12.4& 0.098 & 245& 23& 259& 0.089 & DL($30,12$) &417& 14& 741& 0.062 & 280& 17& 109& 0.009 & 7.18 & 4.03 & 0.83 & 2.21 & Pdirty\\
1181 & 30.0 & 25.4 & 19.9 & 0.217 & 10& 12.4& 0.091 & 245& 23& 259& 0.081 & DL($32,12$) &445& 15& 803& 0.053 & 306& 17& 119& 0.009 & 6.67 & 3.55 & 0.83 & 2.19 & Pdirty\\
1244 & 27.7 & 22.3 & 20.4 & 0.084 & 10& 12.7& 0.029 & 245& 24& 282& 0.024 & DL($32,12$) &443& 14& 843& 0.024 & 306& 17& 119& 0.009 & 6.67 & 3.55 & 0.83 & 2.19 & Pdirty\\
1308 & 26.5 & 20.2 & 18.7 & 0.074 & 11& 11.6& 0.029 & 266& 24& 306& 0.025 & DL($33,12$) &456& 15& 886& 0.014 & 300& 17& 116& 0.009 & 6.53 & 3.40 & 0.83 & 2.19 & Pdirty\\
1384 & 25.4 & 18.4 & 17.1 & 0.069 & 11& 11.6& 0.027 & 266& 24& 306& 0.022 & DL($36,12$) &495& 15& 958& 0.014 & 306& 17& 119& 0.008 & 6.03 & 2.93 & 0.83 & 2.16 & Pdirty\\
1466 & 25.5 & 17.4 & 14.1 & 0.191 & 11& 11.2& 0.073 & 266& 23& 281& 0.061 & DL($45,12$) &612& 14& 1070& 0.064 & 299& 17& 115& 0.007 & 5.26 & 2.45 & 0.57 & 2.14 & Pdirty\\
1547 & 24.5 & 15.9 & 14.7 & 0.074 & 11& 11.6& 0.023 & 266& 24& 306& 0.019 & DL($44,12$) &599& 14& 1130& 0.027 & 299& 17& 115& 0.007 & 5.29 & 2.45 & 0.60 & 2.14 & Pdirty\\
1651 & 24.8 & 15.0 & 12.5 & 0.168 & 11& 11.2& 0.066 & 266& 23& 281& 0.054 & DL($52,12$) &705& 15& 1250& 0.053 & 306& 17& 119& 0.006 & 4.76 & 1.97 & 0.57 & 2.11 & Pdirty\\
1739 & 24.2 & 13.9 & 12.9 & 0.072 & 11& 11.6& 0.021 & 266& 24& 306& 0.016 & DL($52,12$) &703& 14& 1310& 0.030 & 306& 17& 119& 0.006 & 4.76 & 1.97 & 0.57 & 2.11 & Pdirty\\
1841 & 24.3 & 13.2 & 12.4 & 0.062 & 14& 9.1& 0.021 & 329& 24& 379& 0.023 & DL($52,12$) &703& 15& 1340& 0.014 & 306& 17& 119& 0.006 & 4.76 & 1.97 & 0.57 & 2.11 & Pdirty\\
2050 & 25.7 & 12.5 & 10.4 & 0.173 & 14& 8.8& 0.059 & 329& 23& 348& 0.063 & DL($67,12$) &900& 15& 1590& 0.056 & 299& 17& 115& 0.005 & 4.25 & 1.49 & 0.57 & 2.09 & Pdirty\\
2160 & 25.2 & 11.6 & 10.7 & 0.078 & 14& 9.1& 0.019 & 329& 24& 379& 0.019 & DL($67,12$) &898& 14& 1670& 0.037 & 299& 17& 115& 0.005 & 4.25 & 1.49 & 0.57 & 2.09 & Pdirty\\
2366 & 26.1 & 11.0 & 10.4 & 0.061 & 16& 8.0& 0.019 & 371& 24& 427& 0.023 & DL($72,12$) &963& 15& 1820& 0.017 & 299& 17& 115& 0.005 & 4.18 & 1.49 & 0.57 & 2.01 & Pclean\\
2790 & 29.2 & 10.5 & 8.50 & 0.187 & 16& 7.7& 0.052 & 371& 23& 393& 0.061 & DL($50,24$) &1305& 14& 2280& 0.083 & 299& 17& 115& 0.004 & 3.73 & 1.01 & 0.55 & 2.06 & Pdirty\\
2943 & 28.7 & 9.76 & 9.04 & 0.073 & 14& 9.1& 0.017 & 329& 24& 379& 0.015 & DL($50,24$) &1303& 14& 2450& 0.038 & 299& 17& 115& 0.004 & 3.73 & 1.01 & 0.55 & 2.06 & Pdirty\\
3184 & 29.4 & 9.23 & 8.71 & 0.057 & 16& 8.0& 0.016 & 371& 24& 427& 0.018 & DL($53,24$) &1380& 15& 2650& 0.020 & 280& 17& 109& 0.004 & 3.69 & 1.01 & 0.55 & 2.01 & Pclean\\
\hline
293.5 & 210 & 714 & 607 & 0.150 & 2& 65.5& 0.050 & 56& 25& 70.0& 0.059 & L($17,2$) &70& 14& 101& 0.029 & 280& 18& 123& 0.021 & 33.4 & 25.13 & 5.13 & 3.00 & dirty\\
1193 & 28.3 & 23.7 & 21.7 & 0.087 & 10& 12.7& 0.031 & 245& 24& 282& 0.026 & DL($30,12$) &417& 14& 796& 0.023 & 299& 17& 115& 0.010 & 7.03 & 3.88 & 0.83 & 2.21 & Pdirty\\
1763 & 24.2 & 13.7 & 12.7 & 0.073 & 12& 10.6& 0.021 & 287& 24& 331& 0.018 & DL($52,12$) &703& 14& 1310& 0.030 & 306& 17& 119& 0.006 & 4.76 & 1.97 & 0.57 & 2.11 & Pdirty\\

\hline\hline
\end{tabular}
\endgroup
\caption{Selected compilation parameters on space-time pareto frontier for performing ground state phase estimation of utility-scale electronic structure to chemical accuracy (Continued).}
\end{table}